\documentclass[12pt,a4paper,twoside,titlepage]{article}
\usepackage[latin2]{inputenc}
\usepackage[OT4]{fontenc}

\usepackage[titletoc, title, header]{appendix}

\usepackage[sort&compress,numbers]{natbib}

\usepackage{url}
\usepackage{latexsym}
\usepackage{amsmath}
\usepackage{amssymb}
\usepackage{amsfonts}
\usepackage{wasysym}
\usepackage{fancyhdr}

\usepackage[figuresright]{rotating}
\usepackage{graphicx}
\usepackage[german,polish,english]{babel}
\usepackage{subfigure}
\usepackage{multirow}
\makeatletter
  \newcommand\tinyv{\@setfontsize\tinyv{4pt}{6}}
\makeatother
\graphicspath{{images/}}
\newcommand{\emptydoublepage}{\clearpage\thispagestyle{empty}\cleardoublepage}
\newlength{\fullwidth} 
\usepackage{chngpage}
\newcommand{\myFrameBigFigure}[4][image]
{
\begin{figure}[p]
	\framebox{%
		\includegraphics[width=\textwidth]{#2}%
		}
		\caption[#4]{#3}
		\label{#1_#2}
\end{figure}
}

\newcommand{\myFrameHugeFigure}[4][image]
{
\begin{sidewaysfigure}[t!bp]
\framebox{\includegraphics[width=20cm, height=12cm]{#2}}		
\caption[#4]{#3}
		\label{#1_#2}
\end{sidewaysfigure}
}

\newcommand{\myFigure}[4][image]%
{%
\begin{figure}[ht!bp]%
	\begin{center}%
		\includegraphics[width= \textwidth]{#2}%
		\caption[#4]{#3}
		\label{#1_#2}%
	\end{center}%
\end{figure}%
}%

\newcommand{\myFrameFigure}[4][image]%
{
\begin{figure}[ht!bp]
	\begin{center}
	\framebox{\includegraphics[width= \textwidth]{#2}}
		\caption[#4]{#3}
		\label{#1_#2}
	\end{center}
\end{figure}
}

\newcommand{\myFrameFigurer}[4][image]%
{
\begin{figure}[ht!bp]
	\begin{center}
	\framebox{\includegraphics[height= \textwidth, angle=270]{#2}}
		\caption[#4]{#3}
		\label{#1_#2}
	\end{center}
\end{figure}
}

\newcommand{\myFrameSmallFigure}[4][image]%
{
\begin{figure}[ht!bp]
	\begin{center}
	\framebox{\includegraphics[width=0.7\textwidth]{#2}}
		\caption[#4]{#3}
		\label{#1_#2}
	\end{center}
\end{figure}
}
\newcommand{\myFrameSmallFigurer}[4][image]%
{%
\begin{figure}[ht!bp]
	\begin{center}
	\framebox{\includegraphics[height= 0.7\textwidth, angle=270]{#2}}
		\caption[#4]{#3}
		\label{#1_#2}
	\end{center}
\end{figure}
}%

\newcommand{\myImgRef}[2][image]%
{%
	(Fig.~\ref{#1_#2})%
}
\newcommand{\myTable}[3]%
{%
\begin{table}[htdp]%
	\begin{center}%
		#1%
		\caption{#2}%
		\label{#3}%
	\end{center}%
\end{table}%
}%

\newcommand{\myTabRef}[1]%
{%
	Table~\ref{#1}%
}%
\newcommand{\myEqRef}[1]%
{%
	(Eq.~\ref{#1})%
}%
\begin{document}

\pagenumbering{Alph}

\selectlanguage{english}

\begin{titlepage}

\begin{center}
\begin{Huge}Jagiellonian University\\
\end{Huge}
\begin{center}
\textsc{Marian Smoluchowski Institute of Physics}
\end{center}
\end{center}
\vspace{0.5cm}
\begin{center}
\includegraphics[width=4.5cm]{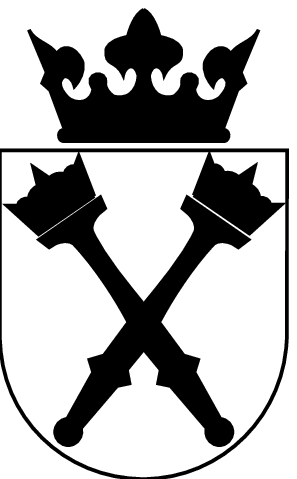}
\end{center}
\vspace{1cm}


\begin{center}
\begin{Large}
{ {Leading Modes of the $3\pi^{0}$ production} \\ {in proton--proton collisions} \\ {at incident proton momentum $3.35\mathrm{~GeV/c}$} }
\end{Large}
\end{center}
\vspace{0.5cm}
\begin{center}
\textbf{\begin{Large}Benedykt R. Jany\end{Large}}
\end{center}

\vspace{1.cm}
\begin{flushright}
Doctoral Dissertation prepared at the Nuclear Physics Department\\
Supervisor: Prof. Dr. hab. Zbigniew Rudy
\end{flushright}
\vspace{2cm}

\begin{center}
\textbf{\begin{large}Cracow 2011\end{large}}\\
\textsc{Poland}
\end{center}

\emptydoublepage
\thispagestyle{empty}

\begin{tiny}$\infty$\end{tiny}\\
\vspace{0.5\textheight}


\begin{center} 
\begin{Huge}\textsc{Ex Nihil Multum}\end{Huge}
\end{center}

\end{titlepage}

\emptydoublepage

\pagenumbering{Roman}
\begin{abstract}
This work deals with the prompt $pp \rightarrow pp 3\pi^{0}$ reaction where the $3\pi^{0}$ do not origin from the decay of narrow resonances like $\eta(547),\omega(782),\eta'(958)$.
The reaction was measured for the proton beam momentum of $3.35\mathrm{~GeV/c}$ with the WASA-at-COSY detector setup.
The dynamics of the reaction is investigated by Dalitz and Nyborg plots studies.
The reaction is described by the model assuming simultaneous excitation of two baryon resonances $\Delta(1232)$ and $N^{*}(1440)$ where
resonances are identified by their unique decays topology on the missing mass of two protons $MM_{pp}$ dependent Dalitz and Nyborg plots.
The ratio $R=\Gamma(N^{*}(1440) \rightarrow N \pi \pi )/\Gamma(N^{*}(1440) \rightarrow \Delta(1232) \pi \rightarrow N \pi \pi)=0.039 \pm 0.011 (stat.)  \pm 0.008 (sys.)$ is measured for the first time in a direct way.
It shows that the $N^{*}(1440) \rightarrow \Delta(1232) \pi \rightarrow N \pi \pi$ decay is a leading mode of $3\pi^{0}$ production.
It is also shown that the $MM_{pp}$ is very sensitive to the structure of the spectral line shape of the $N^{*}(1440)$ resonance as well as on the interaction between
 the  $\Delta(1232)$ and $N^{*}(1440)$ resonances.
The multipion spectroscopy -- a precision tool to directly access the properties of baryon resonances is considered.

The $pp \rightarrow pp \eta(3\pi^{0})$ reaction was also measured simultaneously.
It is shown that the $\eta$ production mechanism via $N^{*}(1535)$ is $43.4 \pm 0.8(stat.) \pm 2.0(sys.)\% $ of the total production,
 for the $\eta$ momentum in the CM system $q_{\eta}^{CM}=0.45-0.7\mathrm{~GeV/c}$.
First time momentum dependence of the $\eta$ angular distribution is seen, the strongest effect is observed for the  $\cos(\theta_{\eta^{CM}})$ distribution.
\end{abstract}

\emptydoublepage
\begin{otherlanguage}{polish}
  \begin{abstract}
Praca ta jest dedykowana reakcji bezpośredniej produkcji $pp \rightarrow pp 3\pi^{0}$ gdzie $3\pi^{0}$ nie pochodzą z rozpadu wąskich rezonansów jak $\eta(547),\omega(782),\eta'(958)$.
Reakcja została zmierzona dla pędu wiązki protonowej $3.35\mathrm{~GeV/c}$ przy pomocy systemu detekcyjnewgo WASA-at-COSY.
Dynamika reakcji jest studiowana przy pomocy wykresów Dalitza i Nyborga.
Reakcja jest opisywana przez model zakładający równoczesne wzbudzenie dwóch rezonansów barionowych  $\Delta(1232)$ i $N^{*}(1440)$ gdzie
rezonansy są identyfikowane dzięki niepowtarzalnej topologii ich rozpadów na zależnych od masy brakującej dwóch protów $MM_{pp}$ wykresach Dalitza i Nyborga.
Stosunek rozgałęzien $R=\Gamma(N^{*}(1440) \rightarrow N \pi \pi )/\Gamma(N^{*}(1440) \rightarrow \Delta(1232) \pi \rightarrow N \pi \pi)=0.039 \pm 0.011 (stat.)  \pm 0.008 (sys.)$
został po raz pierwszy wyznaczony w bezpośredni sposób;
oznacza to że gałąź rozpadu $N^{*}(1440) \rightarrow \Delta(1232) \pi \rightarrow N \pi \pi$ jest członem wiodącym produkcji $3\pi^{0}$.         
Pokazane jest że  $MM_{pp}$ jest bardzo czuła ze względu na strukturę linii spektralnej rezonansu  $N^{*}(1440)$ oraz na oddziaływanie między resonansami $\Delta(1232)$ i $N^{*}(1440)$.
Spektroskopia wielopionowa jako precyzyjne narzędzie do bezpośredniego dostępu do własności rezonansów barionowych jest rozważana.    
Przekrój czynny na reakcję został wyznaczony $\sigma_{pp \rightarrow pp 3\pi^{0}} = 123 \pm 1 (stat.) \pm 8 (sys.) \pm 19 (norm.)  \mu b$.

Równoczesnie została zmierzona reakcja $pp \rightarrow pp \eta(3\pi^{0})$.     
Pokazane zostało, że mechanizm produkcji mezonu $\eta$ przez rezonans $N^{*}(1535)$ jest równy $43.4 \pm 0.8(stat.) \pm 2.0(sys.)\% $ całkowitej produkcji,
został on wyznaczony dla pędu mezonu $\eta$ w środku masy równego $q_{\eta}^{CM}=0.45-0.7\mathrm{~GeV/c}$.
Po raz pierwszy zaobserwowano zależność pędową rozkładów kątowych dla mezonu $\eta$, najsilniejszy efekt jest widoczny dla rozkładu $\cos(\theta_{\eta^{CM}})$.  
  \end{abstract}
\end{otherlanguage}

\emptydoublepage

{
\changepage{+5cm}{}{}{}{}{-5cm}{-5cm}{}{}
\tableofcontents
}
\thispagestyle{empty}
\emptydoublepage

\pagenumbering{arabic}
\setcounter{page}{1}
\section{Introduction}
\thispagestyle{plain}
Nucleon--Nucleon reactions typically lead to the abundant pion production. It is due to the fact that this pseudoscalar meson \cite{PDG2008} has the lowest mass from
all members of pseudoscalar meson nonet and carries no exotic quantum numbers.
Isospin of pion is $1$, this was used in suitable definition of $G$~parity. Unstable particles, which appear at first stage of nuclear reaction,
usually decay into pion channels; it means that investigation of pion spectra is one of the techniques refined for the analysis of the unstable particles or states.

\noindent As an example, so-called ABC~effect \cite{ABCeffect} is observed in the reactions with two pion production.
Similarly, simultaneous detection of three pions was used e.g. in the investigation of $\eta$~meson produced in $pp \rightarrow pp \eta \rightarrow 3\pi^{0}$ reaction;
luckily the $\eta$~meson has narrow width and technically the analysis was not that difficult \cite{Vlasov3pi0}.

In this dissertation the properties of the prompt $3\pi^{0}$ production in the proton-proton collisions at the incident proton momentum of $P_{beam}=3.35\mathrm{~GeV/c}$ where the three
pions do not origin from the decays of the narrow resonances like $\eta(547), \omega(782), \eta'(958)$) are described.
The dynamics of this process was never studied in details neither experimentally nor theoretically, the cross section is also unknown.

The reaction was measured with the WASA-at-COSY detector setup \cite{proposal} located in the Institute f\"ur Kernphysik of the Forschungszentrum J\"ulich Germany
at the Cooler Synchrotron COSY. Using the unique capabilities of the WASA-at-COSY installation to detect the charged and neutral multi-particle coincidences with a large acceptance,
all final state particles were reconstructed from the signals in the detectors.
This provided a data set of high statistics for the later analysis.

The studies presented in this work concentrate on the extraction of the reaction dynamics in the
model independent way using only the basic principles like energy and momentum conservation -- applying kinematic calculations in the framework of Monte-Carlo model simulations.
This is realized in a systematic way by studying the invariant masses of the subsystems using missing mass of two protons dependent Dalitz and Nyborg plots.    
The analysis leads to conclusions that the reaction proceeds via simultaneous excitations of two baryon resonances $\Delta(1232)$ and $N^{*}(1440)$.
The developed approach first time allowed to extract the spectroscopic informations (branching ratio, spectral line shape) of the $N^{*}(1440)$ resonance in a direct way.  

In parallel also the $pp \rightarrow pp \eta(3\pi^{0})$ reaction was measured.
The $\eta$ meson production was successfully described by assuming two mechanisms: the resonant production via excitation of $N^{*}(1535)$ and a non resonant part.
\newpage
For the first time also the momentum dependence of the $\eta$ angular distribution~was~seen.

\medskip
\noindent As a consequence one concludes that the multipion spectroscopy may be treated as a precision tool to directly access the properties of baryon resonances.  

\bigskip
\noindent This dissertation is divided into four main mainstream parts.

First part Section~\ref{sec:setup} describes the experimental setup. The Cooler Synchrotron COSY is described. 
The properties of the WASA-at-COSY detector setup are overviewed. The properties of the detector components (Forward and Central Detector) as well as the pellet target are given. 

Second part Section~\ref{sec:PhysicsMotivation} is dedicated to the physics of $3\pi^{0}$ production.
Theory and data status together with physics motivations are given. Later the selected choice of the observables is presented.

Third part Section~\ref{sec:AnalysisExperimental} deals with the analysis of the experimental data.
The experimental conditions are listed. Next, the selection of events is discussed. Later, the experimental detector resonance is compared with
the detector response simulations.
Finally the procedure of the kinematic fitting of the events together with the used error parametrization is presented.

Forth part Section~\ref{sec:results} contains the results and their errors.

\noindent The $pp \rightarrow pp 3\pi^{0}$ reaction is studied. The Monte-Carlo model description is proposed together with the supporting arguments.
Next, the parameters of the model are derived from the experimental data by studying the missing mass of two protons dependent Dalitz and Nyborg plots, the overall model of the reaction is presented.
The parameters of the model are discussed; the possible interaction between the $\Delta(1232)$ and $N^{*}(1440)$ resonances or
the influence of the $N^{*}(1440)$ spectral line shape is taken into consideration.
The model is validated by detailed statistical analysis and other processes contribution to the model are verified.
Later, the cross section is extracted. The Dalitz and Nyborg plots are corrected for the detection efficiency and geometrical acceptance,
 the absolute normalization of the spectra is performed.

\noindent The $pp \rightarrow pp \eta(3\pi^{0})$ reaction is studied in parallel.
The accessibility of the phase space is checked and the observables are defined.
Next, the production mechanism is studied and described by assuming two mechanisms: the resonant production via excitation of $N^{*}(1535)$ and a non resonant part.
Later, the angular distributions are investigated and the angular anisotropy is extracted.     

A summary with conclusions is given in Section~\ref{sec:SumConclusions}.
The obtained results are discussed and compared with other experimental measurements as well as available theoretical models.
The multipion spectroscopy -- a high precision tool to directly access the properties of baryon resonances is considered.
 
There are also six appendixes \ref{appendix:Kine5part},~\ref{appendix:DetectorCalibration},~\ref{appendix:wmc},~\ref{appendix:TrackReconstruction},~\ref{appendix:kfit},~\ref{appendix:Bayes},
 which are toolboxes, covering the technical aspects behind the data analysis presented in the main parts.
In the Appendix~\ref{appendix:DataTables} the results of this work i.e. the acceptance and efficiency corrected Dalitz and Nyborg Plots
and the angular distributions of the $\eta$ meson are presented as tables of numbers. 

\newpage ~
\thispagestyle{empty}
\emptydoublepage
\newpage

\section{The Experimental Setup \label{sec:setup}}
\thispagestyle{plain}
The WASA detector setup had been operated since 1998 in The Svedberg Laboratory in Uppsala (Sweden).
It has been built with the aim on studying the decays of $\eta$-mesons in nuclear reactions.
In 2003 it has been decided to shut down the CELSIUS ring, to stop the operation of WASA~at~CELSIUS.
Shortly after that announcement the idea came up to continue the operation of WASA~at~COSY.
The reasons are obvious: The combination of WASA and COSY would be of advantages for both communities.
COSY offers a higher energy than CELSIUS, allowing the extension of the studies into the $\eta'$ sector. 
The WASA detector has an electromagnetic calorimeter as a central component and, thus, the ability to detect neutral decay
modes involving photons -- such a device is missing at COSY. The proposal \cite{proposal} for moving WASA to COSY \cite{WASAatCOSY} was accepted by the COSY PAC in 2004. 
The WASA detector was dismounted during summer 2005 and shipped to J\"ulich (Germany). The final installation and commissioning took place in the end of 2006. 

\subsection{Cooler Synchrotron COSY}
\label{subsec:cosy}
\myFrameSmallFigure{ikpcosy}{View at the Cooler Synchrotron COSY.}{View at the  Cooler Synchrotron COSY}
The Cooler Synchrotron COSY \myImgRef{ikpcosy} is located in the Institute f\"ur Kernphysik of the Forschungszentrum J\"ulich Germany. It delivers phase-space cooled polarized or unpolarized protons (deuterons) of momentum from $p=300\mathrm{~MeV/c}$ up to $p=3700\mathrm{~MeV/c}$. The ring has a circumference of $184\mathrm{~m}$ and can be filled with up to $10^{11}$ particles. When using the internal cluster target the luminosity of $10^{31}\mathrm{cm^{-2}s^{-1}}$ can be reached.
Two cooling methods can be applied during accumulation of the beam to reduce the phase-space volume, electron cooling at injection energies and stochastic cooling at higher energies.
In the electron cooling method an electron beam, moving with the same average velocity as proton beam (acting as a cold gas), 
is mixed with the protons (hot gas), by mixing the two gases with different temperatures the average kinetic energy drops.
The stochastic cooling is the process in which the deviations of nominal energy or position of particles in a beam are measured and corrected. The electron cooling system in COSY is applied at injection momentum $p=300\mathrm{~MeV/c}$ and reaches up to $p=600\mathrm{~MeV/c}$, and the stochastic cooling from $p=1500\mathrm{~MeV/c}$ to $3700\mathrm{~MeV/c}$.
Both, proton and deuteron beams, can be provided unpolarized as well as polarized. At COSY internal and external target positions are in operation \myImgRef{cosy}. For further details see \cite{cosy}.

\myFrameBigFigure{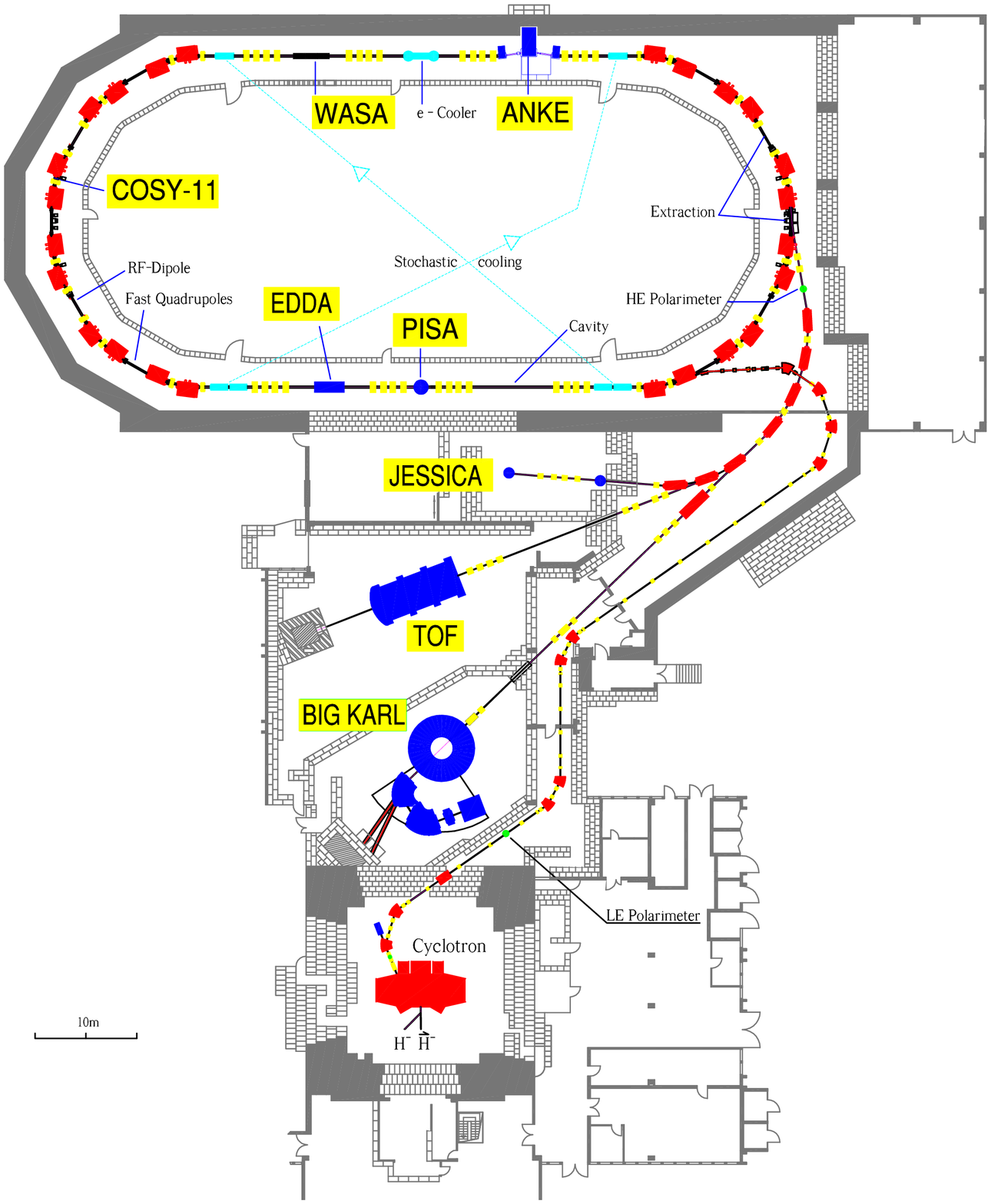}{The accelerator complex with the cyclotron, the COSY ring and the experimental installations. The place of WASA is within one of the straight sections of COSY.
In the presented Figure the beam circulates clockwise.
For more details please visit \url{http://www.fz-juelich.de/ikp/cosy}.}{The accelerator complex COSY}

\subsection{The WASA~at~COSY detector}
\label{subsec:detector}
\myFrameFigure{WASA}{Cross section of the WASA-~at~COSY detector. Beam comes from the left. The Central Detector (\textbf{CD}) build around the interaction point (at the left). The layers of the Forward Detector (\textbf{FD}) are shown on the right. Other symbols (MDC, PSB, FPC, ...) will be explained in text of the thesis.}{Cross section of the WASA detector}
\myFrameSmallFigure{wdetector2}{3D View of the WASA detector\cite{proposal}.}{3D View at the WASA detector}

As mentioned before the \textbf{WASA} detector -- \textit{\textbf{W}ide \textbf{A}ngle \textbf{S}hower \textbf{A}pparatus }  \myImgRef{WASA}, was designed to study various decay modes of the $\eta$-meson. This is reflected in the detector setup.
The $\eta$-mesons are produced in reactions of the type $pp \longrightarrow pp\eta$. Due to the kinematics boost, the two protons are going into the forward direction, while the light decay products of the $\eta$ are distributed into $4\pi$.

In order to detect the protons, a $\phi$-symmetric~($0-360\mathrm{~deg}$) forward detector for $\theta \leq 18^{\circ}$ is installed. The particles are identified and reconstructed by means of dE measurement and track reconstruction using drift chambers.
A trigger, which is set only on the forward detector, can be used to select events independently from the decay mode of the $\eta$-meson registered by the Central Detector.

Particles coming from meson decays ($e^{\pm}, \mu^{\pm}, \pi^{\pm}$ and $\gamma$), are detected in the central part of WASA ($\theta \geq 20^{\circ}$). 
Momentum reconstruction for charged particles is done by tracking in a magnetic field and the energy of the neutral particles is measured using an electromagnetic calorimeter. By including the central detector in the trigger also very rare decay modes can be studied while using a very high luminosity of up to $10^{32}$~cm$^{-2}$s$^{-1}$. A 3D view of the detector setup \myImgRef{wdetector2}.

\subsubsection{The Pellet Target}
\label{subsubsec:pellet}
The pellet target system \myImgRef{pellet1} was a special development for WASA. 
The "pellets" are frozen droplets of hydrogen or deuterium with a diameter between $25\mathrm{~\mu m}$ and $35\mathrm{~\mu m}$. The advantages of using pellet target compared with a standard internal gas target are the following:

\begin{itemize}
\item high target density, allows high luminosities necessary for studying rare decays
\item thin tube delivery through the detector, $4\pi$ detection possible
\item precisely localized target, small probability of secondary interactions inside the target
\end{itemize}

The central part of the system is the pellet generator where a stream of liquid gas (hydrogen or deuterium)
 is broken into droplets by a vibrating nozzle. The droplets freeze by evaporation into a first vacuum chamber
 forming a pellet beam. The beam enters a vacuum-injection capillary where it is collimated and is fed through
 a $2\mathrm{~m}$ long pipe into the scattering chamber \myImgRef{pellet2}.
 An effective beam thickness for hydrogen of $3*10^{15}\mathrm{~atoms/cm^2}$ has been achieved with a beam diameter $2-4\mathrm{~mm}$,
 a frequency of pellets $5-10\mathrm{~kHz}$, and an average distance between the pellets of $9-20\mathrm{~mm}$.

\begin{figure}[ht!bp]%
\vskip 1cm
	\begin{center}%
	\frame{
		\includegraphics[height= 0.7\textwidth, angle=270]{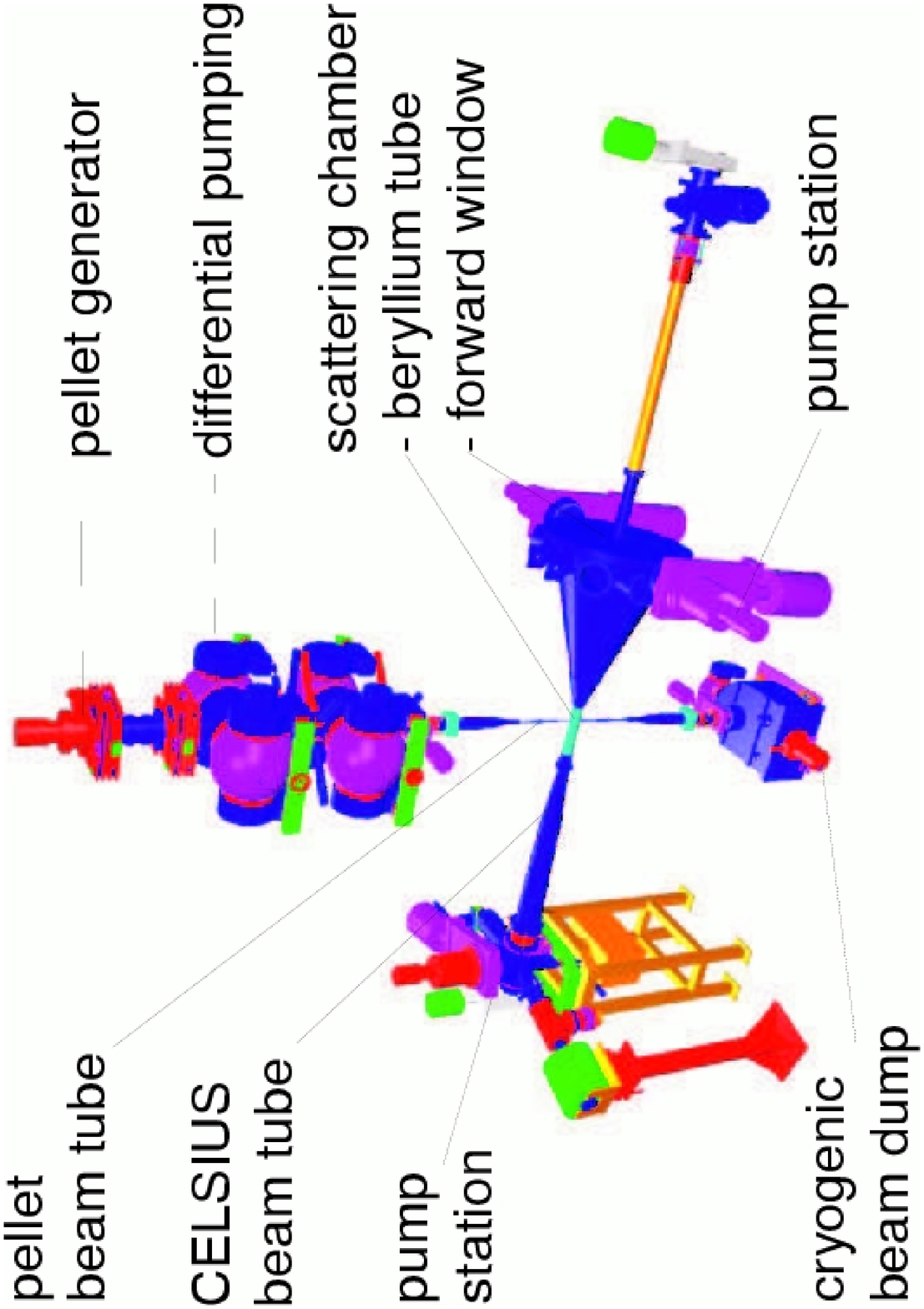}%
		}
		\caption[The Pellet Target system]{The Pellet Target system \cite{proposal}.}
		\label{image_pellet1}%
	\end{center}
\end{figure}
\begin{figure}[ht!bp]
	\begin{center}
	\frame{\includegraphics[width=0.5\textwidth]{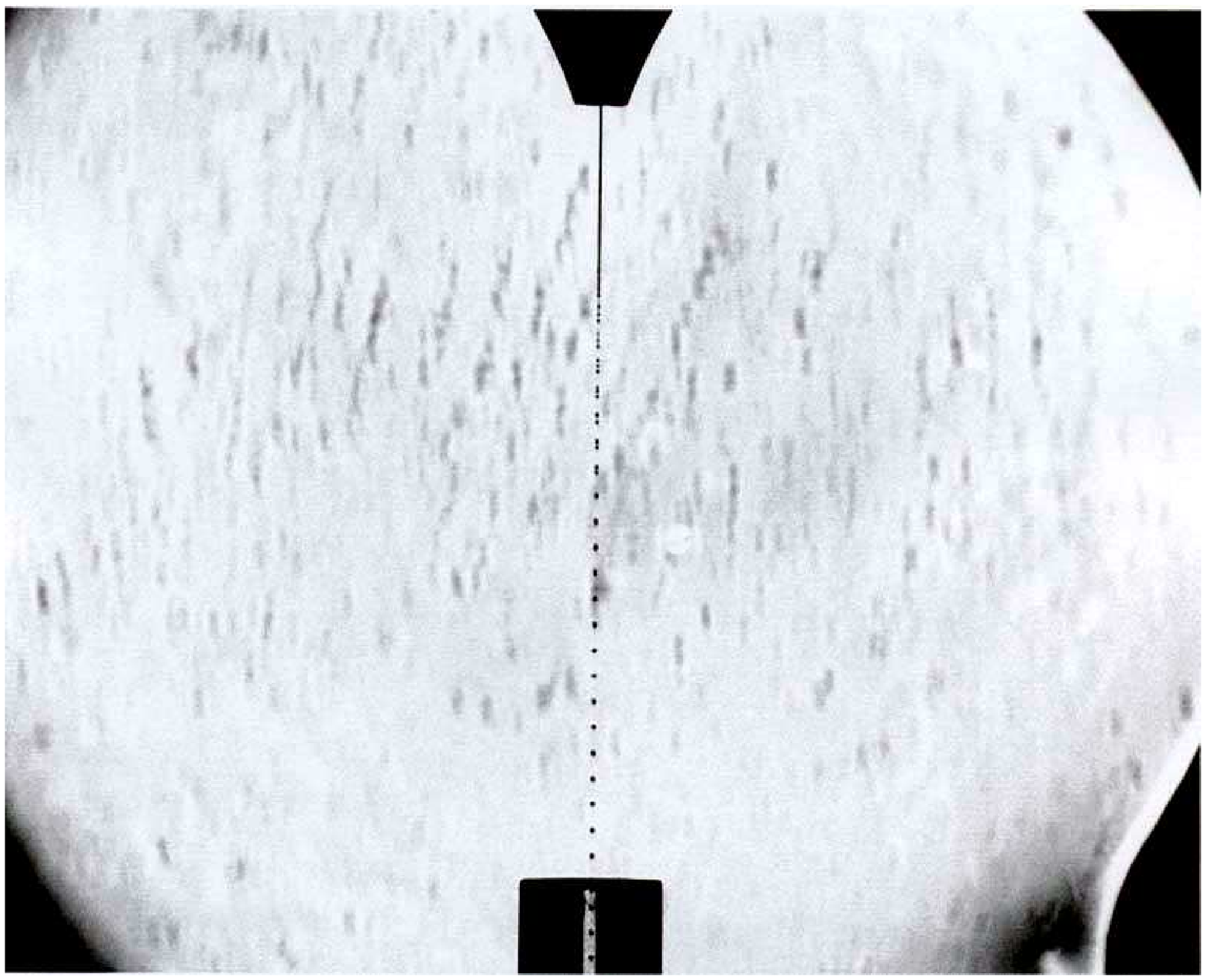}}
		\caption[Photography of generated pellets]{Photography of generated pellets.}
		\label{image_pellet2}
	\end{center}
\end{figure}

\newpage
\subsubsection{The Forward Detector}
\label{subsubsec:fd}
The Forward Detector (FD) \myImgRef{WASA} tags meson production by measuring the energies (dE-E) and angles of forward scattered projectiles like protons, deuterons, neutrons and charged pions.
The produced mesons are then reconstructed using the missing mass technique. Detector covers angles from $3^{\circ}$ to $18^{\circ}$. It consists of several layers of detectors described below: 
\begin{itemize}
\item \textbf{FWC} The Forward Window Counters\\
The FWC is the first detector in the Forward Detector. It consists of $12$ plastic scintillators of $5\mathrm{~mm}$ thickness. It is used to reduce the background from scattered particles originated from the beam pipe or the exit flange.
\item \textbf{FPC} The Forward Proportional Chambers\\
The next detector is the FPC. It consists of $4$ modules each containing $122$ straw tubes detectors. The modules are rotated relatively to each other by $45^{\circ}$. The FPC is used as a precise tracking device.
\item \textbf{FTH} The Forward Trigger Hodoscope\\
Close to FPC the FTH (``J\"ulich Quirl'') is installed. It consists of 3 layers of plastic scintillators, one with straight modules, two with bended ones. Each layer has a thickness of $5\mathrm{~mm}$. It is used for a rough determination of the hit position on the higher level and as a starting value for the track reconstruction, see Appendix~\ref{appendix:TrackReconstructionFD}.
\item \textbf{FRH} The Forward Range Hodoscope\\
The kinetic energy of the particles is measured by the FRH. It consists of $5$ layers of cake-piece shaped plastic scintillators of $11\mathrm{~cm}$ thickness. There are $24$ scintillators pro layer. It is also used for particle identification by the dE-E technique.
\item \textbf{FRI} The Forward Range Interleaving Hodoscope\\
Between third and forth layer of FRH two layers of plastic scintillators are installed (FRI). Each layer is made of $32$ strips of $5.2\mathrm{~mm}$ thickness. The FRI is used to determine the scattering angles of neutrons.
\item \textbf{FVH} The Forward Veto Hodoscope\\
The last layer of FD is FVH. It consists of $12$ horizontally oriented scintillator strips with photomultipliers on both sides. The hit position is determined from the time differences of the signals. It is used to identify particles which are not stopped in the FRH. 
\end{itemize}

\subsubsection{The Central Detector}
\label{subsubsec:cd}
The Central Detector (CD) surrounds the interaction point and is constructed to identify energies and angles of the decay products of $\pi^{0}$ and $\eta$ mesons, with close to $4\pi$ acceptance.  It consists of:
\begin{itemize}
\item \textbf{SCS} The Superconducting Solenoid\\
The SCS produces an axial magnetic field necessary for momentum reconstruction using the inner drift chambers. As superconductor NbTi/Cu is used cooled down by liquid He at $4.5\mathrm{K}$. The maximal central magnetic field is $1.3\mathrm{T}$. The return path for the field is done by a yoke made of $5$ tons of pure iron  with low carbon content.
\item \textbf{MDC} The Mini Drift Chamber\\
The MDC is build around the beam pipe and it is used for momentum and vertex determination \myImgRef{mdc}.  It consists of $17$ layers with in total $1738$ straw tubes detectors. It covers scattering angles from $24^{\circ}$ to $159^{\circ}$ \cite{jacewicz}. For the resolution refer to \myTabRef{tab:mdc}.

\myTable{
\begin{tabular}{|l|c|c|}
\hline
 particle& $p$ $\mathrm{[ MeV/c ]}$ & resolution $\bigtriangleup p/p$ \\ 
\hline
\hline
electrons & $20-600$ & $<1\%$ \\ 
\hline
 pions, muons & $100-600$ & $<4\%$ \\ 
\hline
protons & $200-800$ & $<5\%$\\
\hline
\end{tabular}
}{MDC resolution}{tab:mdc}

\begin{figure}[ht!bp]%
	\begin{center}%
	\frame{
		\includegraphics[height= 0.7\textwidth, angle=270]{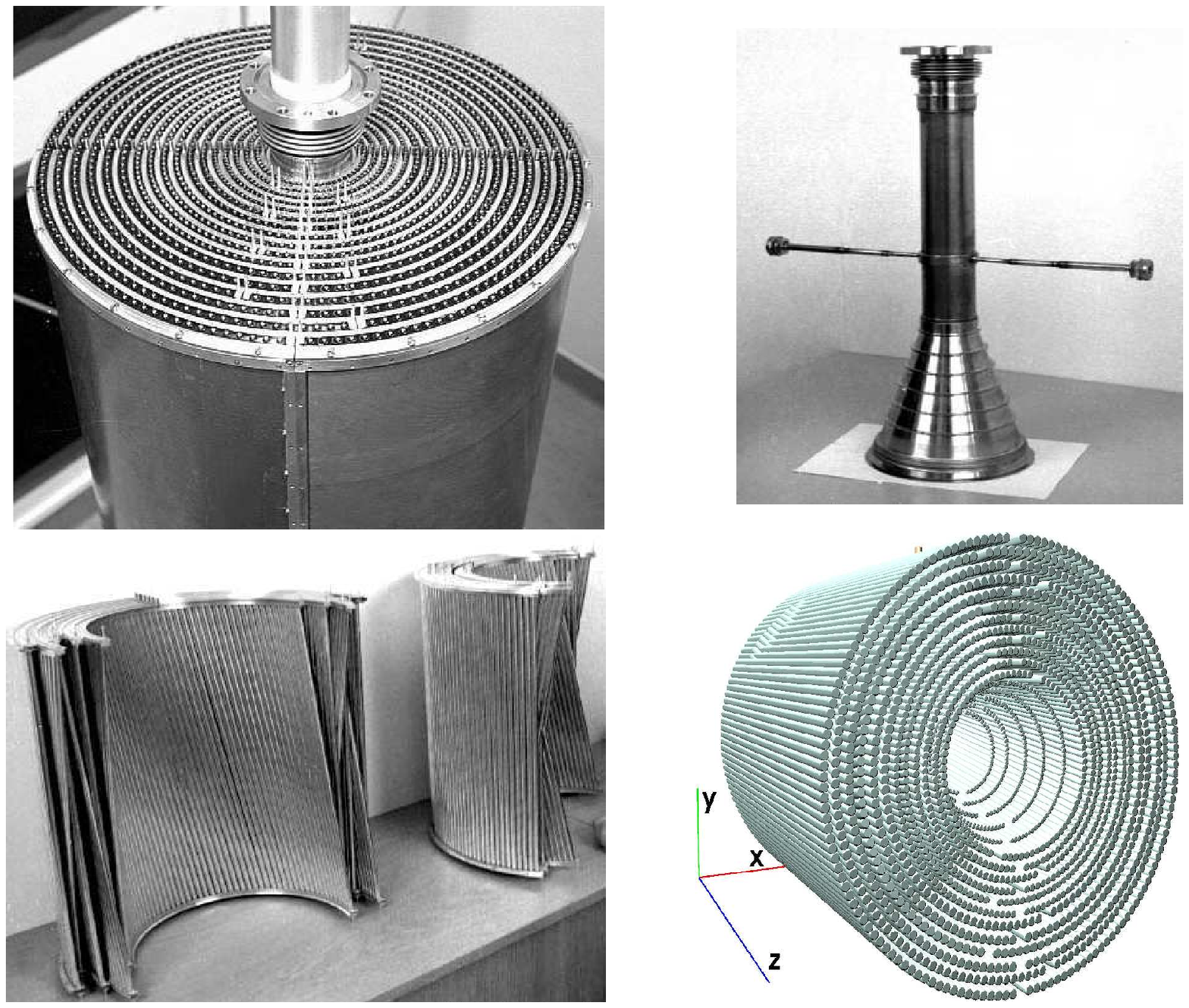}%
		}
		\caption[MDC and Be pipe.]{MDC and Be pipe. The fully assembled MDC inside Al-Be cylinder (upper left)\cite{jacewicz}.}
		\label{image_mdc}%
	\end{center}%
\end{figure}

\item \textbf{PSB} The Plastic Scintillator Barrel\\
The PSB surrounds the MDC inside the SCS. It consists of $146$ pieces of $8\mathrm{~mm}$ thick strips that form a barrel like shape. It is used together with MDC and SEC, and acts as a dE-E and dE-momentum detector and  as well as a veto for photons.

\item \textbf{SEC} The Scintillator Electromagnetic Calorimeter\\
\label{SECofWASA}
The SEC is the heart of the WASA detector and maybe the most important part.
At CELSIUS it has been used to measure electrons and photons up to $800$~MeV.
However, using a different setting the energy range can be extended taking into account the higher energy available at COSY. 
It consists of $1012$ CsI(Na) crystals shaped like a truncated pyramids \myImgRef{crystal}. It covers angles from $20^{\circ}$ to $169^{\circ}$. The crystals are placed in $24$ layers along the beam \myImgRef{sec2}. The lengths of the crystals vary from $30\mathrm{~cm}$ (central part), $25\mathrm{~cm}$ (forward part) to $20\mathrm{~cm}$ (backward part). The forward part consists of $4$ layers each $36$ crystals, covering the range of $20^{\circ}-36^{\circ}$. The central part consists of $17$ layers with $48$ elements each, covering the range between $36^{\circ}-150^{\circ}$, and the backward part with $3$ layers, two with $24$ crystals and one with $12$. The geometrical distribution of the crystals \myImgRef{sec1} and \myImgRef{sec3}.

The calorimeter consists of sodium doped CsI crystals. 
They are painted with transparent varnish for moisture protection and wrapped in $150\mathrm{~\mu m}$ teflon and $25\mathrm{~\mu m}$ aluminized mylar foil \cite{zabierowski}. For more information refer to \myTabRef{tab:sec}.
Detailed properties are described in \cite{BRJmaster}.

More detail information on WASA Detector components and electronics can be found in \cite{WASAcelsius, DAQ1, DAQ2}.\\

\myTable{
\begin{tabular}{|l|c|}
\hline
 amount of active material& $16$~$X_0$ (radiation length \cite{BRJmaster}) \\ 
\hline
geometric acceptance:& $96\%$\\
polar angle:&$20^{\circ}-169^{\circ}$\\
azimuthal angle:&$0^{\circ}-180^{\circ}$\\ 
\hline
 relative energy resolution:& $30\%$ (FWHM)\\ 
$Cs(137) 662\mathrm{keV}$ & \\
\hline
maximal kinetic energy for stopping: &\\
pions/protons/deuterons & $190/400/500 \mathrm{MeV}$ \\
\hline
\end{tabular}
}{SEC parameters}{tab:sec}

\myFrameSmallFigure{crystal}{CsI(Na) crystal fully equipped with light guide, photomultiplier tube and housing \cite{zabierowski}.}{CsI(Na) crystal fully equipped}

\myFrameSmallFigure{sec2}{SEC planar map (arrow indicates beam direction) \cite{proposal}.}{SEC planar map}

\begin{figure}[ht!bp]%
	\begin{center}%
	\frame{
		\includegraphics[height= 0.5\textwidth, angle=270]{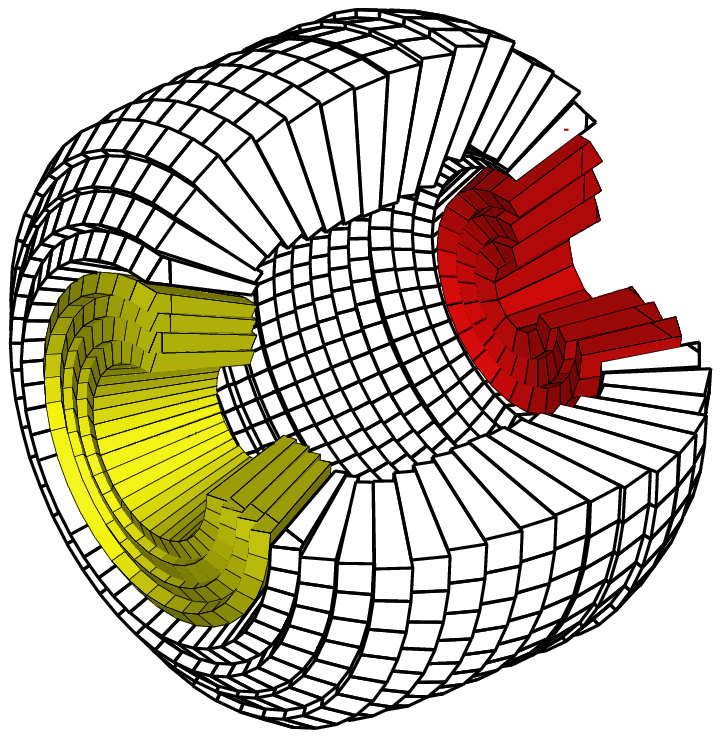}%
		}
		\caption[SEC schematic view.]{Schematic view of the calorimeter layout. It consists of the forward part (yellow, on the left), central part and the backward part (red, on the right) \cite{proposal}.}
		\label{image_sec1}%
	\end{center}%
\end{figure}

\myFrameSmallFigure{sec3}{Photography of the SEC (forward part to the left, beam comes from the right).}{Photography of the SEC}


\end{itemize}

\newpage ~
\thispagestyle{empty}
\emptydoublepage
\newpage

\section{Physics of $3\pi^{0}$ production}\label{sec:PhysicsMotivation}
\thispagestyle{plain}
\subsection{Theory and data status - Physics Motivations}
\label{subsec:TheoryData}

\myFrameFigure{pp3pi0Mesons.eps}{The existing experimental data \cite{ETACS01} for the 
total cross section for $pp \rightarrow pp 3\pi^{0}$ and $pp \rightarrow pp \pi^{+}\pi^{-}\pi^{0}$ reaction versus proton beam kinetic energy $T_{beam}$,
 the proposed models (from \cite{ETACS01,3pichargedPauly1,3pichargedPauly2}) for cross section scaling \cite{3pi0CSmodel2,3pi0CSmodel3, 3pi0CSmodel1}.\newline
The beam kinetic energy at threshold $T_{b}^{Thr.}$ and beam momentum at threshold $P_{b}^{Thr.}$ 
 for $pp \rightarrow pp X$ reaction where\newline
 $X=\eta(547)$~~$(T_{b}^{Thr.}=1.255\mathrm{~GeV},~P_{b}^{Thr.}=2.670\mathrm{~GeV/c})$\newline
 $X=\omega(782)$~~$(T_{b}^{Thr.}=1.892\mathrm{~GeV},~P_{b}^{Thr.}=2.670\mathrm{~GeV/c})$\newline
$X=\eta'(958)$~~$(T_{b}^{Thr.}=2.405\mathrm{~GeV},~P_{b}^{Thr.}=3.208\mathrm{~GeV/c})$\newline
$X=\Phi(1020)$~$(T_{b}^{Thr.}=2.593\mathrm{~GeV},~P_{b}^{Thr.}=3.404\mathrm{~GeV/c})$ respectively indicated.
 }{Cross Section}

The field of $3\pi^{0}$ production in the proton-proton collisions in the proton beam kinetic energy region between $T_{beam}=1-3\mathrm{~GeV}$ where the three
pions do not origin from the decays of the narrow resonances (like $\eta, \omega, \eta'$) is unexplored both
experimentally and theoretically. 
The dynamics of this process was never studied in details.
The total cross section was measured recently for few data points for low energies \myImgRef{pp3pi0Mesons.eps} \cite{ETACS01}.
and the simple models for the cross section scaling were proposed \cite{3pi0CSmodel2, 3pi0CSmodel3, 3pi0CSmodel1}.

The statistical (Phase Space) model prediction is based on the phase space considerations \cite{3pi0CSmodel2}.\newline
The FSI Faeld-Wilkin model prediction is based of the assumption that there exist a proton-proton Final State Interaction (FSI) which could be modeled as discussed in \cite{3pi0CSmodel3}.\newline    
The FSI Delof model prediction is also based on the assumption that there exist a proton-proton FSI which could be modeled as discussed in \cite{3pi0CSmodel1}
(for the $pp$ relative momenta in the Center of Mass greater than $300\mathrm{~MeV/c}$ pure Phase Space is used).

It should be noticed that the two proposed models \cite{3pi0CSmodel3, 3pi0CSmodel1} are consistent with existing $3\pi^{0}$ data,
 but for the higher beam kinetic energy $T_{beam}\sim 2.5\mathrm{~GeV}$ their predictions differ by a factor $2$.

As proposed by \cite{ETACS01} to get some informations about the dynamics of the reaction one can study also the cross section ratio
\begin{equation}
 \frac{\sigma(pp \rightarrow pp \pi^{+}\pi^{-}\pi^{0})}{\sigma(pp \rightarrow pp \pi^{0}\pi^{0}\pi^{0})}
\label{eq:CSratioPions}
\end{equation}

(assuming isospin conservation one can perform calculations for different reaction scenarios see Table.~\ref{tab:ReacScenario}).

\myTable{
\begin{tabular}{|c||c|}
\hline
& \\
Assumed Reaction Scenario &  $\frac{\sigma(pp \rightarrow pp \pi^{+}\pi^{-}\pi^{0})}{\sigma(pp \rightarrow pp \pi^{0}\pi^{0}\pi^{0})}$ \\
\hline
\hline
& \\
Phase Space (statistical model)  &  $8$ \cite{ETACS01,3pi0CSmodel2}\\ 
\hline
& \\
$\Delta(1232)N^{*}(1440)$ & \\
 $\Delta(1232) \rightarrow p \pi$&$4$ \cite{ETACS01} \\
$N^{*}(1440) \rightarrow p \pi \pi$ & \\
\hline
& \\
$\Delta(1232)N^{*}(1440)$ & \\
 $\Delta(1232) \rightarrow p \pi$&$7$ \cite{ETACS01,3pi0Baryons} \\
$N^{*}(1440) \rightarrow \pi \Delta(1232)$ & \\
\hline
& \\
$\Delta(1232)N^{*}(1520)$ & $>>8$ \cite{ETACS01}\\
\hline
\end{tabular}
}{Different scenarios of $pp \rightarrow pp \pi \pi \pi$ reaction. Description in text. }{tab:ReacScenario}

Using the available experimental data (Fig.~\ref{image_pp3pi0Mesons.eps}) it is possible to calculate the ratio (Eq.~\ref{eq:CSratioPions}) only at energy $1.36\mathrm{~GeV}$ \cite{ETACS01}:
\begin{equation}
 \left( \frac{\sigma(pp \rightarrow pp \pi^{+}\pi^{-}\pi^{0})}{\sigma(pp \rightarrow pp \pi^{0}\pi^{0}\pi^{0})}\right)_{T_{beam}^{Exp.}=1.36\mathrm{~GeV}} = 6.3 \pm 0.6 (stat.) \pm 1.0 (syst.) 
\end{equation}
It is seen that the calculated ratio is consistent with the Phase Space (statistical model) value Table~\ref{tab:ReacScenario}.  

But such a studies cannot substitute the full study of the dynamics (which might be very complicated)
in terms of invariant mass studies in the subsystems and assumptions about various baryon resonances excitations (i.e. excited states of the nucleon \cite{PDG2008, BaryonsBook1, BaryonsBook2}).

\begin{sidewaysfigure}
\centering
{
\subfigure[]{\fbox{\includegraphics[width=0.4\textwidth, height=0.3\textwidth]{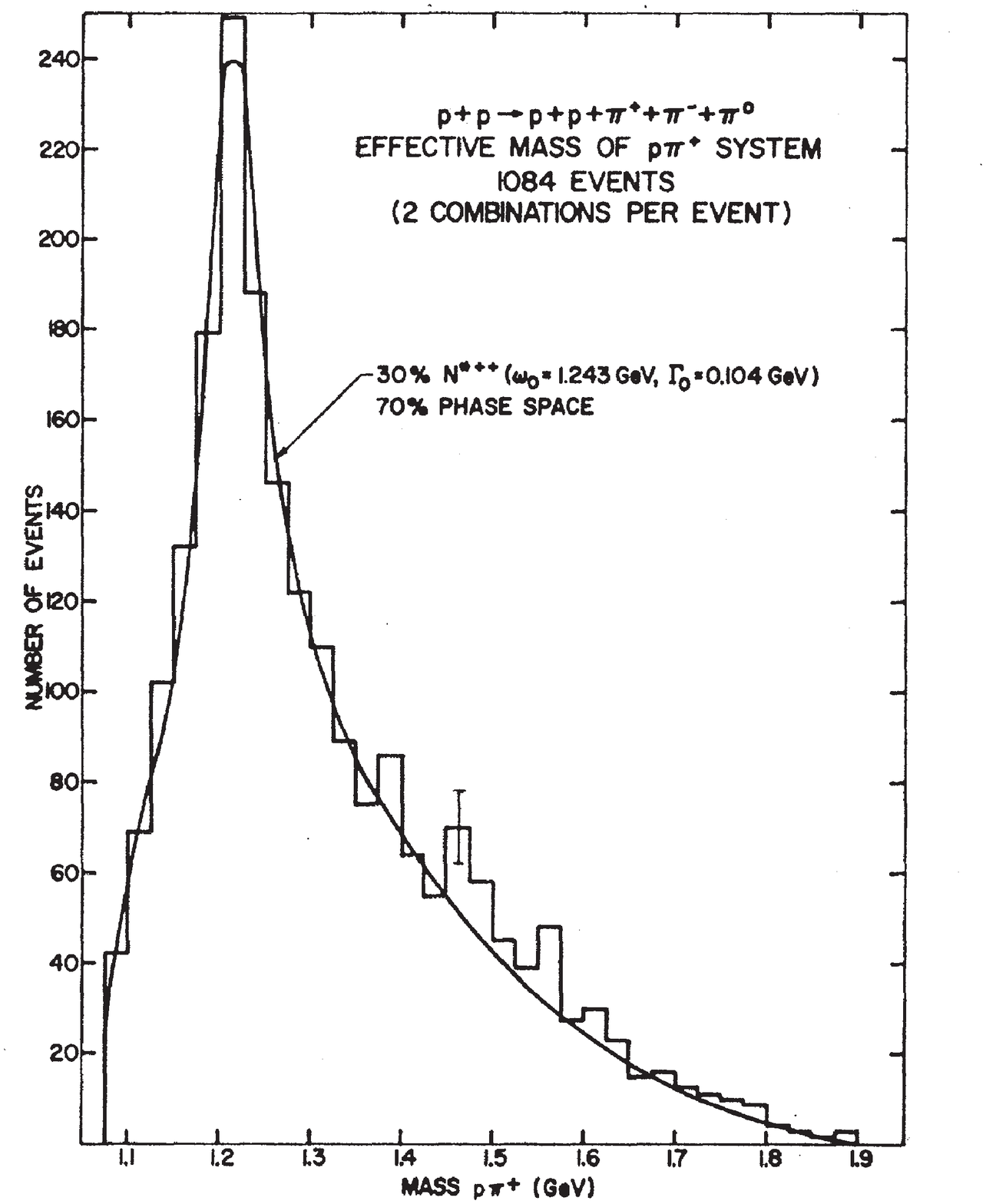}}}\quad
\subfigure[]{\fbox{\includegraphics[width=0.4\textwidth, height=0.3\textwidth]{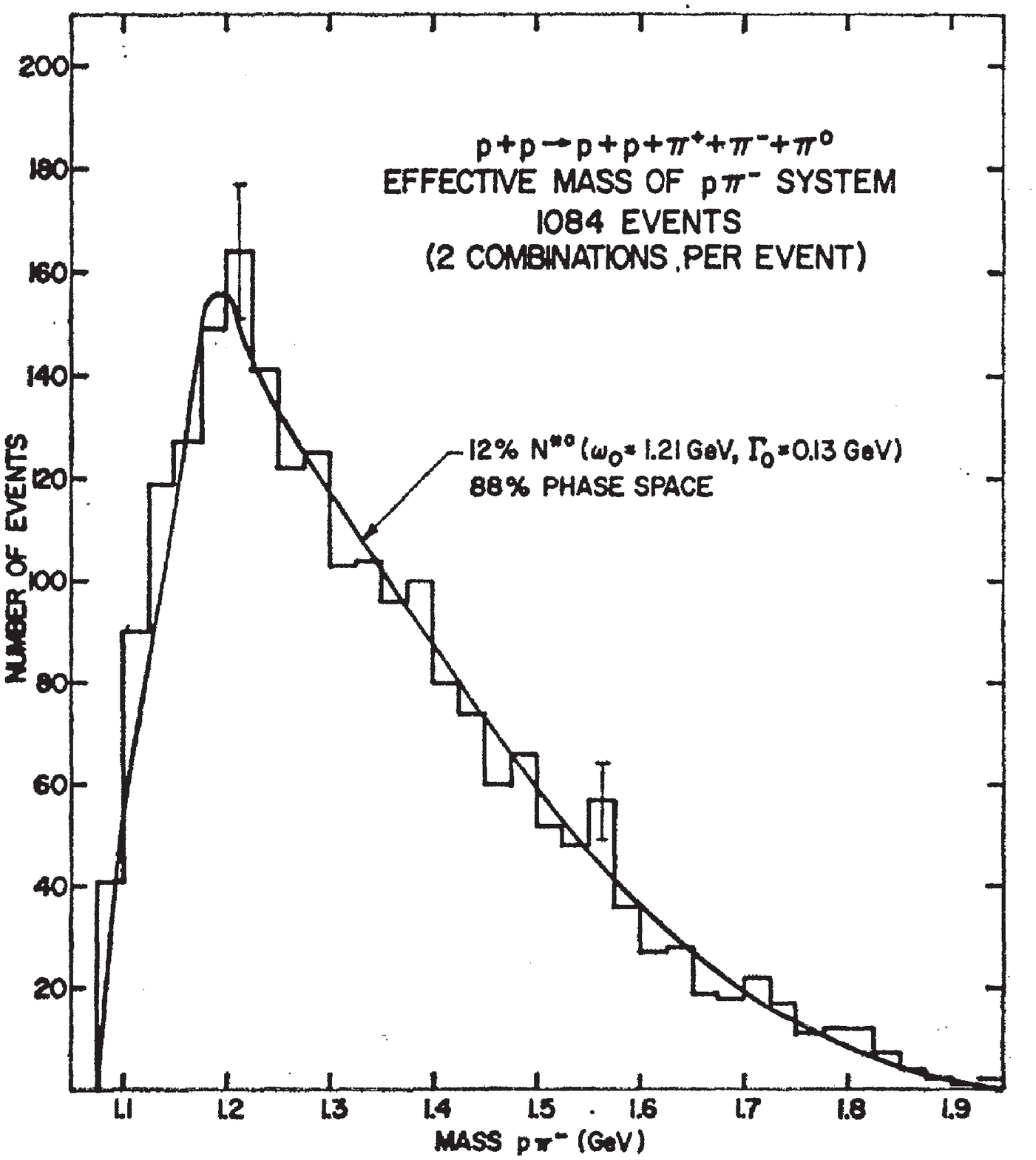}}}\\
\subfigure[]{\fbox{\includegraphics[width=0.4\textwidth, height=0.3\textwidth]{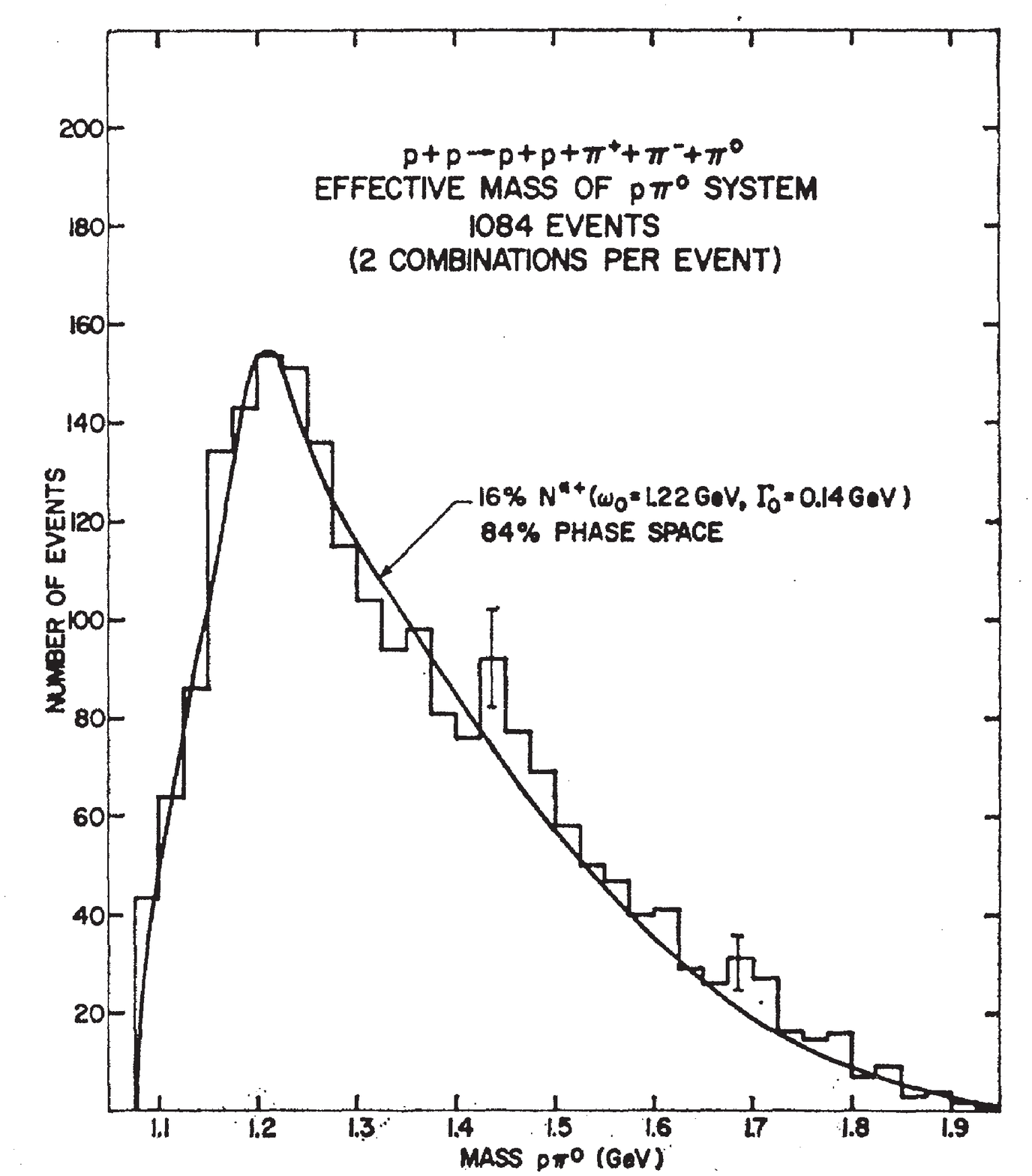}}}
}
\caption{Experimental data, the $pp \rightarrow pp \pi^{+}\pi^{-}\pi^{0}$ reaction at $5\mathrm{~GeV/c}$ proton beam momentum,
 invariant masses for different subsystems, taken from \cite{3pichargedDATA2}.
Description of lines in original papers.
Strong signal from the $\Delta(1232)$ seen in $p\pi$ invariant masses.
}
\label{image_Fig23_3pi0DATA2_PhysRev.161.1387.eps}
\end{sidewaysfigure}

\begin{sidewaysfigure}
\centering
{
\fbox{\includegraphics[width=12cm, height=20cm,angle=270]{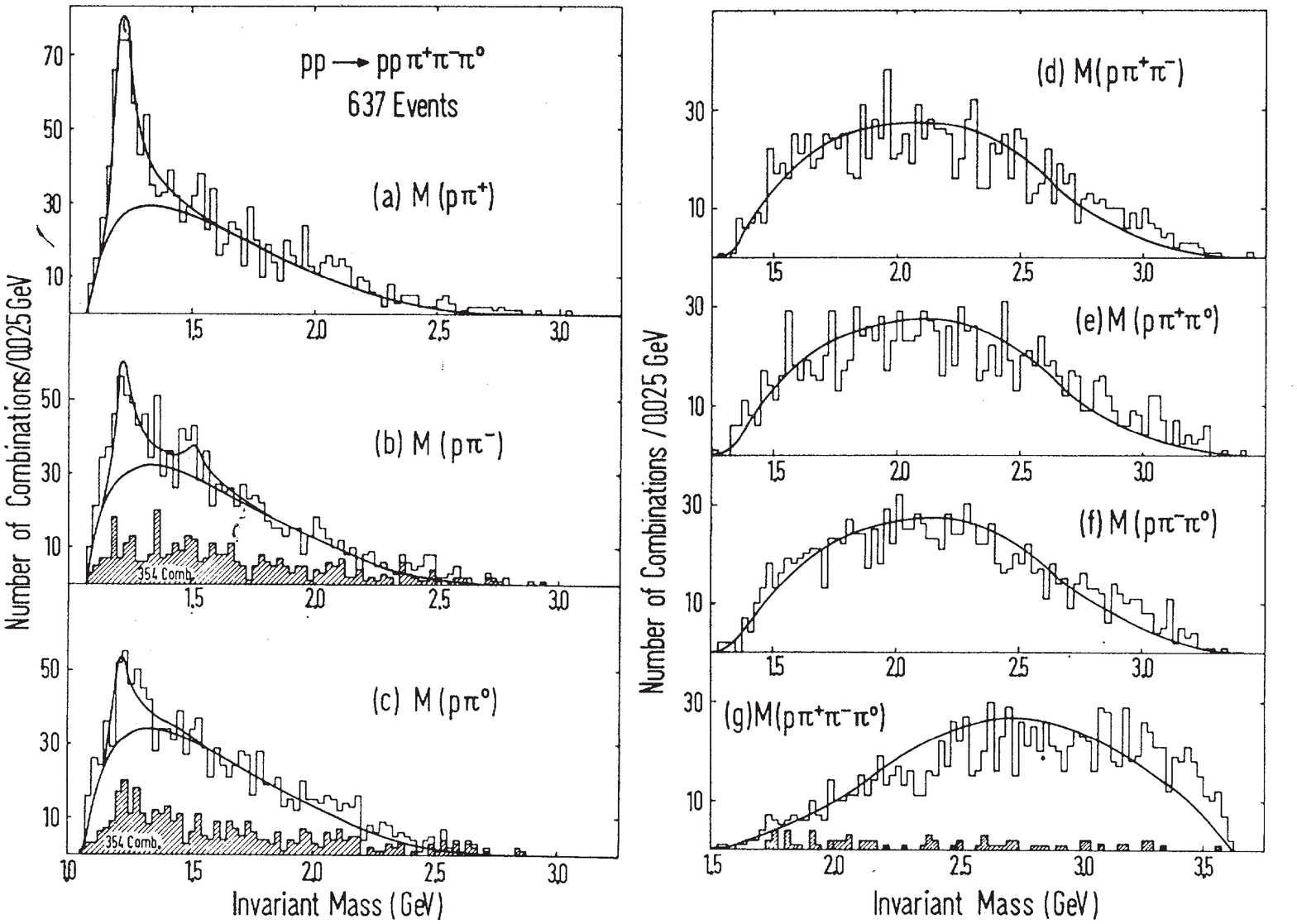}}
}
\caption{Experimental data, the $pp \rightarrow pp \pi^{+}\pi^{-}\pi^{0}$ reaction at $10\mathrm{~GeV/c}$ proton beam momentum, invariant masses for different subsystems,
 taken from \cite{3pichargedDATA3}. Description of lines in original papers.
Strong signal from the $\Delta(1232)$ seen in $p\pi$ invariant masses.
}
\label{image_Fig14_3pi0DATA3_PhysRev.174.1638.eps}
\end{sidewaysfigure}

Such a studies, in a simplified form, were performed for the case of the $pp \rightarrow pp \pi^{+}\pi^{-}\pi^{0}$ reaction \cite{3pichargedDATA2,3pichargedDATA1,3pichargedDATA3}, for much higher beam momenta $5\mathrm{~GeV/c}$ (Fig.~\ref{image_Fig23_3pi0DATA2_PhysRev.161.1387.eps}),
$5.5\mathrm{~GeV/c}$ (Fig.~\ref{image_Fig24_3pi0DATA1_PhysRev.154.1284_Resonances.eps}) and $10\mathrm{~GeV/c}$ (Fig.~\ref{image_Fig14_3pi0DATA3_PhysRev.174.1638.eps}).
The different invariant mass systems were considered, in all of the cases strong signal from the $\Delta(1232)$ was seen in $p\pi$ invariant masses.
Also an evidence for $N^{*}(1440)$ signal was seen in $p\pi^{+}\pi^{-}$ and $p\pi^{+}\pi^{0}$ invariant masses for the case of $5.5\mathrm{~GeV/c}$ beam momentum.

From a pure theoretical point of view,
there exist no microscopic model for the three pion production in contrast to the $NN \rightarrow NN \pi \pi$ reactions 
where complete microscopic model based on the excitations and decays of various baryon resonances exists \cite{ValenciaModel}.

In addition, in case of the reaction at $3.35\mathrm{~GeV/c}$ beam momentum (beam kinetic energy $2.541~\mathrm{~GeV}$) one can calculate the De~Broglie wavelength $\lambda$ of the incoming proton
\cite{QMDavidov,QMSchiff}:

\begin{equation}
 \lambda = \frac{h}{p} = 0.3701 \mathrm{~fm}
\end{equation}

where $h$ is the Planck's constant and $p$ is a particle momentum.

When one compares this value with the proton diameter \cite{ProtonRadius}:

\begin{equation}
1.754 \pm 0.014 \mathrm{~fm}
\end{equation}

One concludes that the $\lambda$ is more than four time smaller i.e. the incoming proton ``feels'' the inner structure of the target proton.
One can also compare this value with the range of the gluon induced interaction (in case of $\eta'$-nucleon interaction), estimated by the two-gluon effective potential,
which is in order of $\sim 0.3\mathrm{~fm}$ \cite{EtaPrimeGluon}.
Probing such small distances, the quark-gluon degrees of freedom may play a significant role in the production dynamics.

This fact will make the difficulty in the interpretation of the particle interactions by common  existing microscopic models \cite{OBE1,OBE2,OBE3,OBE4,OBE5,OBE6,OBE7,OBELast}
 mimic interaction by exchange of various light mesons like $\pi, \eta, \rho, \omega$
and are more applicable for lower energies (below $\sim 2\mathrm{~GeV}$ beam kinetic energy, equivalent to $\sim 2.8\mathrm{~GeV/c}$ beam momentum) \cite{OBElimits}.
  
It might be more plausible to use rather the microscopic approaches based on the Quantum Chromo Dynamics (QCD) (excitations of quark-gluon degrees of freedom) \cite{QCD01,QCD02,HSD0,HSD,HSD1}
 for the future model of $3\pi$ production - which might be very difficult theoretical task. 

Taking all these facts into account, nowadays the experimental analysis of the $3\pi$ production should be concentrated on the extraction of the reaction dynamics in the
model independent way using only the basic principles like energy and momentum conservation.
This might be done is a systematic way by studying the invariant masses of the subsystems.
Such a approach is presented later in this work.

\bigskip
The dynamics of the $pp \rightarrow pp 3\pi^{0}$ reaction is not understood and it was never investigated in details,
this makes the reaction a very interesting object for studies itself and also for the following reasons: 

\medskip
In heavy ion collisions the multiple pion production offers a possibility to look at a properties of the baryon resonances in the nuclear matter \cite{BaryonsHI,BaryonsHI01,BaryonsHI02,BaryonsHI03,BaryonsHI04,BaryonsHI05,BaryonsHI06,BaryonsHI07,BaryonsHI08,BaryonsHI09}.
It is also well established that in these reactions pions are mostly produced by the baryon resonances excitations \cite{BaryonsHI,BaryonsHI01,BaryonsHI02}. 

\noindent As shown above, the physics of $pp \rightarrow pp 3\pi^{0}$ at $T=2.54\mathrm{~GeV}$ is very closely related to the baryon resonances.
There is enough energy to excite various baryon resonances - specially $\Delta(1232)$ and $N^{*}(1440)$  (seen Table~\ref{tab:R1R1Baryons} on page \pageref{tab:R1R1Baryons} and Fig.~\ref{image_BaryonsCS2.eps} on page \pageref{image_BaryonsCS2.eps}).
One can study the dynamics of the produced baryon resonances
in similar way like it was attempted for the $pp \rightarrow pp \pi^{+}\pi^{-}\pi^{0}$ reaction \cite{3pichargedDATA2,3pichargedDATA1,3pichargedDATA3}.
Using this elementary reaction one can also get the spectroscopic informations about baryon resonances states \cite{PDG2008,BaryonsBook1, BaryonsBook2}.
Like in the heavy ion collisions in nuclear medium \cite{BaryonsHI,BaryonsHI01,BaryonsHI02,BaryonsHI03,BaryonsHI04,BaryonsHI05,BaryonsHI06,BaryonsHI07,BaryonsHI08,BaryonsHI09},
one may think about the multipion spectroscopy -- a tool to directly aces the properties of baryon resonances.
\medskip

The knowledge about the reaction is very important for transport codes like BUU/HSD~\cite{BUU1,BUU2,HSD,HSD0, HSD1}, INCL~\cite{INCL}, QMD~\cite{QMD1,QMD2} used intensively in the modeling of
the nucleon--nucleus or nucleus--nucleus  interactions, since they use the elementary reactions as an input for their calculations.
The pions are the most abundantly produced mesons in these reactions and they deliver the information about the first stage of the reaction.
 All of the models fail in the description of the pion spectra \cite{BUUpions,INCLpions,QMDpions}.
Inclusion of the new $pp \rightarrow pp 3\pi^{0}$ reaction channel might help in the better understanding of the pion dynamics in these models. 

\medskip
It is essential also to know the cross section 
and the dynamics of the $pp \rightarrow pp 3\pi^{0}$ reaction, specially in the beam kinetic energy range $1.8-2.8\mathrm{~GeV}$ (see Fig.~\ref{image_pp3pi0Mesons.eps}). 
It forms one of the most severe background for $\omega(782),\eta'(958),\Phi(1020)$ mesons:

\begin{itemize}
 \item hadronic decays e.g. $\eta' \rightarrow 3\pi^{0}$ - isospin forbidden, $\omega \rightarrow 3\pi^{0}$ - $C$ parity forbidden
 \item leptonic and semileptonic decays also, since the $\pi^{0}$ can undergo Dalitz decay  $\pi^{0} \rightarrow e^{+}e^{-}\gamma$ e.g. $\omega \rightarrow e^{+}e^{-}$, $\Phi \rightarrow \pi^{0} e^{+}e^{-}$   
\end{itemize}
 
\noindent These decays are of special interest of
many collaborations like WASA-at-COSY~\cite{proposal,WASAatCOSY} or HADES~\cite{HADES} since using them one can study the symmetries (e.g. isospin symmetry) and symmetry breaking ($C$, $CP$ parity).


\myFrameFigure{Fig24_3pi0DATA1_PhysRev.154.1284_Resonances.eps}{Experimental data, the $pp \rightarrow pp \pi^{+}\pi^{-}\pi^{0}$ reaction at $5.5\mathrm{~GeV/c}$ proton beam momentum,
 invariant masses for different subsystems, taken from \cite{3pichargedDATA1}. Description of lines in original papers.
Strong signal from the $\Delta(1232)$ seen in $p\pi$ invariant masses.
Evidence for $N^{*}(1440)$ signal seen in $p\pi^{+}\pi^{-}$ and $p\pi^{+}\pi^{0}$ invariant masses. 
}{charged pions}

\newpage
\subsection{Choice of the Observables}
\label{sec:DefinitionOfVariables}

One considers reaction

\begin{equation}
 a + b \rightarrow 1 + 2 +3 +4 +5
\end{equation}

 with masses $m_{i}$ and momenta $\overrightarrow{p_{i}}$ in the center of mass frame, where total energy in the center of mass frame is $\sqrt{s}$.
The probability that the momentum of the $i^{th}$ particle will be in the range $d^{3}p_{i}$ can be expressed as \cite{byckling}:

\begin{equation}
  d^{15}\mathcal{P} = d^{15}\mathcal{V}\left|\mathcal{M}\right|^{2}
\label{eq:ProbPhSp1}
\end{equation}

where $\mathcal{M}$ states for the invariant matrix element for the process
and $d^{15}\mathcal{V}$ is the Lorentz invariant phase space element available for the reaction.

One can rewrite (Eq.~\ref{eq:ProbPhSp1}) to the following form:   

{
\footnotesize
\begin{equation}
 d^{15}\mathcal{P} = d^{3}p_{1}d^{3}p_{2}d^{3}p_{3}d^{3}p_{4}d^{3}p_{5} \frac{1}{32 E_{1}E_{2}E_{3}E_{4}E_{5} } \delta^3\left( \sum_{j=1}^{5} \overrightarrow{p_{j}}\right)\delta^{3}\left( \sum_{j=1}^{5} E_{j}-\sqrt{s}\right)\left|\mathcal{M}\right|^{2} 
\label{eq:ProbPhSp2}
\end{equation}
}

where $E_{i}=\sqrt{\overrightarrow{p_{i}}^{2}+m_{i}^{2}}$ are energies in the CM frame (units are chosen so the $c=1$).

The event distribution (Eq.~\ref{eq:ProbPhSp2}) could be expressed in terms of the invariant masses defined by:

\begin{equation}
 M^{2}_{i \ldots j} = \left(E_{i} + \ldots E_{j} \right)^{2} - \left(\overrightarrow{p_{i}} + \ldots \overrightarrow{p_{j}} \right)^{2} 
\end{equation}
 
The three particle invariant masses and the four particle invariant masses are connected with the two particle invariant masses by:
{
\footnotesize
\begin{eqnarray}
 M^{2}_{ijk} &=& M^{2}_{ij} + M^{2}_{ik} + M^{2}_{jk} - \left(m^{2}_{i} + m^{2}_{j} + m^{2}_{k}  \right)\\
M^{2}_{ijkl} &=& M^{2}_{ij} + M^{2}_{ik} + M^{2}_{il} + M^{2}_{jk} + M^{2}_{jl} + M^{2}_{kl} - 2 \left(m^{2}_{i} + m^{2}_{j} + m^{2}_{k} + m^{2}_{l} \right)  
\end{eqnarray}
}

Such a representation is also very convenient to analyze the reaction in terms of the resonances in the subsystems 
 (one assumes that $\mathcal{M}$ depends only on this invariant masses).

\myFrameSmallFigurer{Fig1_nyborg_65_phasespace_165432.eps}{Pentagon of nonplanar vectors lengths describes momentum conservation
for five particles. The pentagon is completely defined by the variables $p_{1}, p_{2}, p_{3}, p_{4}, p_{5}, p_{12}, p_{123}, \phi_{(12)3}, \phi_{(123)4}$
Due to the energy and momentum conservation $p_{5}$ could be eliminated. Taken from \cite{Nyborg5body}}{Definition}

One can integrate (Eq.~\ref{eq:ProbPhSp2}) over spatial orientations of the entire system.
This integration can be done assuming that the created system ``has forgotten'' about direction of the beam particle ($3$ dimensions less).
Next, one can integrate further, leaving explicit dependence only on $E_{1}, E_{2}, E_{3}, E_{4}, p_{12}, p_{123}, \phi_{(12)3}, \phi_{(123)4}$
(this can be done as system has to fulfill energy and momentum conservation, 4 dimensions less).
\newpage
Finally, one is left with $8$ dimensions only \cite{Nyborg5body}:

\begin{equation}
 d^{8}\mathcal{P} = \frac{\pi^{2}}{4} dE_{1} dE_{2} dE_{3} dE_{4} dp_{12} dp_{123} d\phi_{(12)3} d\phi_{(123)4}\left|\mathcal{M}\right|^{2}
\label{eq:ProbPhSp3}
\end{equation}

where $\overrightarrow{p_{12}}=\overrightarrow{p_{1}}+\overrightarrow{p_{2}}$, $\overrightarrow{p_{123}}=\overrightarrow{p_{1}}+\overrightarrow{p_{2}}+\overrightarrow{p_{3}}$, $\phi_{(12)3}$~-~angle between the surface defined by the $\overrightarrow{p_{1}},\overrightarrow{p_{2}}$ vectors and the $\overrightarrow{p_{3}}$ vector,
 $\phi_{(123)4}$~-~angle between the surface defined by the $\overrightarrow{p_{12}},\overrightarrow{p_{3}}$ vectors and the $\overrightarrow{p_{4}}$ vector ( see Fig.~\ref{image_Fig1_nyborg_65_phasespace_165432.eps}).
$p_{i}$ plays for the length of the vector $\overrightarrow{p_{i}}$.

\bigskip 
If one assumes that $\mathcal{M}$ does not depend on $\phi_{(12)3}, \phi_{(123)4}$ one may integrate over it and obtain \cite{Nyborg5body}:

\begin{equation}
 d^{6}\mathcal{P} = \pi^{4} dE_{1} dE_{2} dE_{3} dE_{4} dp_{12} dp_{123}\left|\mathcal{M}\right|^{2}
\label{eq:ProbPhSp4}
\end{equation}

\bigskip
In our case we are studying the reaction
\begin{equation}
p p \rightarrow \underbrace{p}_{3} \underbrace{p}_{4} \underbrace{\pi^{0}}_{1} \underbrace{\pi^{0}}_{2} \underbrace{\pi^{0}}_{5}
\end{equation}

\newpage
One would like to study the dynamics for this reaction:

\begin{itemize}
 \item The correlations between the pions\\
One would need to study the $M^{2}(\pi^{0}\pi^{0})$ i.e $M^{2}_{12}$ and $M^{2}_{25}$

\item The interactions between the protons and between the pions\\
One would need to study the $M^{2}(pp)$ and $M^{2}(3\pi^{0})$ i.e $M^{2}_{34}$ and $M^{2}_{1235}$

\item The possible resonances in proton pion and proton two pions systems\\
One would need to study the $M^{2}(p\pi^{0})$ and $M^{2}(p\pi^{0}\pi^{0})$ i.e $M^{2}_{45}$ and $M^{2}_{123}$ 
\end{itemize}

It would be also very convenient to study in addition the event distribution as function of the missing mass of the two protons

\begin{equation}
 MM_{pp}^{2} = \left(\sqrt{s}-E_{3}-E_{4}\right)^{2}+\left(\overrightarrow{p_{3}} + \overrightarrow{p_{4}}\right) = M^{2}_{125} 
\end{equation}

as a parameter, since the $MM_{pp}$ defines the maximal available kinetic energy for the $3\pi^{0}$ system in its rest frame ($Q_{3\pi^{0}}$) and simultaneously
the maximal available kinetic energy for the $pp$ system in its rest frame ($Q_{pp}^{Max}$):

\begin{eqnarray}
Q_{3\pi^{0}}^{Max} &=& MM_{pp} - m_{1}- m_{2}- m_{5} = MM_{pp} - 3 m_{\pi^{0}}\\
Q_{pp}^{Max} &=& \sqrt{s} - MM_{pp} - m_{3} - m_{4} = \sqrt{s} - MM_{pp} - 2 m_{p} 
\end{eqnarray}
where $m_{\pi^{0}}$~-~ mass of the $\pi^{0}$ meson, $m_{p}$~-~ mass of the proton.
\medskip

\noindent Selecting the missing mass one selects how the energy is distributed between the three pion system and a two proton system. 

\bigskip
It will be shown in Appendix~\ref{appendix:Kine5part} how to express the event distribution in appropriate variables.

\newpage ~
\thispagestyle{empty}
\emptydoublepage
\newpage

\section{Analysis of the experimental data}
\label{sec:AnalysisExperimental}
\thispagestyle{plain}

\myFrameHugeFigure{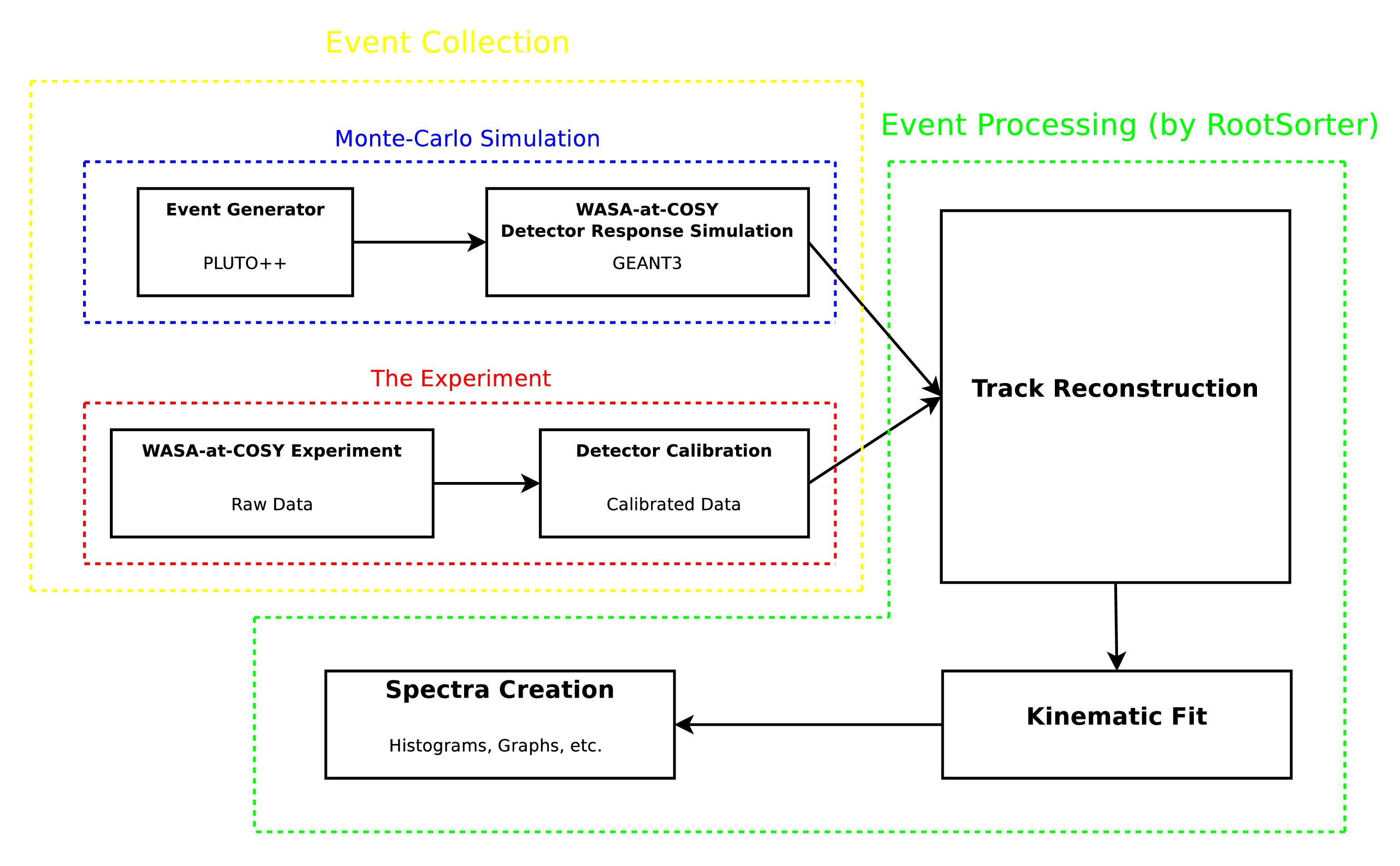}{Schematic view of events flow and the analysis chain. 
First the events are collected in the experiment (Section~\ref{subsec:detector}) and later calibrated (Appendix~\ref{appendix:DetectorCalibration}) or
simulated by the Monte-Carlo which includes the detector response (Appendix~\ref{appendix:wmc}).
Later the events (both from the experiment and Monte-Carlo) are processed by the same common procedure
 i.e. the tracks are built (Appendix~\ref{appendix:TrackReconstruction}) then they are fitted kinematically (Appendix~\ref{appendix:kfit}).}{Events flow.}

The purpose of the analysis is to obtain physical spectra from the collected data and validate it by the Monte-Carlo simulation.
The schematic view of the events flow and analysis chain is presented in \myImgRef{EventReconstruction2}.
One can divided this process into the two phases:

\begin{enumerate}
 \item The events collection phase

\begin{itemize}
 \item The raw data collection\\
The data are collected in the WASA-at-COSY experiment\newline (see Section~\ref{subsec:detector}), later the raw data have to be calibrated\newline (see Appendix~\ref{appendix:DetectorCalibration}).

 \item The Monte-Carlo simulation\\
 First the events are generated and later the detector response is simulated (see Appendix~\ref{appendix:wmc}).

\end{itemize}

\item The data processing phase\\
This part is done using the RootSorter analysis framework \cite{RootSorter} which is based on ROOT \cite{Root}.
First from the collected data the tracks are build, which correspond to the physical particles (see Appendix~\ref{appendix:TrackReconstruction}).
Later the statistical hypothesis test is done on tracks by the kinematic fit procedure (see Appendix~\ref{appendix:kfit}). 

\end{enumerate}

\subsection{The Experimental Conditions}

\begin{figure}[ht!bp]
\centering
{
\subfigure[]{\fbox{\includegraphics[width=0.7\textwidth]{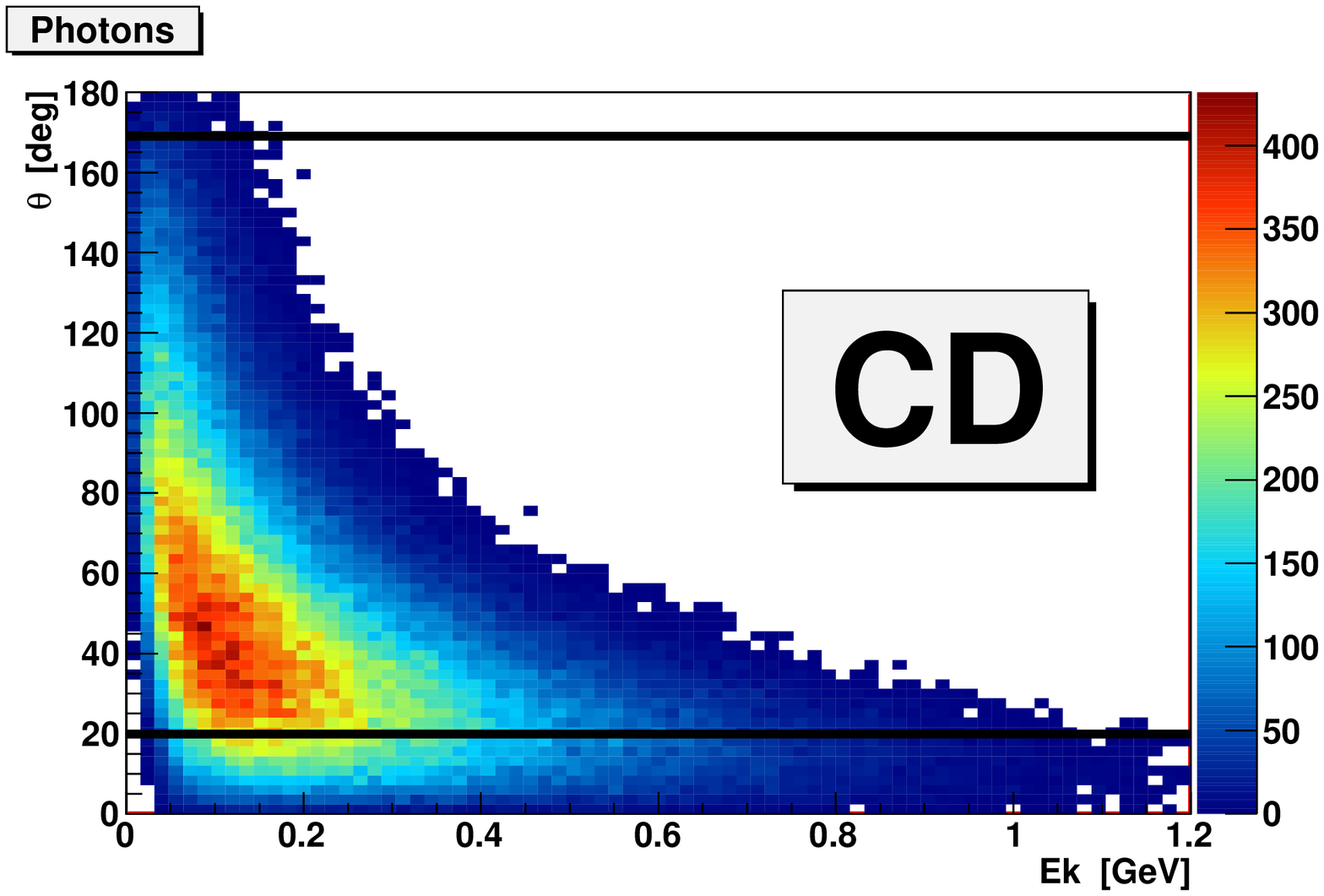}} \label{fig:3pi0PhSpKinematicsPhotons}}\\
\subfigure[]{\fbox{\includegraphics[width=0.7\textwidth]{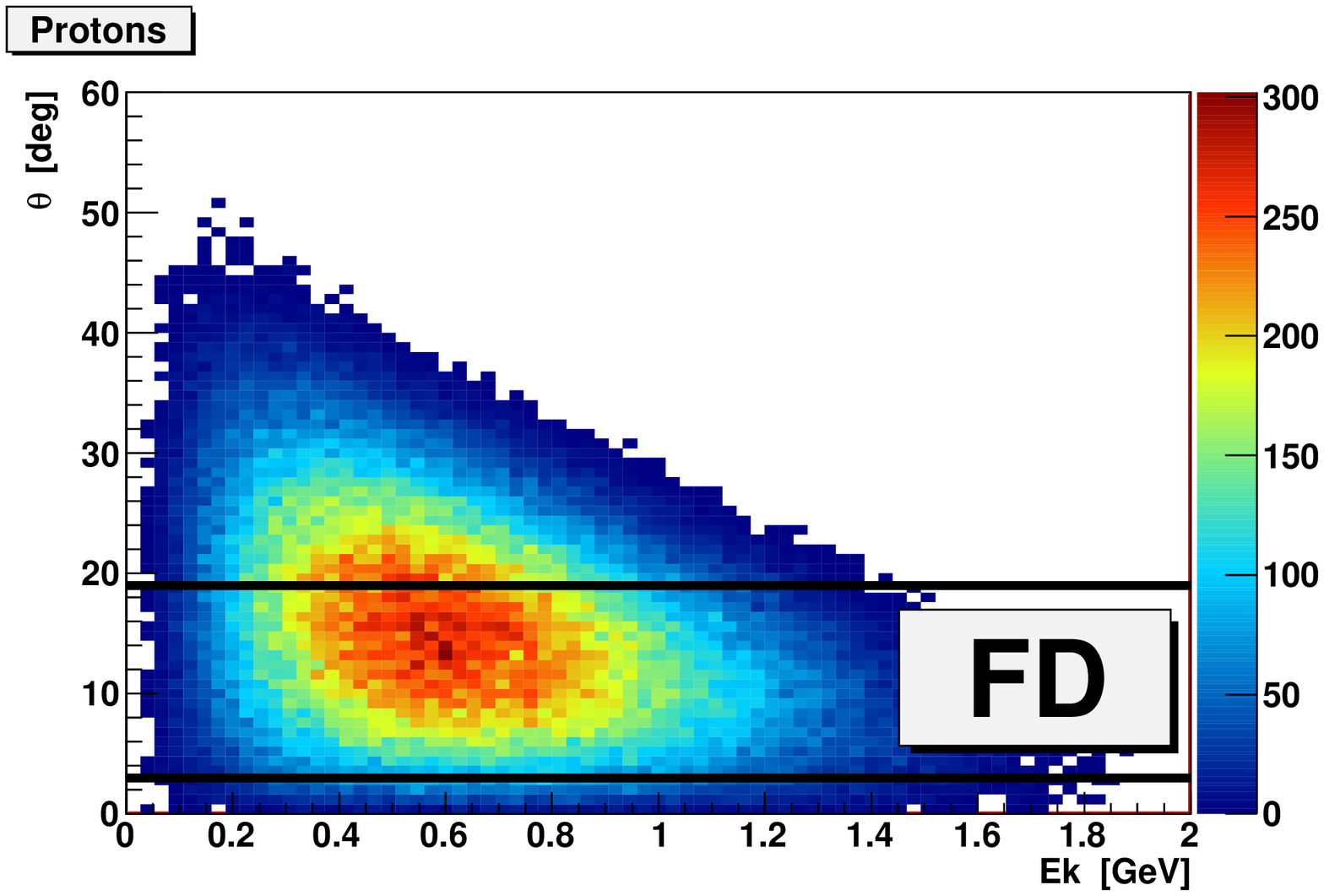}} \label{fig:3pi0PhSpKinematicsProtons}}
}

\caption{The kinematics of the  $pp \rightarrow pp 3\pi^{0} \rightarrow pp 6\gamma$ reaction at $3.35\mathrm{GeV/c}$ - Monte Carlo simulation homogeneously and isotropically populated phase space. Fig.~\ref{fig:3pi0PhSpKinematicsPhotons} 
photon scattering angle versus kinetic energy, geometrical boundaries of the CD detector marked. Fig.~\ref{fig:3pi0PhSpKinematicsProtons}
proton scattering angle versus kinetic energy, geometrical boundaries of the FD detector marked. }
\label{fig:3pi0PhSpKinematics}
\end{figure}

The aim of the experiment was to measure the $pp \rightarrow pp 3\pi^{0} \rightarrow pp 6\gamma$ reaction at proton kinetic energy
$T=2.541\mathrm{~GeV}$, which corresponds to the momentum of $3.350\mathrm{~GeV/c}$ and excess energy $Q=598\mathrm{~MeV}$
(center of mass energy $\sqrt{s}=2.879\mathrm{~GeV/c^{2}}$). 
The Monte Carlo simulation, based on homogeneously and isotropically populated phase space (see Appendix~\ref{appendix:wmc}),
was performed to study the kinematics of the reaction. The results (Fig.~\ref{fig:3pi0PhSpKinematics}) show that most of
the protons are going to the FD detector and most of the photons to the CD detector.
Assuming that two protons are registered in FD detector and six photons in CD detector,using Monte-Carlo simulation, the geometrical acceptance of
\begin{equation}
 Geom.Acc. = 14.24\%
\label{eq:GeomAcc3pi0}
\end{equation}

has been obtained.     

The experimental data were collected during one week run in May 2008, 
using pellet proton target and COSY phase space cooled proton beam at incident momentum of $3.35\mathrm{GeV/c}$. 
The data were collected under following trigger condition:

\begin{itemize}
 \item more then one hit in the FD detector - time overlap with FWC and FTH detectors
(to ensure that the hits come from the same interaction)

 \item more then one hit in the third layer of the FRH detector
(this ensures that one has at least two high energetic particles)

 \item more then one neutral group(veto on overlapped element of PS detector) in the calorimeter with low threshold on the group ($\sim 50\mathrm{~MeV}$)
(this implies that at least two photons are registered)

 \item veto on the signal ``at least one hit in PS detector''
(one ensures that the charged particles are excluded) 
\end{itemize}

the trigger was not prescaled.

\subsection{The Event Selection}

\begin{figure}[ht!bp]
\centering
{
\subfigure[Experimental Data.]{\fbox{\includegraphics[width=0.7\textwidth]{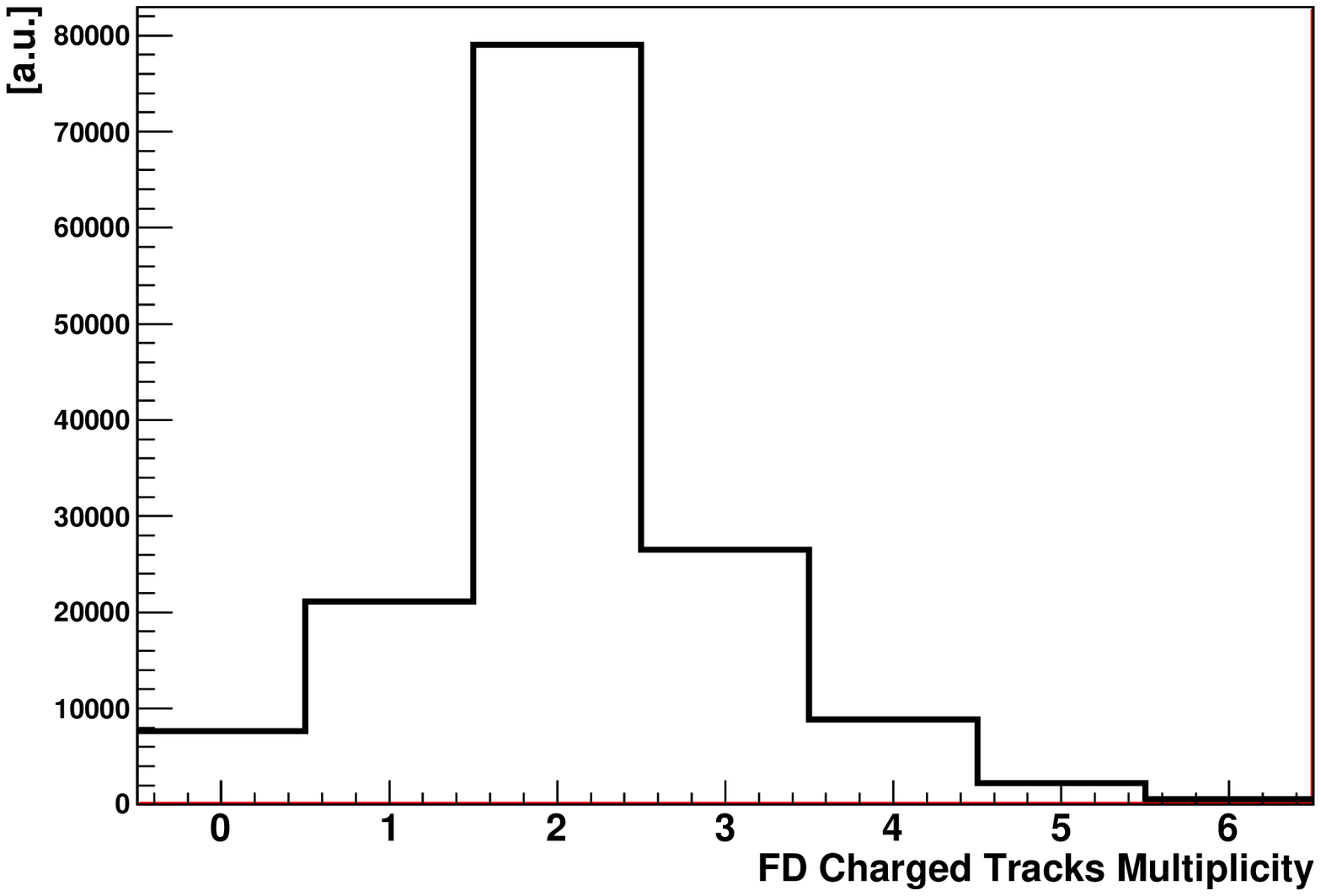}} \label{fig:FDChargedMultiplicityData}}\\
\subfigure[Monte-Carlo simulation]{\fbox{\includegraphics[width=0.7\textwidth]{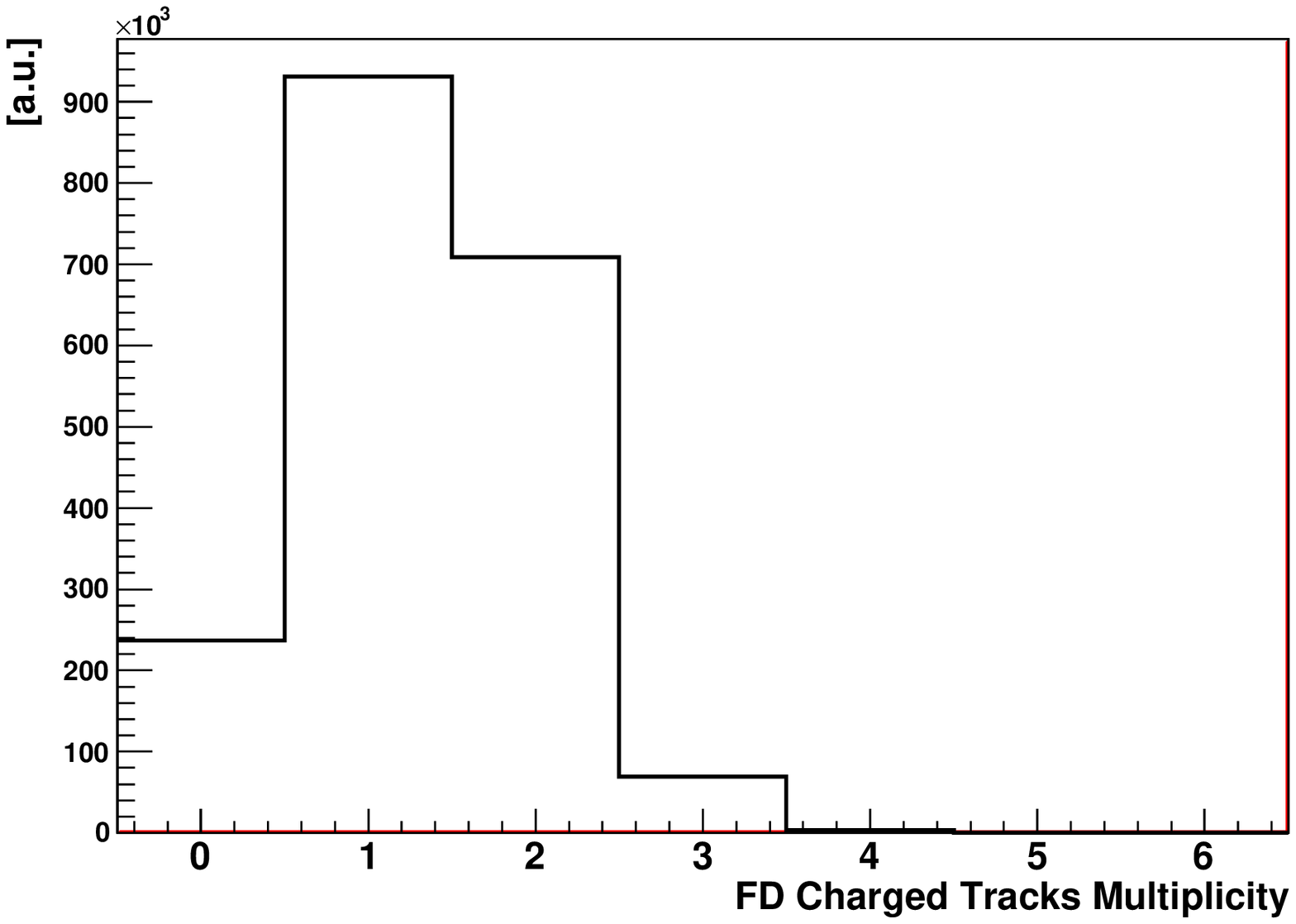}} \label{fig:FDChargedMultiplicityMC}}
}
\caption{Charged Tracks multiplicity FD detector.}
\label{fig:FDChargedMultiplicity}
\end{figure}

The event selection was optimized to select from the data $p p 3\pi^{0}[6\gamma]$ final state.
First the FD tracks (see Appendix~\ref{appendix:TrackReconstructionFD}) with a minimum energy deposit of the track of $20\mathrm{~MeV}$ were selected to suppress detector noises.
After this selection charged tracks multiplicity in FD detector was compared with 
Monte-Carlo simulation $pp \rightarrow pp 3\pi^{0}$ assuming homogeneously and isotropically populated phase space (Fig.~\ref{fig:FDChargedMultiplicity}).

\begin{figure}[ht!bp]
\centering
{
\subfigure[Time of the first charged track in FD versus the second one.]{\fbox{\includegraphics[width=0.7\textwidth]{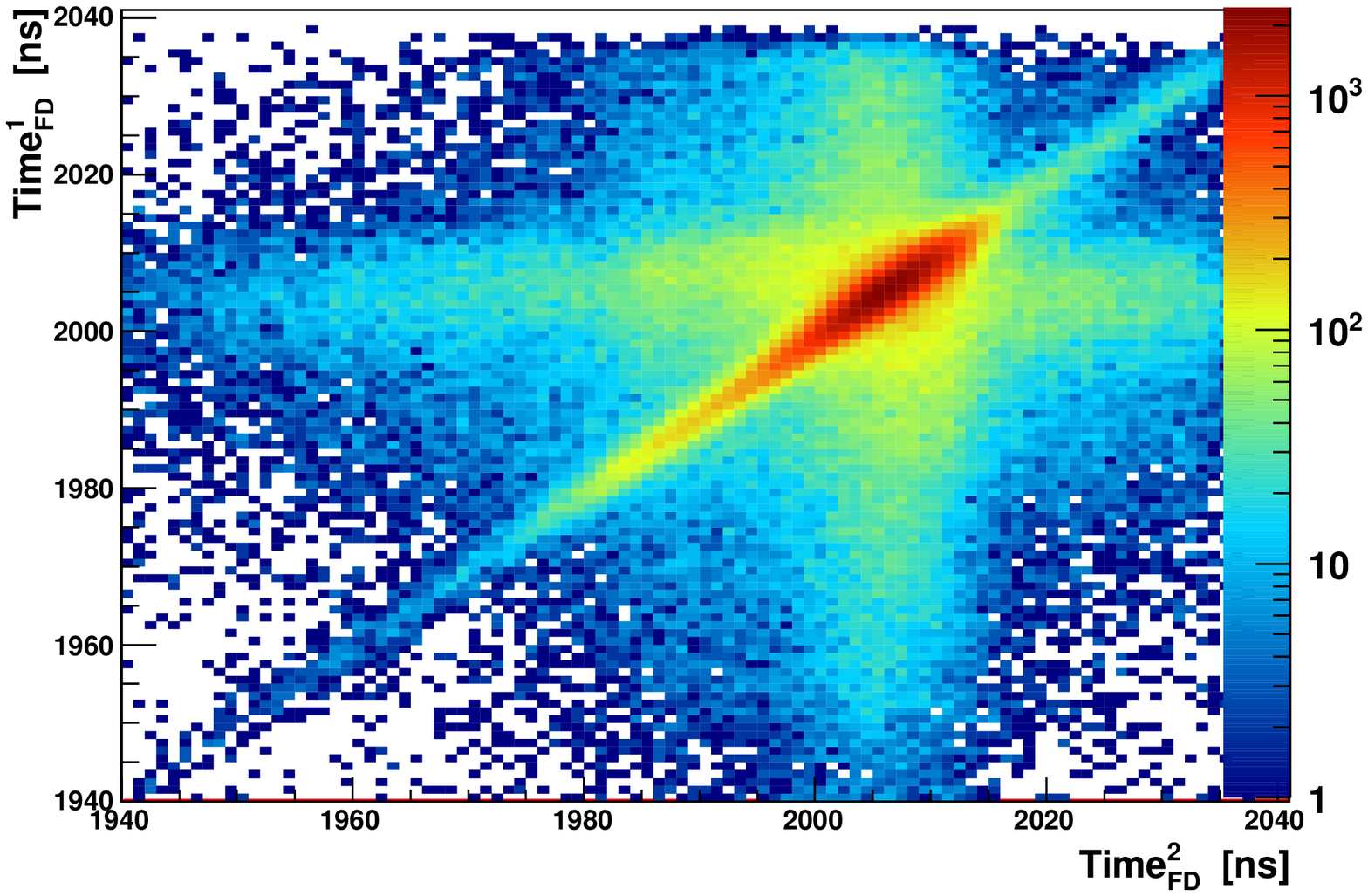}} \label{fig:TimeFDvsFD2D}}\\
\subfigure[Time difference between first charged track in FD and the second one.]{\fbox{\includegraphics[width=0.7\textwidth]{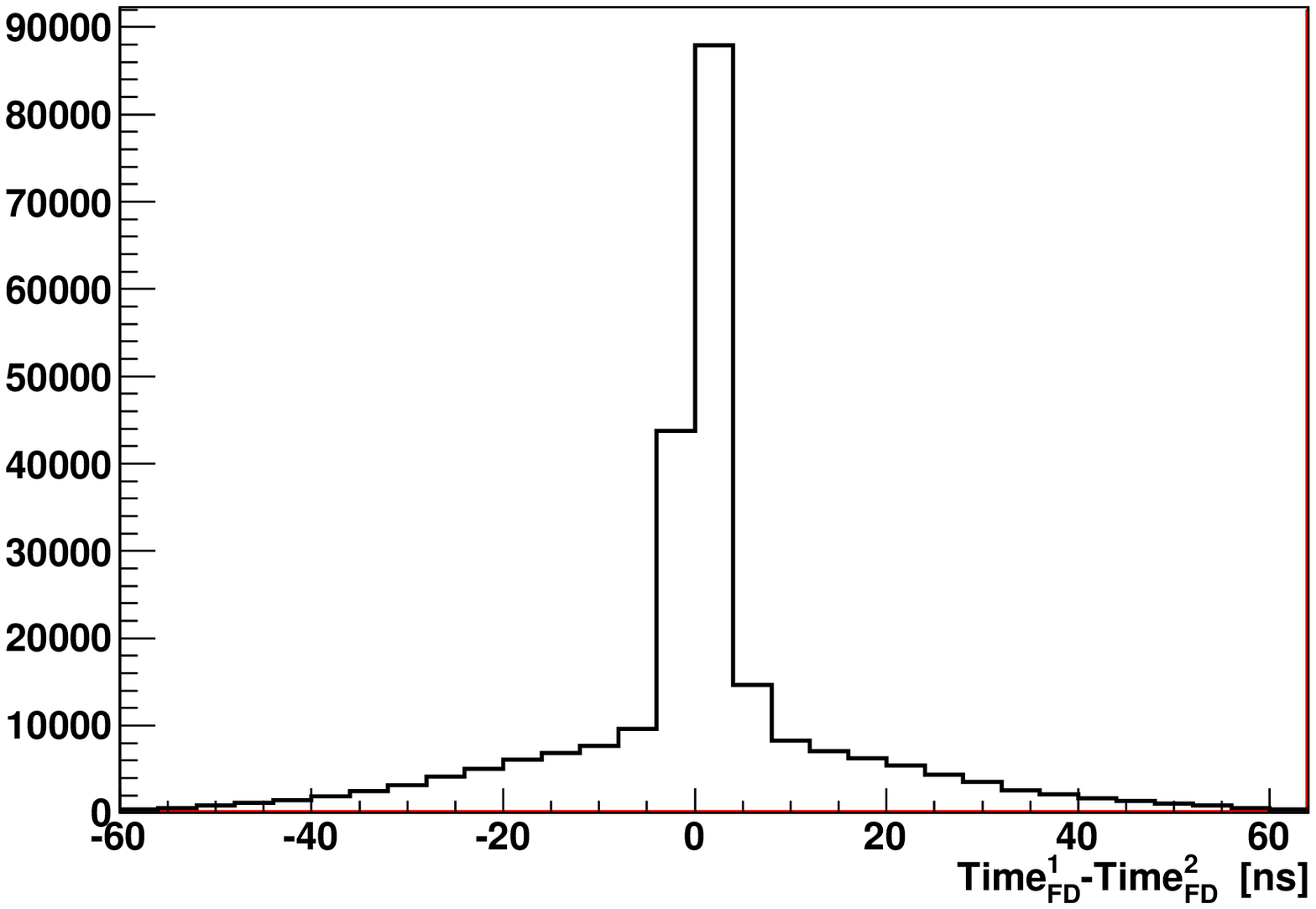}} \label{fig:TimeDiffFDFD}}
}
\caption{Time dependences charged tracks FD detector.}
\label{fig:TimeFDvsFD}
\end{figure}

\begin{figure}[ht!bp]
\centering
{
\subfigure[Time of the first charged track in FD versus the second one after the cut.]{\fbox{\includegraphics[width=0.7\textwidth]{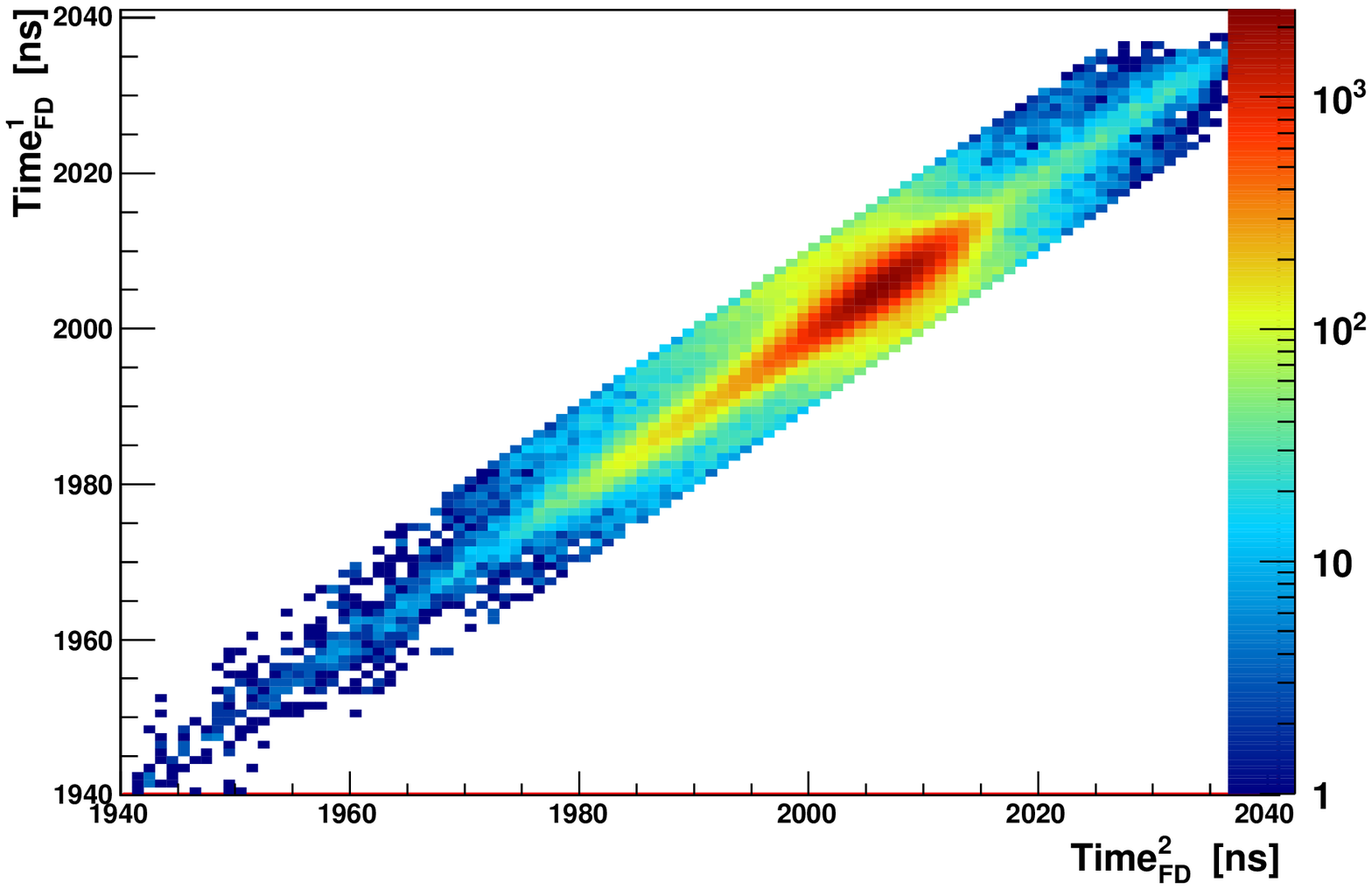}} \label{fig:TimeFDvsFD2Dcut}}\\
\subfigure[Time difference between first charged track in FD and the second one after the cut.]{\fbox{\includegraphics[width=0.7\textwidth]{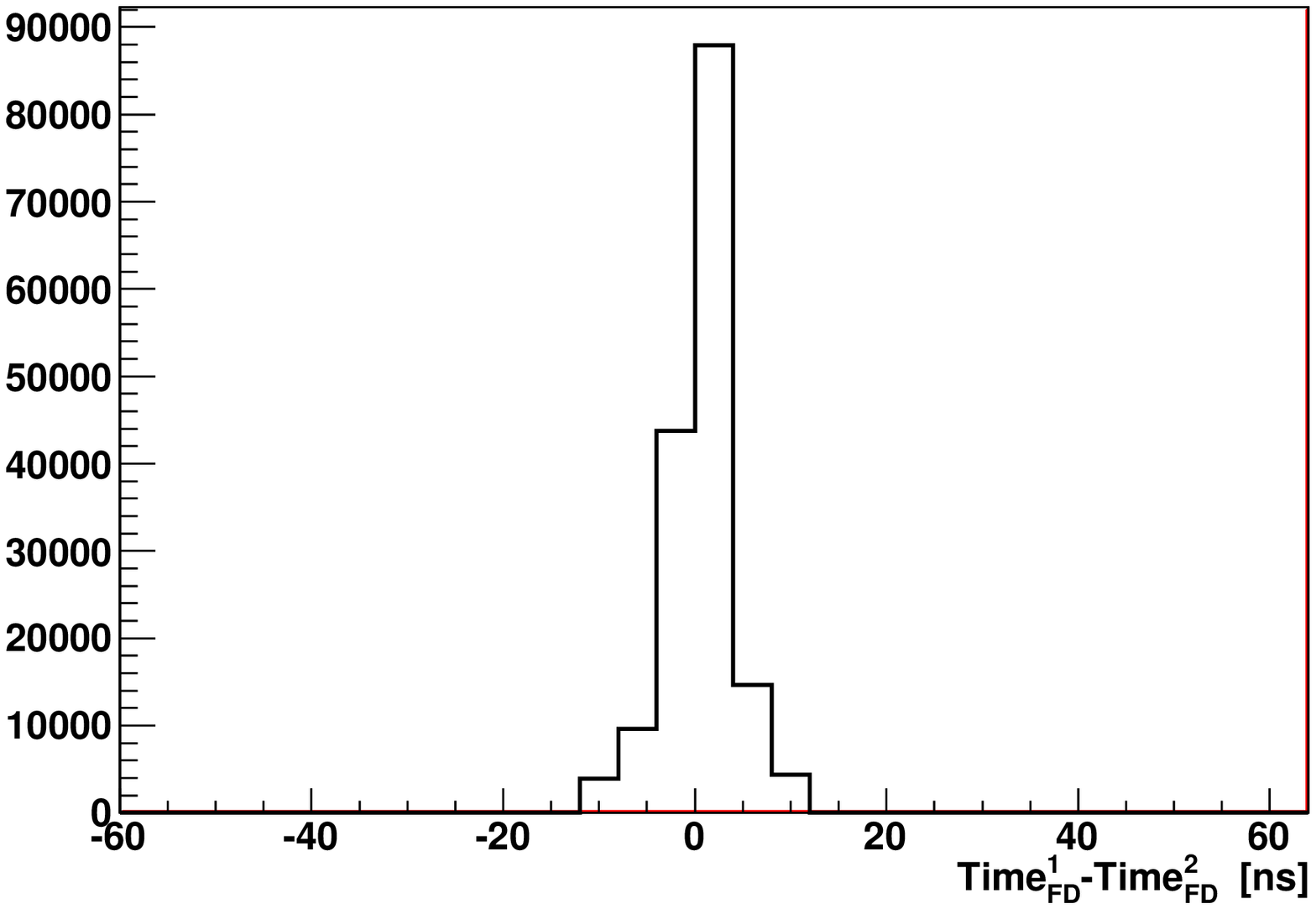}} \label{fig:TimeDiffFDFDcut}}
}
\caption{Time dependences charged tracks FD detector with the cut on the difference between two charged tracks.}
\label{fig:TimeFDvsFDcut}
\end{figure}

Next the events with two or more charged tracks tracks in FD detector were chosen, time correlation between the tracks pairs was checked (Fig.~\ref{fig:TimeFDvsFD}).
The cut on the time difference between two charged tracks was performed, the events in prompt peak were selected and accepted for Time~Difference~from~$-4\mathrm{~ns}$ to $10\mathrm{~ns}$) (Fig.~\ref{fig:TimeFDvsFDcut}), to create a pair of two charged tracks.  

%

\begin{sidewaysfigure}[t!bp]
\centering
{
\subfigure[Charged Tracks Multiplicity, Experimental Data.]{\fbox{\includegraphics[width=0.45\textwidth]{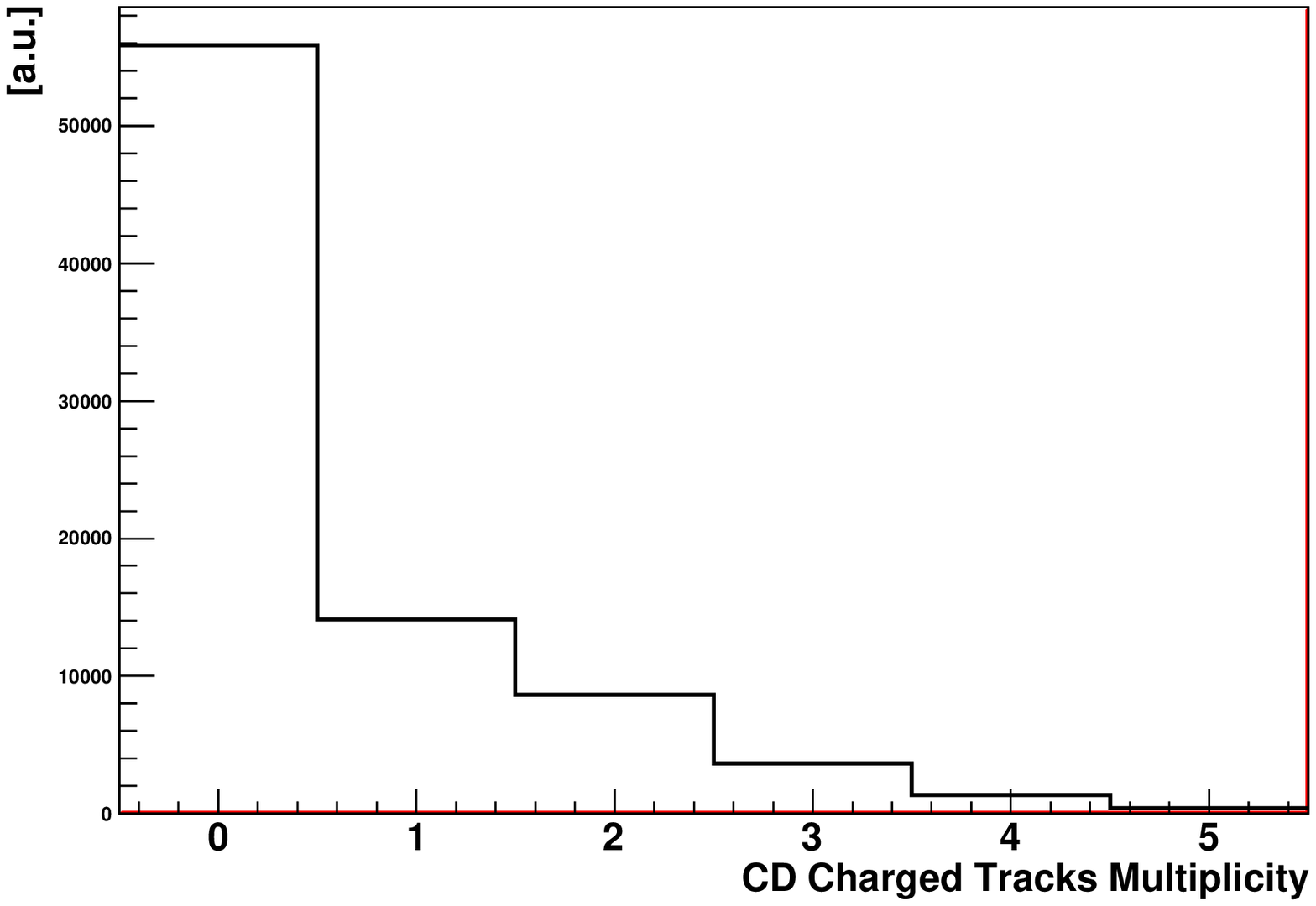}} \label{fig:CDnMulti}}\quad
\subfigure[Neutral Tracks Multiplicity, Experimental Data.]{\fbox{\includegraphics[width=0.45\textwidth]{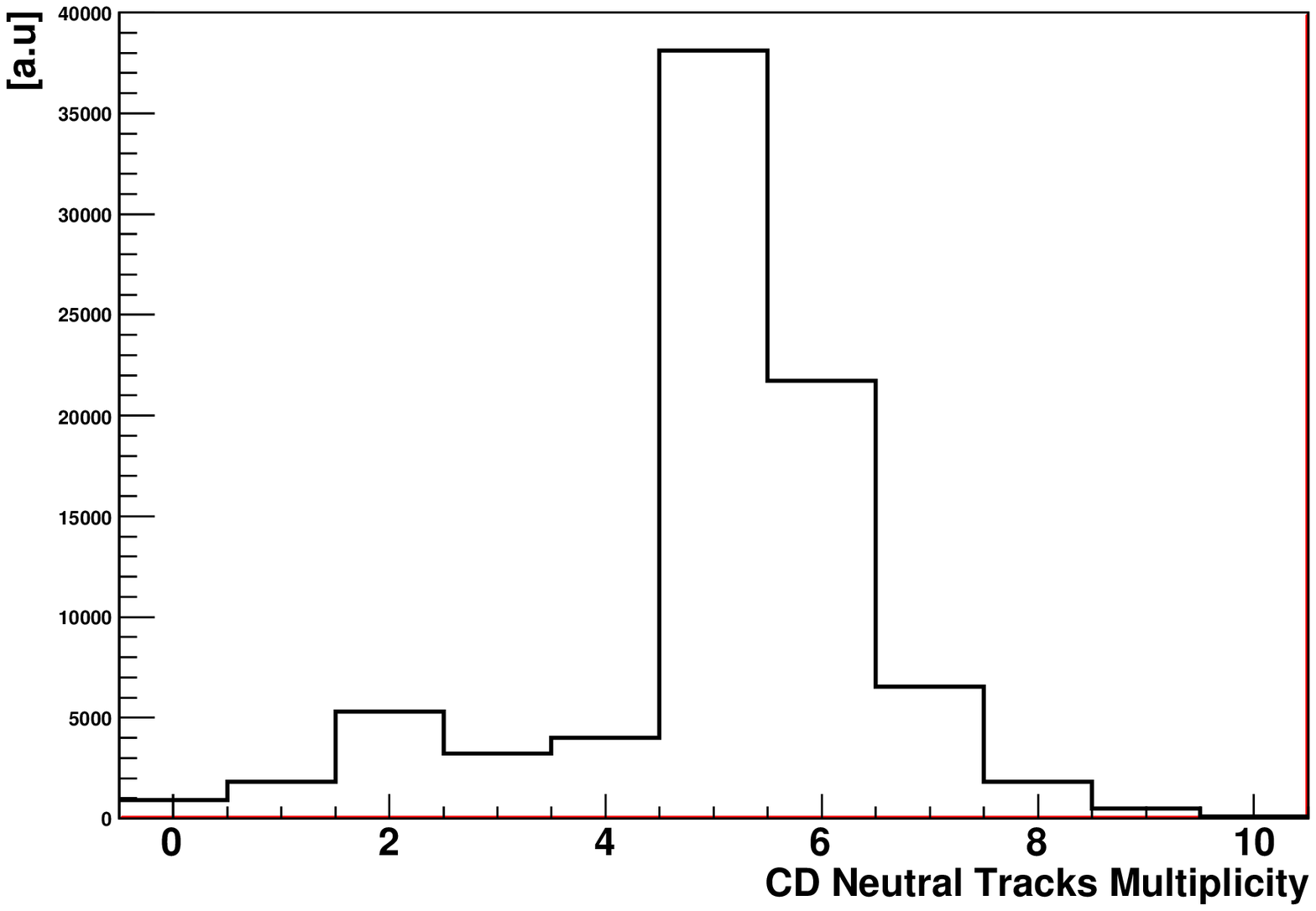}} \label{fig:CDcMulti}}\\
\subfigure[Charged Tracks Multiplicity, Monte-Carlo simulation.]{\fbox{\includegraphics[width=0.45\textwidth]{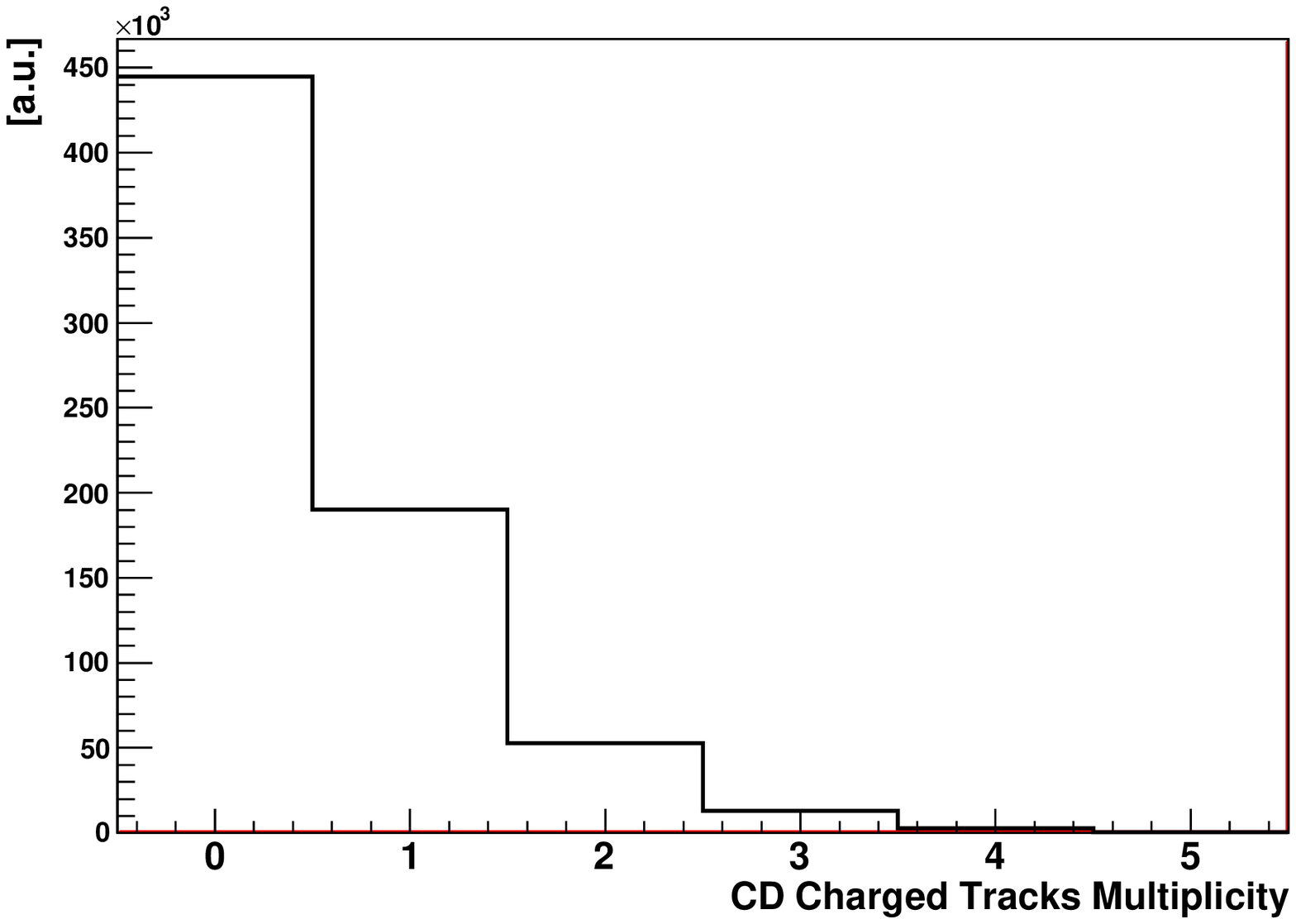}} \label{fig:CDnMultiMC}}\quad
\subfigure[Neutral Tracks Multiplicity, Monte-Carlo simulation.]{\fbox{\includegraphics[width=0.45\textwidth]{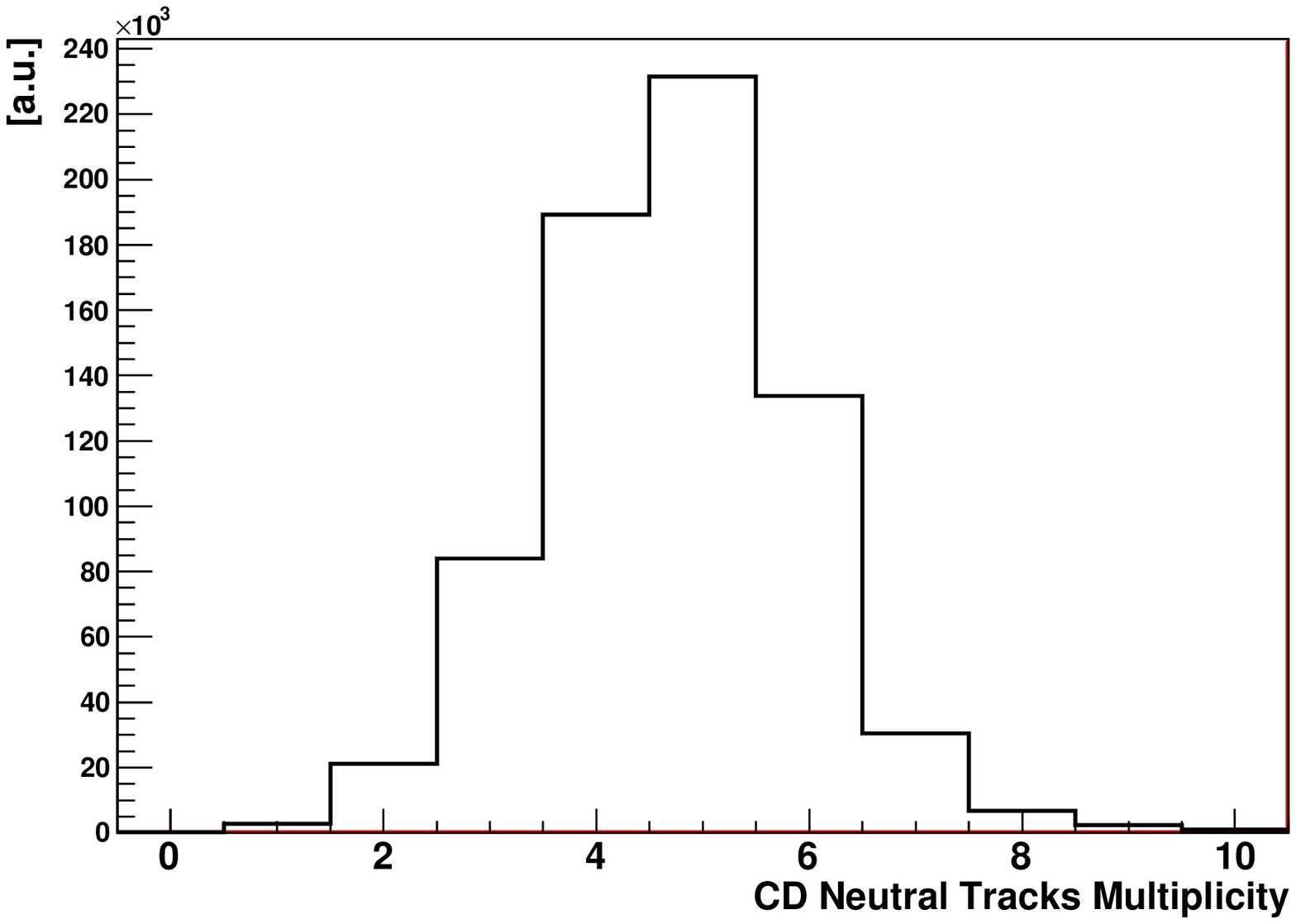}} \label{fig:CDcMultiMC}}
}
\caption{Tracks Multiplicity in CD detector. Only events without charged tracks in CD will be finally accepted.}
\label{fig:CDMulti}
\end{sidewaysfigure}

\begin{figure}[ht!bp]
\centering
{
\subfigure[Mean time of the two charged tracks in FD versus the time of the neutral track in CD.]{\fbox{\includegraphics[width=0.7\textwidth]{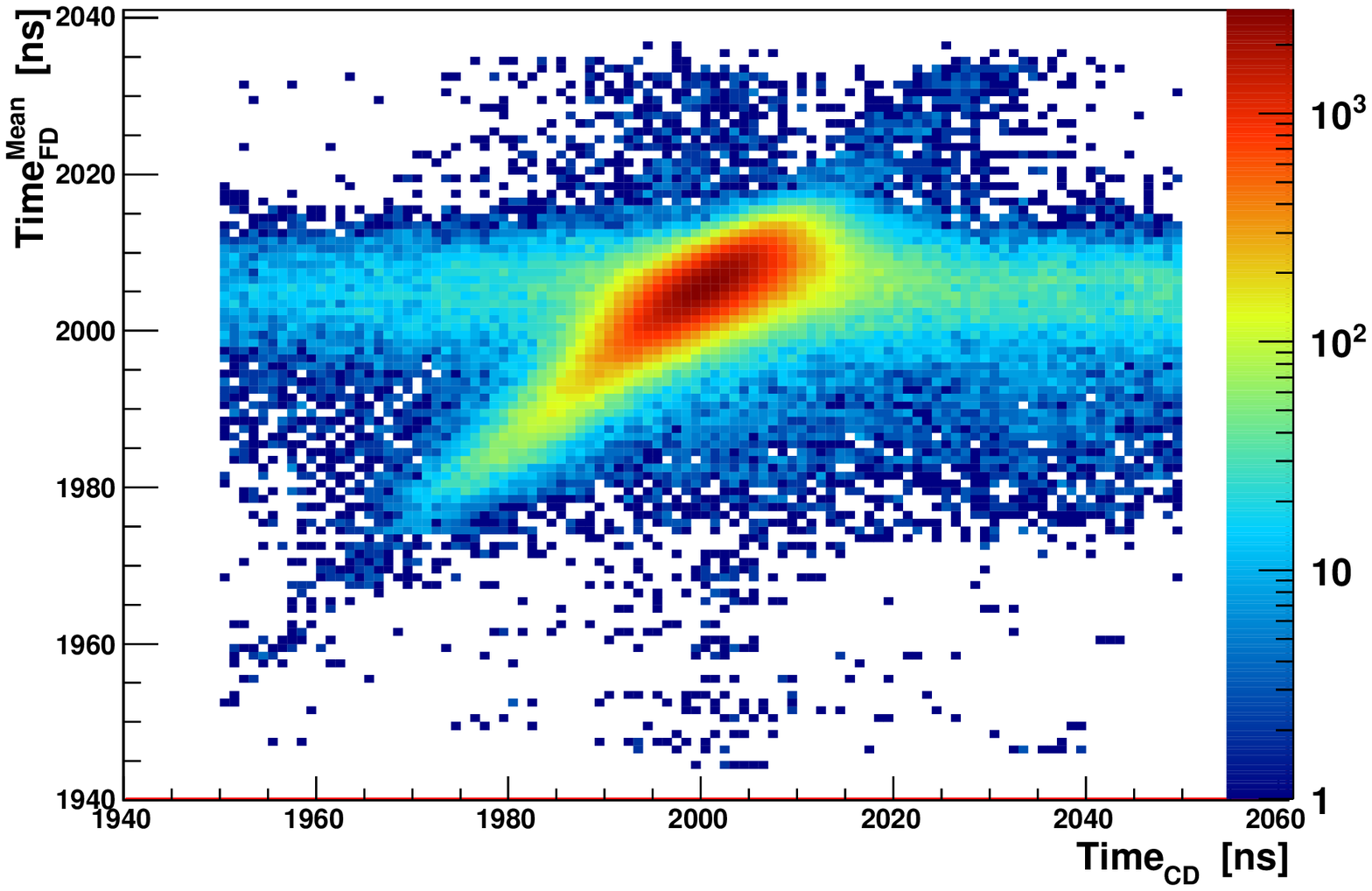}} \label{fig:TimeFDvsCD2D}}\\
\subfigure[Time difference mean time of the two charged tracks in FD and the time of the neutral track in CD.]{\fbox{\includegraphics[width=0.7\textwidth]{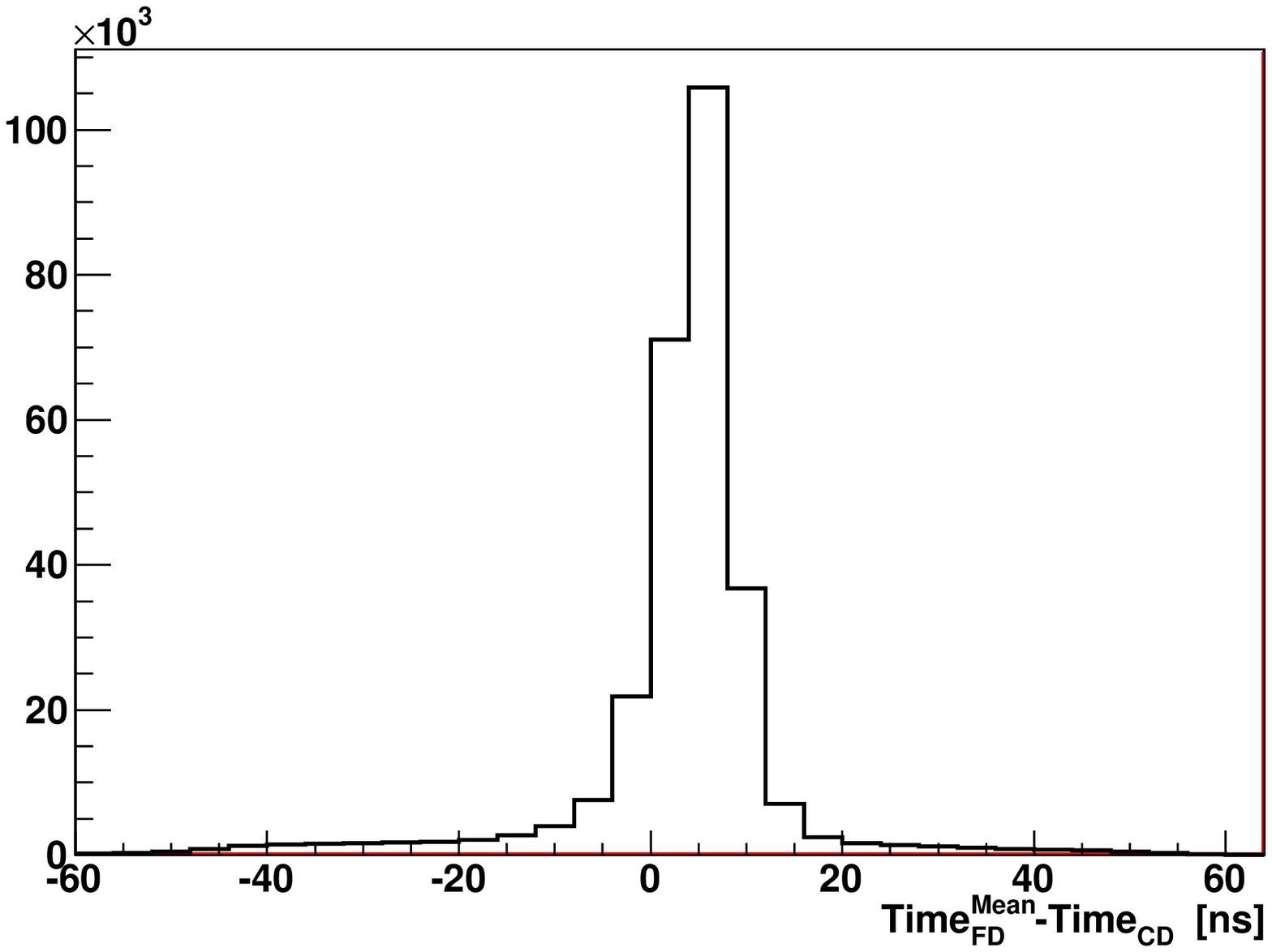}} \label{fig:TimeDiffFDCD}}
}
\caption{Time dependences between mean time of the two charged tracks in FD and the time of the neutral track in CD.}
\label{fig:TimeFDvsCD}
\end{figure}

\begin{figure}[ht!bp]
\centering
{
\subfigure[Mean time of the two charged tracks in FD versus the time of the neutral track in CD after the cut.]{\fbox{\includegraphics[width=0.7\textwidth]{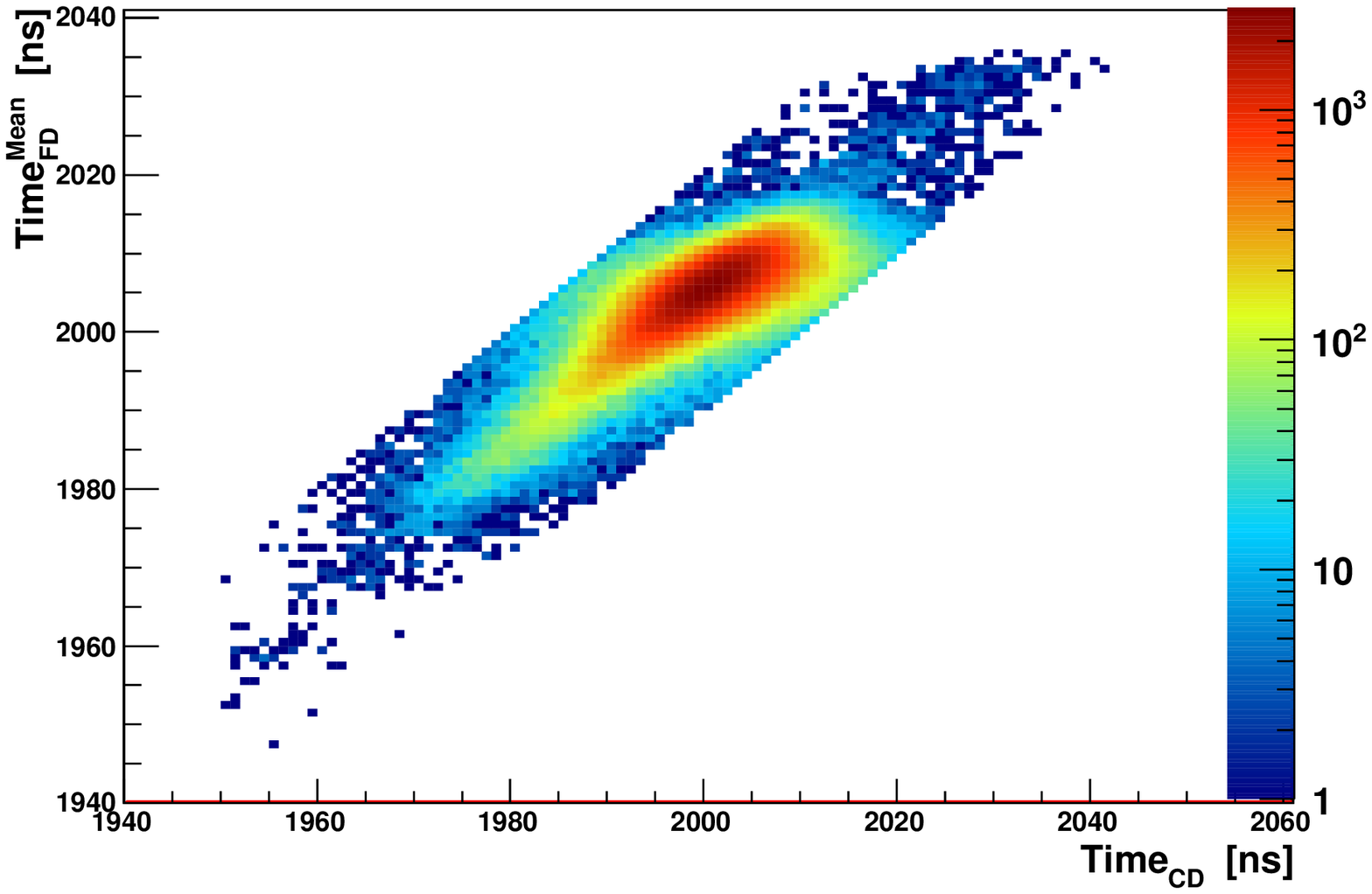}} \label{fig:TimeFDvsCD2Dcut}}\\
\subfigure[Time difference mean time of the two charged tracks in FD and the time of the neutral track in CD after the cut.]{\fbox{\includegraphics[width=0.7\textwidth]{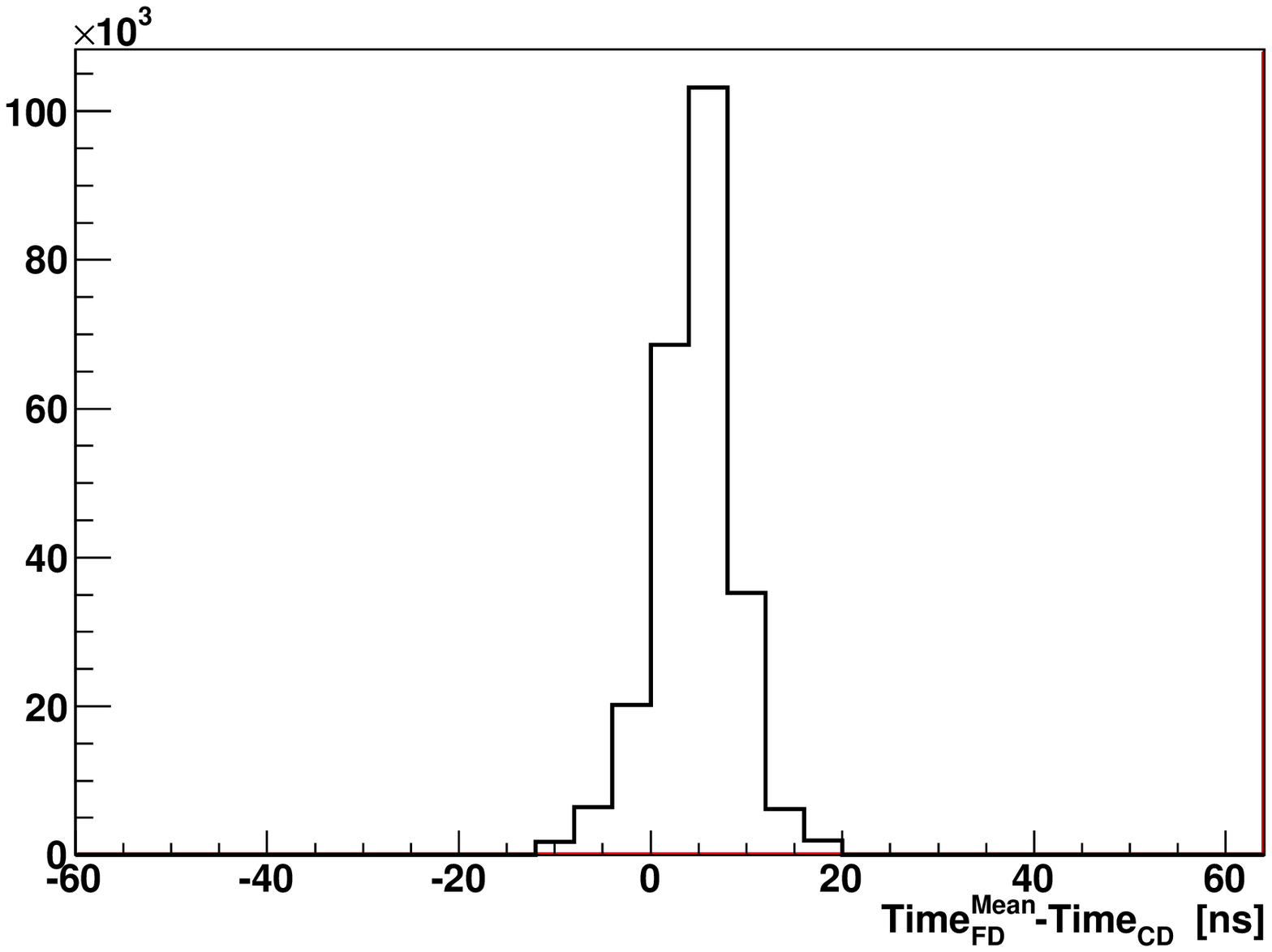}} \label{fig:TimeDiffFDCDcut}}
}
\caption{Time dependences between mean time of the two charged tracks in FD and the time of the neutral track in CD with the cut on the time difference.}
\label{fig:TimeFDvsCDcut}
\end{figure}

\begin{figure}[ht!bp]
\centering
{
\subfigure[Experimental Data.]{\fbox{\includegraphics[width=0.7\textwidth]{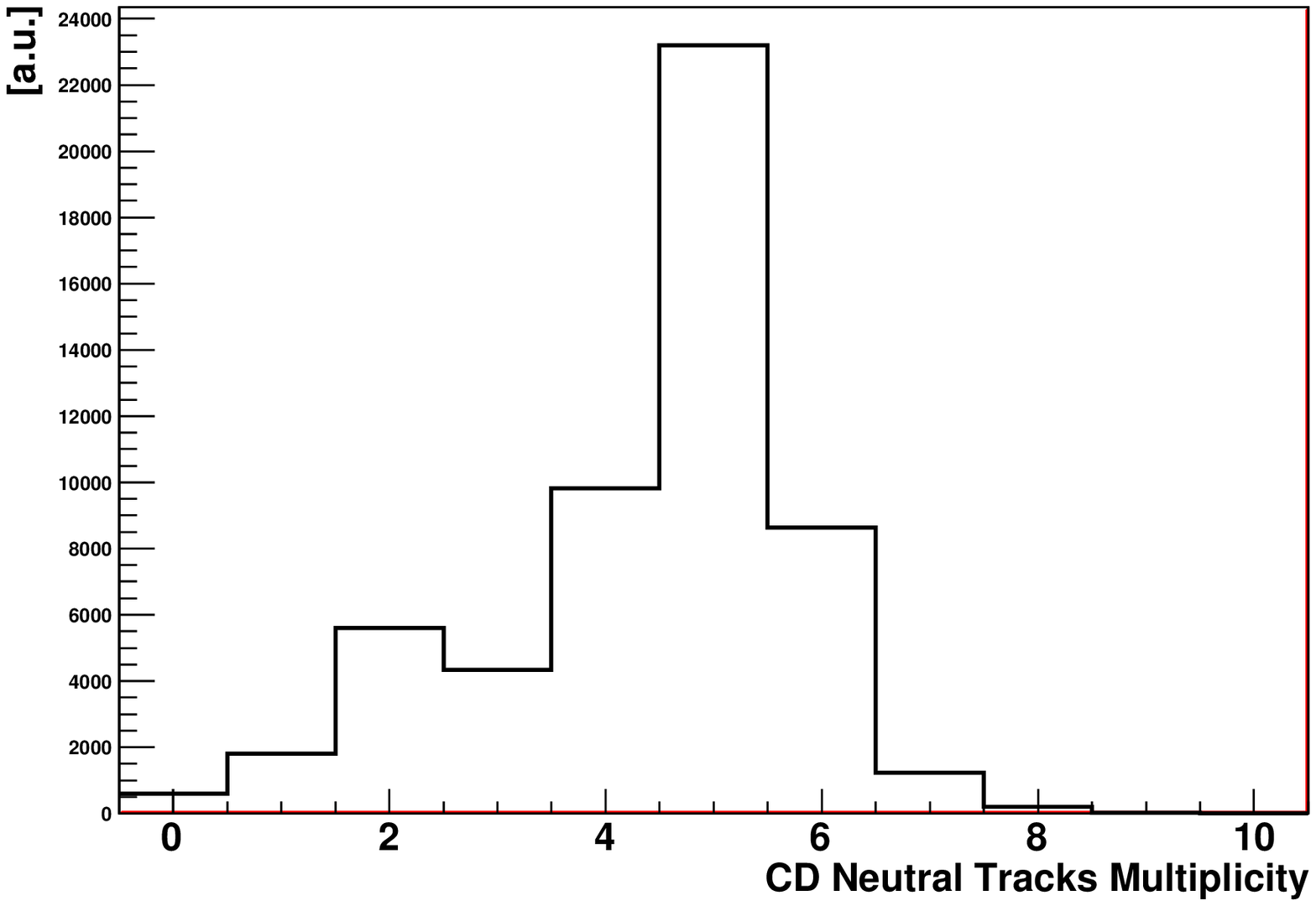}} \label{fig:CDNeutralMultiplicityCutData}}\\
\subfigure[Monte-Carlo simulation.]{\fbox{\includegraphics[width=0.7\textwidth]{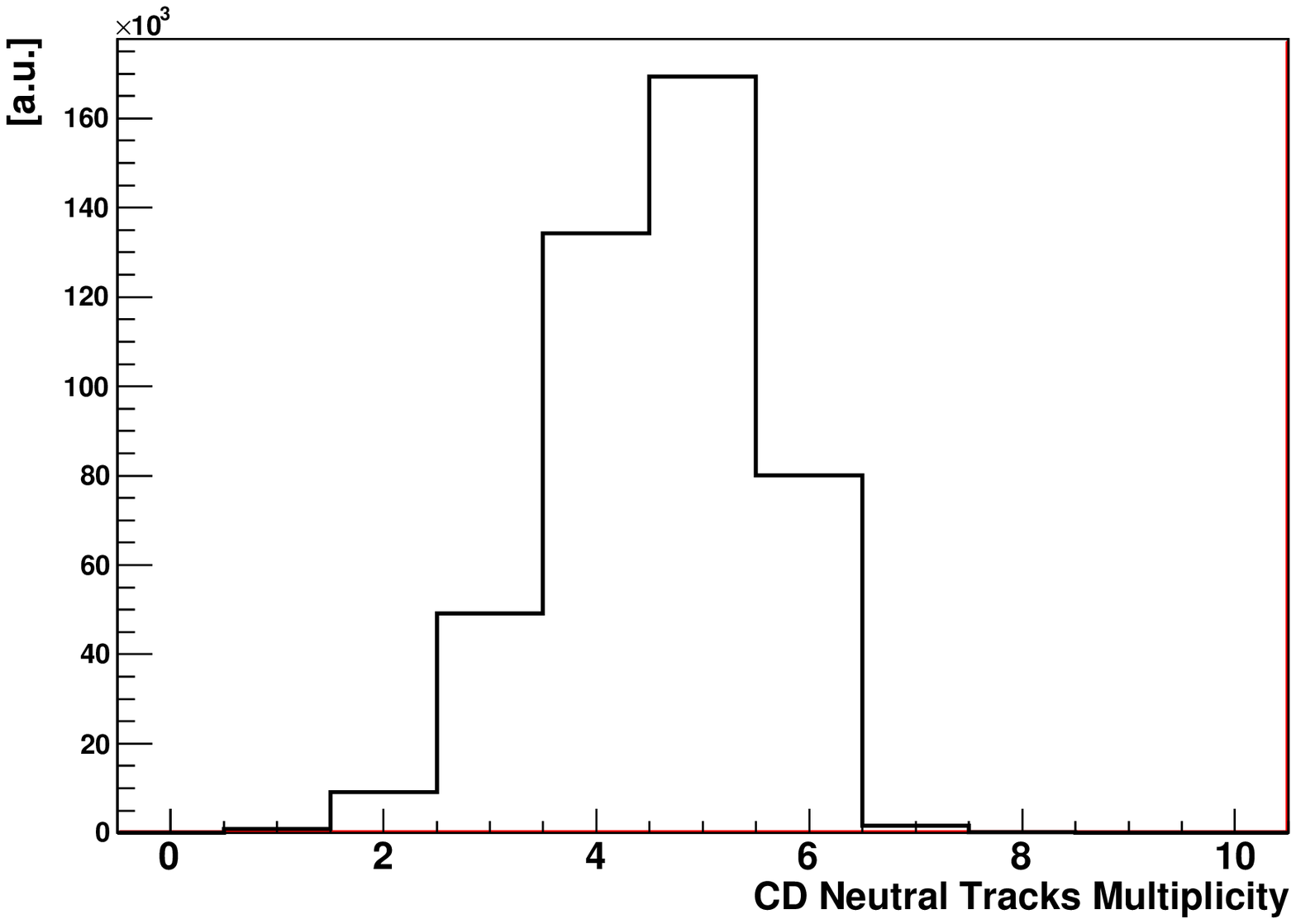}} \label{fig:CDNeutralMultiplicityCutMC}}
}
\caption{Multiplicity of neutral CD detector tracks, after time difference cut.}
\label{fig:CDNeutralMultiplicityCut}
\end{figure}

After selection of two charged tracks in FD correlated in time, event was analyzed further. The tracks multiplicities in CD detector were examined and compared with Monte-Carlo simulation (Fig.~\ref{fig:CDMulti}).
In the CD track finding procedure (see Appendix~\ref{appendix:TrackReconstructionCD}) the signals of PS detector were used as a selection criteria for charged(neutral) tracks, 
the $1\mathrm{~MeV}$ values was chosen as a PS cluster threshold.
The veto on charged CD tracks was put.
Time correlation was checked between mean time of two charged tracks in FD and a neutral CD track (Fig.~\ref{fig:TimeFDvsCD}).
The cut on the time difference between mean time of two charged tracks in FD and a neutral CD track was performed, the events
 in prompt peak were selected (time~difference~from~$-10\mathrm{~ns}$ to $20\mathrm{~ns}$ ) (Fig.~\ref{fig:TimeFDvsCDcut}).
The neutral tracks multiplicity was checked Fig.~\ref{fig:CDNeutralMultiplicityCut} and compared with Monte-Carlo simulation.
The events with six neutral CD tracks correlated with mean time of the two charged tracks in FD were chosen. 
Around $4.4$ millions of such events were selected for later analysis.
Total reconstruction efficiency defined as
\begin{equation}
 Tot.Rec.Eff=Rec.Eff*Geom.Acc
\label{eq:TotRecEffDefinition}
\end{equation}
where $Rec.Eff$~-~reconstruction efficiency of the applied cuts, $Geom.Acc$~-~geometrical acceptance (Eq.~\ref{eq:GeomAcc3pi0}).
It was estimated via Monte-Carlo, based on homogeneously and isotropically populated phase space, to be:

\begin{equation}
 Tot.Rec.Eff = 3.90\%
\label{eq:TotRecEffeventselection}
\end{equation}

\newpage
\subsection{The Detector Response}\label{subsec:DetResp}
After the selection of two charged tracks in FD (protons) correlated in time with six neutral tracks in CD (photons) the detector response
was compared with the Monte-Carlo simulation of $pp \rightarrow pp 3\pi^{0}$ assuming homogeneously and isotropically populated phase space.

\begin{figure}[ht!bp]
\centering
{
\subfigure[Experimental Data.]{\fbox{\includegraphics[width=0.7\textwidth]{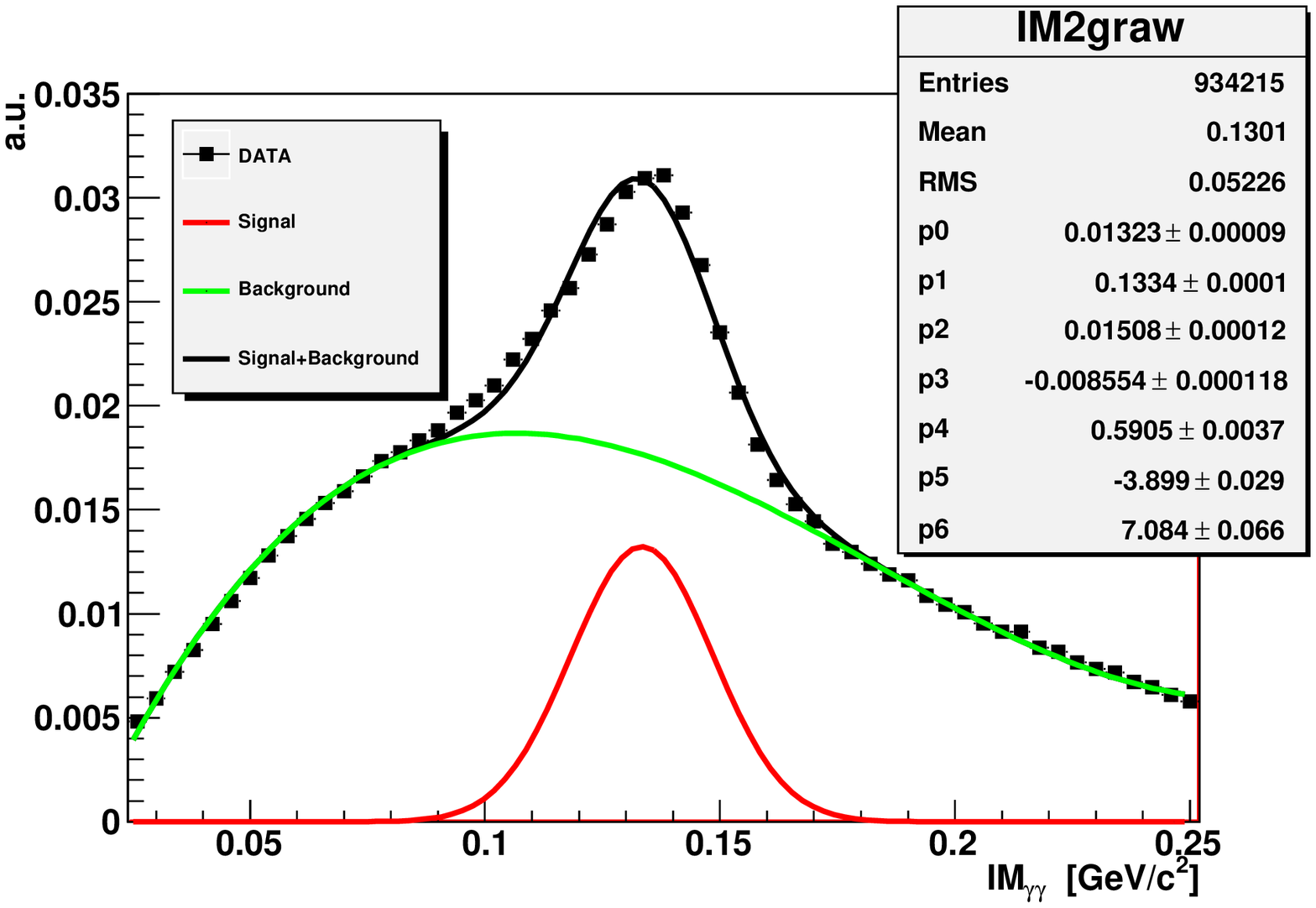}} \label{fig:IMggRAW}}\\
\subfigure[Monte-Carlo simulation.]{\fbox{\includegraphics[width=0.7\textwidth]{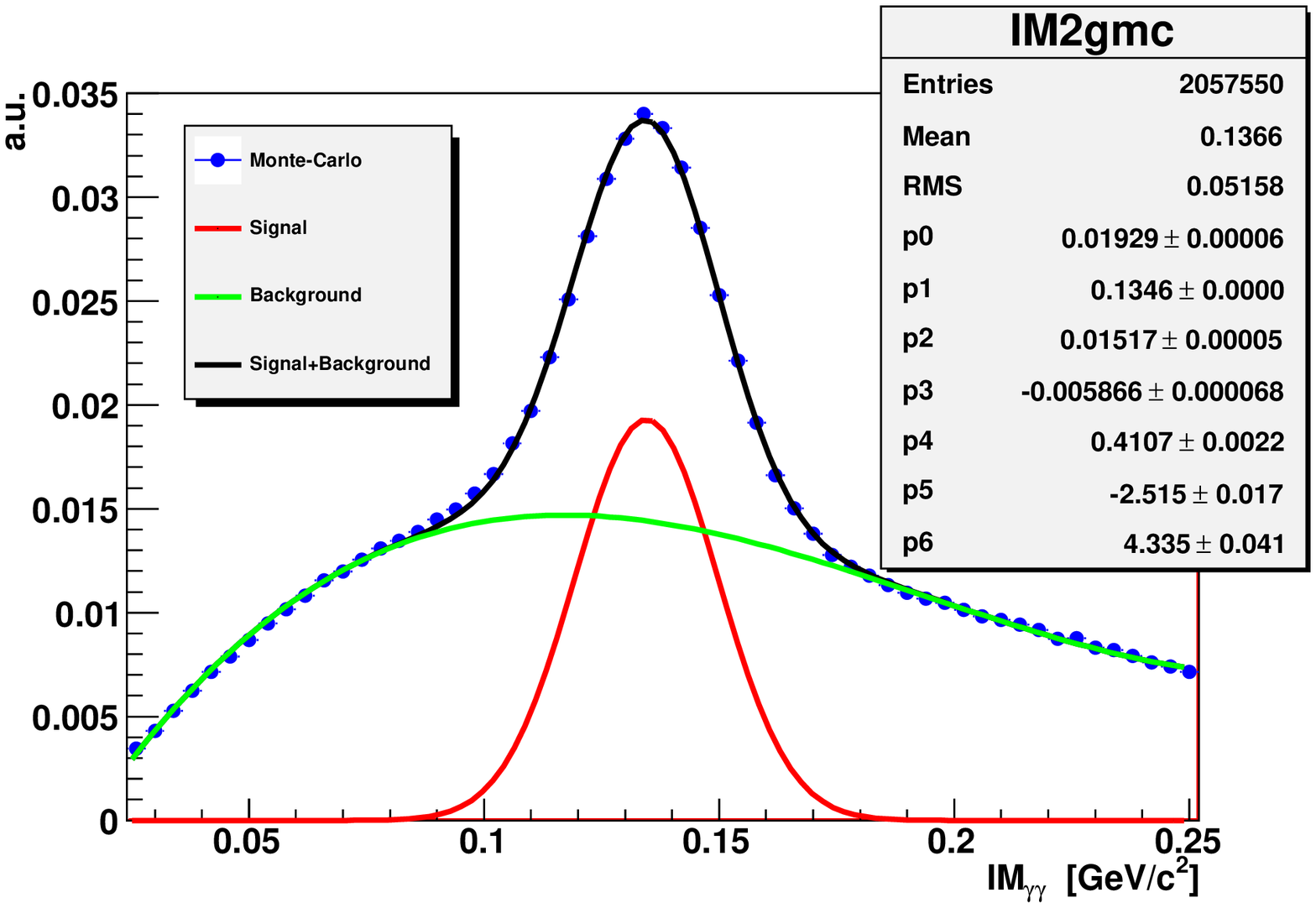}} \label{fig:IMggMC}}
}
\caption{Invariant Mass of two photon pairs, $\pi^{0}$ peak visible on a combinatorial background.}
\label{fig:IMgg}
\end{figure}

The response of the CD detector was checked by plotting invariant mass of two photon pairs (Fig.~\ref{fig:IMgg}).
The $\pi^{0}$ signal is seen on the combinatorial background. The spectra were fitted with a sum of Gaussian (describing the peak) and
the polynomial of the forth order(describing the combinatorial background). 
The fitted curve describing the background was subtracted from the data point and simulation, the spectra were 
normalized to the same area and compared \myImgRef{IMggComp}. It is seen that the Monte-Carlo simulation describes the experimental data very well.
The maximum of the peak is at $M_\pi^{0}\approx0.135\mathrm{~GeV/c^{2}}$ indicated by red line in \myImgRef{IMggComp}.

\myFrameSmallFigure{IMggComp}{Invariant Mass of two photon pairs, combinatorial background subtracted. Experimental data and Monte-Carlo comparison, line indicates the $\pi^{0}$ mass.}{Invariant Mass of two photon pairs, combinatorial background subtracted.}

\begin{sidewaysfigure}[t!bp]
\centering
{
\subfigure[Experimental Data.]{\fbox{\includegraphics[width=0.45\textwidth]{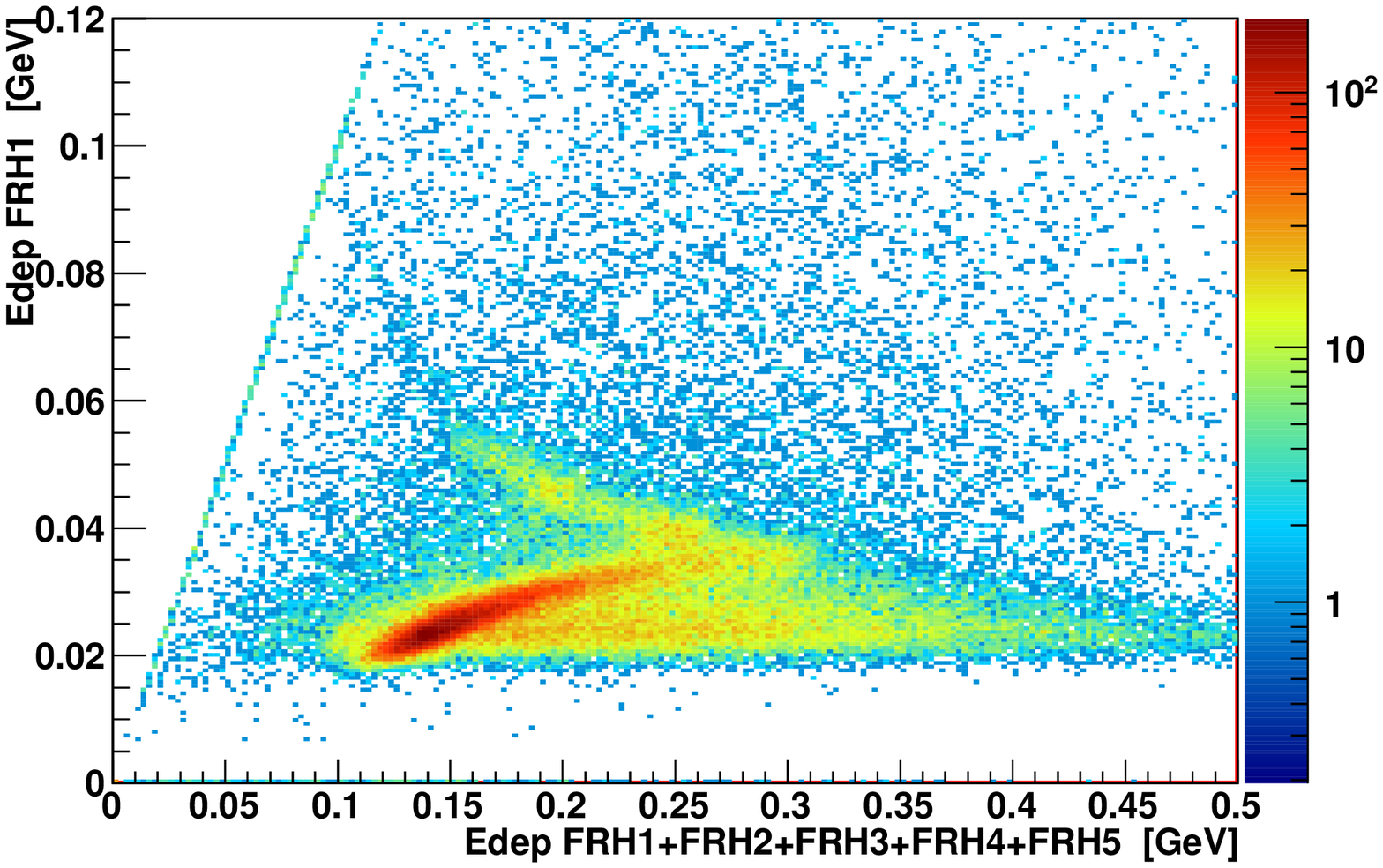}} \label{fig:EdepFRHcompRAW1}}\\

\subfigure[Monte-Carlo simulation.]{\fbox{\includegraphics[width=0.45\textwidth]{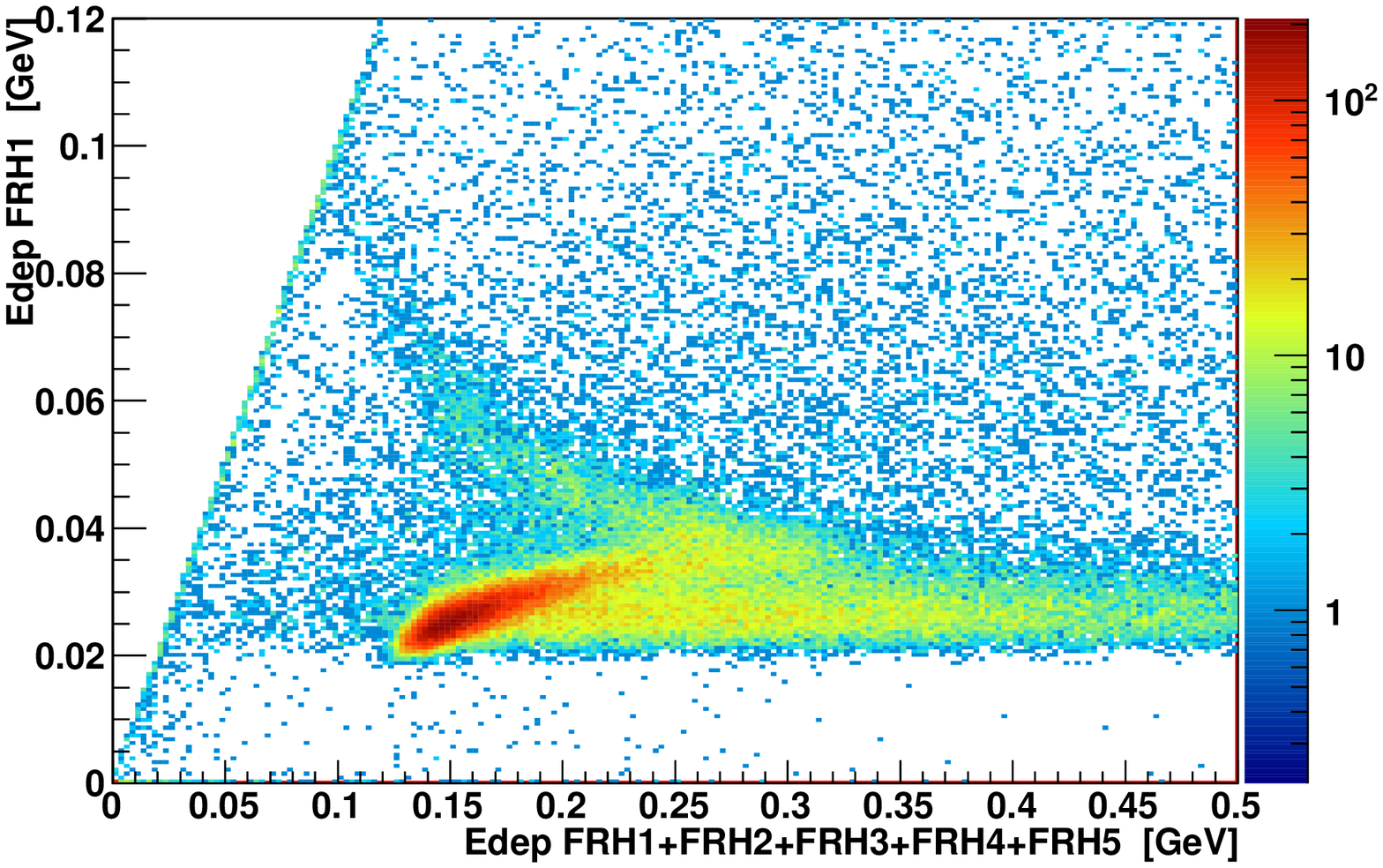}} \label{fig:EdepFRHcompMC1}}\quad
\subfigure[Monte-Carlo simulation, with no hadronic interactions.]{\fbox{\includegraphics[width=0.45\textwidth]{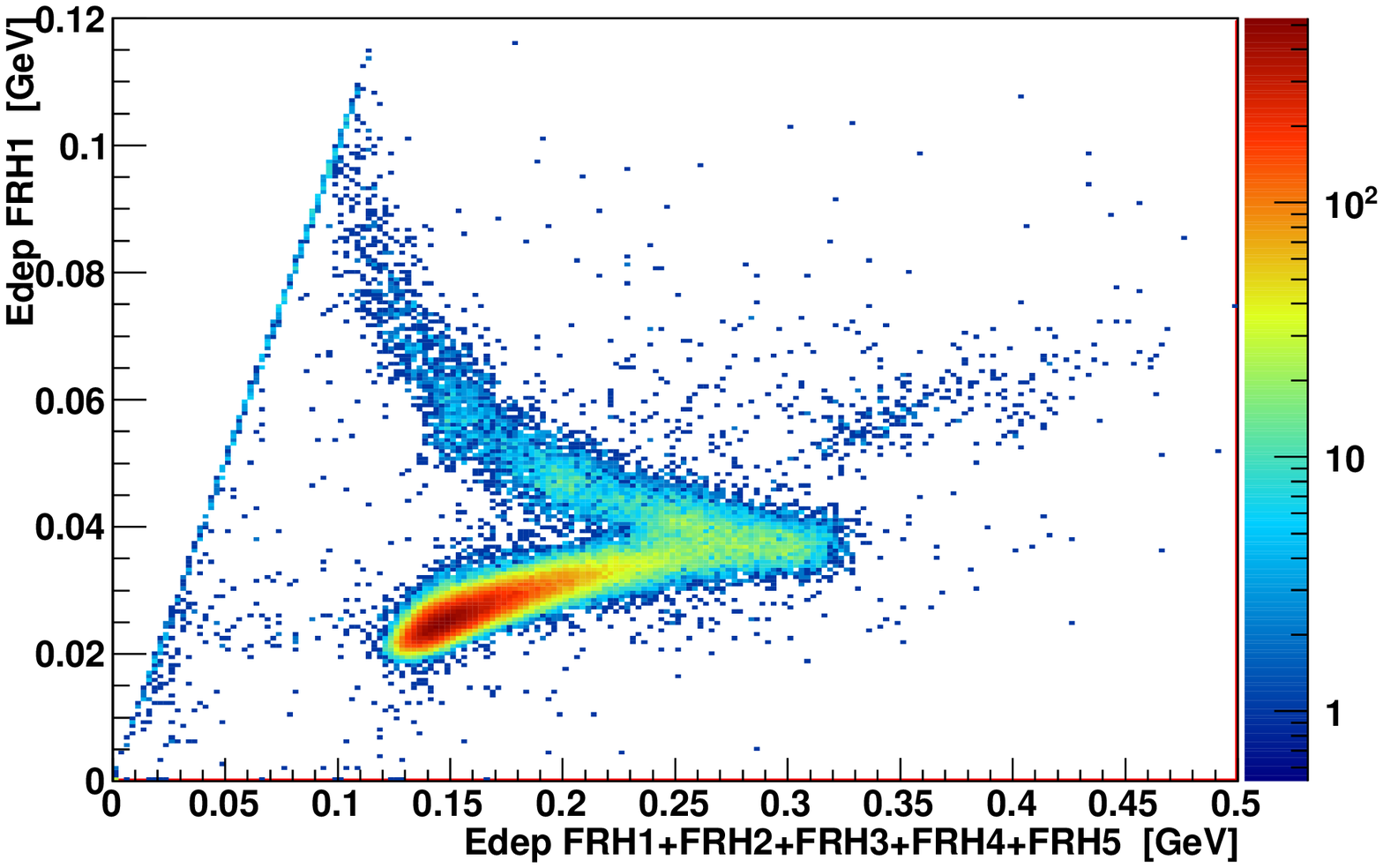}} \label{fig:EdepFRHcompMC1NoHadr}}

}
\caption{Energy deposits in FRH detector, layer 1 versus whole FRH, $dE-E$ plots, of two charged tracks. Good agreement between experimental data and Monte-Carlo visible. It is seen that the most of the charged tracks travel through the whole FD detector.}
\label{fig:EdepFRHcomp1}
\end{sidewaysfigure}

\begin{sidewaysfigure}[t!bp]
\centering
{
\subfigure[Experimental Data.]{\fbox{\includegraphics[width=0.45\textwidth]{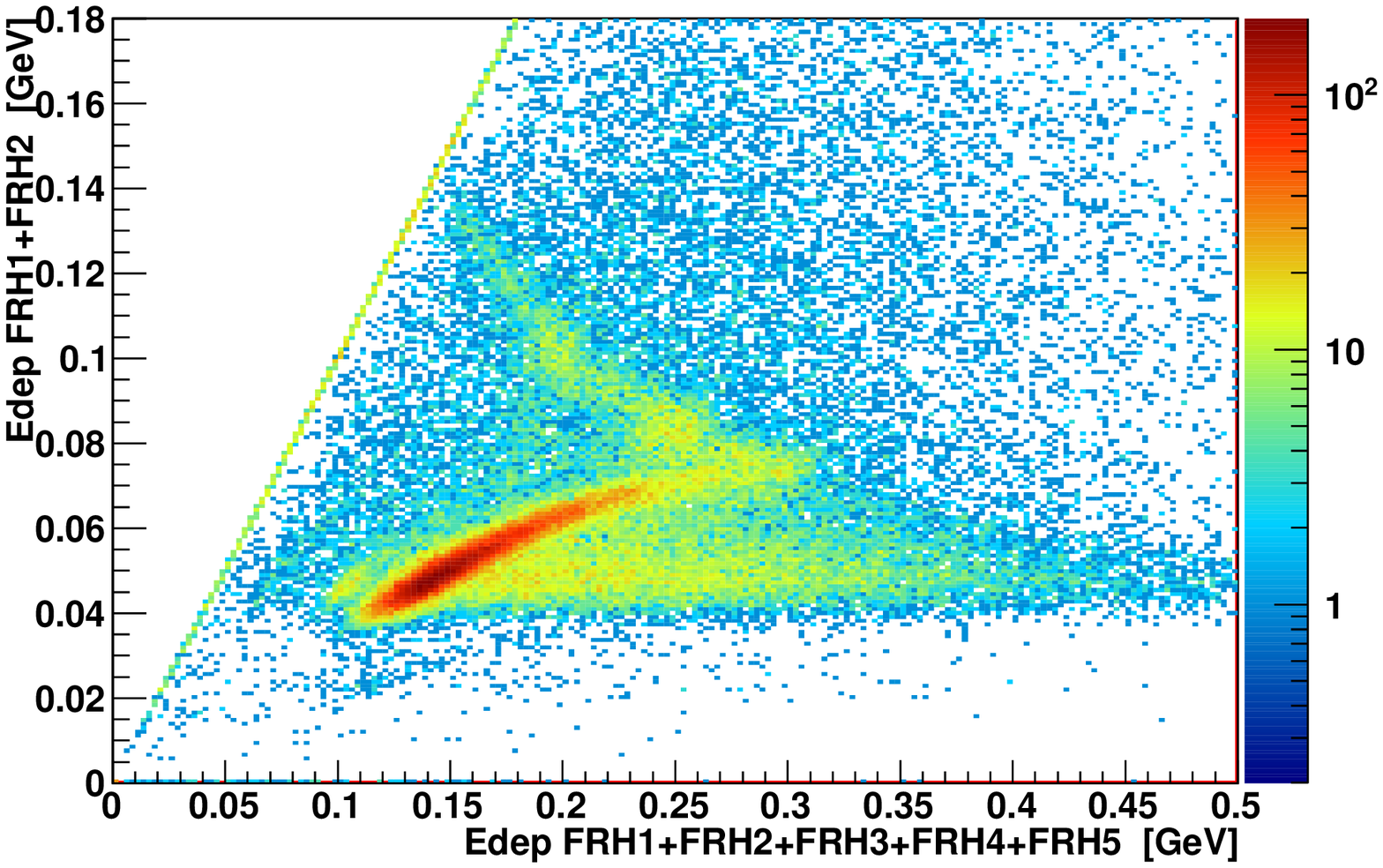}} \label{fig:EdepFRHcompRAW2}}\\

\subfigure[Monte-Carlo simulation.]{\fbox{\includegraphics[width=0.45\textwidth]{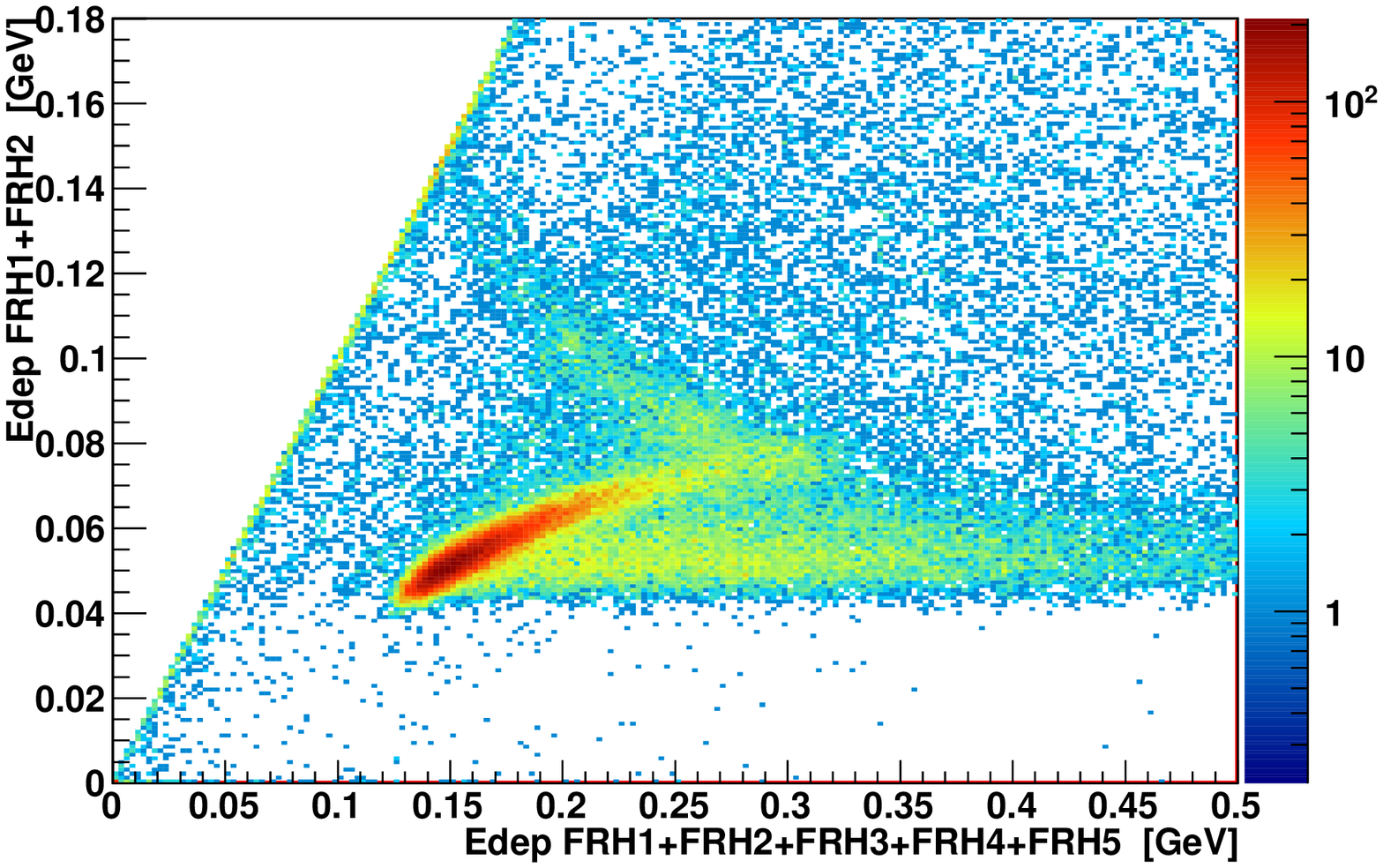}} \label{fig:EdepFRHcompMC2}}\quad
\subfigure[Monte-Carlo simulation, with no hadronic interactions.]{\fbox{\includegraphics[width=0.45\textwidth]{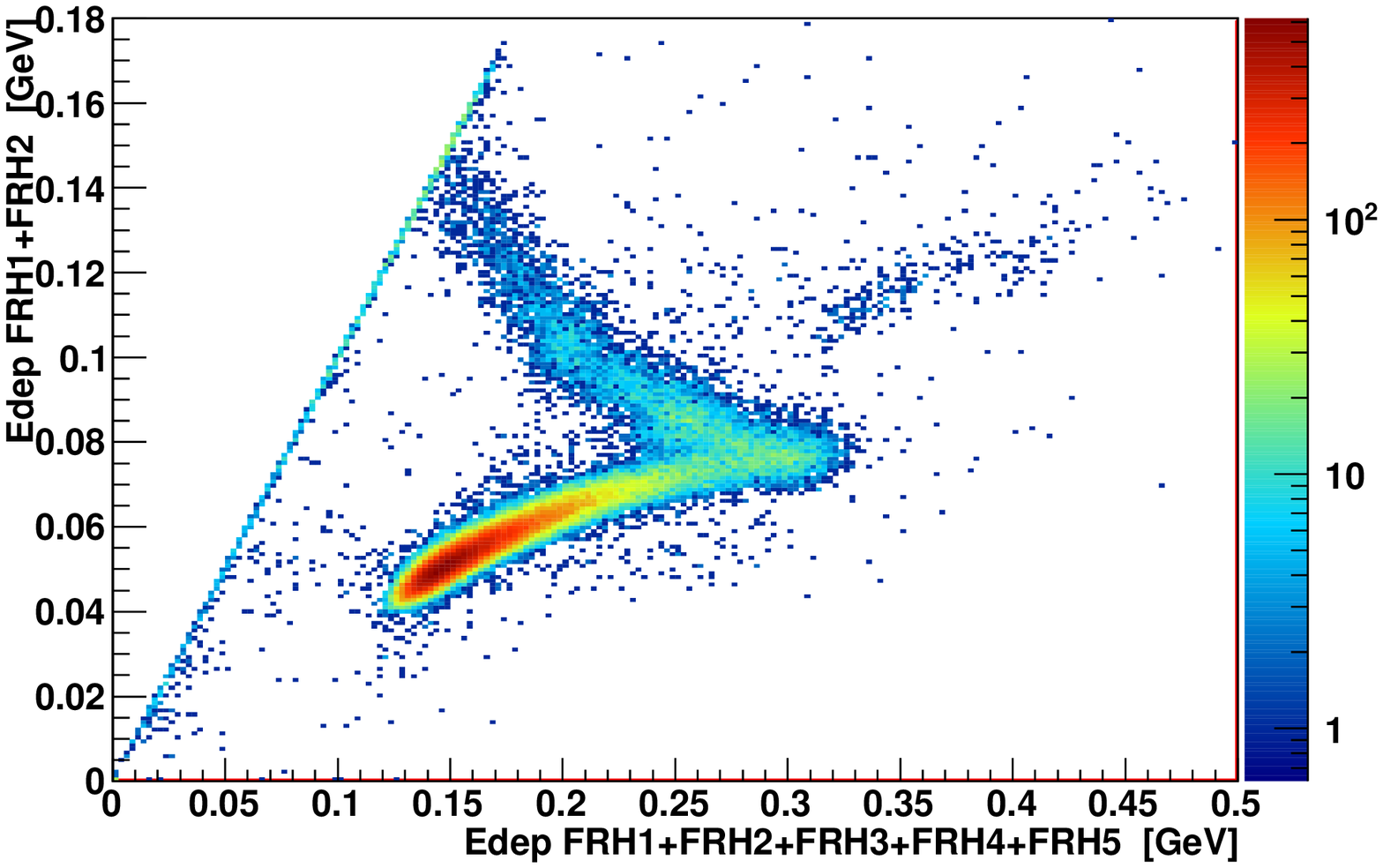}} \label{fig:EdepFRHcompMC2NoHadr}}

}
\caption{Energy deposits in FRH detector, layer from 1 to 2 versus whole FRH, $dE-E$ plots, of two charged tracks. Good agreement between experimental data and Monte-Carlo visible. It is seen that the most of the charged tracks travel through the whole FD detector.}
\label{fig:EdepFRHcomp2}
\end{sidewaysfigure}

\begin{sidewaysfigure}[t!bp]
\centering
{
\subfigure[Experimental Data.]{\fbox{\includegraphics[width=0.45\textwidth]{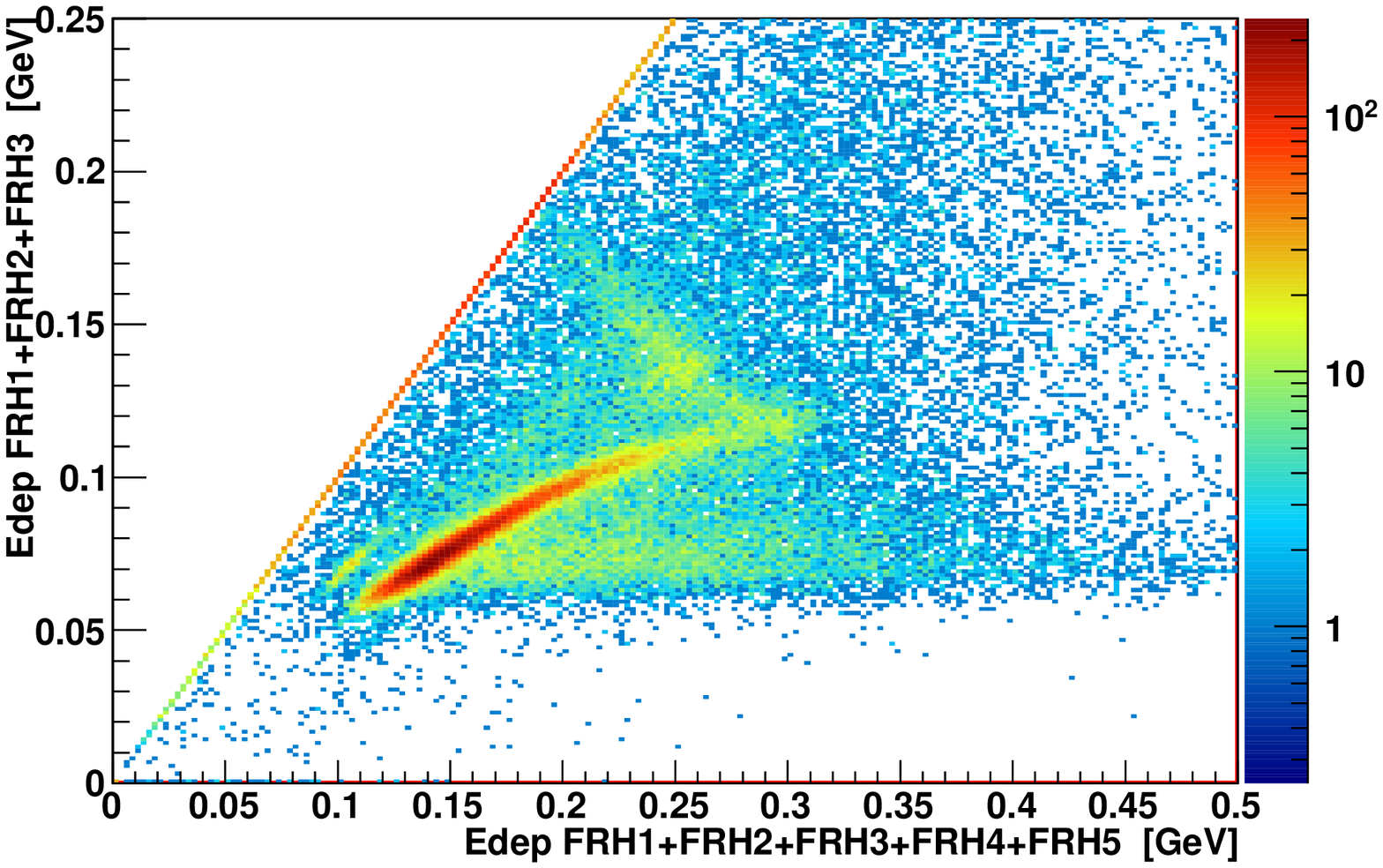}} \label{fig:EdepFRHcompRAW3}}\\

\subfigure[Monte-Carlo simulation.]{\fbox{\includegraphics[width=0.45\textwidth]{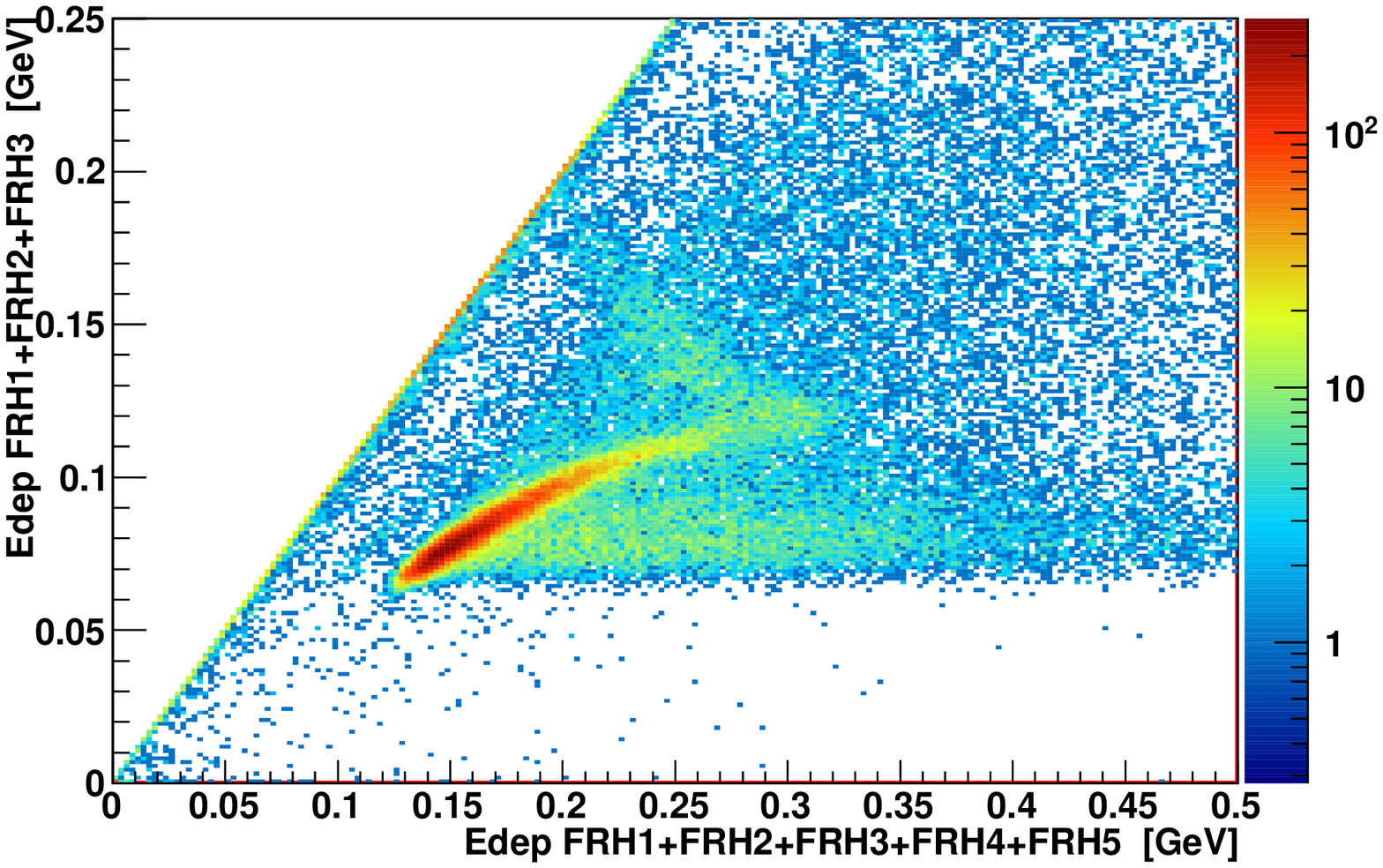}} \label{fig:EdepFRHcompMC3}}\quad
\subfigure[Monte-Carlo simulation, with no hadronic interactions.]{\fbox{\includegraphics[width=0.45\textwidth]{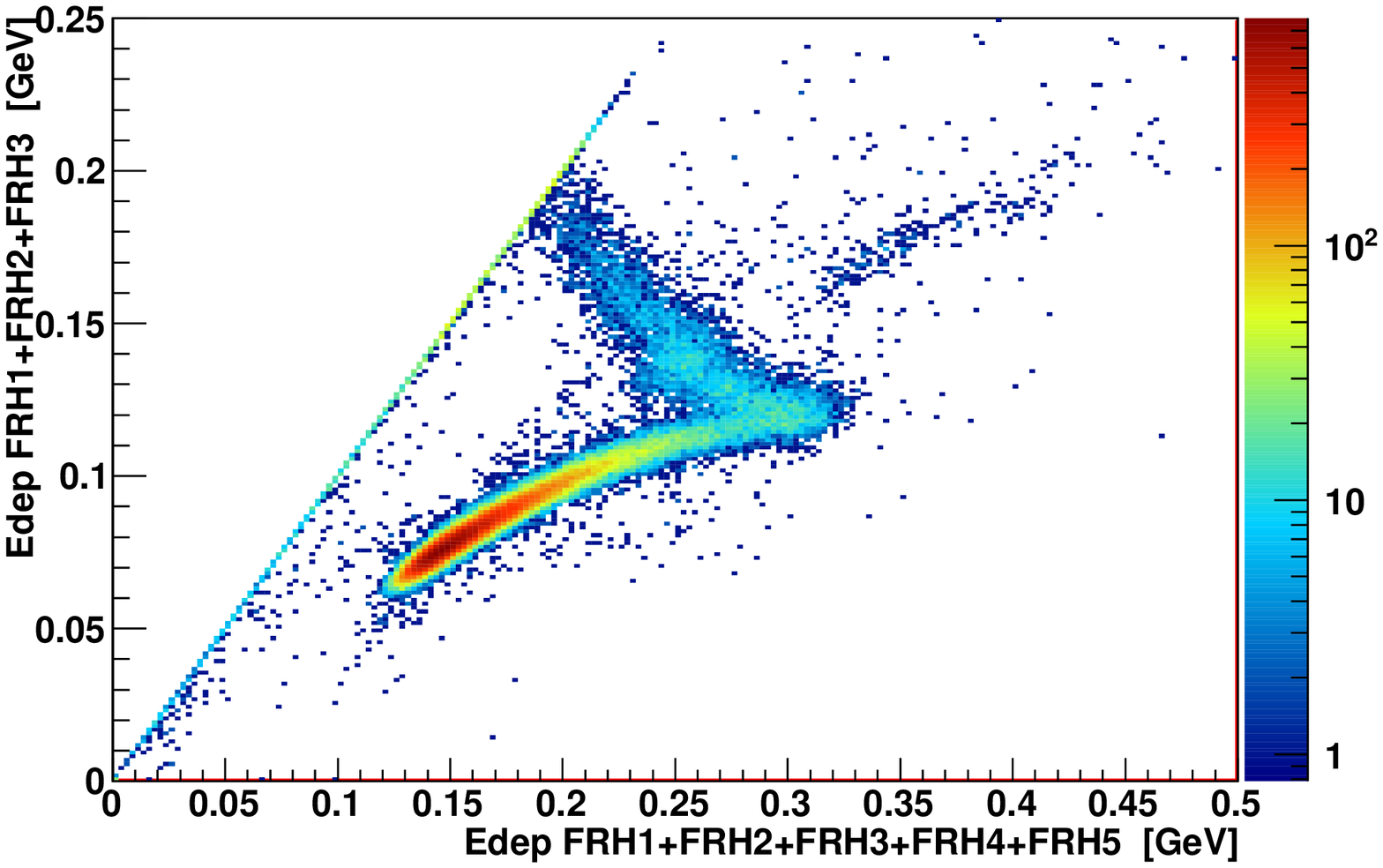}} \label{fig:EdepFRHcompMC3NoHadr}}

}
\caption{Energy deposits in FRH detector, layer from 1 to 3 versus whole FRH, $dE-E$ plots, of two charged tracks. Good agreement between experimental data and Monte-Carlo visible. It is seen that the most of the charged tracks travel through the whole FD detector.}
\label{fig:EdepFRHcomp3}
\end{sidewaysfigure}

\begin{sidewaysfigure}[t!bp]
\centering
{
\subfigure[Experimental Data.]{\fbox{\includegraphics[width=0.45\textwidth]{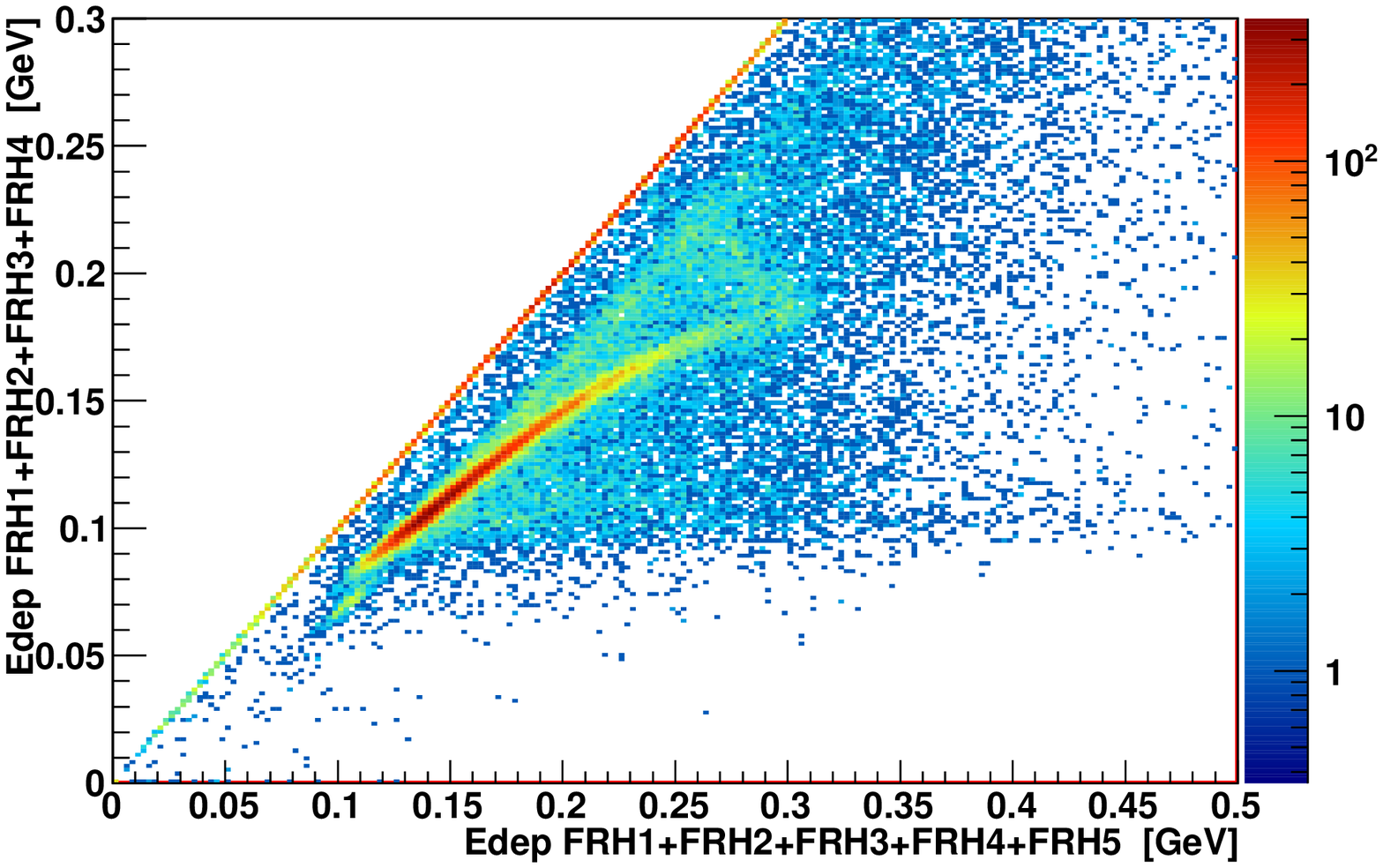}} \label{fig:EdepFRHcompRAW4}}\\

\subfigure[Monte-Carlo simulation.]{\fbox{\includegraphics[width=0.45\textwidth]{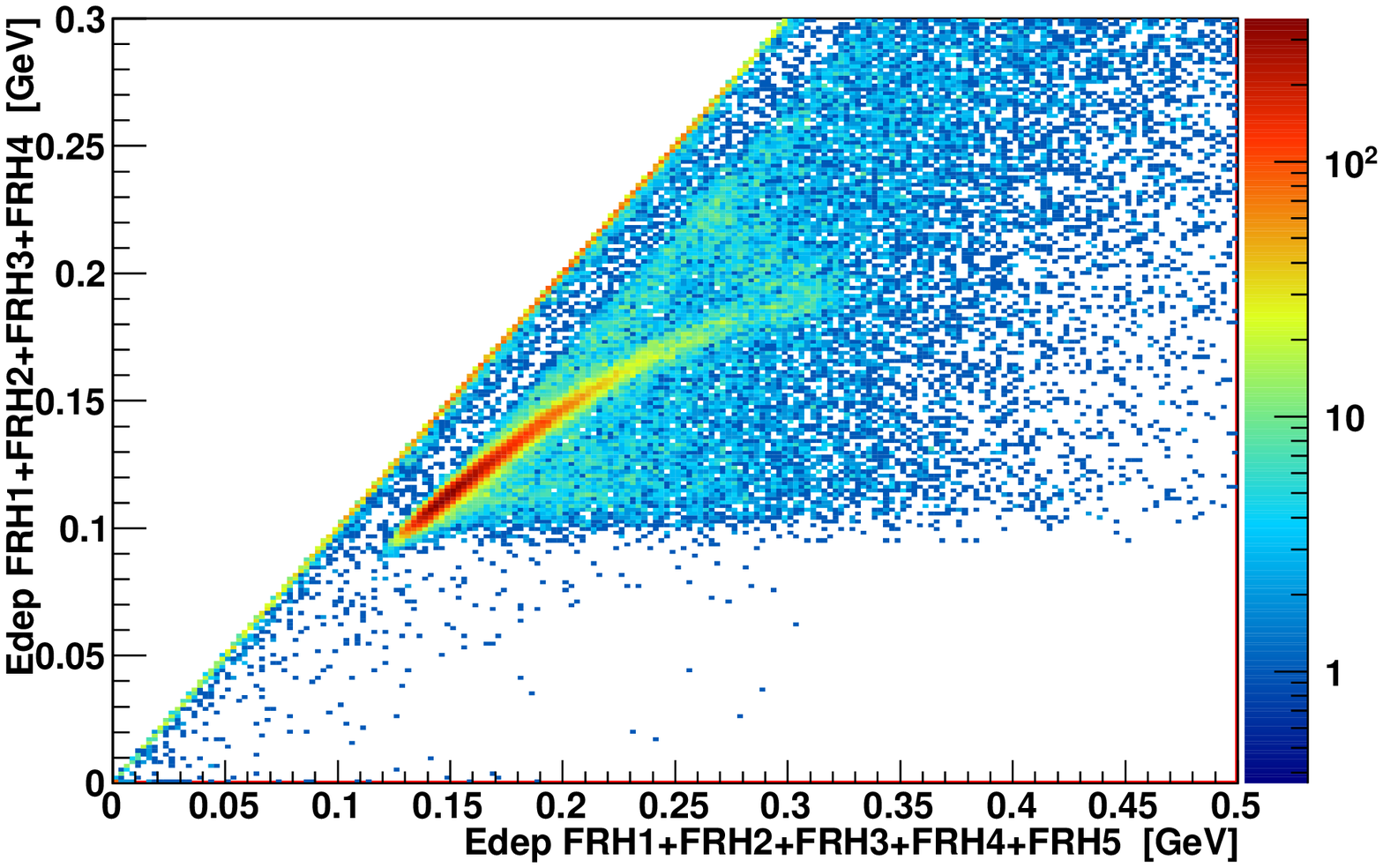}} \label{fig:EdepFRHcompMC4}}\quad
\subfigure[Monte-Carlo simulation, with no hadronic interactions.]{\fbox{\includegraphics[width=0.45\textwidth]{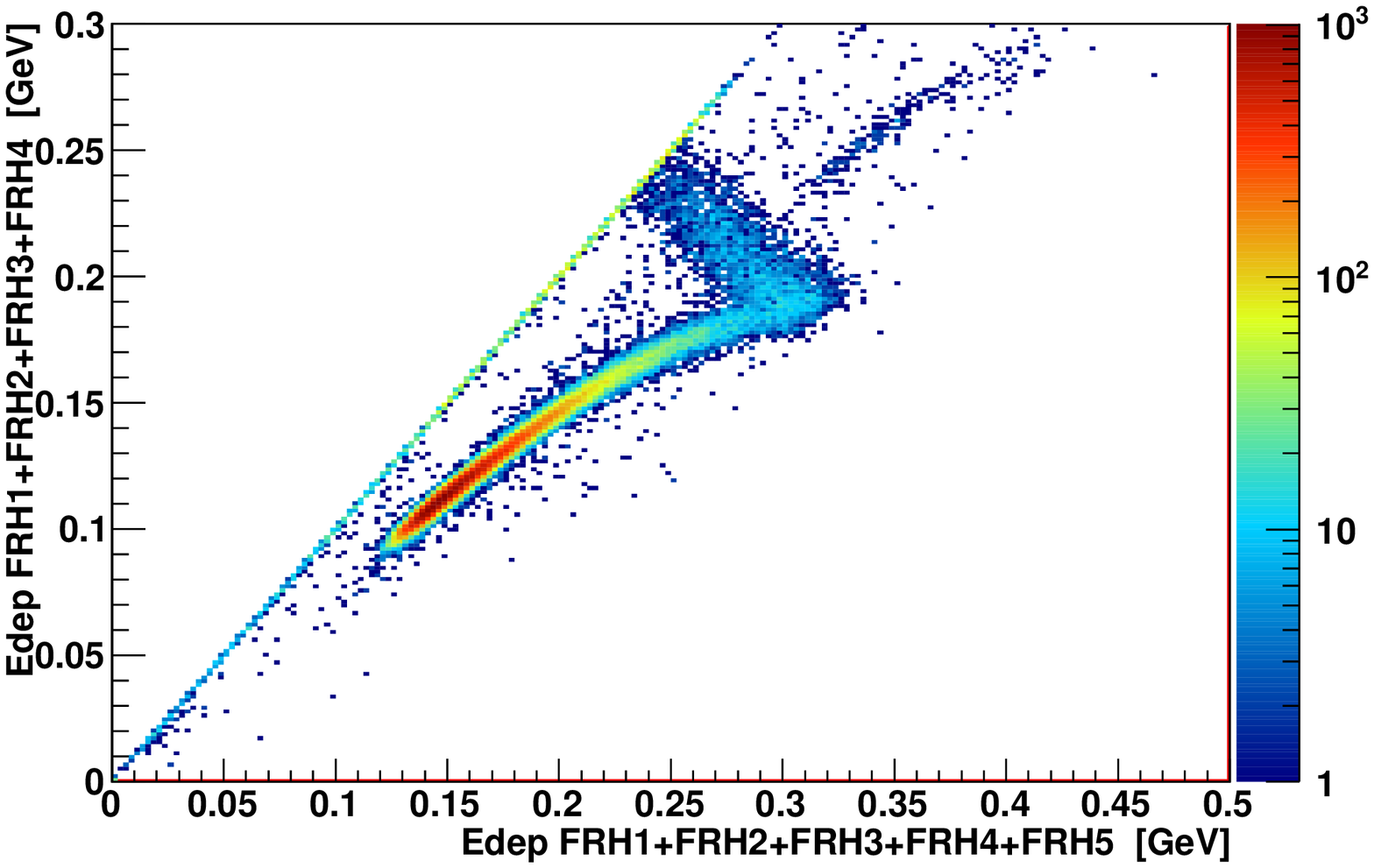}} \label{fig:EdepFRHcompMC4NoHadr}}

}
\caption{Energy deposits in FRH detector, layer from 1 to 4 versus whole FRH, $dE-E$ plots, of two charged tracks. Good agreement between experimental data and Monte-Carlo visible. It is seen that the most of the charged tracks travel through the whole FD detector.}
\label{fig:EdepFRHcomp4}
\end{sidewaysfigure}
The response of the FD detector was also checked by comparing different $dE-E$ plots using different layers of the FRH detector (Figs.~\ref{fig:EdepFRHcomp1}~\ref{fig:EdepFRHcomp2}~\ref{fig:EdepFRHcomp3}~\ref{fig:EdepFRHcomp4}).
It is seen that the most of the charged tracks travel through the whole FD detector.
The experimental $dE-E$ plots are good described by the Monte-Carlo simulation, which proves that the measured energy deposits are consistent with protons energy deposits. 

\newpage
\subsection{The Kinematic Fitting}
\label{subsec:AnaChainKinFit}
The Monte-Carlo simulations (see Appendix~\ref{appendix:wmc}) are extremely useful tool for understanding of complex data.
The following convention concerning Monte-Carlo simulation will be used:

\begin{itemize}
 \item The true value: value assumed in the simulation, used as an input
 \item The reconstructed(measured) value: value after tracks propagation and reconstruction (see Appendix~\ref{appendix:TrackReconstruction})
 \item The fitted value: value after the kinematic fitting
\end{itemize}

The kinematic fitting (see Appendix~\ref{appendix:kfit}) was used to improve the resolution of the variables and to balance the overall four-momentum to keep only the kinematic complete events.

The hypothesis of the reaction $ pp \rightarrow pp 6\gamma \rightarrow pp 3\pi^{0}$ was tested on the experimental data using kinematic fit.
The overall four momentum was balanced. The $6\gamma$ were combined into the $3\pi^{0}$, the $\pi^{0}$ mass constraint was used.
The energy, polar angle $\theta$ and an azimuthal angle $\phi$ of the neutral tracks in CD detector (the photons) was fitted.
For the two tracks in FD detector (the proton candidates) the polar angle $\theta$ and an azimuthal angle $\phi$ was fitted. 
The energy of the track was put as an unknown for the fit, since the protons energies are to high to reconstruct them via $dE-E$ energy losses in FRH detector
in a standard way as well as in the developed Bayesian~Likelihood approach (see Appendix~\ref{appendix:Bayes}).
Most of proton tracks are passing through the whole FRH detector (see Section~\ref{subsec:DetResp} Figs.~\ref{fig:EdepFRHcomp1}~\ref{fig:EdepFRHcomp2}~\ref{fig:EdepFRHcomp3}~\ref{fig:EdepFRHcomp4}) .

\subsubsection{The error parametrization}

\begin{figure}[ht!bp]
\centering
{
\subfigure[True minus reconstructed value of the $\theta_{Rec}$ angle of the photon as a function of the photon energy $E_{Rec}$.]{\fbox{\includegraphics[width=0.7\textwidth]{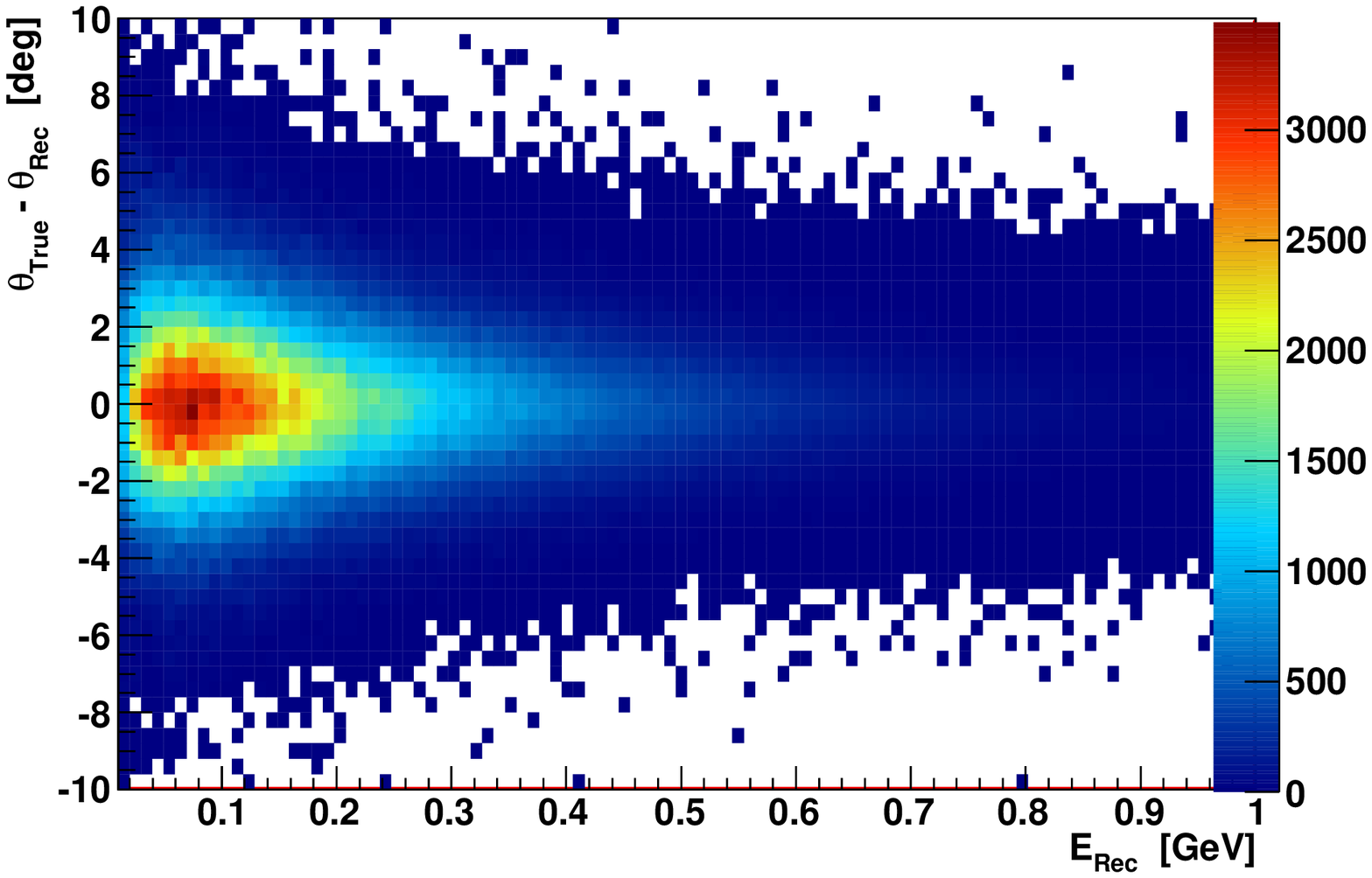}} \label{fig:ErrorParam2D}}\\
\subfigure[True minus reconstructed value of the $\theta_{Rec}$ angle of the photon for the particular photon energy. Line indicates the Gaussian fit.]{\fbox{\includegraphics[width=0.7\textwidth]{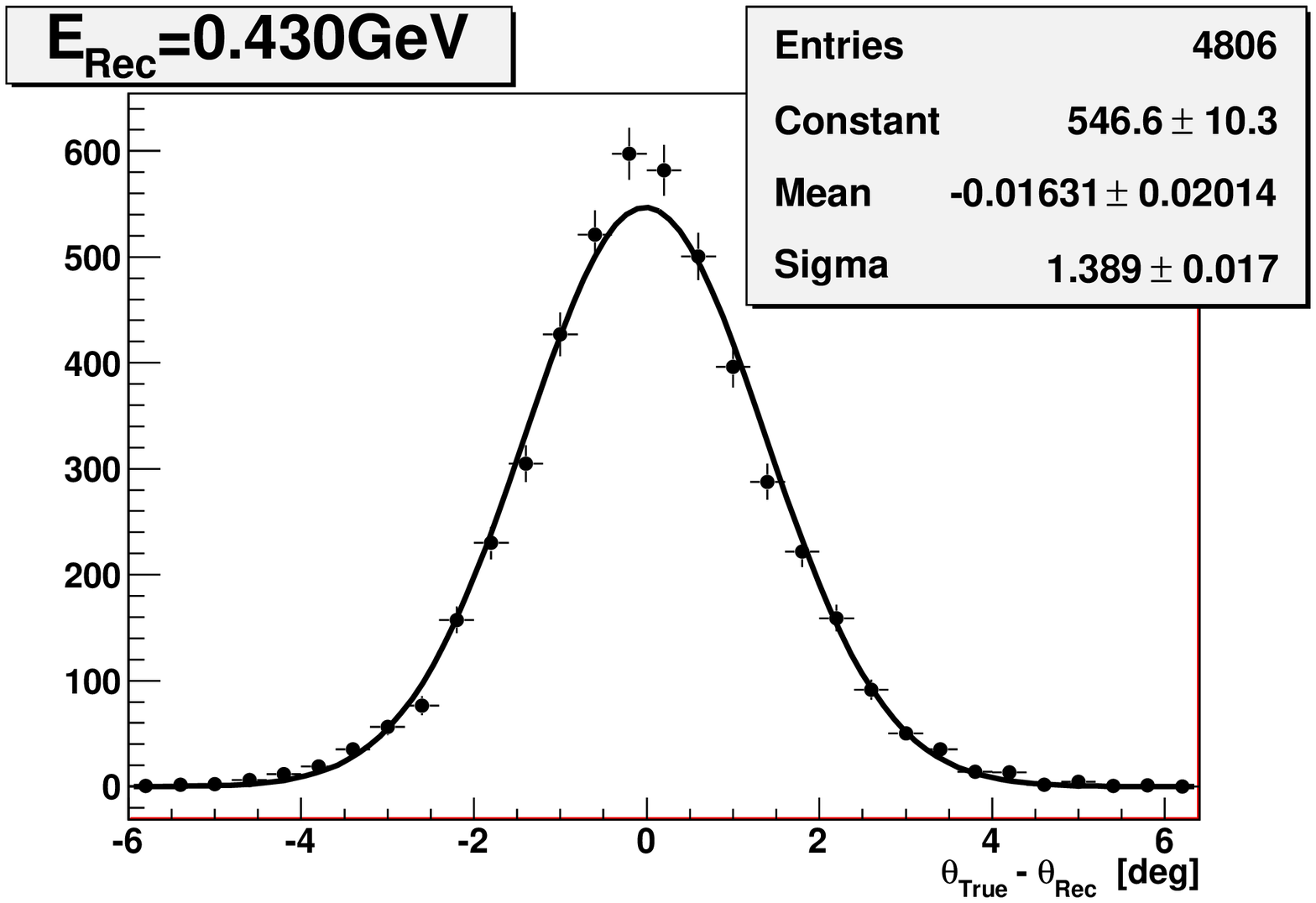}} \label{fig:ErrorParamFit}}
}
\caption{Example of the error parametrization technique.}
\label{fig:ErrorParam}
\end{figure}

The errors of the parameters were determined using WASA-at-COSY Monte-Carlo simulation,
 by simulating single photon tracks in CD detector and single proton tracks in FD detector respectively. 
The errors of the variables were derived by the fitting of the Gaussian function to the true minus reconstructed(measured) values
in steps of the variable from which this error depends. The discrete values were put into the histograms, later the linear interpolation between the values was used to get the error value.
Example of the procedure is presented in (Fig.~\ref{fig:ErrorParam}).

\begin{sidewaysfigure}[t!bp]
\centering
{
\subfigure[Relative Error of energy versus energy.]{\fbox{\includegraphics[width=0.28\textwidth]{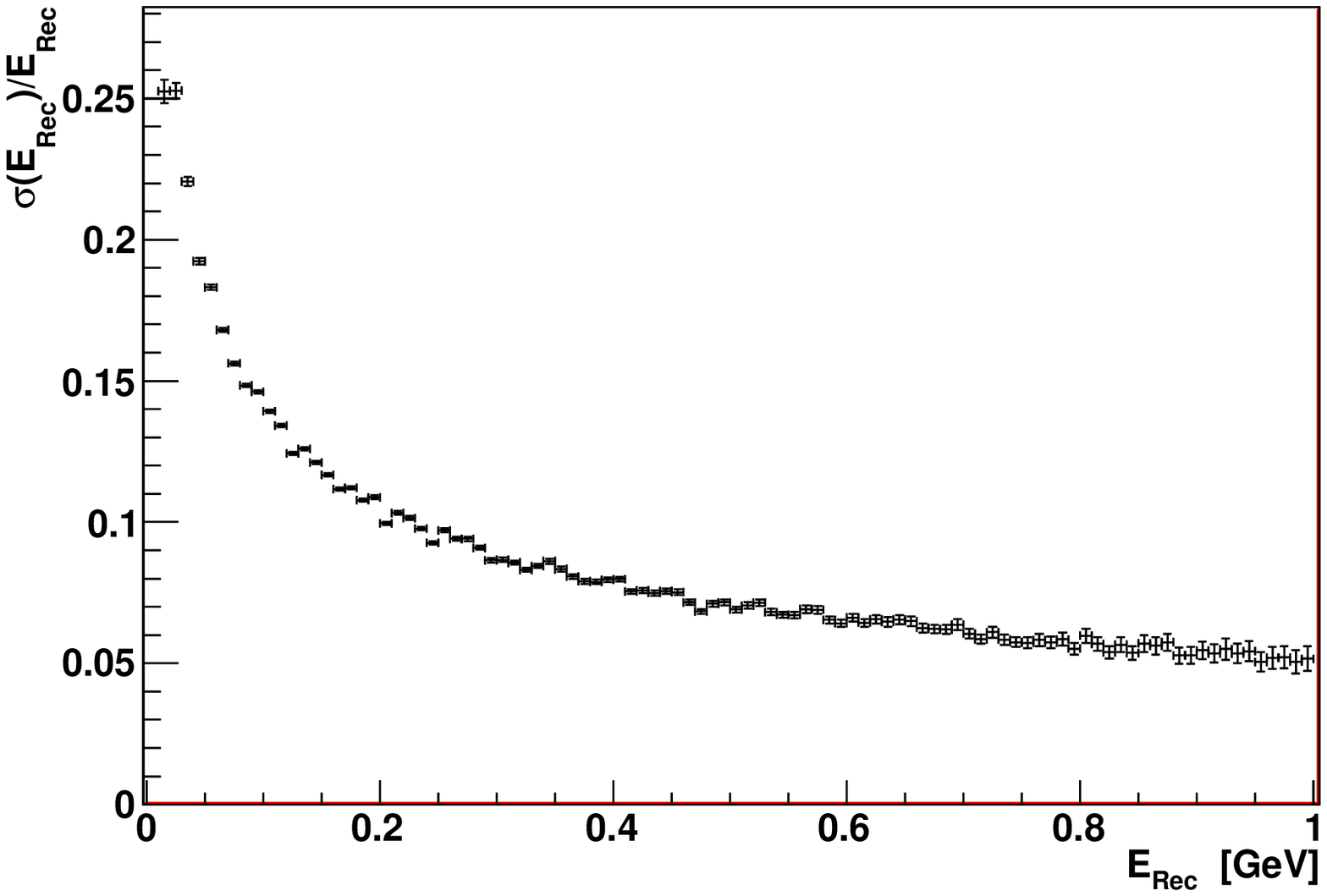}} \label{fig:CDErrorCheckEA}}\quad
\subfigure[Relative Error of energy versus $\theta_{Rec}$.]{\fbox{\includegraphics[width=0.28\textwidth]{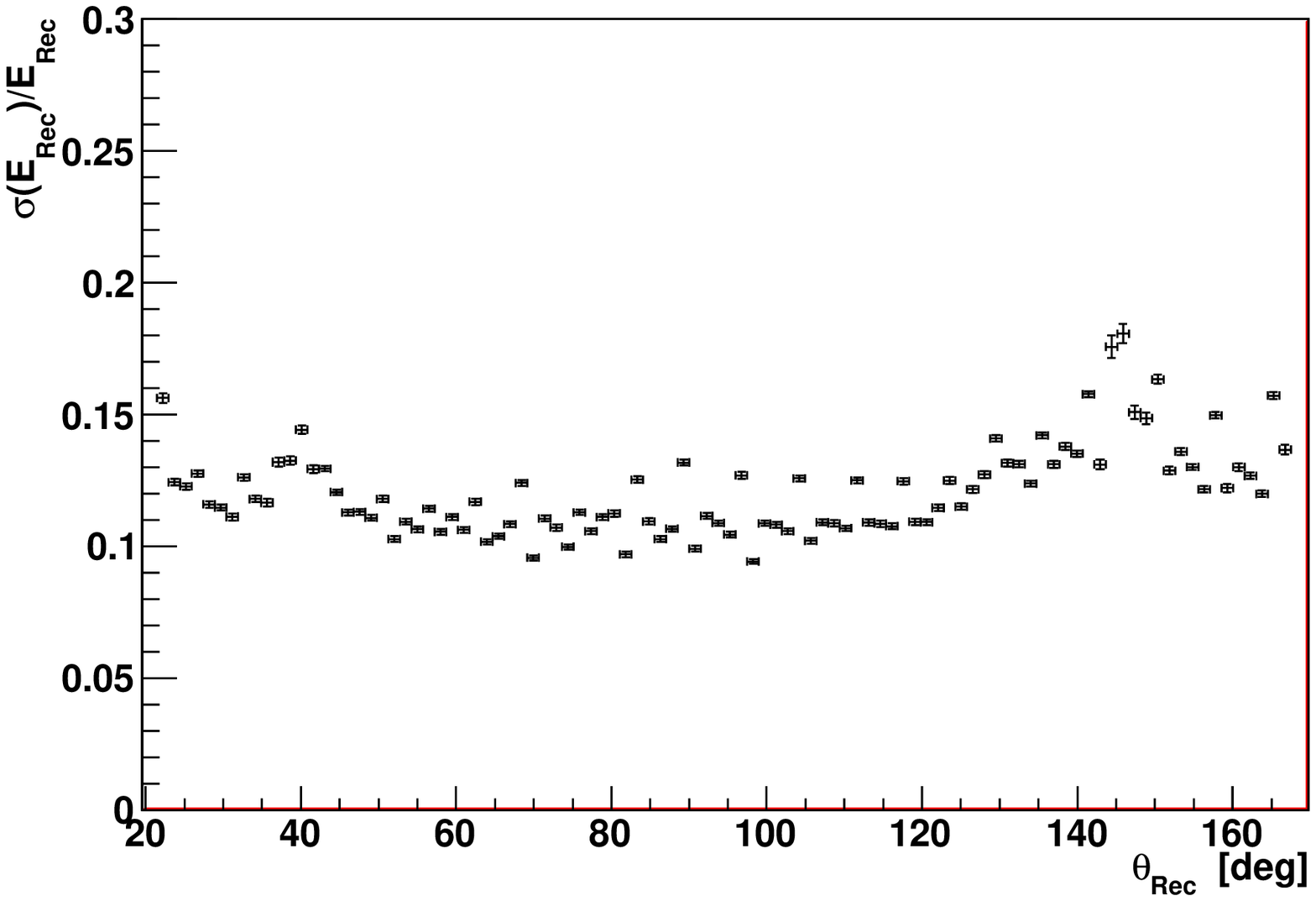}} \label{fig:CDErrorCheckEB}}\quad
\subfigure[Relative Error of energy versus $\phi_{Rec}$.]{\fbox{\includegraphics[width=0.28\textwidth]{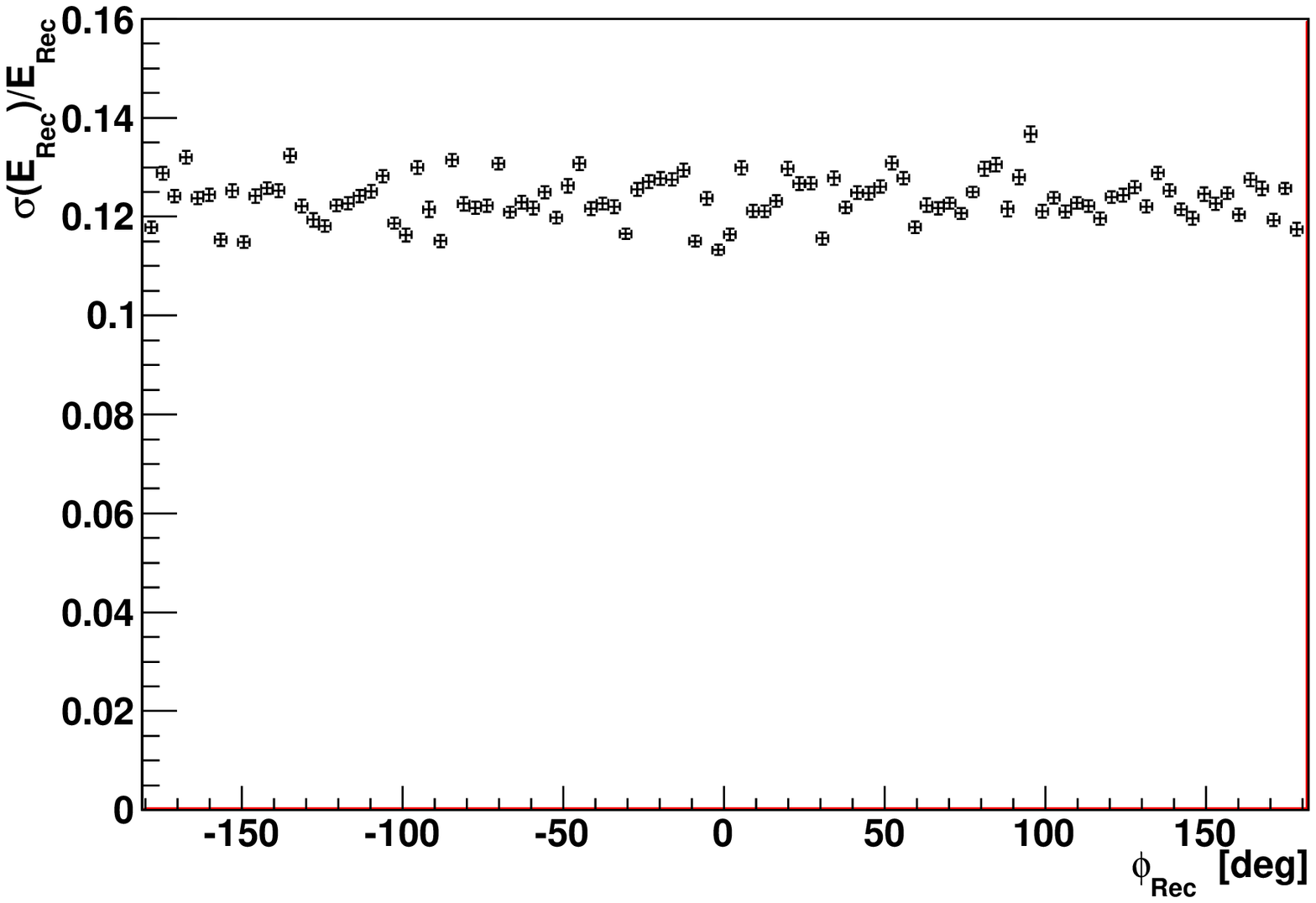}} \label{fig:CDErrorCheckEC}}\\

\subfigure[Error of $\theta_{Rec}$ versus energy.]{\fbox{\includegraphics[width=0.28\textwidth]{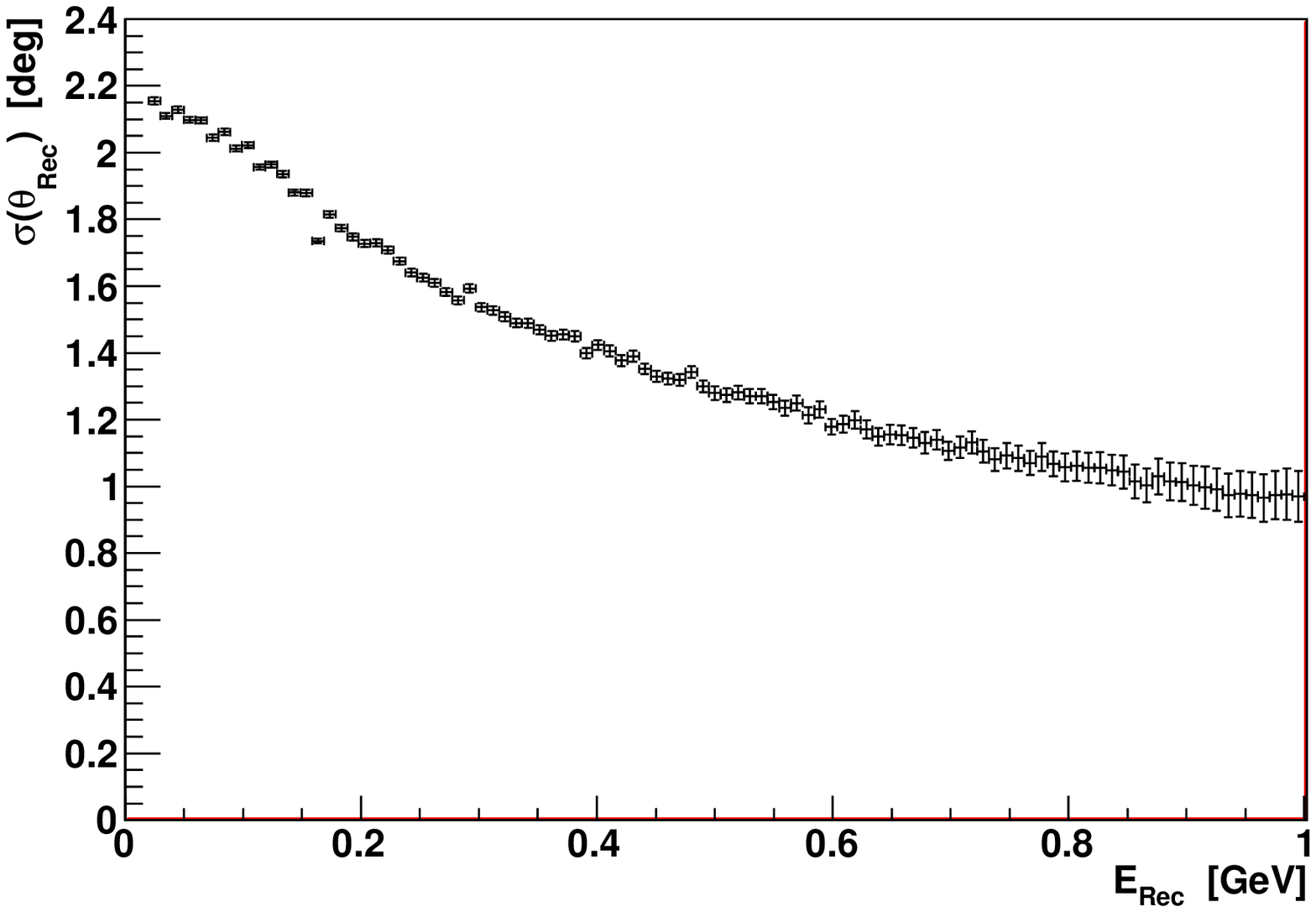}} \label{fig:CDErrorCheckTA}}\quad
\subfigure[Error of $\theta_{Rec}$ versus $\theta_{Rec}$.]{\fbox{\includegraphics[width=0.28\textwidth]{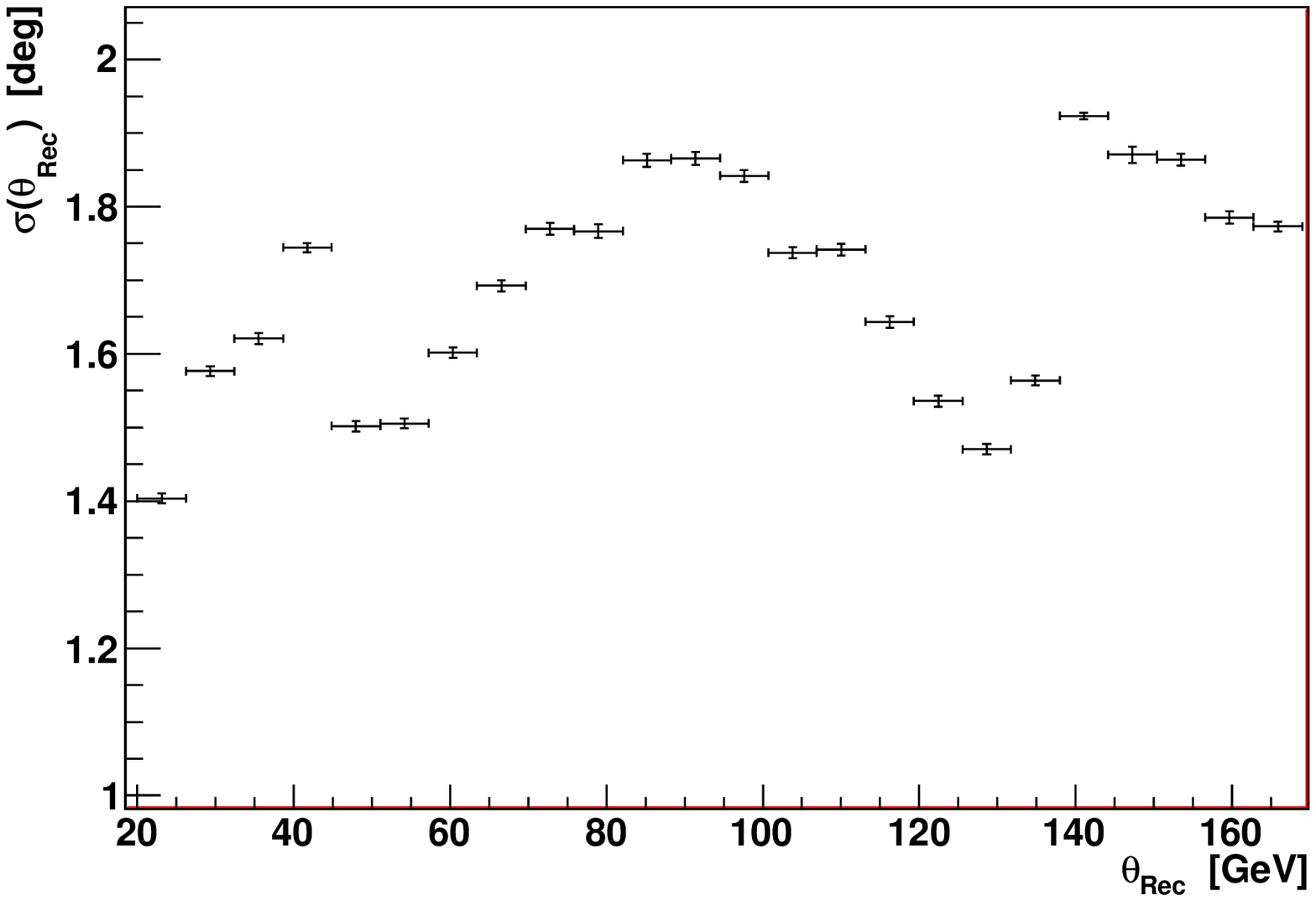}} \label{fig:CDErrorCheckTB}}\quad
\subfigure[Error of $\theta_{Rec}$ versus $\phi_{Rec}$.]{\fbox{\includegraphics[width=0.28\textwidth]{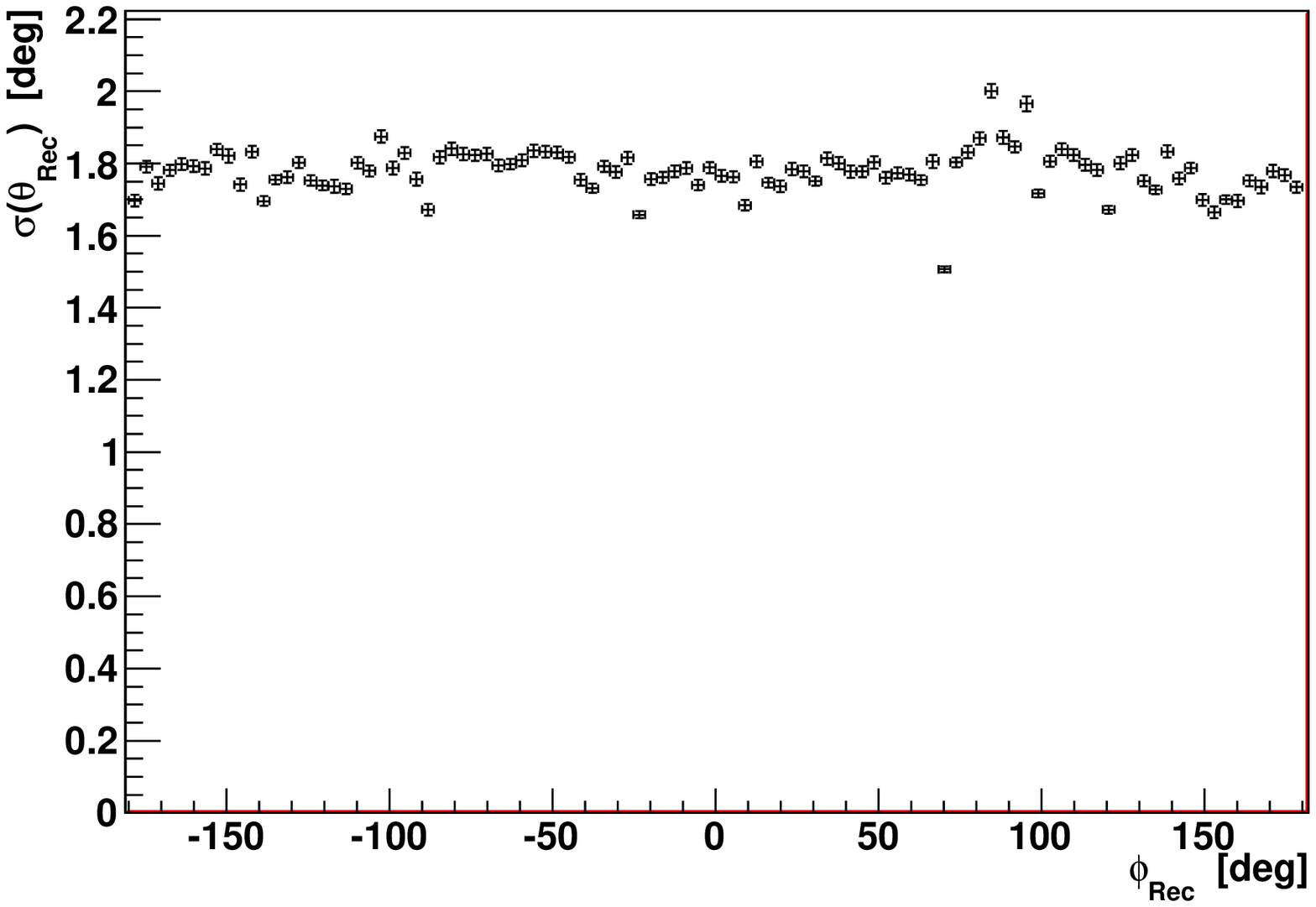}} \label{fig:CDErrorCheckTC}}\\

\subfigure[Error of $\phi_{Rec}$ versus energy.]{\fbox{\includegraphics[width=0.28\textwidth]{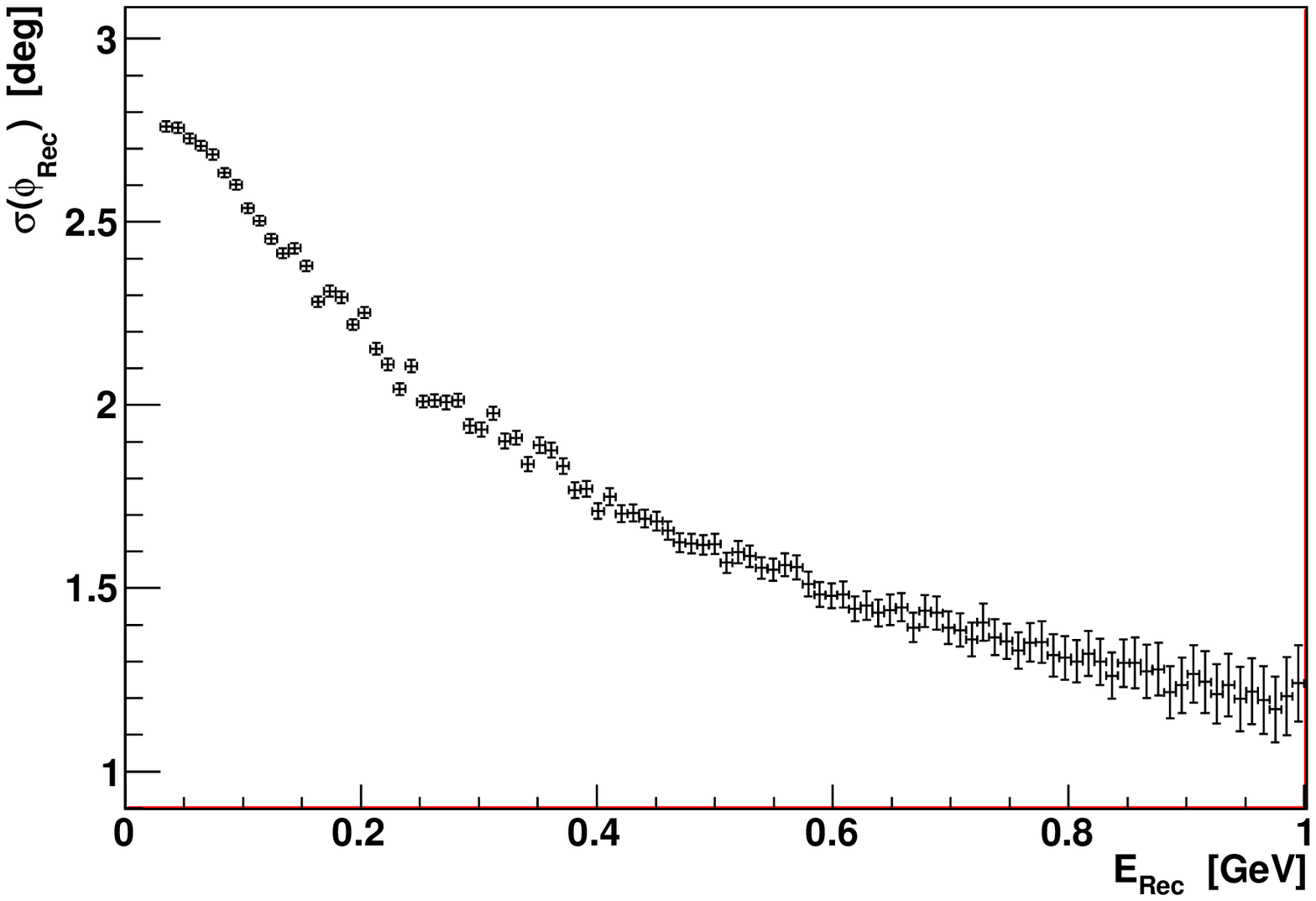}} \label{fig:CDErrorCheckPA}}\quad
\subfigure[Error of $\phi_{Rec}$ versus $\theta_{Rec}$.]{\fbox{\includegraphics[width=0.28\textwidth]{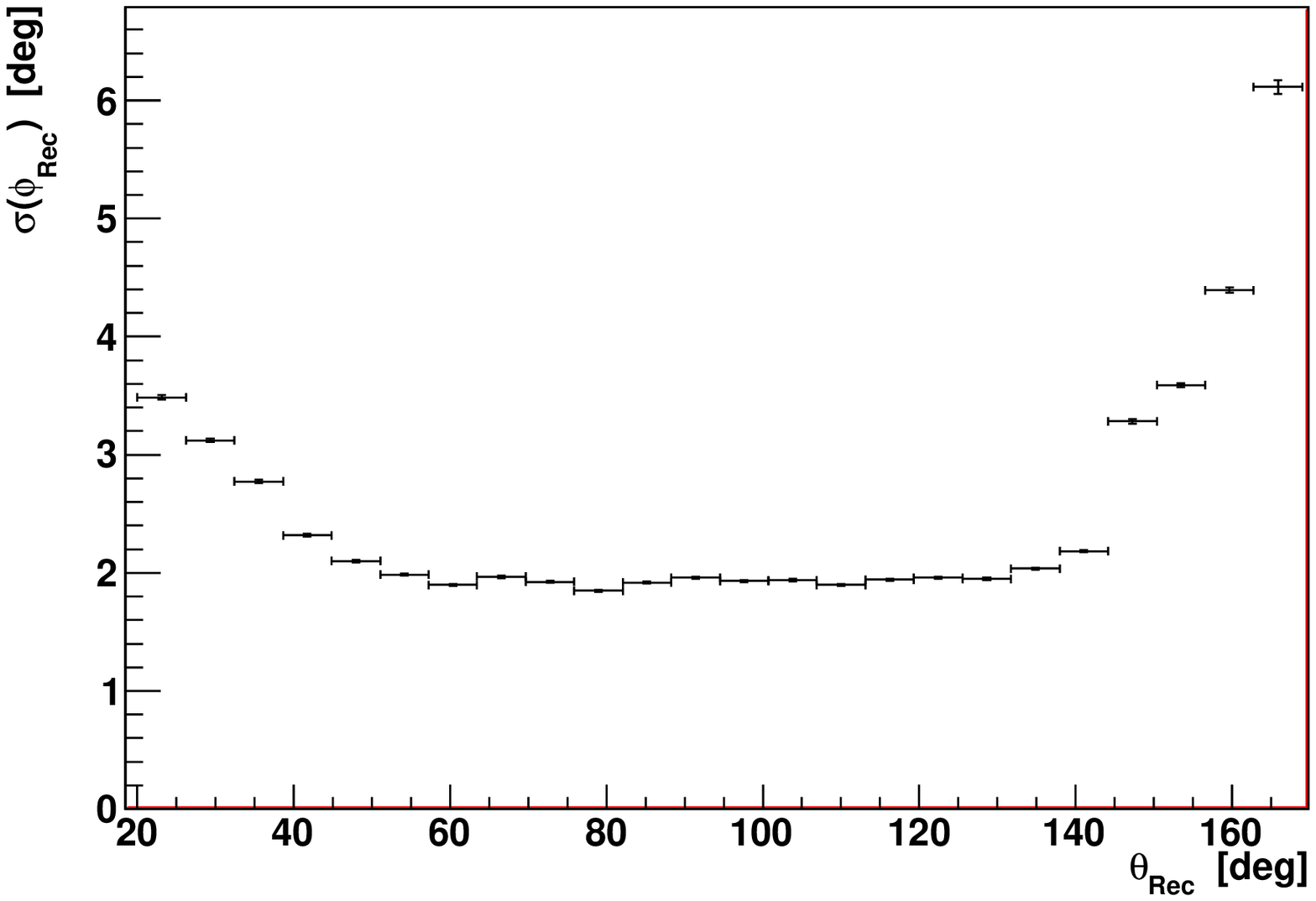}} \label{fig:CDErrorCheckPB}}\quad
\subfigure[Error of $\phi_{Rec}$ versus $\phi_{Rec}$.]{\fbox{\includegraphics[width=0.28\textwidth]{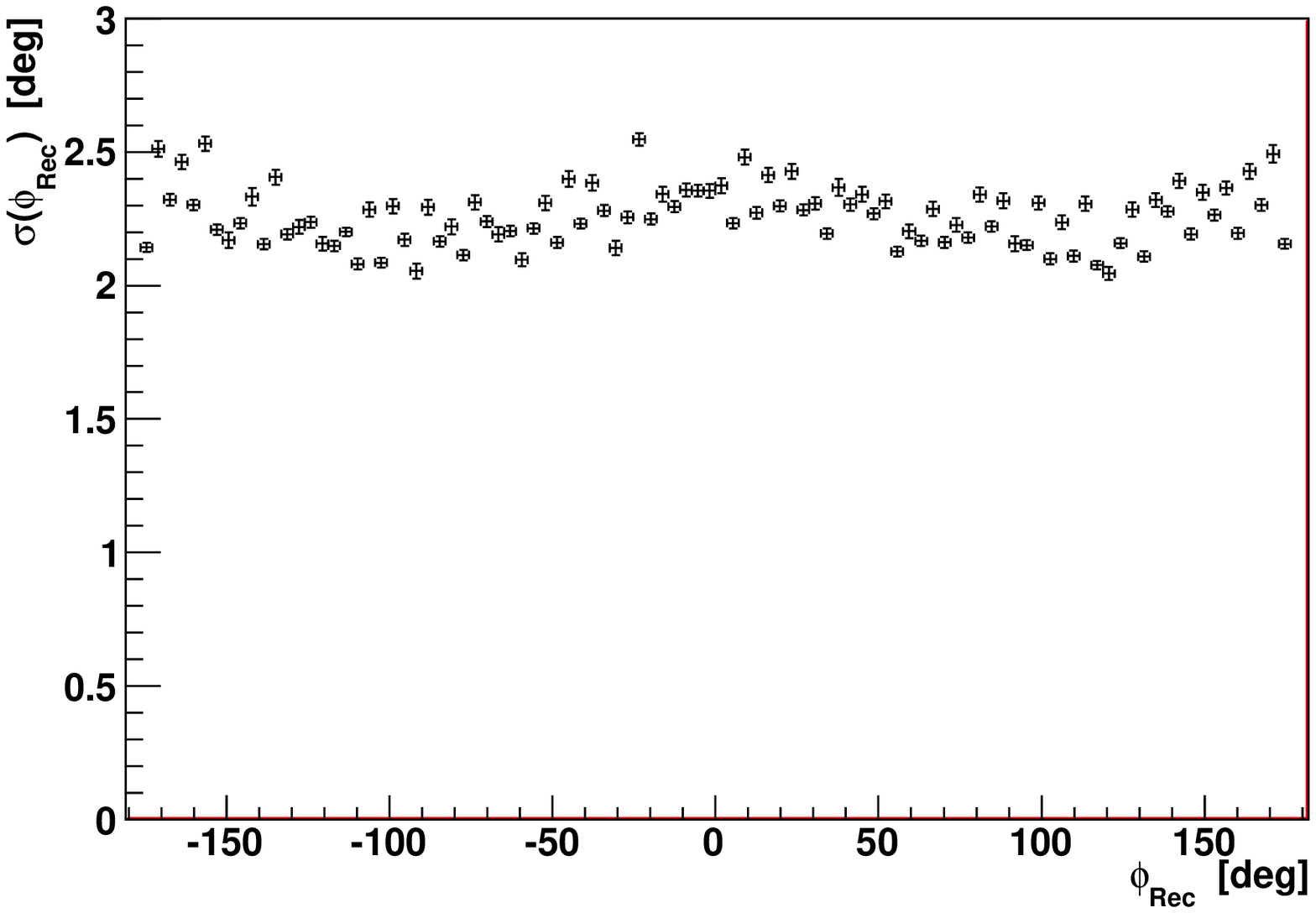}} \label{fig:CDErrorCheckPC}}
}
\caption{Dependences of the errors for the photons in the CD detector. It is see that the errors do not depend of the azimuthal angle $\phi_{Rec}$. Results of the Monte-Carlo simulations. }
\label{fig:CDErrorCheck}
\end{sidewaysfigure}

\begin{sidewaysfigure}[t!bp]
\centering
{
\subfigure[Error of photon energy.]{\fbox{\includegraphics[width=0.45\textwidth]{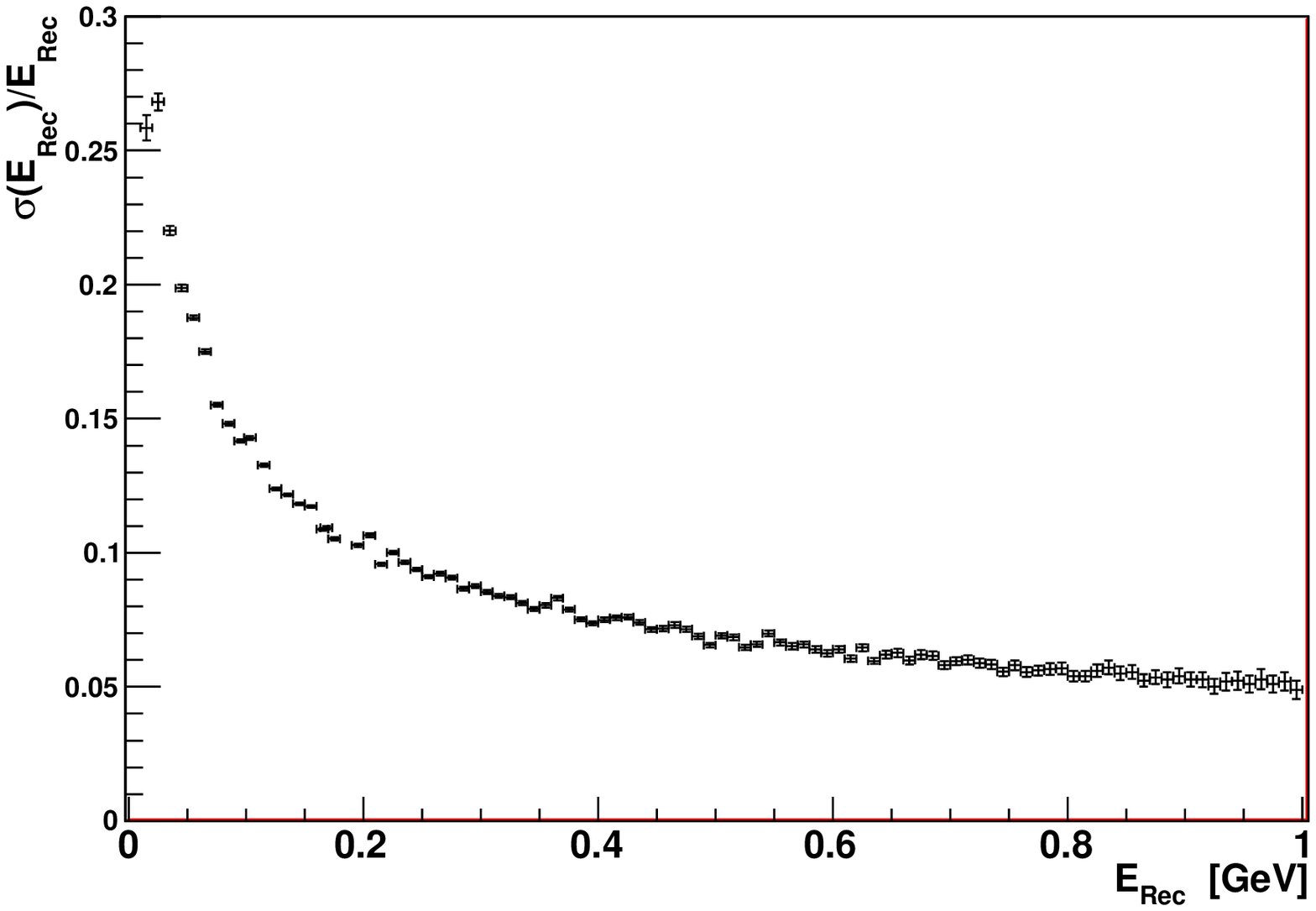}} \label{fig:CDErrorParamEnergy}}\\
\subfigure[Error of photon $\theta_{Rec}$ angle.]{\fbox{\includegraphics[width=0.45\textwidth]{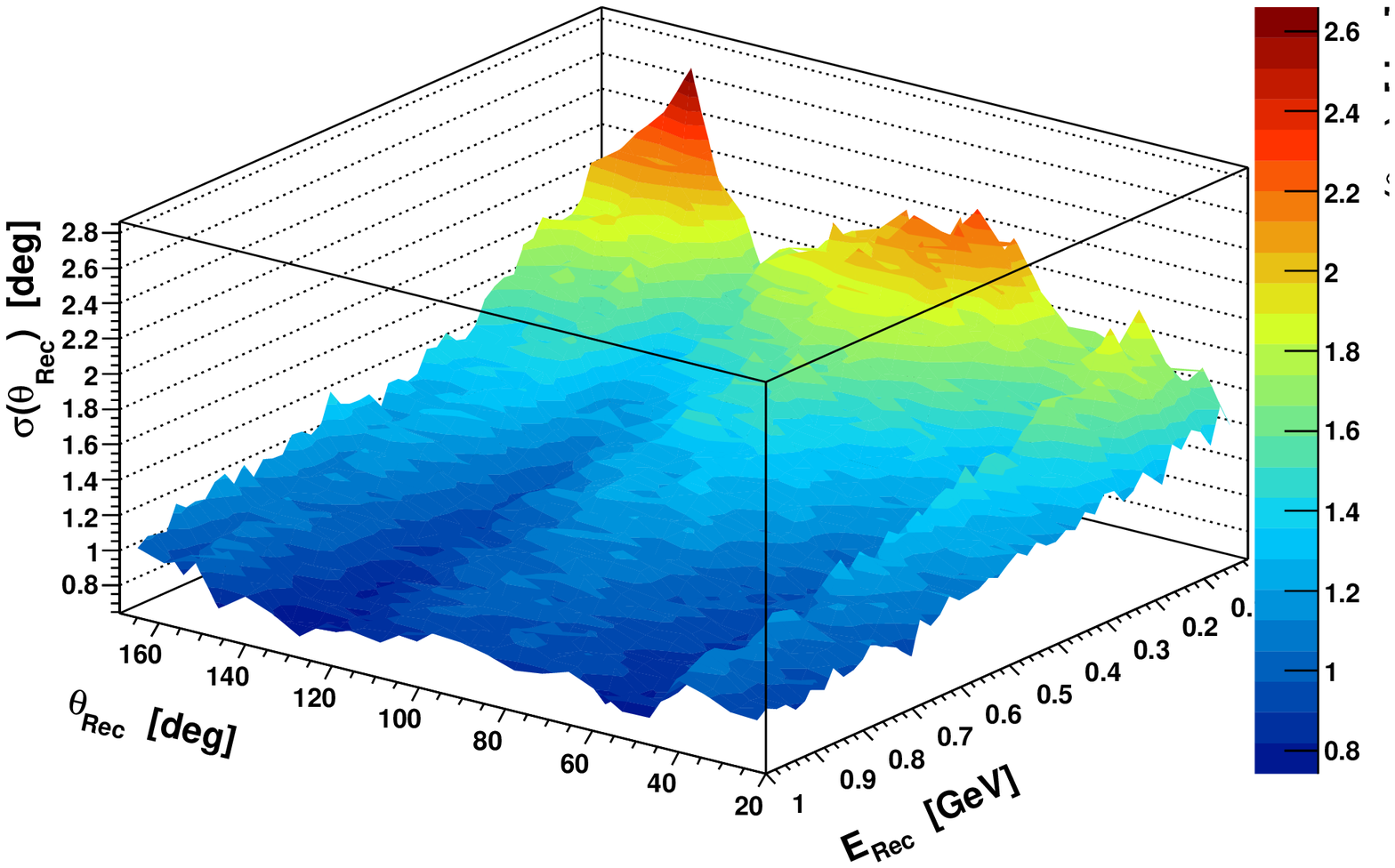}} \label{fig:CDErrorParamTheta}}\quad
\subfigure[Error of photon $\phi_{Rec}$ angle.]{\fbox{\includegraphics[width=0.45\textwidth]{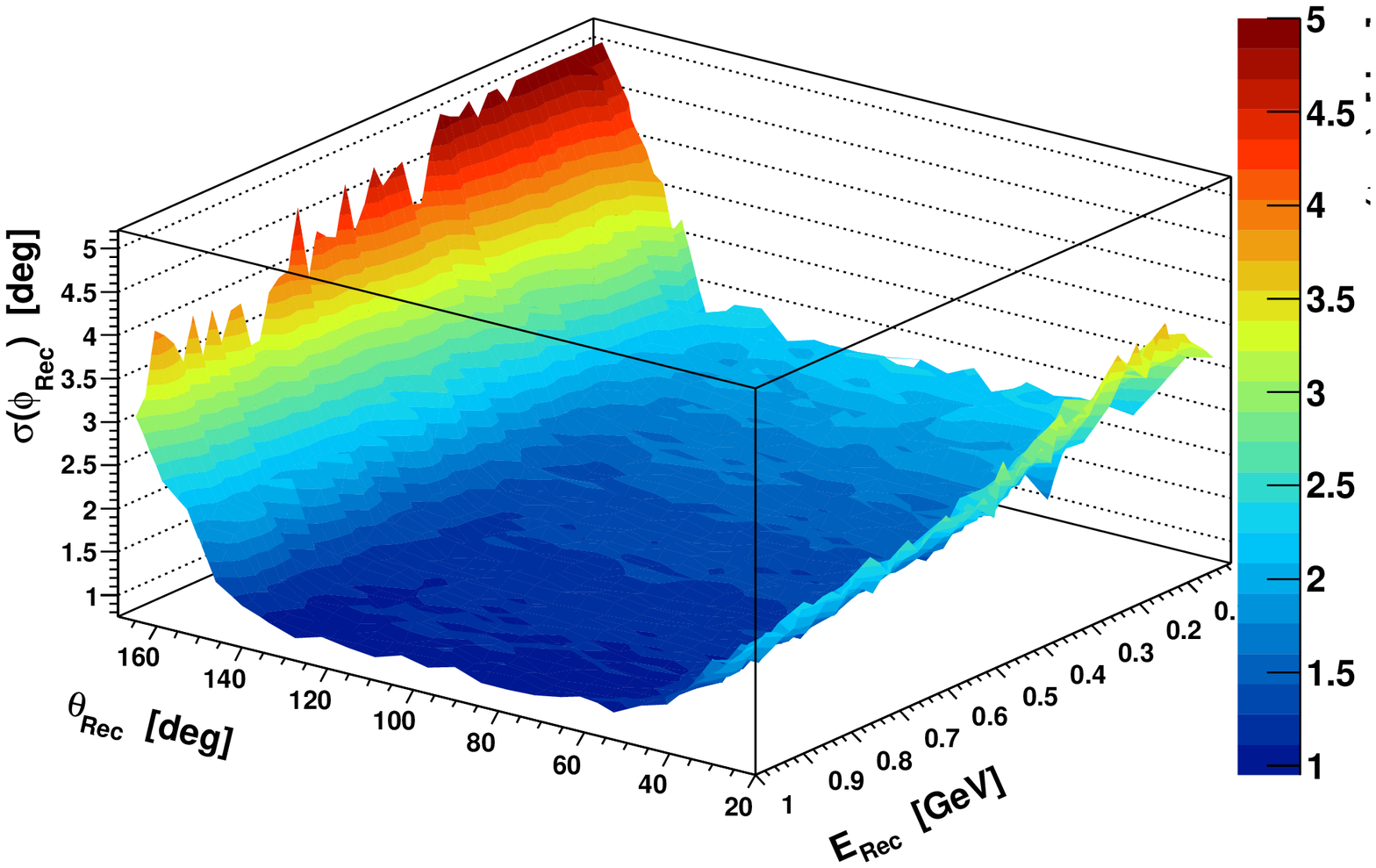}} \label{fig:CDErrorParamPhi}}
}
\caption{The error parametrization for photons in the CD detector used in the kinematic fit. Results of the Monte-Carlo simulations. }
\label{fig:CDErrorParam}
\end{sidewaysfigure}

The variable dependencies of the error for the photons in the CD detector were checked (Fig.~\ref{fig:CDErrorCheck}).
One sees that the errors do not depend on the azimuthal angle $\phi_{Rec}$. 
It is seen that the error of the photon energy depends only on the energy. The error of the photon polar angle $\theta_{Rec}$ depends on the photon energy (the cluster size dependence)
 and on the $\theta_{Rec}$ (the effect of the detector), the two dimensional error parametrization is needed. The error of the photon azimuthal angle $\phi_{Rec}$ depends on the photon energy (the cluster size dependence) and
on the $\theta_{Rec}$ (the effect of the detector), the two dimensional error parametrization is needed.   
The appropriate error parametrization for photons, which were needed for kinematic fit, were prepared (Fig.~\ref{fig:CDErrorParam}).

\begin{figure}[ht!bp]
\centering
{
\subfigure[Error of proton $\theta$ angle.]{\fbox{\includegraphics[width=0.7\textwidth]{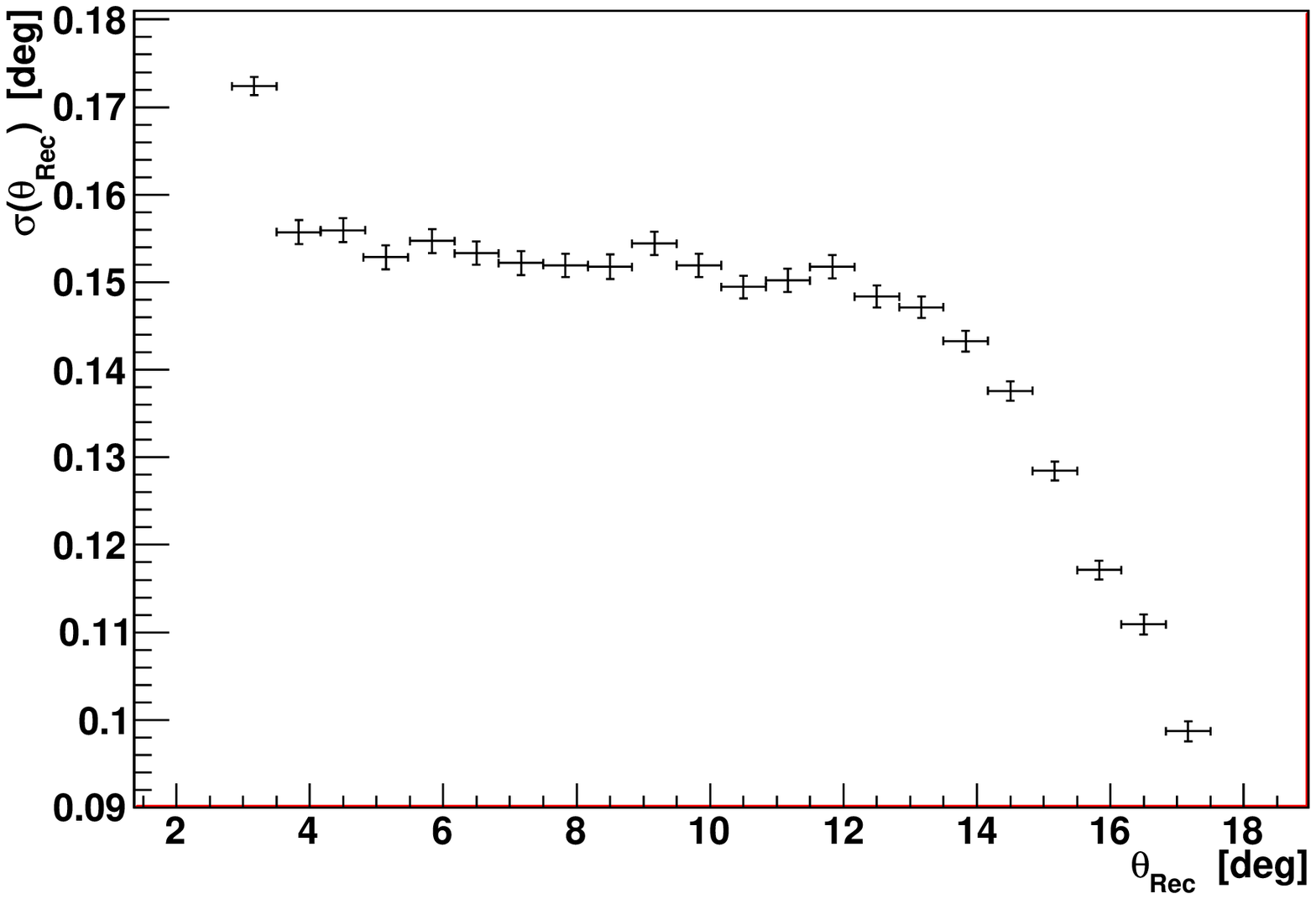}} \label{fig:FDErrorParamTheta}}\\
\subfigure[Error of proton $\phi$ angle.]{\fbox{\includegraphics[width=0.7\textwidth]{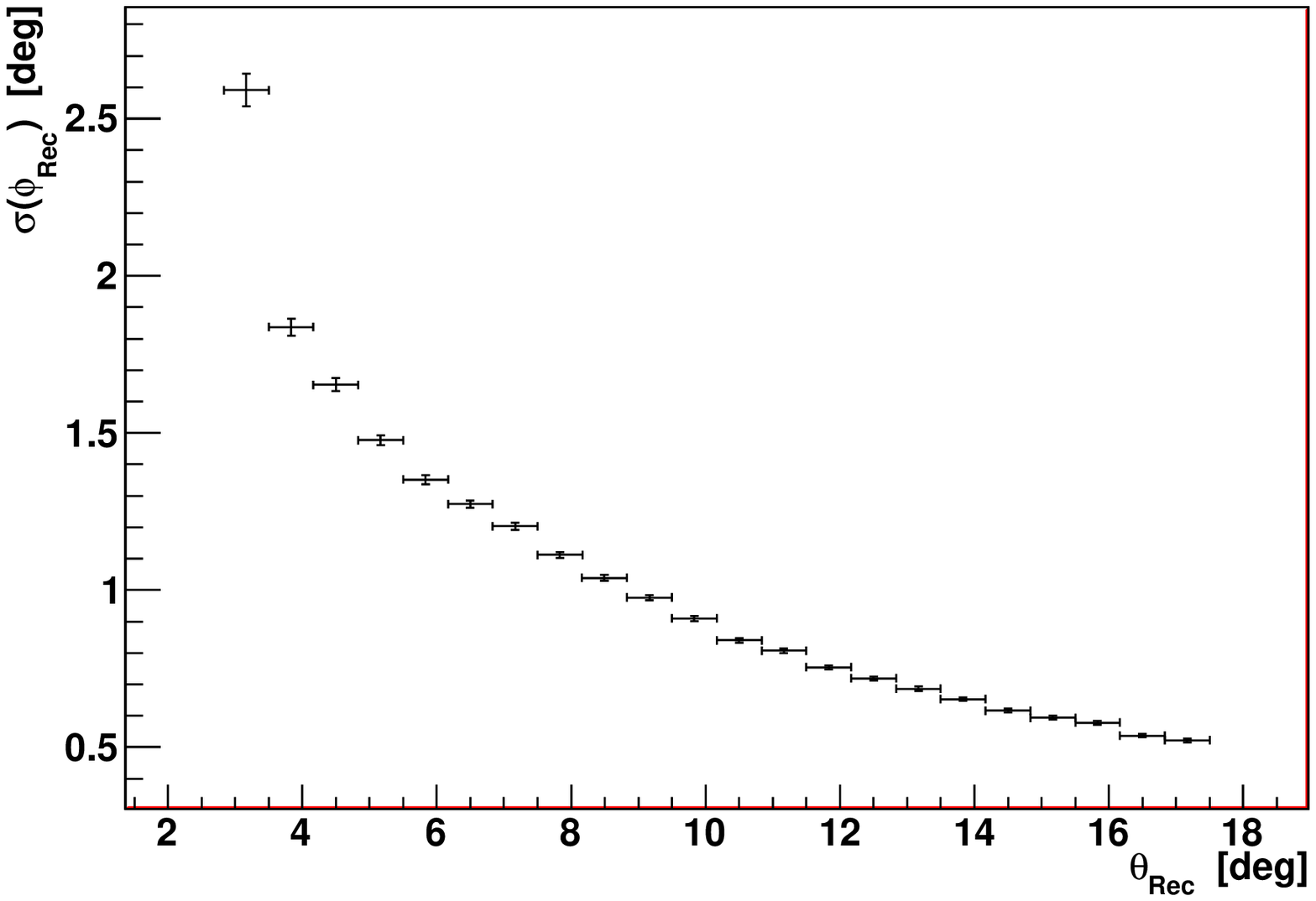}} \label{fig:FDErrorParamPhi}}
}
\caption{The error parametrization for protons in the FD detector used in the kinematic fit. Results of the Monte-Carlo simulations.}
\label{fig:FDErrorParam}
\end{figure}

For the proton in FD detector the error of the proton polar angle $\theta_{Rec}$ as a function of the $\theta_{Rec}$ 
and the error of the proton azimuthal angle $\phi_{Rec}$ as a function of the $\theta_{Rec}$ were used for kinematic fit as an error parametrization (Fig.~\ref{fig:FDErrorParam}). 

The used error parametrization (Figs.~\ref{fig:CDErrorCheck},~\ref{fig:CDErrorParam},~\ref{fig:FDErrorParam}) are results of the Monte-Carlo simulations using single tracks.

\subsubsection{The diagnostics after the Kinematic Fit}

\paragraph{The statistical relations\\}

\begin{figure}[ht!bp]
\centering
{
\subfigure[Logarithmic scale.]{\fbox{\includegraphics[width=0.7\textwidth]{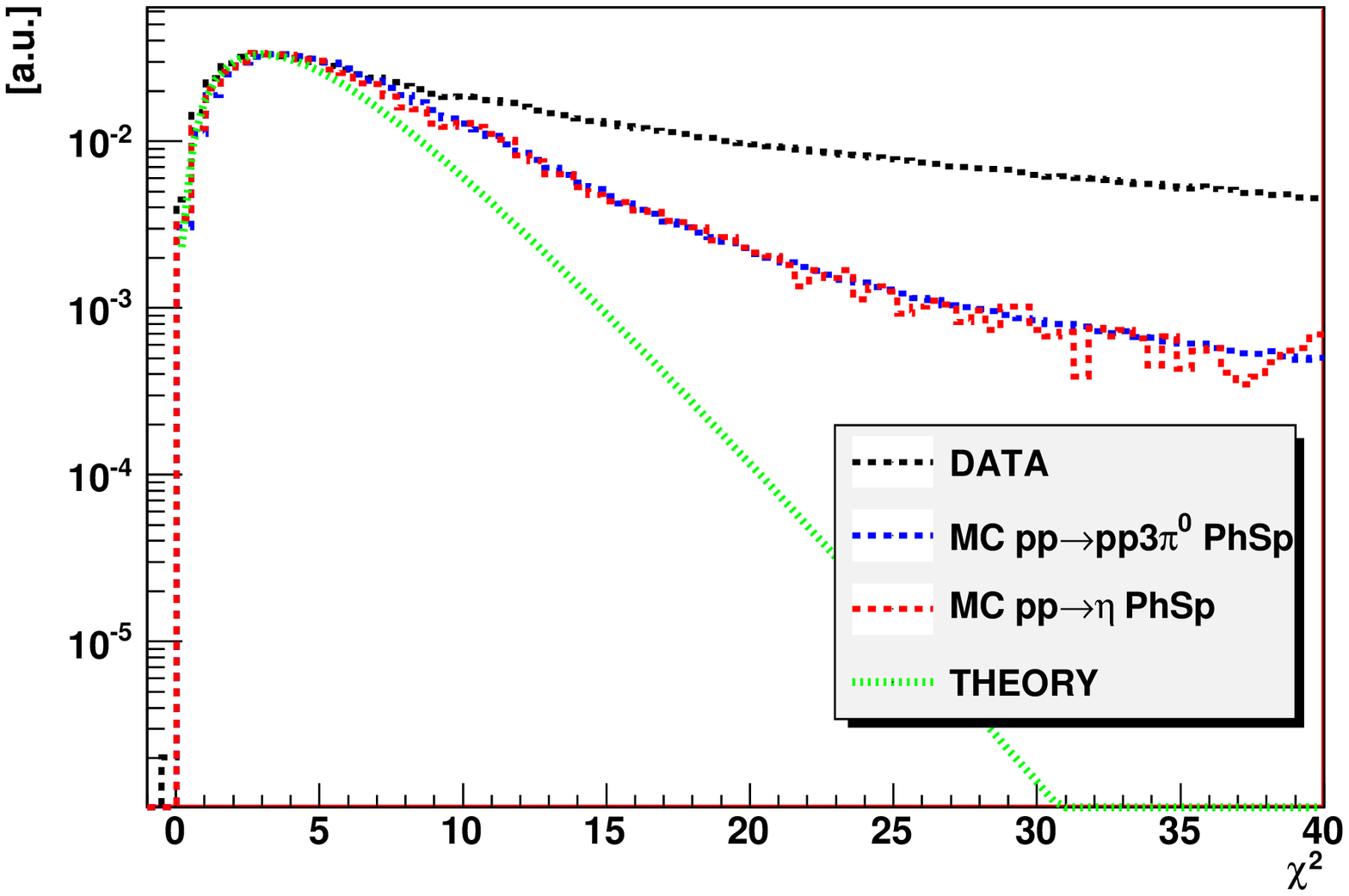}} \label{fig:KFitChi2Log}}\\
\subfigure[Linear scale.]{\fbox{\includegraphics[width=0.7\textwidth]{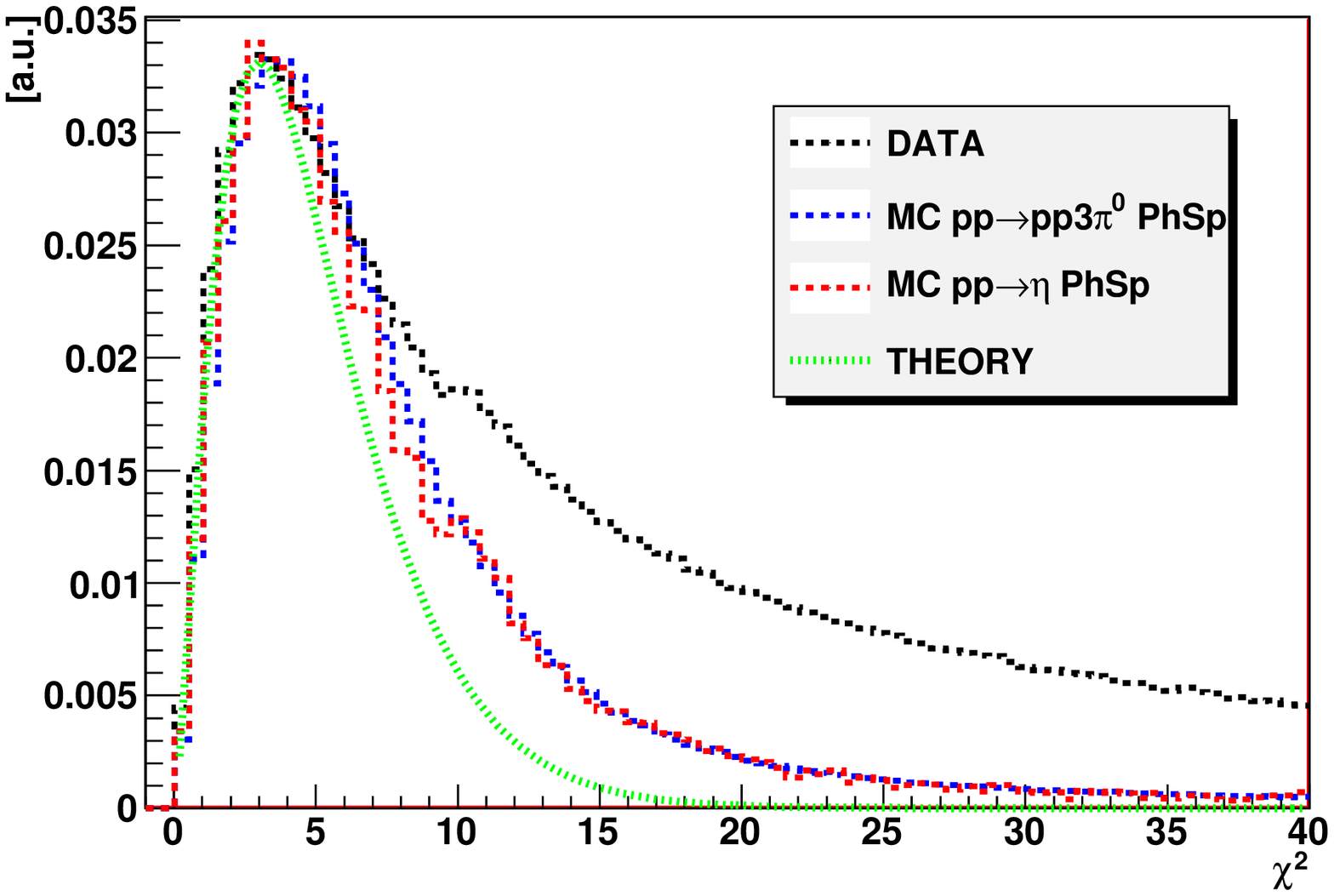}} \label{fig:KFitChi2Lin}}
}
\caption{The $\chi^{2}$ distribution for the kinematic fit of  $ pp \rightarrow pp 3\pi^{0} \rightarrow pp 6\gamma$ reaction.}
\label{fig:KFitChi2}
\end{figure}

\begin{figure}[ht!bp]
\centering
{
\subfigure[Logarithmic scale.]{\fbox{\includegraphics[width=0.7\textwidth]{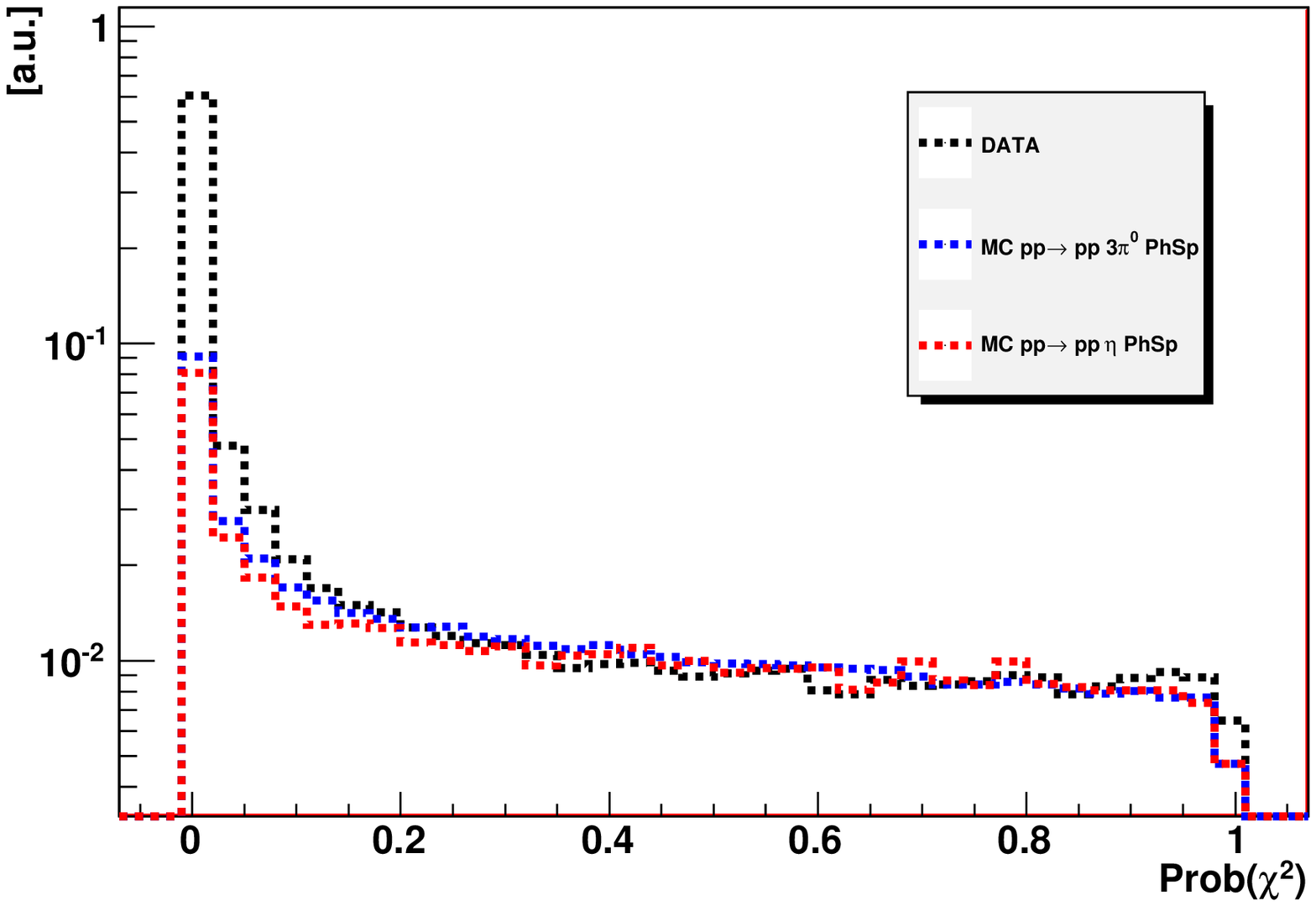}} \label{fig:KFitProbLog}}\\
\subfigure[Linear scale.]{\fbox{\includegraphics[width=0.7\textwidth]{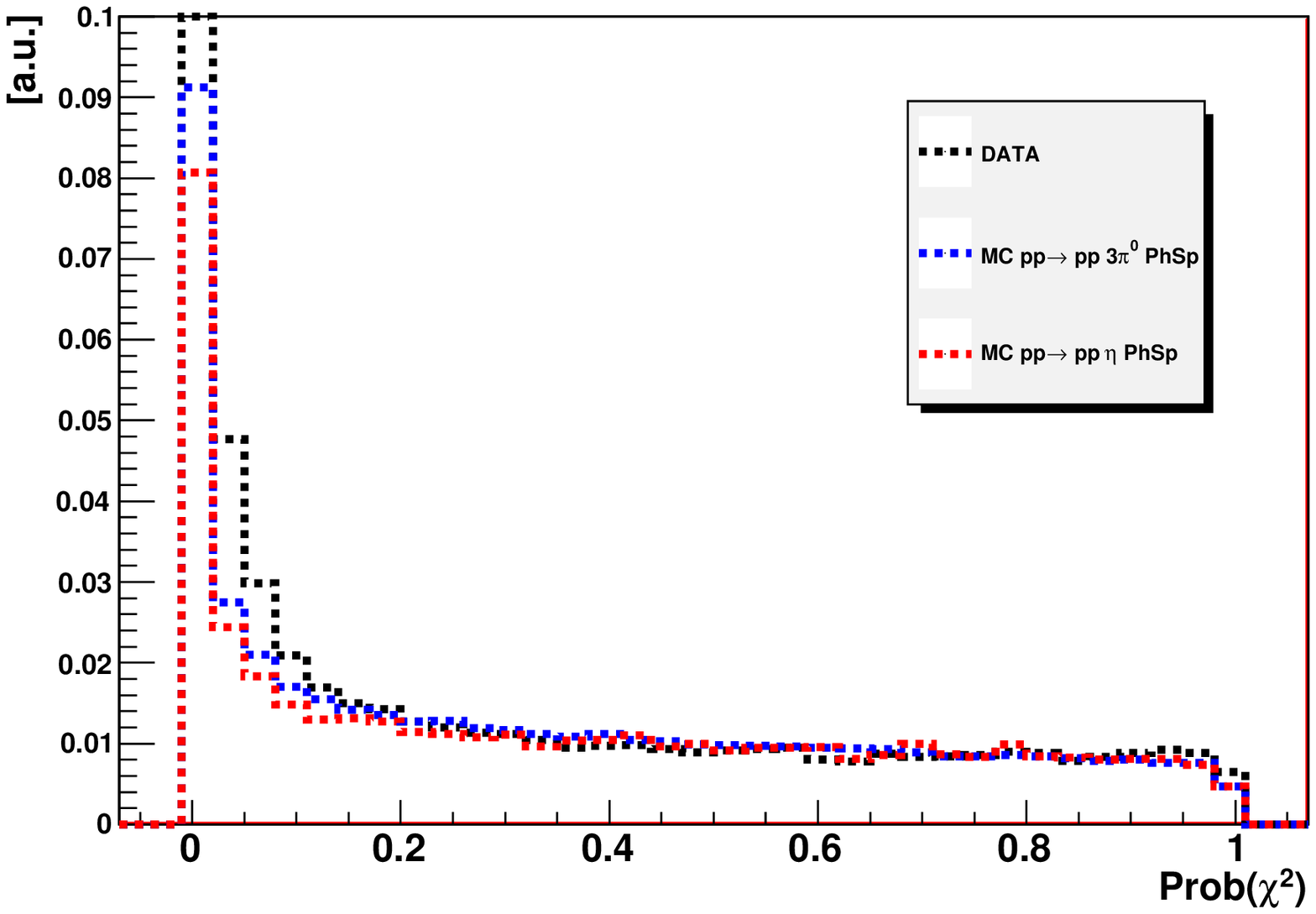}} \label{fig:KFitProbLin}}
}
\caption{The probability distribution for the kinematic fit of  $ pp  \rightarrow pp 3\pi^{0} \rightarrow pp 6\gamma$ reaction.}
\label{fig:KFitProb}
\end{figure}

The $\chi^{2}$ distribution (Fig.~\ref{fig:KFitChi2}) and corresponding probability distribution (Fig.~\ref{fig:KFitProb}) (see Appendix~\ref{appendix:kfit}) 
for the kinematic fit of  $ pp \rightarrow pp 3\pi^{0} \rightarrow pp 6\gamma $ reaction were checked and compared with the Monte-Carlo predictions
of the $ pp \rightarrow pp 3\pi^{0}$(blue curve) and $ pp \rightarrow pp \eta$ (red curve) reaction, also the theoretical $\chi^{2}$ distribution for the (Number of Degrees of Freedom) $NDF=5$ is presented. The difference between the data and Monte-Carlo is visible
 and it is related to the contribution in the data of different kind of processes then $ pp \rightarrow pp 3\pi^{0}$ reaction also the
 non Gaussian error response of the detector plays here an important role. Nevertheless all of the events related to these factors populate
low probability area (near $0$) and they could be rejected by selecting higher probability region where the probability function is getting flat and is consistent with Monte-Carlo.    

The region of the $Prob>0.2$ where the probability function (Fig.~\ref{fig:KFitProb}) is getting flat was selected for the later analysis.
This additional cut ($Prob>0.2$), introduced in order to purify the data, lowers the Total Reconstruction Efficiency (as defined in Eq.~\ref{eq:TotRecEffDefinition}).
Via Monte-Carlo simulations with homogeneously and isotropically populated phase space one can calculate Total Reconstruction Efficiency
including the above cut:  

\begin{equation}
 Tot.Rec.Eff = 2.26\%
\label{eq:TotRecEffKFprob02}
\end{equation}

The following statistical relation for the kinematic fit based on the $\chi^{2}$ method for the residuals should hold (see Appendix~\ref{appendix:kfit}) : 
\begin{equation}
 \sigma_{residual} = \sqrt{\sigma_{m}^{2} - \sigma_{f}^{2}}
\label{eq:residualRelation}
\end{equation}
and
\begin{equation}
 residual = m-f
\end{equation}

where $m$~-~measured value(before kinematic fit) of energy, polar angle $\theta$, azimuthal angle $\phi$; $f$~-~fitted value(after kinematic fit) of energy, polar angle $\theta$, azimuthal angle $\phi$; 
$\sigma_{residual}$~-~the standard deviation of the residual value distribution;$\sigma_{m}$~-~the standard deviation of the measured value (the error of the measured value);
$\sigma_{f}$~-~the standard deviation of the fitted value (the error of the fitted value).
The only possibility to calculate the $\sigma_{m}$ and $\sigma_{f}$ is via Monte-Carlo simulations.
Usually $\sigma_{m}$ and $\sigma_{f}$ might be very similar, the correction of the kinematic fit could be very small. This could lead to the very high
numerical inaccuracy of the difference $\sigma_{m}^{2} - \sigma_{f}^{2}$. To avoid this problem instead of calculation directly this difference and comparing it with the
$\sigma_{residual}$ it could be done other way by changing the (Eq.~\ref{eq:residualRelation}) to the form:
\begin{equation}
 \sigma_{f} = \sqrt{\sigma_{m}^{2} - \sigma_{residual}^{2}}
\label{eq:residualRelationChanged}
\end{equation}

Now one compares the $\sqrt{\sigma_{m}^{2} - \sigma_{residual}^{2}}$ with the $\sigma_{f}$, here always $\sigma_{m}$ will be different
 from $\sigma_{residual}$.

The relation (Eq.~\ref{eq:residualRelationChanged}) was checked for the kinematic fit of the data.
The $\sigma_{m}$ was extracted by fitting true minus measured(reconstructed) value distribution with the Gaussian function using Monte-Carlo simulation.
The $\sigma_{f}$ was extracted by fitting true minus fitted value distribution with the Gaussian function using Monte-Carlo simulation.
The $\sigma_{residual}$ was derived from the experimental data by fitting Gaussian function to the measured(reconstructed) minus fitted value distribution.
The estimation of the parameters was done for all fitted variables i.e. energy of the photons, polar angle $\theta$ of photons and protons and  
 azimuthal angle $\phi$ of photons and protons. This task was done by means of the Monte-Carlo simulations. 
The distributions used for the parameter extraction (Figs.~\ref{image_PullsEnergyCD},~\ref{image_PullsThetaCD},~\ref{image_PullsPhiCD},~\ref{image_PullsThetaFD},~\ref{image_PullsPhiFD}) are all centered at $0$ value.
The measured minus fitted distribution, (right plots on Figs.~\ref{image_PullsEnergyCD},~\ref{image_PullsThetaCD},~\ref{image_PullsPhiCD},~\ref{image_PullsThetaFD},~\ref{image_PullsPhiFD}), shows good agreement between experimental data and Monte-Carlo simulation.
   
\myTable{
\begin{tabular}{|r||c|c|c|c|c|c|}
\hline
& from MC & from MC & from Exp. & \multirow{2}{*}{ $\sqrt{\sigma_{m}^{2} - \sigma_{residual}^{2}}$} &\multirow{2}{*}{$\frac{\sigma_{f}}{\sqrt{\sigma_{m}^{2} - \sigma_{residual}^{2}}}$} \\
 & $\sigma_{m}$ & $\sigma_{f}$ & $\sigma_{residual}$ & & \\ 
\hline
\hline
$E_{photon}$~~~~~~~~~& 0.1081 & 0.0770& 0.0744 &0.0785 & \textbf{0.9812} \\
\hline
$\theta_{photon}$~~[deg] & 1.5270 & 1.4995& 0.2030 & 1.5138& \textbf{0.9906} \\
\hline
$\phi_{photon}$~~[deg] & 1.8920 & 1.8200& 0.2198& 1.8792& \textbf{0.9685} \\
\hline
$\theta_{proton}$~~[deg] & 0.0996 & 0.0999& 0.0087 & 0.0992& \textbf{1.0070} \\
\hline
$\phi_{proton}$~~[deg] & 0.4908 & 0.5068& 0.1143 & 0.4773& \textbf{1.0618} \\
\hline
\end{tabular}
}{The check of the statistical relation for the kinematic fit.}{tab:residualRelation}

The results of the comparison are presented in \myTabRef{tab:residualRelation}. It is seen, in the last column of the table, that the $\sigma_{f}$
differ from the calculated one $\sqrt{\sigma_{m}^{2} - \sigma_{residual}^{2}}$ by a few percent, the ratio is almost equal $1$.
This proves the correctness of the kinematic fit procedure as well as the properly derived error parameterization.   
\newline

\myFrameHugeFigure{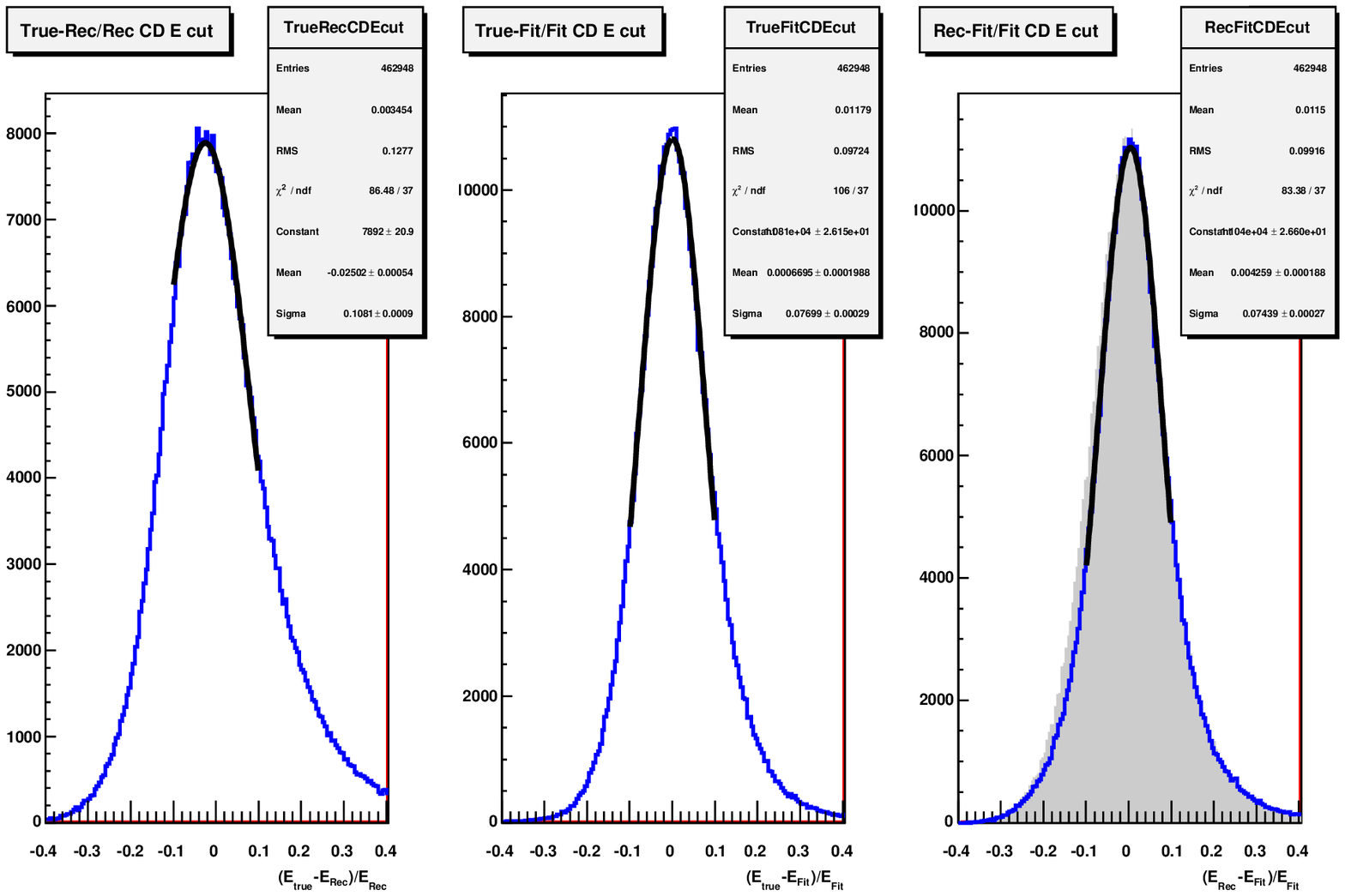}{Photon Energy. \textit{left}: Monte-Carlo Simulation, true minus reconstructed value divided
 by the reconstructed value.\textit{center}: Monte-Carlo Simulation, true minus fitted value divided by the fitted value.
\textit{right}: Reconstructed minus fitted value divided by the fitted value,gray area corresponds to the experimental data,
 blue line to the Monte-Carlo simulation. Black line corresponds to the fit of the Gaussian function.}{Photon Energy, Kinematic Fit Check.}

\myFrameHugeFigure{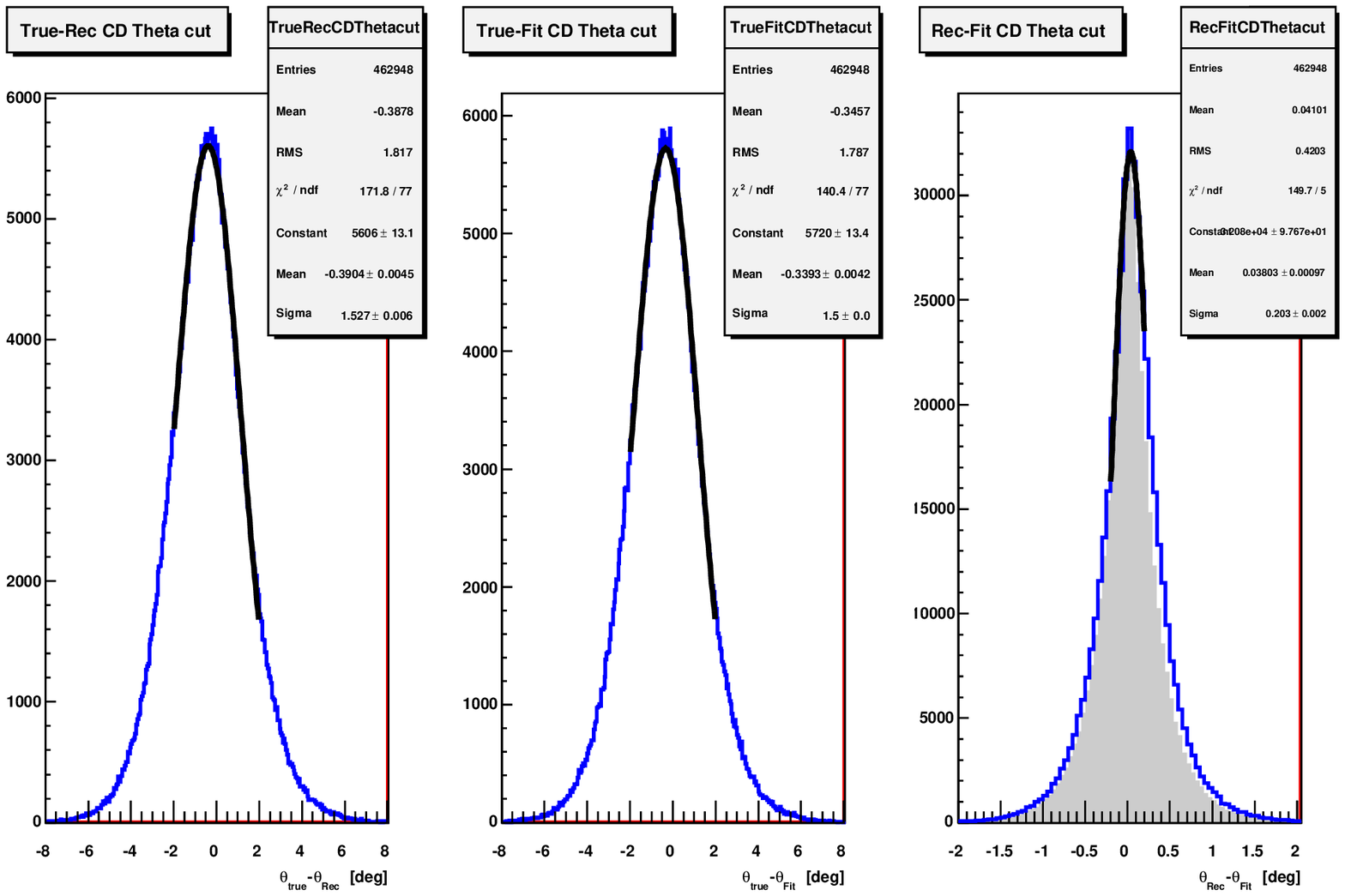}{Photon polar angle $\theta$. \textit{left}: Monte-Carlo Simulation, true minus reconstructed value.\textit{center}: Monte-Carlo Simulation, true minus fitted value.
\textit{right}: Reconstructed minus fitted value,gray area corresponds to the experimental data,
 blue line to the Monte-Carlo simulation. Black line corresponds to the fit of the Gaussian function.}{Photon polar angle, Kinematic Fit Check.}

\myFrameHugeFigure{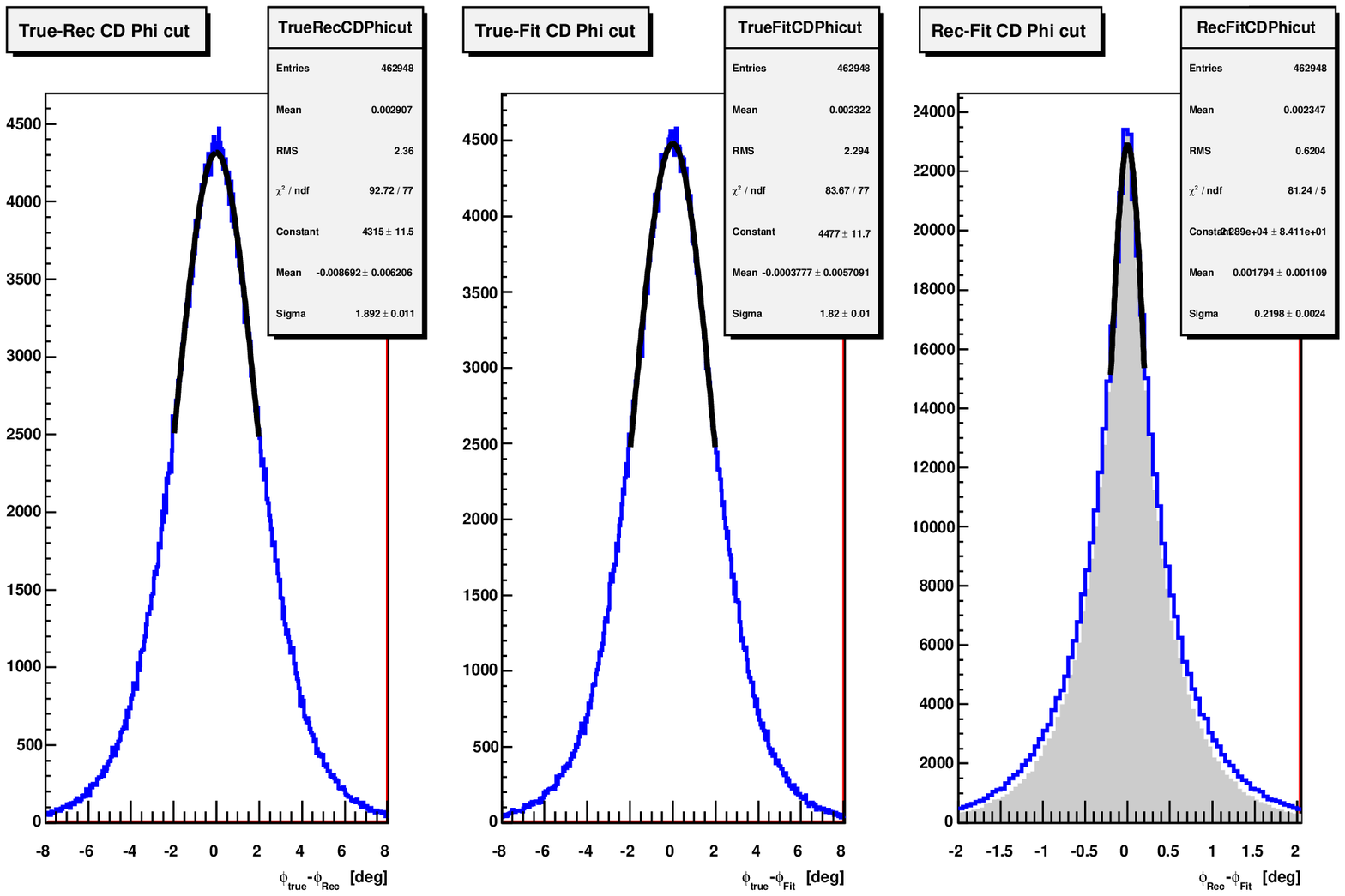}{Photon azimuthal angle $\phi$. \textit{left}: Monte-Carlo Simulation, true minus reconstructed value.\textit{center}: Monte-Carlo Simulation, true minus fitted value.
\textit{right}: Reconstructed minus fitted value,gray area corresponds to the experimental data,
 blue line to the Monte-Carlo simulation. Black line corresponds to the fit of the Gaussian function.}{Photon azimuthal angle, Kinematic Fit Check.}

\myFrameHugeFigure{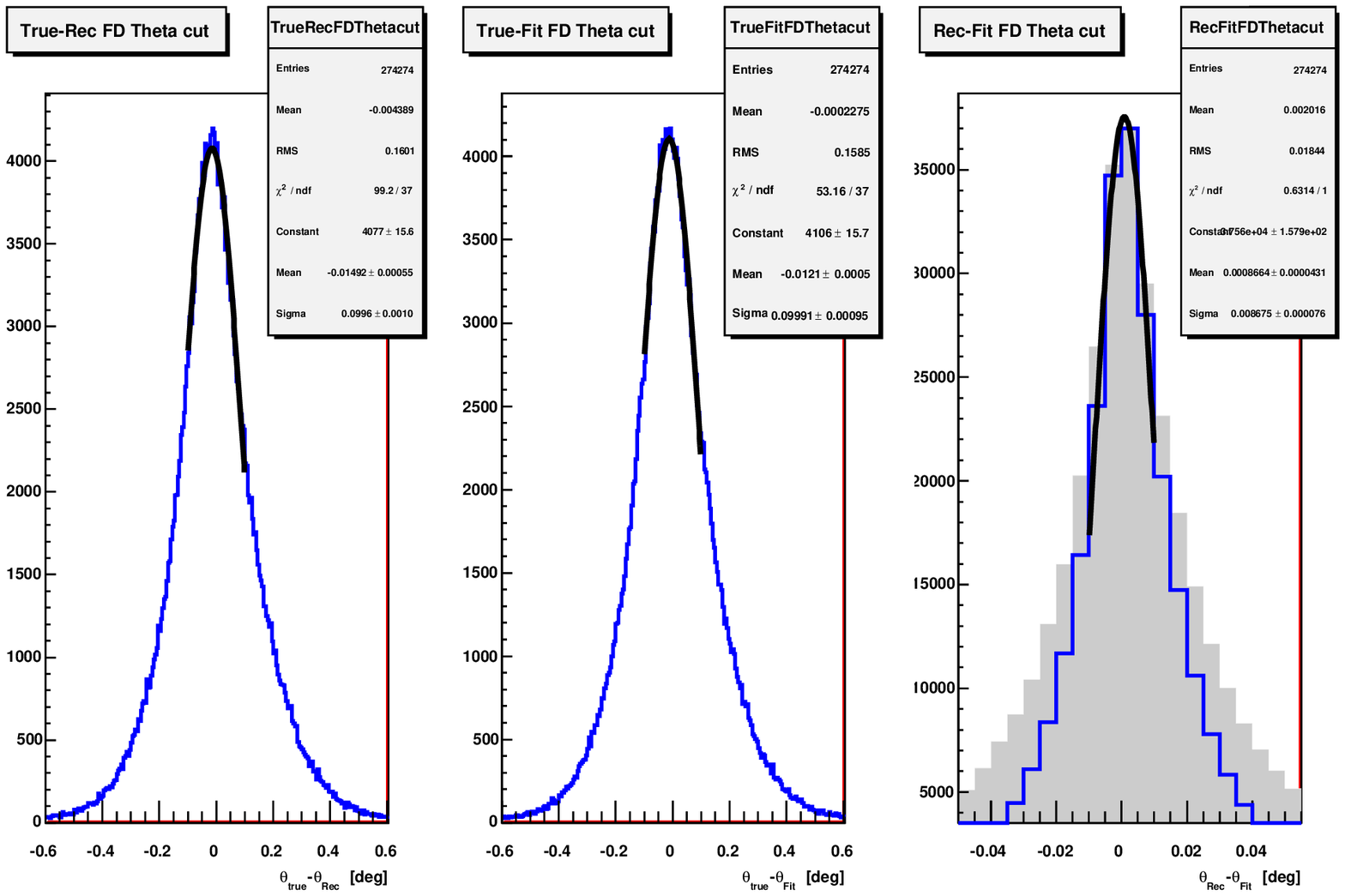}{Proton polar angle $\theta$. \textit{left}: Monte-Carlo Simulation, true minus reconstructed value.\textit{center}: Monte-Carlo Simulation, true minus fitted value.
\textit{right}: Reconstructed minus fitted value,gray area corresponds to the experimental data,
 blue line to the Monte-Carlo simulation. Black line corresponds to the fit of the Gaussian function.}{Proton polar angle, Kinematic Fit Check.}

\myFrameHugeFigure{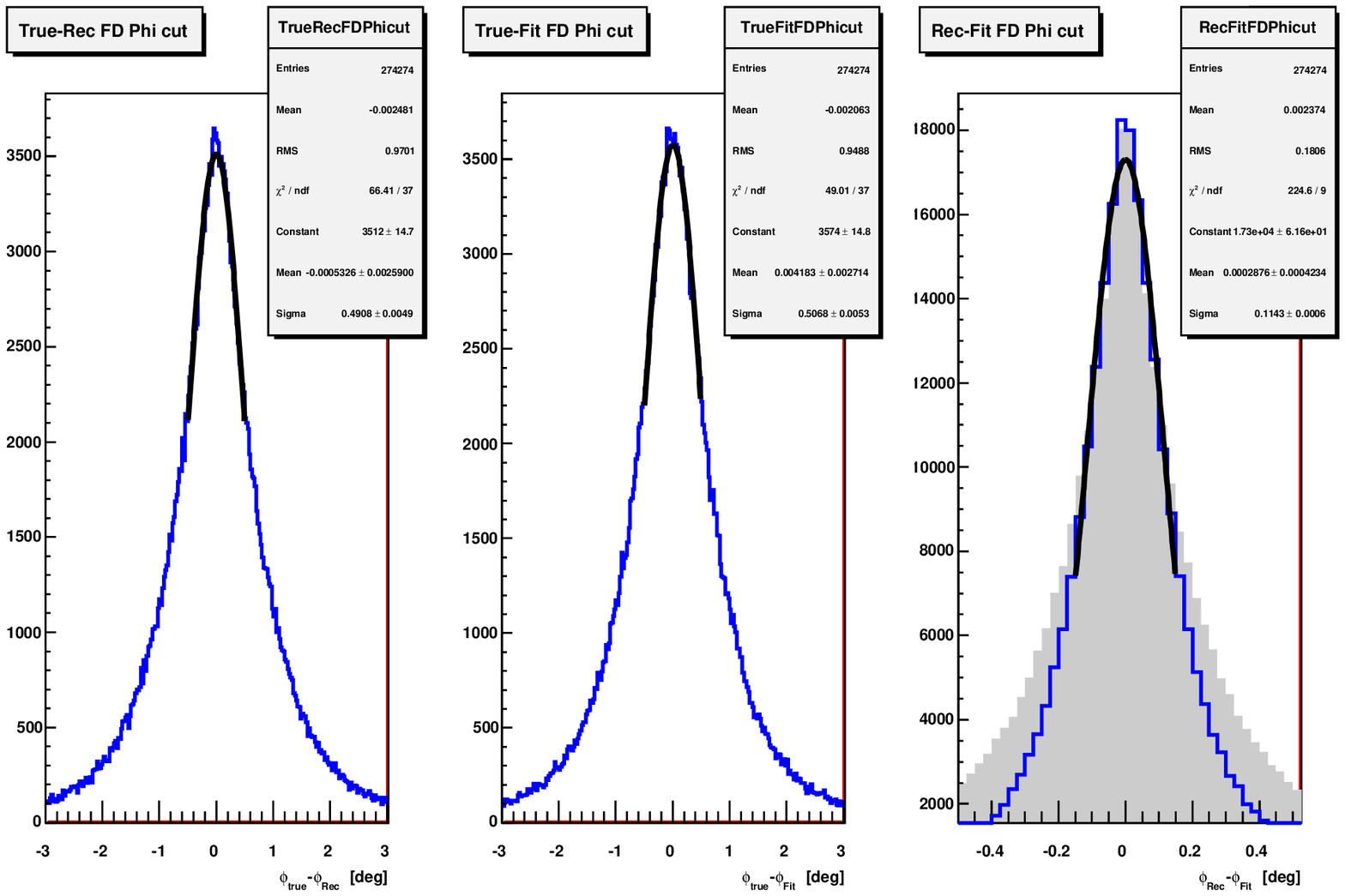}{Proton azimuthal angle $\phi$. \textit{left}: Monte-Carlo Simulation, true minus reconstructed value.\textit{center}: Monte-Carlo Simulation, true minus fitted value.
\textit{right}: Reconstructed minus fitted value,gray area corresponds to the experimental data,
 blue line to the Monte-Carlo simulation. Black line corresponds to the fit of the Gaussian function.}{Proton azimuthal angle, Kinematic Fit Check.}

\newpage
\paragraph{The $\pi^{0}$'s and photons\\}

\begin{sidewaysfigure}[t!bp]
\centering
{
\subfigure[$Prob<0.2$, Experimental Data.]{\fbox{\includegraphics[width=0.45\textwidth]{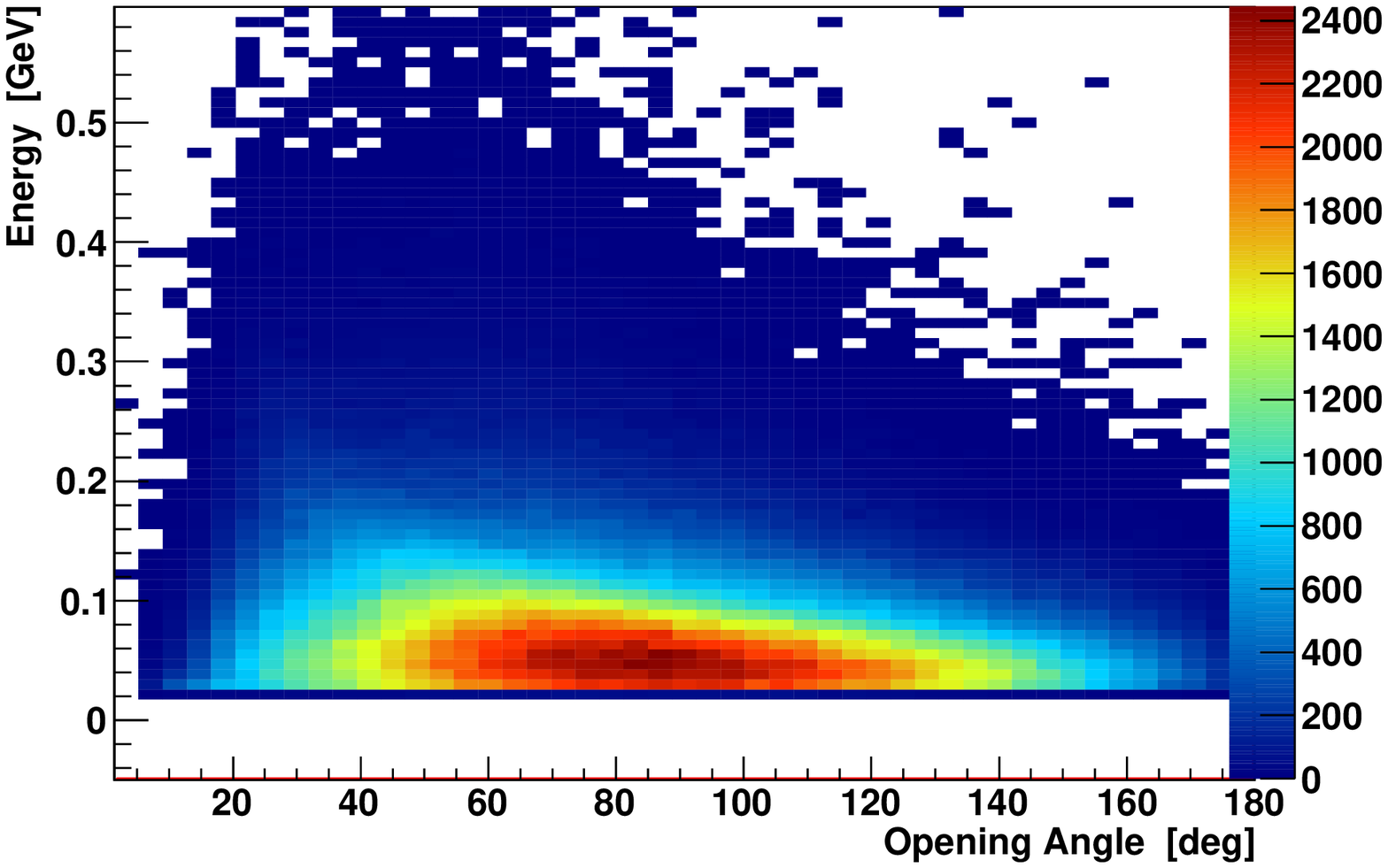}} \label{fig:CDSplitOffsSmall}}\quad
\subfigure[$Prob>0.2$, Experimental Data.]{\fbox{\includegraphics[width=0.45\textwidth]{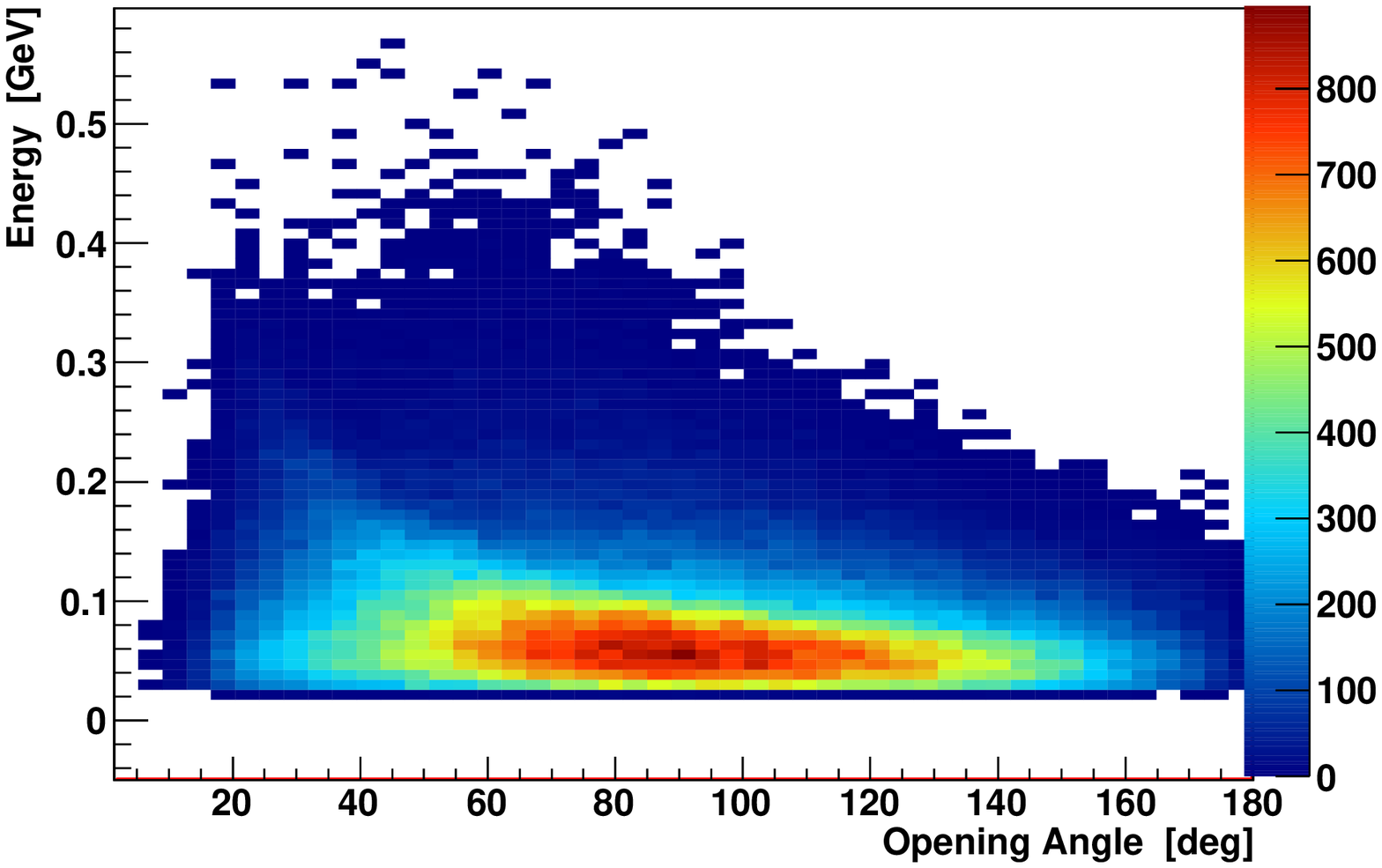}} \label{fig:CDSplitOffsBig}}\\
\subfigure[$Prob<0.2$, Monte-Carlo simulation.]{\fbox{\includegraphics[width=0.45\textwidth]{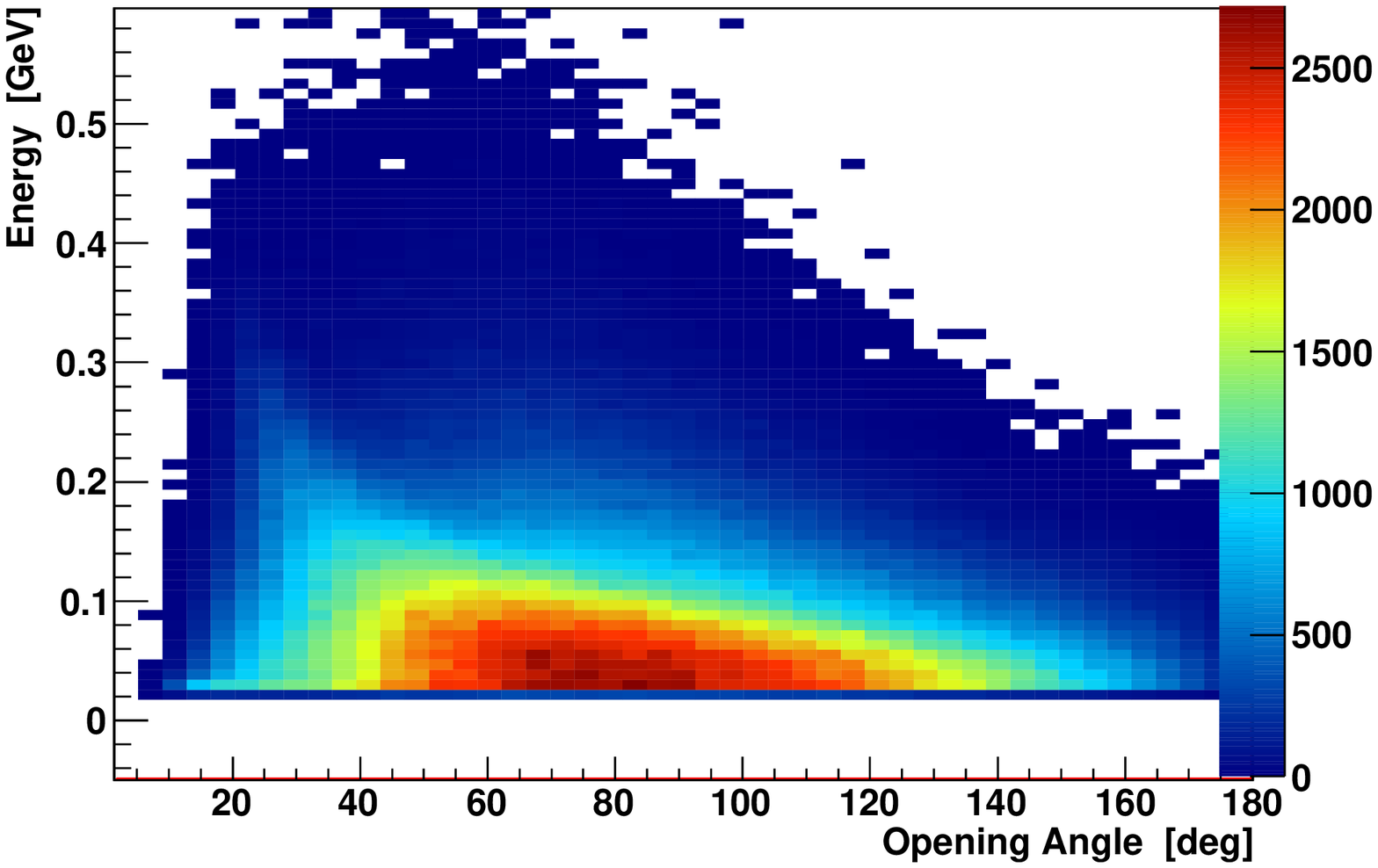}} \label{fig:CDSplitOffsSmallMC}}\quad
\subfigure[$Prob>0.2$, Monte-Carlo simulation.]{\fbox{\includegraphics[width=0.45\textwidth]{OpeningAngleProbBig01.eps}} \label{fig:CDSplitOffsBigMC}}

}
\caption{Energy of the lowest energetic photon from a photon pair versus opening angle between the two photons. The area of the small energies and small angles,
 where one would expect the hadronic split-offs, is not populated. The experimental data are well described by the Monte-Carlo simulation.}
\label{fig:CDSplitOffs}
\end{sidewaysfigure}

To see what is the hadronic split off level for the photon in the electromagnetic calorimeter in the CD detector,
 the energy of the lowest energetic photon from a photon pair versus opening angle between the two photons were plotted for two probability regions $Prob<0.2$ and $Prob>0.2$,
the experimental data were compared with a Monte-Carlo simulation of $ pp \rightarrow pp 3\pi^{0}$ (Fig.~\ref{fig:CDSplitOffs}). The area of the small energies and small angles,
 where one should expect the hadronic split-offs\footnote{secondary interaction of the photon leading to spurious cluster in calorimeter}, is not populated. This means that there is low level of the hadronic split offs for both probability cut regions. 
The Monte-Carlo describes well the experimental data.

\myFrameSmallFigure{IMggAfterKFIT}{Invariant Mass of a photon pair forming $\pi^{0}$ selected by the combinatorics for $Prob>0.2$. Grey area corresponds to the experimental data, blue line to the Monte-Carlo simulation.}{Invariant Mass of a photon pair forming $\pi^{0}$ selected by the Kinematic Fit for $Prob>0.2$.}

The identification and reconstruction of $\pi^{0}$ mesons from photons in CD detector by the kinematic fit was checked. The invariant mass of two photons forming $\pi^{0}$ which were selected
by the combinatorics was plotted, for the probability region $Prob>0.2$ both for experimental data and Monte-Carlo \myImgRef{IMggAfterKFIT}. The peak position corresponds to the theoretical
 $\pi^{0}$ mass. The experimental spectrum agrees with the Monte-Carlo simulation. 

\begin{figure}[ht!bp]
\centering
{
\subfigure[Experimental Data.]{\fbox{\includegraphics[width=0.7\textwidth]{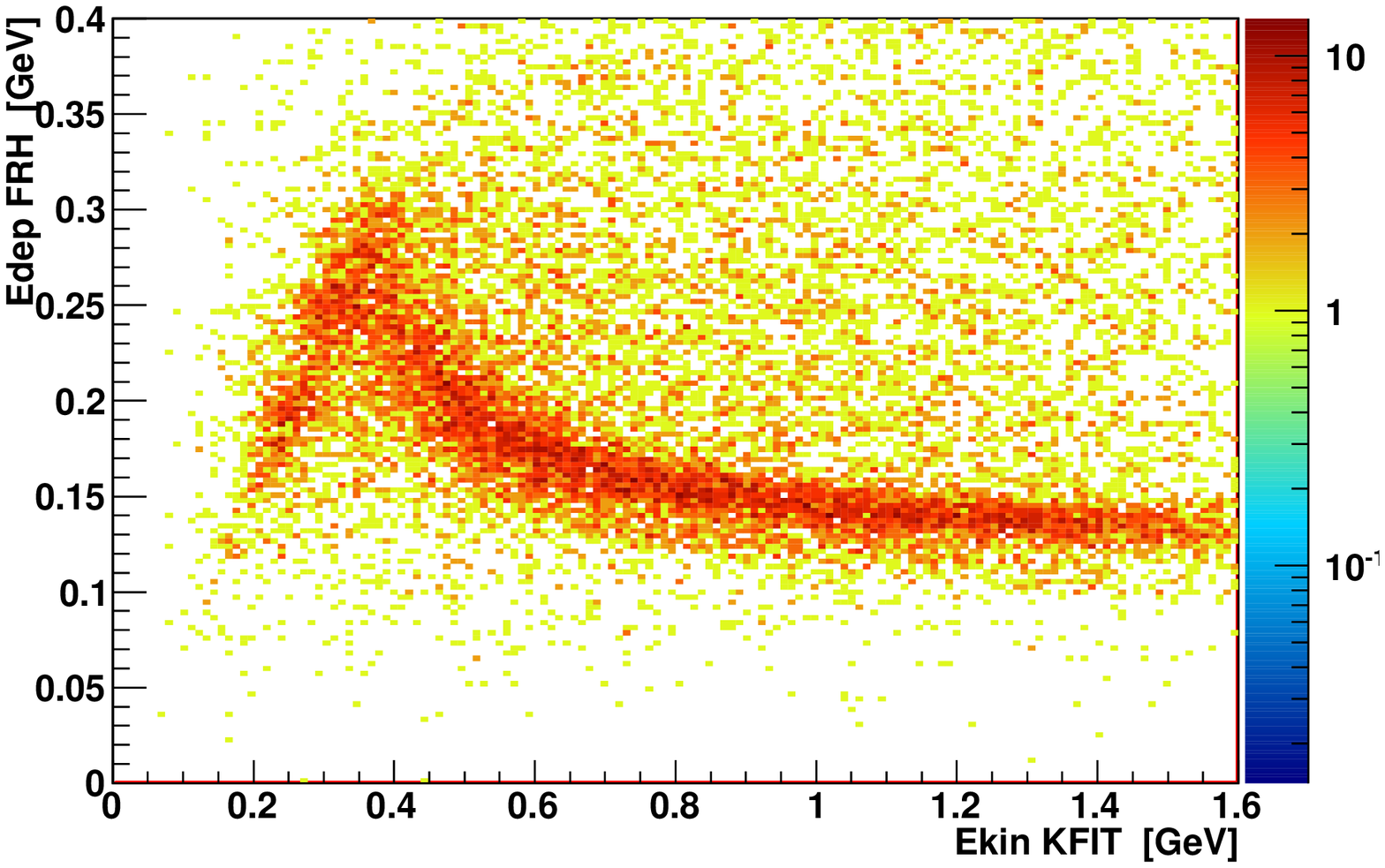}} \label{fig:KFitEkinEdepData}}\\
\subfigure[Monte-Carlo Simulation.]{\fbox{\includegraphics[width=0.7\textwidth]{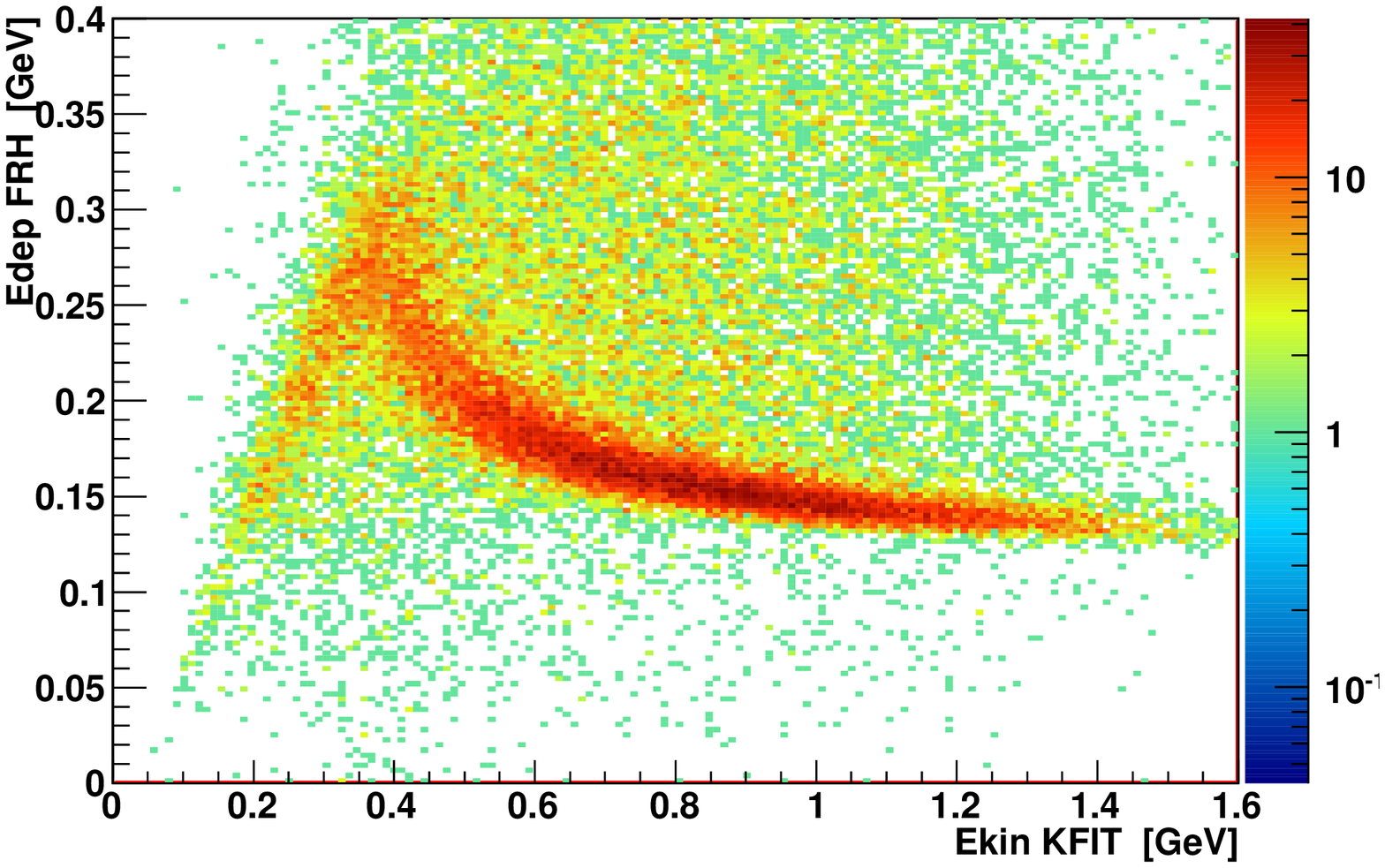}} \label{fig:KFitEkinEdepMC}}
}
\caption{Energy loss in FHR detector versus the kinetic energy derived from the kinematic fit for the proton for $Prob>0.2$. The position of the proton band is the same for both above cases.}
\label{fig:KFitEkinEdep}
\end{figure}

\paragraph{The protons\\}
The identification of protons in FD detector by the kinematic fit was also checked.
The energy loss in FHR detector versus the kinetic energy derived from the kinematic fit for the proton for $Prob>0.2$ was plotted both for
 experimental data and Monte-Carlo (Fig.~\ref{fig:KFitEkinEdep}). The experimental data are consistent with the Monte-Carlo simulation, which proves that
the measured particles are protons.
\begin{figure}[ht!bp]
\centering
{
\subfigure[True kinetic energy versus the kinetic energy derived from the kinematic fit for proton, correlation line visible.]{\fbox{\includegraphics[width=0.7\textwidth]{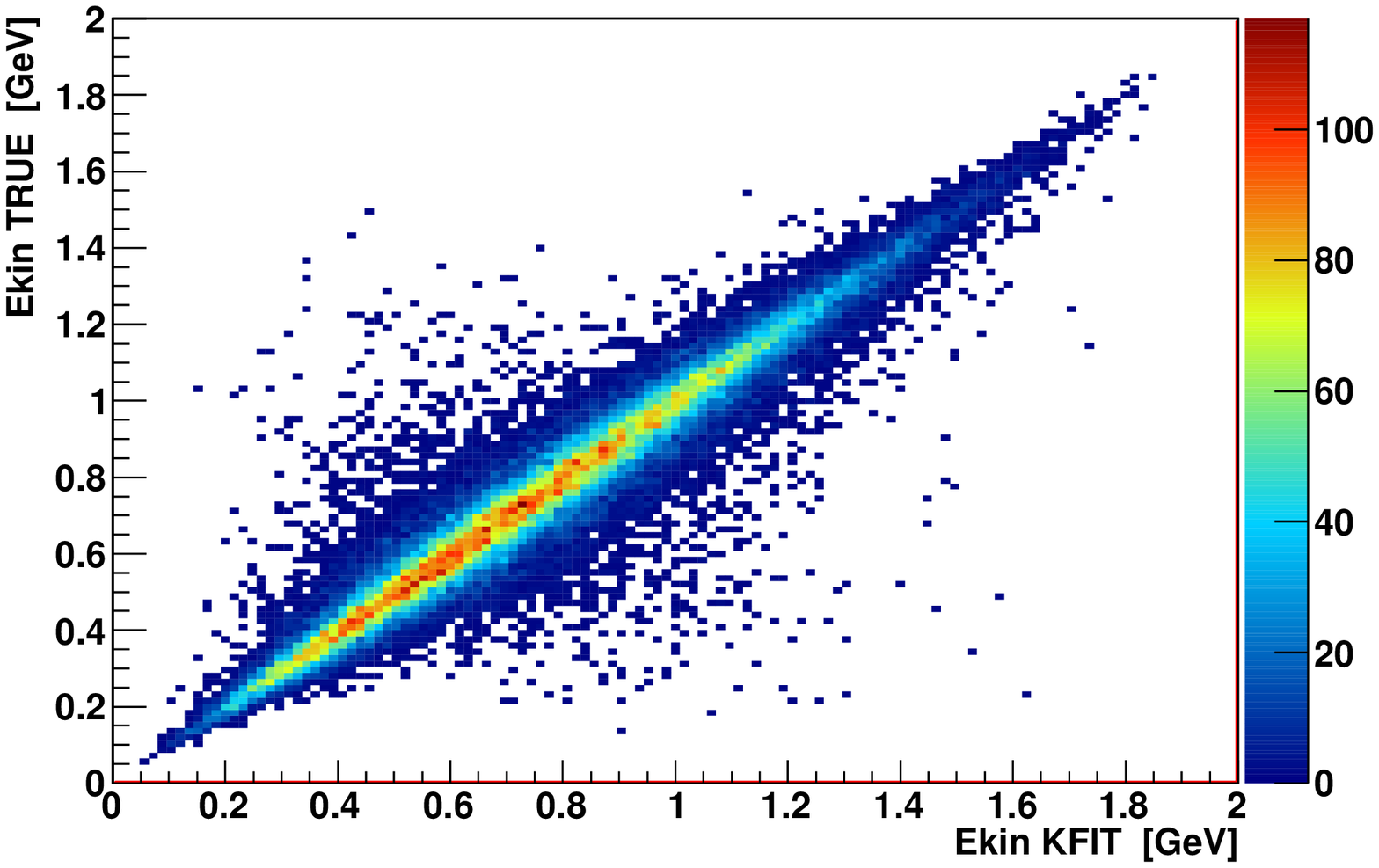}} \label{fig:KFitEkinCorr2D}}\\
\subfigure[Difference between the true kinetic energy and the kinetic energy derived from the kinematic fit divided by true kinetic energy for proton.]{\fbox{\includegraphics[width=0.7\textwidth]{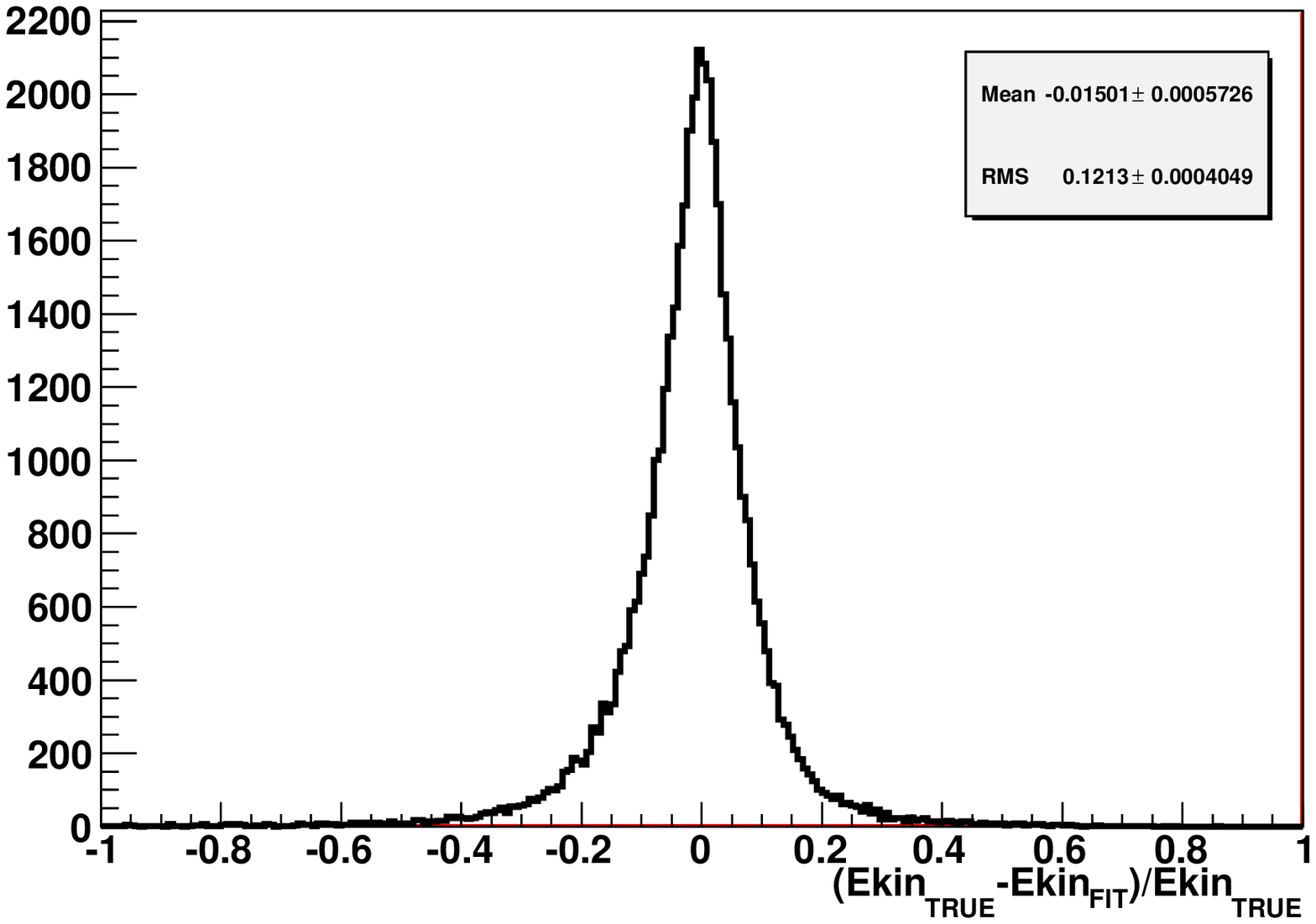}} \label{fig:KFitEkinCorrDiff}}
}
\caption{Monte-Carlo simulation of $pp \rightarrow pp 3\pi^{0}$ test of the proton kinetic energy reconstruction by the kinematic fit for $Prob>0.2$.}
\label{fig:KFitEkinCorr}
\end{figure}

The reconstruction of proton kinetic energy by the kinematic fit was found out using Monte-Carlo simulation by plotting true kinetic energy versus the kinetic
 energy derived from the kinematic fit for proton (Fig.~\ref{fig:KFitEkinCorr2D}), which shows the correlation line. From the width of the difference between the true kinetic energy 
 and the kinetic energy derived from the kinematic fit divided by true kinetic energy (Fig.~\ref{fig:KFitEkinCorrDiff}) the resolution of the reconstructed kinetic energy
was estimated to be $\sim12\%$.

\begin{figure}[ht!bp]
\centering
{
\subfigure[Kinetic energy reconstructed using FRH detector versus the kinetic energy derived from the kinematic fit for proton, correlation line visible.]{\fbox{\includegraphics[width=0.7\textwidth]{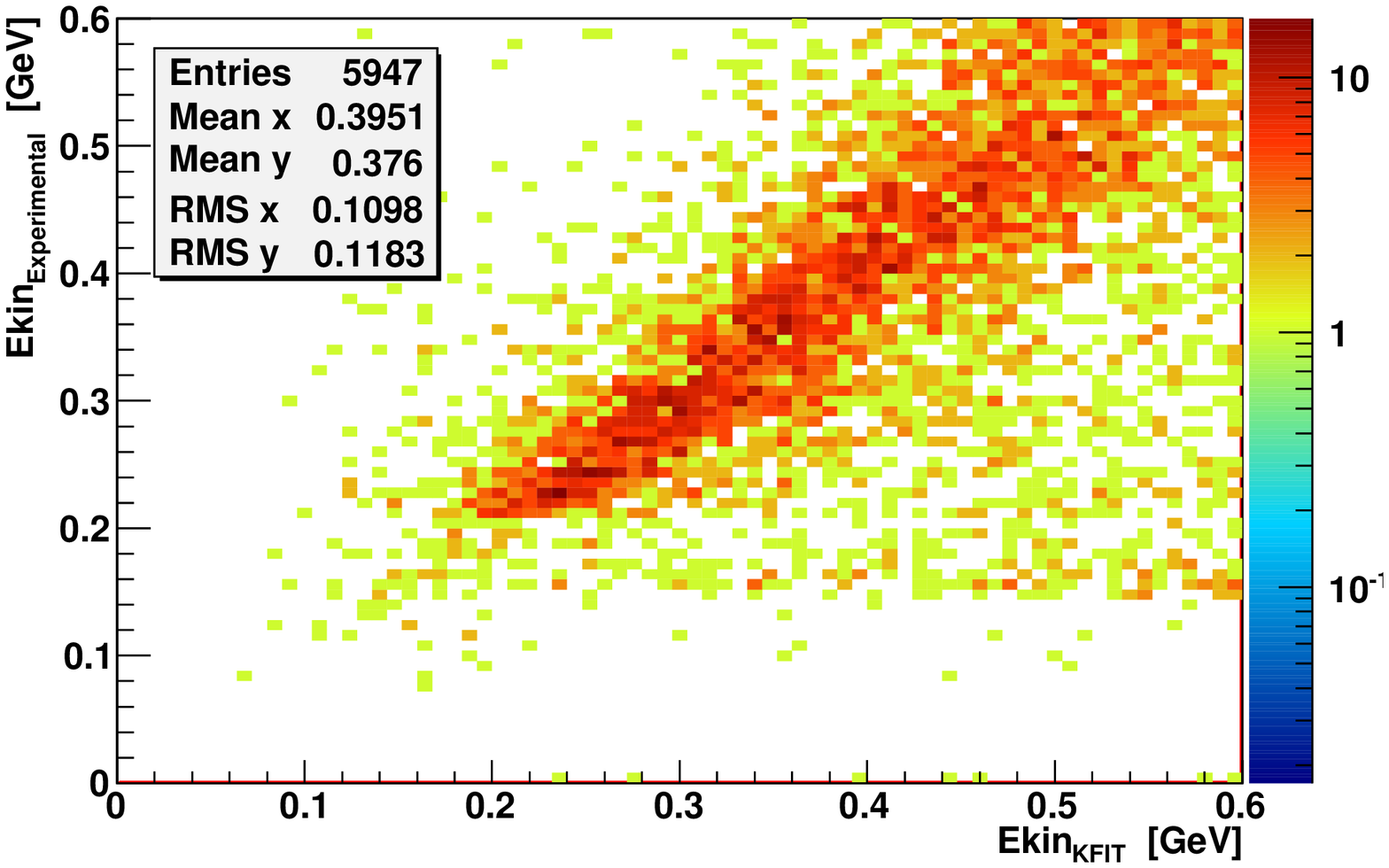}} \label{fig:KFitEkinEkin2D}}\\
\subfigure[Difference between the kinetic energy reconstructed using FRH detector and the kinetic energy derived from the kinematic fit for proton.]{\fbox{\includegraphics[width=0.7\textwidth]{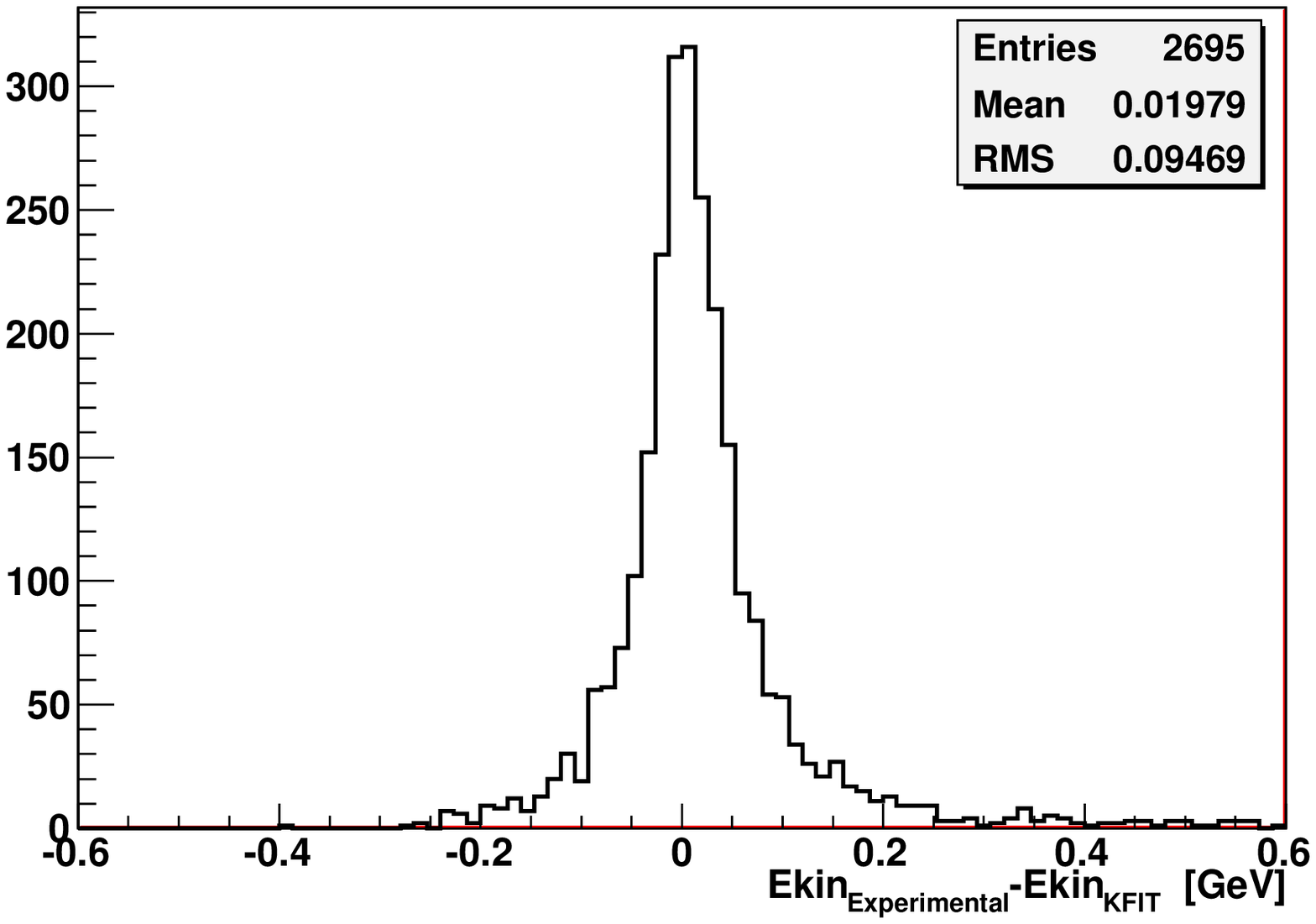}} \label{fig:KFitEkinEkinDiff}}
}
\caption{Experimental data, small kinetic energies of the protons in FD detector selected $E_{Kin}<0.6\mathrm{GeV}$, $Prob>0.2$.}
\label{fig:KFitEkinEkin}
\end{figure}

\begin{figure}[ht!bp]
\centering
{
\subfigure[Energy loss in FRH detector versus the energy loss recalculated from the kinetic energy from the kinematic fit for proton, correlation line visible.]{\fbox{\includegraphics[width=0.7\textwidth]{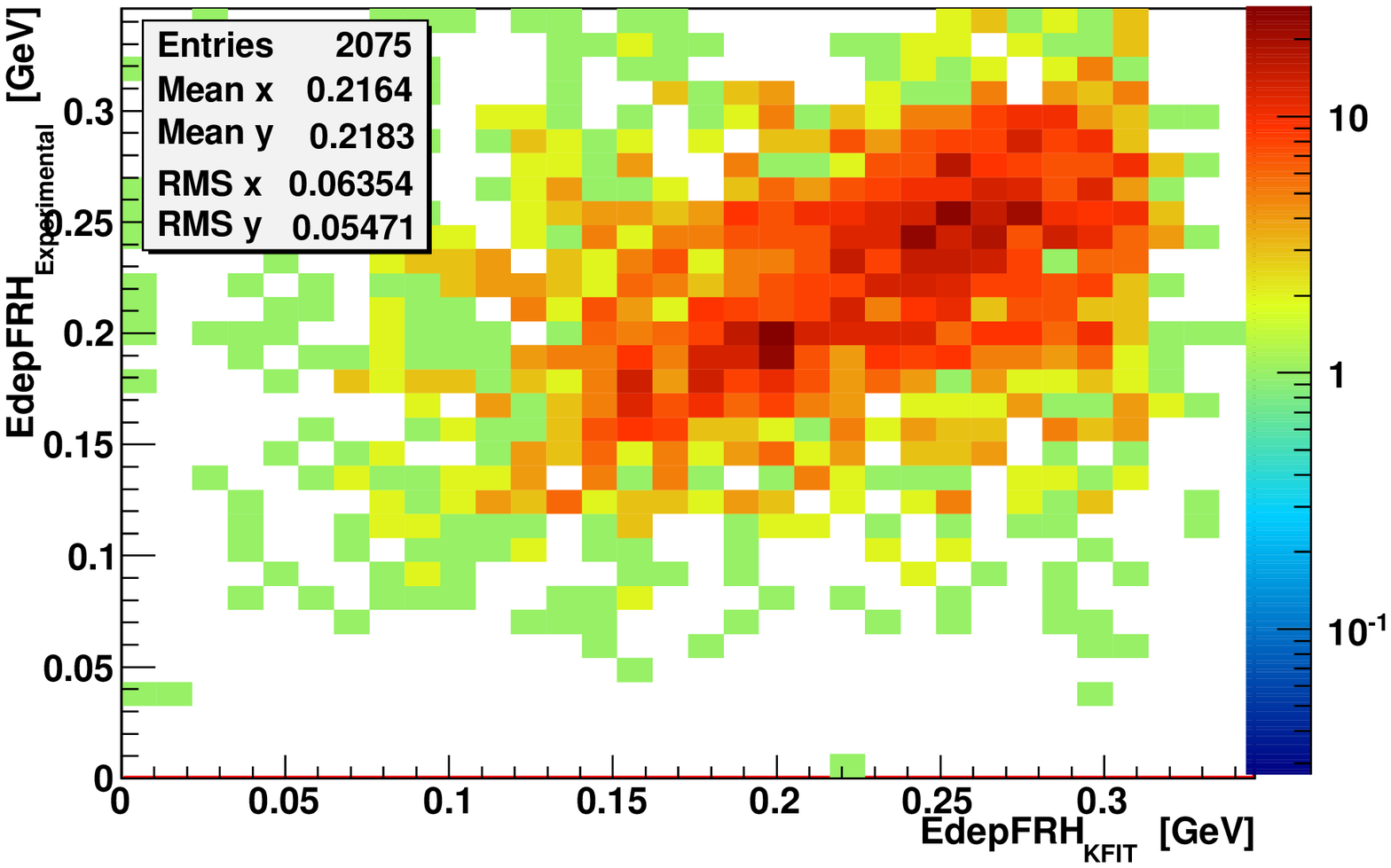}} \label{fig:KFitEdepEdep2D}}\\
\subfigure[Difference between energy loss in FRH detector and the energy loss recalculated from the kinetic energy from the kinematic fit for proton.]{\fbox{\includegraphics[width=0.7\textwidth]{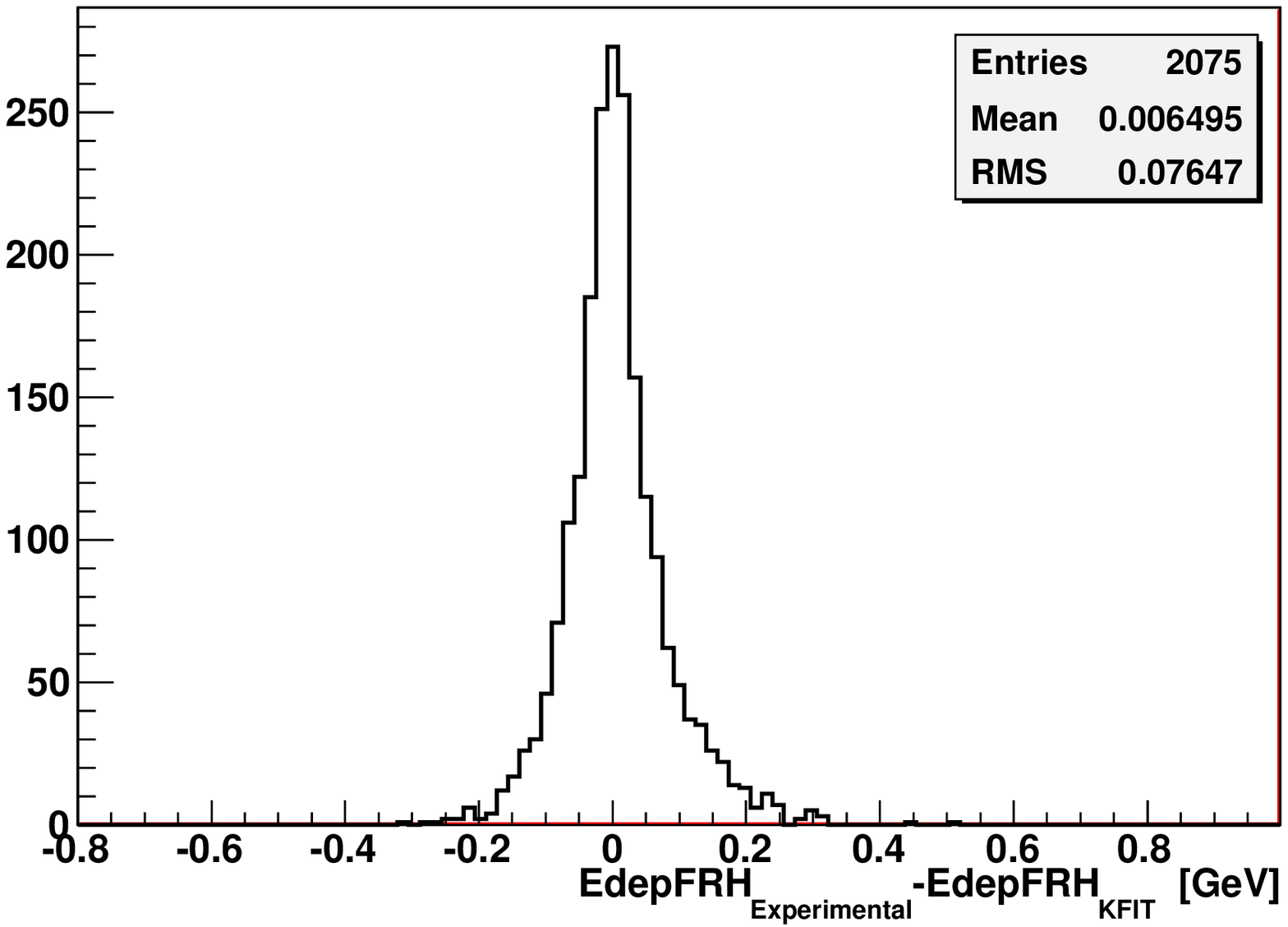}} \label{fig:KFitEdepEdepDiff}}
}
\caption{Experimental data, small kinetic energies of the protons in FD detector selected $E_{Kin}<0.6\mathrm{GeV}$, $Prob>0.2$.}
\label{fig:KFitEdepEdep}
\end{figure}

The checks on the experimental data were also performed. Small kinetic energies of the protons in FD detector reconstructed by the kinematic fit were selected $E_{Kin}<0.6\mathrm{~GeV}$ for $Prob>0.2$
and compared with the kinetic energy reconstructed using FRH detector. The correlation is visible (Fig.~\ref{fig:KFitEkinEkin2D}), the peak in difference (Fig.~\ref{fig:KFitEkinEkinDiff}) confirms it.        

For the consistency, additional cross check was performed for protons of kinetic energy $E_{Kin}<0.6\mathrm{~GeV}$. 
The kinetic energy derived from the kinematic fit was also recalculated to the energy loss in FRH
 detector by the Monte-Carlo method and compared with the measured energy loss in FRH, the correlation and a peak in difference is visible (Fig.~\ref{fig:KFitEdepEdep}).
Moreover the plot of difference of energy loss in FRH and the energy loss recalculated using obtained from kinematic fit proton kinetic energy is centered at $0$ value.
It confirms the correctness of kinetic energy calculation by the kinematic fit.

\newpage ~
\thispagestyle{empty}
\emptydoublepage
\newpage

\section{Results and error discussion}\label{sec:results}
\thispagestyle{plain}
The selection involving kinematic fitting of the events was performed. The complete $pp3\pi^{0}$ final state events are selected.
The missing mass of the two protons \myImgRef{MMpp02PhSp.eps} shows the $\eta\rightarrow3\pi^{0}$ meson decay peak and a prompt $3\pi^{0}$ production.
The Monte-Carlo simulation of $pp\rightarrow pp3\pi^{0}$ assuming homogeneously and isotropically populated phase space does not describe the experimental data with excluded $\eta$ meson production - the non resonant part. 

\myFrameFigure{MMpp02PhSp.eps}{Spectrum of Missing Mass of two protons. Black histogram corresponds to the experimental data, the blue one to
 the Monte-Carlo phase space simulation of $pp \rightarrow pp 3\pi^{0}$. The $\eta\rightarrow 3\pi^{0}$ meson decay peak visible as well as the
 prompt $3\pi^{0})$ production.
Vertical axes - number of events (in given bin) is shown.
}{Missing Mass of two protons.}

\subsection{The $pp \rightarrow pp 3\pi^{0}$ reaction}
\label{sec:PP3pi0reaction}

\subsubsection{The model description}
\label{subsec:3pi0MCmodel}
\paragraph{Introduction - the idea of the Model\\}
\myFrameHugeFigure{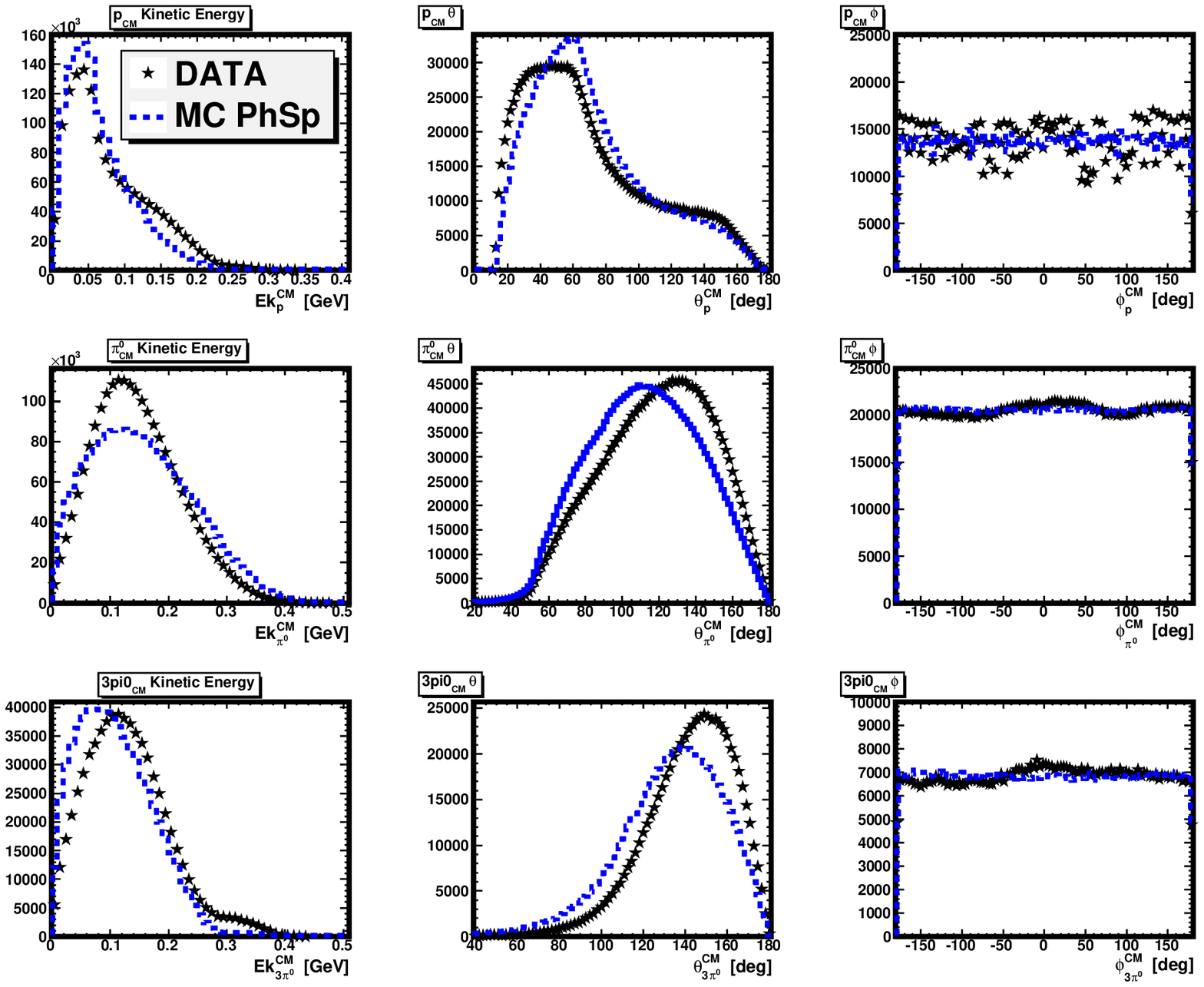}{Comparison of the experimental data with the Monte-Carlo phase space simulation for the
 $MM_{pp}<0.5\mathrm{~GeV/c^{2}}$ and $MM_{pp}>0.6\mathrm{~GeV/c^{2}}$.
\textit{Upper row}: proton in the center of mass frame,\textit{Middle row}: pion in the center of mass frame, \textit{Lower row}: $3\pi^{0}$ system in the center of mass frame. From left kinetic energy, 
the polar angle and the azimuthal angle distribution. The experimental data are shown as a black marker, the Monte-Carlo simulation as a blue line.
Vertical axes - number of events (in given bin) is shown.
}{Comparison of the experimental data with the Monte-Carlo phase space simulation for the $MM_{pp}<0.5\mathrm{GeV/c^{2}}$ and $MM_{pp}>0.6\mathrm{GeV/c^{2}}$.}

The $pp \rightarrow pp 3\pi^{0}$ reaction was separated from $pp \rightarrow pp \eta(3\pi^{0})$ reaction by introducing a cut on the Missing Mass
 of two protons \myImgRef{MMpp02PhSp.eps}. The region without the $\eta$ meson signal was selected as $MM_{pp}<0.5\mathrm{~GeV/c^{2}}$ and $MM_{pp}>0.6\mathrm{~GeV/c^{2}}$. 

Next the distribution of kinetic energy, the polar angle $\theta$ and the azimuthal angle $\phi$ for the protons, pions and $3\pi^{0}$ system,
 in the center of mass system was compared to the Monte-Carlo simulation of $pp\rightarrow pp3\pi^{0}$ assuming homogeneously and isotropically populated phase space \myImgRef{ComparisonPhSp.eps}.
The kinetic energy distributions and the polar angle distributions differ strongly from the simulation: the phase space simulation does not describe this process. 

The following scheme of the dynamics is proposed:
the $pp \rightarrow pp 3\pi^{0}$ reaction is sequential two-body process

\begin{equation}
 pp \rightarrow R_{1}R_{2} \rightarrow pp 3\pi^{0}
\label{eq:3pi0ModelR1R1}  
\end{equation}

$R_{1,2}$~-~unstable particle (Baryon Resonance).

\begin{figure}[ht!bp]
\centering
{
\subfigure[$Model_{1}$]{\fbox{\includegraphics[width=0.45\textwidth, height=0.25\textwidth]{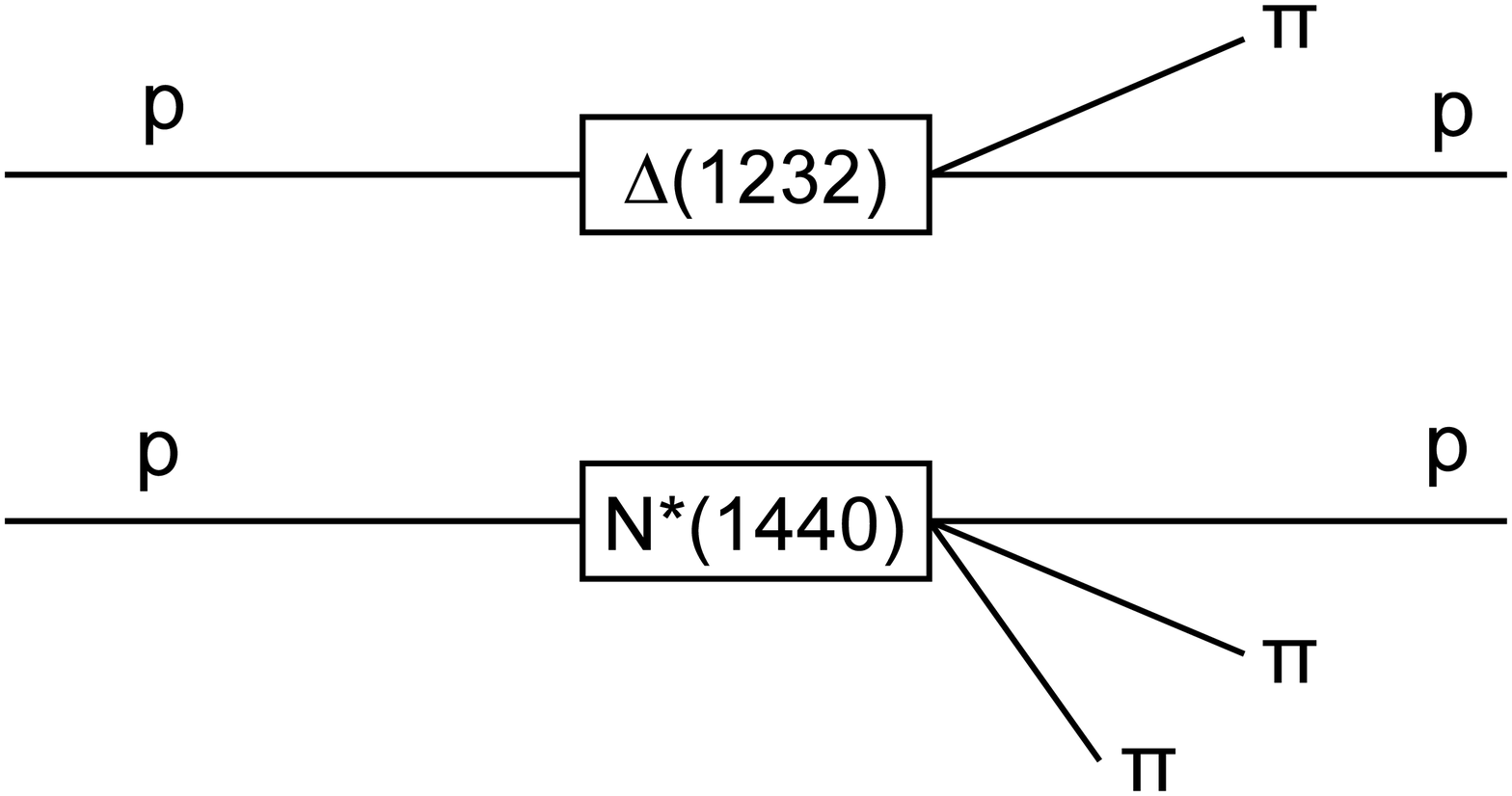}} \label{fig:3pi0Model1}}\quad
\subfigure[$Model_{2}$]{\fbox{\includegraphics[width=0.45\textwidth, height=0.25\textwidth]{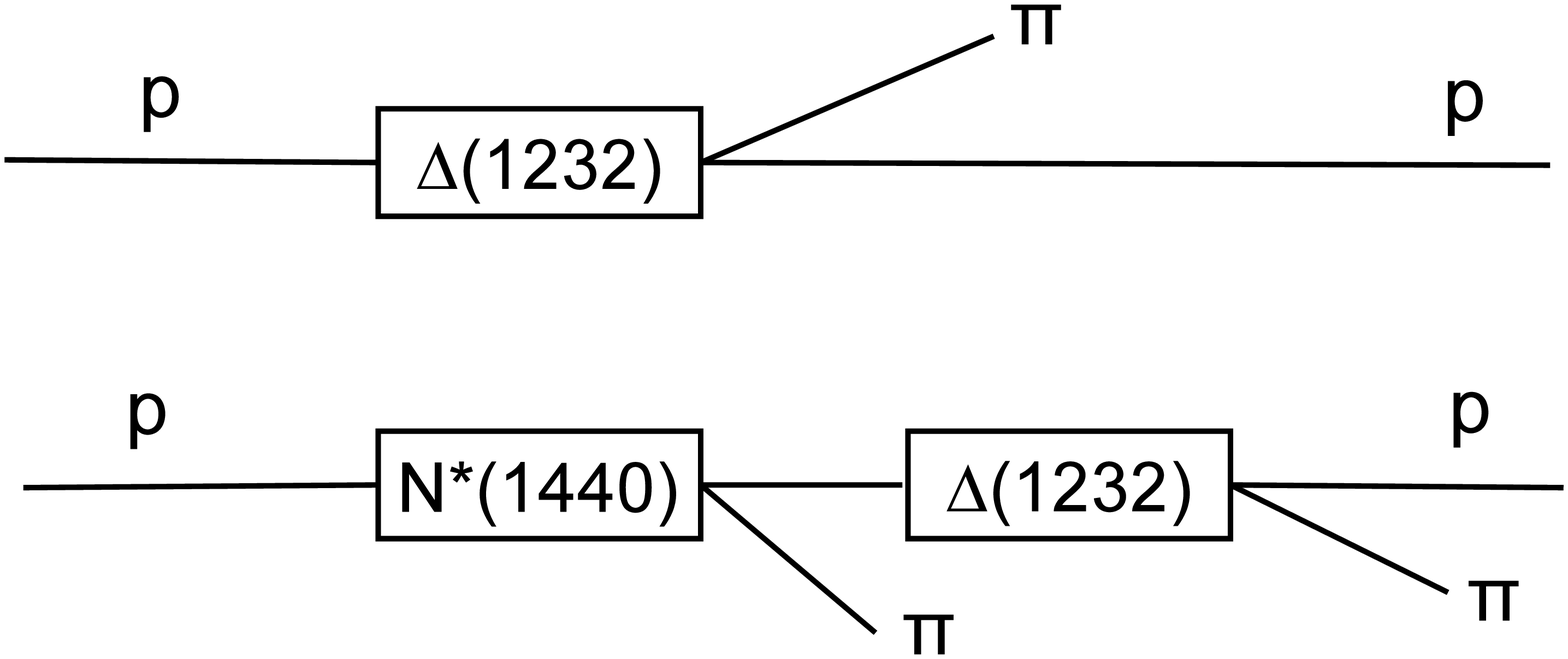}} \label{fig:3pi0Model2}}
}
\caption{Two model possibilities of $pp \rightarrow pp 3\pi^{0}$ production.}
\label{fig:3pi0Models}
\end{figure}

Searching for the candidates for $R_{1,2}$ in the most simple form one may take the two lowest lying resonances like $\Delta(1232)$ and $N^{*}(1440)$, 
since the available phase space volume for them will be the biggest (later other arguments supporting feasibility of the choice will be presented).
The $\Delta(1232)$ can decay $\Delta(1232) \rightarrow p \pi^{0}$ and the $N^{*}(1440)$ has two dominant decay branches, directly $N^{*}(1440) \rightarrow p \pi^{0}\pi^{0}$
or sequentially $N^{*}(1440) \rightarrow \pi^{0} \Delta(1232) \rightarrow \pi^{0} p \pi^{0}$.
The joint reaction mechanism:
       
\begin{equation}
 pp \rightarrow \Delta(1232)N^{*}(1440) \rightarrow pp 3\pi^{0}
\label{eq:3pi0ModelDN} 
\end{equation}

could be written as a two Feynman graphs (Fig.~\ref{fig:3pi0Models}).

\newpage
\paragraph{Supporting arguments\\}

\myTable{
\footnotesize
\begin{adjustwidth}{-2cm}{-3cm}
\begin{tabular}{|l|l||c||p{4cm}|}
\hline
& & & \\
        &         & $T_{Beam}^{Thr.}\mathrm{~[GeV]}$   &            \\
$R_{1}$ & $R_{2}$ &  $P_{Beam}^{Thr.}\mathrm{~[GeV/c]}$& properties \\ 
        &         &$\mathbf{Q}\mathrm{~[MeV]}$                    &             \\
\hline
\hline
& & & \\
$\Delta(1232)P_{33}$ & $N^{*}(1440)P_{11}$          & $1.930$ & \multirow{3}{4cm}{will be strongly excited} \\
$BR(N\pi)=100\%$       & a)~$BR(N\pi\pi)=5\%-10\%$     & $2.710$      &                \\
                     & b)~$BR(\pi\Delta(1232))=20\%-30\%$ & $\mathbf{207}$       &                \\
\hline
& & & \\
$\Delta(1232)P_{33}$ & $N^{*}(1520)D_{13}$          & $2.161$ & \multirow{3}{4cm}{will be excited} \\
$BR(N\pi)=100\%$      & a)~$BR(N\pi\pi)<8\%$     &  $2.954$     &                \\
                     & b)~$BR(\pi\Delta(1232))=15\%-25\%$ & $\mathbf{127}$       &                \\
\hline
& & & \\
$\Delta(1232)P_{33}$ & $N^{*}(1535)S_{11}$          & & \multirow{4}{4cm}{$R_{2}$ spin flip excitation(weak), will be weakly excited~\cite{HPMorsch,PDG2008}} \\
$BR(N\pi)=100\%$       & a)~$BR(N\pi\pi)<3\%$     &   $2.205$    &                \\
                     & b)~$BR(\pi\Delta(1232))<1\%$ &  $3.000$     &                \\
                    &  c)~$BR(\pi N^{*}(1440))<7\%$ &   $\mathbf{112}$     &                \\
\hline
& & & \\
$\Delta(1232)P_{33}$ & $\Delta(1600)P_{33}$          & $2.400$ & \multirow{3}{4cm}{$R_{2}$ monopole excitation of $\Delta(1232)P_{33}$, will be weakly excited~\cite{HPMorsch,PDG2008}} \\
$BR(N\pi)=100\%$       & a)~$BR(\pi\Delta(1232))=40\%-70\%$     & $3.203$      &                \\
                     & b)~$BR(\pi N^{*}(1440))=10\%-35\%$ & $\mathbf{47}$      &                \\
\hline
& & & \\
$\Delta(1232)P_{33}$ & $\Delta(1620)S_{13}$          & $2.460$ & \multirow{3}{4cm}{$R_{2}$ monopole excitation of $\Delta(1232)P_{33}$, will be weakly excited~\cite{HPMorsch,PDG2008}}\\
$BR(N\pi)=100\%$        & a)~$BR(\pi\Delta(1232))=30\%-60\%$     & $3.270$      &                \\
                     & b)~$BR(\pi N^{*}(1440))=11\%$ & $\mathbf{27}$       &                \\
\hline
& & & \\
$\Delta(1232)P_{33}$ & $N^{*}(1680)F_{15}$          & $2.644$ & \multirow{2}{4cm}{Kinematically not possible}\\
$BR(N\pi)=100\%$        & a)~$BR(\pi\Delta(1232))=30\%-40\%$     & $3.457$      &                \\
                                                             & & $\mathbf{-33}$& \\
\hline
& & & \\
$\Delta(1232)P_{33}$ & $\Delta (1700)D_{33}$          & $2.706$ & \multirow{2}{4cm}{Kinematically not possible} \\
$BR(N\pi)=100\%$       & a)~$BR(N\pi\pi)=80\%-90\%$     &   $3.522$    &                \\
                                                         & & $\mathbf{-53}$& \\
\hline
\hline
& & & \\
$N^{*}(1440)P_{11}$ & $N^{*}(1440)P_{11}$          & $2.545$ & \multirow{3}{4cm}{Kinematically not possible} \\
$BR(N\pi)=55\%-75\%$       & a)~$BR(N\pi\pi)=5\%-10\%$     & $3.355$      &                \\
                     & b)~$BR(\pi\Delta(1232))=20\%-30\%$ &   $\mathbf{-1}$     &                \\
\hline
\end{tabular}
\end{adjustwidth}
}{Possible excitations of baryon resonances($R_{1,2}$) in the reaction $p p \rightarrow R_{1} R_{2}$, which give $pp3\pi^{0}$ final state, at incident proton momentum 
of $3.35\mathrm{~GeV/c}$. The excess energy $Q$,the beam kinetic energy at threshold  $T_{Beam}^{Thr.}$ and also the beam momentum at threshold  $P_{Beam}^{Thr.}$ is given. 
}{tab:R1R1Baryons}

\myFrameBigFigure{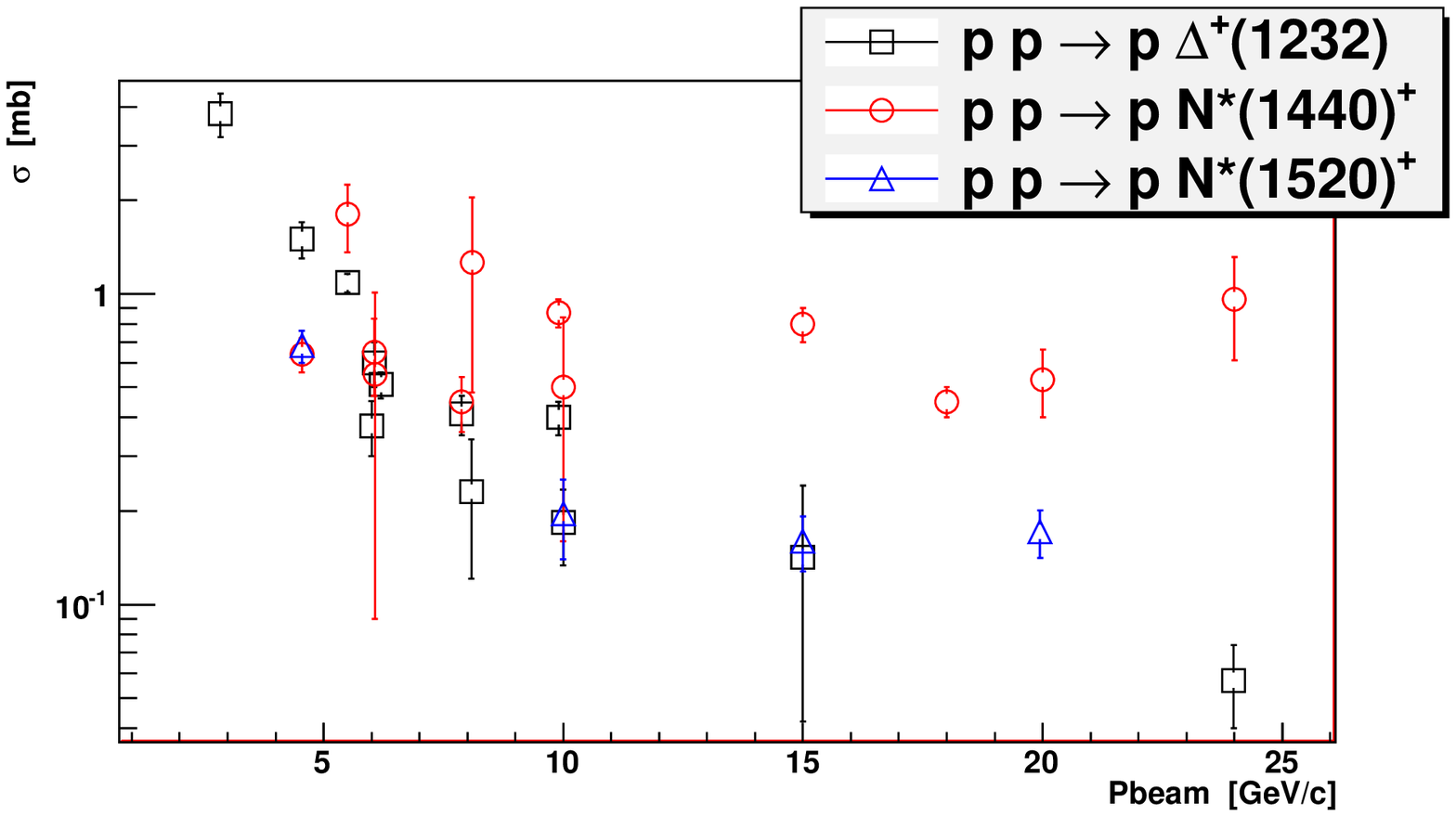}{Total Cross Section for the reactions $p p \rightarrow p \Delta^{+}(1232)$, $p p \rightarrow p N^{*}(1440)^{+}$, $p p \rightarrow p N^{*}(1520)^{+}$
 as a function of beam momentum. The beam momentum at threshold $1.262\mathrm{~GeV/c}$, $1.852\mathrm{~GeV/c}$, $2.081\mathrm{~GeV/c}$
and beam kinetic energy at threshold $0.634\mathrm{~GeV}$, $1.138\mathrm{~GeV}$, $1.345\mathrm{~GeV}$  for $\Delta^{+}(1232)$, $N^{*}(1440)^{+}$ and $N^{*}(1520)^{+}$ respectively.     
The highest cross section in case of $\Delta(1232)$ and $N^{*}(1440)$ visible.
Data come from \cite{BaryonsCS}}{Cross Section.}

The arguments that support the idea of lowest available baryon resonances playing role of $R_{1}$ and $R_{2}$ in scheme (Eq.~\ref{eq:3pi0ModelR1R1})
are both qualitative and quantitative.
The resonances $R_{1}, R_{2}$ should excite with large probability in pp reaction and decay predominantly into channels with one or two pions.
$\Delta(1232)$ decays into nucleon and pion ($\sim 100\%$) while $N^{*}(1440)$ decay sequence almost always leads to states with two pions.
In Table~\ref{tab:R1R1Baryons} branching ratios of decays of different baryon resonances that could play role of $R_{2}$ are presented, together with $Q$ value
of considered $R_{1}R_{2}$ channels. It is seen that $N^{*}(1680)$ and $\Delta (1700)$ are kinematically not possible.   
One concludes that the branching is considerably highest for $N^{*}(1440)$, $\Delta(1600)$, $\Delta(1620)$.
Moreover, the last two are known to be monopole excitations of $\Delta(1232)$ that excite extremely weakly in pp reaction \cite{HPMorsch,PDG2008}.
Also available phase space would be lower for $\Delta(1600)$, $\Delta(1620)$ as seen from the Table~\ref{tab:R1R1Baryons}.

\myFrameFigurer{Fig4_BlairNuovo.eps}{A sample of the inelastically scattered proton spectra at $2.85\mathrm{~GeV/c}$ incident proton 
momentum, a) $2.292$~deg; b) $2.911$~deg; c) $3.845$~deg; d) $5.288$~deg; e) $6.503$~deg f) $7.758$~deg. 
The strong $\Delta(1232)$ signal visible.
The plot comes from \cite{BaryonsProd2}.}{}

\myFrameFigurer{Fig5_BlairNuovo.eps}{A sample of the inelastically scattered proton spectra at $4.55\mathrm{~GeV/c}$ incident proton 
momentum,  a) $2.292$~deg; b) $2.911$~deg; c) $3.845$~deg; d) $5.288$~deg; e) $6.503$~deg f) $7.758$~deg. 
The strong $\Delta(1232)$ signal and clear evidence of $N^{*}(1440)$ visible.
The plot comes from \cite{BaryonsProd2}.}{}

\myFrameSmallFigure{Fig5P5_PhysRevD.5.1073.eps}{Total Cross section as a function of the Missing Mass of the proton for the $p p \rightarrow p  X$ reaction at incident proton momenta range $6.2-29.7\mathrm{~GeV/c}$.
The spectra are presented for the four momentum transfer $-t=0.044\mathrm{~GeV^{2}/c^{2}}$(Upper Plot) and $-t=0.88\mathrm{~GeV^{2}/c^{2}}$(Lower Plot). The plot comes from \cite{BaryonsProd1}.}{Cross section as a function of the Missing Mass of the proton for the $p p \rightarrow p  X$ reaction.}

Total Cross Section for $pp \rightarrow p X$ is presented \myImgRef{BaryonsCS2.eps}; one concludes that the highest cross section is
in case of  $\Delta(1232)$ and $N^{*}(1440)$.
Similar conclusion can be drawn from literature data (Figs.~\ref{image_Fig4_BlairNuovo.eps},~\ref{image_Fig5_BlairNuovo.eps},~\ref{image_Fig5P5_PhysRevD.5.1073.eps}).
For the lower proton momentum $2.85\mathrm{~GeV/c}$ (Fig.~\ref{image_Fig4_BlairNuovo.eps}) than this work (i.e.$3.35\mathrm{~GeV/c}$) only strong $\Delta(1232)$ signal is seen.
In case of higher momentum $4.55\mathrm{~GeV/c}$ (Fig.~\ref{image_Fig5_BlairNuovo.eps}) the strong $\Delta(1232)$ signal and clear evidence of $N^{*}(1440)$ is seen.
When the momentum of the proton increases $6.2-29.7\mathrm{~GeV/c}$ (Fig.~\ref{image_Fig5P5_PhysRevD.5.1073.eps}) the $\Delta(1232)$ excitation looses its importance,
higher baryon resonances appear with much bigger probability.
Since experiment considered in this work was performed at $3.35\mathrm{~GeV/c}$ proton momentum, from the above data (Figs.~\ref{image_Fig4_BlairNuovo.eps},~\ref{image_Fig5_BlairNuovo.eps},~\ref{image_Fig5P5_PhysRevD.5.1073.eps})
one can conclude that the most strongly excited baryon resonances will be $\Delta(1232)$ and $N^{*}(1440)$.

One can present as well qualitative arguments that the simultaneous excitation of two Roper($N^{*}(1440)$) resonances (playing role of $R_{1}R_{2}$) is also
unfeasible. In the last row of Table~\ref{tab:R1R1Baryons} this case is considered; the $Q$ value is negative which makes the process
kinematically unfavored. In addition it is seen that the excitation of the Roper resonance is considerably lower than that of $\Delta(1232)$
(Figs.~\ref{image_BaryonsCS2.eps},~\ref{image_Fig4_BlairNuovo.eps},~\ref{image_Fig5_BlairNuovo.eps}).
This makes simultaneous excitation of two Roper resonances rather unprobable in comparison to the $\Delta(1232)N^{*}(1440)$ system.

The production of $3\pi^{0}$ via simultaneous excitation of $\Delta(1232)$ and $N^{*}(1440)$ is considered as a predominant and also supported by \cite{ETACS01,3pi0Baryons}
(see Section~\ref{sec:PhysicsMotivation} on page \pageref{sec:PhysicsMotivation}).

\newpage
\paragraph{The derivation of the model parameters from experimental data\\}\label{par:modelparam}
To check the validity of the hypothesis that the $pp \rightarrow pp 3\pi^{0}$ reaction follows via simultaneous excitation of two baryon resonances 
The $pp \rightarrow \Delta(1232) N^{*}(1440)$ reaction was simulated using kinematic calculations by PLUTO++ event generator \cite{Pluto}
, without any interaction between the $\Delta(1232)$ and $N^{*}(1440)$ - homogeneously and isotropically populated phase space was assumed (see Appendix~\ref{appendix:wmc}).
Two decay paths of the $N^{*}(1440)$ the direct decay $Model_{1}$~(Fig.~\ref{fig:3pi0Model1})
and sequential decay $Model_{2}$~(Fig.~\ref{fig:3pi0Model2}) were taken into account.
   
The experimental data were compared with these two simulations by comparing the following distributions:

\begin{itemize}
 \item Dalitz Plot $ppX$ ($M^{2}(pp)$ versus $M^{2}(p3\pi^{0})$) \cite{DalitzPlot,DalitzPlot1, DalitzPlot2,DalitzPlot3}\\
The interactions between the protons and the three pion system could be examined.

\item Dalitz Plot $3\pi^{0}$ ($M^{2}(2\pi^{0})$ versus $M^{2}(2\pi^{0})$) \cite{DalitzPlot, DalitzPlot1, DalitzPlot2,DalitzPlot3}\\
The correlations between the pions could be examined.

\item Nyborg Plot ($M(p\pi^{0})$ versus $M(p\pi^{0}\pi^{0}$) \cite{NyborgPlot}\\
The resonances in the $\pi^{0}-p$ and $\pi^{0}\pi^{0}-p$ system could be studied.
\end{itemize}

Here $M$ states for the invariant mass.
The distributions (Figs.~\ref{image_Dal1New2},~\ref{image_Dal2New2},~\ref{image_Dal3New2}) were studied 
for five different regions of the missing mass of two protons $MM_{pp}$, excluding the region of the $\eta$ meson ($MM_{pp}=0.5-0.6\mathrm{~GeV/c^{2}}$) (Fig.\ref{image_MMpp02PhSp.eps} and Table~\ref{tab:MMQ}). 
The selection of the $MM_{pp}$ ranges implies the boundaries on the maximal available kinetic energy for the $3\pi^{0}$ system in its rest frame ($Q_{3\pi^{0}}^{Max}$) and simultaneously
on the maximal available kinetic energy for the $pp$ system in its rest frame ($Q_{pp}^{Max}$) Table~\ref{tab:MMQ}.
More details about the variables choice can be found in Section~\ref{sec:DefinitionOfVariables} on page \pageref{sec:DefinitionOfVariables} and in Appendix~\ref{appendix:Kine5part}.   

\myTable{
\begin{tabular}{|c||c|c|c|c|c|}
\hline
$MM_{pp}\mathrm{~GeV/c^{2}}$ &  $0.4-0.5$ & $0.6-0.7$ & $0.7-0.8$ & $0.8-0.9$ & $0.9-1.0$ \\
\hline
\hline
$Q_{3\pi^{0}}^{Max}\mathrm{~MeV}$ &$\sim 100$ & $\sim 300$ & $\sim 400$ & $\sim 500$ & $\sim 600$ \\
\hline
$Q_{pp}^{Max}\mathrm{~MeV}$ & $\sim 600$ & $\sim 400$ & $\sim 300$ & $\sim 200$ & $\sim 100$ \\   
\hline
\end{tabular}
}{Selected five different regions of the missing mass of the two protons $MM_{pp}$. Description in text.}{tab:MMQ}

\myFrameHugeFigure{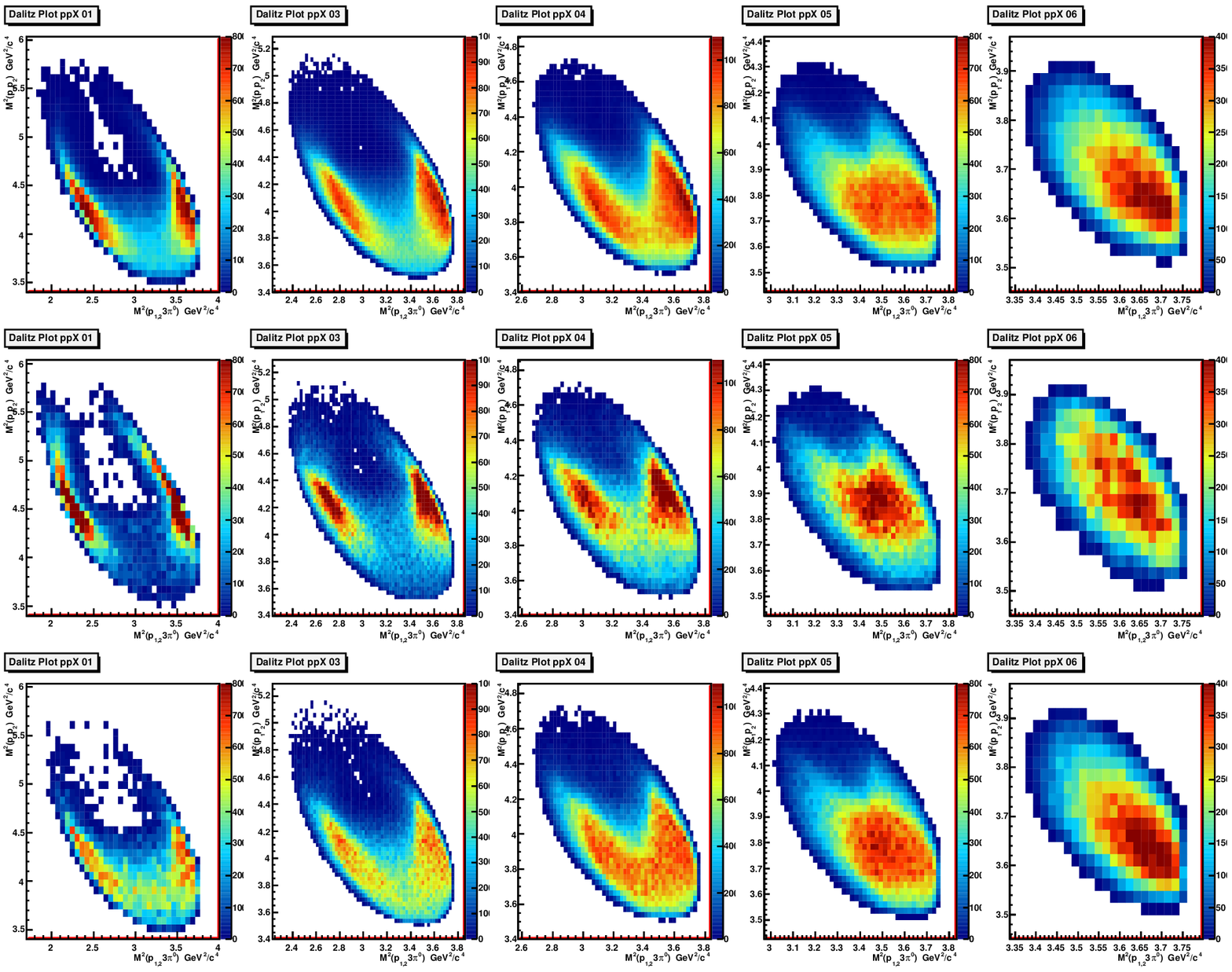}{Dalitz Plot $ppX$. $M^{2}(pp)$ versus $M^{2}(p3\pi^{0})$. The upper row corresponds to the experimental data,
the middle row to the Monte-Carlo $Model_{1}$, the lower row to the Monte-Carlo $Model_{2}$ (Fig.~\ref{fig:3pi0Models}). The columns from left to right correspond
to the following Missing Mass of two protons bins, column~1~$MM_{pp}=0.4-0.5\mathrm{~GeV/c^{2}}$,column~2~$MM_{pp}=0.6-0.7\mathrm{~GeV/c^{2}}$,
column~3~$MM_{pp}=0.7-0.8\mathrm{~GeV/c^{2}}$, column~4~$MM_{pp}=0.8-0.9\mathrm{~GeV/c^{2}}$, column~5~$MM_{pp}=0.9-1.0\mathrm{~GeV/c^{2}}$.
 The plots are symmetrized against two protons - each event is filled two times.
Fully expandable and colored version of the figure is available in the attached electronic version of the thesis. 
}{Dalitz Plot $ppX$.}

\myFrameHugeFigure{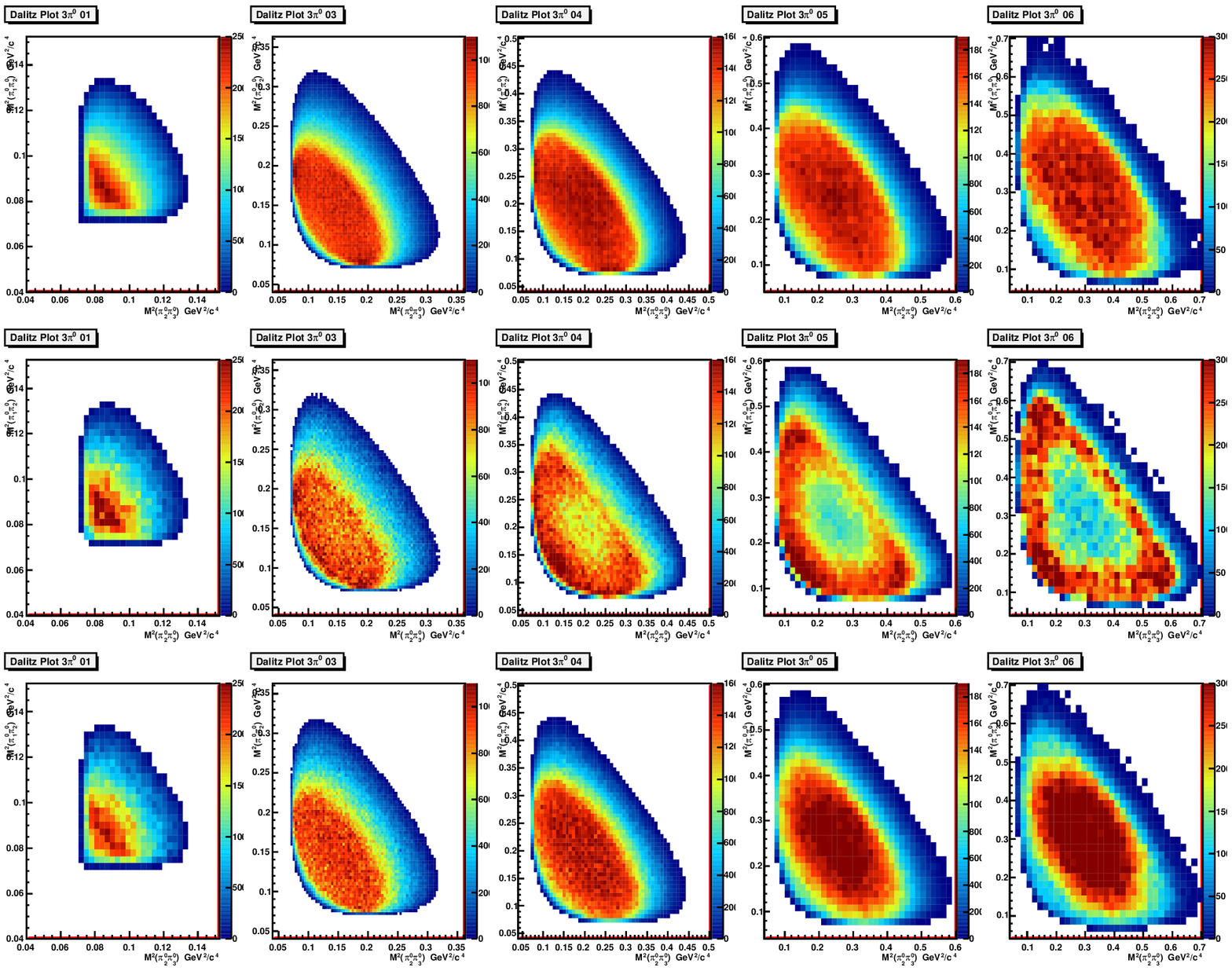}{Dalitz Plot $3\pi^{0}$. $M^{2}(2\pi^{0})$ versus $M^{2}(2\pi^{0})$.The upper row corresponds to the experimental data,
the middle row to the Monte-Carlo $Model_{1}$, the lower row to the Monte-Carlo $Model_{2}$ (Fig.~\ref{fig:3pi0Models}). The columns from left to right correspond
to the following Missing Mass of two protons bins, column~1~$MM_{pp}=0.4-0.5\mathrm{~GeV/c^{2}}$,column~2~$MM_{pp}=0.6-0.7\mathrm{~GeV/c^{2}}$,
column~3~$MM_{pp}=0.7-0.8\mathrm{~GeV/c^{2}}$, column~4~$MM_{pp}=0.8-0.9\mathrm{~GeV/c^{2}}$, column~5~$MM_{pp}=0.9-1.0\mathrm{~GeV/c^{2}}$.
 The plots are symmetrized against three pions - each event is filled six times.
Fully expandable and colored version of the figure is available in the attached electronic version of the thesis. 
}{Dalitz Plot $3\pi^{0}$.}

\myFrameHugeFigure{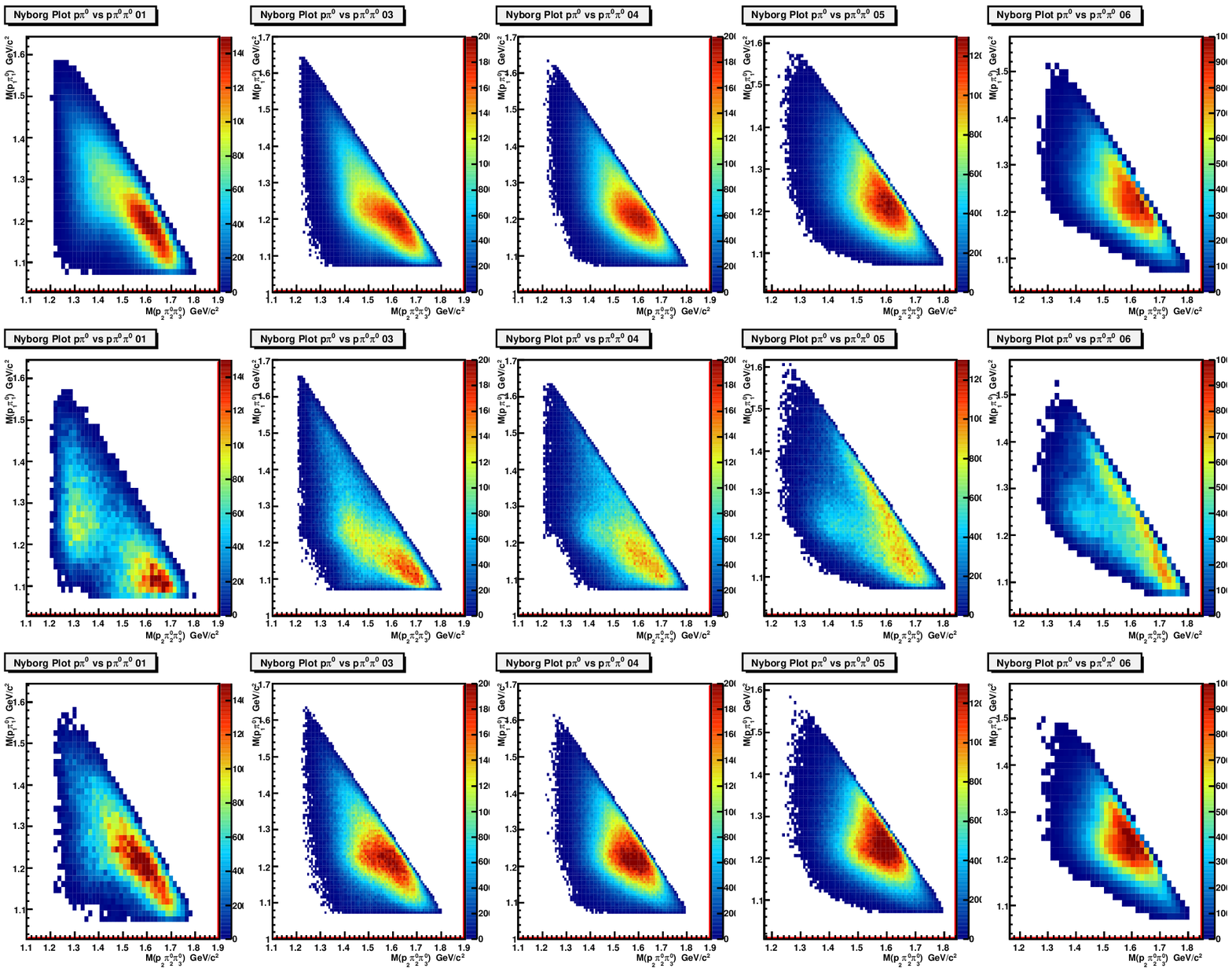}{Nyborg Plot. $M(p\pi^{0})$ versus $M(p\pi^{0}\pi^{0})$.The upper row corresponds to the experimental data,
the middle row to the Monte-Carlo $Model_{1}$, the lower row to the Monte-Carlo $Model_{2}$ (Fig.~\ref{fig:3pi0Models}). The columns from left to right correspond
to the following Missing Mass of two protons bins, column~1~$MM_{pp}=0.4-0.5\mathrm{~GeV/c^{2}}$,column~2~$MM_{pp}=0.6-0.7\mathrm{~GeV/c^{2}}$,
column~3~$MM_{pp}=0.7-0.8\mathrm{~GeV/c^{2}}$, column~4~$MM_{pp}=0.8-0.9\mathrm{~GeV/c^{2}}$, column~5~$MM_{pp}=0.9-1.0\mathrm{~GeV/c^{2}}$.
 The plots are symmetrized against two protons and three pions - each event is filled six times.
Fully expandable and colored version of the figure is available in the attached electronic version of the thesis. 
}{Nyborg Plot.}

One concludes that the two models populate different areas on the plots.
To check if one could describe the experimental data with a sum of the two models:

\begin{equation}
 \beta Model_{1} + (1-\beta)Model_{2}
\label{eq:Model3pi0First}
\end{equation}

one needs to know the fraction of $Model_{1}$  to the $Model_{1}+Model_{2}$ i.e. the $\beta$ parameter - which actually is the fraction of the direct decay to the sum of direct and sequential decay of the $N^{*}(1440)$.

To estimate the $\beta$ from the experimental data one needs to fit the shape of the plots (Figs.~\ref{image_Dal1New2},~\ref{image_Dal2New2},~\ref{image_Dal3New2})
 for mixture of $Model_{1}$ and $Model_{2}$ to the experimental data.
The population of the events in missing mass of two protons is not a purpose of the fit, and it will be extracted after the shape fit.
Fit was performed using the chi-square method; one defines the $\chi^{2}$ function to minimize:
\begin{equation}
 \chi^{2} = \sum \dfrac{\left[ Data - \left(\beta Model_{1} + (1-\beta)Model_{2}\right)\right]^{2} }{\sigma^{2}_{Data} + \beta^{2}\sigma^{2}_{Model_{1}} + (1-\beta)^{2}\sigma^{2}_{Model_{2}}}
\label{eq:chi2DalitzPlotFit}
\end{equation}
where the sum goes over each bin of the five Dalitz Plots $ppX$, five Dalitz Plots $3\pi^{0}$ and five Nyborg Plots i.e. $15$ plots,
 all in all $19752$ data point are fitted simultaneously. The $\sigma_{Data}$ is the error of the point for experimental data,
 $\sigma_{Model_{1}}$~-~ is the error of the point for $Model_{1}$ and $\sigma_{Model_{2}}$~-~ is the error of the point for $Model_{2}$.

For the numerical purpose of doing the fit, the $\chi^{2}$ function (Eq.~\ref{eq:chi2DalitzPlotFit}) was redefined to:
\begin{equation}
 \chi^{2} = \sum \dfrac{\left[ Data - \left(\alpha Model_{1}^{*} + (1-\alpha)Model_{2}^{*}\right)\right]^{2} }{\sigma^{2}_{Data} + \alpha^{2}\sigma^{2}_{Model_{1}^{*}} + (1-\alpha)^{2}\sigma^{2}_{Model_{2}^{*}}}
\label{eq:chi2DalitzPlotFitNorm}
\end{equation}     
where the models were normalized to the experimental data:
\begin{eqnarray}
  Model_{1}^{*} &=&\dfrac{\int Data}{\int Model_{1}} Model_{1}\nonumber \\
  Model_{2}^{*} &=&\dfrac{\int Data}{\int Model_{2}} Model_{2}
\label{eq:ModelsNormalization}
\end{eqnarray}

Now parameter $\alpha$ is the fraction of $Model_{1}^{*}$ to the $Model_{1}^{*}+Model_{2}^{*}$, so one can write:

\begin{equation}
\dfrac{Model_{1}^{*}}{Model_{2}^{*}} = \dfrac{\alpha}{1-\alpha}
\end{equation}

to get the ratio of the $Model_{1}$ to $Model_{1}+Model_{2}$ one just redoes the normalization (Eq.~\ref{eq:ModelsNormalization}),
 no other factor is needed since the same amount of events for $Model_{1}$ and $Model_{2}$ was generated:

\begin{eqnarray}
 \dfrac{Model_{1}}{Model_{2}} &=& \dfrac{Model_{1}^{*}}{Model_{2}^{*}} \dfrac{\int Model_{1}}{\int Model_{2}} = \nonumber \\
&=&  \dfrac{\alpha}{1-\alpha} \dfrac{\int Model_{1}}{\int Model_{2}} = \nonumber \\
&=&  \dfrac{\beta}{1-\beta}
\label{eq:ModelRatio}
\end{eqnarray}

\begin{figure}[ht!bp]
\centering
{
\subfigure[]{\fbox{\includegraphics[width=0.7\textwidth]{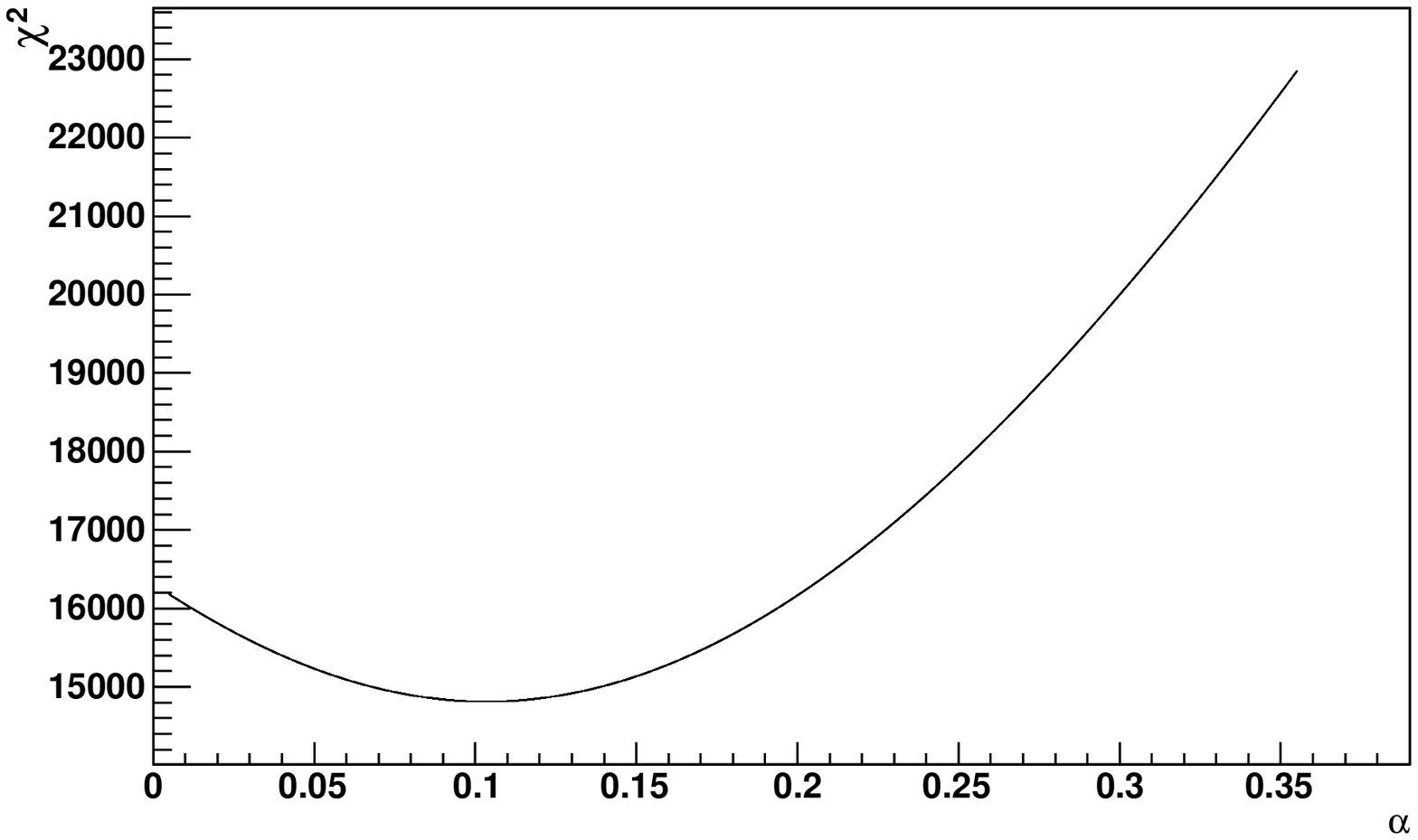}} \label{fig:chi2DalitzPlotFit1}}\\
\subfigure[]{\fbox{\includegraphics[width=0.7\textwidth]{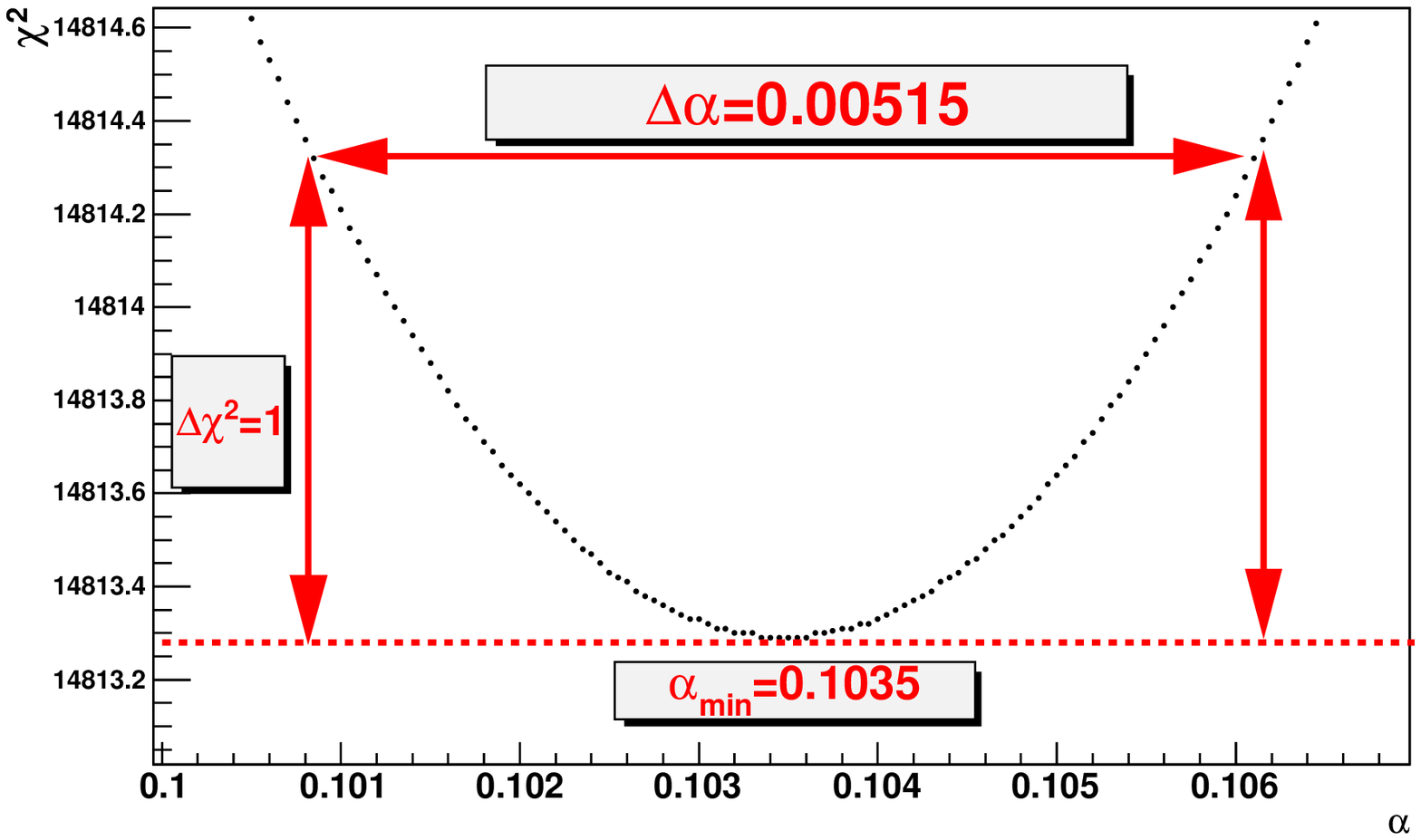}} \label{fig:chi2DalitzPlotFit2}}
}
\caption{$\chi^{2}$ versus the searched parameter $\alpha$ for the sum of $Model_{1}^{*}$ and $Model_{2}^{*}$ fit.}
\label{fig:chi2DalitzPlotFit}
\end{figure}

The $\chi^{2}$ function (Eq.~\ref{eq:chi2DalitzPlotFitNorm}) was minimize in respect to parameter $\alpha$.
The $\chi^{2}$ versus the searched parameter $\alpha$  (Fig.~\ref{fig:chi2DalitzPlotFit}), the function has one minimum.
The estimated value of the $\alpha$ parameter is the one which gives the smallest $\chi^{2}$, the error of the parameter is half of
 the distance for which the $\chi^{2}$ function changes by $1$, this gives:

\begin{equation}
 \frac{\chi^{2}_{min}}{NDF} = \frac{14813.29}{19751} =  0.750 \pm 0.010 
\end{equation}
where $\chi^{2}_{min}$ is the $\chi^{2}$ value at minimum, and $NDF$ is the Number of Degrees of Freedom. 

\begin{equation}
 \alpha = 0.1035 \pm 0.0026~.
\label{eq:alpha3pi0Model}
\end{equation}
It corresponds (Eq.~\ref{eq:ModelRatio}) to ratio:
\begin{equation}
\dfrac{Model_{1}}{Model_{2}} = 0.0580 \pm 0.0016
\end{equation}
giving the
\begin{equation}
 \beta = 0.0548 \pm 0.0015
\end{equation}

The same fit was performed where the spectra for the missing mass region $0.4-0.5\mathrm{~GeV/c^{2}}$ were excluded,
to check the sensitivity of the of the procedure,
 this gave the following result;

\begin{equation}
 \left(\frac{\chi^{2}_{min}}{NDF}\right)_{2} = \frac{14166.21}{18308} =  0.774 \pm 0.011 
\end{equation}

\begin{equation}
 \alpha_{2} = 0.1032 \pm 0.0028
\end{equation}

\begin{equation}
\left(\dfrac{Model_{1}}{Model_{2}}\right)_{2} = 0.0568 \pm 0.0017
\end{equation}

\begin{equation}
 \beta_{2} = 0.0537 \pm 0.0015
\end{equation}

The difference between these two results was used to estimate the systematic errors:
\begin{equation}
 \Delta\alpha_{sys.} = |\alpha-\alpha_{2}|=0.0003
\end{equation}
\begin{equation}
\Delta\left(\dfrac{Model_{1}}{Model_{2}}\right)_{sys.} = \left|\dfrac{Model_{1}}{Model_{2}}-\left(\dfrac{Model_{1}}{Model_{2}}\right)_{2}\right|=0.0012
\end{equation}
\begin{equation}
 \Delta\beta_{sys.} = |\beta - \beta_{2}| = 0.0011
\end{equation}

The final values with the systematic errors are:
\begin{equation}
 \alpha =  0.1035 \pm 0.0026 (stat.) \pm 0.0003 (sys.)
\end{equation}

\begin{equation}
\dfrac{Model_{1}}{Model_{2}} = 0.0580 \pm 0.0016 (stat.) \pm 0.0012 (sys.)
\label{eq:Model12Final}
\end{equation}

\begin{equation}
 \beta =  0.0548 \pm 0.0015 (stat.) \pm 0.0011 (sys.)
\label{eq:ModelBetaValue}
\end{equation}

To verify how the Monte-Carlo simulation based on the sum of two models, with the fitted parameter, describes the experimental data,  
the comparison was done showing the models sum for all the Dalitz and Nyborg plots (Figs.~\ref{image_Dal1FitNew3}~\ref{image_Dal2FitNew3}~\ref{image_Dal3FitNew3}).
It it seen that sum of the models describes very good the event populations on the Dalitz and Nyborg plots.

It is now also possible to calculate the ration $R$ of the partial decay widths for the decay of the Roper resonance $N^{*}(1440)$:
\begin{eqnarray}
 R &=& \frac{\Gamma(N^{*}(1440) \rightarrow N \pi \pi )}{\Gamma(N^{*}(1440) \rightarrow \Delta(1232) \pi \rightarrow N \pi \pi )} = \label{eq:RatioRoper}\\
   &=& \frac{4}{6}\frac{Model_{1}}{Model_{2}} =\nonumber\\
   &=& 0.039 \pm 0.011 (stat.)  \pm 0.008 (sys.)\nonumber
\end{eqnarray}
where $4/6$ comes from the summing up over all isospin states \cite{PDG2008} and $\frac{Model_{1}}{Model_{2}}$ (Eq.~\ref{eq:Model12Final}).

\myFrameHugeFigure{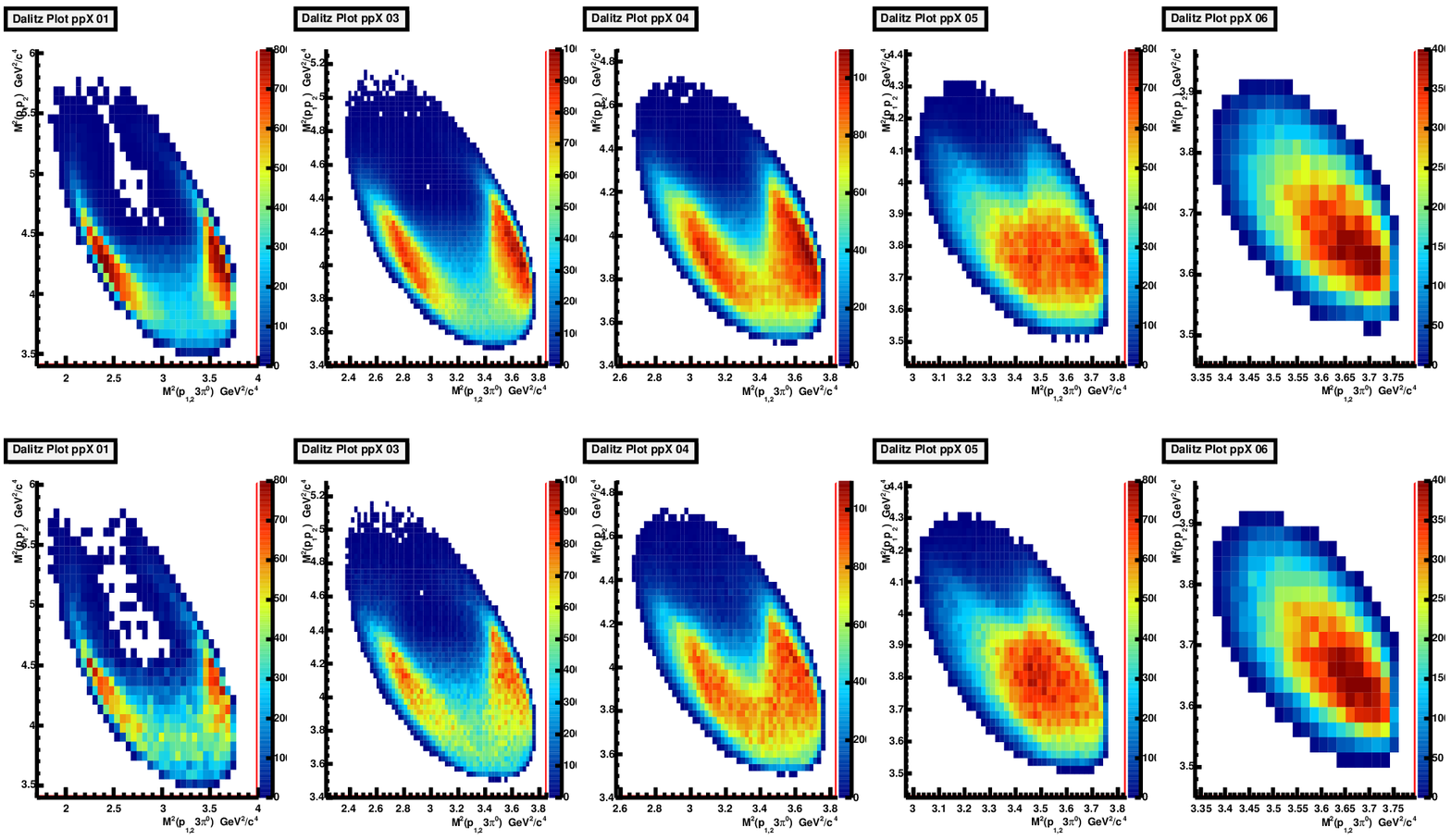}{Dalitz Plot $ppX$. $M^{2}(pp)$ versus $M^{2}(p3\pi^{0})$. The upper row corresponds to the experimental data,
the lower row to the Monte-Carlo model sum of $Model_{1}$ and $Model_{2}$ (Fig.~\ref{fig:3pi0Models}).
The columns from left to right correspond
to the following Missing Mass of two protons bins, column~1~$MM_{pp}=0.4-0.5\mathrm{~GeV/c^{2}}$,column~2~$MM_{pp}=0.6-0.7\mathrm{~GeV/c^{2}}$,
column~3~$MM_{pp}=0.7-0.8\mathrm{~GeV/c^{2}}$, column~4~$MM_{pp}=0.8-0.9\mathrm{~GeV/c^{2}}$, column~5~$MM_{pp}=0.9-1.0\mathrm{~GeV/c^{2}}$.
 The plots are symmetrized against two protons - each event is filled two times.
Fully expandable and colored version of the figure is available in the attached electronic version of the thesis. 
}{Dalitz Plot $ppX$.}

\myFrameHugeFigure{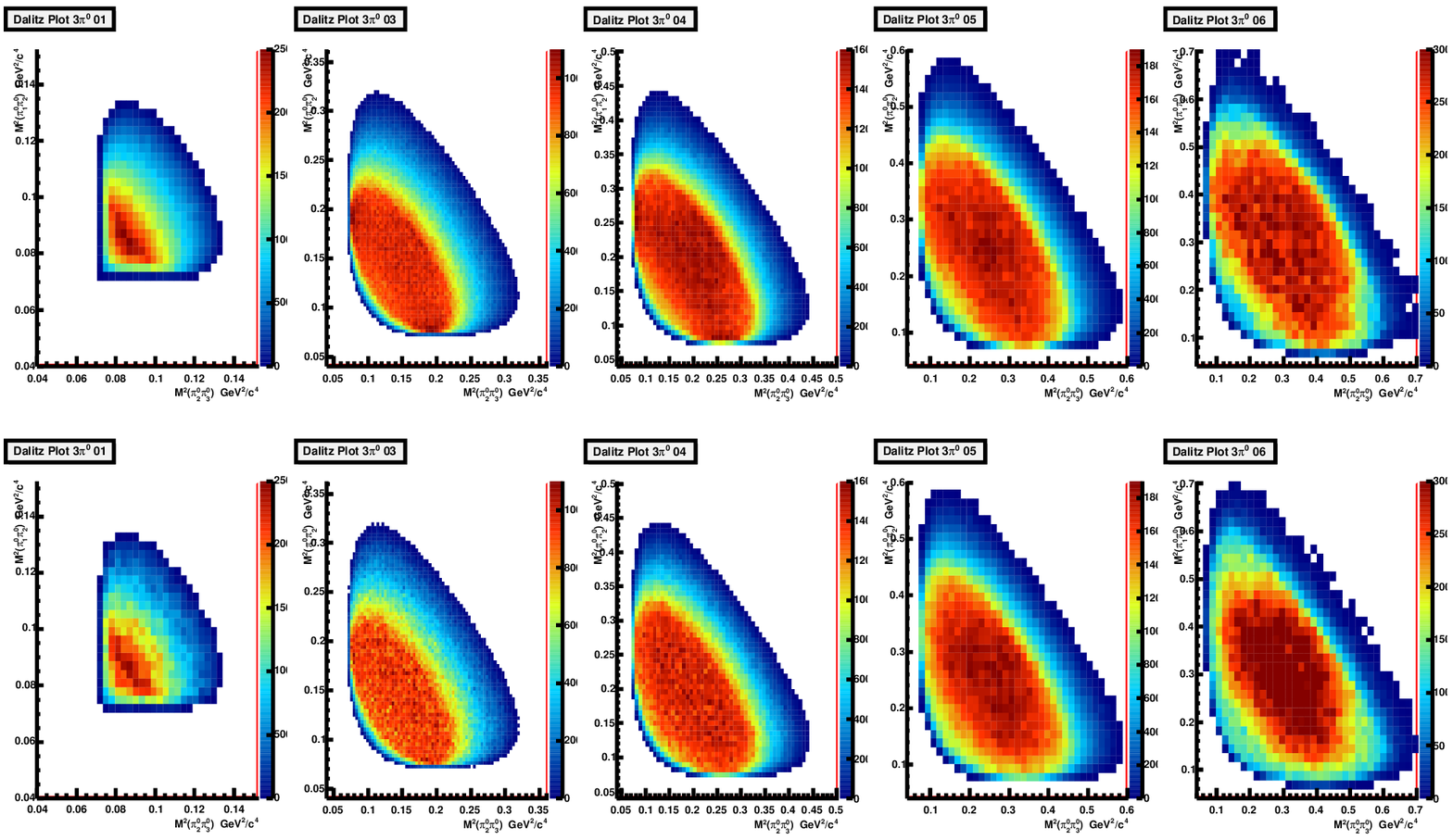}{Dalitz Plot $3\pi^{0}$. $M^{2}(2\pi^{0})$ versus $M^{2}(2\pi^{0})$.The upper row corresponds to the experimental data,
the lower row to the Monte-Carlo model sum of $Model_{1}$ and $Model_{2}$ (Fig.~\ref{fig:3pi0Models}).
 The columns from left to right correspond
to the following Missing Mass of two protons bins, column~1~$MM_{pp}=0.4-0.5\mathrm{~GeV/c^{2}}$,column~2~$MM_{pp}=0.6-0.7\mathrm{~GeV/c^{2}}$,
column~3~$MM_{pp}=0.7-0.8\mathrm{~GeV/c^{2}}$, column~4~$MM_{pp}=0.8-0.9\mathrm{~GeV/c^{2}}$, column~5~$MM_{pp}=0.9-1.0\mathrm{~GeV/c^{2}}$.
 The plots are symmetrized against three pions - each event is filled six times.
Fully expandable and colored version of the figure is available in the attached electronic version of the thesis. 
}{Dalitz Plot $3\pi^{0}$.}

\myFrameHugeFigure{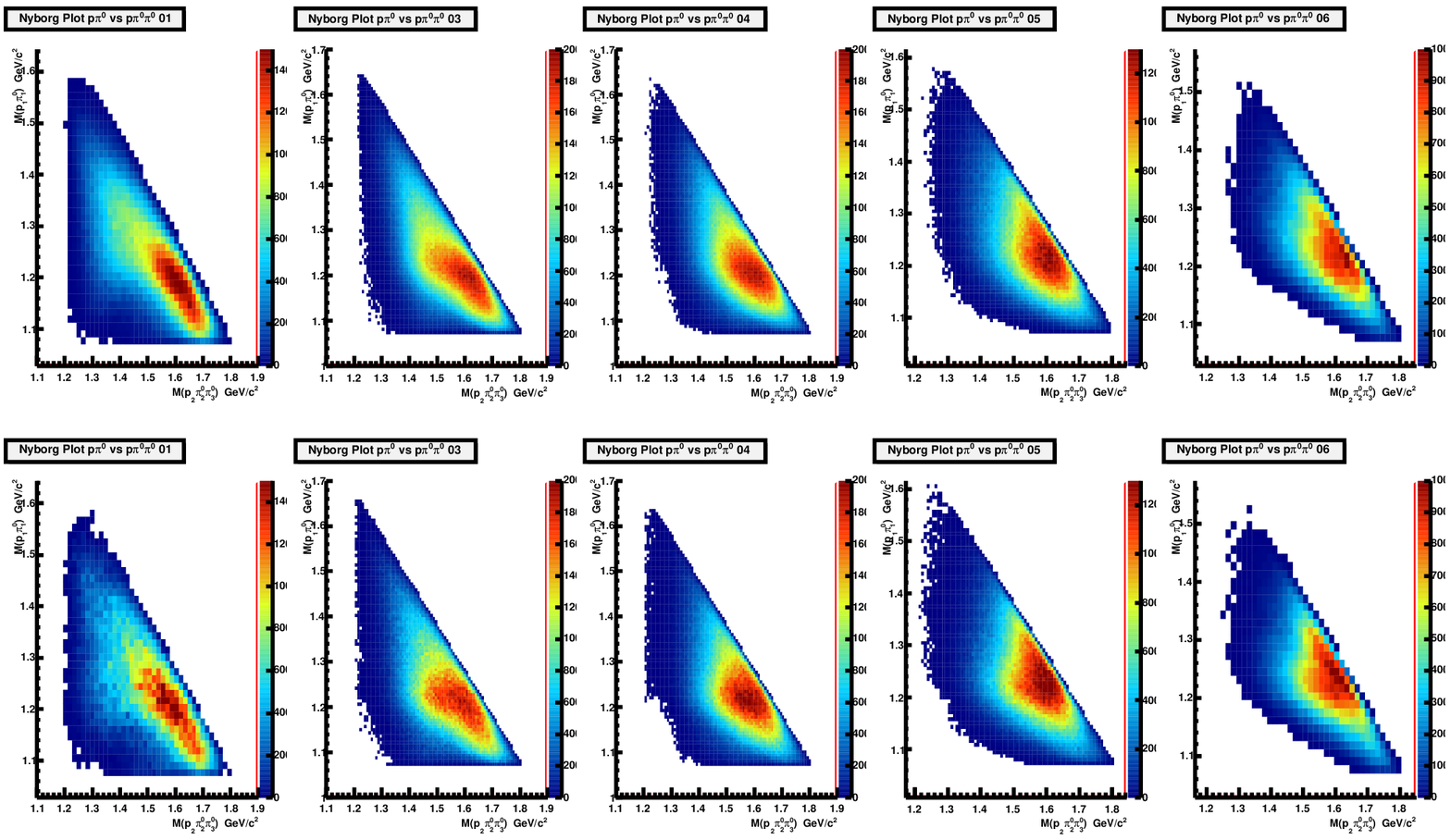}{Nyborg Plot. $M(p\pi^{0})$ versus $M(p\pi^{0}\pi^{0})$.The upper row corresponds to the experimental data,
the lower row to the Monte-Carlo model sum of $Model_{1}$ and $Model_{2}$ (Fig.~\ref{fig:3pi0Models}).
The columns from left to right correspond
to the following Missing Mass of two protons bins, column~1~$MM_{pp}=0.4-0.5\mathrm{~GeV/c^{2}}$,column~2~$MM_{pp}=0.6-0.7\mathrm{~GeV/c^{2}}$,
column~3~$MM_{pp}=0.7-0.8\mathrm{~GeV/c^{2}}$, column~4~$MM_{pp}=0.8-0.9\mathrm{~GeV/c^{2}}$, column~5~$MM_{pp}=0.9-1.0\mathrm{~GeV/c^{2}}$.
 The plots are symmetrized against two protons and three pions - each event is filled six times.
Fully expandable and colored version of the figure is available in the attached electronic version of the thesis. 
}{Nyborg Plot.}

\myFrameSmallFigure{MMpp02ModelSum.eps}{Missing Mass of two protons. Comparison of the experimental data (black marker) with the Monte-Carlo simulation phase space (blue line), the $Model_{1}$ (Fig.~\ref{fig:3pi0Model1})
 (yellow line) , the $Model_{2}$ (Fig.~\ref{fig:3pi0Model2}) (green line), the model sum (Eq.~\ref{eq:3pi0Model}) (red line).
Vertical axes - number of events (in given bin) is shown.
}{Missing Mass of two protons.}

\begin{figure}[ht!bp]
\centering
{
\subfigure[Fit of the fourth order polynomial outside the $\eta$ meson peak.]{\fbox{\includegraphics[width=0.7\textwidth]{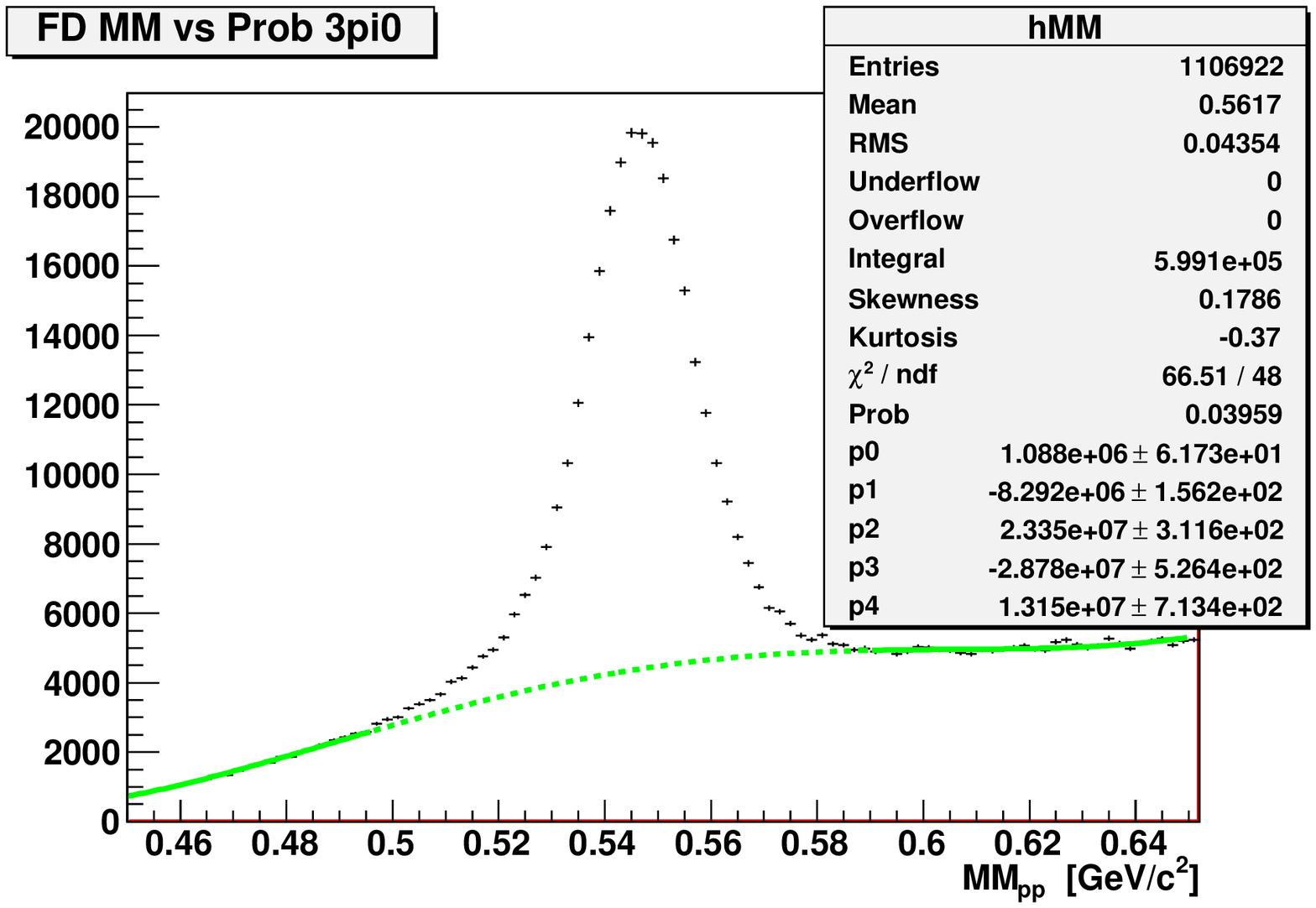}} \label{fig:MMppEtaBGSubtraction1}}\\
\subfigure[Subtracted $\eta$ meson signal, blue markers correspond to the Monte-Carlo simulation.]{\fbox{\includegraphics[width=0.7\textwidth]{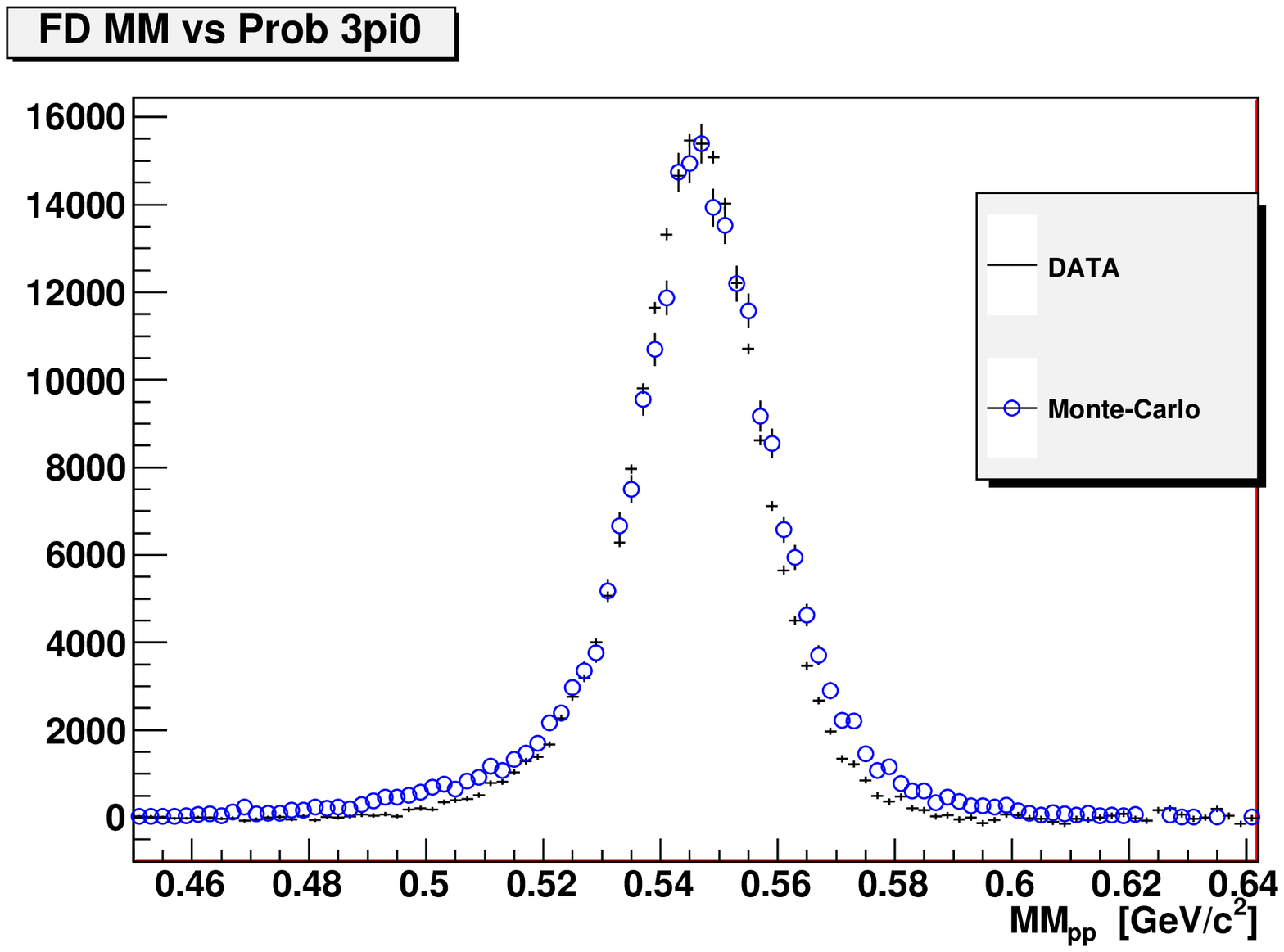}} \label{fig:MMppEtaBGSubtraction2}}
}
\caption{Subtraction of the background under the $\eta$ peak from $MM_{pp}$ distribution.}
\label{fig:MMppEtaBGSubtraction}
\end{figure}

\begin{figure}[ht!bp]
\centering
{
\subfigure[black: subtracted experimental data, blue: the true value of the Monte-Carlo simulation composed of the sum of $Model_{1}$ and $Model_{2}$.]{\fbox{\includegraphics[width=0.7\textwidth]{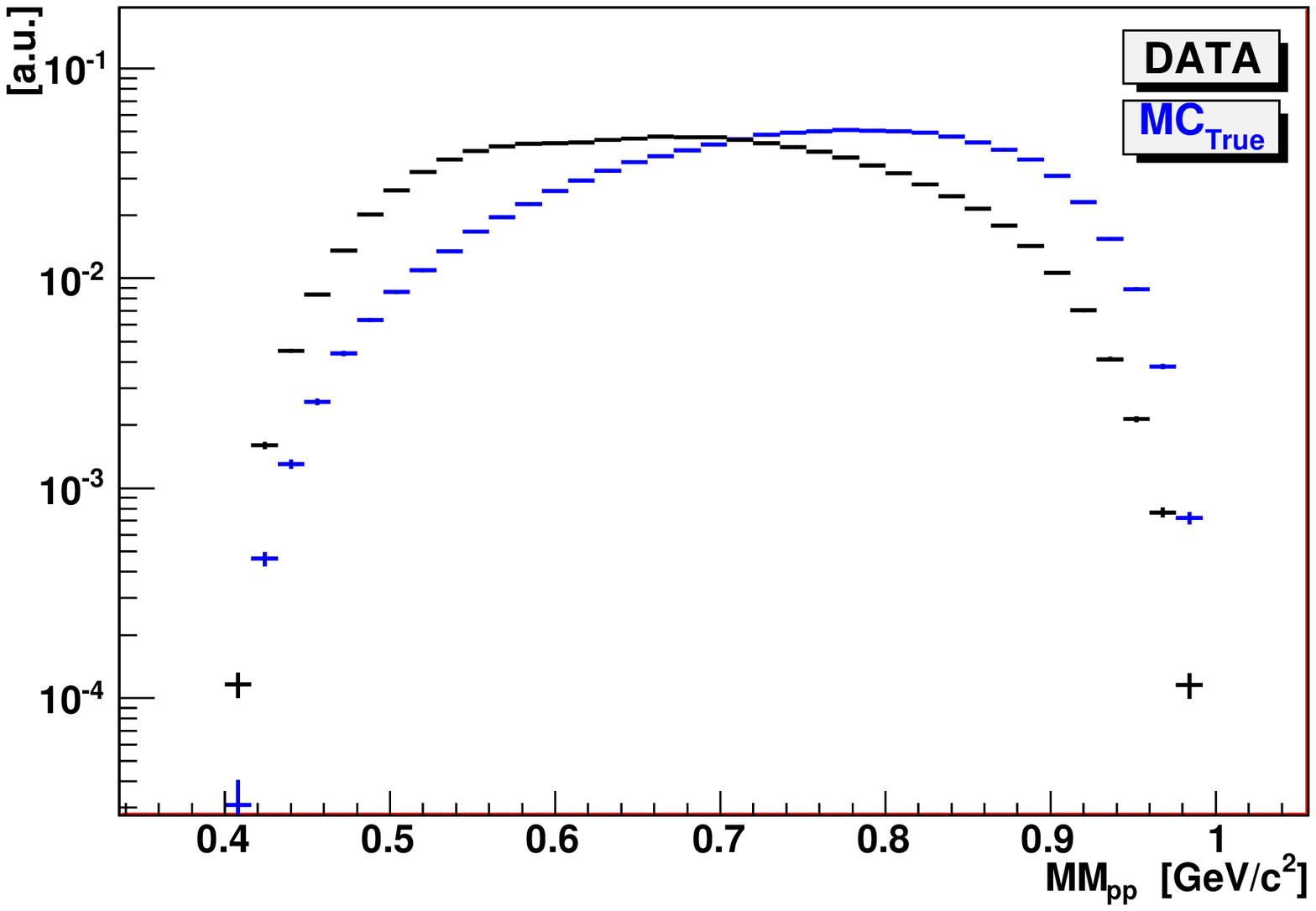}} \label{fig:MMfacor1}}\\
\subfigure[The missing mass population function $f(MM_{pp})$.]{\fbox{\includegraphics[width=0.7\textwidth]{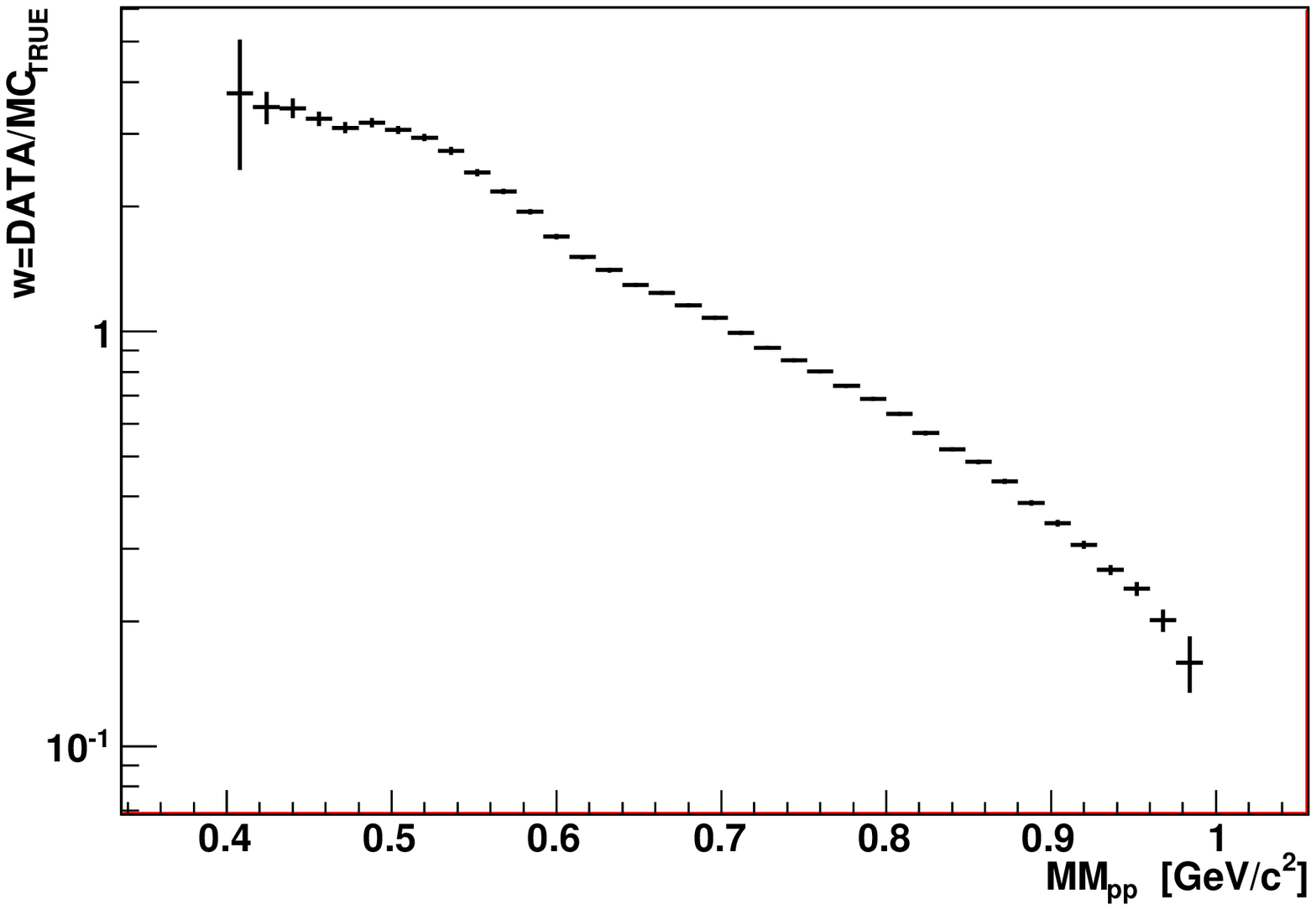}} \label{fig:MMfacor2}}
}
\caption{Missing Mass dependence of event population.}
\label{fig:MMfacor}
\end{figure}

\bigskip
The $\Delta(1232)$ and $N^{*}(1440)$ were identified by their unique topologies of the events on Dalitz and Nyborg plots which were mimic by the the sum of $Model_{1}$ and $Model_{2}$ (Fig.~\ref{fig:3pi0Models}),
which were fitted to the experimental data.
The population of the events as a function of the missing mass of two protons was not a purpose of the fit.
It is seen (Fig.~\ref{image_MMpp02ModelSum.eps}) that the proposed process $Model_{1}$ and $Model_{2}$ (Fig.~\ref{fig:3pi0Models}) as well as the homogeneously and isotropically populated phase space
populate almost the same area on the missing mass of the two protons $MM_{pp}$, which is different from the population of the experimental data.  
In order to describe the event population as a function of $MM_{pp}$, 
the missing mass population function $f(MM_{pp})$ was derived from the experimental data.
First the $\eta$ meson signal was subtracted from the data by fitting outside the $\eta$ meson peak the fourth order polynomial (Fig.~\ref{fig:MMppEtaBGSubtraction}).
Next the subtracted experimental data were compared with the true value of the Monte-Carlo simulation composed of the sum of $Model_{1}$ and $Model_{2}$ (Fig.~\ref{fig:3pi0Models}); 
the experimental data histogram was divided by the Monte-Carlo model histogram to obtain the missing mass population function $f(MM_{pp})$ Table~\ref{tab:MMfactor} (Fig.~\ref{fig:MMfacor2}).

\myTable{
\footnotesize
\begin{tabular}{|c|c|}
\hline
$MM_{pp}\mathrm{~GeV/c^{2}}$ & $f(MM_{pp})$ \\ 
\hline
\hline
$0.408 \pm 0.008$	& $3.8 \pm 1.3$ \\
\hline
 $0.424 \pm 0.008$	& $3.47 \pm 0.31$ \\
\hline
 $0.440 \pm 0.008$	& $3.46 \pm 0.19$\\
\hline
 $0.456 \pm 0.008$	& $3.26 \pm 0.13$\\
\hline
 $0.472 \pm 0.008$	& $3.102 \pm 0.092$\\
\hline
$ 0.488 \pm 0.008$	& $3.191 \pm 0.079$\\
\hline
 $0.504 \pm 0.008$	& $3.062 \pm 0.066$\\
\hline
 $0.520 \pm 0.008$	& $2.936 \pm 0.058$\\
\hline
 $0.536 \pm 0.008$	& $2.729 \pm 0.057$\\
\hline
 $0.552 \pm 0.008$	& $2.419 \pm 0.048$\\
\hline
 $0.568 \pm 0.008$	& $2.178 \pm 0.034$\\
\hline
$ 0.584 \pm 0.008$	& $1.944 \pm 0.027$\\
\hline
$ 0.600 \pm 0.008$	& $1.695 \pm 0.022$\\
\hline
$ 0.616 \pm 0.008$	& $1.513 \pm 0.019$\\
\hline
 $0.632 \pm 0.008$	& $1.407 \pm 0.017$\\
\hline
$ 0.648 \pm 0.008$	& $1.297 \pm 0.015$\\
\hline
 $0.664 \pm 0.008$	& $1.240 \pm 0.014$\\
\hline
 $0.680 \pm 0.008$	& $1.156 \pm 0.013$\\
\hline
 $0.696 \pm 0.008$	& $1.079 \pm 0.012$\\
\hline
$ 0.712 \pm 0.008$	& $0.995 \pm 0.011$\\
\hline
 $0.728 \pm 0.008$	& $0.9131 \pm 0.0099$\\
\hline
$ 0.744 \pm 0.008$	& $0.8540 \pm 0.0093$\\
\hline
$ 0.760 \pm 0.008$	& $0.8012 \pm 0.0088$\\
\hline
 $0.776 \pm 0.008$	& $0.7396 \pm 0.0082$\\
\hline
$ 0.792 \pm 0.008$	& $0.6877 \pm 0.0078$\\
\hline
 $0.808 \pm 0.008$	& $0.6330 \pm 0.0074$\\
\hline
 $0.824 \pm 0.008$	& $0.5690 \pm 0.0069$\\
\hline
 $0.840 \pm 0.008$	& $0.5207 \pm 0.0066$\\
\hline
$ 0.856 \pm 0.008$	& $0.4851 \pm 0.0065$\\
\hline
$ 0.872 \pm 0.008$	& $0.4354 \pm 0.0063$\\
\hline
$ 0.888 \pm 0.008$	&$ 0.3862 \pm 0.0061$\\
\hline
 $0.904 \pm 0.008$	&$ 0.3456 \pm 0.0062$\\
\hline
$ 0.920 \pm 0.008$	& $0.3060 \pm 0.0066$\\
\hline
$ 0.936 \pm 0.008$	& $0.2665 \pm 0.0073$\\
\hline
$ 0.952 \pm 0.008$	& $0.2400 \pm 0.0091$\\
\hline
$ 0.968 \pm 0.008$	&$ 0.202 \pm 0.012$\\
\hline
 $0.984 \pm 0.008$	&$ 0.159 \pm 0.025$\\
\hline
\end{tabular} 
}{The missing mass population function $f(MM_{pp})$.}{tab:MMfactor}

\paragraph{The Overall Model\\}

\myFrameFigure{MMpp02ModelSumFactor}{Missing Mass of two protons. Comparison of the experimental data (black marker) with the Monte-Carlo simulation phase space (blue line), the $Model_{1}$ (Fig.~\ref{fig:3pi0Model1})
 (yellow line) , the $Model_{2}$ (Fig.~\ref{fig:3pi0Model2}) (green line), the model sum (Eq.~\ref{eq:3pi0Model}) (red line).
The derived $f(MM_{pp})$ (see Table~\ref{tab:MMfactor} and Fig.~\ref{fig:MMfacor2}) was used for the models. 
Vertical axes - number of events (in given bin) is shown.
}{Missing Mass of two protons.}

\myFrameHugeFigure{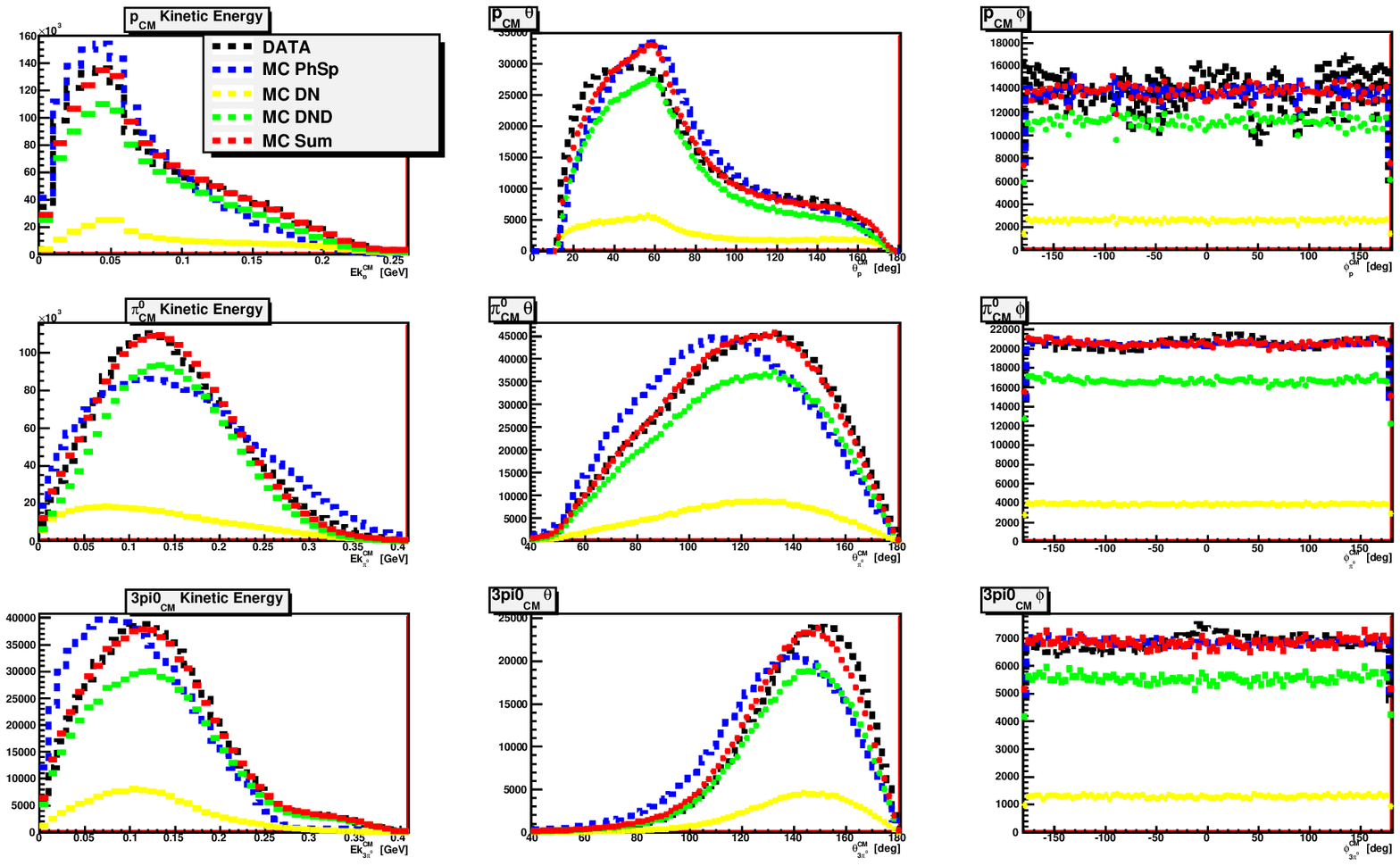}{Comparison of the experimental data with the Monte-Carlo simulation for the $MM_{pp}<0.5\mathrm{~GeV/c^{2}}$ and $MM_{pp}>0.6\mathrm{~GeV/c^{2}}$.
\textit{Upper row}: proton in the center of mass frame,\textit{Middle row}: pion in the center of mass frame, \textit{Lower row}: $3\pi^{0}$ system in the center of mass frame. From left kinetic energy, 
the polar angle and the azimuthal angle distribution. The experimental data are shown as a black marker, the Monte-Carlo simulation phase space (blue line), the $Model_{1}$ (Fig.~\ref{fig:3pi0Model1}) (yellow line)
 , the $Model_{2}$ (Fig.~\ref{fig:3pi0Model2}) (green line), the model sum (Eq.~\ref{eq:3pi0Model}) (red line).
The derived $f(MM_{pp})$ (see Table~\ref{tab:MMfactor} and Fig.~\ref{fig:MMfacor2}) was used for the models. 
Vertical axes - number of events (in given bin) is shown.
}{Comparison of the experimental data with the Monte-Carlo
 simulation for the $MM_{pp}<0.5\mathrm{GeV/c^{2}}$ and $MM_{pp}>0.6\mathrm{GeV/c^{2}}$.}
 
One can write the overall proposed model of the $pp \rightarrow pp 3\pi^{0}$ production, extracted by the Monte-Carlo comparison from experimental data:

\begin{equation}
 Model = \left[ \beta Model_{1} + (1-\beta) Model_{2}\right] \times f(MM_{pp})
\label{eq:3pi0Model}
\end{equation}

The $Model_{1}$ corresponds to the process (Fig.~\ref{fig:3pi0Model1}) and the $Model_{2}$ to the process (Fig.~\ref{fig:3pi0Model2}), which assumes no interaction between the baryons - homogeneously and isotropically populated phase space.

The $\beta$ parameter is the fraction of $Model_{1}$ to the $Model_{1}+Model_{2}$ obtain from the chi-square fit to the experimental data (Eq.~\ref{eq:ModelBetaValue}). 

The $f(MM_{pp})$ corresponds to the missing mass population function Table~\ref{tab:MMfactor} (Fig.~\ref{fig:MMfacor2}), which is derived from the experimental data.

It might be related to the possible interaction between the baryons not included in the Monte-Carlo model and the internal properties of the $N^{*}(1440)$, which is discussed below (see page~\pageref{par:fMMppOrigin}).

The (Figs.~\ref{image_MMpp02ModelSumFactor},~\ref{image_ComparisonModelSumFactor}) shows the comparison of the proposed model (Eq.~\ref{eq:3pi0Model}) with the experimental data. 
It is seen that the model describes very good the experimental data.
More detail quantitative discussion on the validation of the developed model will be presented in Section~\ref{subsec:MCModelValidation} on page \pageref{subsec:MCModelValidation}.

\paragraph{The origin of the missing mass population function\\}\label{par:fMMppOrigin}
Searching for the origin of the missing mass population function $f(MM_{pp})$ (Table~\ref{tab:MMfactor}, (Fig.~\ref{fig:MMfacor2}))   
the kinematics of the $pp \rightarrow \Delta(1232) N^{*}(1440)$ reaction was studied in details by \textbf{Pluto++} calculations, via Monte-Carlo method (see Appendix~\ref{appendix:wmc}).
The following explanations are considered.

\subparagraph{The Possibility of $\Delta(1232) N^{*}(1440)$ interaction\\}\label{par:DNinteraction}

\begin{sidewaysfigure}[t!bp]
\centering
{
\subfigure[Four momentum transfer to the $\Delta(1232)$ as a function of the
missing mass of the two protons.]{\fbox{\includegraphics[width=0.45\textwidth]{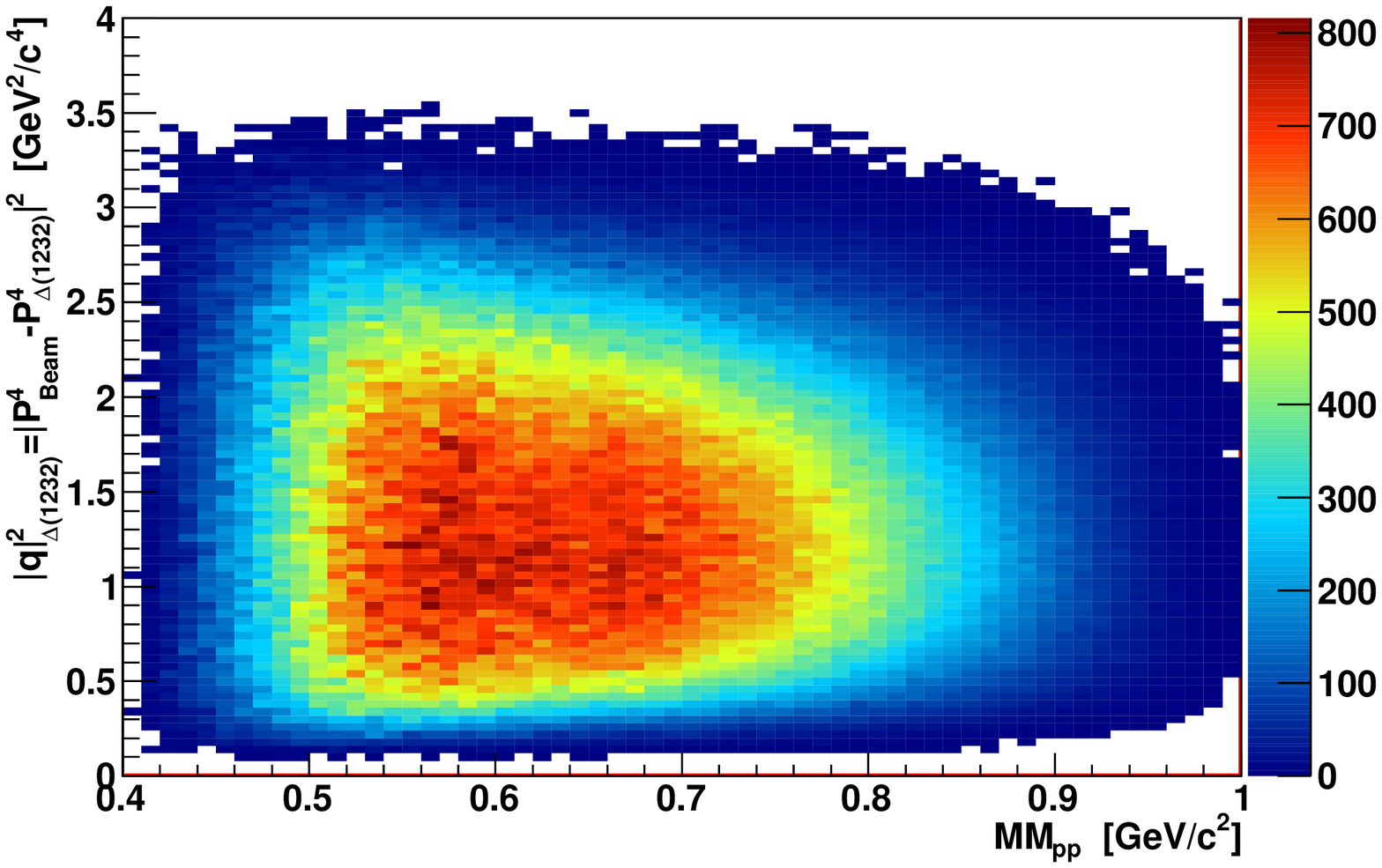}} \label{fig:MomTransferDN1}}\quad
\subfigure[Four momentum transfer to the $N^{*}(1440)$ as a function of the
missing mass of the two protons.]{\fbox{\includegraphics[width=0.45\textwidth]{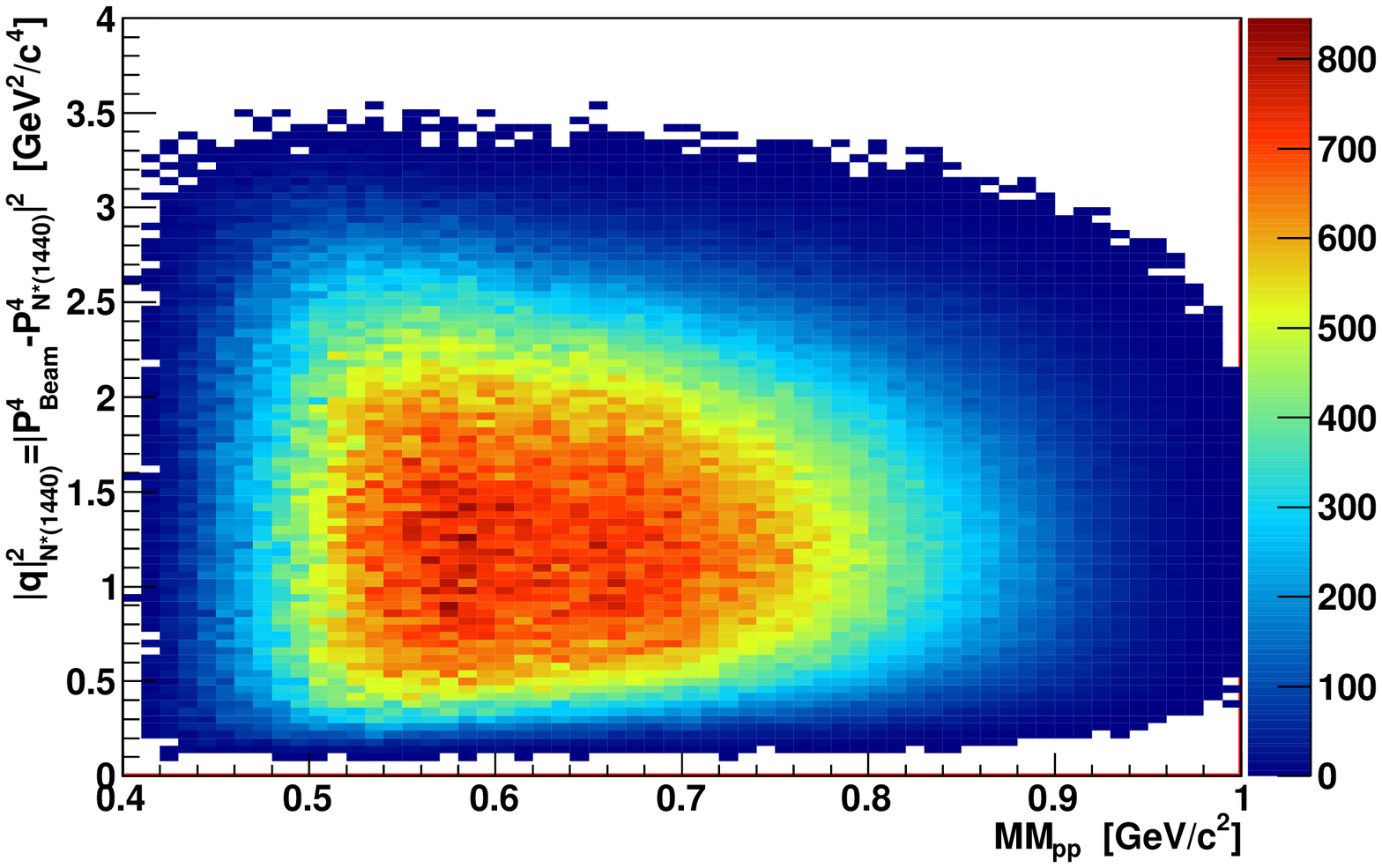}} \label{fig:MomTransferDN2}}\\
\subfigure[Four momentum transfer to the $N^{*}(1440)$ versus the four momentum transfer to the $\Delta(1232)$.]{\fbox{\includegraphics[width=0.45\textwidth]{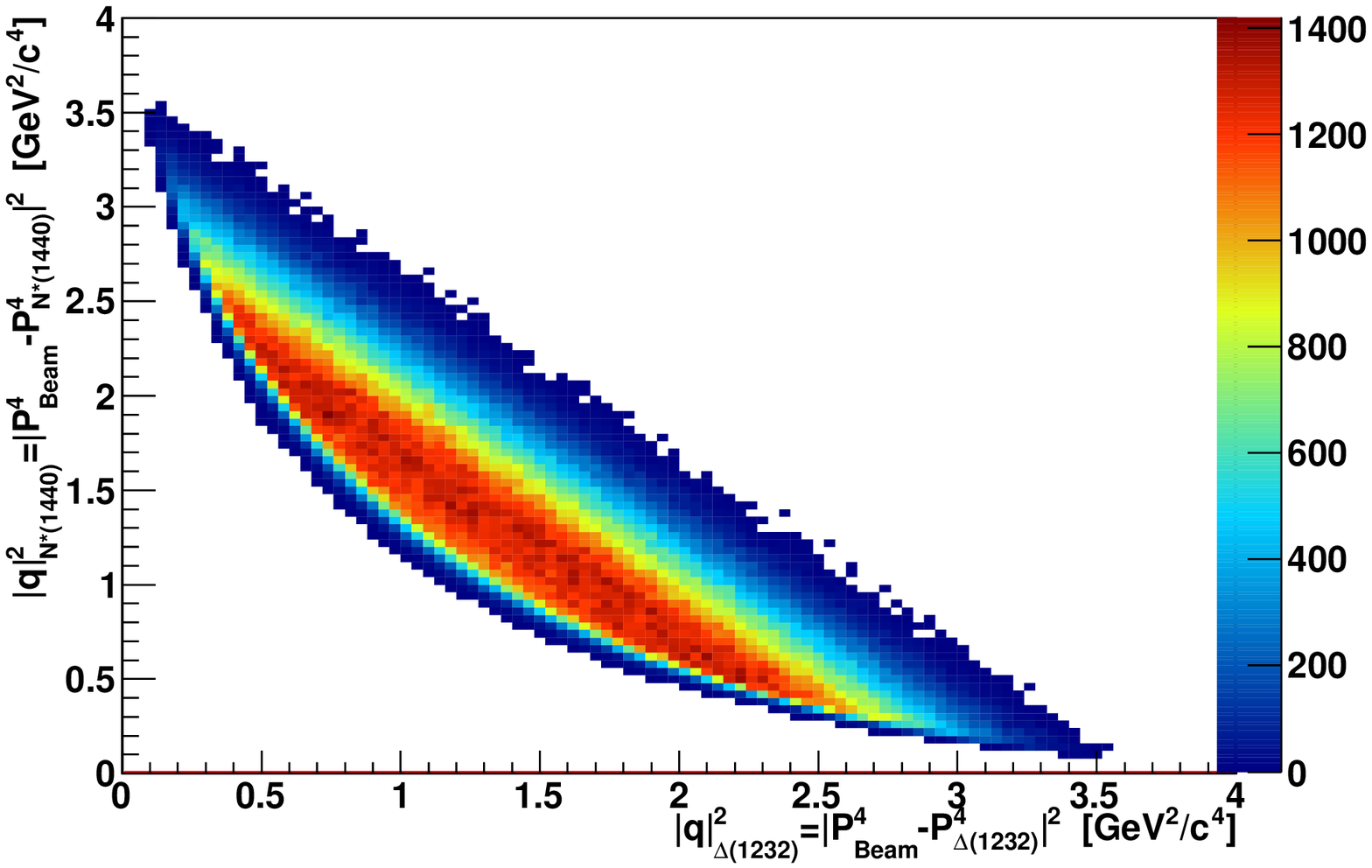}} \label{fig:MomTransferDN3}}
}
\caption{Calculations by \textbf{Pluto++} via Monte-Carlo simulations. 
}
\label{fig:MomTransferDN}
\end{sidewaysfigure}

To check the possibility of $\Delta(1232) N^{*}(1440)$ interaction and its influence of the $MM_{pp}$,
it is very convenient to study the four momentum transfers $q$ dependences since it gives the informations about the distance resolution of the reaction $\Delta r$ (interaction distance)
accordingly to the relation \cite{ReactionsLipperheide,QMDavidov,QMSchiff}:

\begin{equation}
 \Delta r = \frac{\hbar}{|q|}
\end{equation}

where $\hbar$ is a Planck's constant.
 
\bigskip

First the four momentum transfer to the $\Delta(1232)$ as a function of the
missing mass of the two protons and four momentum transfer to the $N^{*}(1440)$ as a function of the
missing mass of the two protons was examined (Fig.~\ref{fig:MomTransferDN}).
It is seen that there is no dependence between these four momentum transfers and the $MM_{pp}$.

\myFrameSmallFigure{DNMomTransfer2D.eps}{Monte-Carlo simulation,difference of four momentum between the $\Delta(1232)$ and $N^{*}(1440)$ as a function of the
missing mass of the two protons. Calculations by \textbf{Pluto++}. }{fMMpopulation Q}

\myFrameSmallFigure{DNMomTransfer1D.eps}{Monte-Carlo simulation,average difference of four momentum between the $\Delta(1232)$ and $N^{*}(1440)$ as a function of the
missing mass of the two protons. Calculations by \textbf{Pluto++}. }{fMMpopulation Q}

\myFrameSmallFigure{MMFactorDNMomTransfer1D.eps}{The $f(MM_{pp})$ as a function of the average difference of four momentum between the $\Delta(1232)$ and $N^{*}(1440)$}{fmmpp}

One can also try to relate the $MM_{pp}$ with the difference of four momentum between the $\Delta(1232)$ and $N^{*}(1440)$ in the center of mass frame $|q|^{2}$ \myImgRef{DNMomTransfer2D.eps}.
It is seen that the variables are correlated. One can calculate also the average $q$ as a function of $MM_{pp}$ \myImgRef{DNMomTransfer1D.eps}.
It is now possible using the relation from \myImgRef{DNMomTransfer1D.eps} to recalculate the $f(MM_{pp})$ - the missing mass population function Table~\ref{tab:MMfactor} (Fig.~\ref{fig:MMfacor2})
to dependence on average difference of four momentum between the $\Delta(1232)$ and $N^{*}(1440)$ $|q|^{2}$ \myImgRef{MMFactorDNMomTransfer1D.eps}.
It is seen that when the average difference of four momentum increases the $f(MM_{pp})$ increases exponentially.
If the possible interaction would be interpreted as the One Boson Exchange (i.e. by exchange of the $\pi^{0}$, $f_{0}\diagup\sigma(600)$) than the behavior of the interaction
would be $\sim \frac{1}{|q|^{2}}$ \cite{OBE1,OBE2,OBE3,OBE4,OBE5,OBE6,OBE7,OBELast}. When the difference of four momentum increases the interaction should strongly decrease.
Since the $f(MM_{pp})$ behaves completely differently with an increase of the $|q|^{2}$, as shown above,
the interaction between the $\Delta(1232) N^{*}(1440)$ system understood as the One Boson Exchange is excluded.           

\newpage
\subparagraph{The influence of the $N^{*}(1440)$ line shape\\}\label{par:Nlineshape}

\begin{sidewaysfigure}[t!bp]
\centering
{
\subfigure[$\Delta(1232)$ line shape versus $MM_{pp}$,\newline where $N^{*}(1440) \rightarrow p \pi^{0}\pi^{0}$ (stable particles)]{\fbox{\includegraphics[width=0.45\textwidth]{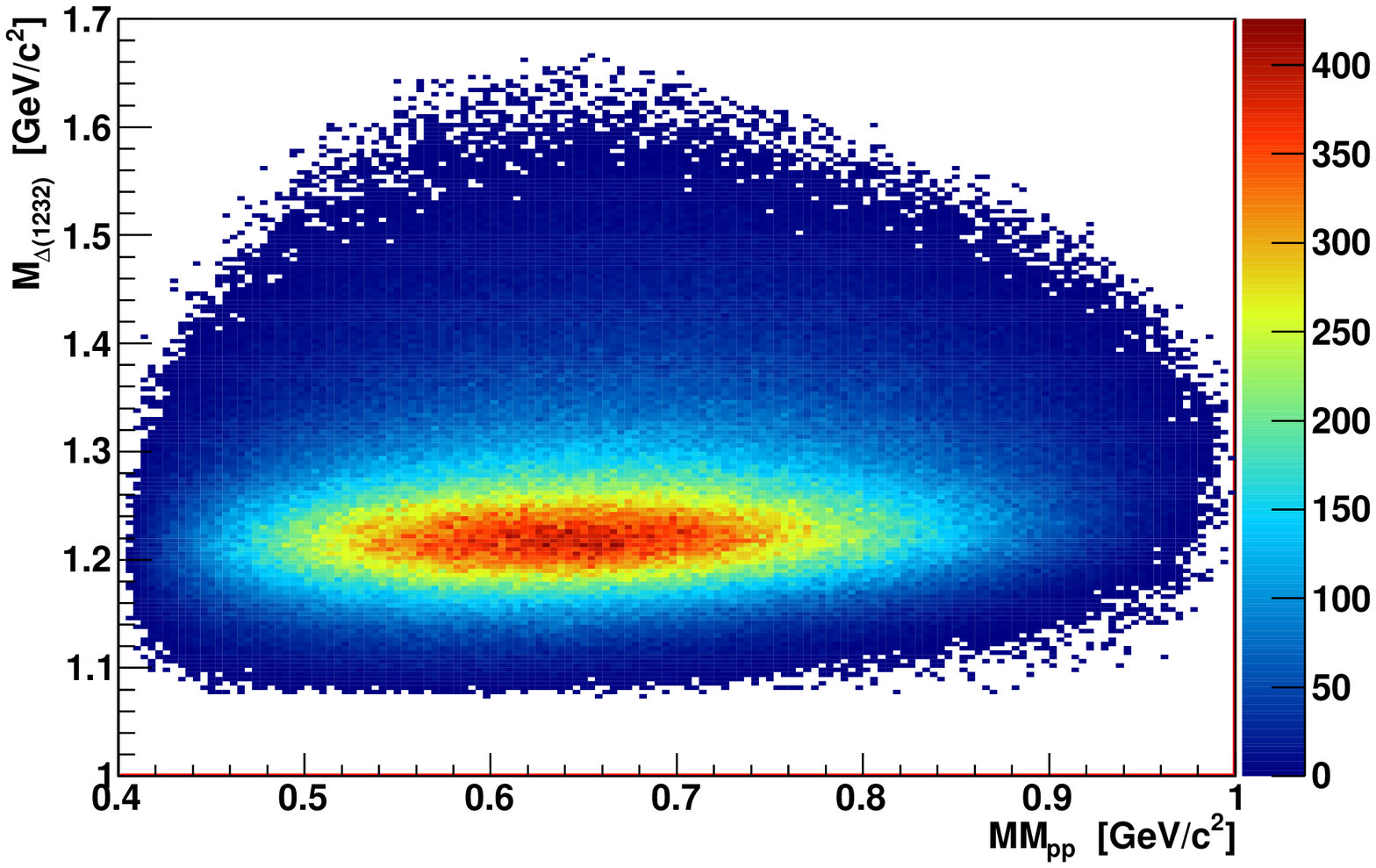}} \label{fig:D1232N1440lineshapeMod2StableD}}\quad
\subfigure[$N^{*}(1440)$ line shape versus $MM_{pp}$,\newline where $N^{*}(1440) \rightarrow p \pi^{0}\pi^{0}$ (stable particles)]{\fbox{\includegraphics[width=0.45\textwidth]{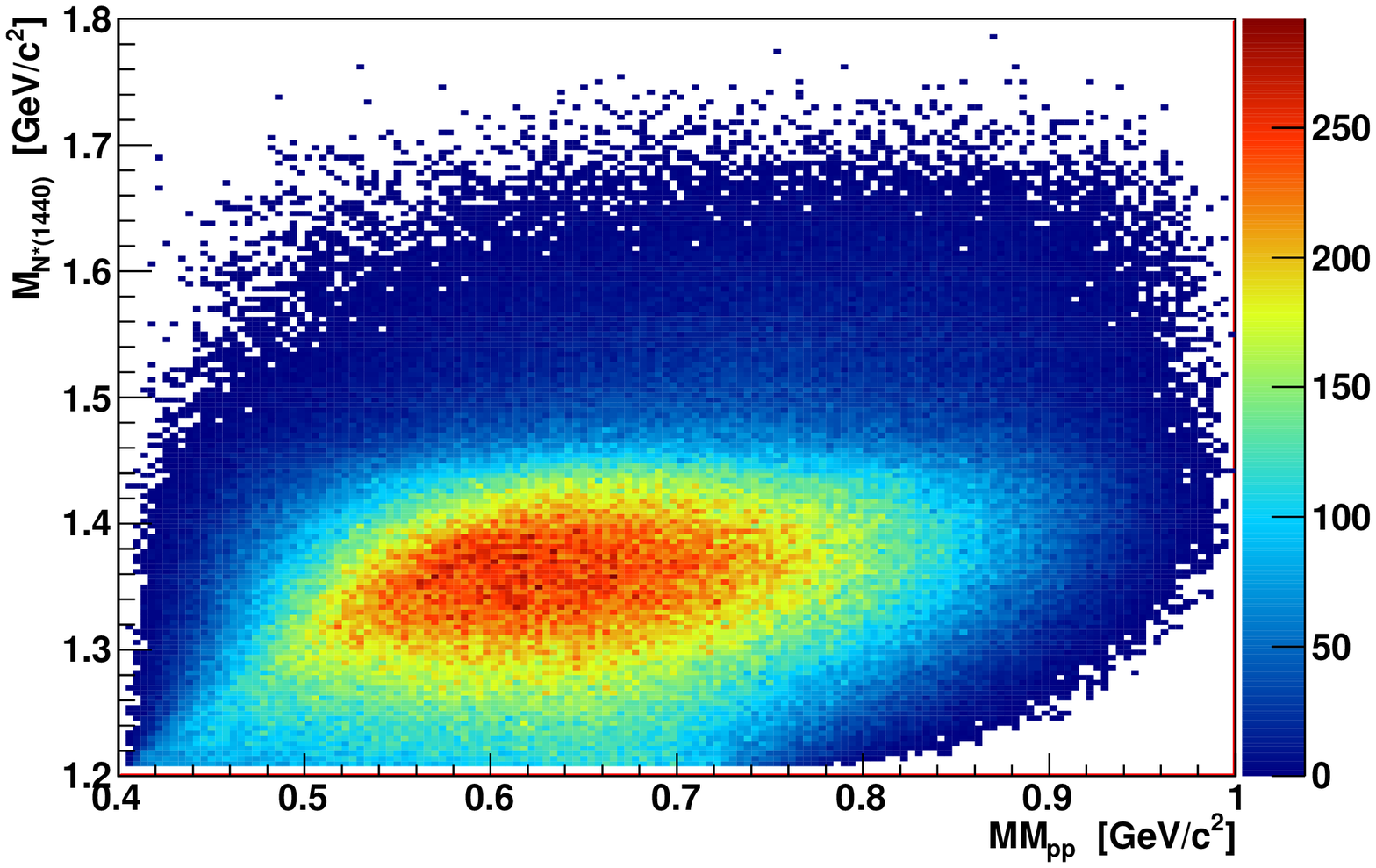}} \label{fig:D1232N1440lineshapeMod2StableN}}\\
\subfigure[$\Delta(1232)$ line shape versus $MM_{pp}$,\newline where $N^{*}(1440) \rightarrow \pi^{0} \Delta(1232)$ (unstable particle)]{\fbox{\includegraphics[width=0.45\textwidth]{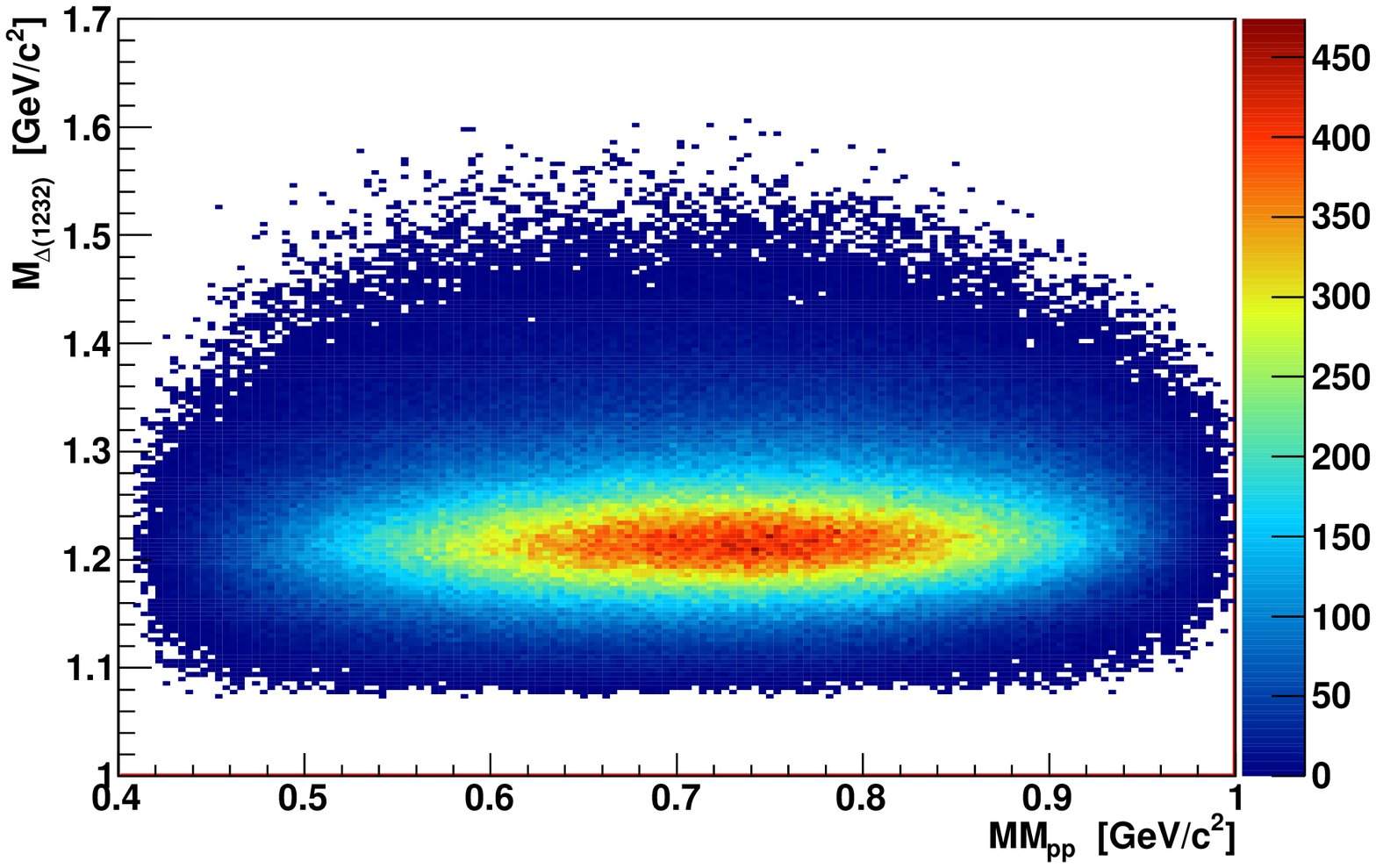}} \label{fig:D1232N1440lineshapeMod2UnStableD}}\quad
\subfigure[$N^{*}(1440)$ line shape versus $MM_{pp}$,\newline where $N^{*}(1440) \rightarrow \pi^{0} \Delta(1232)$ (unstable particle)]{\fbox{\includegraphics[width=0.45\textwidth]{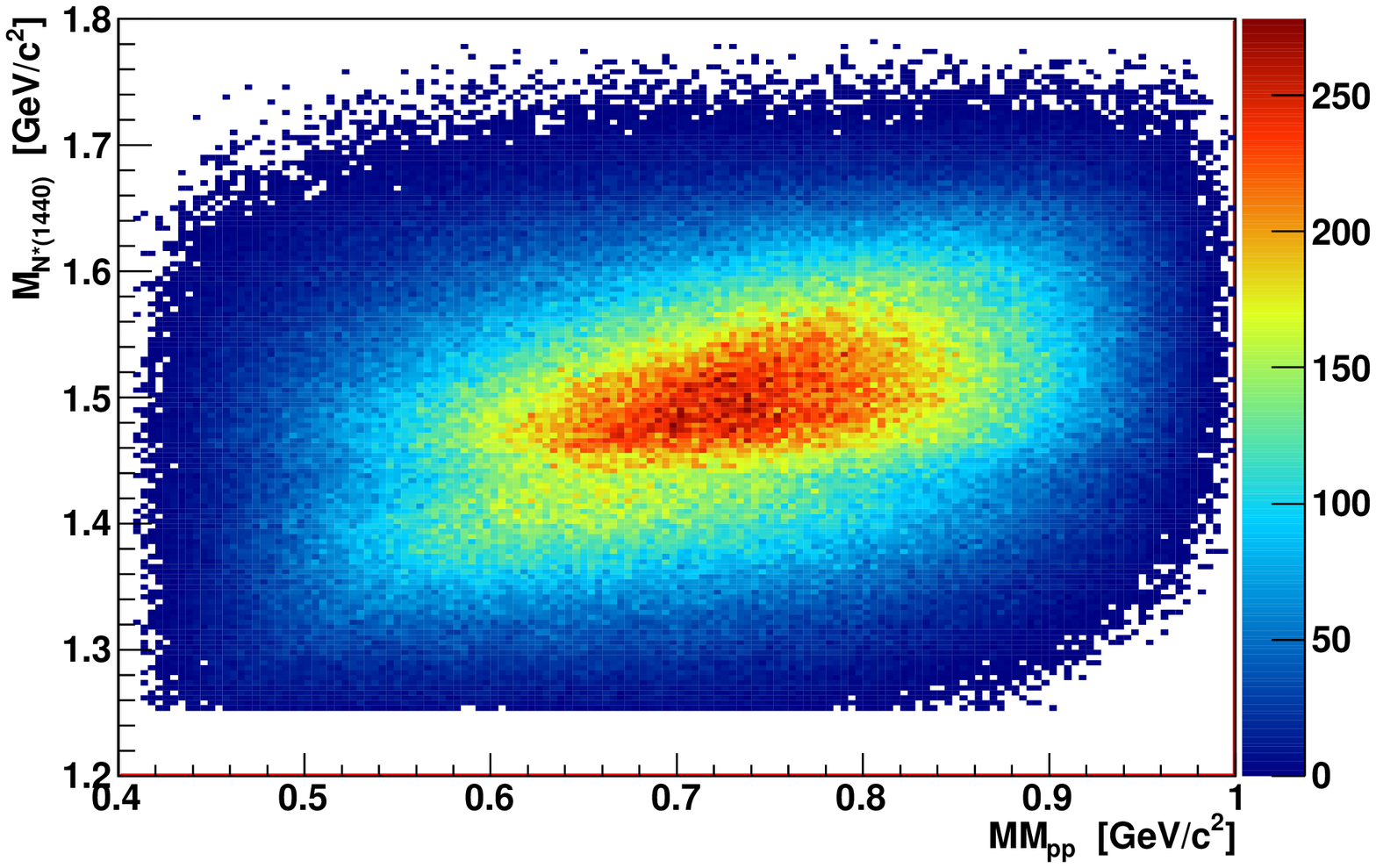}} \label{fig:D1232N1440lineshapeMod2UnStableN}}
}
\caption{Realistic effective spectral line shape $g(m)$ of $\Delta(1232)$ and $N^{*}(1440)$
in the $pp \rightarrow \Delta(1232) N^{*}(1440)$ reaction at incident proton momentum of $3.350\mathrm{~GeV/c^{2}}$ versus $MM_{pp}$.
The $\Delta(1232)$ decays into $p\pi^{0}$ (stable particles), the $N^{*}(1440)$ decays into $p\pi^{0}\pi^{0}$ (stable particles) or into $\pi^{0}\Delta(1232)$ (unstable particle),
when later $\Delta(1232)$ decays into $p\pi^{0}$. 
The correlation between the $N^{*}(1440)$ line shape and $MM_{pp}$ visible, no correlation in case of $\Delta(1232)$ seen. 
Calculations by \textbf{Pluto++} (via Monte-Carlo method).  
}
\label{fig:D1232N1440lineshapeMod2}
\end{sidewaysfigure}

\begin{sidewaysfigure}[t!bp]
\centering
{
\subfigure[Average $N^{*}(1440)$ mass versus $MM_{pp}$,\newline where $N^{*}(1440) \rightarrow p \pi^{0}\pi^{0}$ (stable particles)]{\fbox{\includegraphics[width=0.45\textwidth]{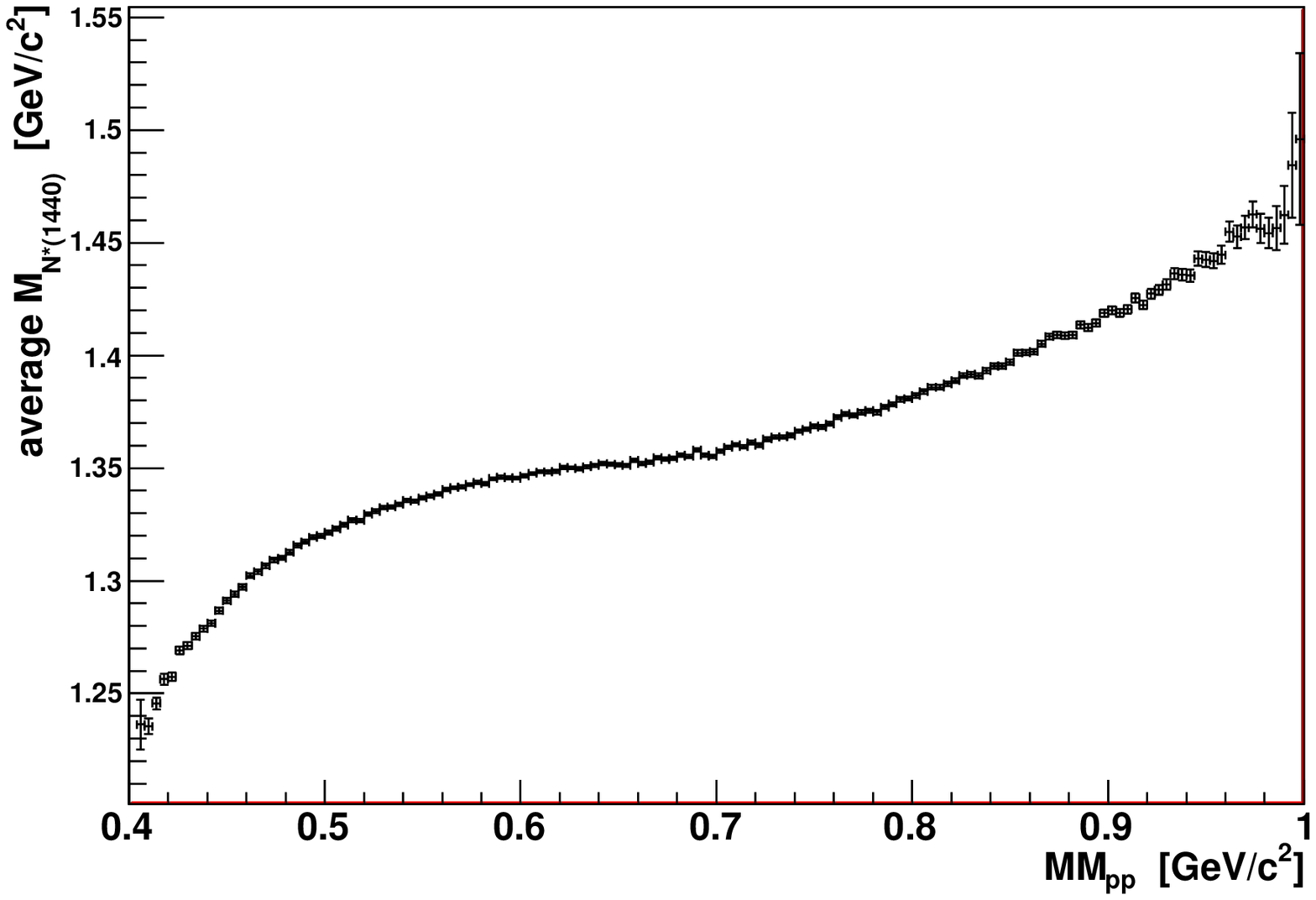}} \label{fig:D1232N1440lineshapeMod3StableD}}\quad
\subfigure[The $f(MM_{pp})$ as a function of the average $N^{*}(1440)$ mass,\newline where $N^{*}(1440) \rightarrow p \pi^{0}\pi^{0}$ (stable particles)]{\fbox{\includegraphics[width=0.45\textwidth]{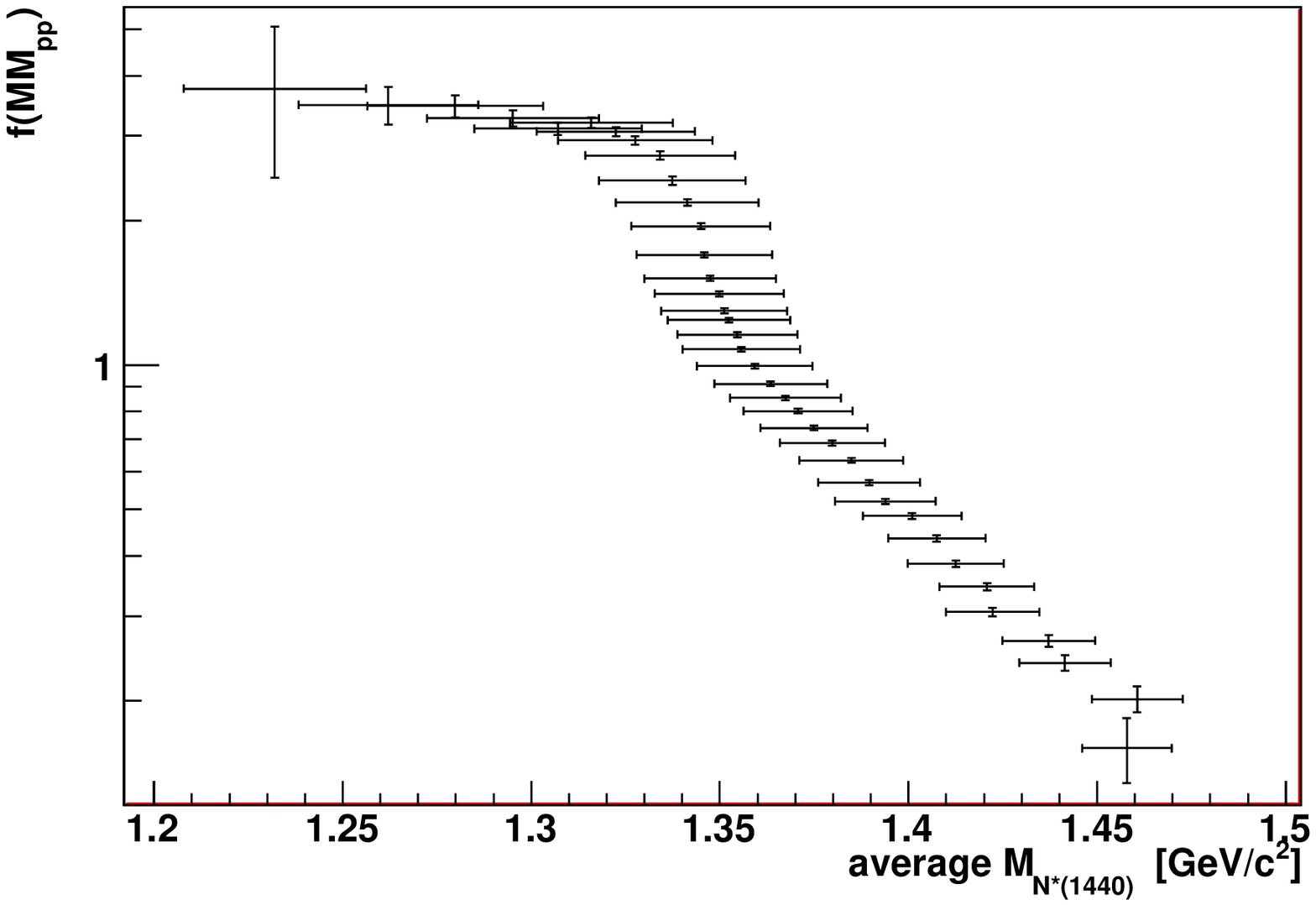}} \label{fig:D1232N1440lineshapeMod3StableN}}\\
\subfigure[Average $N^{*}(1440)$ mass versus $MM_{pp}$,\newline where $N^{*}(1440) \rightarrow p \Delta(1232)$ (unstable particle)]{\fbox{\includegraphics[width=0.45\textwidth]{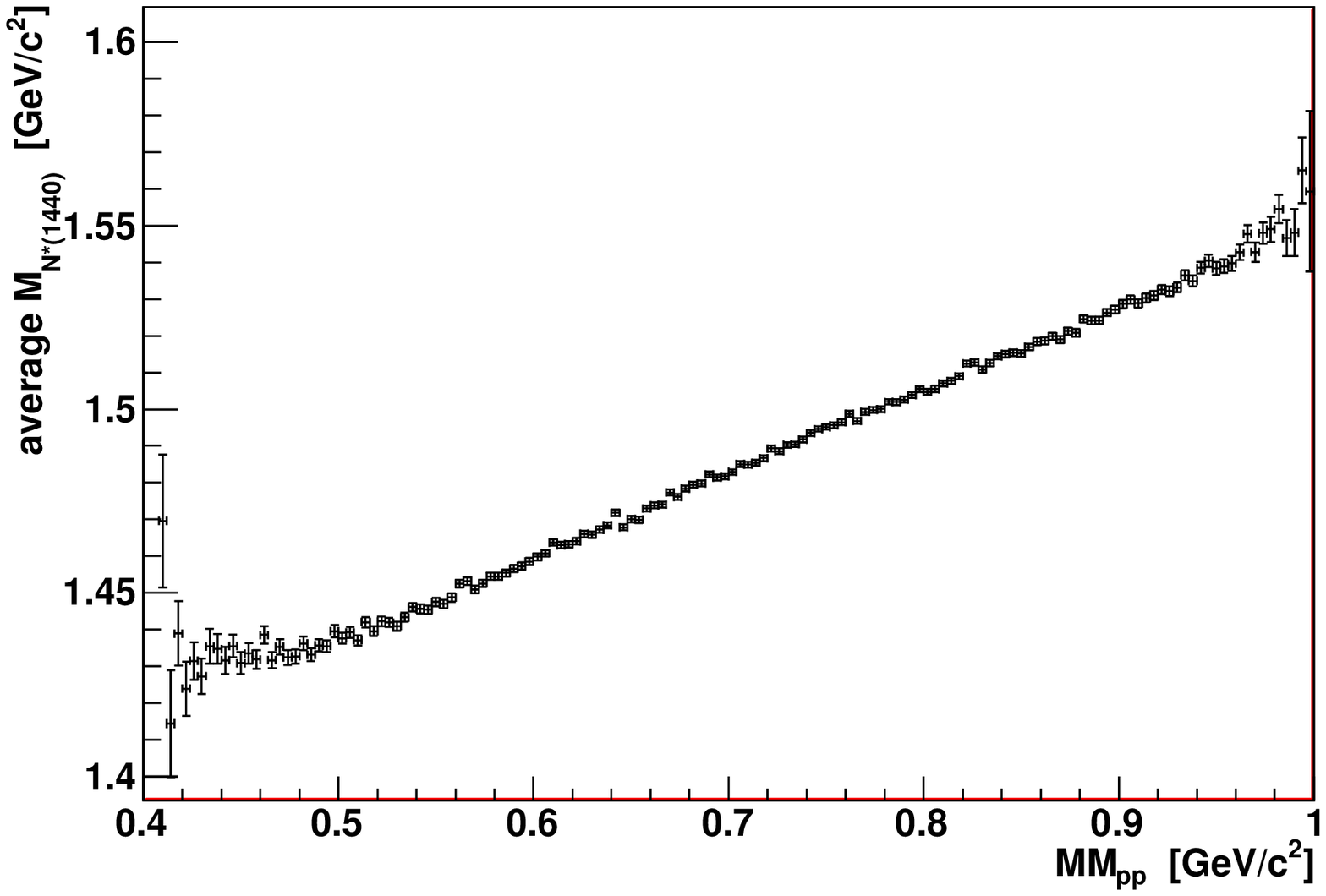}} \label{fig:D1232N1440lineshapeMod3UnStableD}}\quad
\subfigure[The $f(MM_{pp})$ as a function of the average $N^{*}(1440)$ mass,\newline where $N^{*}(1440) \rightarrow p \Delta(1232)$ (unstable particle)]{\fbox{\includegraphics[width=0.45\textwidth]{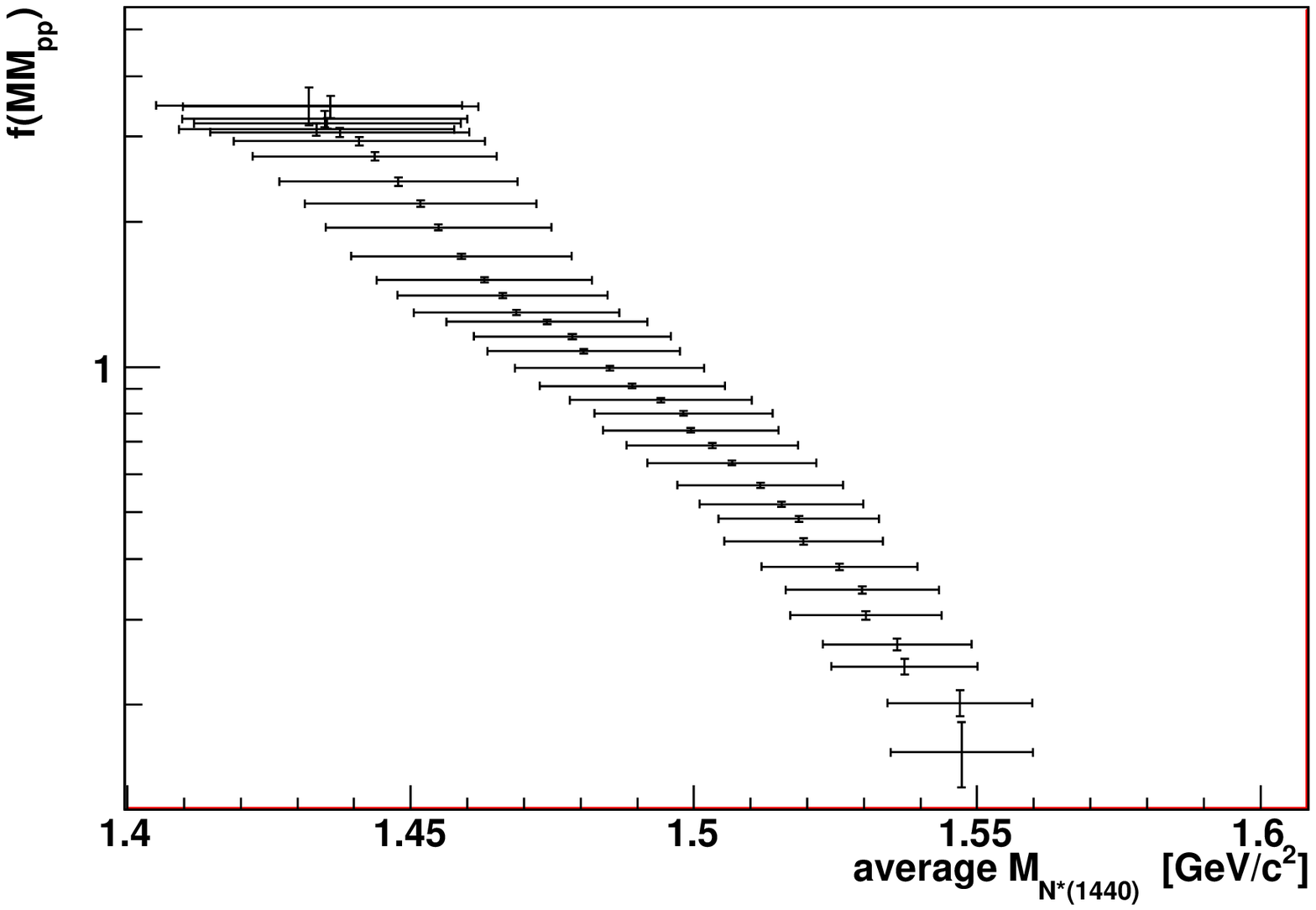}} \label{fig:D1232N1440lineshapeMod3UnStableN}}
}
\caption{Calculations of $pp \rightarrow \Delta(1232) N^{*}(1440)$ reaction kinematics done by \textbf{Pluto++} via Monte-Carlo method.  
}
\label{fig:D1232N1440lineshapeMod3}
\end{sidewaysfigure}

One can try to relate the $MM_{pp}$ with the internal properties of the $\Delta(1232)$ and $N^{*}(1440)$; for this purpose the
mass of the resonances (the realistic effective spectral line shape) for different decay modes of the $N^{*}(1440)$ is plotted against the $MM_{pp}$ (Fig.~\ref{fig:D1232N1440lineshapeMod2}).  
It is seen that the $N^{*}(1440)$ mass is correlated with the $MM_{pp}$ (Figs.~\ref{fig:D1232N1440lineshapeMod2StableD},~\ref{fig:D1232N1440lineshapeMod2UnStableN}) -- the correlation is seen in both cases independently on the spectral line shape.
 in case of the $\Delta(1232)$ (Figs.~\ref{fig:D1232N1440lineshapeMod2StableD},\ref{fig:D1232N1440lineshapeMod2UnStableD}) no correlation is visible.
To see it in details, one can calculate also the average mass of the $N^{*}(1440)$ as a function of the $MM_{pp}$ (Figs.~\ref{fig:D1232N1440lineshapeMod3StableD},~\ref{fig:D1232N1440lineshapeMod3UnStableD}).  
It is seen that when the missing mass $MM_{pp}$ of the two protons increases one selects in average higher mass of the $N^{*}(1440)$ from the spectral line shape.
One can now transform the $f(MM_{pp})$ - the missing mass population function Table~\ref{tab:MMfactor} (Fig.~\ref{fig:MMfacor2})
to dependence on the average mass of the $N^{*}(1440)$ (Figs.~\ref{fig:D1232N1440lineshapeMod3StableN},~\ref{fig:D1232N1440lineshapeMod3UnStableN}).
It is seen that $f(MM_{pp})$ decreases with the average mass of the $N^{*}(1440)$.
One can now use the obtained $f(MM_{pp})$ versus average mass of the $N^{*}(1440)$ relation (Figs.~\ref{fig:D1232N1440lineshapeMod3StableN},~\ref{fig:D1232N1440lineshapeMod3UnStableN})  
 and use it as an correction to the realistic effective spectral line shape of the $N^{*}(1440)$ calculated by \textbf{Pluto++} (see Appendix~\ref{appendix:wmc}).
It is seen (Figs.~\ref{fig:D1232N1440lineshapeModStableN},~\ref{fig:D1232N1440lineshapeModUnStableN}) that the proposed modified spectral line shapes of the $N^{*}(1440)$
are significantly different from the one obtained originally from the \textbf{Pluto++}.

In particular interesting effect is seen in case of $N^{*}(1440)$ decaying into $\pi^{0}\Delta(1232)$ (Fig.~\ref{fig:D1232N1440lineshapeModUnStableN}).
The calculated by \textbf{Pluto++} realistic effective spectral line shape of the $N^{*}(1440)$ changes, when applying correction (Figs.~\ref{fig:D1232N1440lineshapeMod3StableN},~\ref{fig:D1232N1440lineshapeMod3UnStableN}), to the shape very similar to the Breit-Wigner distribution (see Fig.~\ref{fig:D1232N1440lineshapeModUnStableN}).

This might indicate that the proposed by \textbf{Pluto++} modification of the $N^{*}(1440)$ spectral line caused by decay to $\Delta(1232)$ is not so strong as proposed.

It is also seen that when modifying the $N^{*}(1440)$ spectral line the $\Delta(1232)$ spectral line is not disturbed (Figs.~\ref{fig:D1232N1440lineshapeModStableD},~\ref{fig:D1232N1440lineshapeModStableD}).

Now one can use the modified line shapes of $N^{*}(1440)$ by $f(MM_{pp})$ (Figs.~\ref{fig:D1232N1440lineshapeModStableN},~\ref{fig:D1232N1440lineshapeModUnStableN})
and see how the data are described by the model (Eq.~\ref{eq:3pi0Model}), now without explicit $f(MM_{pp})$.          
It is seen (Figs.~\ref{image_MMpp_FSIDNDComp.eps},~\ref{image_CompAll_FSIDND.eps}) that the proposed modification of the model by changing the spectral line shape of the $N^{*}(1440)$ results
 in almost the same behavior as the one with the $f(MM_{pp})$ included explicitly (see for comparison Figs.~\ref{image_MMpp02ModelSumFactor},~\ref{image_ComparisonModelSumFactor}).    
It is also seen that in this case the model describes the data significantly better than the homogeneously and isotropically populated phase space.

\bigskip
Concluding, two possible explanations of the origin of $f(MM_{pp})$ - the missing mass population function were considered.

\begin{itemize}
 \item The possibility of $\Delta(1232) N^{*}(1440)$ interaction via One Boson Exchange (OBE), which was excluded due to completely
 different behavior of the $f(MM_{pp})$ as a function of the difference of four momentum than in OBE \cite{OBE1,OBE2,OBE3,OBE4,OBE5,OBE6,OBE7,OBELast};

 \item Next, it was shown that the $MM_{pp}$ is very sensitive to the structure of the spectral line shape of the $N^{*}(1440)$.
The proposed modification of the spectral line mainly of the $N^{*}(1440)\rightarrow\pi^{0}\Delta(1232)$ accomplishes  the explicitly added $f(MM_{pp})$;
since the main effect influencing the description of the data is due to the modification of the $N^{*}(1440)\rightarrow\pi^{0}\Delta(1232)$ line shape -- it is the
leading mode of $3\pi^{0}$ production $\sim 95\%$ (see Eq.~\ref{eq:3pi0Model}).  
In particular the proposed modification of this spectral line is very similar to the Breit-Wigner distribution.
This could indicate that the proposed by \textbf{Pluto++} \cite{Pluto} modification of the $N^{*}(1440)$ spectral line caused by decay to $\Delta(1232)$ (unstable hadron) is not as prominent as proposed.
For instance, it remains inconclusive whether the spectral line shape of $N^{*}(1440)$ remains the same in case of $N^{*}(1440) \rightarrow p\pi^{0}\pi^{0}$
since the contribution of this branch consists of only $\sim 5\%$ (see Eq.~\ref{eq:3pi0Model}).  
\end{itemize}

\noindent The possibility of molecule or bound state creation of $\Delta(1232)N^{*}(1440)$ system
as well as the excitations of the quark-gluon degrees of freedom is not excluded.        

\begin{sidewaysfigure}[t!bp]
\centering
{
\subfigure[$\Delta(1232)$ line shape,\newline where $N^{*}(1440) \rightarrow p \pi^{0}\pi^{0}$ (stable particles)]{\fbox{\includegraphics[width=0.45\textwidth]{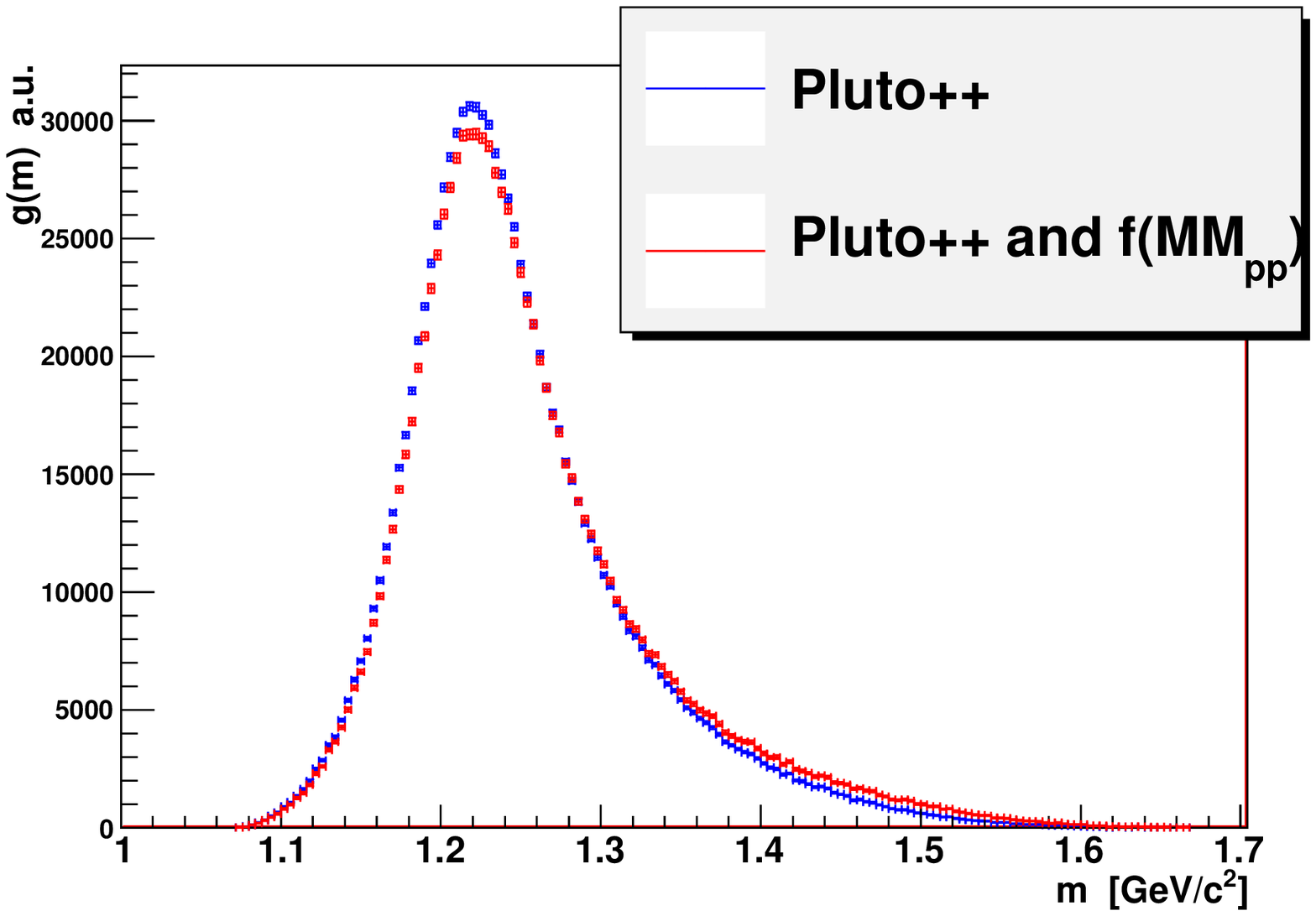}} \label{fig:D1232N1440lineshapeModStableD}}\quad
\subfigure[$N^{*}(1440)$ line shape,\newline where $N^{*}(1440) \rightarrow p \pi^{0}\pi^{0}$ (stable particles)]{\fbox{\includegraphics[width=0.45\textwidth]{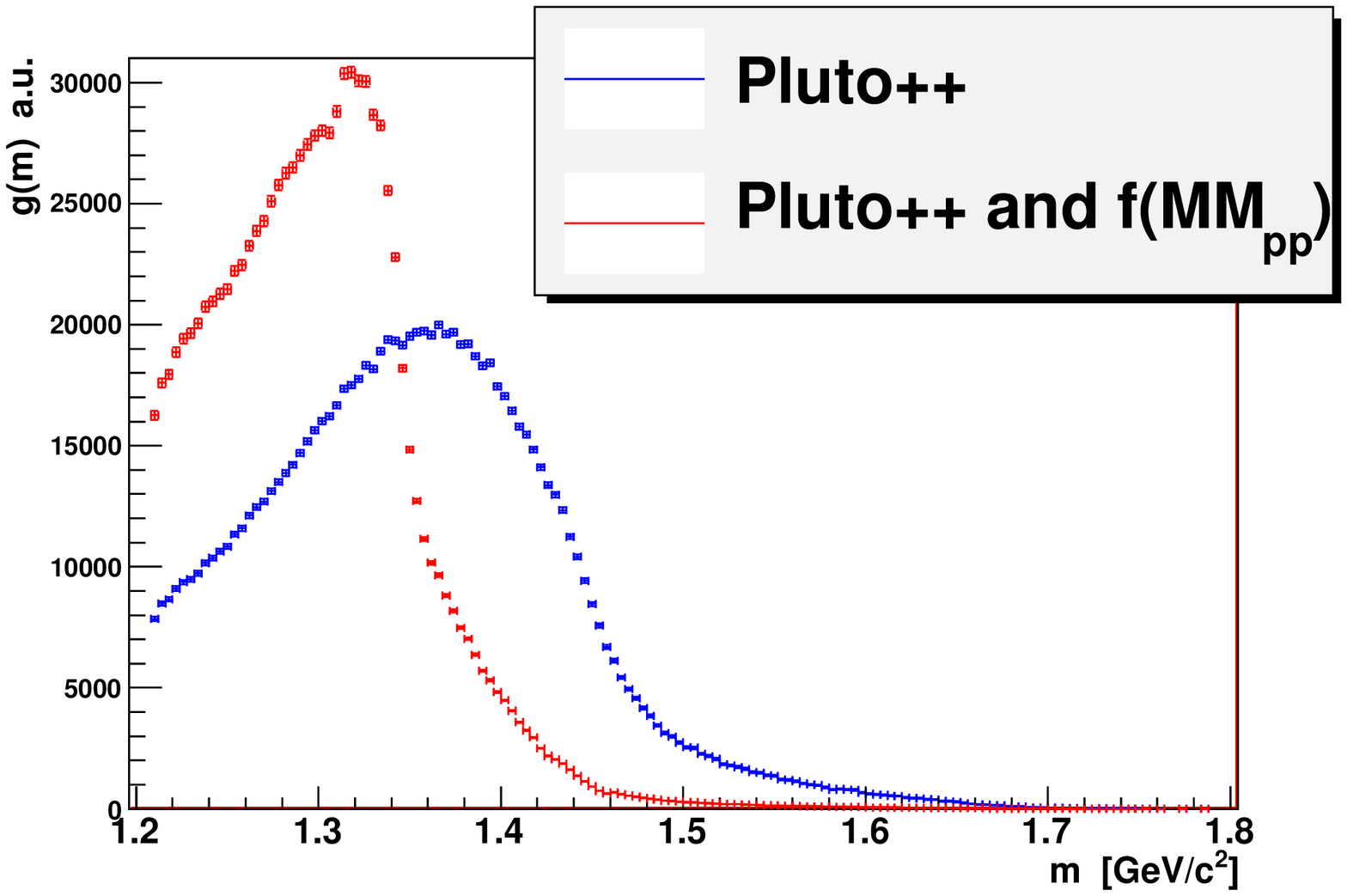}} \label{fig:D1232N1440lineshapeModStableN}}\\
\subfigure[$\Delta(1232)$ line shape,\newline where $N^{*}(1440) \rightarrow \pi^{0} \Delta(1232)$ (unstable particle)]{\fbox{\includegraphics[width=0.45\textwidth]{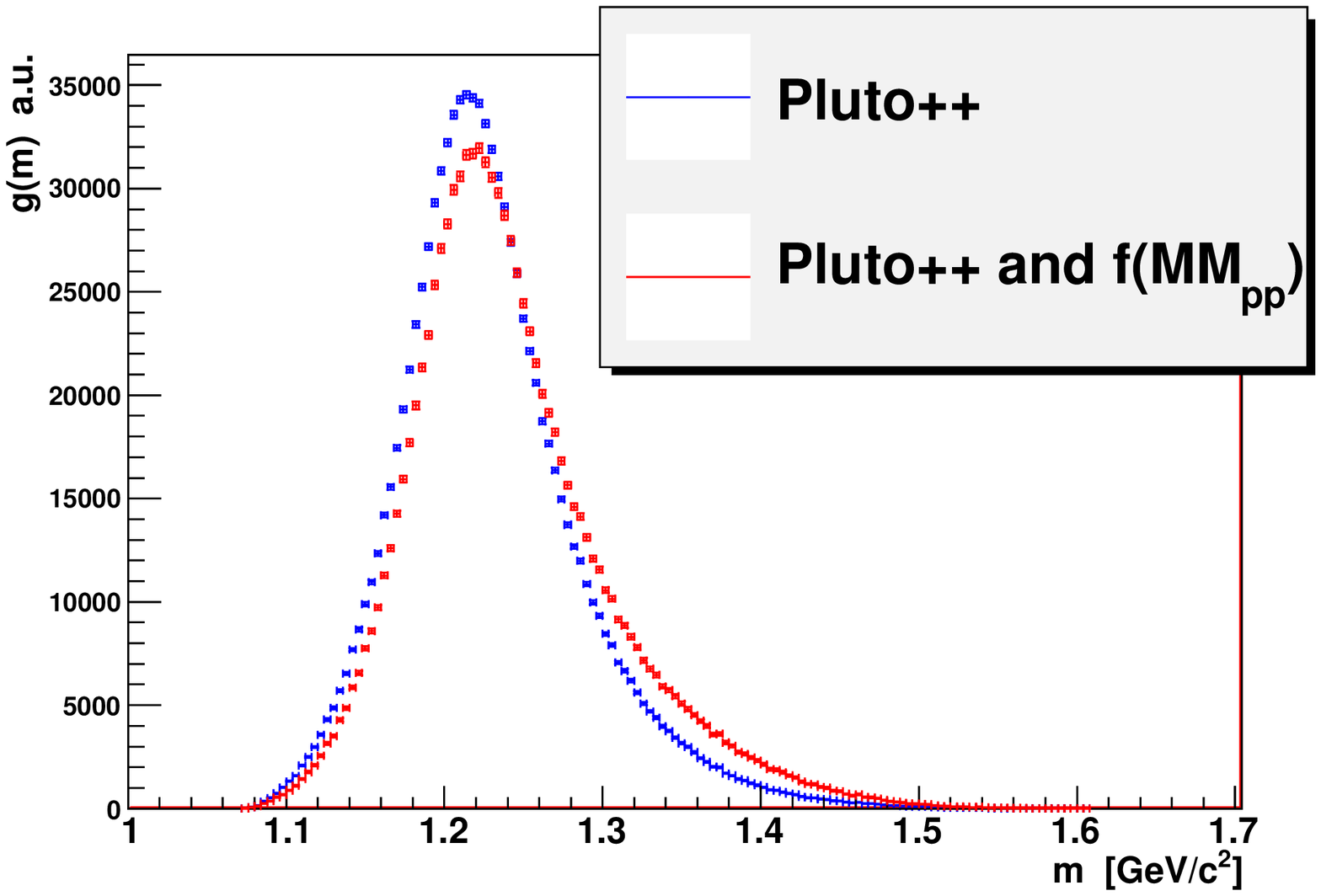}} \label{fig:D1232N1440lineshapeModUnStableD}}\quad
\subfigure[$N^{*}(1440)$ line shape,\newline where $N^{*}(1440) \rightarrow \pi^{0} \Delta(1232)$ (unstable particle)]{\fbox{\includegraphics[width=0.45\textwidth]{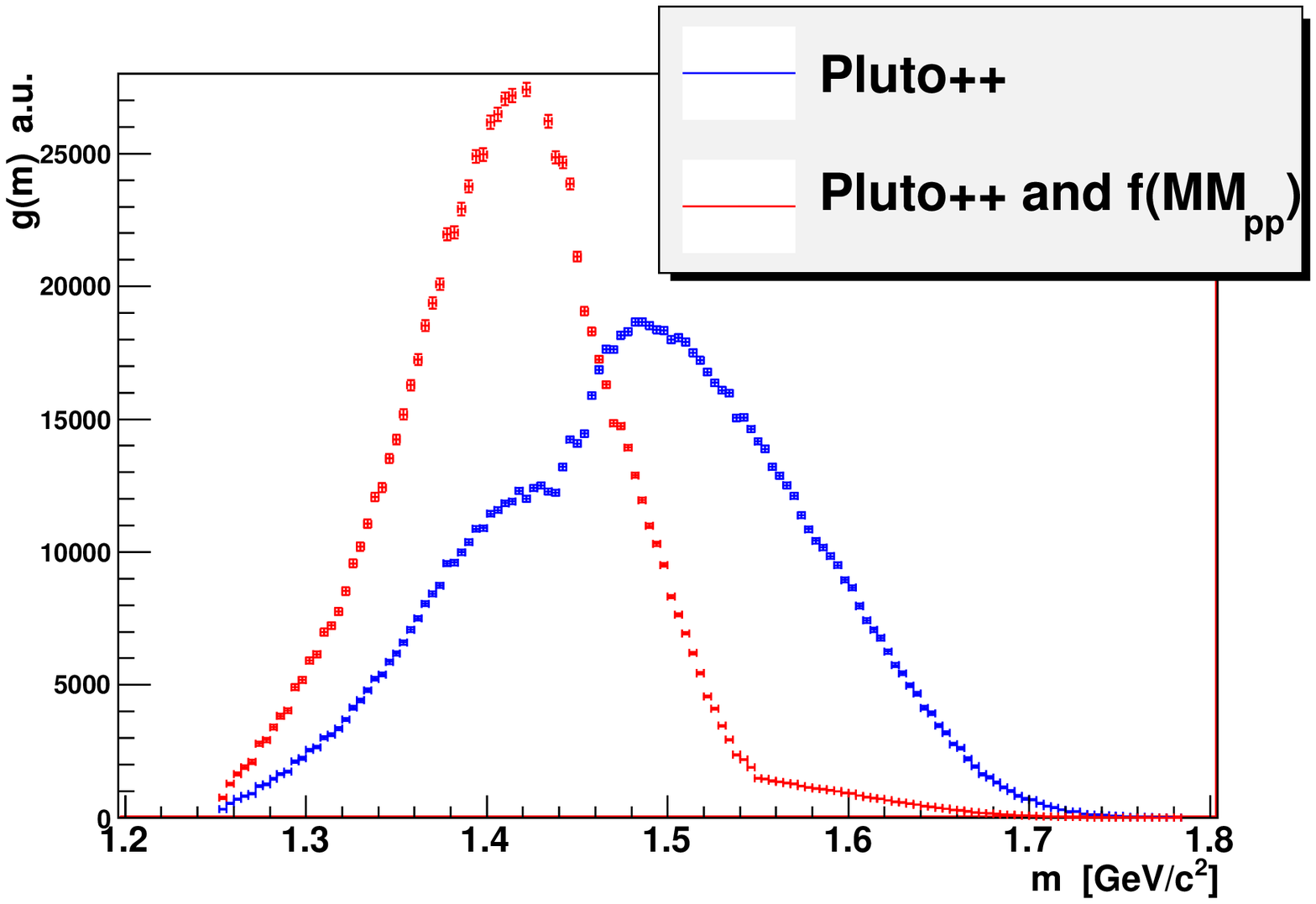}} \label{fig:D1232N1440lineshapeModUnStableN}}
}
\caption{Realistic effective spectral line shape $g(m)$ of $\Delta(1232)$ and $N^{*}(1440)$
in the $pp \rightarrow \Delta(1232) N^{*}(1440)$ reaction at incident proton momentum of $3.350\mathrm{~GeV/c^{2}}$.
The $\Delta(1232)$ decays into $p\pi^{0}$ (stable particles), the $N^{*}(1440)$ decays into $p\pi^{0}\pi^{0}$ (stable particles) or into $p\Delta(1232)$ (unstable particle),
when later $\Delta(1232)$ decays into $p\pi^{0}$. 
Spectral line shapes calculated by \textbf{Pluto++} (blue), proposed modified line shapes by $f(MM_{pp})$ (red) (via Monte-Carlo method).  
}
\label{fig:D1232N1440lineshapeMod}
\end{sidewaysfigure}

\myFrameFigure{MMpp_FSIDNDComp.eps}{Spectrum of the Missing Mass of two protons. Comparison of the experimental data (black marker) with the Monte-Carlo simulation phase space (blue line), the $Model_{1}$ (Fig.~\ref{fig:3pi0Model1})
 (yellow line) , the $Model_{2}$ (Fig.~\ref{fig:3pi0Model2}) (green line), the model sum (Eq.~\ref{eq:3pi0Model}) (red line).
The model is calculated without the $f(MM_{pp})$ population function but with proposed modified spectral line shape of the $N^{*}(1440)$ (see Figs.~\ref{fig:D1232N1440lineshapeModStableN},~\ref{fig:D1232N1440lineshapeModUnStableN}.)
The proposed modification of the model results in almost the same behavior as the one with the $f(MM_{pp})$ included explicitly (see Fig.~\ref{image_MMpp02ModelSumFactor}).    
It is seen that the model describes the data significantly better than the homogeneously and isotropically populated phase space.
Vertical axis - number of events (in given bin) is shown.
}{Missing Mass of two protons.}{}

\myFrameHugeFigure{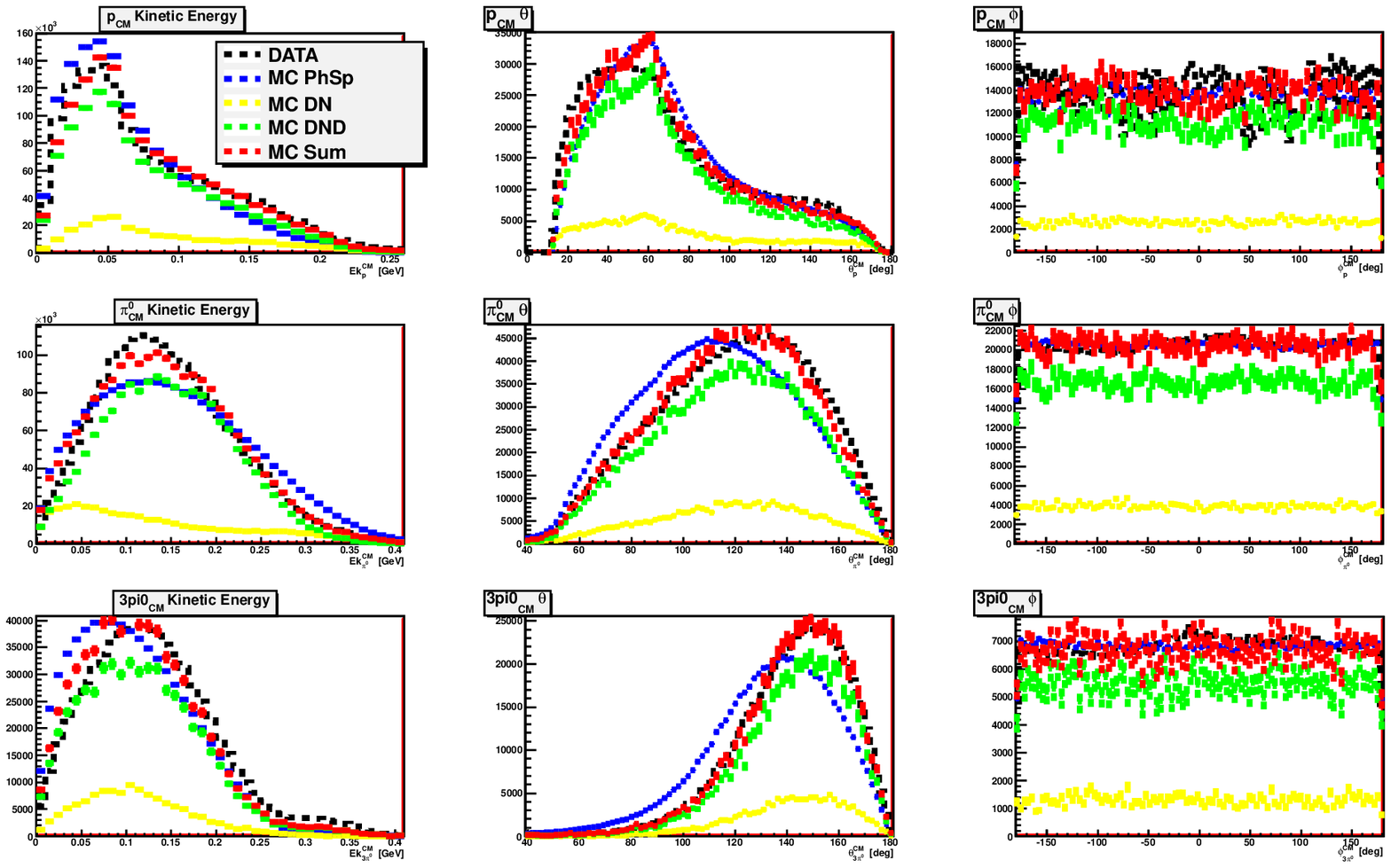}{Comparison of the experimental data with the Monte-Carlo simulation for the $MM_{pp}<0.5\mathrm{~GeV/c^{2}}$ and $MM_{pp}>0.6\mathrm{~GeV/c^{2}}$.
\textit{Upper row}: proton in the center of mass frame,\textit{Middle row}: pion in the center of mass frame, \textit{Lower row}: $3\pi^{0}$ system in the center of mass frame. From left kinetic energy, 
the polar angle and the azimuthal angle distribution. The experimental data are shown as a black marker, the Monte-Carlo simulation phase space (blue line), the $Model_{1}$ (Fig.~\ref{fig:3pi0Model1}) (yellow line)
 , the $Model_{2}$ (Fig.~\ref{fig:3pi0Model2}) (green line), the model sum (Eq.~\ref{eq:3pi0Model}) (red line).
The model is calculated without the $f(MM_{pp})$ population function but with proposed modified spectral line shape of the $N^{*}(1440)$ (see Figs.~\ref{fig:D1232N1440lineshapeModStableN},~\ref{fig:D1232N1440lineshapeModUnStableN}.)
The proposed modification of the model results in almost the same behavior as the one with the $f(MM_{pp})$ included explicitly (see Fig.~\ref{image_ComparisonModelSumFactor}).    
It is seen that the model describes the data significantly better than the homogeneously and isotropically populated phase space.
On vertical axes - number of events (in given bin) is shown.}{}

%

\emptydoublepage
\subsubsection{The model validation}
\label{subsec:MCModelValidation}
\thispagestyle{fancy}

\paragraph{The model consistency\\}\label{par:ModelConsistency}

To check the Monte Carlo developed Model (Eq.~\ref{eq:3pi0Model}) consistency with experimental data, 
the model was compared with the experimental data for all considered spectra 
(i.e. Dalitz Plot $ppX$, Dalitz Plot $3\pi^{0}$, Nyborg Plot for different missing mass ranges) by 
constructing the ratio:
\begin{equation}
 ration = \frac{Experimental~Data}{The~Model}
\label{eq:ValidationRatio}
\end{equation}

In case of data and model agreement the ration should be equal to the value $1$.
The spectra of the $ratio$ (Eq.~\ref{eq:ValidationRatio})  were prepared (Upper row of Figs.~\ref{image_Dal1CompCol2.eps},~\ref{image_Dal2CompCol2.eps},~\ref{image_Dal3CompCol2.eps})  
The color contours correspond to relative deviation from the constant value $1$.
Also the relative statistical error of the $ratio$ (Eq.~\ref{eq:ValidationRatio}) was computed (Lower row of Figs.~\ref{image_Dal1CompCol2.eps},~\ref{image_Dal2CompCol2.eps},~\ref{image_Dal3CompCol2.eps}).  
The color contours were selected to reflect the contours coloring of the relative deviation.
It is seen that the color contours structure of the relative deviation from the constant value $1$ follows the statistical error distribution.
One can conclude that in the range of the statistical error the ratio is equal to constant value $1$.    
The model is consistent with the experimental data.
\bigskip

In addition the difference $x$ between the experimental data and the model normalized by data errors ($\sigma_{Data}$) and the model errors ($\sigma_{Data}$)

\begin{equation}
 x=\frac{Data-Model}{\sqrt{\sigma^{2}_{Data} + \sigma^{2}_{Model}}}
\end{equation}

for all data points included in the $\chi^{2}$ fit (Eq.~\ref{eq:chi2DalitzPlotFitNorm}) was studied in details.

\begin{sidewaysfigure}
\centering
{
\subfigure[Probability Density Function]{\fbox{\includegraphics[width=0.4\textwidth]{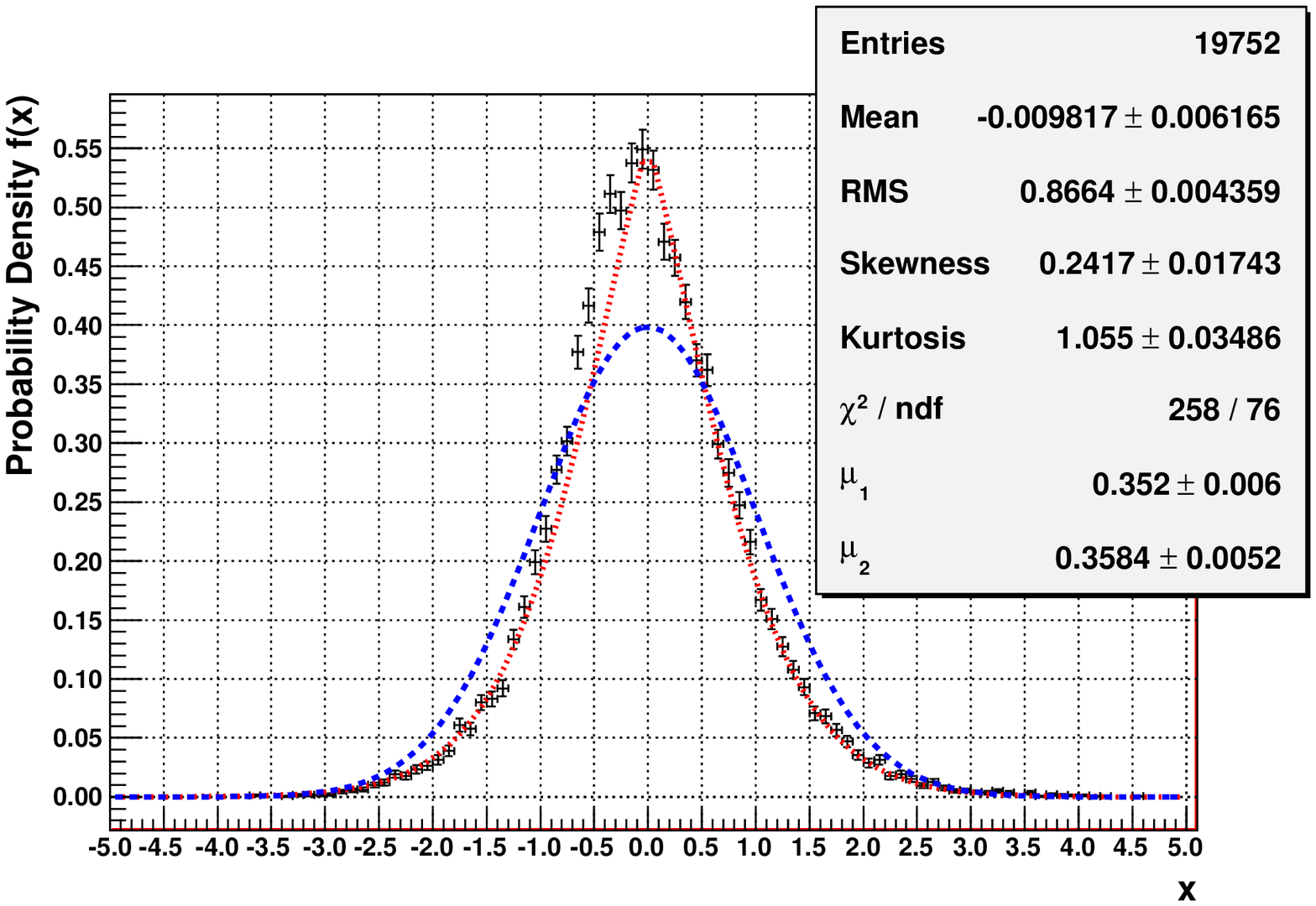}} \label{fig:ModelValidationPDF}}\quad
\subfigure[Cumulative Distribution Function ]{\fbox{\includegraphics[width=0.4\textwidth]{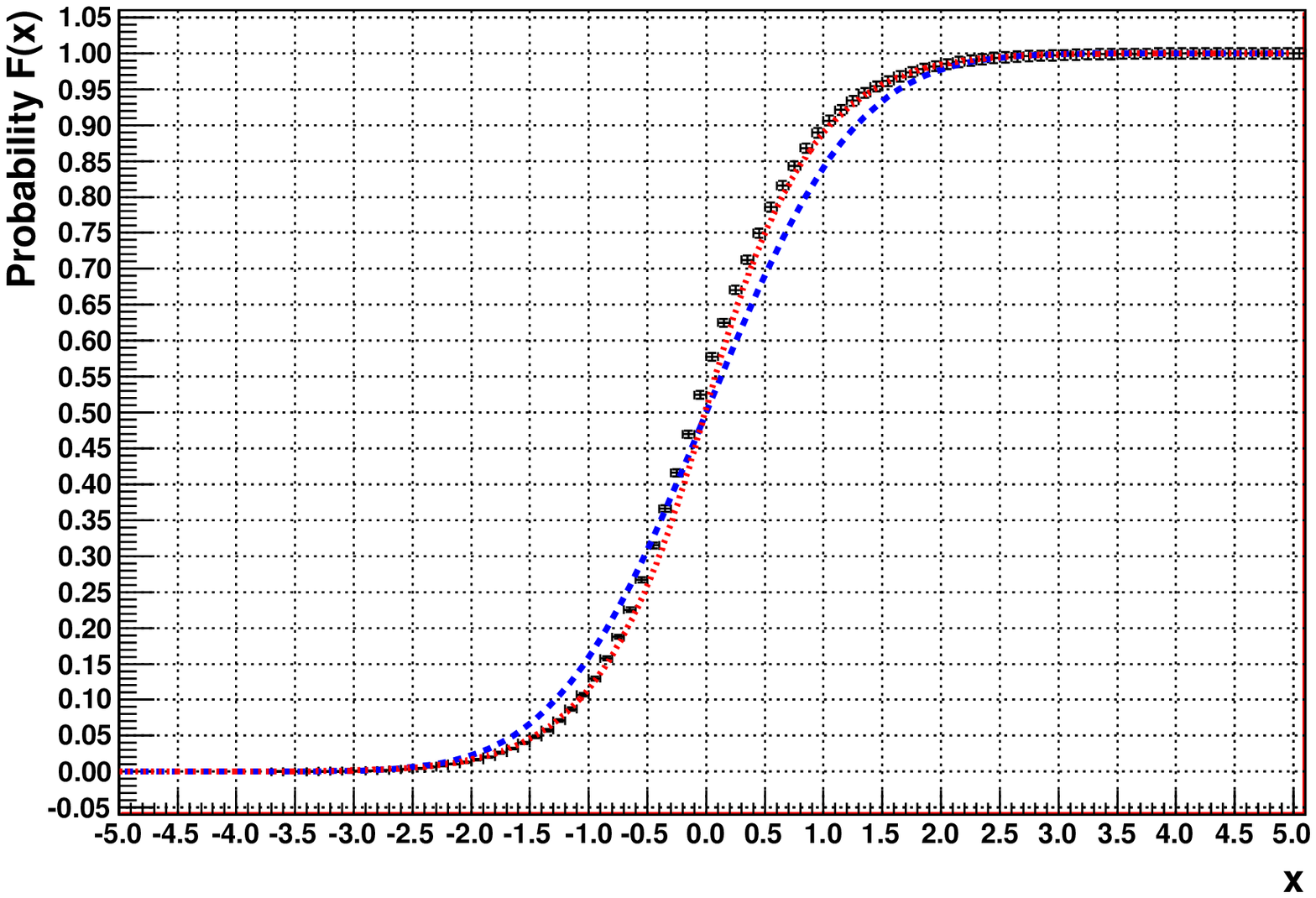}} \label{fig:ModelValidationCDF}}\\
\subfigure[Probability to find events in range $(-x,x)$]{\fbox{\includegraphics[width=0.4\textwidth]{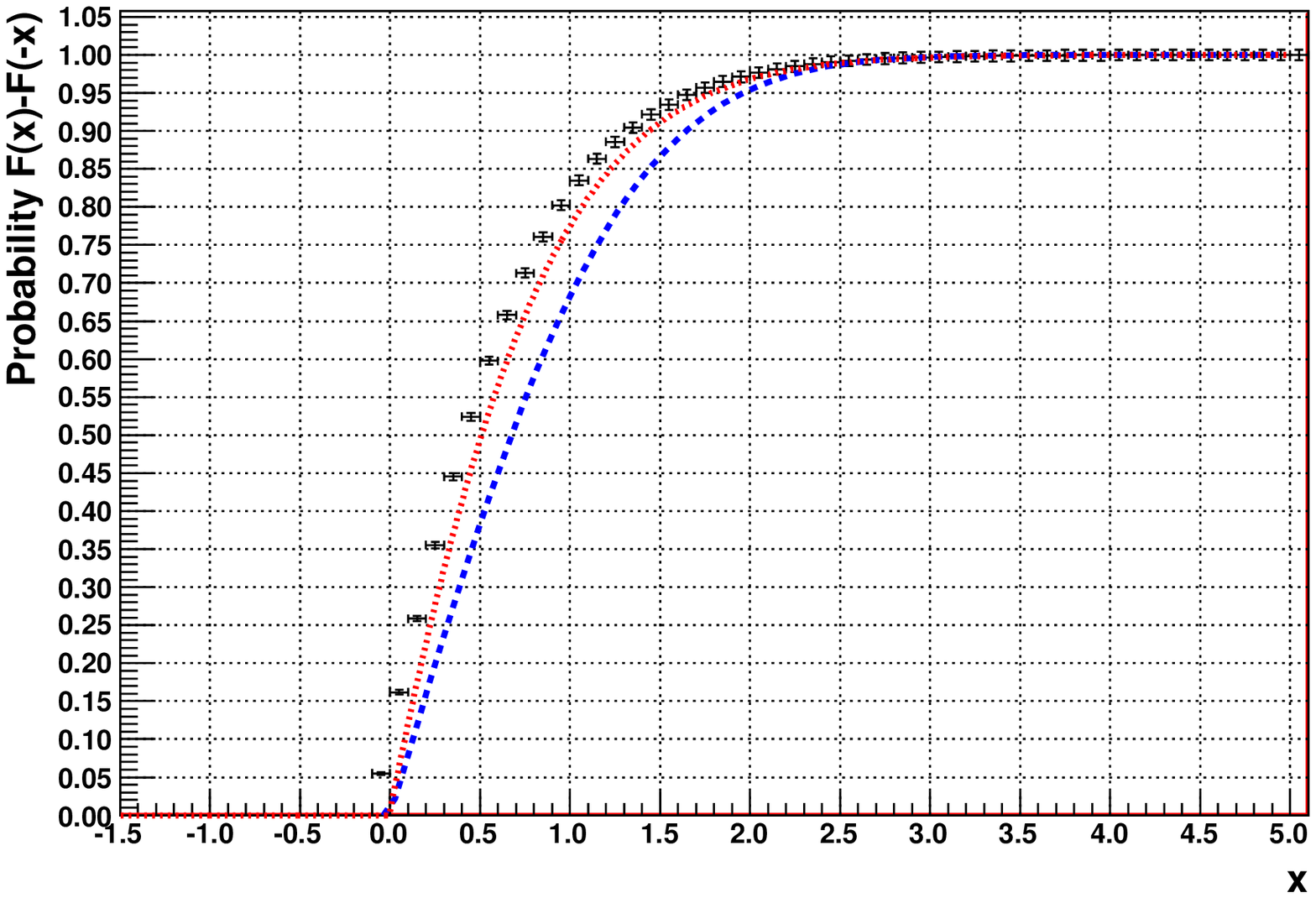}} \label{fig:ModelValidationProb}}
}
\caption{The derived distributions for the $x=\frac{Data-Model}{\sqrt{\sigma^{2}_{Data} + \sigma^{2}_{Model}}}$ for the Monte Carlo developed Model (Eq.~\ref{eq:3pi0Model}). 
The $x$ value was calculated for all data points included in the $\chi^{2}$ fit (Eq.~\ref{eq:chi2DalitzPlotFitNorm}).
The Probability Density Function was fitted by the Skellam distribution (Eq.~\ref{eq:skellamPDF}).
The experimental data distributions are plotted as a black histogram, the standard normal distribution is plotted as a blue dashed line,
the Skellam distribution is shown a red dotted line.
The Skellam distribution describes the difference distributions significantly better than the standard normal distribution.
It is seen that the probability to find the events in the range of one standard deviation $x\in(-1,1)$ is equal to $\sim0.8$ i.e
the Monte Carlo developed Model describes $\sim80\%$ of the experimental data within the statistical errors of the data and the model.}
\label{fig:ModelValidationAll}
\end{sidewaysfigure}

The Probability Density Function~(PDF), and the Cumulative Distribution Function~(CDF) were calculated (Figs.~\ref{fig:ModelValidationPDF},~\ref{fig:ModelValidationCDF}).
It is seen that the PDF is centered around mean value of $0$ and it is in good agreement symmetric (the skewness parameter is close to $0$ value). 
There is no systematic shift effect.

Since the statistical fluctuations of the Data and Model points are of the Poissonian behavior 
,one can show that
the difference $x=x_{1}-x_{2}$ of two statistically independently random variables $x_{1}, x_{2}$ having Poissonian distribution $f_{1}, f_{1}$

\begin{eqnarray}
 f_{1}(x_{1},\mu_{1}) &=& \frac{\mu_{1}^{x_{1}} e^{-\mu_{1}}}{x_{1}!},\nonumber \\
 f_{2}(x_{2},\mu_{2}) &=& \frac{\mu_{2}^{x_{2}} e^{-\mu_{2}}}{x_{2}!}
\end{eqnarray}
where $\mu_{1,2}$ is the expected value of the distribution,
is distributed like Skellam distribution \cite{skellam}:

\begin{equation}
 f(x,\mu_{1},\mu_{2}) = e^{-(\mu_{1}+\mu_{2})} \left( \frac{ \mu_{1} } { \mu_{2} } \right)^{x/2} I_{|x|}( 2 \sqrt{\mu_{1}\mu_{2}} )
\label{eq:skellamPDF}
\end{equation}
where $I_{|x|}$ is modified Bessel function of the first kind.
\bigskip

The PDF (Fig.~\ref{fig:ModelValidationPDF}) was fitted with the Skellam distribution (Eq.~\ref{eq:skellamPDF}).
The following values were obtained:

\begin{eqnarray}
 \mu_{1} &=&  0.3520 \pm 0.0059\\
 \mu_{2} &=& 0.3584 \pm 0.0052
\end{eqnarray}

The standard normal distribution was plotted for the comparison.
The Skellam distribution describes the PDF and CDF spectra significantly better than the standard normal distribution.
Also the probability to find events in range $(-x,x)$ was computed (Fig.~\ref{fig:ModelValidationProb}).   
It is seen that the probability to find the events in the range of one standard deviation $x\in(-1,1)$ is equal to $\sim0.8$ i.e
the Monte Carlo developed Model describes $\sim80\%$ of the experimental data within the statistical errors of one standard deviation.
It is also seen that the probability value $\sim 1.0$ is reached for the range $x\in(-2.5,2.5)$ - around $100\%$ of the experimental data
is described by the Monte Carlo Model withing the statistical error of $\sim 2.5$ standard deviations.
Concluding the Monte Carlo developed model fully describes the data within the statistical precision of data and model.

\myFrameHugeFigure{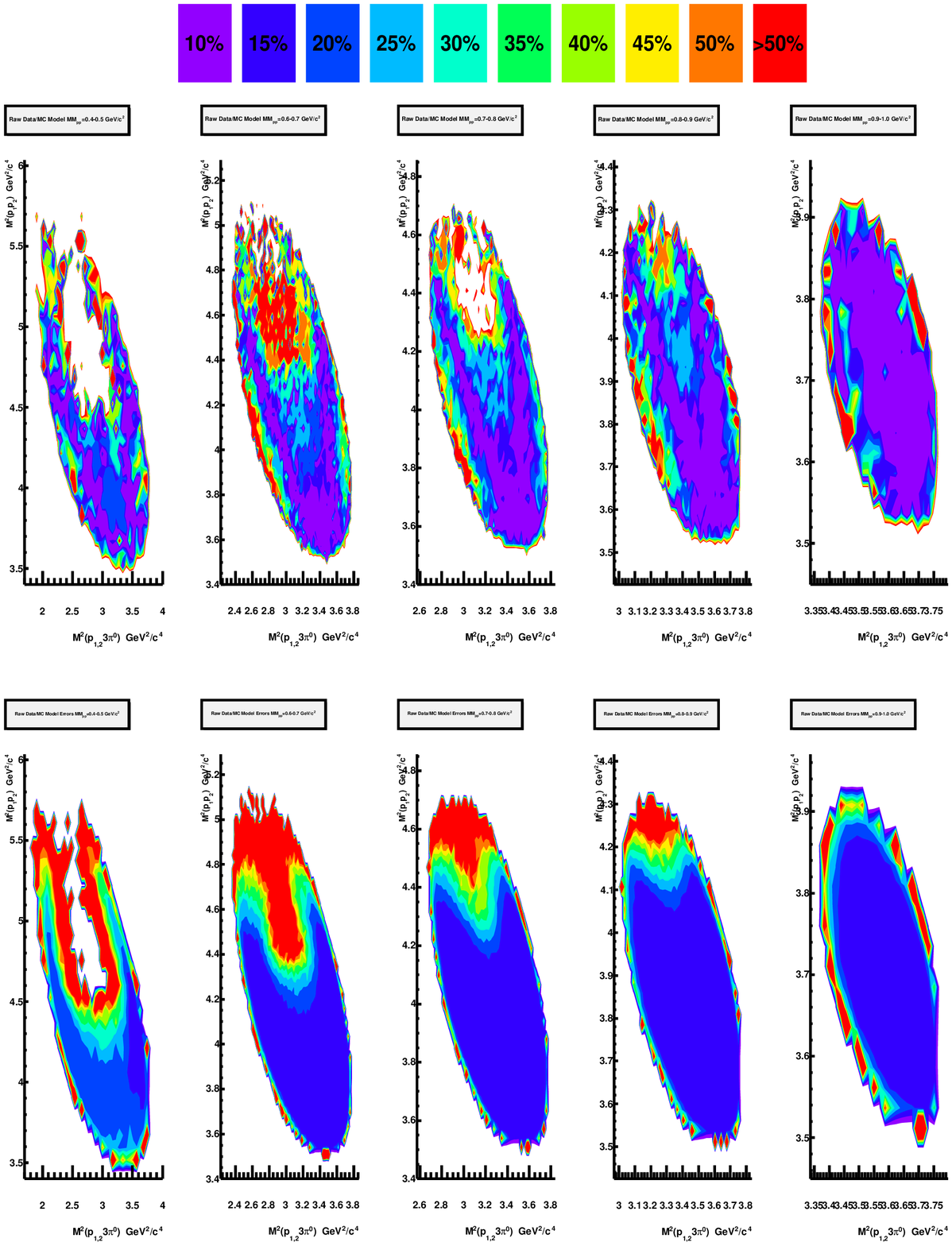}{Dalitz Plot $ppX$. $M^{2}(pp)$ versus $M^{2}(p3\pi^{0})$. The upper row corresponds to ratio of the experimental data divided by
 the Monte-Carlo model sum of $Model_{1}$ and $Model_{2}$ (Fig.~\ref{fig:3pi0Models}), the contour colors correspond to the relative deviation from constant value $1$.
The lower row corresponds to the relative statistical error of the ratio.
The columns from left to right correspond
to the following Missing Mass of two protons bins, column~1~$MM_{pp}=0.4-0.5\mathrm{~GeV/c^{2}}$,column~2~$MM_{pp}=0.6-0.7\mathrm{~GeV/c^{2}}$,
column~3~$MM_{pp}=0.7-0.8\mathrm{~GeV/c^{2}}$, column~4~$MM_{pp}=0.8-0.9\mathrm{~GeV/c^{2}}$, column~5~$MM_{pp}=0.9-1.0\mathrm{~GeV/c^{2}}$.
 The plots are symmetrized against two protons - each event is filled two times.
The color contours structure of the relative deviation from the constant value $1$ follows the statistical error distribution.
It is seen that in the range of the statistical error the ratio is equal to constant value $1$.
The model is consistent with the experimental data.
Fully expandable and colored version of the figure is available in the attached electronic version of the thesis.
}{Dalitz Plot $ppX$.}

\myFrameHugeFigure{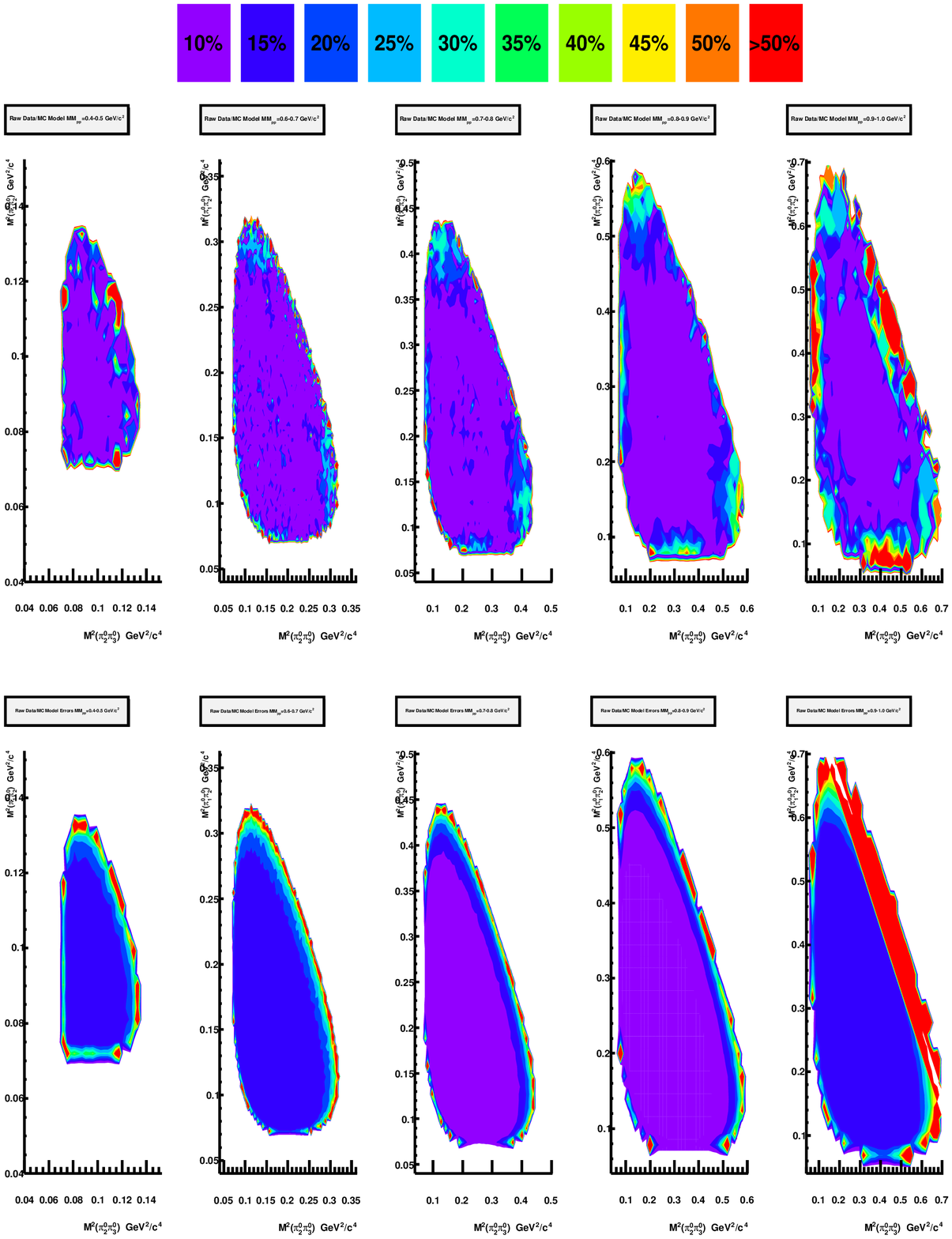}{Dalitz Plot $3\pi^{0}$. $M^{2}(2\pi^{0})$ versus $M^{2}(2\pi^{0})$. The upper row corresponds to ratio of the experimental data divided by
 the Monte-Carlo model sum of $Model_{1}$ and $Model_{2}$ (Fig.~\ref{fig:3pi0Models}), the contour colors correspond to the relative deviation from constant value $1$.
The lower row corresponds to the relative statistical error of the ratio.
The columns from left to right correspond
to the following Missing Mass of two protons bins, column~1~$MM_{pp}=0.4-0.5\mathrm{~GeV/c^{2}}$,column~2~$MM_{pp}=0.6-0.7\mathrm{~GeV/c^{2}}$,
column~3~$MM_{pp}=0.7-0.8\mathrm{~GeV/c^{2}}$, column~4~$MM_{pp}=0.8-0.9\mathrm{~GeV/c^{2}}$, column~5~$MM_{pp}=0.9-1.0\mathrm{~GeV/c^{2}}$.
 The plots are symmetrized against three pions - each event is filled six times.
The color contours structure of the relative deviation from the constant value $1$ follows the statistical error distribution.
It is seen that in the range of the statistical error the ratio is equal to constant value $1$.
The model is consistent with the experimental data.
Fully expandable and colored version of the figure is available in the attached electronic version of the thesis.
}{Dalitz Plot $3\pi^{0}$.}

\myFrameHugeFigure{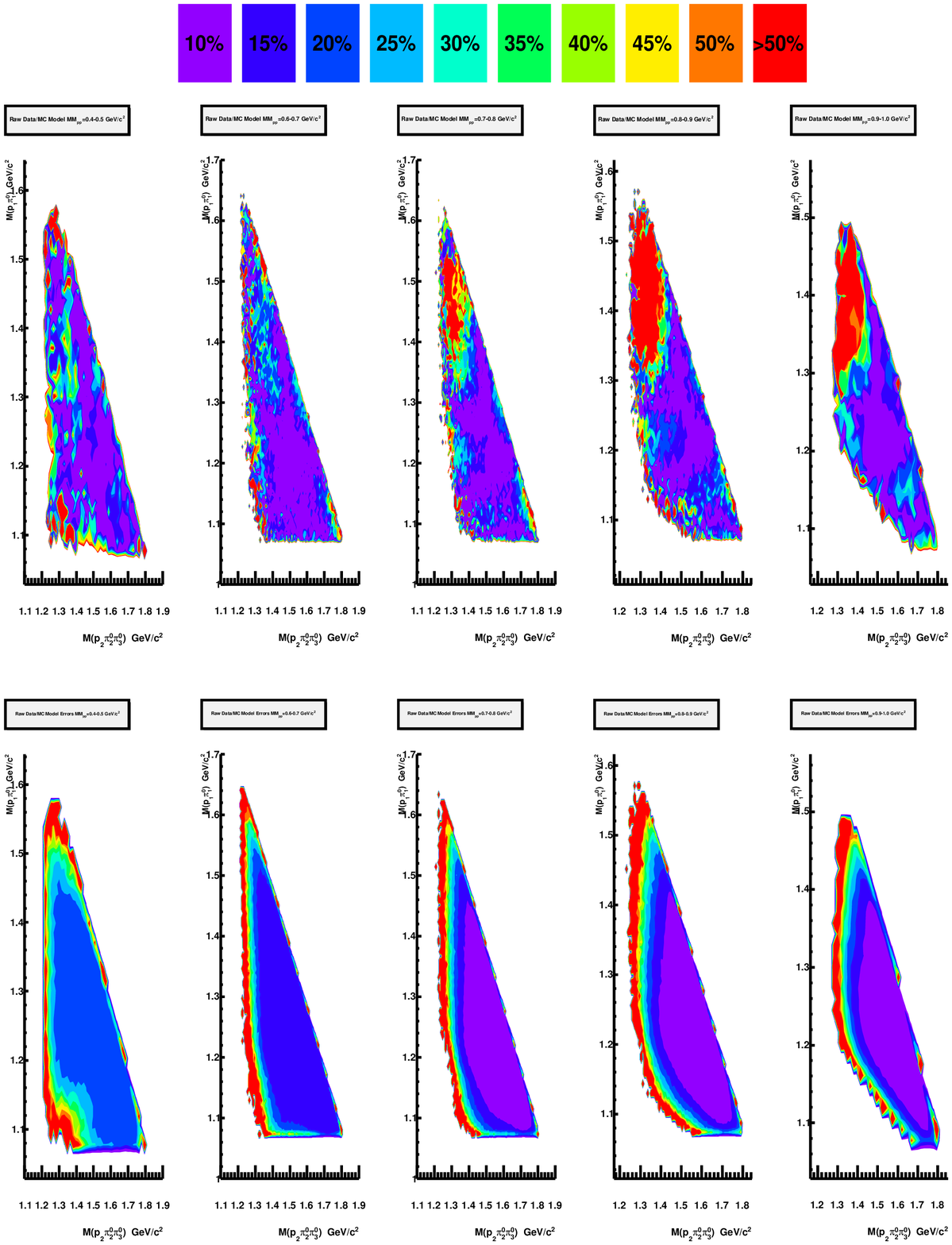}{Nyborg Plot. $M(p\pi^{0})$ versus $M(p\pi^{0}\pi^{0})$. The upper row corresponds to ratio of the experimental data divided by
 the Monte-Carlo model sum of $Model_{1}$ and $Model_{2}$ (Fig.~\ref{fig:3pi0Models}), the contour colors correspond to the relative deviation from constant value $1$.
The lower row corresponds to the relative statistical error of the ratio.
The columns from left to right correspond
to the following Missing Mass of two protons bins, column~1~$MM_{pp}=0.4-0.5\mathrm{~GeV/c^{2}}$,column~2~$MM_{pp}=0.6-0.7\mathrm{~GeV/c^{2}}$,
column~3~$MM_{pp}=0.7-0.8\mathrm{~GeV/c^{2}}$, column~4~$MM_{pp}=0.8-0.9\mathrm{~GeV/c^{2}}$, column~5~$MM_{pp}=0.9-1.0\mathrm{~GeV/c^{2}}$.
 The plots are symmetrized against two protons and three pions - each event is filled six times.
The color contours structure of the relative deviation from the constant value $1$ follows the statistical error distribution.
It is seen that in the range of the statistical error the ratio is equal to constant value $1$.
The model is consistent with the experimental data.
Fully expandable and colored version of the figure is available in the attached electronic version of the thesis.
}{Nyborg Plot.}

\newpage
\paragraph{Verification of other processes contribution to the model\\}\label{par:ModelOtherProc}

To verify a possibility of other processes contribution ($Model_{3}$) to the Monte Carlo developed Model (Eq.~\ref{eq:3pi0Model}) ($Model_{Tot.}$)
the following procedure was applied.

One may assume that in order to describe the experimental data in addition to the developed Model ($Model_{Tot.}$) one has other processes that could contribute ($Model_{3}$),
where other processes one may try to mimic by the Monte Carlo simulation of $pp \rightarrow pp 3\pi^{0}$ reaction assuming homogeneously and isotropically populated phase space.
This could be written as:
  
\begin{equation}
 (1-\epsilon) Model_{Tot.} + \epsilon Model_{3}
\end{equation}

one needs to know the fraction of $Model_{3}$  to the $Model_{3}+Model_{Tot.}$ i.e. the $\epsilon$ parameter - which actually is the fraction of the other processes to the sum of the developed Model and other processes.

To estimate the $\epsilon$ from the experimental data one needs to fit once more the shape of the plots (Figs.~\ref{image_Dal1New2},~\ref{image_Dal2New2},~\ref{image_Dal3New2})
by the mixture of $Model_{3}$ and $Model_{Tot.}$ to the experimental data.  

The population of the events in missing mass of two protons was not a purpose of the fit, it was previously derived from the data Table~\ref{tab:MMfactor} (Fig.~\ref{fig:MMfacor2}).
Fit was performed using the chi-square method; one defines the $\chi^{2}$ function to minimize:
\begin{equation}
 \chi^{2} = \sum \dfrac{\left[ Data - \left((1-\epsilon) Model_{Tot.} + \epsilon Model_{3}\right)\right]^{2} }{\sigma^{2}_{Data} + (1-\epsilon)^{2}\sigma^{2}_{Model_{Tot.}} + \epsilon^{2}\sigma^{2}_{Model_{3}}}
\label{eq:chi2DalitzPlotFitNew4}
\end{equation}
where the sum goes over each bin of the five Dalitz Plots $ppX$, five Dalitz Plots $3\pi^{0}$ and five Nyborg Plots i.e. $15$ plots.
The $\sigma_{Data}$ is the error of the point for experimental data,
 $\sigma_{Model_{Tot.}}$~-~ is the error of the point for the Monte Carlo developed Model (Eq.~\ref{eq:3pi0Model}) and $\sigma_{Model_{3}}$~-~ is the error of the point for $Model_{3}$.

For the numerical purpose of doing the fit, the $\chi^{2}$ function (Eq.~\ref{eq:chi2DalitzPlotFitNew4}) was redefined to:
\begin{equation}
 \chi^{2} = \sum \dfrac{\left[ Data - \left((1-e) Model_{Tot.}^{*} + e Model_{3}^{*}\right)\right]^{2} }{\sigma^{2}_{Data} + (1-e)^{2}\sigma^{2}_{Model_{Tot.}^{*}} + e^{2}\sigma^{2}_{Model_{3}^{*}}}
\label{eq:chi2DalitzPlotFitNormNew4}
\end{equation}    
The same as in the Section~\ref{subsec:3pi0MCmodel} on page \pageref{subsec:3pi0MCmodel}.

\begin{figure}[ht!bp]
\centering
{
\subfigure[]{\fbox{\includegraphics[width=0.7\textwidth]{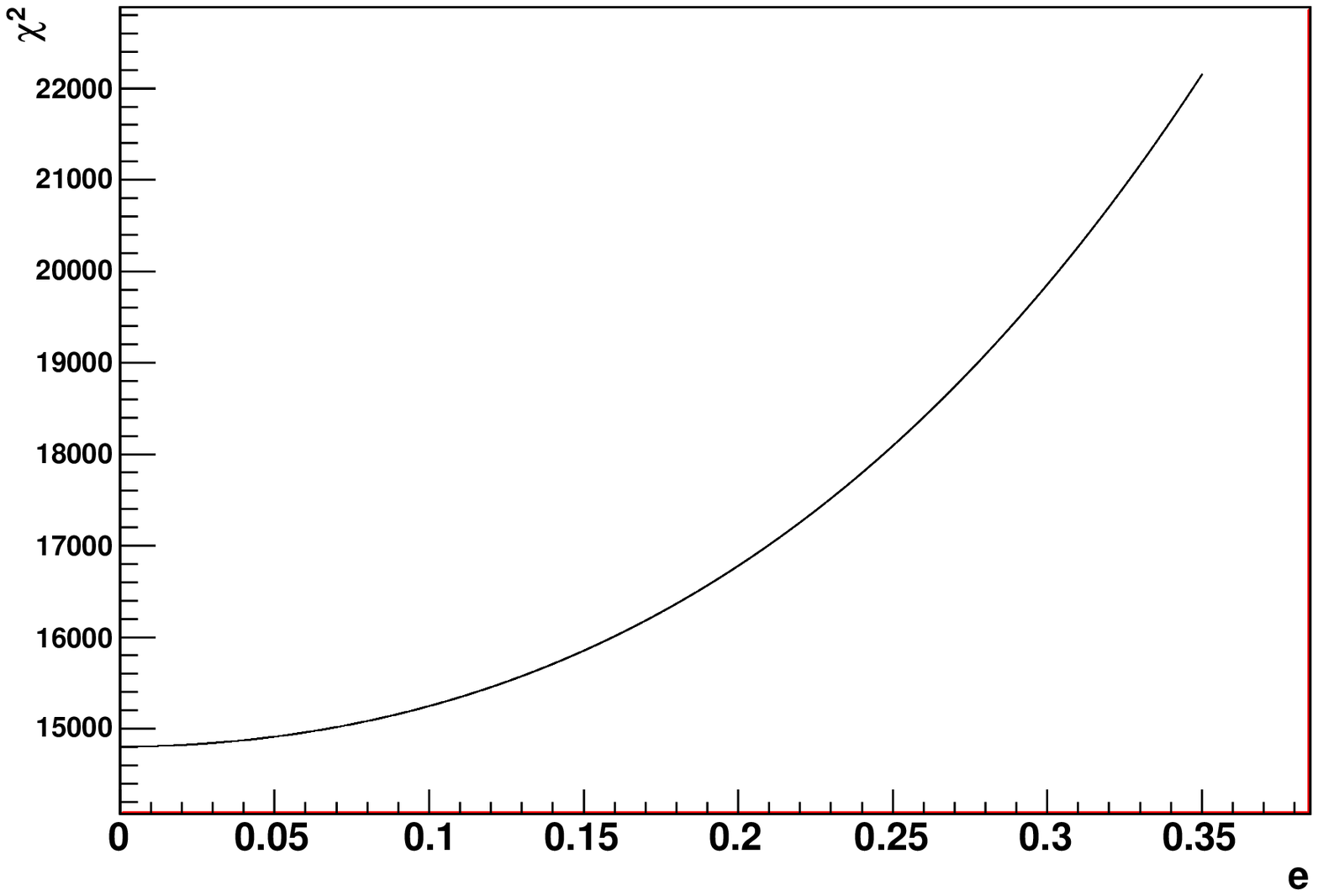}} \label{fig:chi2New41}}\\
\subfigure[]{\fbox{\includegraphics[width=0.7\textwidth]{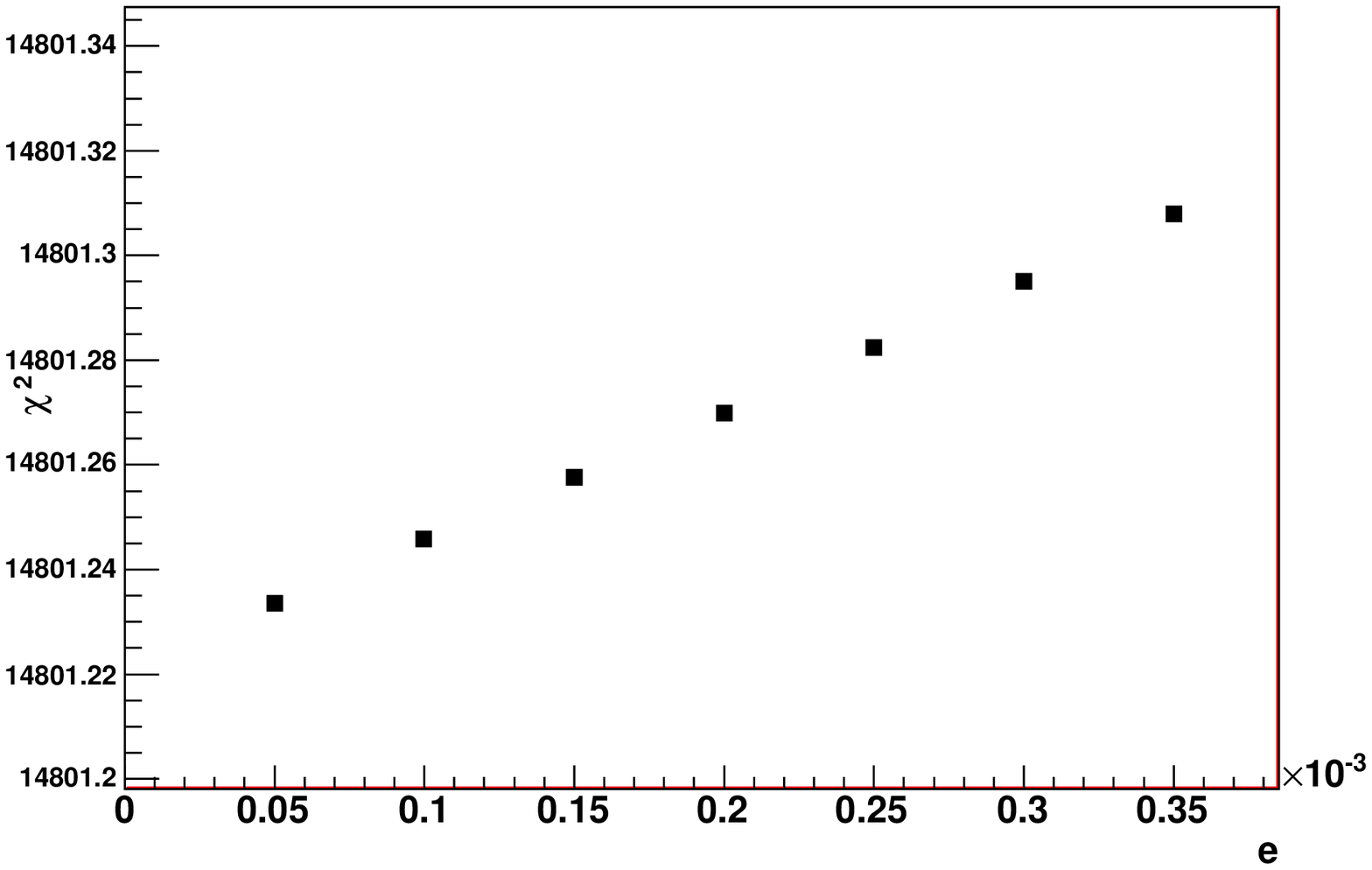}} \label{fig:chi2New42}}
}
\caption{$\chi^{2}$ versus the search parameter $e$ for the sum of $Model_{Tot.}^{*}$ and $Model_{3}^{*}$ fit.
The searched parameter $e$ approaches asymptotically to $0$ for minimal $\chi^{2}$ value.}
\label{fig:chi2New4}
\end{figure}

The $\chi^{2}$ function (Eq.~\ref{eq:chi2DalitzPlotFitNormNew4}) was minimize in respect to parameter $e$.
The (Fig.~\ref{fig:chi2New4}) shows the $\chi^{2}$ versus the searched parameter $e$.
The $\chi^{2}$ function does not have a minimum. 
It is seen that the searched parameter $e$ approaches asymptotically to $0$ for minimal $\chi^{2}$ value.
The contribution of other processes estimated by the above procedure would be equal to $0$ value.

\newpage
To check in addition the sensitivity of the estimate of other processes contribution in respect to the Monte Carlo developed model parameter $\beta$ (Eq.~\ref{eq:3pi0Model} on page \pageref{eq:3pi0Model}).
The estimate was repeated with the $\beta$ treated as a free parameter.   
This could be written as:

\begin{equation}
 (1-\epsilon)\beta Model_{1} + (1-\beta)Model_{2} + \epsilon Model_{3}
\end{equation}
where processes $Model_{1,2}$ are defined by (Eq.~\ref{eq:3pi0Model}). 

Fit was performed using the chi-square method; one defines the $\chi^{2}$ function to minimize:
\begin{equation}
 \chi^{2} = \sum \dfrac{\left[ Data - \left( (1-\epsilon) \left[ \beta Model_{1} + (1-\beta)Model_{2} \right] + \epsilon Model_{3}\right)\right]^{2} }{\sigma^{2}_{Data} + (1-\epsilon)^{2}\left[ \beta^{2}\sigma^{2}_{Model_{1}}+(1-\beta)^{2}\sigma^{2}_{Model_{2}}\right] + \epsilon^{2}\sigma^{2}_{Model_{3}}}
\label{eq:chi2DalitzPlotFitNew5}
\end{equation}
where the sum goes over each bin of the five Dalitz Plots $ppX$, five Dalitz Plots $3\pi^{0}$ and five Nyborg Plots i.e. $15$ plots.
The $\sigma_{Data}$ is the error of the point for experimental data,
 $\sigma_{Model_{1,2}}$~-~ is the error of the point for the $Model_{1,2}$ (Eq.~\ref{eq:3pi0Model}) and $\sigma_{Model_{3}}$~-~ is the error of the point for $Model_{3}$.

For the numerical purpose of doing the fit, the $\chi^{2}$ function (Eq.~\ref{eq:chi2DalitzPlotFitNew5}) was redefined to:
\begin{equation}
 \chi^{2} = \sum \dfrac{\left[ Data - \left( (1-e) \left[ \alpha Model_{1}^{*} + (1-\beta)Model_{2}^{*} \right] + e Model_{3}^{*}\right)\right]^{2} }{\sigma^{2}_{Data} + (1-e)^{2}\left[ \alpha^{2}\sigma^{2}_{Model_{1}^{*}}+(1-\alpha)^{2}\sigma^{2}_{Model_{2}^{*}}\right] + e^{2}\sigma^{2}_{Model_{3}^{*}}}
\label{eq:chi2DalitzPlotFitNormNew5}
\end{equation}
The same as in the Section~\ref{subsec:3pi0MCmodel} on page \pageref{subsec:3pi0MCmodel}.

\begin{figure}[ht!bp]
\centering
{
\subfigure[]{\fbox{\includegraphics[width=0.7\textwidth]{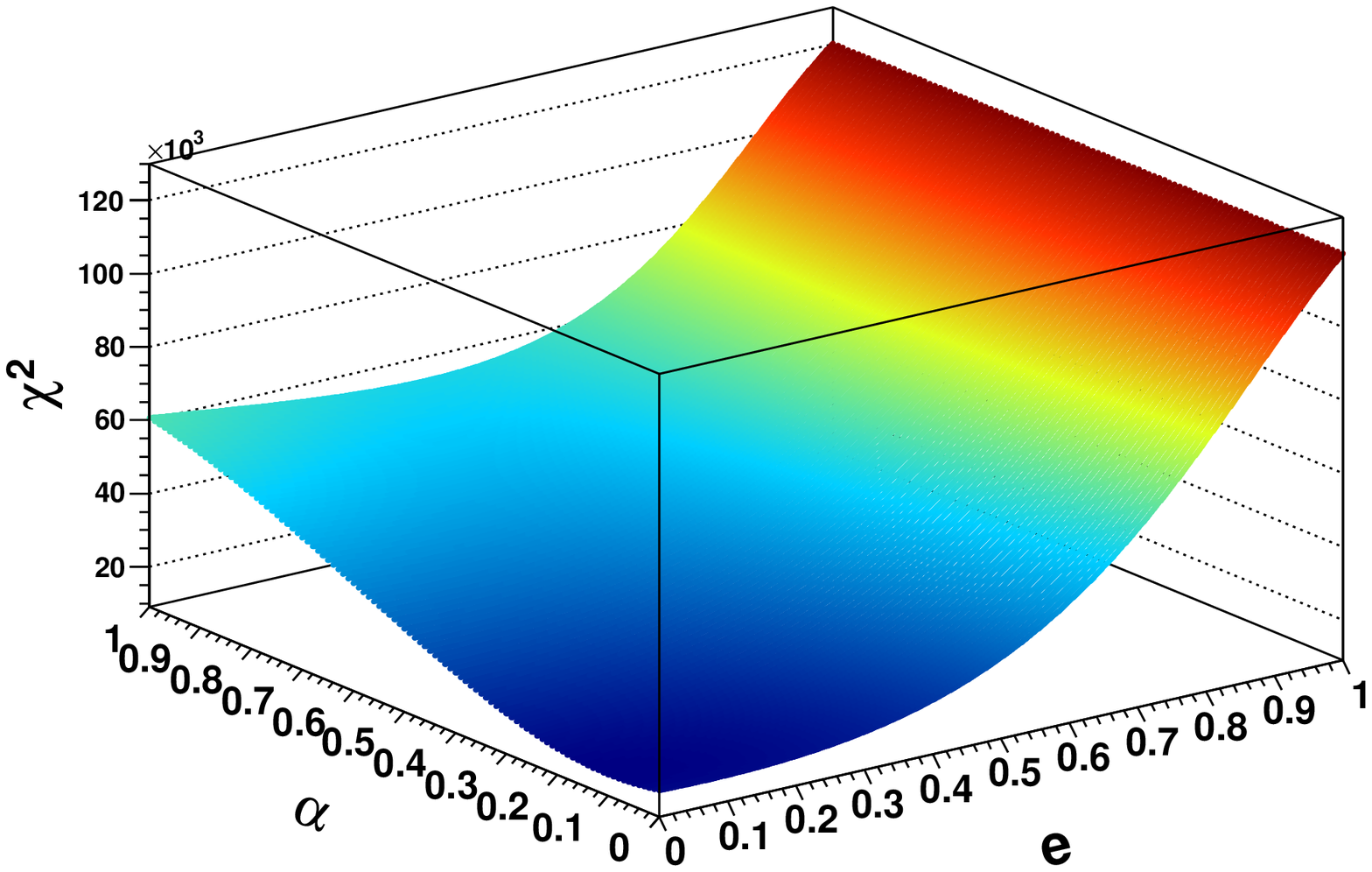}} \label{fig:chi2New51}}\\
\subfigure[]{\fbox{\includegraphics[width=0.7\textwidth]{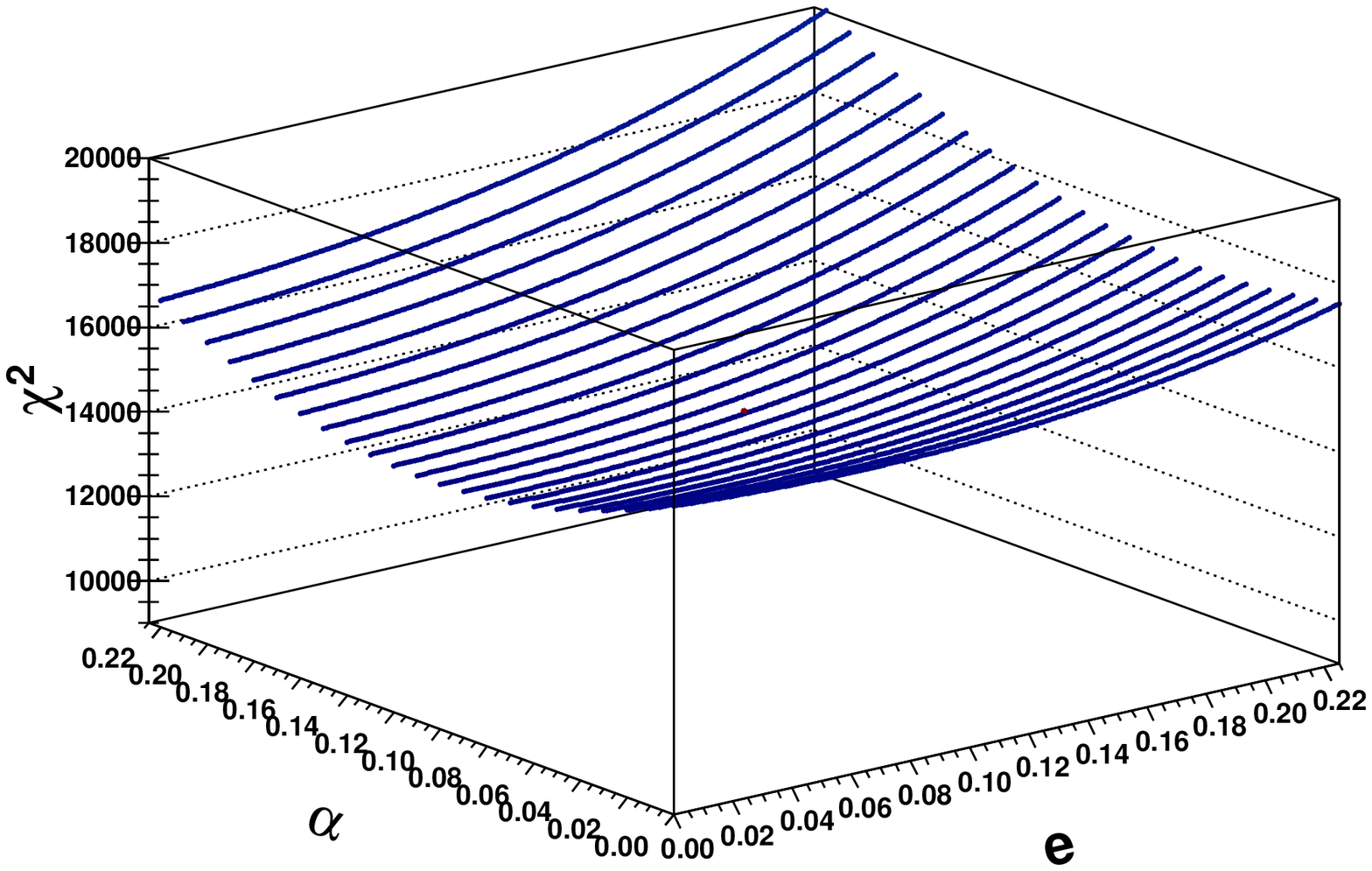}} \label{fig:chi2New52}}
}
\caption{$\chi^{2}$ versus the search parameter $e$ and $\alpha$ for the sum of $Model_{1}^{*}$, $Model_{2}^{*}$ and $Model_{3}^{*}$ fit simultaneously.
The $\chi^{2}$ function has a global minimum for the $e=0$ and $\alpha=0.1$ and $\chi^{2}=14803.3$.}
\label{fig:chi2New5}
\end{figure}

The $\chi^{2}$ function (Eq.~\ref{eq:chi2DalitzPlotFitNormNew5}) was minimize in respect to parameter $e$ and $\alpha$.
The (Fig.~\ref{fig:chi2New5}) shows the $\chi^{2}$ versus the searched parameter $e$ and $\alpha$.
It is seen that the $\chi^{2}$ has a global minimum for the following values:

\begin{equation}
 \frac{\chi^{2}}{NDF} = \frac{14803.3}{19715} = 0.751 \pm 0.010
\end{equation}

\begin{equation}
\alpha = 0.10 \pm 0.01
\label{eq:alpha3pi0ModelNew5}
\end{equation}

\begin{equation}
e = 0.000 \pm 0.001
\end{equation}

It is seen that the searched parameter $e$ approaches asymptotically to $0$ for the $\chi^{2}$ approaching to the global minimum.
The obtained $\alpha$ (Eq.~\ref{eq:alpha3pi0ModelNew5}) value is consistent within the reached precision with the previously obtained result (Eq.~\ref{eq:alpha3pi0Model} on page \pageref{eq:alpha3pi0Model}). 
If one considers change of the $\chi^{2}$ by $1$ from the minimum at $0$, one gets the  $\epsilon \approx 2\%$.  

Concluding, the contribution of other processes estimated here falls to the value $\sim 2\%$.


%
%
%
%

\emptydoublepage
\subsubsection{The Cross Section extraction}\label{subsec:CrossSection}
\thispagestyle{fancy}
In order to obtain the cross section for the $pp \rightarrow pp 3\pi^{0}$ one needs some suitable reaction of well-known cross section to which one can do the normalization.
The good choice is the $pp \rightarrow pp \eta$ where the $\eta$ meson decays into the same channel $\eta \rightarrow 3\pi^{0}$, this reaction
is nicely seen in the missing mass of the two protons \myImgRef{MMpp02PhSp.eps}.

The cross section for the $pp \rightarrow pp 3\pi^{0}$ one can write as:

\begin{equation}
 \sigma_{3\pi^{0}} = \dfrac{N_{3\pi^{0}}}{Tot.Eff._{3\pi^{0}}} \dfrac{1}{L}
\label{eq:CS3pi01}
\end{equation}
 $N_{3\pi^{0}}$~-~number of the identified $3\pi^{0}$ events(from experimental data), $Tot.Eff._{3\pi^{0}}$~-~total reconstruction efficiency (Eq.~\ref{eq:TotRecEffDefinition}) for the $pp \rightarrow pp 3\pi^{0}$ reaction 
(from Monte-Carlo simulation, since one knows the model of the reaction (Eq.~\ref{eq:3pi0Model})~), $L$~-~the integrated luminosity.
From $pp \rightarrow pp \eta$ reaction one gets:

\begin{equation}
 L =\dfrac{N_{\eta}}{Tot.Eff._{\eta}} \dfrac{1}{\sigma_{\eta}}
\label{eq:CS3pi02}
\end{equation}
$N_{\eta}$~-~number of the identified $\eta$ events(from experimental data), $Tot.Eff._{\eta}$~-~total
 reconstruction efficiency (Eq.~\ref{eq:TotRecEffDefinition}) of the  $pp \rightarrow pp \eta$ reaction, 
from the Monte-Carlo simulation,$\sigma_{\eta}$~-cross section for the $pp \rightarrow pp \eta$ reaction
 for the beam kinetic energy $T=2.54\mathrm{GeV}$.

When combining the two equation (Eq.~\ref{eq:CS3pi01}) and (Eq.~\ref{eq:CS3pi02}), it gives:
\begin{equation}
 \sigma_{3\pi^{0}} = \dfrac{N_{3\pi^{0}}}{Tot.Eff._{3\pi^{0}}} \dfrac{Tot.Eff._{\eta}}{N_{\eta}} \sigma_{\eta}
\label{eq:CS3pi03}
\end{equation}


\begin{figure}[ht!bp]
\centering
{
\subfigure[]{\fbox{\includegraphics[width=0.7\textwidth]{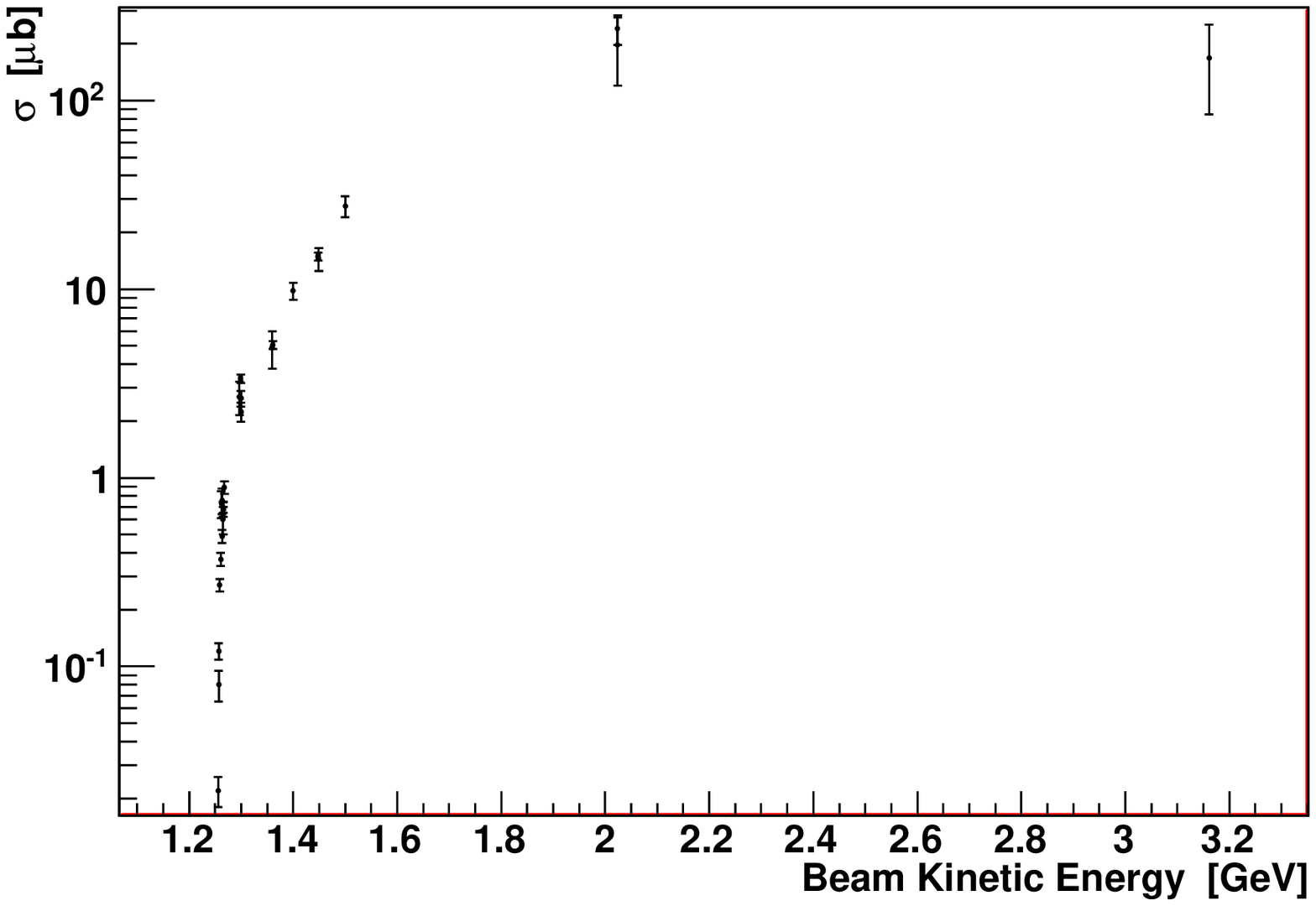}} \label{fig:EtaCrossSection1}}\\
\subfigure[]{\fbox{\includegraphics[width=0.7\textwidth]{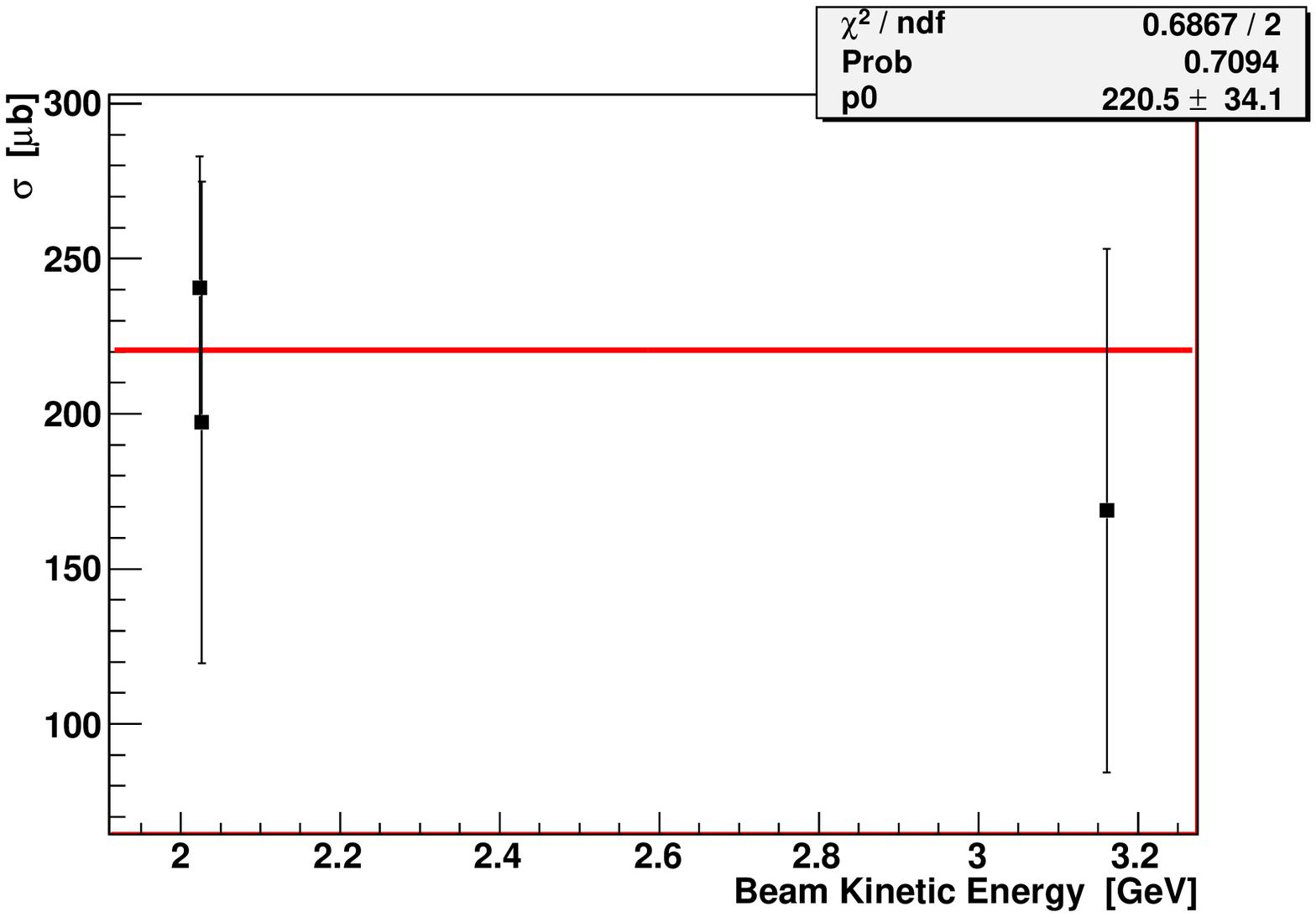}} \label{fig:EtaCrossSection2}}
}
\caption{$pp \rightarrow pp \eta$ cross section as a function of the proton beam kinetic energy \cite{ETACS01,ETACS02,ETACS03,ETACS04,ETACS05}.}
\label{fig:EtaCrossSection}
\end{figure}

The cross section for the $pp \rightarrow pp \eta$ at $T=2.54\mathrm{GeV}$ was evaluated from the existing data \cite{ETACS01,ETACS02,ETACS03,ETACS04,ETACS05}.
One see that the cross section saturates at high kinetic energy energy(around $2\mathrm{~GeV}$) (Fig.~\ref{fig:EtaCrossSection1}).  
The constant was fitted to the values above $2\mathrm{~GeV}$; the extracted cross section used for calculation of $L$ is:
\begin{equation}
 \sigma_{\eta} = 220 \pm 34 \mu b
\label{eq:EtaCrossSectionValue}
\end{equation}
The total efficiency for the $pp \rightarrow pp 3\pi^{0}$ was calculated using Monte-Carlo simulation, using developed model (Eq.~\ref{eq:3pi0Model}),
from the true number of events $N_{3\pi^{0}}^{True}$ (i.e. generated events) and the number of events reconstructed $N_{3\pi^{0}}^{Reco}$ (see Section~\ref{subsec:AnaChainKinFit}, page~\pageref{subsec:AnaChainKinFit}):

\begin{equation}
 Tot.Eff._{3\pi^{0}} = \dfrac{N_{3\pi^{0}}^{Reco}}{N_{3\pi^{0}}^{True}}
\label{eq:TotalEfficiency3pi0}
\end{equation}

The total efficiency for the $pp \rightarrow pp \eta$ reaction was calculated in the similar way, using the Monte-Carlo simulation, assuming
that the $\eta$ meson is produced via $N^{*}(1535)$ excitation and phase space (see Section~\ref{sec:etaprod}, page~\pageref{sec:etaprod} \myImgRef{EtaIMEtaPFitresultNew2}~):
\begin{equation}
 Tot.Eff._{\eta} = \dfrac{ N_{\eta}^{Reco}}{N_{\eta}^{True}}
\label{eq:TotalEfficiencyEta}
\end{equation}

The error of the total efficiency $Error(Tot.Eff)$ was calculated assuming binomial errors distribution \cite{BinominalErrors}:
\begin{equation}
 Error(Tot.Eff) = \dfrac{1}{N_{True}}\sqrt{N_{Reco}\left(1-\dfrac{N_{Reco}}{N_{True}}\right)}
\end{equation}

\myFrameSmallFigure{MMpp02PhSpAreas.eps}{Missing Mass of two protons, the areas indicated. On vertical axis the number of events is shown.}{Missing Mass of two protons.}

To extract from the experimental data the number of the $\eta$ mesons and the number of the prompt $3\pi^{0}$,
the distribution of the missing mass of two protons was divided into four eras $A,B,C,D$ \myImgRef{MMpp02PhSpAreas.eps}.
Areas $A,B$ are without any $\eta$ meson content. Area $C$ is the area where the $\eta$ and $3\pi^{0}$ mesons contribute,
the $D$ area is the area below the $\eta$ meson peak (only $3\pi^{0}$ contribute).
Defining it like that leads to the following relations:
\begin{equation}
 N_{\eta} = N_{C}-N_{D}
\label{eq:NumberOfEtas}
\end{equation}
\begin{equation}
 N_{3\pi^{0}} = N_{A}+N_{B}+N_{D}
\label{eq:NumberOf3pi0}
\end{equation}

$N_{A,B,C,D}$~-~number of events in area $A,B,C,D$ respectively.

\begin{figure}[ht!bp]
\centering
{
\subfigure[Fit of the fourth order polynomial outside the $\eta$ meson peak.]{\fbox{\includegraphics[width=0.7\textwidth]{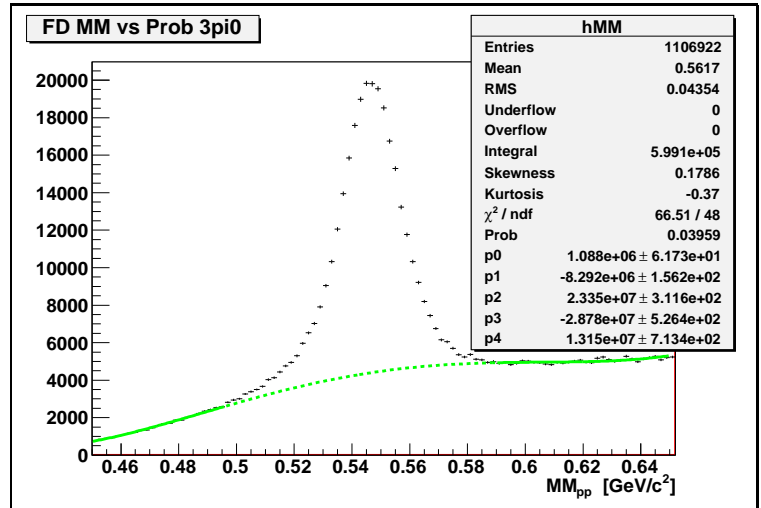}} \label{fig:MMppEtaBGSubtractionB1}}\\
\subfigure[Subtracted $\eta$ meson signal, blue markers correspond to the Monte-Carlo simulation.]{\fbox{\includegraphics[width=0.7\textwidth]{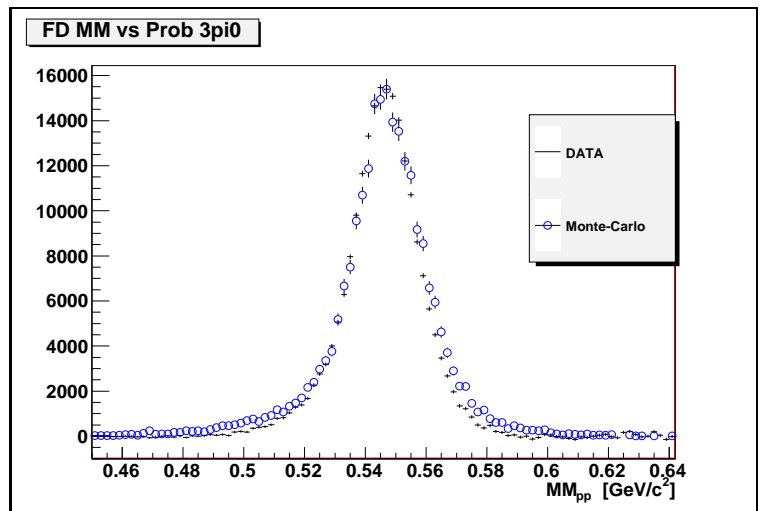}} \label{fig:MMppEtaBGSubtractionB2}}
}
\caption{Subtraction of the background under the $\eta$ peak from $MM_{pp}$ distribution. On vertical axis the number of events is shown.}
\label{fig:MMppEtaBGSubtractionB}
\end{figure}

The only unknown is $N_{D}$ the number of events in area $D$, which is the number of events below the $\eta$ peak.
It was evaluated numerically by fitting the fourth order polynomial outside the $\eta$ meson peak range and prolonging it below the peak (Fig.~\ref{fig:MMppEtaBGSubtractionB}).
Then the $N_{D}$ is the integral of the fitted function in the $\eta$ meson peak range, evaluated numerically.
Also the errors of the number of $\eta$ mesons and $3\pi^{0}$ was calculated:
\begin{equation}
 Error(N_{\eta}) = \sqrt{ Error(N_{C})^{2} + Error(N_{D})^{2} }
\end{equation}
\begin{equation}
Error(N_{3\pi^{0}}) = \sqrt{ Error(N_{A})^{2} + Error(N_{B})^{2} + Error(N_{D})^{2} }
\end{equation}

where $Error(N_{A,B,C,D})$ is the error of the number of events in different area:
\begin{eqnarray}
Error(N_{A}) &=& \sqrt{N_{A}} \\
Error(N_{B}) &=& \sqrt{N_{B}} \\
Error(N_{C}) &=& \sqrt{N_{C}} 
\end{eqnarray}
the $Error(N_{D})$ is calculated numerically as an error of the integral of the mentioned above function.

Now the statistical error of the $pp \rightarrow pp 3\pi^{0}$ cross section $Error(\sigma_{3\pi^{0}})_{stat.}$
was calculated using the standard error propagation procedure:

\begin{eqnarray}
 Error(\sigma_{3\pi^{0}})_{stat.}^{2} &=& Error( \sigma_{3\pi^{0}})_{N_{3\pi^{0}}}^{2} + Error( \sigma_{3\pi^{0}})_{Tof.Eff._{3\pi^{0}}}^{2} + \\
&+& Error( \sigma_{3\pi^{0}})_{N_{\eta}}^{2} + Error( \sigma_{3\pi^{0}})_{Tot.Eff._{\eta}}^{2}\nonumber
\end{eqnarray}

where

\begin{eqnarray}
 Error( \sigma_{3\pi^{0}})_{N_{3\pi^{0}}} &=& \left | \dfrac{\partial \sigma_{3\pi^{0}}}{\partial N_{3\pi^{0}} } \right| Error(N_{3\pi^{0}})\\
 Error( \sigma_{3\pi^{0}})_{Tot.Eff._{3\pi^{0}}} &=& \left | \dfrac{\partial \sigma_{3\pi^{0}}}{\partial Tot.Eff._{3\pi^{0}} } \right| Error(Tot.Eff._{3\pi^{0}})\\
 Error( \sigma_{3\pi^{0}})_{ N_{\eta} } &=& \left | \dfrac{\partial \sigma_{3\pi^{0}}}{\partial N_{\eta} } \right| Error(N_{\eta})\\
 Error( \sigma_{3\pi^{0}})_{Tot.Eff._{\eta}} &=& \left | \dfrac{\partial \sigma_{3\pi^{0}}}{\partial Tot.Eff._{\eta} } \right| Error(Tot.Eff._{\eta})\\
\end{eqnarray}

The normalization error of the $3\pi^{0}$ cross section $Error(\sigma_{3\pi^{0}})_{norm.}$, related to the to error of the $\eta$ meson cross section, 
 was separated since it is not related to the statistical precision of this experiment:

\begin{equation}
 Error(\sigma_{3\pi^{0}})_{norm.} = Error( \sigma_{3\pi^{0}})_{\sigma_{\eta}}
\end{equation}
where
\begin{equation}
 Error( \sigma_{3\pi^{0}})_{\sigma_{\eta}} = \left | \dfrac{\partial \sigma_{3\pi^{0}}}{\partial \sigma_{\eta} } \right| Error(\sigma_{\eta})\\
\end{equation}

\myTable{
\footnotesize
\begin{adjustwidth}{-3cm}{-3cm}
\begin{tabular}{|l|c|c|c|c|c|c|c|c||c|}
\hline
 Prob. Cut (see Fig.~\ref{fig:KFitProb}) & $0.2$ & $0.3$ & $0.4$ & $0.5$ & $0.6$ & $0.7$ & $0.8$ &$0.9$ & Mean  \\ 
\hline
\hline
$\sigma_{3\pi^{0}}\mathrm{[\mu b]}$& $138.51$ & $132.30$ & $128.19$ & $124.49$ & $122.51$ & $119.97$ & $112.48$ & $110.15$&$123.57$\\
\hline
$ Error( \sigma_{3\pi^{0}})_{N_{3\pi^{0}}}\mathrm{[\mu b]}$& $0.14$ & $0.15$ & $0.16$ & $0.17$ & $0.19$ & $0.21$ & $0.24$ & $0.33$ & $0.20$\\
\hline
$Error( \sigma_{3\pi^{0}})_{Tot.Eff._{3\pi^{0}}}\mathrm{[\mu b]}$& $0.27$ & $0.28$ & $0.30$ & $0.32$ & $0.36$ & $0.41$ & $0.48$ & $0.68$ & $0.39$\\
\hline
$Error( \sigma_{3\pi^{0}})_{ N_{\eta} }\mathrm{[\mu b]}$& $0.47$ & $0.47$ & $0.48$ & $0.50$ & $0.55$ & $0.60$ & $0.65$ & $0.87$ & $0.57$\\
\hline
$Error( \sigma_{3\pi^{0}})_{Tot.Eff._{\eta}}\mathrm{[\mu b]}$& $0.47$ & $0.49$ & $0.51$ & $0.55$ & $0.61$ & $0.70$ & $0.81$ & $1.13$ & $0.66$\\
\hline
$Error(\sigma_{3\pi^{0}})_{stat.}\mathrm{[\mu b]}$& $0.73$ & $0.75$ & $0.78$ & $0.83$ & $0.91$ & $1.03$ & $1.17$ & $1.62$ & $0.98$\\
\hline
$Error(\sigma_{3\pi^{0}})_{norm.}\mathrm{[\mu b]}$& $21.41$ & $20.45$ & $19.81$ & $19.24$ & $18.93$ & $18.54$ & $17.38$ & $17.02$ & $19.10$\\
\hline
\end{tabular}
\end{adjustwidth}

}{Results of the $pp \rightarrow pp 3\pi^{0}$ cross section studies.}{tab:CrossSection3pi0}

The $pp \rightarrow pp 3\pi^{0}$ cross section and equivalent errors were evaluated for the different probability cut, the results area
presented in Tab.~\ref{tab:CrossSection3pi0}.
It is seen that the cross section changes systematically with the probability cut.
As the estimator of the cross section the mean value for the probability cut has been taken with the corresponding statistical and normalization error:

\begin{equation}
\sigma_{3\pi^{0}} = 123 \pm 1 (stat.) \pm 19 (norm.) \mu b
\end{equation}

To estimate the systematic error of the cross section $Error(\sigma_{3\pi^{0}})_{sys.}$, related to the changes of the probability cut, the difference between the
maximal cross section value $\sigma_{3\pi^{0}}^{Max}$ and the minimal cross section $\sigma_{3\pi^{0}}^{Min}$ divided by the $2\sqrt{3}$ has been taken \cite{GUM}.
\begin{equation}
 Error(\sigma_{3\pi^{0}})_{sys.} = \dfrac{\sigma_{3\pi^{0}}^{Max}-\sigma_{3\pi^{0}}^{Min}}{2\sqrt{3}} = 8.19 \mu b
\end{equation}

The cross section value for the $pp \rightarrow pp 3\pi^{0}$ reaction at beam kinetic energy $T=2.54\mathrm{~GeV}$ is equal to: 

\begin{equation}
\sigma_{3\pi^{0}} = 123 \pm 1 (stat.) \pm 19 (norm.) \pm 8 (sys.) \mu b
\label{eq:CrossSection3pi0Value}
\end{equation}

Now adding the errors in quadrature this gives:

\begin{equation}
\sigma_{3\pi^{0}} = 123 \pm 21 \mu b
\label{eq:CrossSection3pi0ValueTot}
\end{equation}
 
this corresponds to the integrated luminosity during the whole experiment of;
\begin{equation}
 L=4.30 \pm 0.83~10^{5} {\mu b}^{-1}
\label{eq:IntegratedLuminosityValue}
\end{equation}
 
\myFrameFigure{CS3pi0MyPoint}{$pp \rightarrow pp 3\pi^{0}$ cross section as a function of a proton beam kinetic energy, the experimental data \cite{ETACS01} and available models \cite{3pi0CSmodel1, 3pi0CSmodel2, 3pi0CSmodel3}.}
{$pp \rightarrow pp 3\pi^{0}$ cross section.}

The result was compared with the available data \cite{ETACS01} and models for $pp \rightarrow pp 3\pi^{0}$ production \cite{3pi0CSmodel1, 3pi0CSmodel2, 3pi0CSmodel3}, \myImgRef{CS3pi0MyPoint}.
It is see that the data confirms the the cross section scaling model based on Delof FSI \cite{3pi0CSmodel1}. 
\newpage 

\subsubsection{The Acceptance and Efficiency Correction}\label{subsec:AccEff}

\paragraph{Additional checks\\}

Additional checks were performed, before doing the Acceptance and Efficiency Correction, to see what is the influence of the probability cut on the $\eta$ meson and on the $3\pi^{0}$ system
and on the measured distributions.

\myFrameFigure{EtaProbSummary}{Experimental data - missing mass of the two protons: the subtracted by the polynomial fit $\eta$ meson peak for different regions of the probability. On vertical axis number of events is shown.
The spectra are normalized to the same number of events.}{Missing Mass of the two protons.}

First $\eta$ meson peak was extracted from experimental data, by fitting the polynomial of the forth order to the ranges outside the peak,
for different regions of the probability $Prob=0.2-0.4$, $Prob=0.5-0.7$, $Prob=0.8-1.0$ \myImgRef{EtaProbSummary}.
It is seen that the $\eta$ meson peaks for different probability regions are almost the same shape(the width and the peak position is almost the same),
this proofs the correctness of the analysis as well as the stability of the background extraction technique.       

Next the non resonant $3\pi^{0}$ system was studied and compared for probability cut $Prob>0.2$ and $Prob>0.9$ for experimental data
 (Figs.~\ref{image_CompAllComp09}~\ref{image_Dal1Comp09}~\ref{image_Dal2Comp09}~\ref{image_NyborgComp09}).
It is seen that the distributions are the same independently on the selected probability.

The chosen probability cut does not introduce any systematic effect in the shape of the $\eta$ and $3\pi^{0}$ distributions.

\myFrameHugeFigure{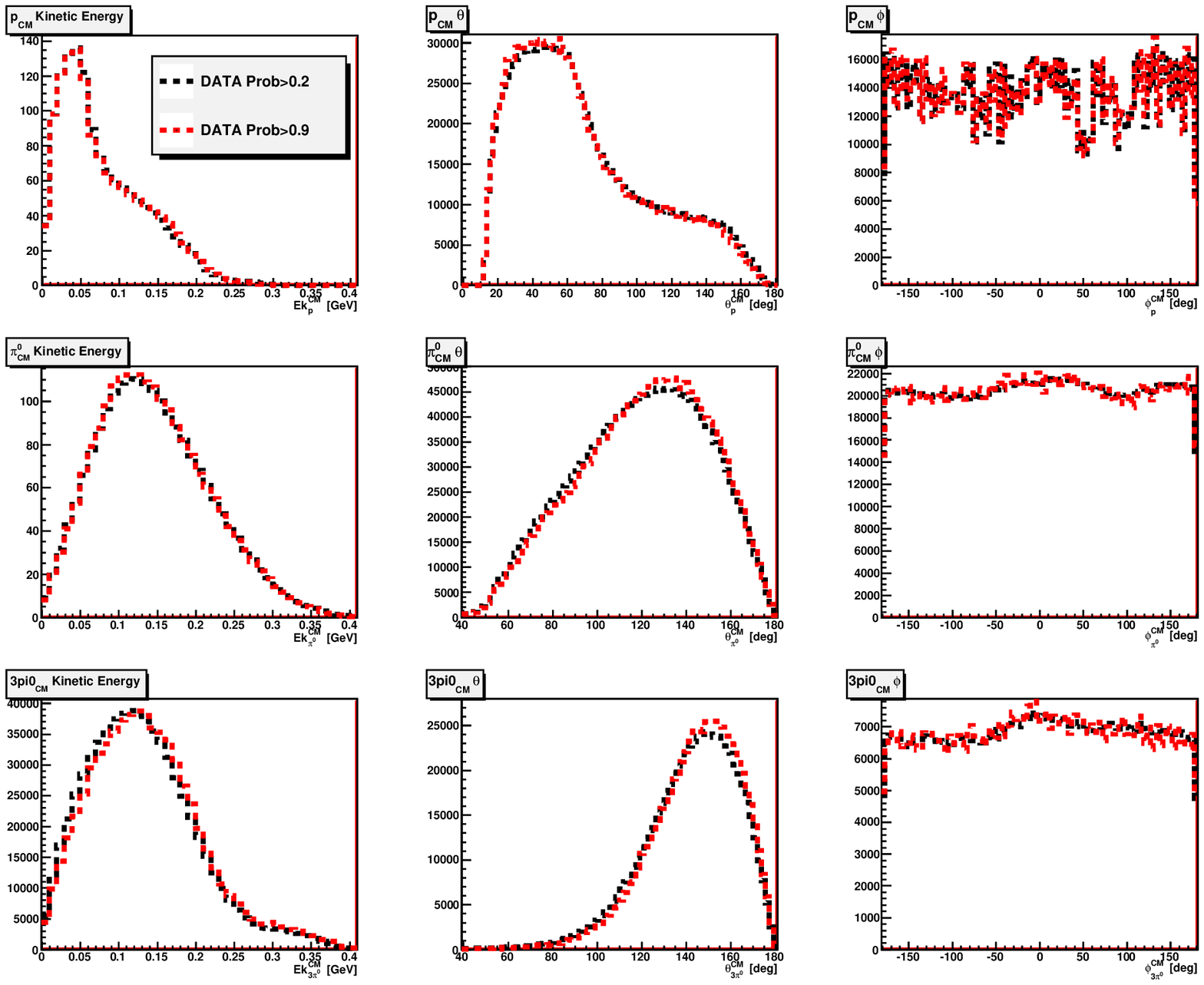}{Comparison of the experimental data with $Prob>0.2$ (black) and $Prob>0.9$ (red), for the $MM_{pp}<0.5\mathrm{GeV/c^{2}}$ and $MM_{pp}>0.6\mathrm{GeV/c^{2}}$.
\textit{Upper row}: proton in the center of mass frame,\textit{Middle row}: pion in the center of mass frame, \textit{Lower row}: $3\pi^{0}$ system in the center of mass frame. From left kinetic energy, 
the polar angle and the azimuthal angle distribution. The spectra are normalized to the same number of events.}{Comparison of the experimental data with with $Prob>0.2$ (black) and $Prob>0.9$ (red), for the $MM_{pp}<0.5\mathrm{GeV/c^{2}}$ and $MM_{pp}>0.6\mathrm{GeV/c^{2}}$.}

\myFrameHugeFigure{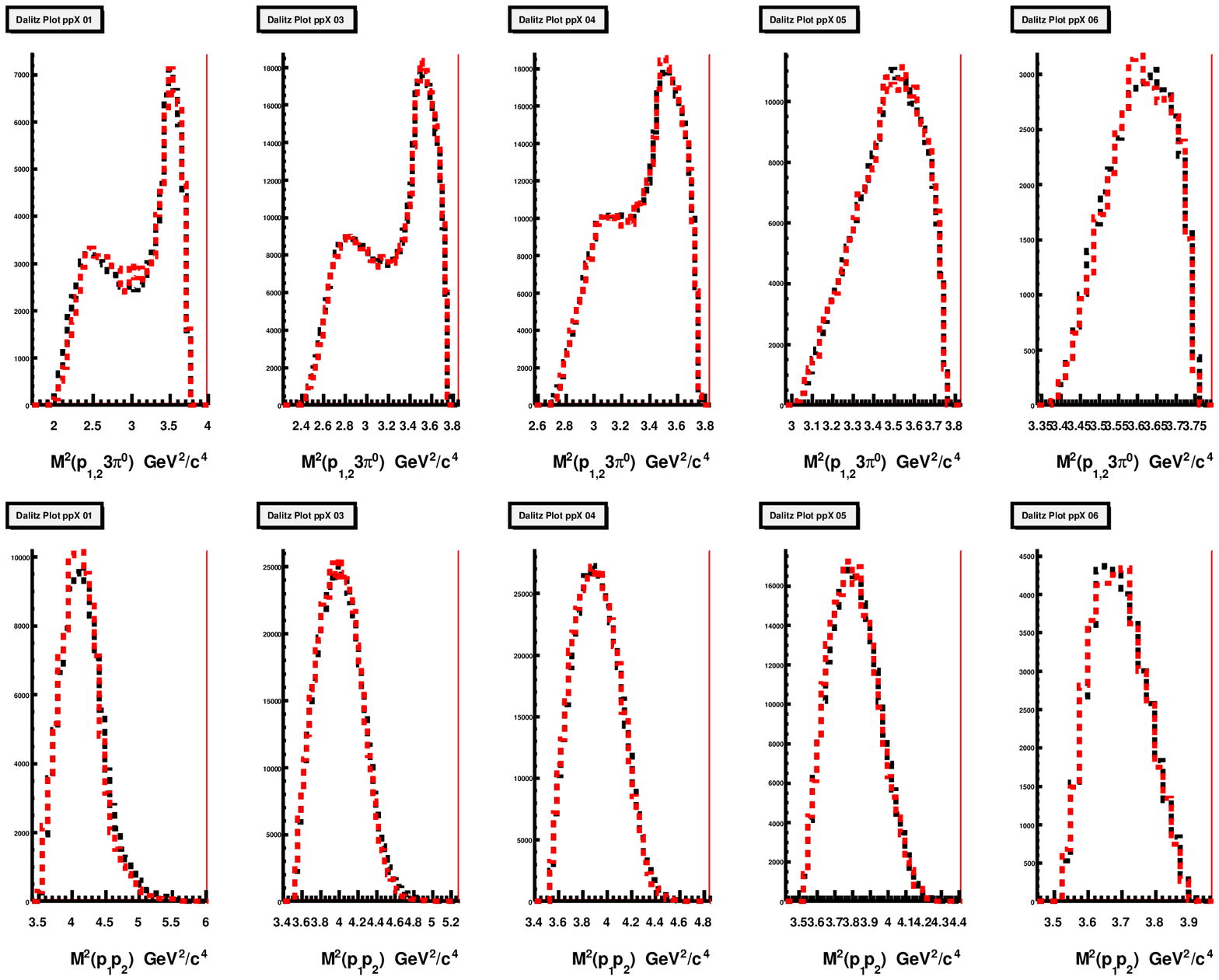}{Comparison of the Dalitz Plot $ppX$ projections
 for the experimental data with $Prob>0.2$ (black) and $Prob>0.9$ (red).
 The upper row corresponds to the projection to the $IM^{2}(p3\pi^{0})$ axis,
the lower row to the $IM^{2}(pp)$ axis. The columns from left to right correspond
to the following missing mass of two protons bins, column~1~$MM_{pp}=0.4-0.5\mathrm{GeV/c^{2}}$,
column~2~$MM_{pp}=0.6-0.7\mathrm{GeV/c^{2}}$, column~3~$MM_{pp}=0.7-0.8\mathrm{GeV/c^{2}}$, 
column~4~$MM_{pp}=0.8-0.9\mathrm{GeV/c^{2}}$, column~5~$MM_{pp}=0.9-1.0\mathrm{GeV/c^{2}}$.
 The plots are symmetrized against two protons - each event is filled two times.
The spectra are normalized to the same number of events.}
{Comparison of the Dalitz Plot $ppX$ projections for the experimental data with $Prob>0.2$ (black) and $Prob>0.9$ (red).}

\myFrameHugeFigure{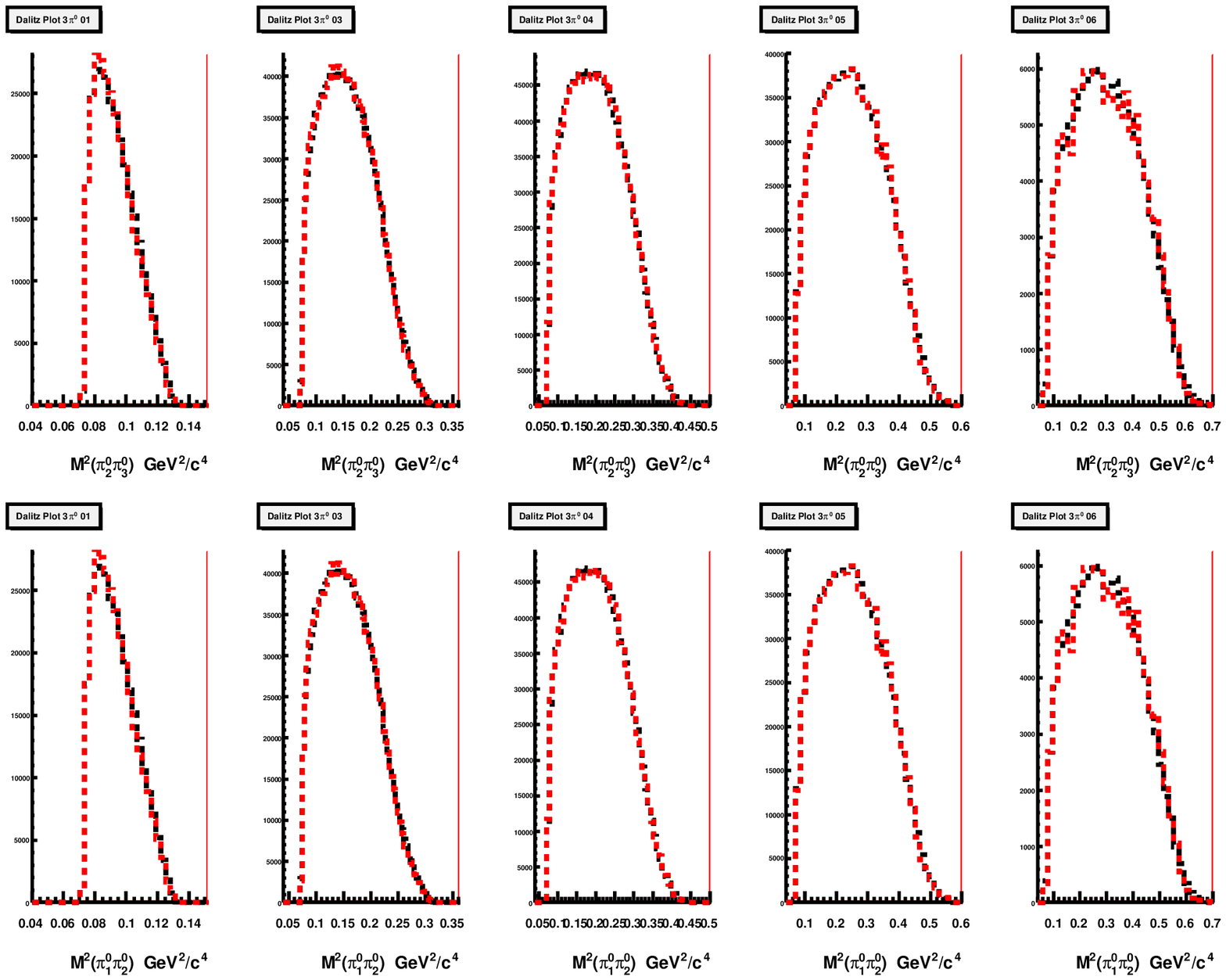}{Comparison of the Dalitz Plot $3\pi^{0}$ projections
 for the experimental data with $Prob>0.2$ (black) and $Prob>0.9$ (red).
 The upper and lower row corresponds to the projection to the $IM^{2}(2\pi^{0})$ axis, $x$ and $y$ axis of the Dalitz Plot,
they are identical.
The columns from left to right correspond
to the following missing mass of two protons bins, column~1~$MM_{pp}=0.4-0.5\mathrm{GeV/c^{2}}$,
column~2~$MM_{pp}=0.6-0.7\mathrm{GeV/c^{2}}$, column~3~$MM_{pp}=0.7-0.8\mathrm{GeV/c^{2}}$, 
column~4~$MM_{pp}=0.8-0.9\mathrm{GeV/c^{2}}$, column~5~$MM_{pp}=0.9-1.0\mathrm{GeV/c^{2}}$.
 The plots are symmetrized against two protons and three pions - each event is filled six times.
The spectra are normalized to the same number of events.}
{Comparison of the Dalitz Plot $ppX$ projections for the experimental data with $Prob>0.2$ (black) and $Prob>0.9$ (red).}

\myFrameHugeFigure{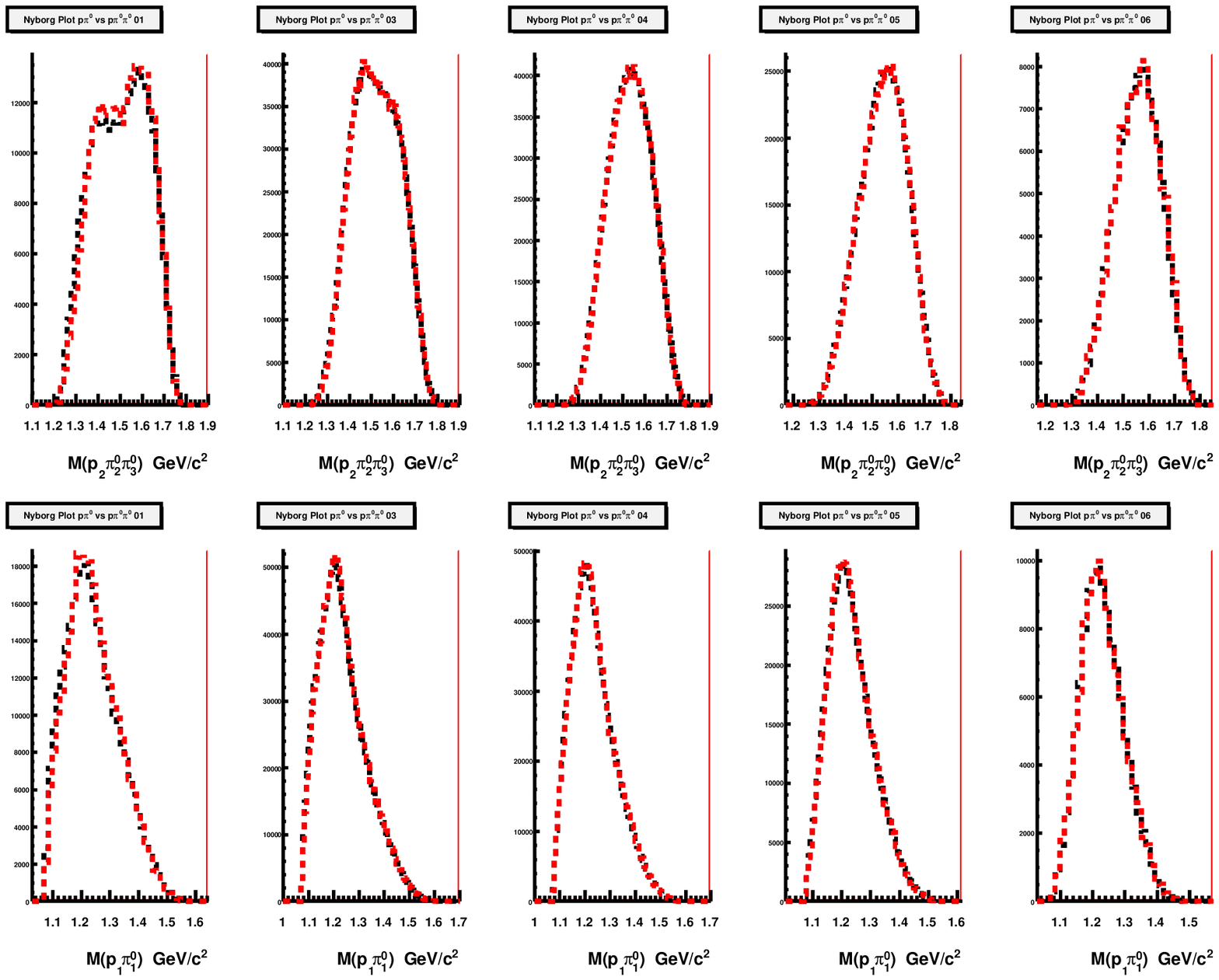}{Comparison of the Nyborg Plot projections
 for the experimental data with $Prob>0.2$ (black) and $Prob>0.9$ (red).
 The upper row corresponds to the projection to the $IM(p2\pi^{0})$ axis,
the lower row to the $IM(p\pi^{0})$ axis. The columns from left to right correspond
to the following missing mass of two protons bins, column~1~$MM_{pp}=0.4-0.5\mathrm{GeV/c^{2}}$,
column~2~$MM_{pp}=0.6-0.7\mathrm{GeV/c^{2}}$, column~3~$MM_{pp}=0.7-0.8\mathrm{GeV/c^{2}}$, 
column~4~$MM_{pp}=0.8-0.9\mathrm{GeV/c^{2}}$, column~5~$MM_{pp}=0.9-1.0\mathrm{GeV/c^{2}}$.
 The plots are symmetrized against two protons and three pions - each event is filled six times.
The spectra are normalized to the same number of events.}
{Comparison of the Dalitz Plot $ppX$ projections for the experimental data with $Prob>0.2$ (black) and $Prob>0.9$ (red).}

\newpage
\paragraph{The acceptance and efficiency correction\\}

The acceptance and efficiency correction was taken into account to remove the bias of the detector acceptance and the reconstruction efficiency.
The aim is to correct the reconstructed variable $V_{Reco}$(it could be one or more dimensional) to obtain the corrected one $V_{Corr}$ (free of the bias). 

First the total efficiency (Eq.~\ref{eq:TotRecEffDefinition}) was calculated as a function of the $V_{Reco}$.
The $pp \rightarrow pp 3\pi^{0}$ using Monte-Carlo simulation was used, using developed model (Eq.~\ref{eq:3pi0Model}),
by comparing the true number of events $N_{3\pi^{0}}^{True}$ (generated events) with the number of events obtained from the reconstruction $N_{3\pi^{0}}^{Reco}$:
\begin{equation}
 Tot.Eff._{3\pi^{0}}(V_{Reco}) = \dfrac{N_{3\pi^{0}}^{Reco}}{N_{3\pi^{0}}^{True}}(V_{Reco})
\label{eq:TotalEfficiencyFunction3pi0}
\end{equation}
The error of the total efficiency $Error(Tot.Eff_{3\pi^{0}})(V_{Reco})$ was calculated assuming binomial errors distribution \cite{BinominalErrors}:
\begin{equation}
 Error(Tot.Eff_{3\pi^{0}})(V_{Reco}) = \dfrac{1}{N_{3\pi^{0}}^{True}}\sqrt{N_{3\pi^{0}}^{Reco}\left(1-\dfrac{N_{3\pi^{0}}^{Reco}}{N_{3\pi^{0}}^{True}}\right)}(V_{Reco})
\label{eq:TotalEfficiencyFunctionError3pi0}
\end{equation}

the derived $Tot.Eff._{3\pi^{0}}(V_{Reco})$ (Eq.~\ref{eq:TotalEfficiencyFunction3pi0}) was applied to correct the reconstructed variable $V_{Reco}$:
\begin{equation}
 V_{Corr} = \dfrac{V_{Reco}}{Tot.Eff._{3\pi^{0}}(V_{Reco})}
\end{equation}

To obtain the absolute normalization, the corrected variable $V_{Corr}$ was normalized to the extracted cross section (Eq.~\ref{eq:CrossSection3pi0ValueTot}),
dividing the spectra by the integrated luminosity value (Eq.~\ref{eq:IntegratedLuminosityValue}).

This procedure was applied to the all Dalitz Plots $ppX$, Dalitz Plots $3\pi^{0}$, Nyborg Plots and their projections,
(see Figs.~\ref{image_Dalitz1AccCor2D}~\ref{image_Dalitz1AccCorPx}~\ref{image_Dalitz1AccCorPy}~\ref{image_Dalitz2AccCor2D}~\ref{image_Dalitz2AccCorPx}~\ref{image_Dalitz3AccCor2D}~\ref{image_Dalitz3AccCorPx}~\ref{image_Dalitz3AccCorPy}).

Later the corrected and normalized data were compared with Monte-Carlo Simulation of homogeneous and isotropic populated phase space
 and the developed Monte-Carlo Model (Eq.~\ref{eq:3pi0Model}), which assumes
simultaneous excitation of two baryons $\Delta(1235)$ and $N^{*}(1440)$ and their decay into $pp3\pi^{0}$ final state (See Figs.~\ref{image_Dalitz1AccCor2D}~\ref{image_Dalitz1AccCorPx}~\ref{image_Dalitz1AccCorPy}~\ref{image_Dalitz2AccCor2D}~\ref{image_Dalitz2AccCorPx}~\ref{image_Dalitz3AccCor2D}~\ref{image_Dalitz3AccCorPx}~\ref{image_Dalitz3AccCorPy}).
It is seen that the Monte-Carlo developed Model (Eq.~\ref{eq:3pi0Model}) describes the data significantly better than the phase space model.  

The acceptance and efficiency corrected Dalitz and Nyborg Plots are available as tables of numbers in Appendix~\ref{appendix:DataTables}.

\myFrameHugeFigure{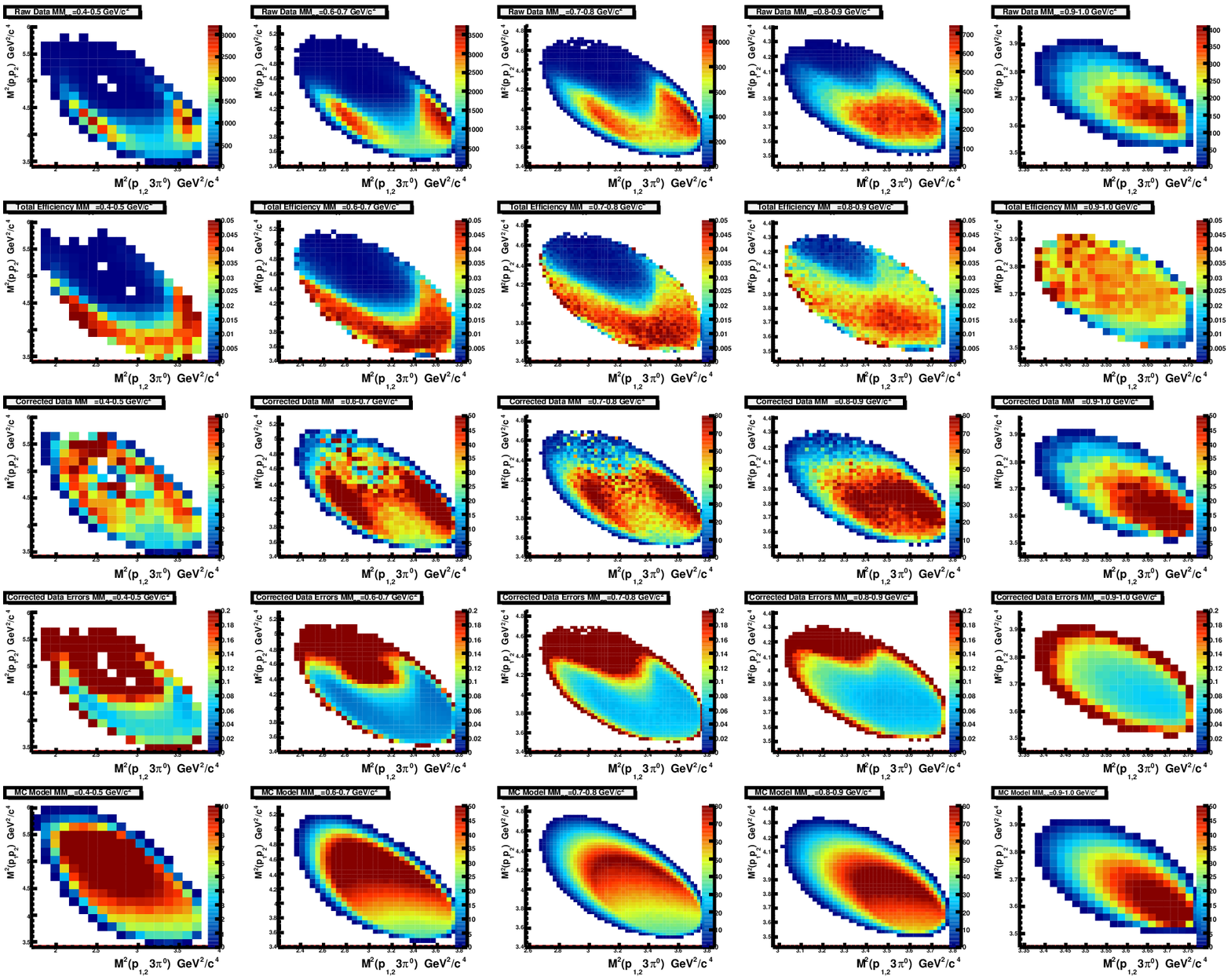}{Dalitz Plot $ppX$. $M^{2}(pp)$ versus $M^{2}(p3\pi^{0})$. The rows from up to down correspond to:
row~1: the experimental data, row~2: the Total Efficiency function (Eq.~\ref{eq:TotalEfficiencyFunction3pi0}),
row~3: the corrected experimental data, row~4: the statistical error of the corrected data,
row~5: the Monte-Carlo developed model (Eq.~\ref{eq:3pi0Model}) . The columns from left to right correspond
to the following missing mass of two protons bins, column~1~$MM_{pp}=0.4-0.5\mathrm{~GeV/c^{2}}$,column~2~$MM_{pp}=0.6-0.7\mathrm{~GeV/c^{2}}$,
column~3~$MM_{pp}=0.7-0.8\mathrm{~GeV/c^{2}}$, column~4~$MM_{pp}=0.8-0.9\mathrm{~GeV/c^{2}}$, column~5~$MM_{pp}=0.9-1.0\mathrm{~GeV/c^{2}}$.
 The plots are symmetrized against two protons - each event is filled two times.
The model is normalized to the same number of events as in the experimental data.
It is seen that the Monte-Carlo developed model (Eq.~\ref{eq:3pi0Model}) row~5 describes the data significantly better than the phase space model (homogenous Dalitz plot).
Fully expandable and colored version of the figure is available in the attached electronic version of the thesis.
}{Dalitz Plot $ppX$.}

\myFrameHugeFigure{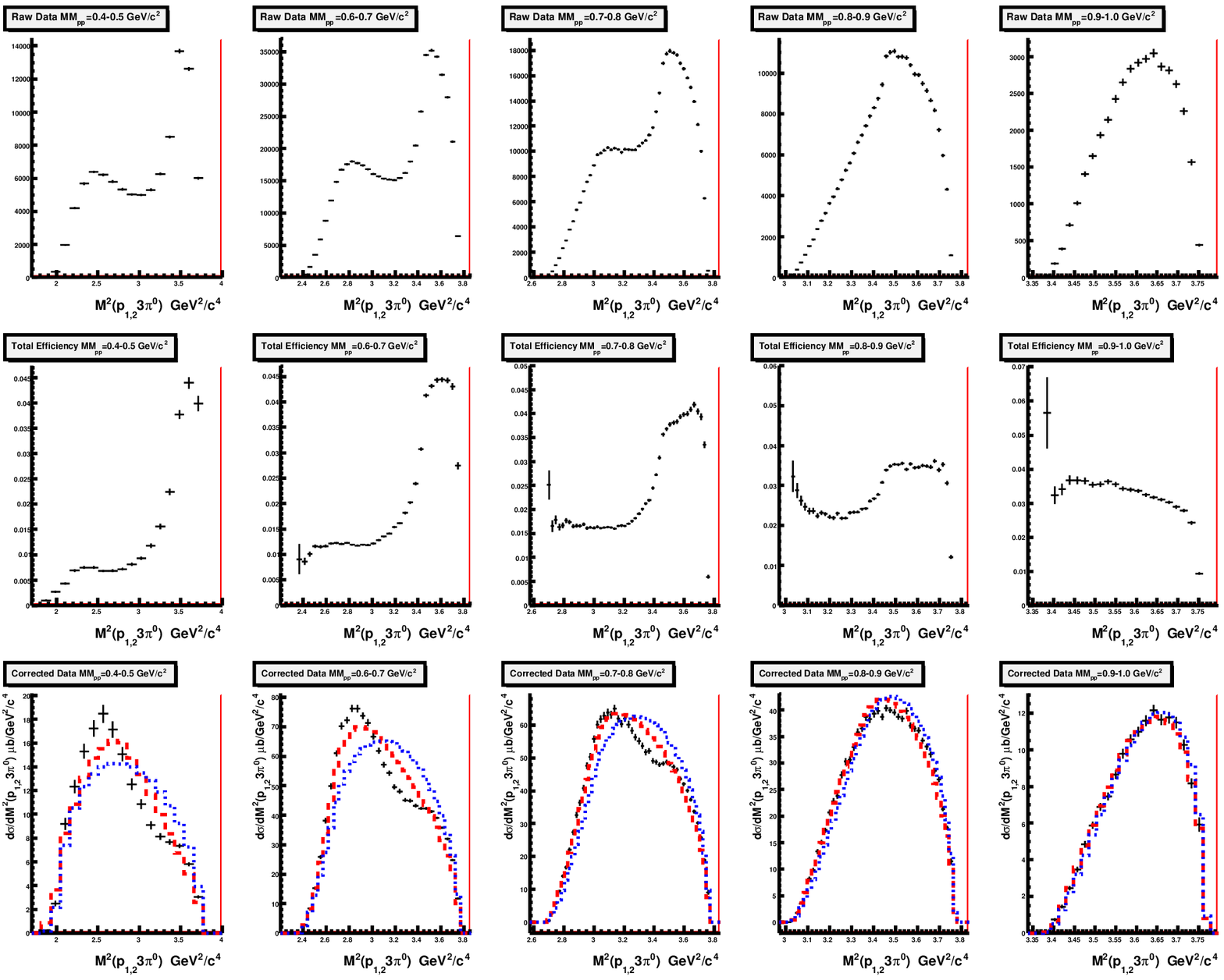}{Dalitz Plot $ppX$ projection to the $M^{2}(p3\pi^{0})$ axis.
The rows from up to down correspond to:
row~1: the experimental data, row~2: the Total Efficiency function (Eq.~\ref{eq:TotalEfficiencyFunction3pi0}),
row~3: the corrected experimental data (black),the Monte-Carlo developed model (Eq.~\ref{eq:3pi0Model}) (red),
the Monte-Carlo phase space (blue).
The columns from left to right correspond
to the following missing mass of two protons bins, column~1~$MM_{pp}=0.4-0.5\mathrm{~GeV/c^{2}}$,
column~2~$MM_{pp}=0.6-0.7\mathrm{~GeV/c^{2}}$, column~3~$MM_{pp}=0.7-0.8\mathrm{~GeV/c^{2}}$, 
column~4~$MM_{pp}=0.8-0.9\mathrm{~GeV/c^{2}}$, column~5~$MM_{pp}=0.9-1.0\mathrm{~GeV/c^{2}}$.
 The plots are symmetrized against two protons - each event is filled two times.
The models are normalized to the same number of events as in the experimental data.
It is seen that the Monte-Carlo developed model (Eq.~\ref{eq:3pi0Model}) row~3 describes the data significantly better than the phase space model.
Fully expandable and colored version of the figure is available in the attached electronic version of the thesis.
}
{Dalitz Plot $ppX$ projection to the $IM^{2}(p3\pi^{0})$ axis.}

\myFrameHugeFigure{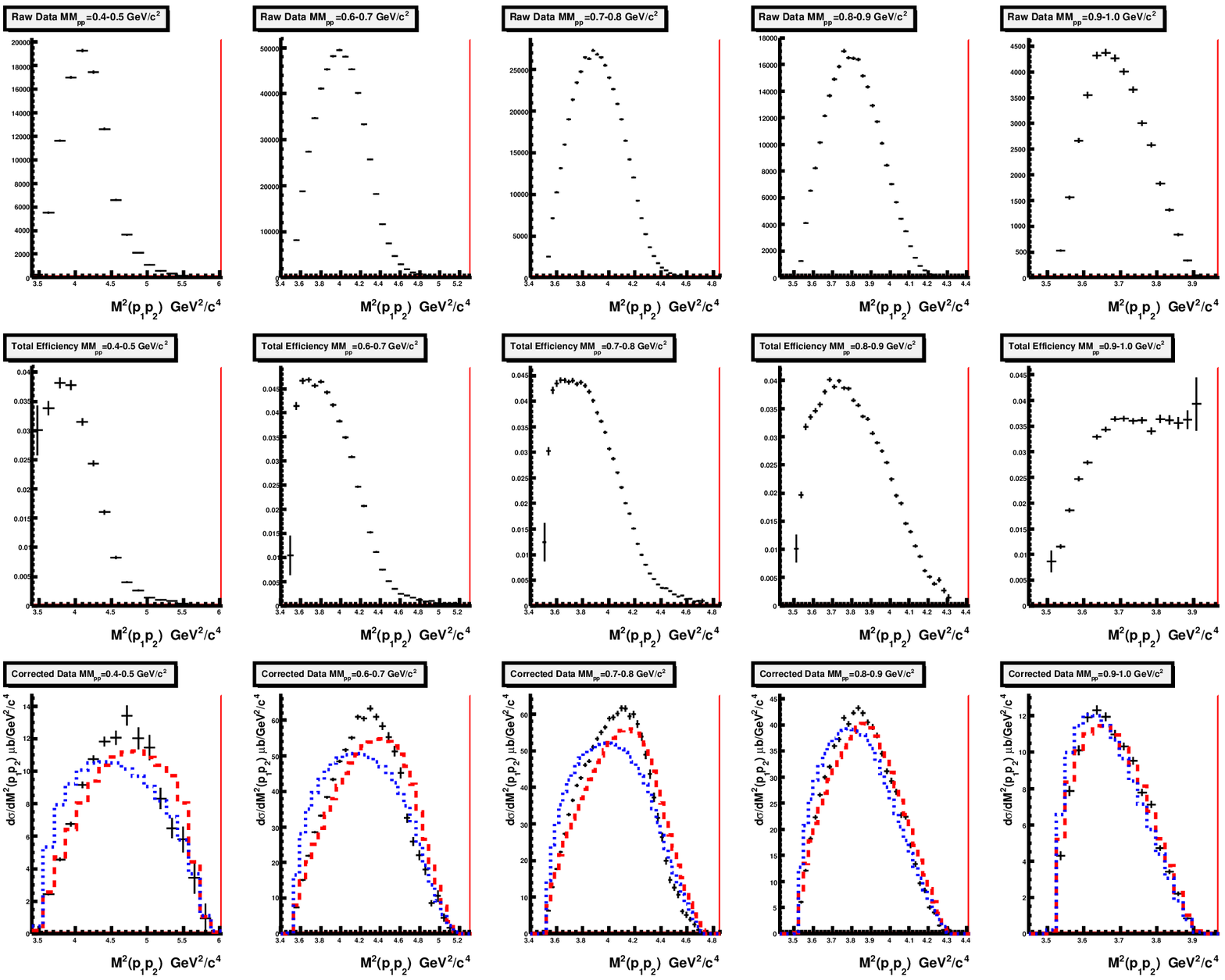}{Dalitz Plot $ppX$ projection to the $M^{2}(pp)$ axis.
The rows from up to down correspond to:
row~1: the experimental data, row~2: the Total Efficiency function (Eq.~\ref{eq:TotalEfficiencyFunction3pi0}),
row~3: the corrected experimental data (black),the Monte-Carlo developed model (Eq.~\ref{eq:3pi0Model}) (red),
the Monte-Carlo phase space (blue).
The columns from left to right correspond
to the following missing mass of two protons bins, column~1~$MM_{pp}=0.4-0.5\mathrm{~GeV/c^{2}}$,
column~2~$MM_{pp}=0.6-0.7\mathrm{~GeV/c^{2}}$, column~3~$MM_{pp}=0.7-0.8\mathrm{~GeV/c^{2}}$, 
column~4~$MM_{pp}=0.8-0.9\mathrm{~GeV/c^{2}}$, column~5~$MM_{pp}=0.9-1.0\mathrm{~GeV/c^{2}}$.
 The plots are symmetrized against two protons - each event is filled two times.
The models are normalized to the same number of events as in the experimental data.
It is seen that the Monte-Carlo developed model (Eq.~\ref{eq:3pi0Model}) row~3 describes the data significantly better than the phase space model.
Fully expandable and colored version of the figure is available in the attached electronic version of the thesis.
}
{Dalitz Plot $ppX$ projection to the $IM^{2}(pp)$ axis.}

\myFrameHugeFigure{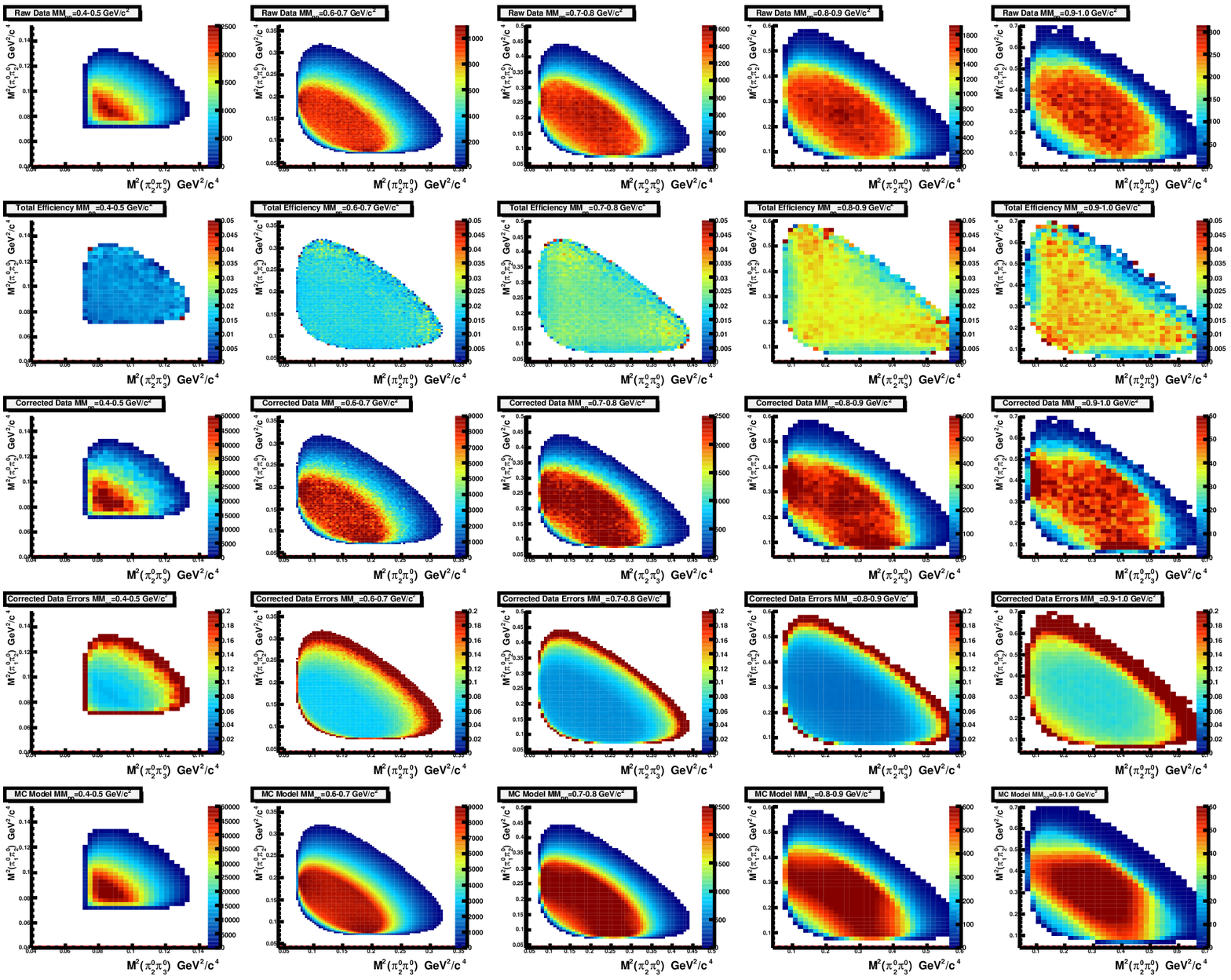}{Dalitz Plot $3\pi^{0}$. $M^{2}(2\pi^{0})$ versus $M^{2}(2\pi^{0})$. The rows from up to down correspond to:
row~1: the experimental data, row~2: the Total Efficiency function (Eq.~\ref{eq:TotalEfficiencyFunction3pi0}),
row~3: the corrected experimental data, row~4: the statistical error of the corrected data,
row~5: the Monte-Carlo developed model (Eq.~\ref{eq:3pi0Model}) . The columns from left to right correspond
to the following missing mass of two protons bins, column~1~$MM_{pp}=0.4-0.5\mathrm{~GeV/c^{2}}$,column~2~$MM_{pp}=0.6-0.7\mathrm{~GeV/c^{2}}$,
column~3~$MM_{pp}=0.7-0.8\mathrm{~GeV/c^{2}}$, column~4~$MM_{pp}=0.8-0.9\mathrm{~GeV/c^{2}}$, column~5~$MM_{pp}=0.9-1.0\mathrm{~GeV/c^{2}}$.
 The plots are symmetrized against tree pions - each event is filled six times.
The model is normalized to the same number of events as in the experimental data.
It is seen that the Monte-Carlo developed model (Eq.~\ref{eq:3pi0Model}) row~5 describes the data significantly better than the phase space model (homogeneous Dalitz plot).
Fully expandable and colored version of the figure is available in the attached electronic version of the thesis.
}{Dalitz Plot $3\pi^{0}$.}

\myFrameHugeFigure{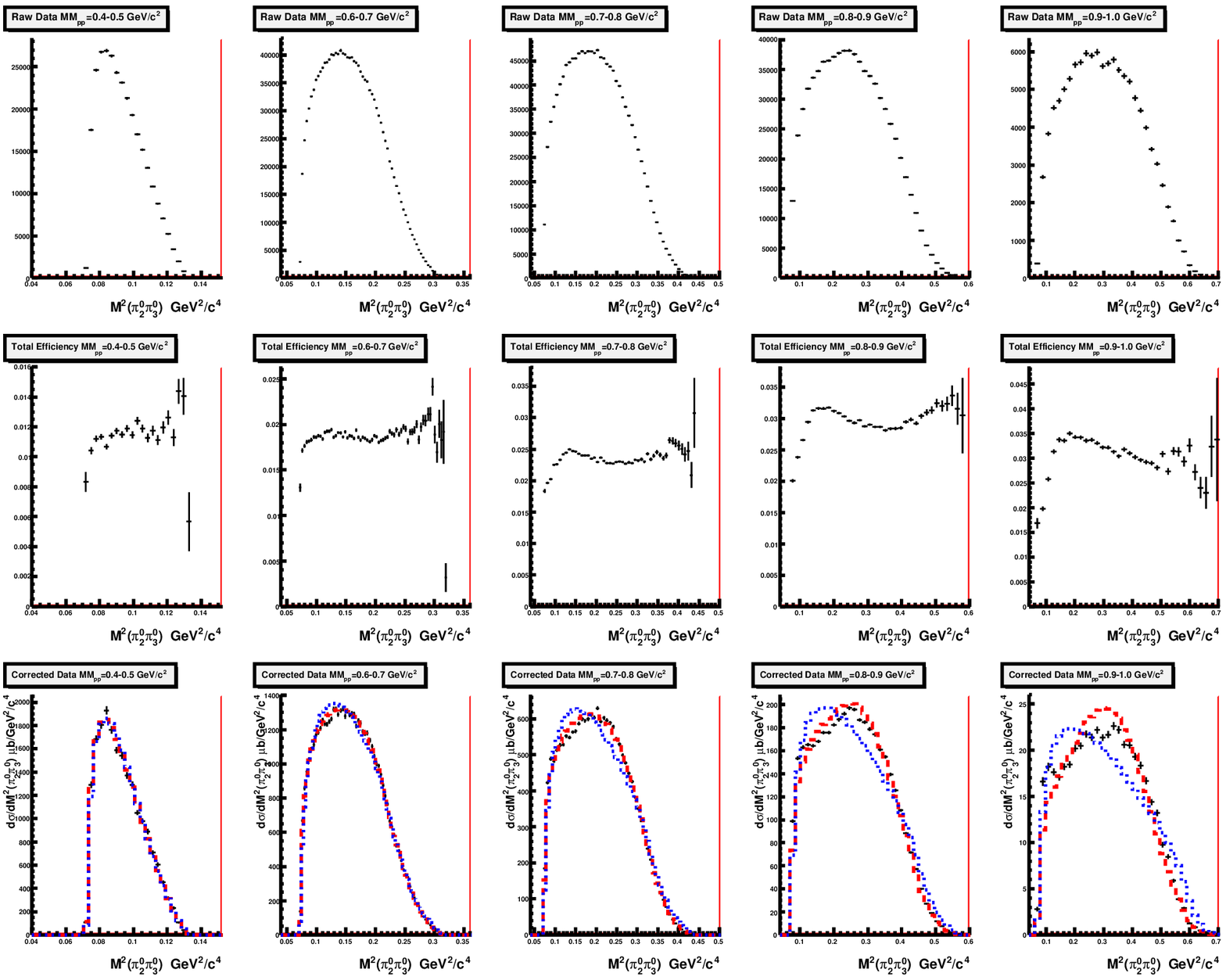}{Dalitz Plot $3\pi^{0}$ projection to the $M^{2}(2\pi^{0})$ axis.
The rows from up to down correspond to:
row~1: the experimental data, row~2: the Total Efficiency function (Eq.~\ref{eq:TotalEfficiencyFunction3pi0}),
row~3: the corrected experimental data (black),the Monte-Carlo developed model (Eq.~\ref{eq:3pi0Model}) (red),
the Monte-Carlo phase space (blue).
The columns from left to right correspond
to the following missing mass of two protons bins, column~1~$MM_{pp}=0.4-0.5\mathrm{~GeV/c^{2}}$,
column~2~$MM_{pp}=0.6-0.7\mathrm{~GeV/c^{2}}$, column~3~$MM_{pp}=0.7-0.8\mathrm{~GeV/c^{2}}$, 
column~4~$MM_{pp}=0.8-0.9\mathrm{~GeV/c^{2}}$, column~5~$MM_{pp}=0.9-1.0\mathrm{~GeV/c^{2}}$.
 The plots are symmetrized against tree pions - each event is filled six times.
The models are normalized to the same number of events as in the experimental data.
It is seen that the Monte-Carlo developed model (Eq.~\ref{eq:3pi0Model}) row~3 describes the data significantly better than the phase space model.
Fully expandable and colored version of the figure is available in the attached electronic version of the thesis.
}
{Dalitz Plot $3\pi^{0}$ projection to the $IM^{2}(2\pi^{0})$ axis.}

\myFrameHugeFigure{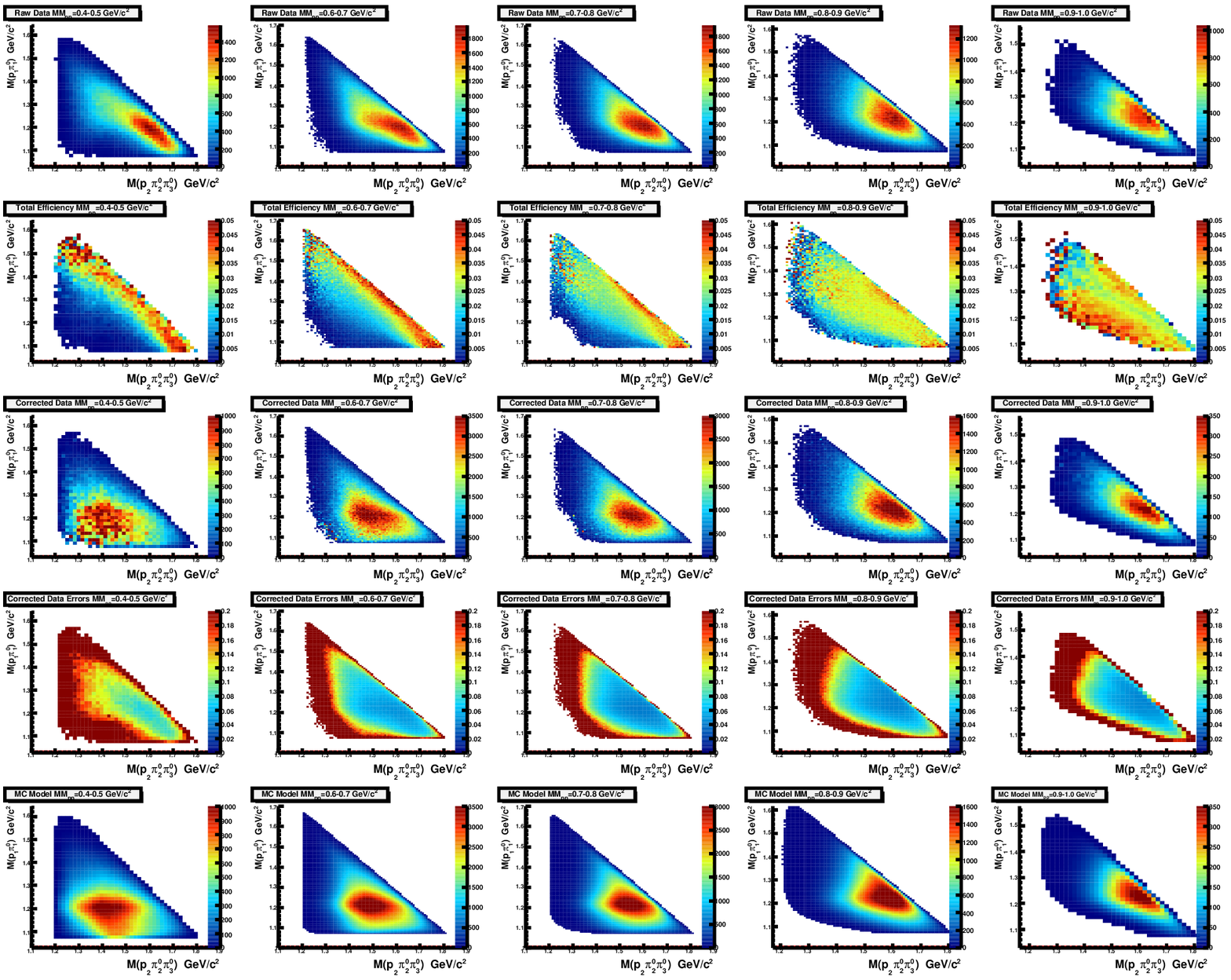}{Nyborg Plot. $M^{2}(p_{1}\pi^{0}_{1})$ versus $M(p_{2}\pi^{0}_{2}\pi^{0}_{3})$. The rows from up to down correspond to:
row~1: the experimental data, row~2: the Total Efficiency function (Eq.~\ref{eq:TotalEfficiencyFunction3pi0}),
row~3: the corrected experimental data, row~4: the statistical error of the corrected data,
row~5: the Monte-Carlo developed model (Eq.~\ref{eq:3pi0Model}). The columns from left to right correspond
to the following missing mass of two protons bins, column~1~$MM_{pp}=0.4-0.5\mathrm{~GeV/c^{2}}$,column~2~$MM_{pp}=0.6-0.7\mathrm{~GeV/c^{2}}$,
column~3~$MM_{pp}=0.7-0.8\mathrm{~GeV/c^{2}}$, column~4~$MM_{pp}=0.8-0.9\mathrm{~GeV/c^{2}}$, column~5~$MM_{pp}=0.9-1.0\mathrm{~GeV/c^{2}}$.
 The plots are symmetrized against tree pions and two protons - each event is filled six times.
The model is normalized to the same number of events as in the experimental data.
Fully expandable and colored version of the figure is available in the attached electronic version of the thesis.
}{Nyborg Plot.}

\myFrameHugeFigure{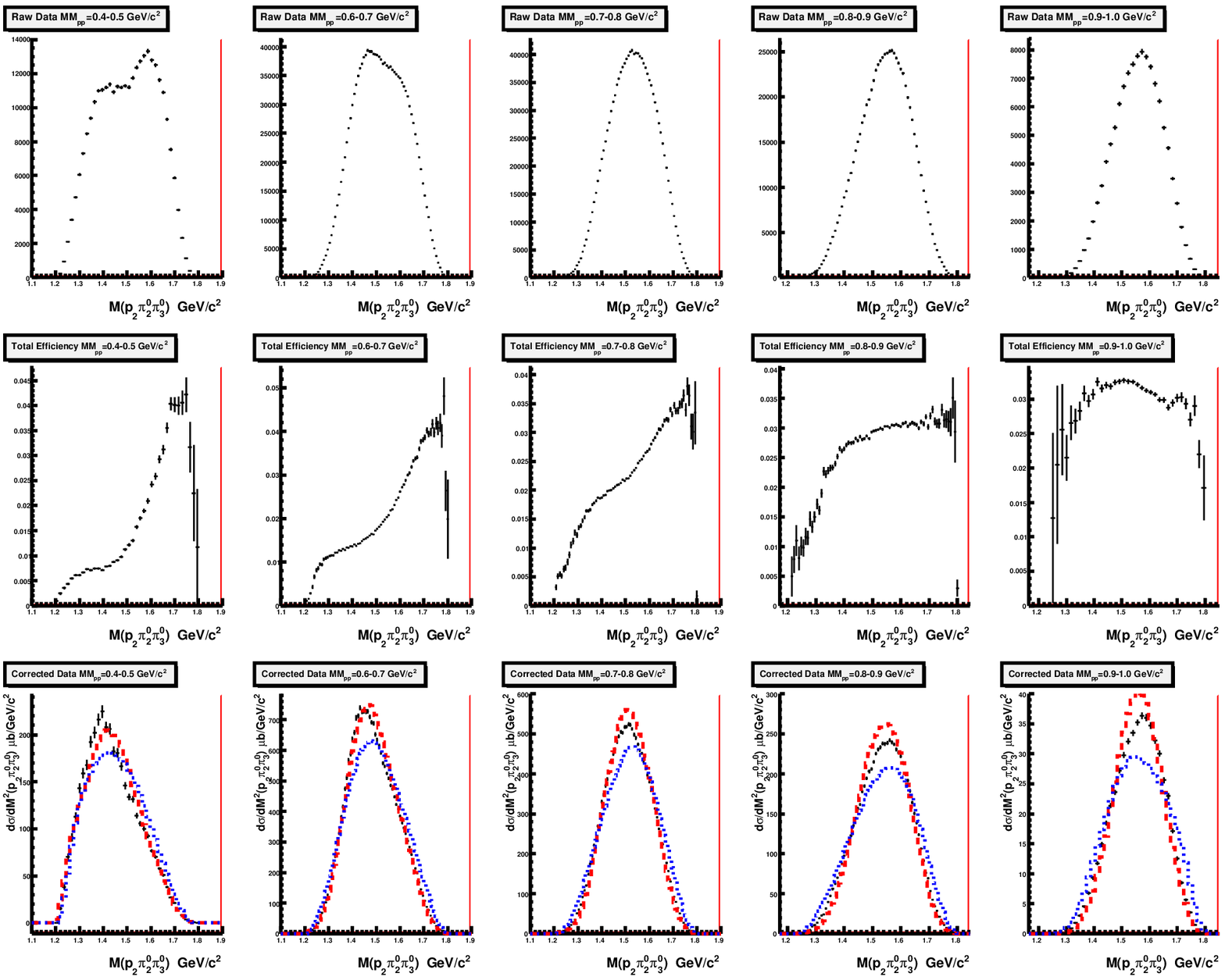}{Nyborg Plot projection to the $M(p_{2}\pi^{0}_{2}\pi^{0}_{3})$ axis.
The rows from up to down correspond to:
row~1: the experimental data, row~2: the Total Efficiency function (Eq.~\ref{eq:TotalEfficiencyFunction3pi0}),
row~3: the corrected experimental data (black),the Monte-Carlo developed model (Eq.~\ref{eq:3pi0Model}) (red),
the Monte-Carlo phase space (blue).
The columns from left to right correspond
to the following missing mass of two protons bins, column~1~$MM_{pp}=0.4-0.5\mathrm{~GeV/c^{2}}$,
column~2~$MM_{pp}=0.6-0.7\mathrm{~GeV/c^{2}}$, column~3~$MM_{pp}=0.7-0.8\mathrm{~GeV/c^{2}}$, 
column~4~$MM_{pp}=0.8-0.9\mathrm{~GeV/c^{2}}$, column~5~$MM_{pp}=0.9-1.0\mathrm{~GeV/c^{2}}$.
 The plots are symmetrized against tree pions and two protons - each event is filled six times.
The models are normalized to the same number of events as in the experimental data.
It is seen that the Monte-Carlo developed model (Eq.~\ref{eq:3pi0Model}) row~3 describes the data significantly better than the phase space model.
Fully expandable and colored version of the figure is available in the attached electronic version of the thesis.
}
{Nyborg Plot projection to the $IM(p_{2}\pi^{0}_{2}\pi^{0}_{3})$ axis.}

\myFrameHugeFigure{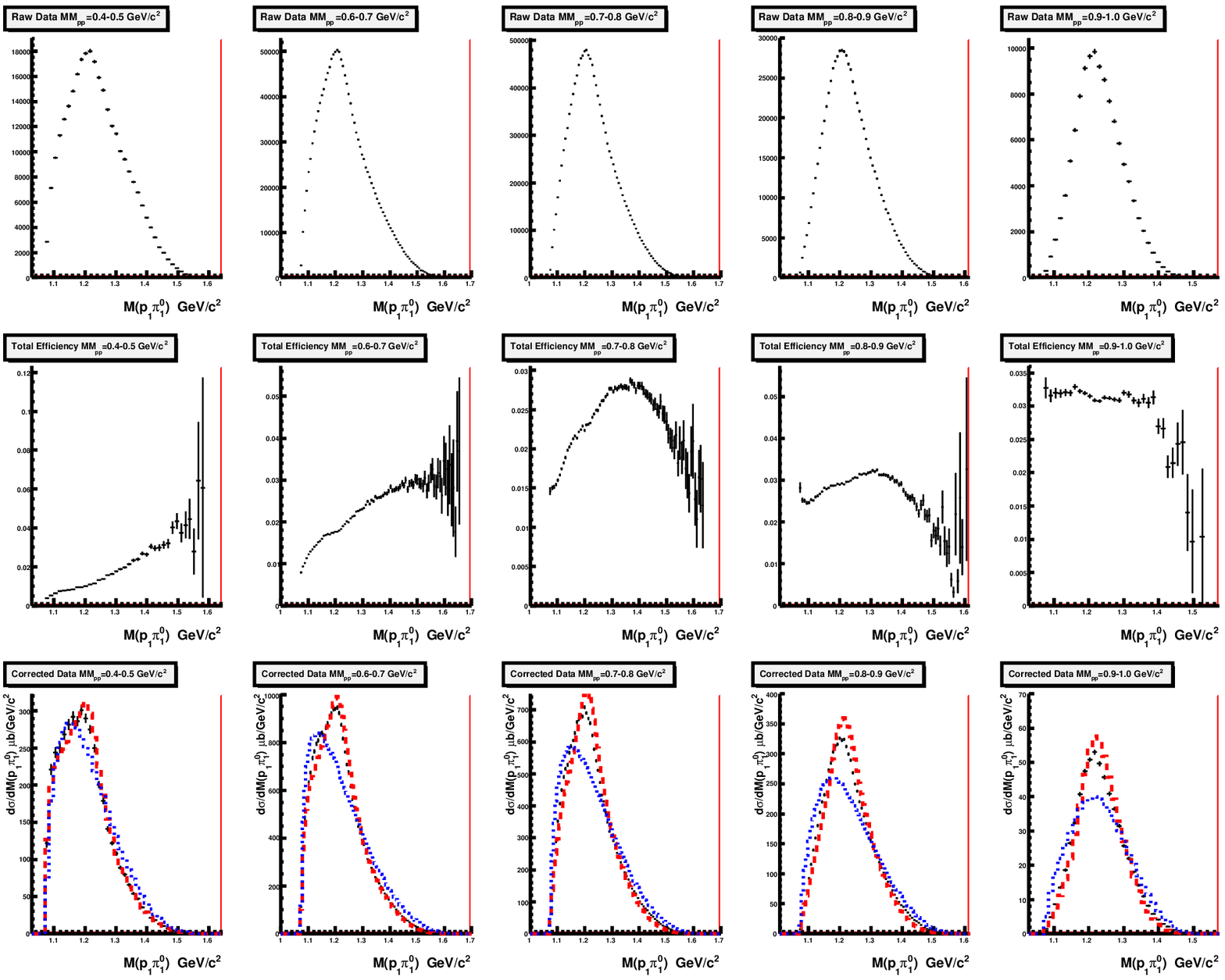}{Nyborg Plot projection to the $M^{2}(p_{1}\pi^{0}_{1})$ axis.
The rows from up to down correspond to:
row~1: the experimental data, row~2: the Total Efficiency function (Eq.~\ref{eq:TotalEfficiencyFunction3pi0}),
row~3: the corrected experimental data (black),the Monte-Carlo developed model (Eq.~\ref{eq:3pi0Model}) (red),
the Monte-Carlo phase space (blue).
The columns from left to right correspond
to the following missing mass of two protons bins, column~1~$MM_{pp}=0.4-0.5\mathrm{~GeV/c^{2}}$,
column~2~$MM_{pp}=0.6-0.7\mathrm{~GeV/c^{2}}$, column~3~$MM_{pp}=0.7-0.8\mathrm{~GeV/c^{2}}$, 
column~4~$MM_{pp}=0.8-0.9\mathrm{~GeV/c^{2}}$, column~5~$MM_{pp}=0.9-1.0\mathrm{~GeV/c^{2}}$.
 The plots are symmetrized against tree pions and two protons - each event is filled six times.
The models are normalized to the same number of events as in the experimental data.
It is seen that the Monte-Carlo developed model (Eq.~\ref{eq:3pi0Model}) row~3 describes the data significantly better than the phase space model.
Fully expandable and colored version of the figure is available in the attached electronic version of the thesis.
}
{Nyborg Plot projection to the $IM^{2}(p_{1}\pi^{0}_{1})$ axis.}

\newpage ~
\thispagestyle{empty}
\emptydoublepage
\newpage
\subsection{The $pp \rightarrow pp \eta(3\pi^{0})$ reaction}
The $pp \rightarrow pp \eta$ reaction at an incident proton momentum $3.35\mathrm{~GeV/c}$ ($T=2.541\mathrm{~GeV}$), which corresponds to the excess energy $Q=455\mathrm{~MeV}$,
was also measured via $\eta$ meson decay into three neutral pions.
All final state particles were detected, the signatures of the two protons were registered in the Forward Detector of the WASA, while the three pions were reconstructed from the decay into six photons 
in the Electromagnetic Calorimeter (see Section~\ref{sec:AnalysisExperimental}) the same way as in the case of the $pp \rightarrow pp 3\pi^{0}$ reaction (see Section~\ref{sec:PP3pi0reaction}).

Many previous high statistics experimental studies of the reaction dynamics in the threshold region \cite{ETACS02,ETACS03,ETACS04,EtaCosy11}, for beam kinetic energies less than $2\mathrm{~GeV}$, show an important role of the $N^{*}(1535)$ baryon resonance.
The near threshold data were interpreted mostly in the framework of the one-boson exchange models \cite{OBE1,OBE2,OBE3,OBE4,OBE5,OBE6,OBE7,OBELast} (by exchange of various light mesons like $\pi, \eta, \rho, \omega$) and a dominant role of the resonance $N^{*}(1535)S_{11}$. 
In the threshold region also the Final State Interaction plays an important role \cite{3pi0CSmodel3, 3pi0CSmodel2}.

For higher energies above $2\mathrm{~GeV}$ beam kinetic energy, there are only few studies  \cite{DISTO, HadesData}
which consider  two dominant production mechanisms: the resonant production (via excitation of $N^{*}(1535)$) and the non resonant production.
%

\paragraph{Available phase space and the observables\\}

\label{sec:etaprod}
\begin{figure}[ht!bp]
\centering
{
\subfigure[Experimental Data (black), Monte-Carlo Phase Space (blue), the Model (Eq.~\ref{eq:3pi0Model}) (red).]{\fbox{\includegraphics[width=0.7\textwidth]{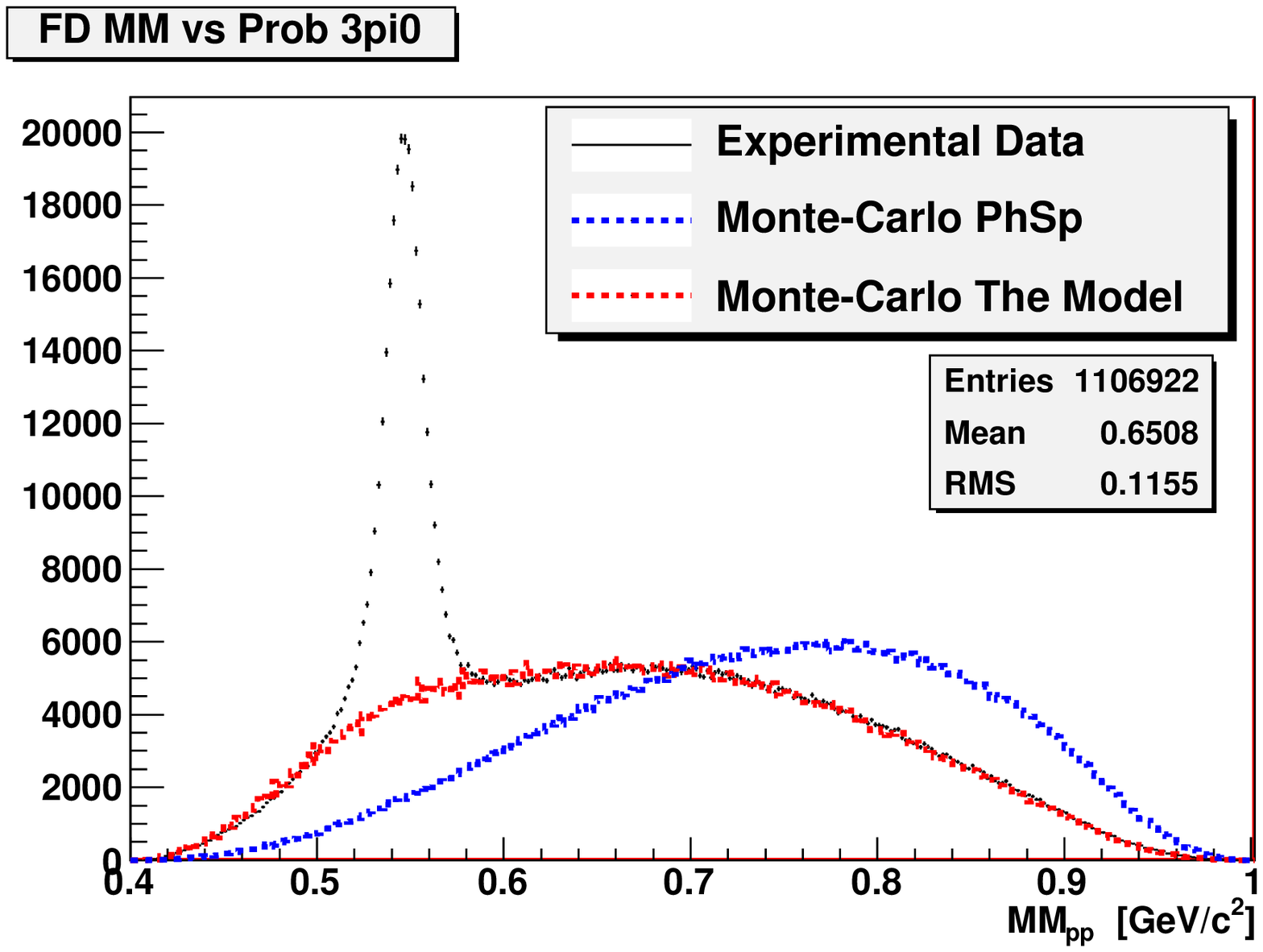}} \label{fig:EtaEvents1}}\\
\subfigure[Subtracted prompt background. Experimental Data (black), Monte-Carlo simulation (blue).]{\fbox{\includegraphics[width=0.7\textwidth]{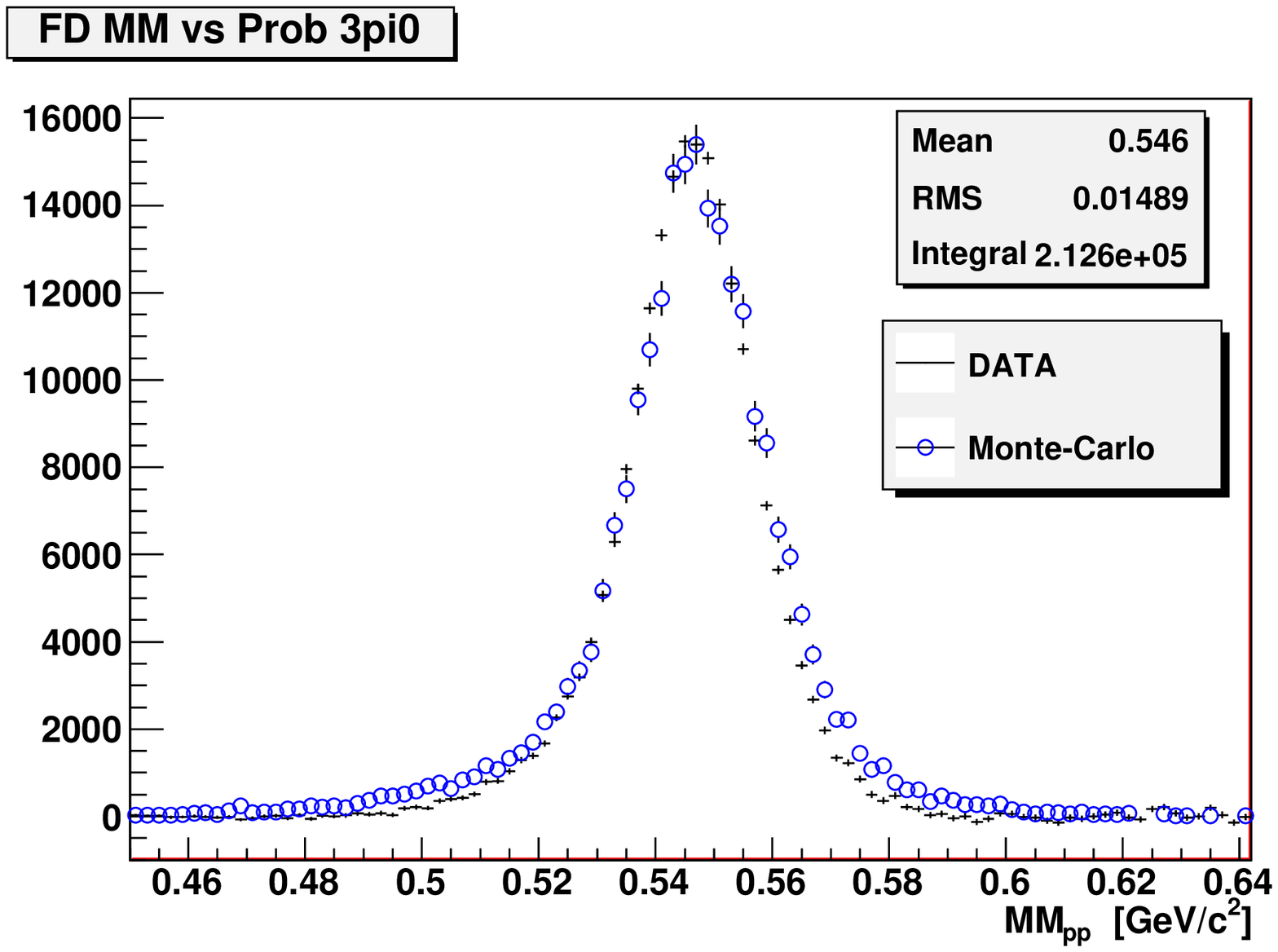}} \label{fig:EtaEvents2}}
}
\caption{Missing Mass of the two protons.}
\label{fig:EtaEvents}
\end{figure}

After the background subtraction around $200k$ events of the $pp \rightarrow pp \eta$ are available (Fig.~\ref{fig:EtaEvents}).  
The data analysis was performed as follows.

\begin{sidewaysfigure}
\centering
{
\subfigure[$N^{*}(1535)$ true events]{\fbox{\includegraphics[width=0.45\textwidth]{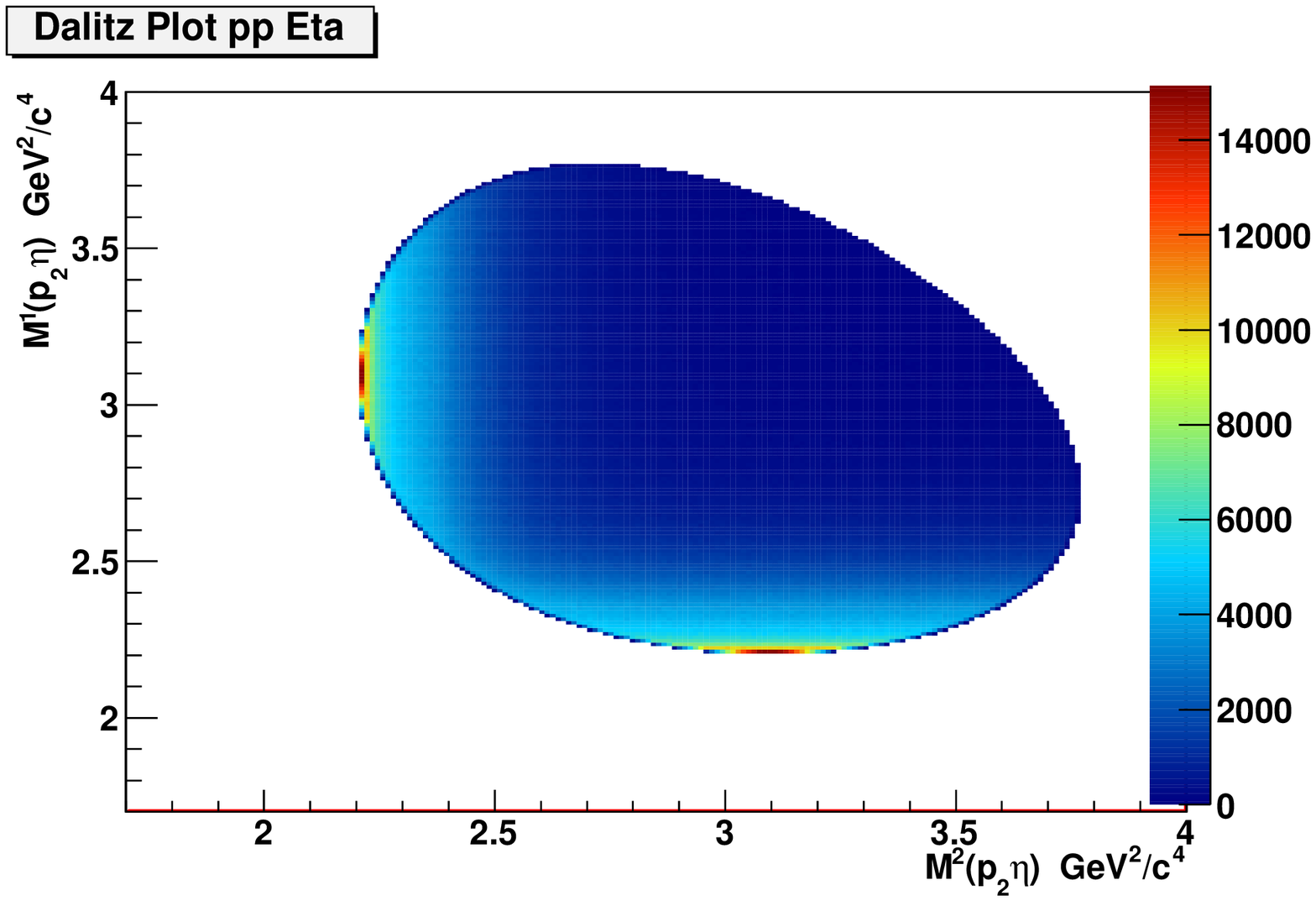}} \label{fig:DalppEtaMC1}}\quad
\subfigure[Phase Space true events]{\fbox{\includegraphics[width=0.45\textwidth]{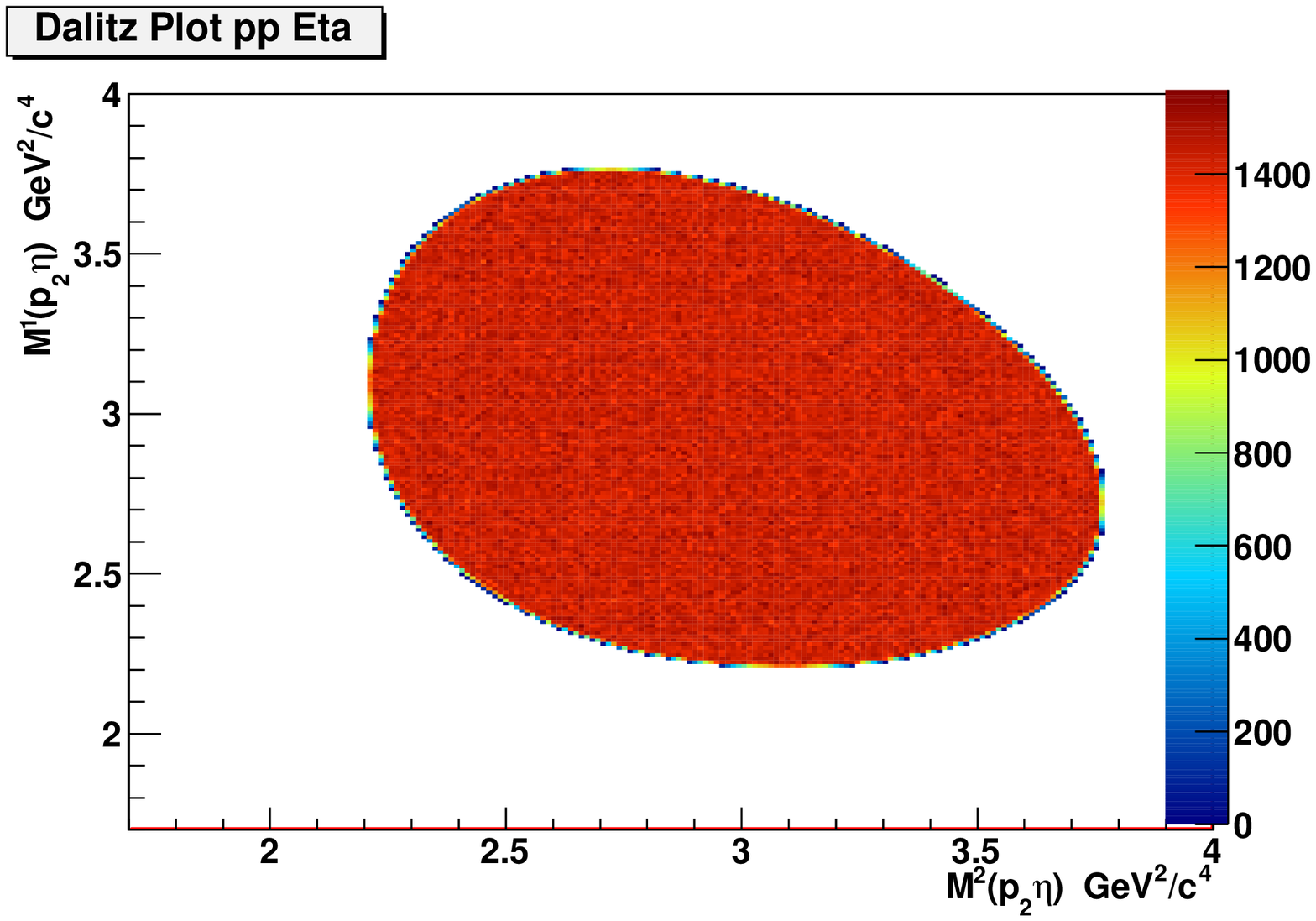}} \label{fig:DalppEtaMC2}}\\
\subfigure[$N^{*}(1535)$ reconstructed events]{\fbox{\includegraphics[width=0.45\textwidth]{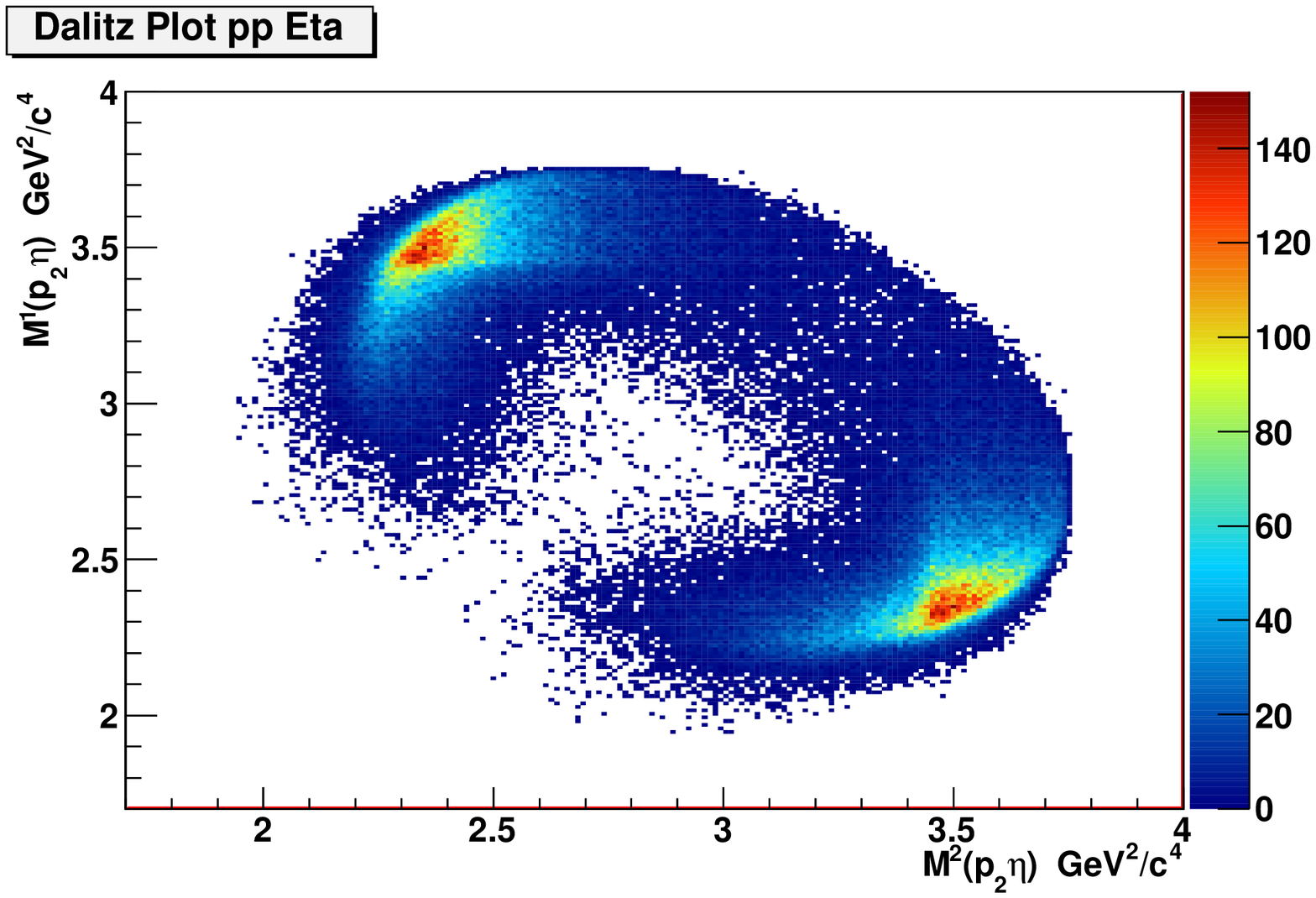}} \label{fig:DalppEtaMC3}}\quad
\subfigure[[Phase Space reconstructed events]{\fbox{\includegraphics[width=0.45\textwidth]{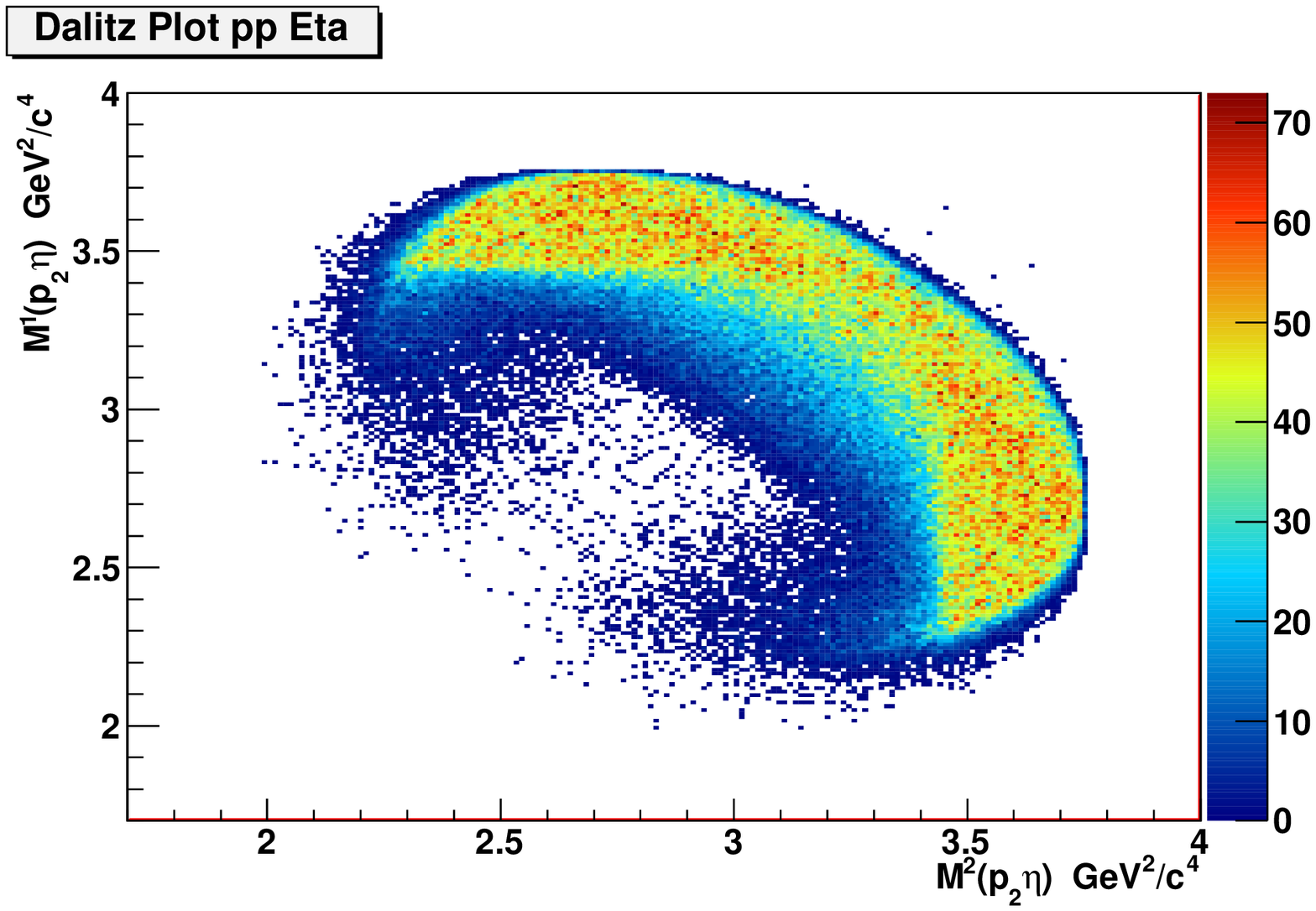}} \label{fig:DalppEtaMC4}}

}
\caption{Monte-Carlo, Dalitz Plot $pp\eta$ - $IM^{2}(p_{1}\eta)$ versus $IM^{2}(p_{2}\eta)$.
The plot is symmetrized against two protons - each event is filled two times.}
\label{fig:DalppEtaMC}
\end{sidewaysfigure}

First the experimental accessibility of the phase space area was studied. Dalitz plot $pp\eta$ for the Monte-Carlo simulation based
on a homogeneous and isotropic phase space population and a production via excitation of $N^{*}(1535)$, were prepared. 
One finds that the acceptance for this reaction is limited (Fig.~\ref{fig:DalppEtaMC}).

\begin{figure}[ht!bp]
\centering
{
\subfigure[true events]{\fbox{\includegraphics[width=0.7\textwidth]{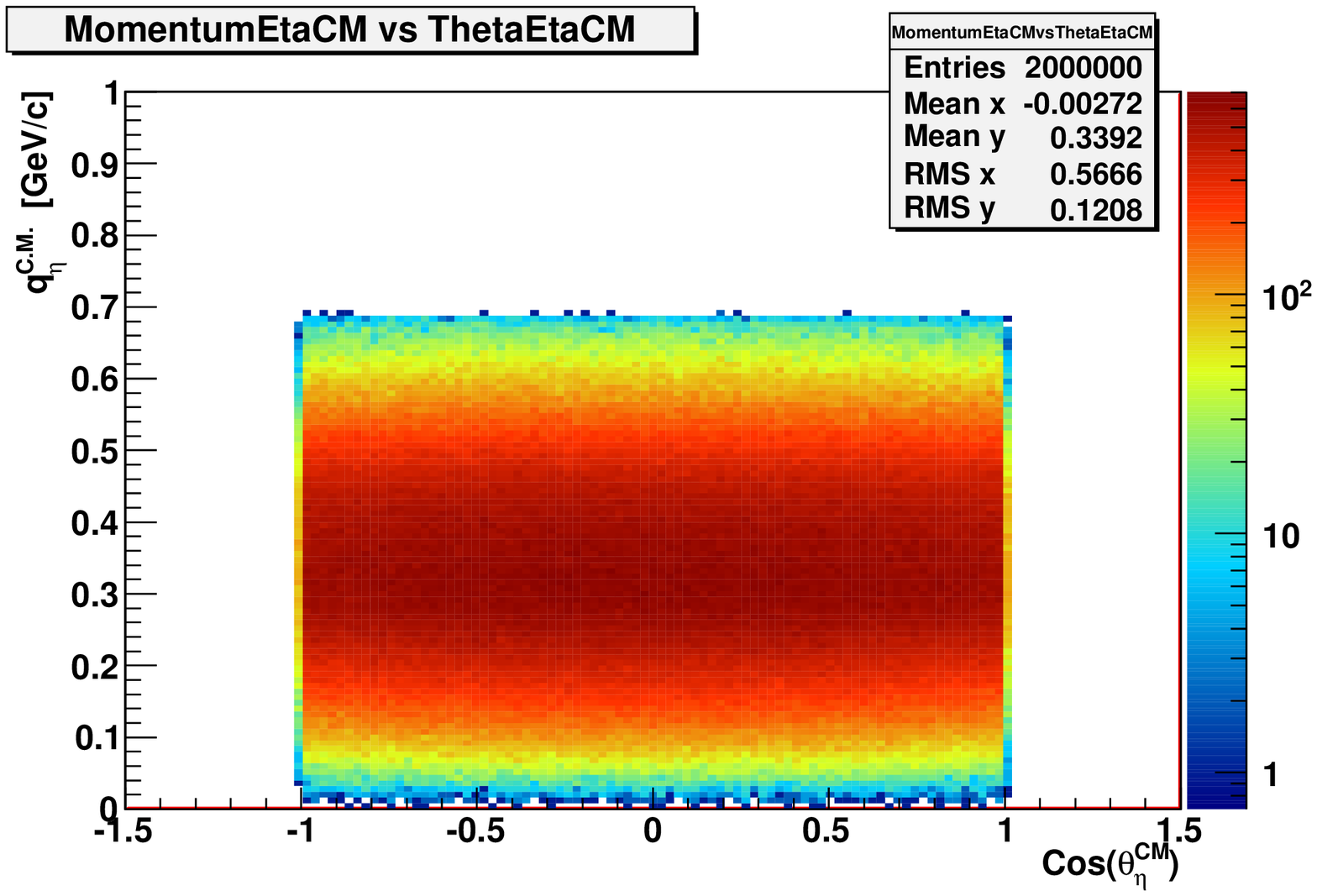}} \label{fig:EtaCMsystemMC1}}\\
\subfigure[reconstructed events]{\fbox{\includegraphics[width=0.7\textwidth]{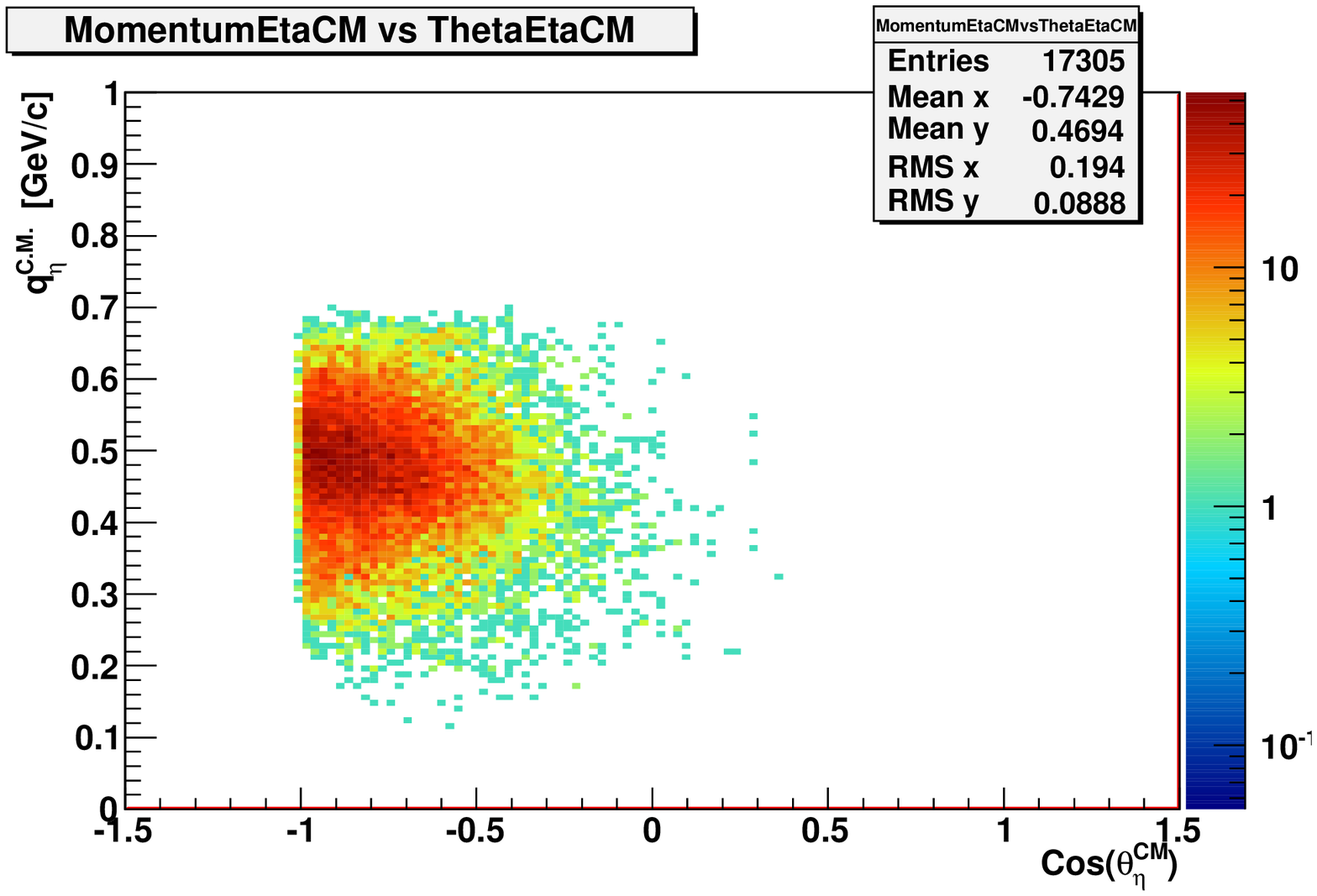}} \label{fig:EtaCMsystemMC2}}
}
\caption{Monte-Carlo simulation, $\eta$ momentum versus $\cos(\theta_{\eta^{CM}})$ in the Center of Mass system.}
\label{fig:EtaCMsystemMC}
\end{figure}

\begin{figure}[ht!bp]
\centering
{
\subfigure[true events]{\fbox{\includegraphics[width=0.7\textwidth]{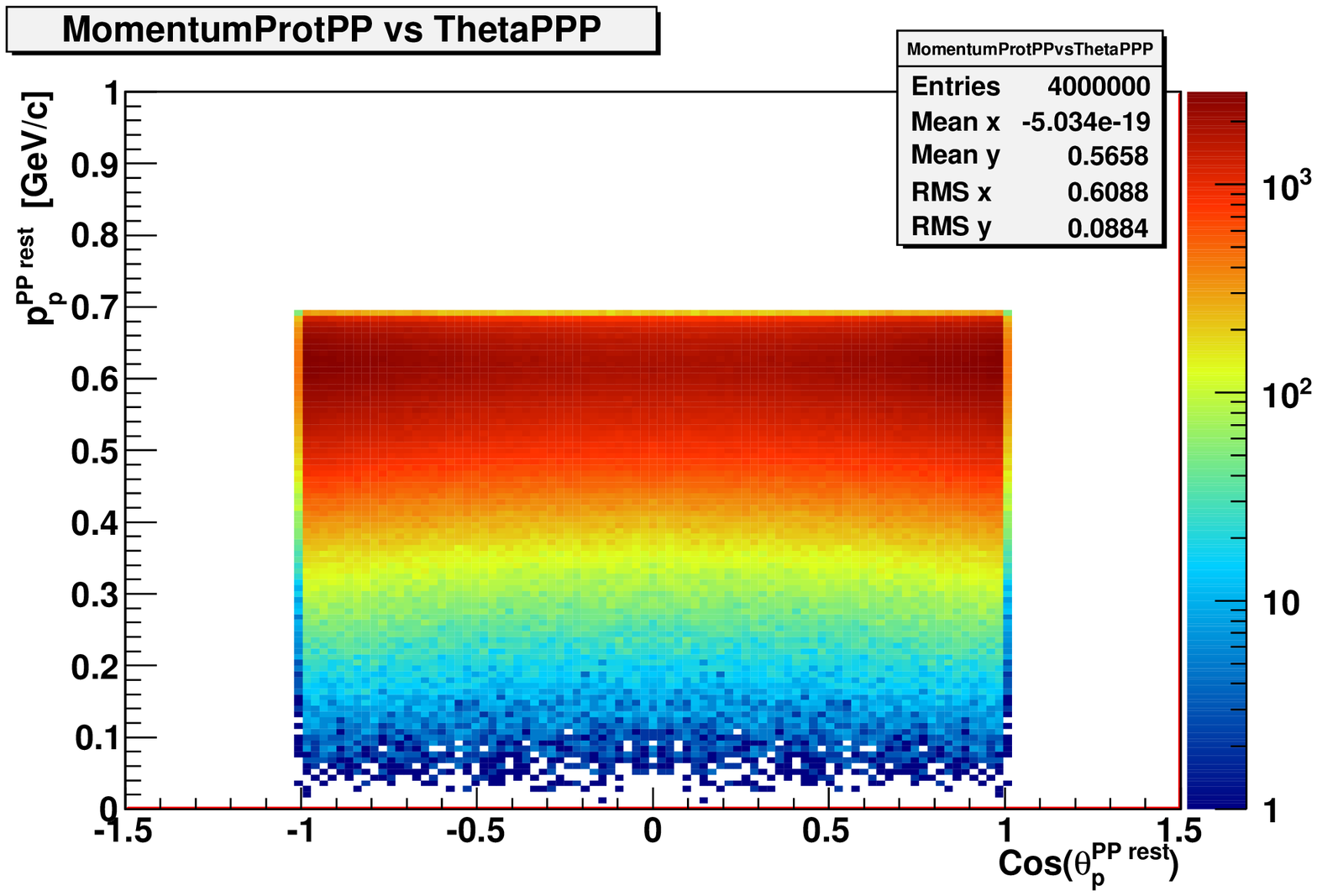}} \label{fig:EtaPPsystemMC1}}\\
\subfigure[reconstructed events]{\fbox{\includegraphics[width=0.7\textwidth]{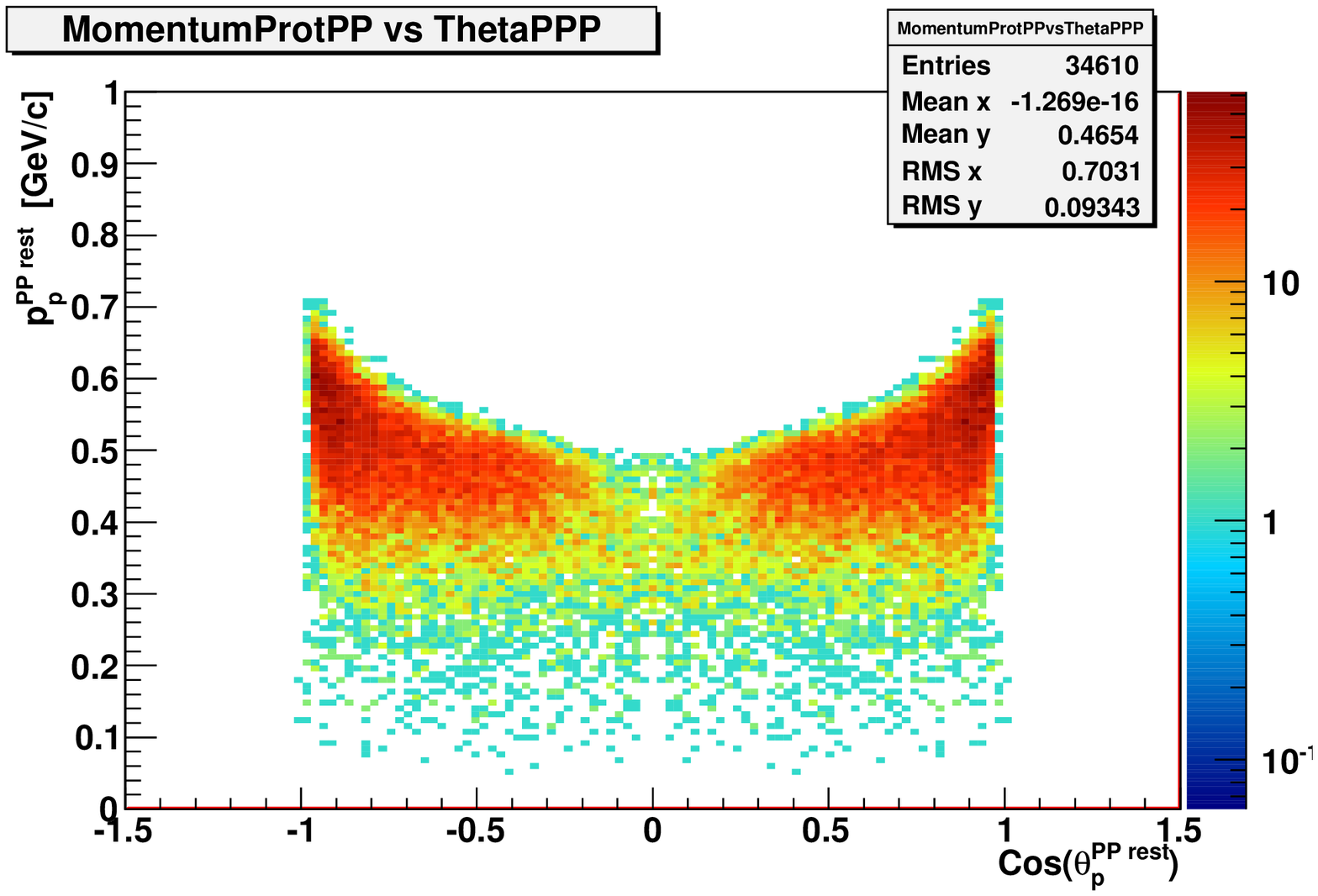}} \label{fig:EtaPPsystemMC2}}
}
\caption{Monte-Carlo simulation, proton momentum versus $\cos(\theta_{p^{pp}})$ in the proton-proton rest frame.}
\label{fig:EtaPPsystemMC}
\end{figure}

\myFrameSmallFigure{EtaMomentumAcceptance}{Monte-Carlo Simulation Phase Space coverage, $\eta$ momentum
 in the CM system, true events (blue), reconstructed events (red)}{Monte-Carlo Simulation Phase Space coverage.}

Since one wants to study the dynamics of this process it is convenient to introduce the following variables,
 accordingly to the \cite{DISTO}.
The momentum of the $\eta$ meson in the Center of Mass $q_{\eta}^{CM}$; the angle between the beam direction and the $\eta$ meson
in the Center of Mass $\theta_{\eta}^{CM}$;
the momentum of the proton in proton-proton rest frame $p_{p}^{PP}$, the angle between the beam direction and the proton direction
in the proton-proton rest frame $\theta_{p}^{PP}$.
Looking into the acceptance in four new defined variables, by comparing the Monte-Carlo simulation for true events with
 the reconstructed ones (Figs.~\ref{fig:EtaCMsystemMC},~\ref{fig:EtaPPsystemMC},~\ref{image_EtaMomentumAcceptance}), one concludes that only high $\eta$
 momenta could be measured. For the further studies the following region was selected:

\begin{eqnarray}
 &q_{\eta}^{CM}&=0.45-0.7~GeV/c\nonumber \\
&\cos(\theta_{\eta^{CM}})&=-1.0-0.0
\label{eq:EtaAcceptanceCut}
\end{eqnarray}

\myFrameFigure{EtacorrelationMomentum}{Monte-Carlo simulation, reconstructed events. Correlation between $\eta$ momentum in CM system $q_{\eta}^{CM}$
and proton momentum in proton-proton rest frame $p_{p}^{PP}$. The broadening of the line corresponds to the detector resolution 
effect.}{Monte-Carlo simulation momentum correlations.}

The proton momentum in proton-proton rest frame $p_{p}^{PP}$ is correlated with the $\eta$
 momentum in CM system $q_{\eta}^{CM}$ \myImgRef{EtacorrelationMomentum}. The selection of the range in $q_{\eta}^{CM}$ implies automatically
the selection of the range in $p_{p}^{PP}$.   

As the initial state is symmetric i.e. two protons, the resulting angular distributions
 should be symmetric around $90\mathrm{~deg}$. To get the full information one needs to measure only the half of the distribution, the other part is a reflection.
For this studies the range from $(-1,0)$ in the $\cos(\theta_{\eta^{CM}})$ and the $\cos(\theta_{p^{pp}})$ were selected.

\begin{figure}[ht!bp]
\centering
{
\subfigure[True minus reconstructed value as a function of the reconstructed value of the $\cos(\theta_{\eta^{CM}})$.]{\fbox{\includegraphics[width=0.7\textwidth]{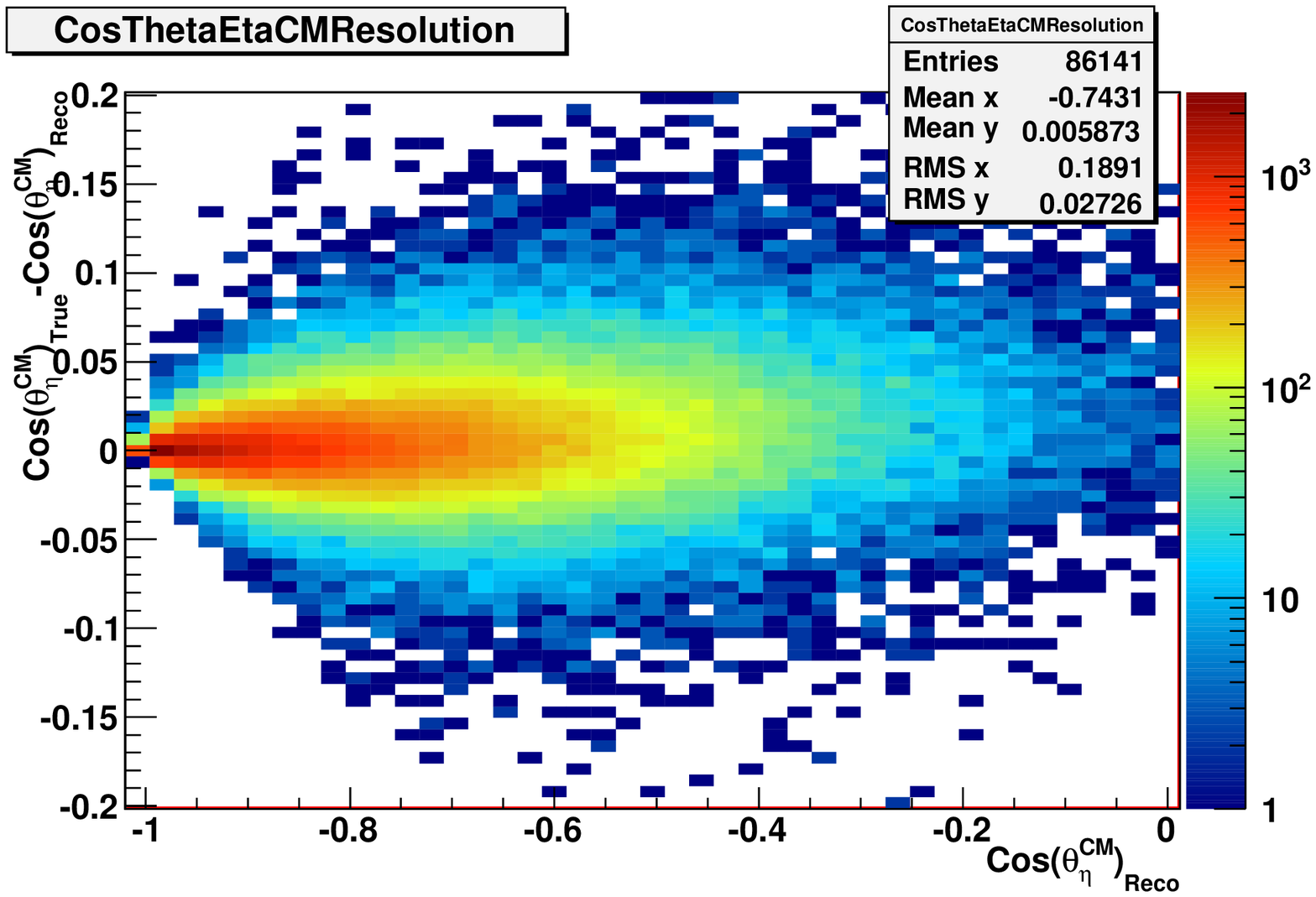}} \label{fig:EtaCMsystemMCresolution1}}\\
\subfigure[The $\sigma$ parameter of the fitted Gaussian peak as a function of the reconstructed $\cos(\theta_{\eta^{CM}})$.]{\fbox{\includegraphics[width=0.7\textwidth]{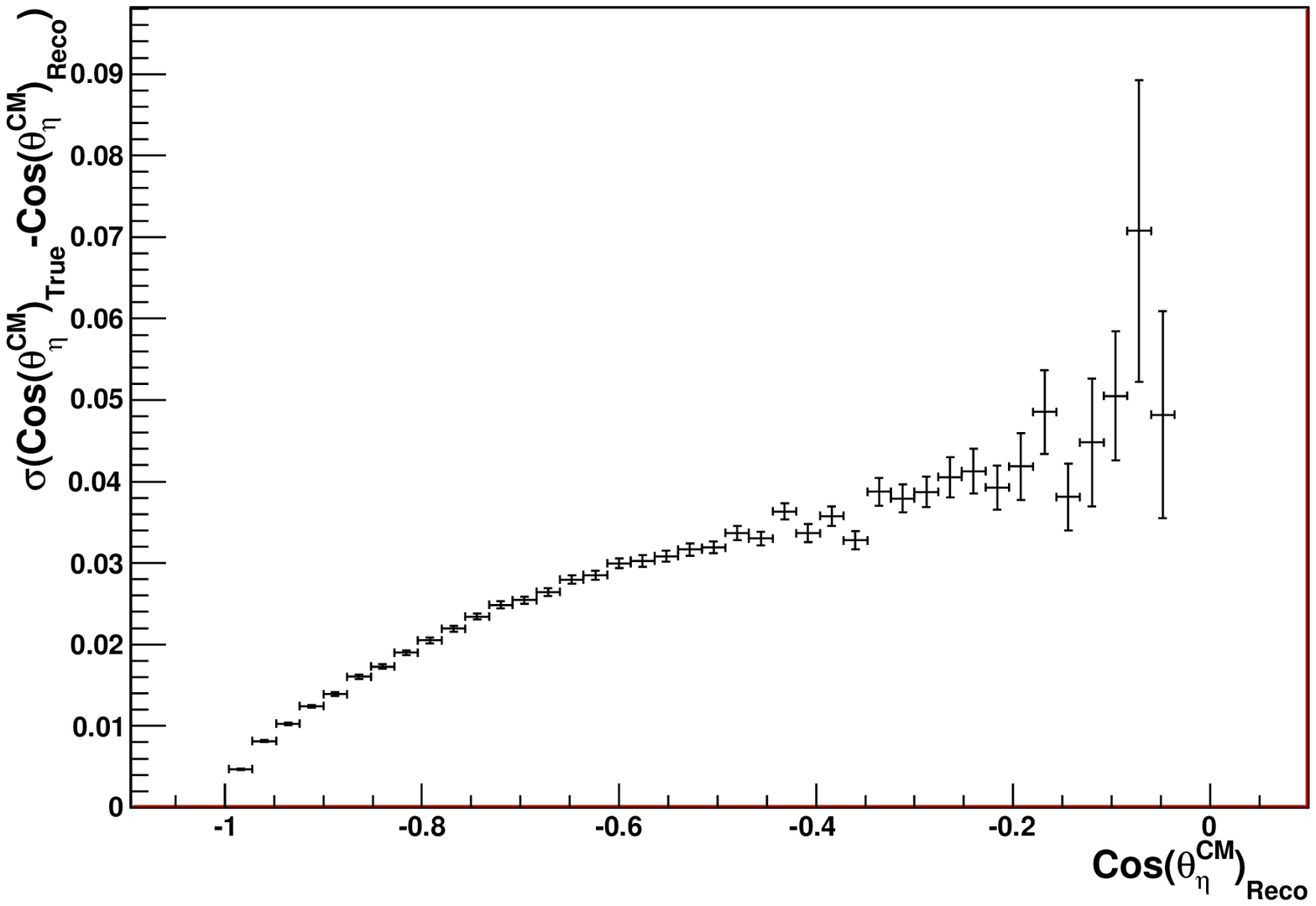}} \label{fig:EtaCMsystemMCresolution2}}
}
\caption{Monte-Carlo simulation, resolution studies of the $\cos(\theta_{\eta^{CM}})$.}
\label{fig:EtaCMsystemMCresolution}
\end{figure}

\begin{figure}[ht!bp]
\centering
{
\subfigure[True minus reconstructed value as a function of the reconstructed value of the $\cos(\theta_{p^{pp}})$.]{\fbox{\includegraphics[width=0.7\textwidth]{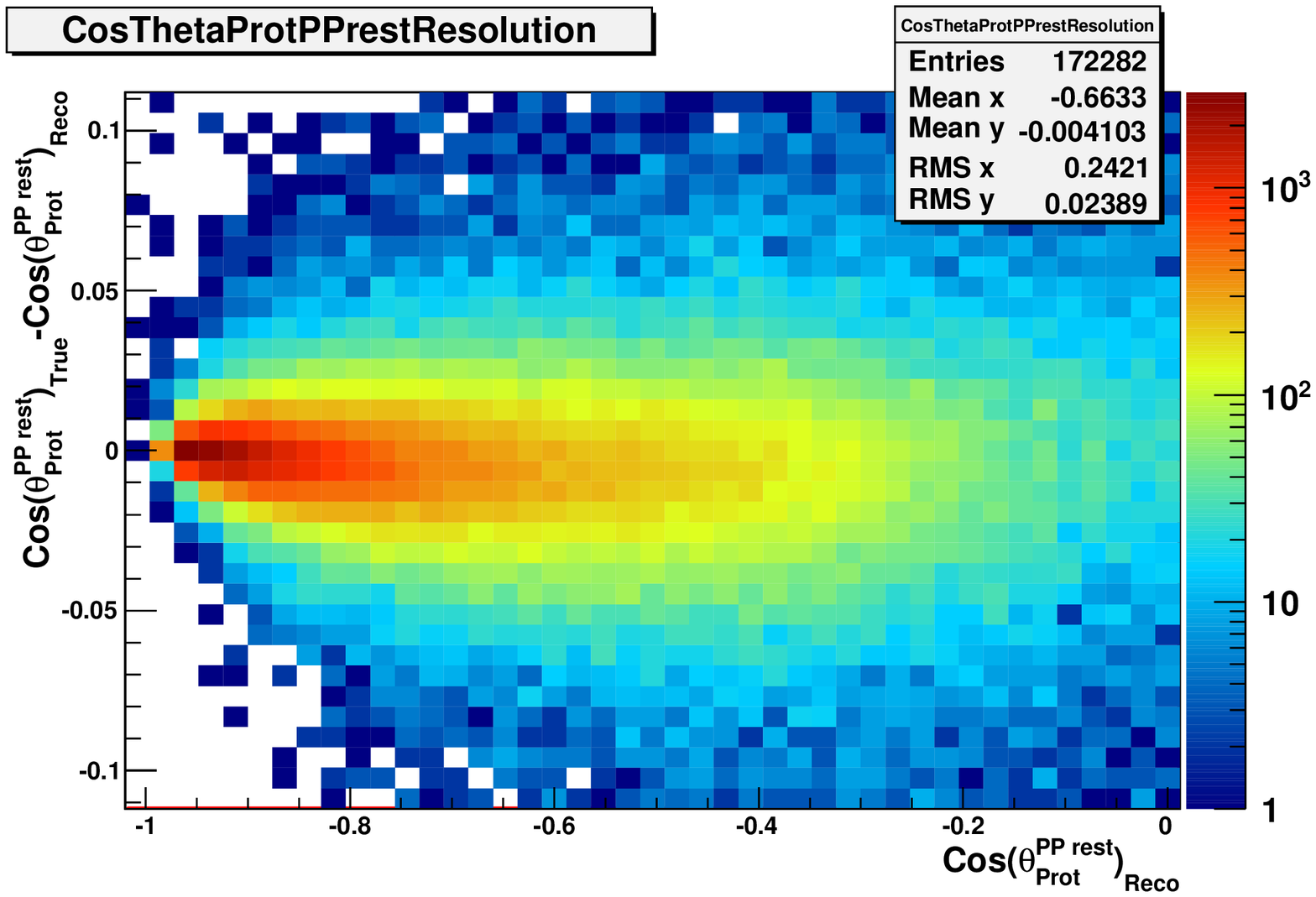}} \label{fig:EtaPPsystemMCresolution1}}\\
\subfigure[The $\sigma$ parameter of the fitted Gaussian peak as a function of the reconstructed $\cos(\theta_{p^{pp}})$.]{\fbox{\includegraphics[width=0.7\textwidth]{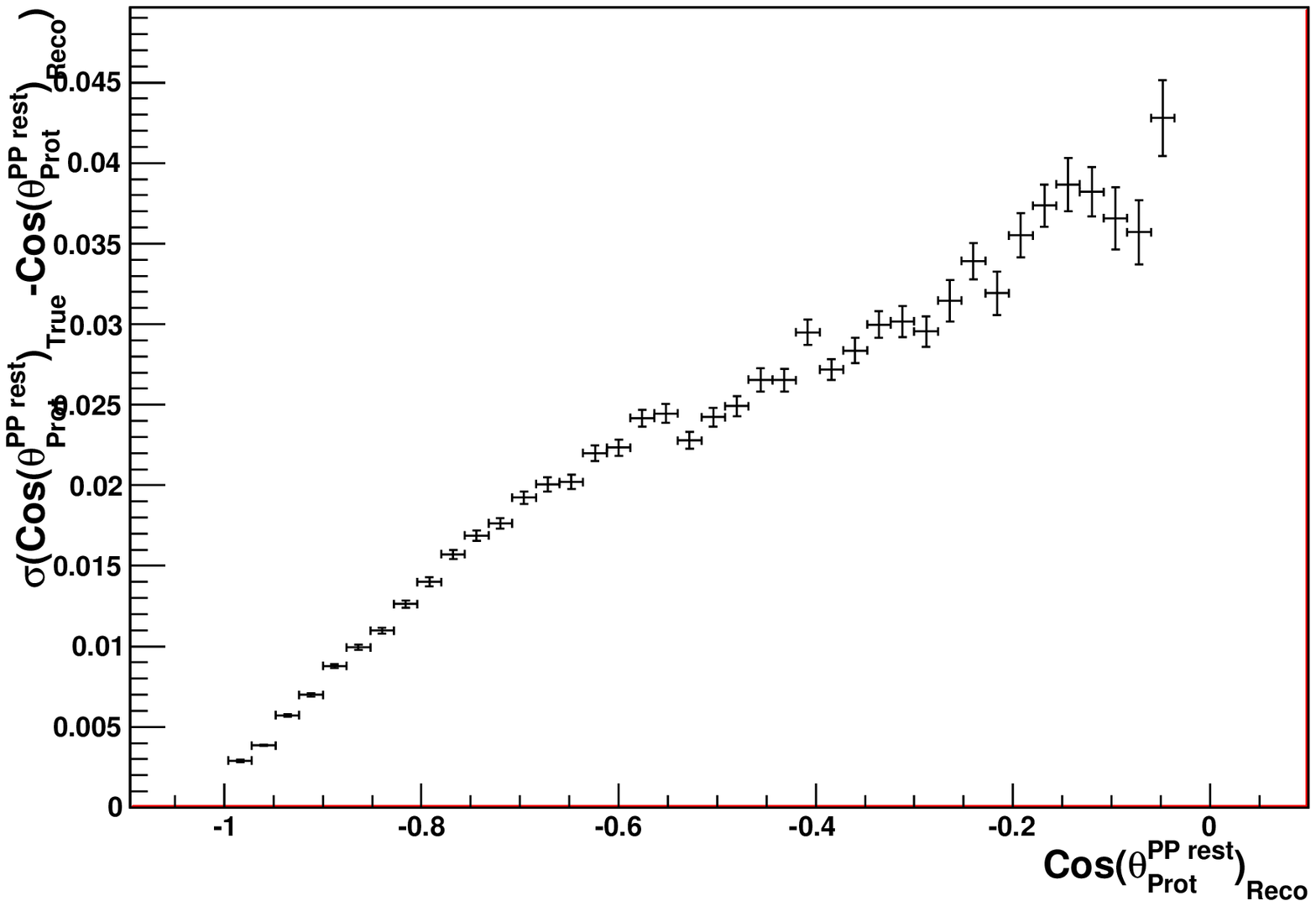}} \label{fig:EtaPPsystemMCresolution2}}
}
\caption{Monte-Carlo simulation, resolution studies of the $\cos(\theta_{p^{pp}})$.}
\label{fig:EtaPPsystemMCresolution}
\end{figure}

Also the resolution for the $\cos(\theta_{\eta^{CM}})$ and the $\cos(\theta_{p^{pp}})$ was studied, to properly select width of the bins in those variables.
Using Monte-Carlo simulation one compares the $\sigma$ parameter of the fitted Gaussian of the true minus reconstructed distribution as a function of the reconstructed value (Figs.~\ref{fig:EtaCMsystemMCresolution},~\ref{fig:EtaPPsystemMCresolution}).
The maximal bin size was chosen.

\newpage
\paragraph{The production mechanism\\}\label{par:etaprod}

\myFrameSmallFigure{EtaMMppVSIMEtaPNew2}{Experimental Data, Missing Mass of two protons versus $M^{2}(p\eta)$.
The signal from the $\eta$ meson as well as the background is seen.}{Experimental Data, Missing Mass of two protons versus $M^{2}(p\eta)$}

\begin{sidewaysfigure}

\centering
{
\subfigure[]{\fbox{\includegraphics[width=0.45\textwidth]{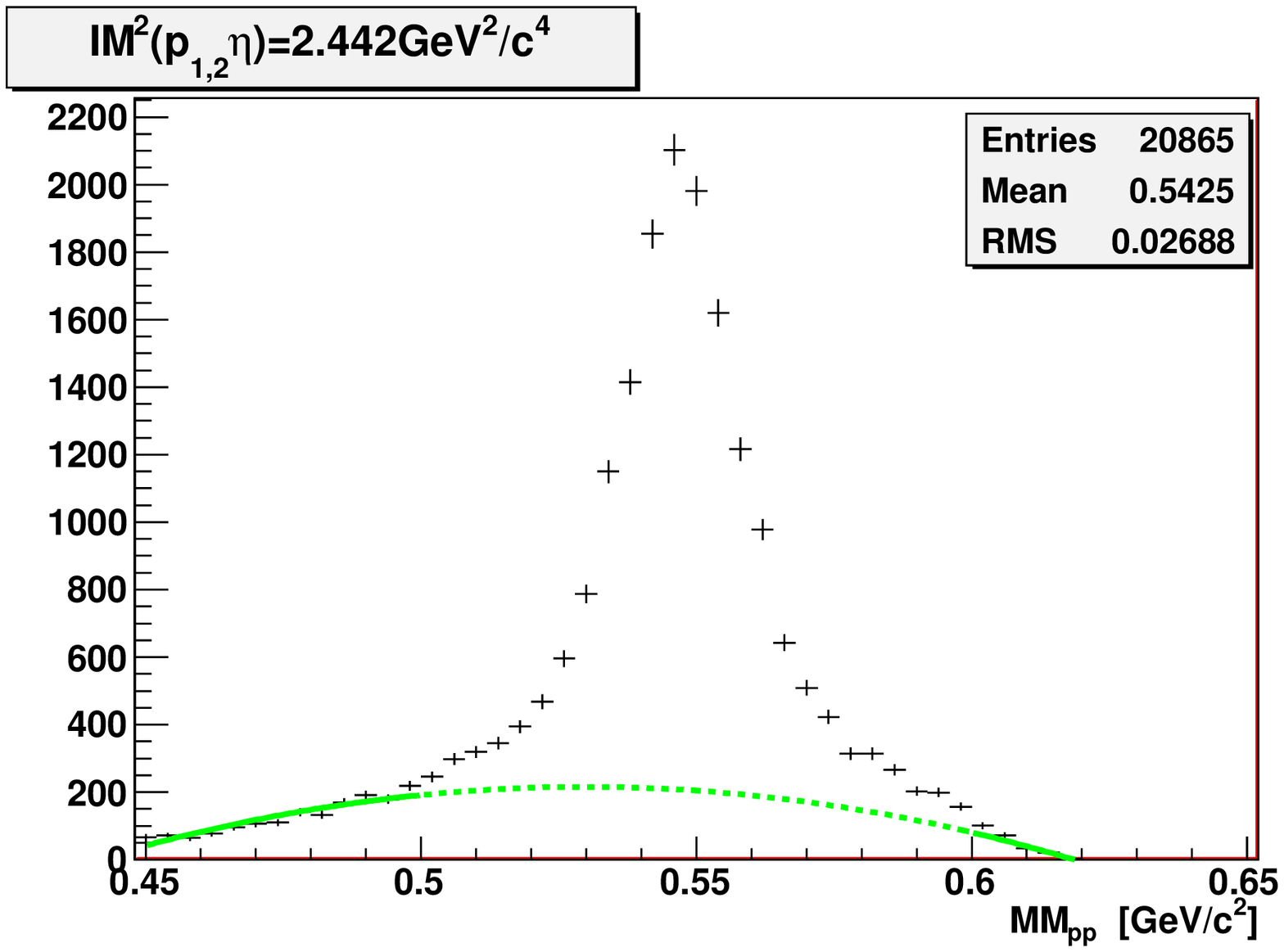}} \label{fig:EtaMMppBGSubtractionA}}\quad
\subfigure[]{\fbox{\includegraphics[width=0.45\textwidth]{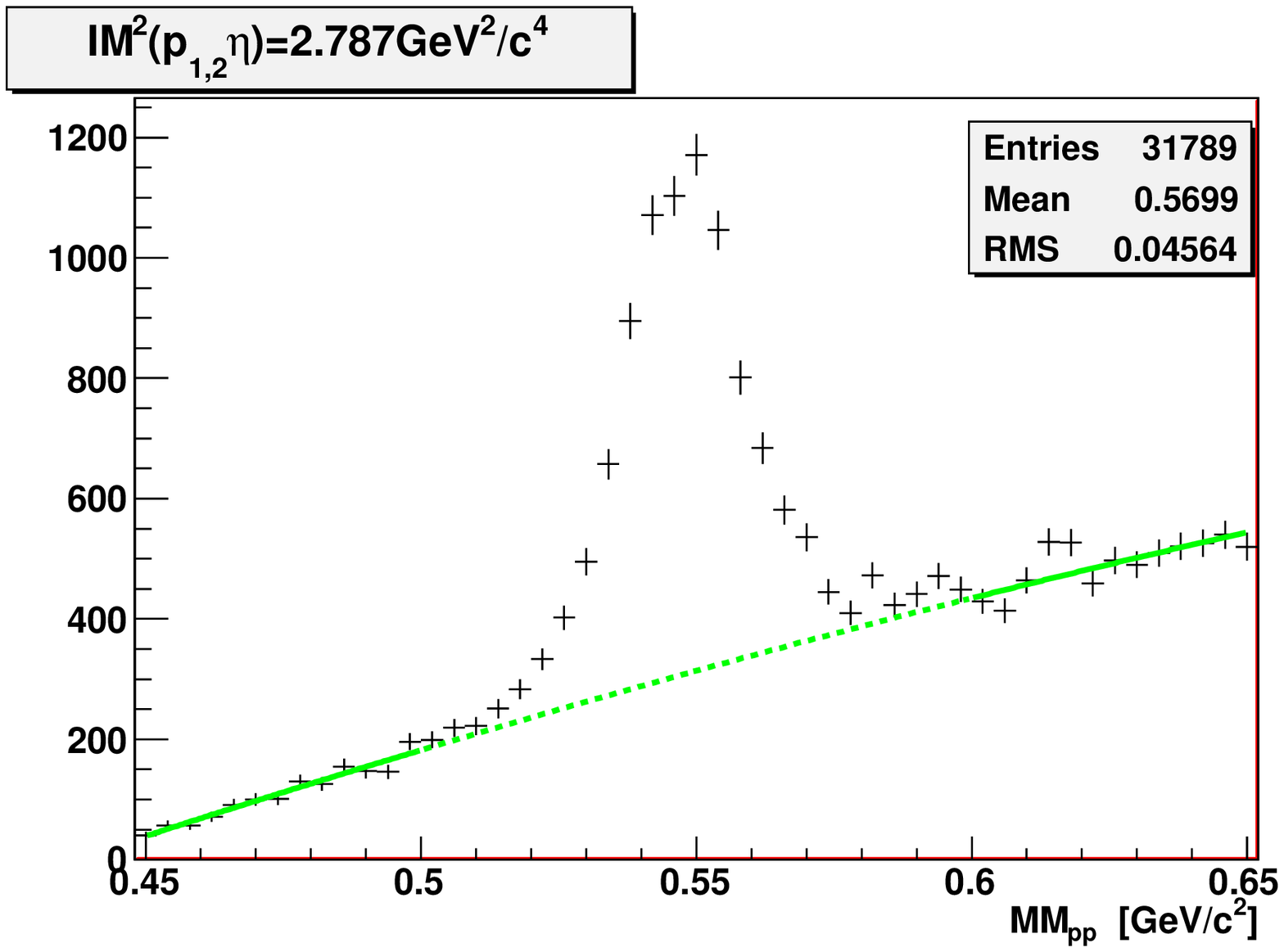}} \label{fig:EtaMMppBGSubtractionB}}\\
\subfigure[]{\fbox{\includegraphics[width=0.45\textwidth]{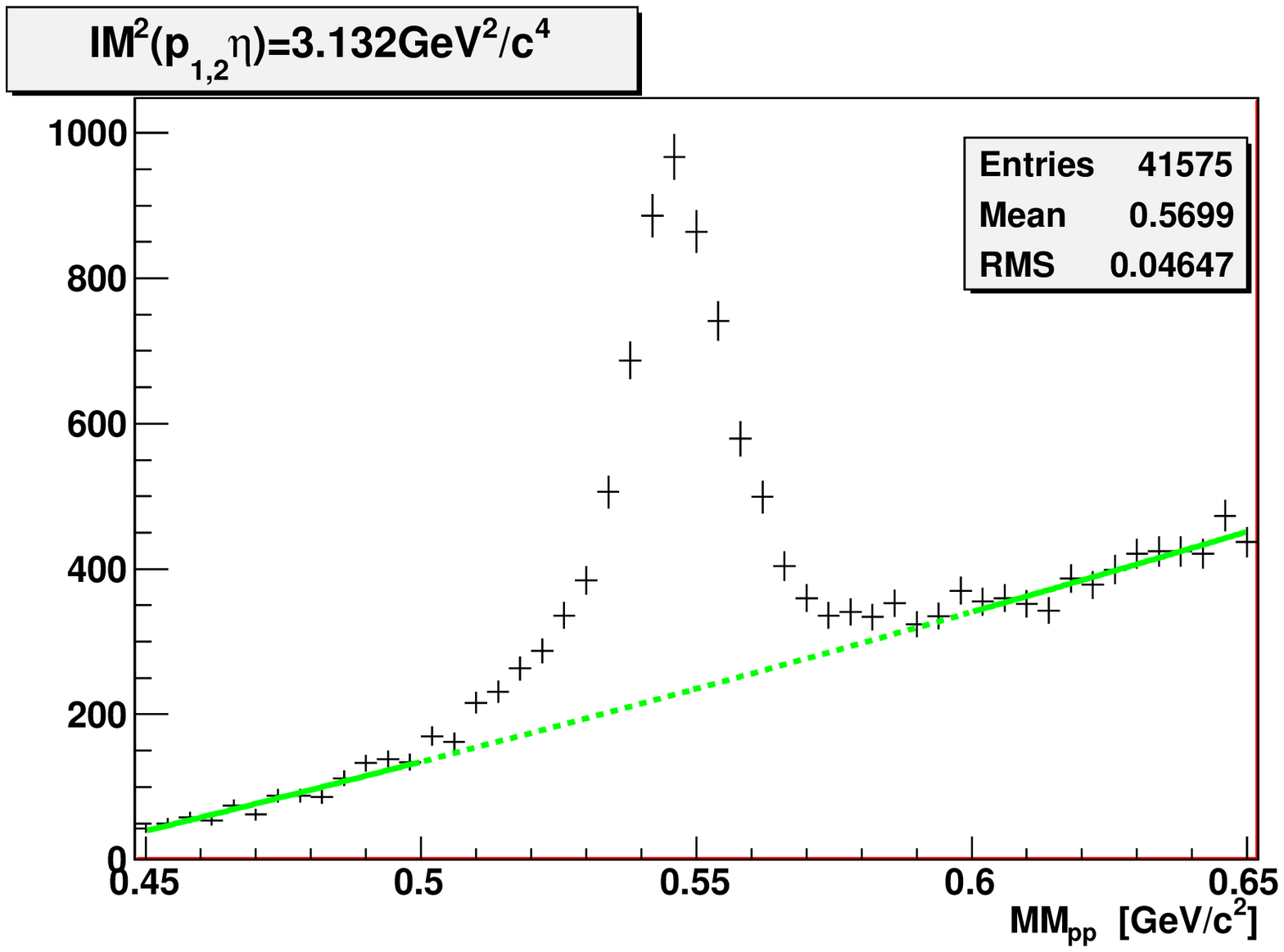}} \label{fig:EtaMMppBGSubtractionC}}\quad
\subfigure[]{\fbox{\includegraphics[width=0.45\textwidth]{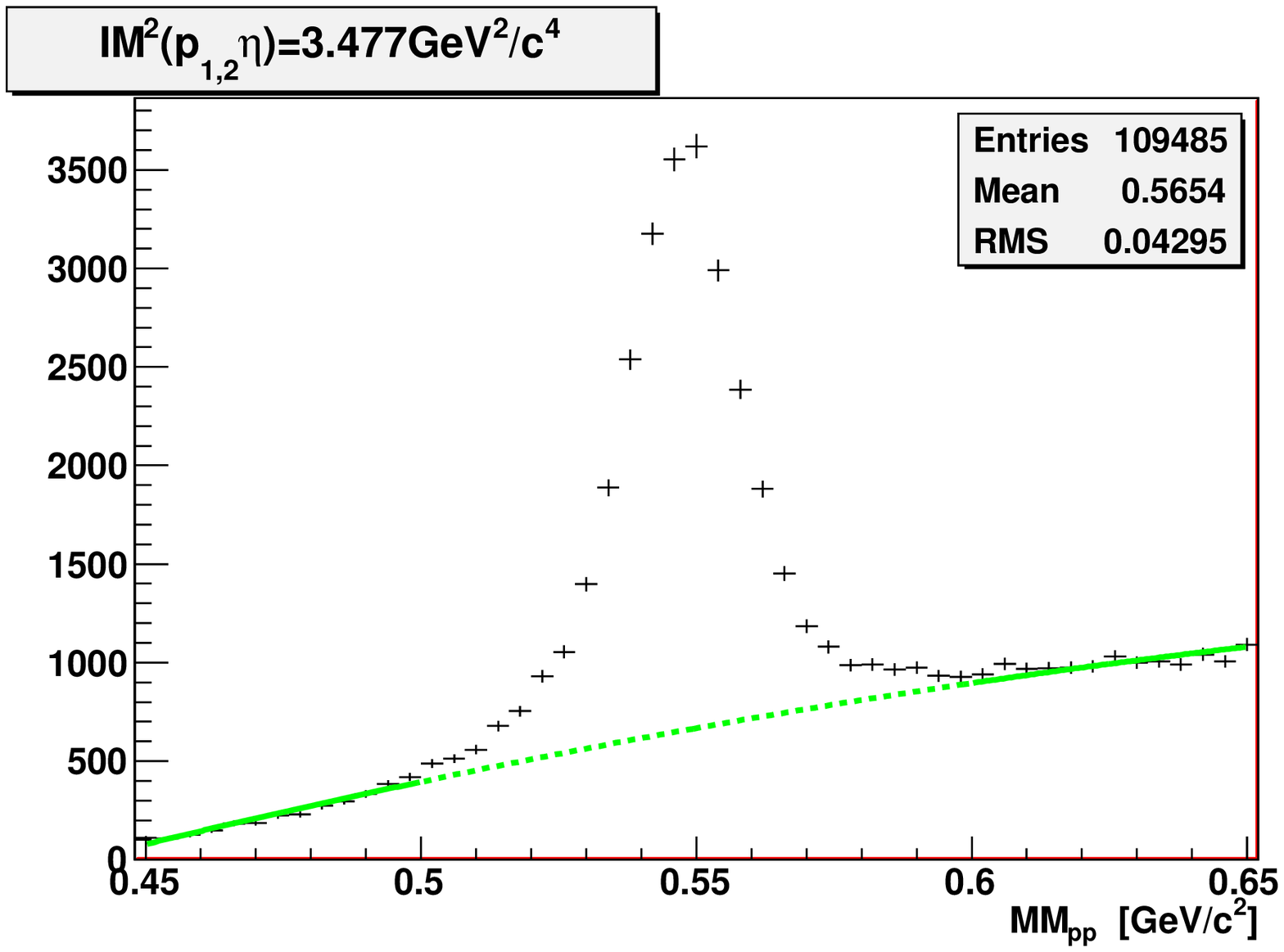}} \label{fig:EtaMMppBGSubtractionD}}
}
\caption{Examples of bin by bin background subtraction from the experimental $M^{2}(p\eta)$ distribution.
The background was approximated by the second order polynomial (green line).}
\label{fig:EtaMMppBGSubtraction}
\end{sidewaysfigure}

\myFrameFigure{EtaIMEtaPFitbeforefitNew2}{$M^{2}(p\eta)$ distribution. The experimental data are black line,
 the Monte-Carlo phase space blue line, the Monte-Carlo simulation assuming excitation of $N^{*}(1535)$ red line.
 The plot is symmetrized against two protons - each event is filled two times. The simulations are normalized to the same number of events as the experimental data.}{$IM^{2}(p\eta)$ distribution.}

Before studying the angular distributions, the $M^{2}(p\eta)$(invariant mass of the proton $eta$ system) distribution was checked. To obtain the background free $M^{2}(p\eta)$
the missing mass of the two protons was plotted against the $M^{2}(p\eta)$ \myImgRef{EtaMMppVSIMEtaPNew2}. Then the background was fitted outside the $\eta$ meson
peak by the second order polynomial and subtracted from the data, this was done for every  $M^{2}(p\eta)$ bin (Fig.~\ref{fig:EtaMMppBGSubtraction}).   
Next the background subtracted experimental data were compared with the Monte-Carlo simulation for two assumed production mechanisms \cite{DISTO},
the phase space production and the production via excitation of the $N^{*}(1532)$ \myImgRef{EtaIMEtaPFitbeforefitNew2} (see Appendix~\ref{appendix:wmc}).
It is seen that non of the Monte-Carlo models describes the experimental data. If one assumes that only those two mechanisms can contribute
to the $\eta$ production, one can perform a fit of the two production models to the experimental data.
Such a fit was performed by using the $\chi^{2}$ method. The $\chi^{2}$ function was minimized:
 \begin{equation}
 \chi^{2} = \sum \dfrac{\left[ Data - \left(b Model_{1} + (1-b)Model_{2}\right)\right]^{2} }{\sigma^{2}_{Data} + b^{2}\sigma^{2}_{Model_{1}} + (1-b)^{2}\sigma^{2}_{Model_{2}}}
\label{eq:chi2EtaFit}
\end{equation}    
where $Model_{1}$~-~is the $\eta$ meson production via $N^{*}(1532)$, $Model_{2}$~-~is the $\eta$ meson phase space production.  
The $\sigma_{Data}$ is the error of the point for experimental data,
 $\sigma_{Model_{1}}$~-~ is the error of the point for $Model_{1}$ and $\sigma_{Model_{2}}$~-~ is the error of the point for $Model_{2}$.
 The $b$ parameter is the fraction of the production via $N^{*}(1532)$ to the sum of phase space production and via $N^{*}(1532)$.
For the numerical purpose of doing the fit algorithm, the $\chi^{2}$ function (Eq.~\ref{eq:chi2EtaFit}) was redefined to:
\begin{equation}
 \chi^{2} = \sum \dfrac{\left[ Data - \left(c Model_{1}^{*} + (1-c)Model_{2}^{*}\right)\right]^{2} }{\sigma^{2}_{Data} + c^{2}\sigma^{2}_{Model_{1}^{*}} + (1-c)^{2}\sigma^{2}_{Model_{2}^{*}}}
\label{eq:chi2EtaFitNorm}
\end{equation}

The same as in the Section~\ref{subsec:3pi0MCmodel} on page \pageref{subsec:3pi0MCmodel}.
%
%
%
%
%

\begin{figure}[ht!bp]
\centering
{
\subfigure[]{\fbox{\includegraphics[width=0.7\textwidth]{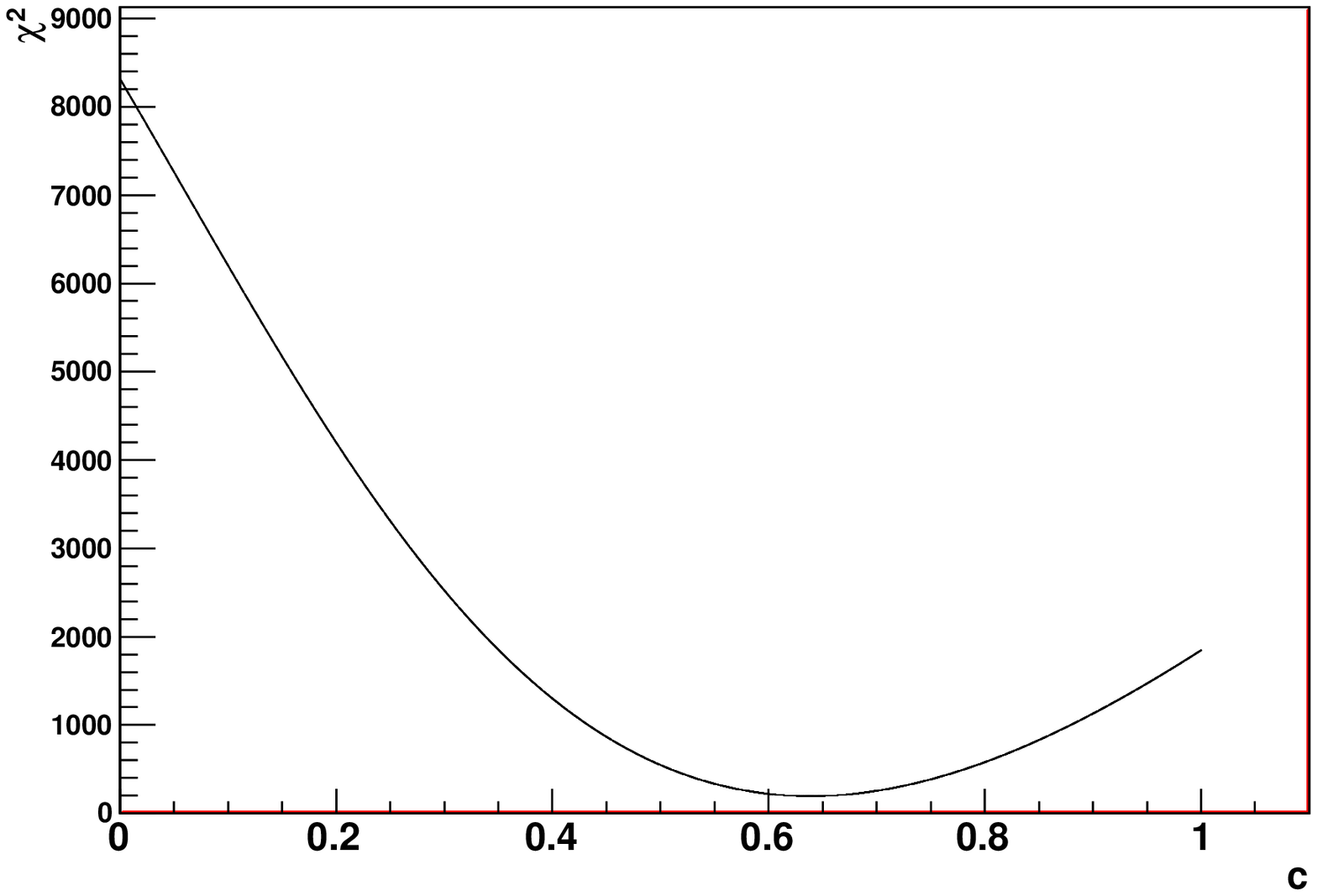}} \label{fig:chi2DalitzPlotFitEta1}}\\
\subfigure[]{\fbox{\includegraphics[width=0.7\textwidth]{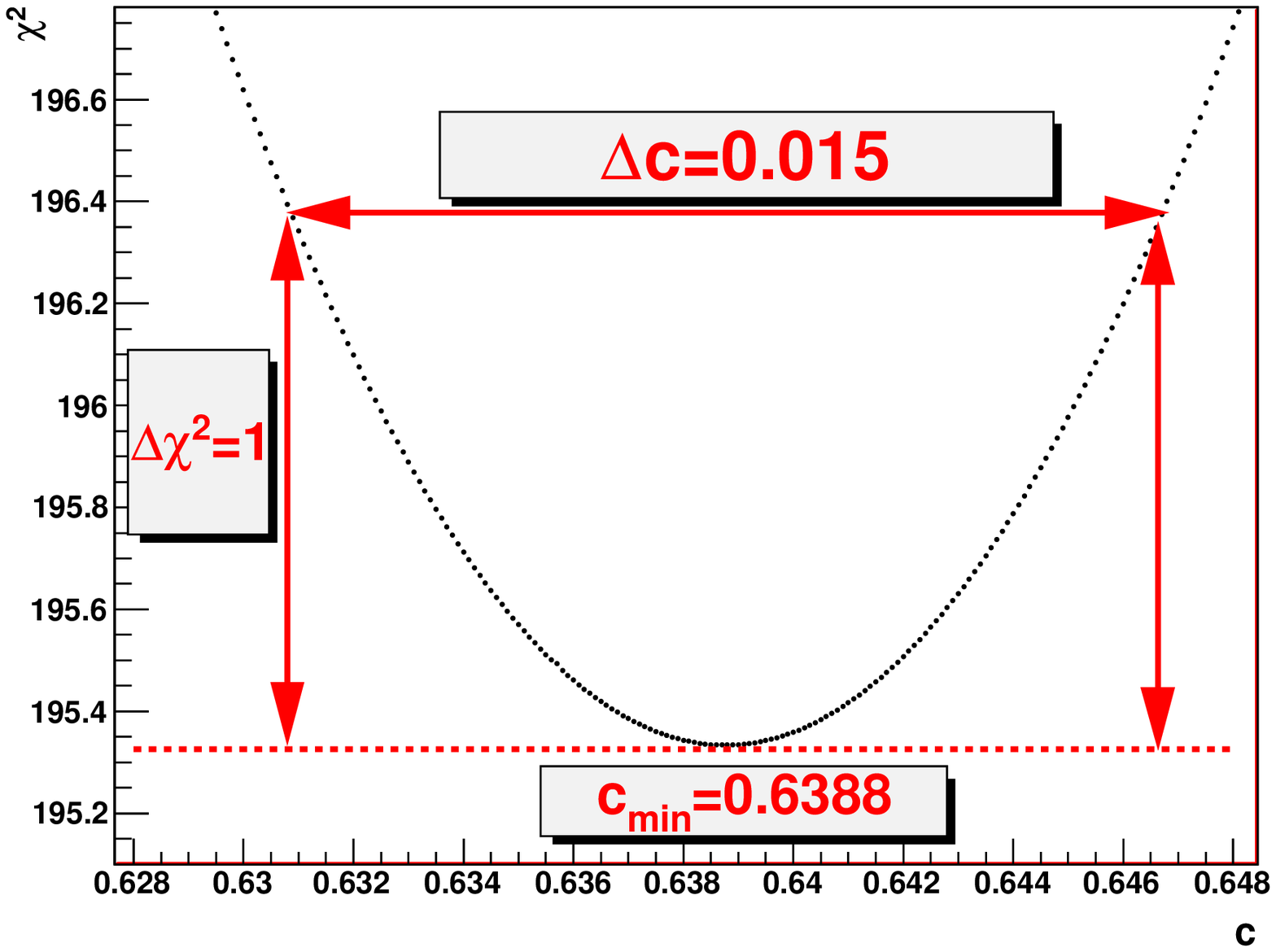}} \label{fig:chi2DalitzPlotFitEta2}}
}
\caption{$\chi^{2}$ versus the searched parameter $c$ for the sum of phase space and $N^{*}(1535)$ fit.}
\label{fig:chi2DalitzPlotFitEta}
\end{figure}

The $\chi^{2}$ function (Eq.~\ref{eq:chi2EtaFitNorm}) was minimize in respect to parameter $c$.
The (Fig.~\ref{fig:chi2DalitzPlotFitEta}) shows the $\chi^{2}$ versus the searched parameter $c$, the function has one minimum.
The estimated value of the $c$ parameter is the one which gives the smallest $\chi^{2}$, the error of the parameter is half of
 the distance for which the $\chi^{2}$ function changes by $1$, this gives:

\begin{equation}
 \frac{\chi^{2}_{min}}{NDF} = \frac{195.3337}{43} =  4.54 \pm 0.22
\end{equation}
where $\chi^{2}_{min}$ is the $\chi^{2}$ value at minimum, and $NDF$ is the Number of Degrees of Freedom.

\begin{equation}
 c =0.6388 \pm 0.0075
\end{equation}
and corresponds to the:
\begin{equation}
\dfrac{Model_{1}}{Model_{2}} = 0.767 \pm 0.011
\end{equation}
giving the
\begin{equation}
 b = 0.4342 \pm 0.0084
\end{equation}

\myFrameFigure{EtaIMEtaPFitresultNew2}{$M^{2}(p\eta)$ distribution. The experimental data are black line,
 the Monte-Carlo phase space blue line, the Monte-Carlo simulation assuming excitation of $N^{*}(1535)$ red line,
the results of the fit - sum of phase space ($56.58\%$) and $N^{*}(1535)$($43.4\%$) production green line.
 The plot is symmetrized against two protons - each event is filled two times. The result of the fit was normalized to the same amount of events as the experimental data.}{$IM^{2}(p\eta)$ distribution.}

To check how the Monte-Carlo simulation based on the sum of two models, with the fitted parameter, describes the experimental data,  
the comparison was done showing the models sum \myImgRef{EtaIMEtaPFitresultNew2}.
It it seen that sum of the models describes very good the event populations on the invariant mass spectrum. 

The same procedure was repeated for the background subtraction using polynomial of the first order, after the fit the results this gives:
\begin{equation}
 b_{pol1} = 0.4141 \pm 0.0084
\end{equation}
The difference between the two background subtraction methods was used to estimate the systematic error:
\begin{equation}
 \Delta b_{sys.}= |b-b_{pol1}| = 0.0201
\end{equation}

The final value with the systematic error is
 
\begin{equation}
 b = 0.434 \pm 0.008(stat.) \pm 0.020(sys.)
\label{eq:MyN1535contrib}
\end{equation}

As a final results one gets that the production mechanism via $N^{*}(1535)$ is the $43.4\%$ of the total $\eta$ production,
 at this energy ($T=2.54\mathrm{GeV}$) for the covered part of the phase space (Eq.~\ref{eq:EtaAcceptanceCut}). 

\bigskip
\paragraph{The angular distributions\\}\label{par:etaDistr}
Having now described the production mechanism (phase space and $N^{*}(1535)$ excitation), one can start to look into the angular
distributions for accessible via Wasa-at-Cosy part of the phase space (Eq.~\ref{eq:EtaAcceptanceCut}) (the conditional distributions).

\begin{sidewaysfigure}

\centering
{
\subfigure[$q_{\eta}^{CM}=0.45-0.475\mathrm{~GeV/c}$]{\fbox{\includegraphics[width=0.45\textwidth]{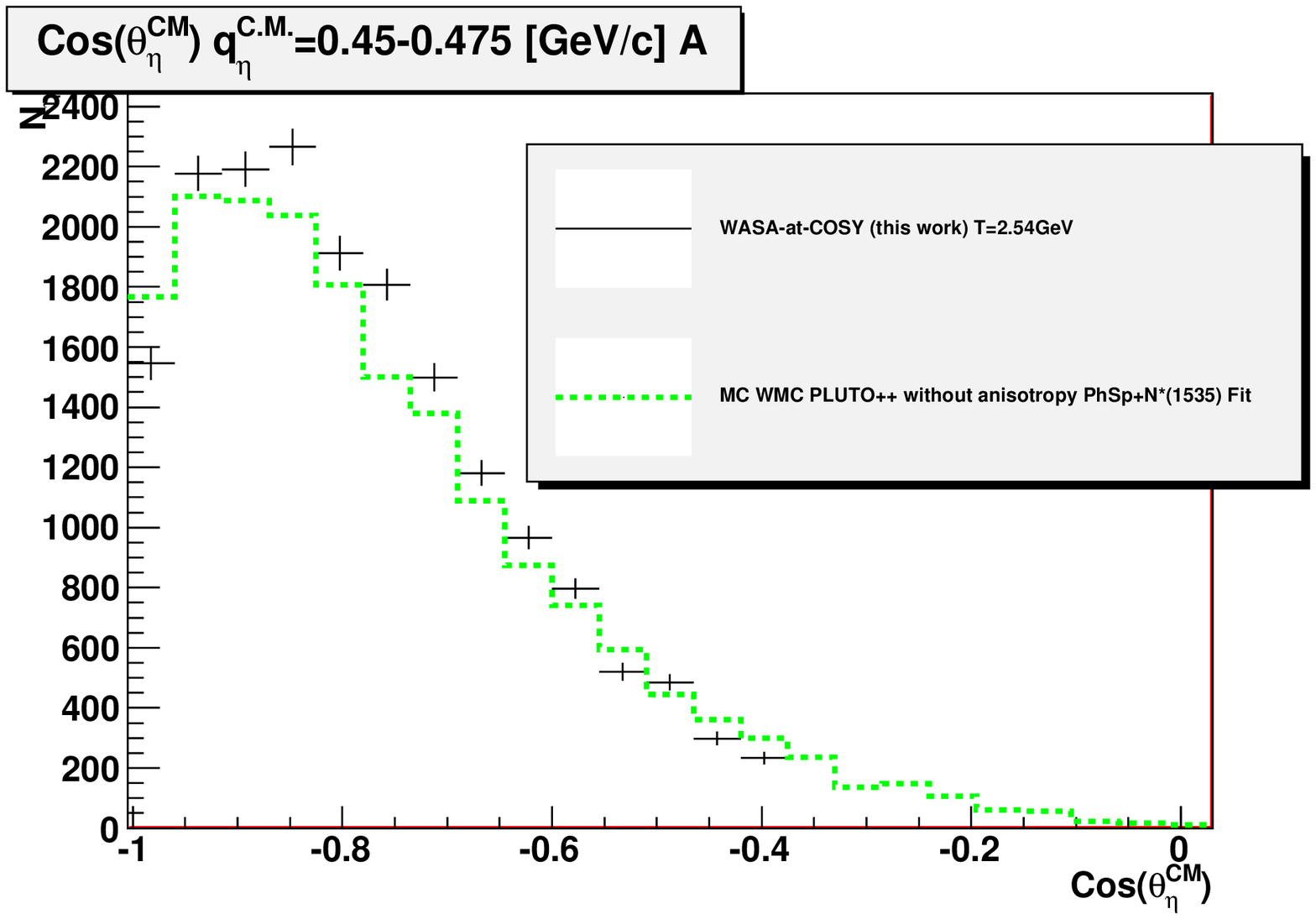}} \label{fig:EtaCMMomDepNA}}\quad
\subfigure[$q_{\eta}^{CM}=0.475-0.5\mathrm{~GeV/c}$]{\fbox{\includegraphics[width=0.45\textwidth]{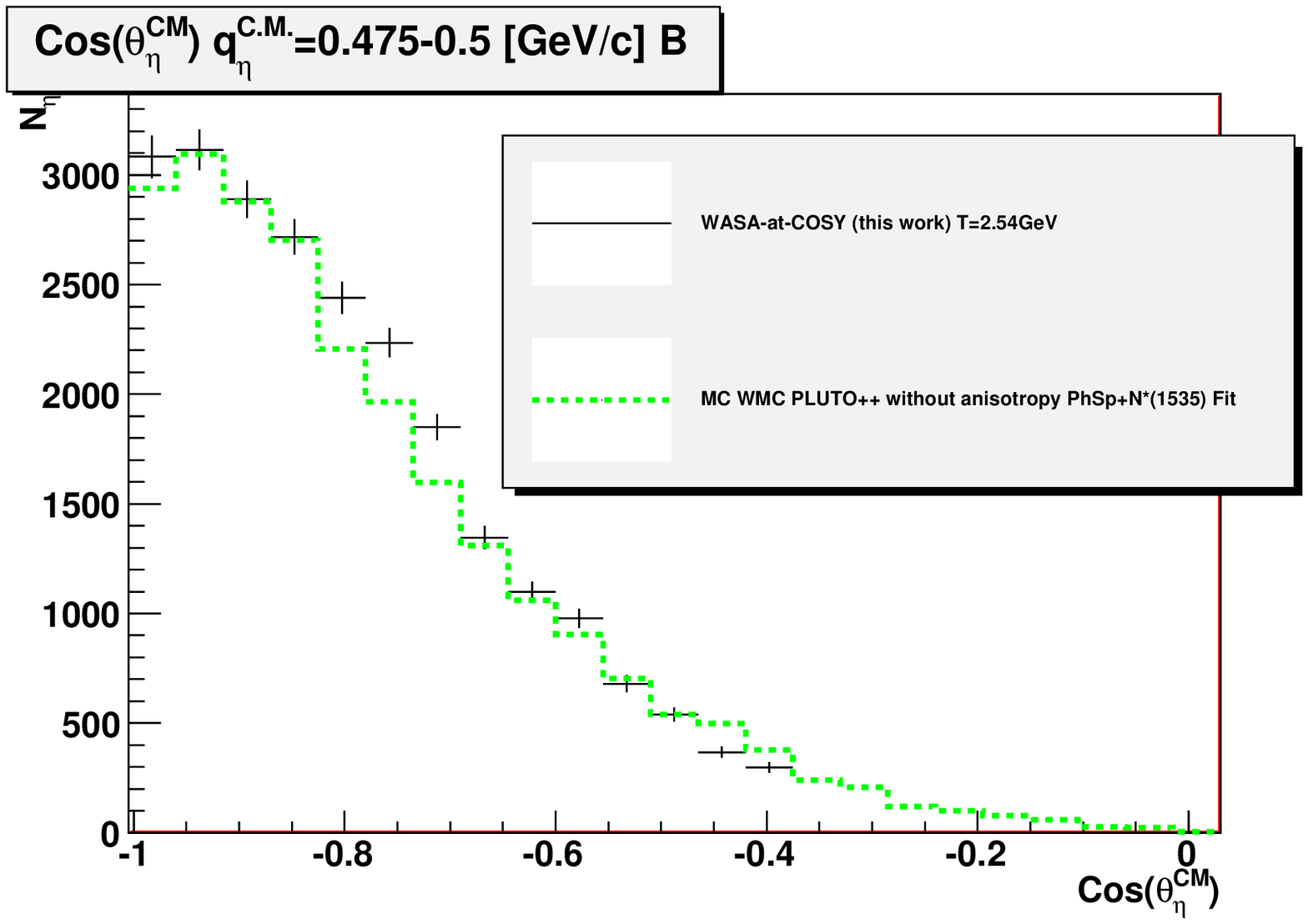}} \label{fig:EtaCMMomDepNB}}\\
\subfigure[$q_{\eta}^{CM}=0.5-0.55\mathrm{~GeV/c}$]{\fbox{\includegraphics[width=0.45\textwidth]{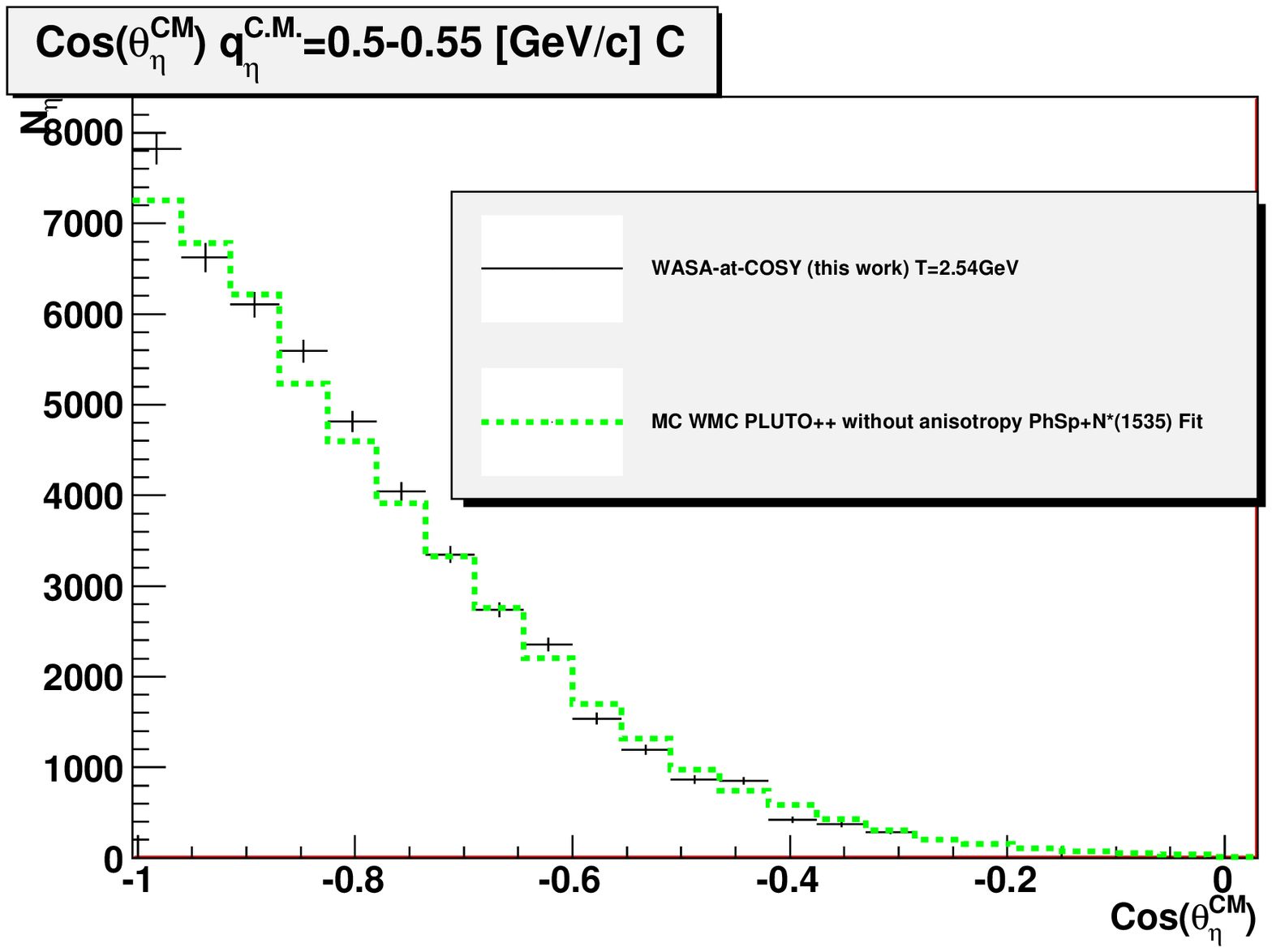}} \label{fig:EtaCMMomDepNC}}\quad
\subfigure[$q_{\eta}^{CM}=0.55-0.7\mathrm{~GeV/c}$]{\fbox{\includegraphics[width=0.45\textwidth]{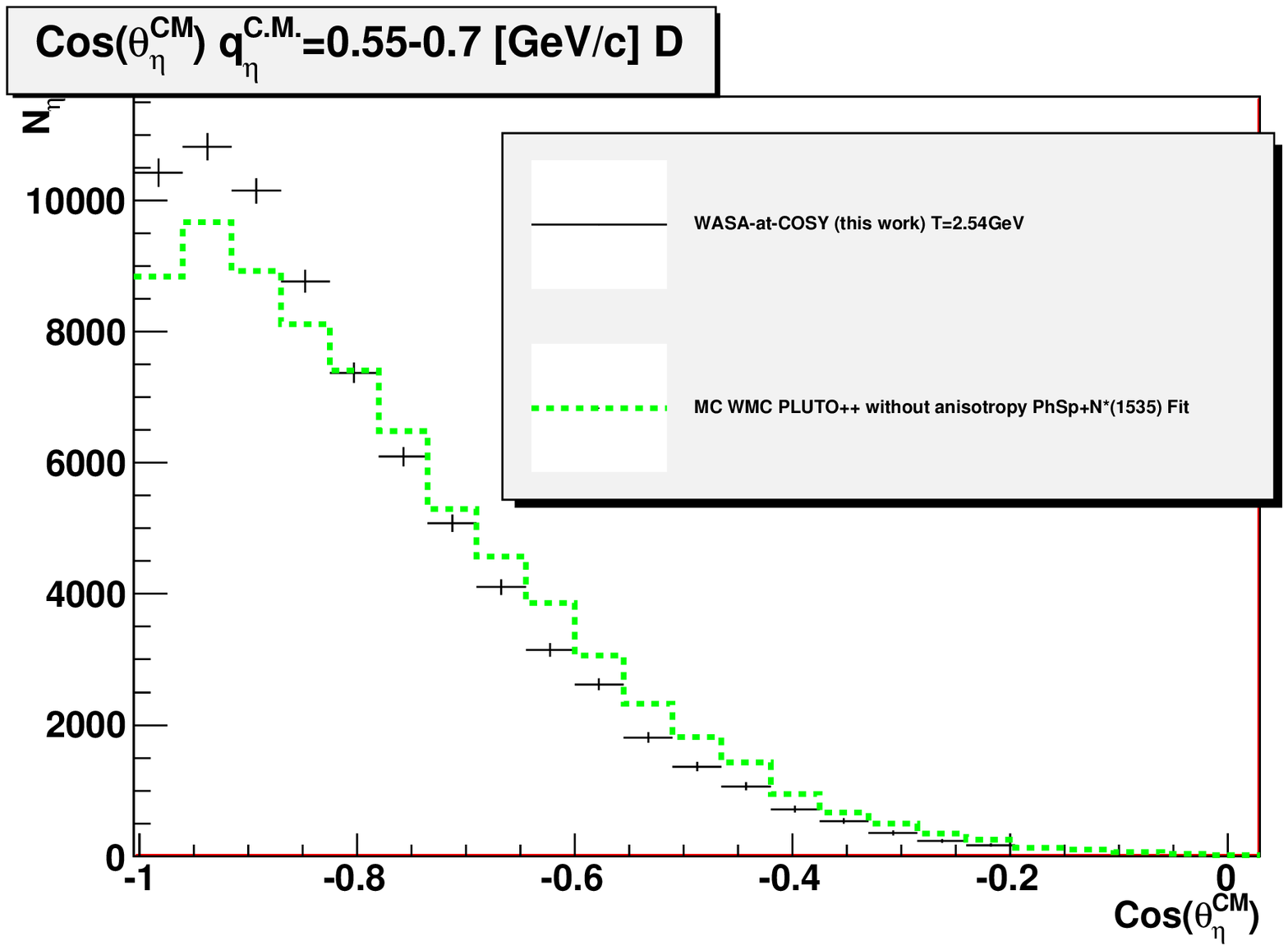}} \label{fig:EtaCMMomDepND}}
}
\caption{$\cos(\theta_{\eta^{CM}})$ distribution for different momenta of the $\eta$ in CM system $q_{\eta}^{CM}$.
Experimental data (black marker), Monte-Carlo simulation without any $\eta$ angle anisotropy ($N^{*}(1535)$ and Phase Space) (green line)}
\label{fig:EtaCMMomDepN}
\end{sidewaysfigure}

\begin{sidewaysfigure}
\centering
{
\subfigure[$q_{\eta}^{CM}=0.45-0.475\mathrm{~GeV/c}$]{\fbox{\includegraphics[width=0.45\textwidth]{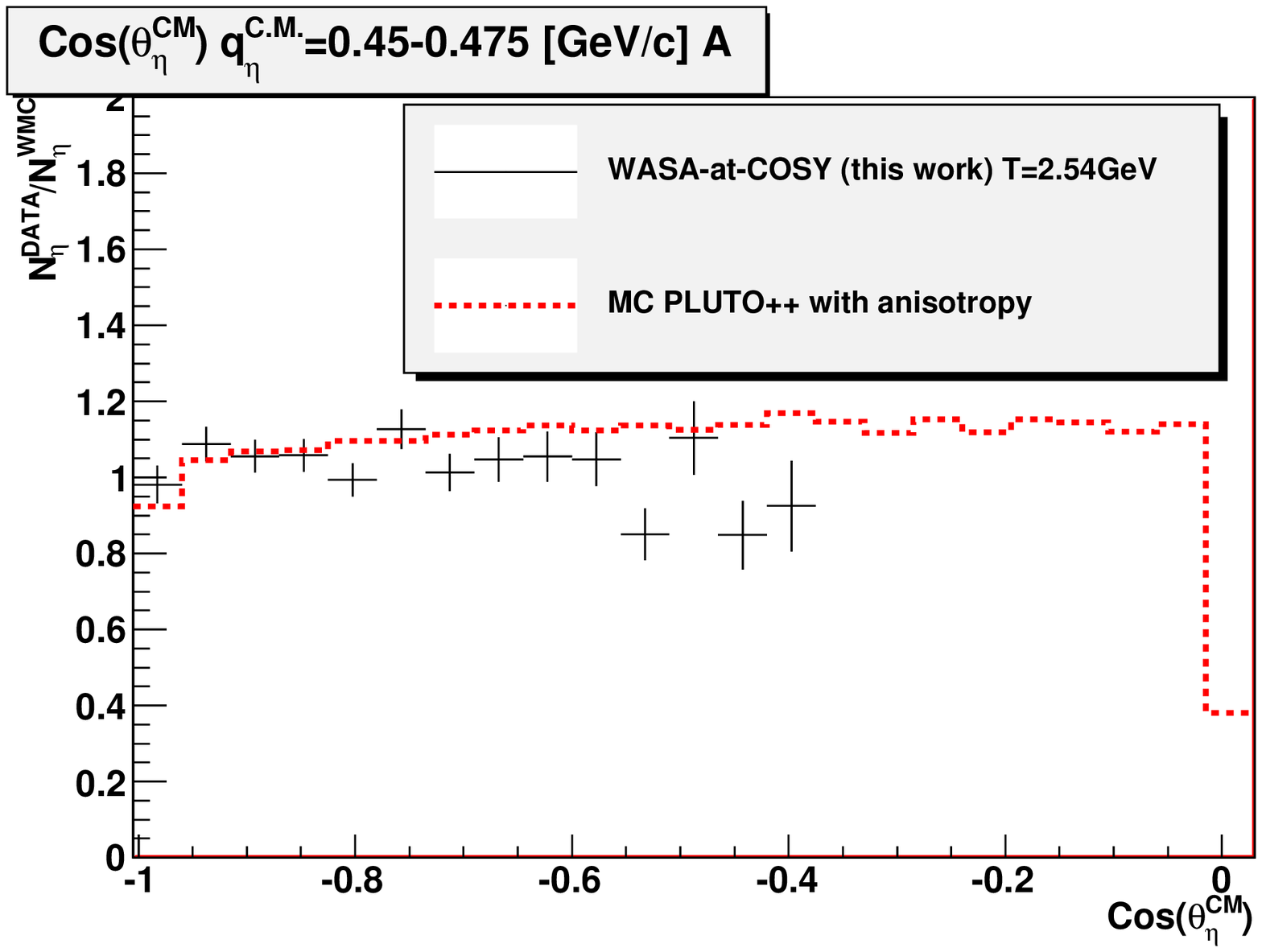}} \label{fig:EtaCMMomDiffNA}}\quad
\subfigure[$q_{\eta}^{CM}=0.475-0.5\mathrm{~GeV/c}$]{\fbox{\includegraphics[width=0.45\textwidth]{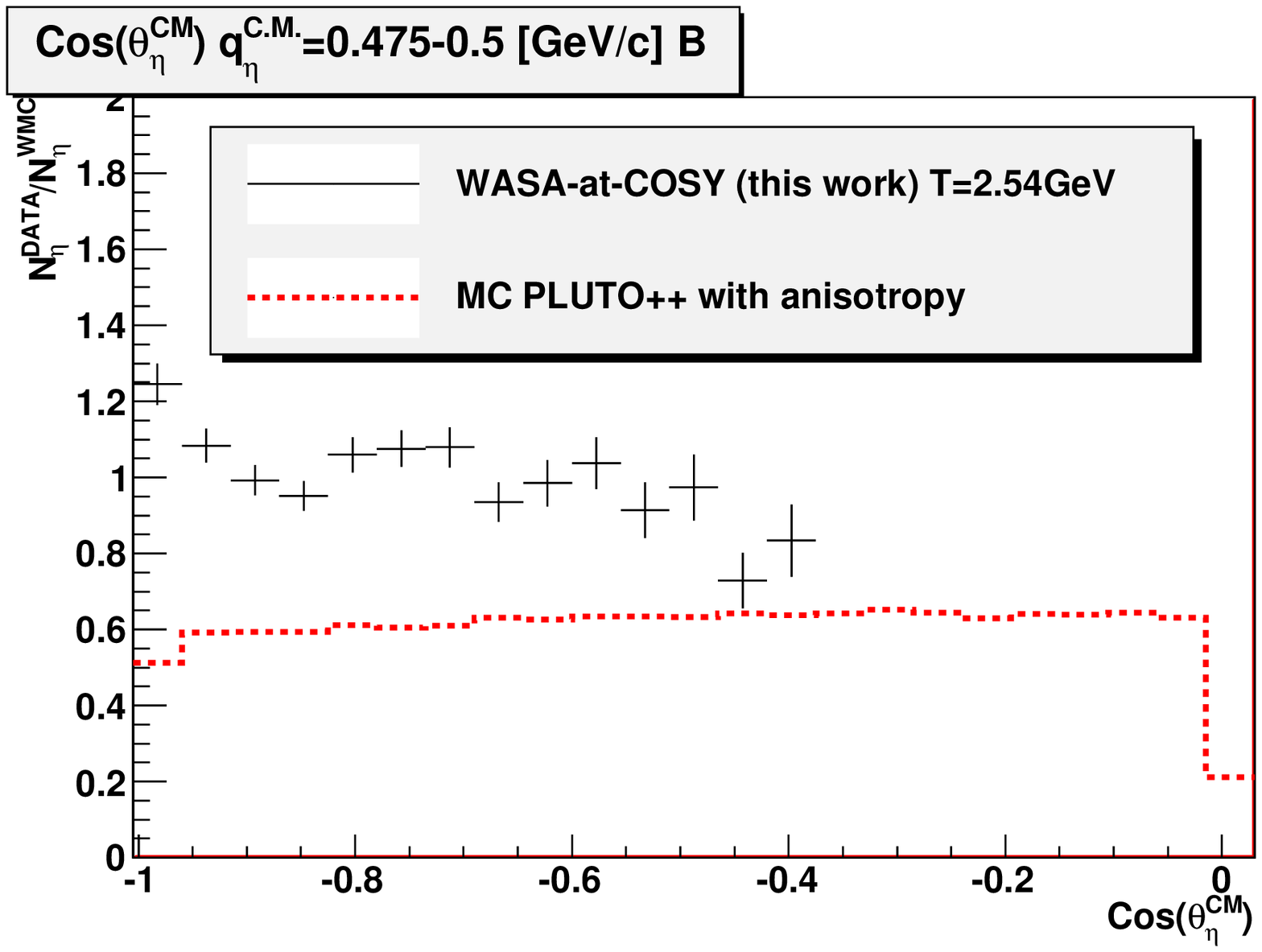}} \label{fig:EtaCMMomDiffNB}}\\
\subfigure[$q_{\eta}^{CM}=0.5-0.55\mathrm{~GeV/c}$]{\fbox{\includegraphics[width=0.45\textwidth]{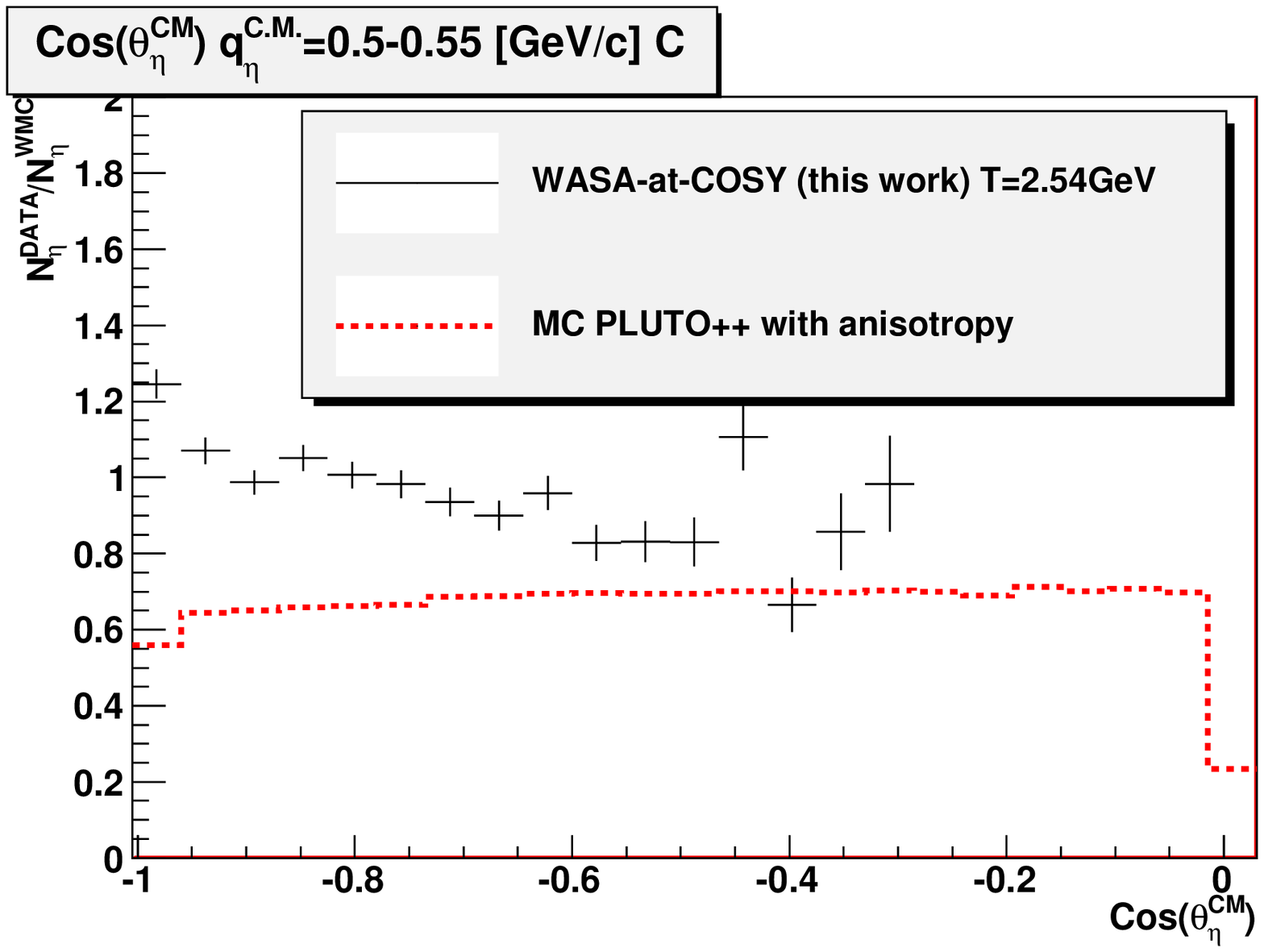}} \label{fig:EtaCMMomDiffNC}}\quad
\subfigure[$q_{\eta}^{CM}=0.55-0.7\mathrm{~GeV/c}$]{\fbox{\includegraphics[width=0.45\textwidth]{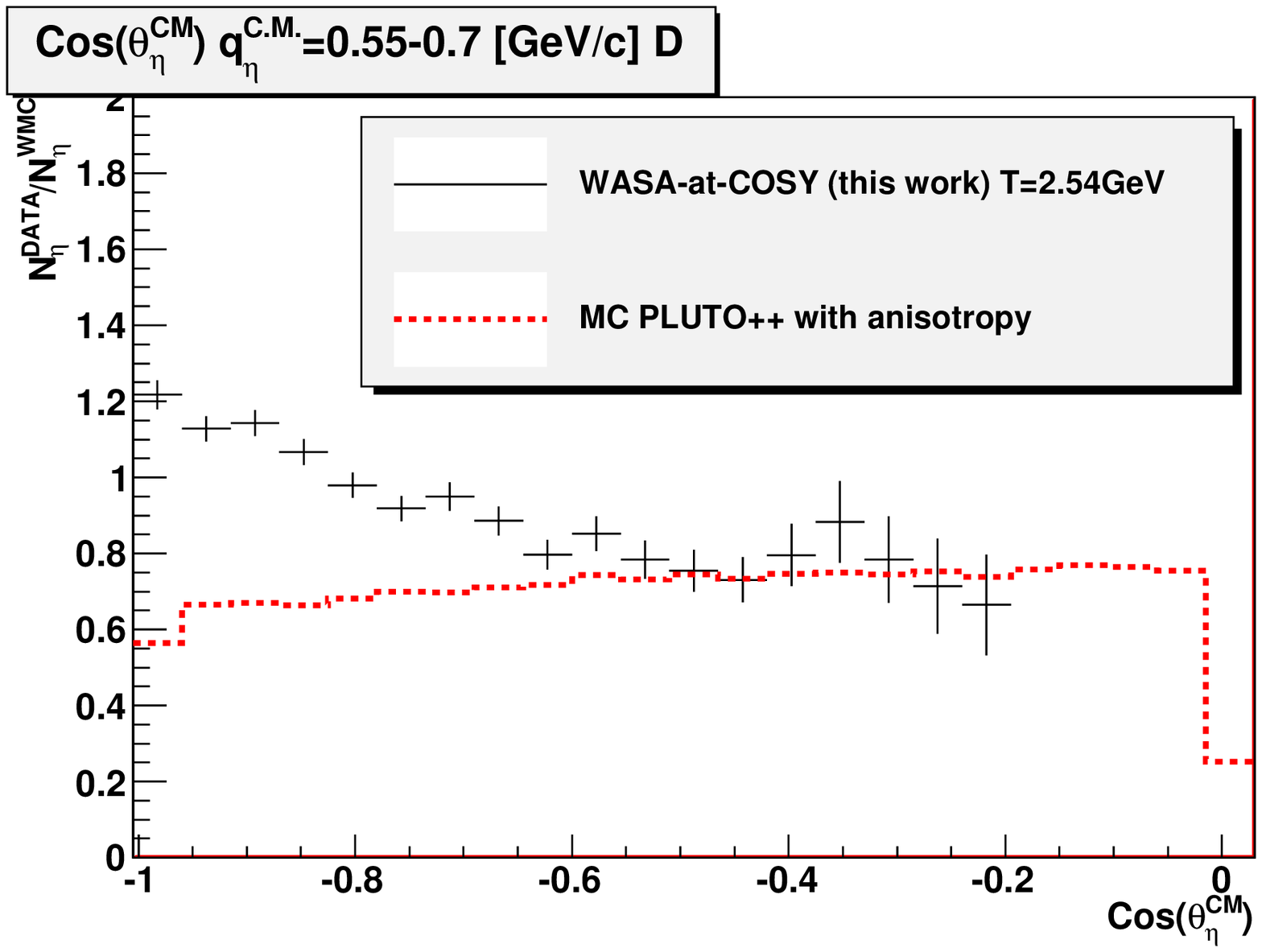}} \label{fig:EtaCMMomDiffND}}
}
\caption{The extracted anisotropy (experimental data divided by the Monte-Carlo simulation of the $\eta$ meson production \myImgRef{EtaIMEtaPFitresultNew2} without any $\eta$ angle anisotropy) versus $\cos(\theta_{\eta^{CM}})$ for different momenta of the $\eta$ in CM system $q_{\eta}^{CM}$.
Experimental data (black marker), Monte-Carlo simulation with anisotropy (arbitrary normalization - to guide the eye) - PLUTO++ parametrization (red line)}
\label{fig:EtaCMMomDiffN}
\end{sidewaysfigure}

First the $\cos(\theta_{\eta^{CM}})$ was studied for the four different regions of the $q_{\eta}^{CM}$.
To obtain the background free $\cos(\theta_{\eta^{CM}})$
the missing mass of the two protons was plotted against the $\cos(\theta_{\eta^{CM}})$. Then the background was fitted outside the $\eta$ meson
peak by the second order polynomial and subtracted from the data, this was done bin by bin.
The background subtracted experimental data were compared with a Monte-Carlo simulation (phase space and $N^{*}(1535)$ excitation)
 without any assumed  $\eta$ angle anisotropy (Fig.~\ref{fig:EtaCMMomDepN}). It is seen that for four different ranges in the $\eta$ momentum
the shape of Monte-Carlo and experimental data changes differently. One sees the momentum dependence of the angular distribution.
To extract this effect and correct for acceptance and efficiency bias, the experimental data were divided by this Monte-Carlo simulation (Fig.~\ref{fig:EtaCMMomDiffN}).
The distribution was compared with a Monte-Carlo simulation with  $\eta$ angle anisotropy model in PLUTO++ \cite{Pluto} (Integrated over all $\eta$ momentum), which is based on \cite{DISTO}.
One sees the systematic changes of the angular distribution when the $\eta$ momentum increases.

The extracted anisotropy (experimental data divided by the Monte-Carlo simulation of the $\eta$ meson production \myImgRef{EtaIMEtaPFitresultNew2} without any $\eta$ angle anisotropy) 
 distributions for four different region of the $\eta$ meson momentum in the CM system
were compared between each other and with the Monte-Carlo simulation with a anisotropy parametrization based on Pluto++ \myImgRef{EtaCMDiffComparison2New2}.
The distributions were normalized so that for the $\cos(\theta_{\eta^{CM}})=-0.67$ all have the value $1$.    
If one looks into that angular distributions, 
 one sees that for the small momentum the data are consistent with this Monte-Carlo model. When the momentum increases the discrepancy between the model and the data
is getting bigger. The model is almost flat, where the data are more and more curved. One sees that the shape of the $\cos(\theta_{\eta^{CM}})$
changes with the $\eta$ meson momentum in the CM.

\myFrameFigure{EtaCMDiffComparison2New2}{The extracted anisotropy (experimental data divided by the Monte-Carlo simulation of the $\eta$ meson production \myImgRef{EtaIMEtaPFitresultNew2} without any $\eta$ angle anisotropy) as a function of $\cos(\theta_{\eta^{CM}})$ for four different
 regions of the $\eta$ momentum in the CM system, compared with the Monte-Carlo simulation with a  $\eta$ angle anisotropy parametrization
 from Pluto++ \cite{Pluto}. The shapes of the experimental distributions changes with the $\eta$ momentum. The normalization of the histograms is described in text.}{Experimental data, $\cos(\theta_{\eta^{CM}})$ comparison.}

\bigskip

\begin{sidewaysfigure}

\centering
{
\subfigure[$q_{\eta}^{CM}=0.45-0.475\mathrm{~GeV/c}$]{\fbox{\includegraphics[width=0.45\textwidth]{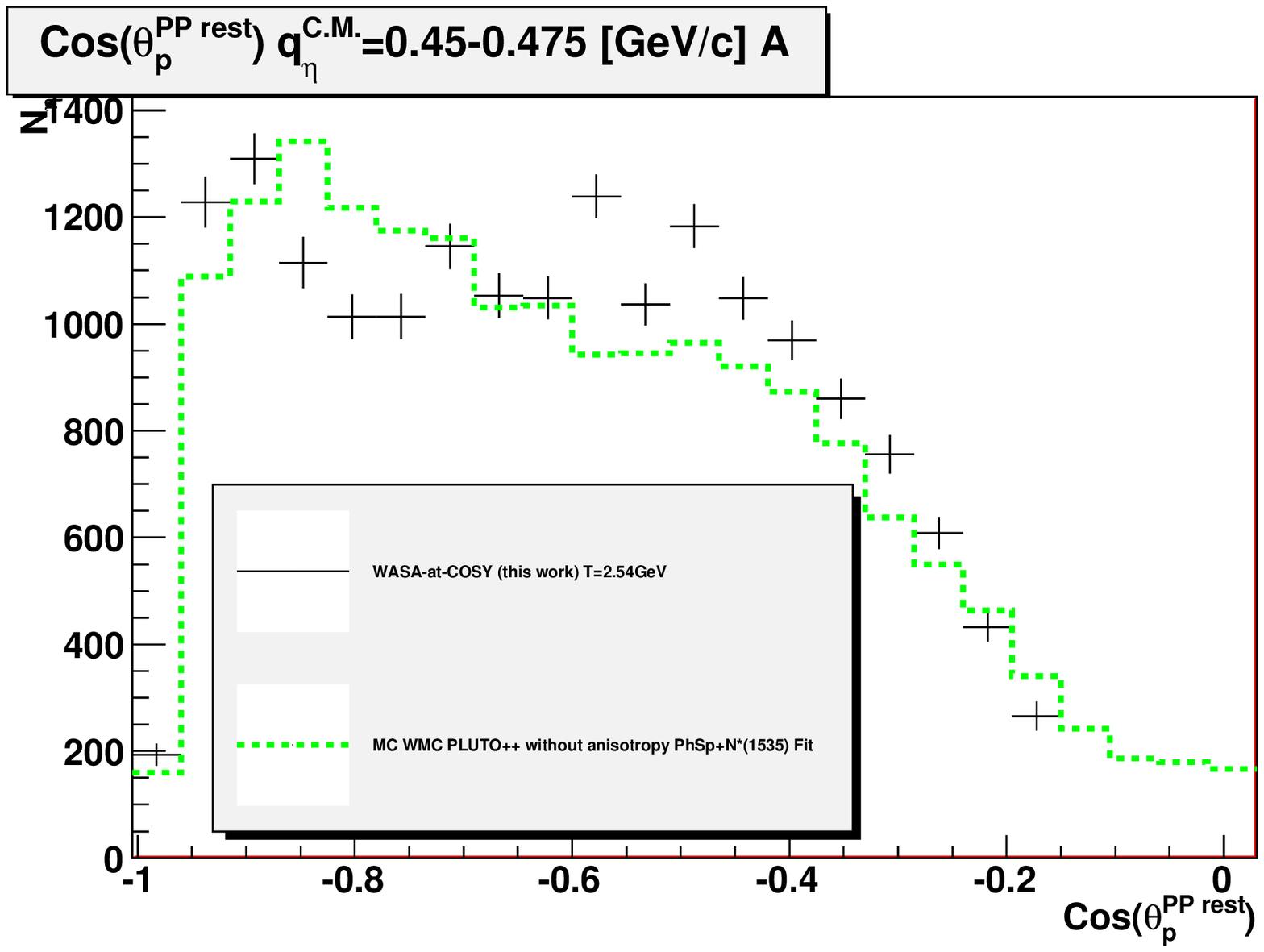}} \label{fig:EtaPPMomDepNA}}\quad
\subfigure[$q_{\eta}^{CM}=0.475-0.5\mathrm{~GeV/c}$]{\fbox{\includegraphics[width=0.45\textwidth]{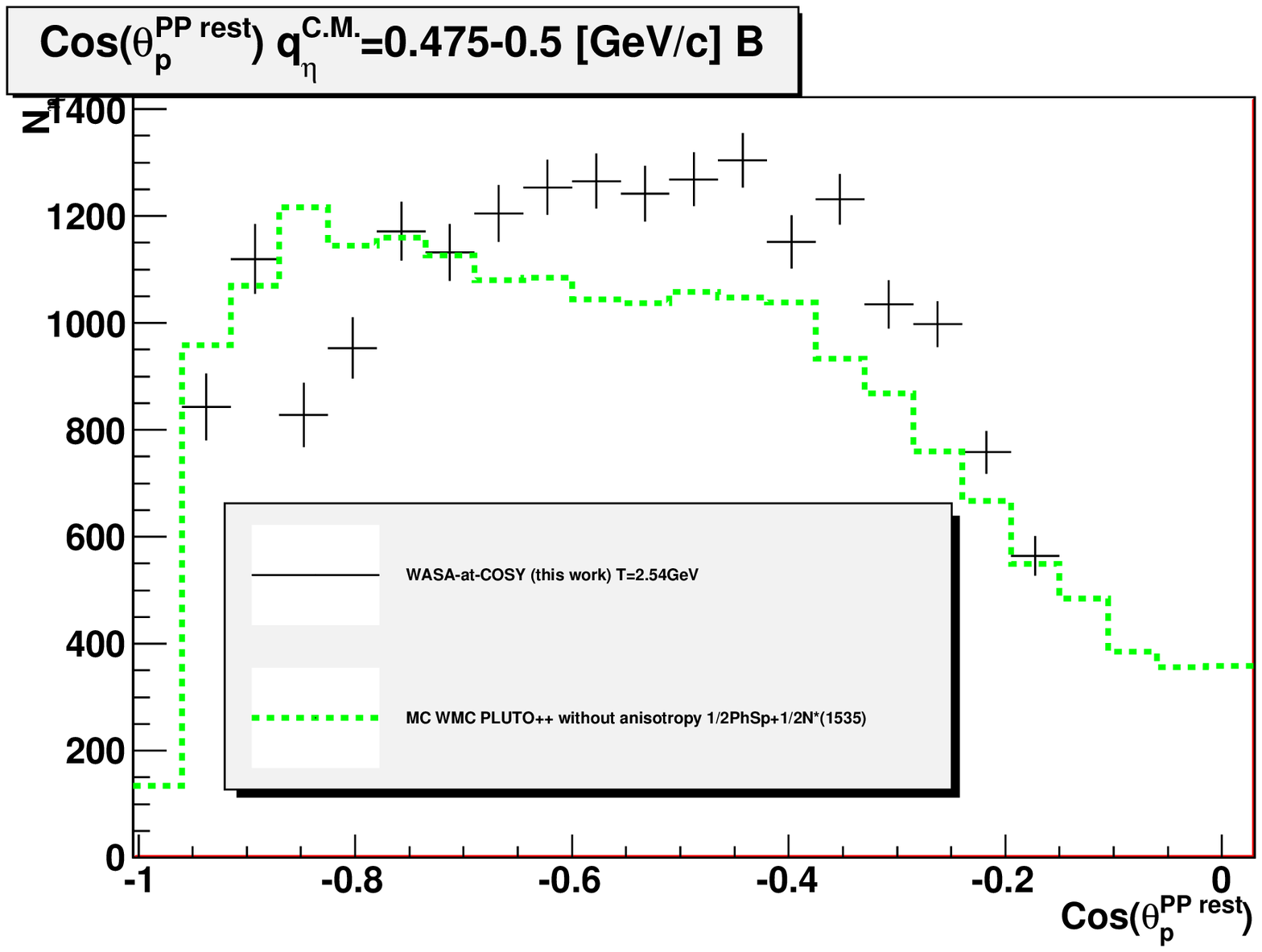}} \label{fig:EtaPPMomDepNB}}\\
\subfigure[$q_{\eta}^{CM}=0.5-0.55\mathrm{~GeV/c}$]{\fbox{\includegraphics[width=0.45\textwidth]{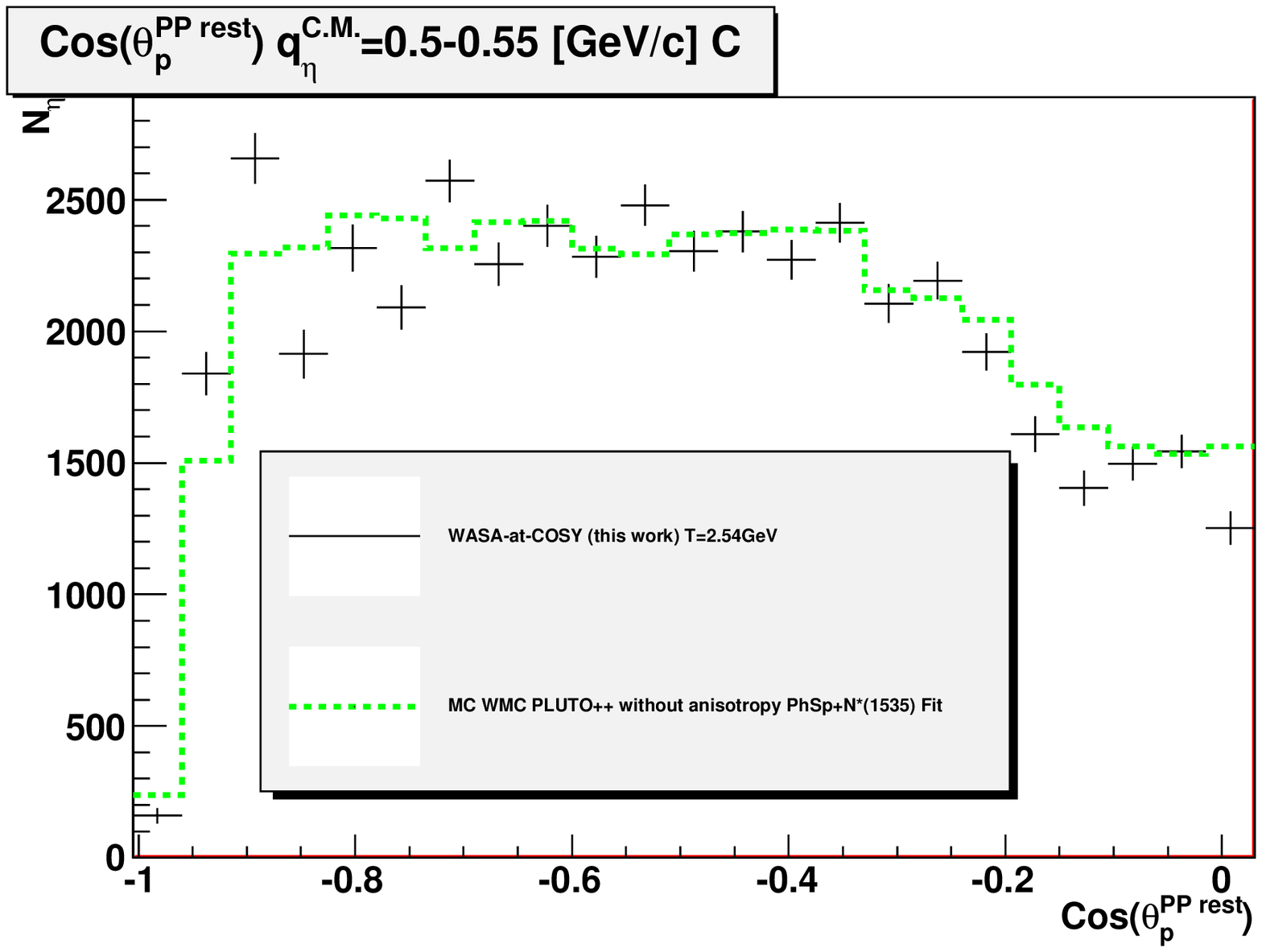}} \label{fig:EtaPPMomDepNC}}\quad
\subfigure[$q_{\eta}^{CM}=0.55-0.7\mathrm{~GeV/c}$]{\fbox{\includegraphics[width=0.45\textwidth]{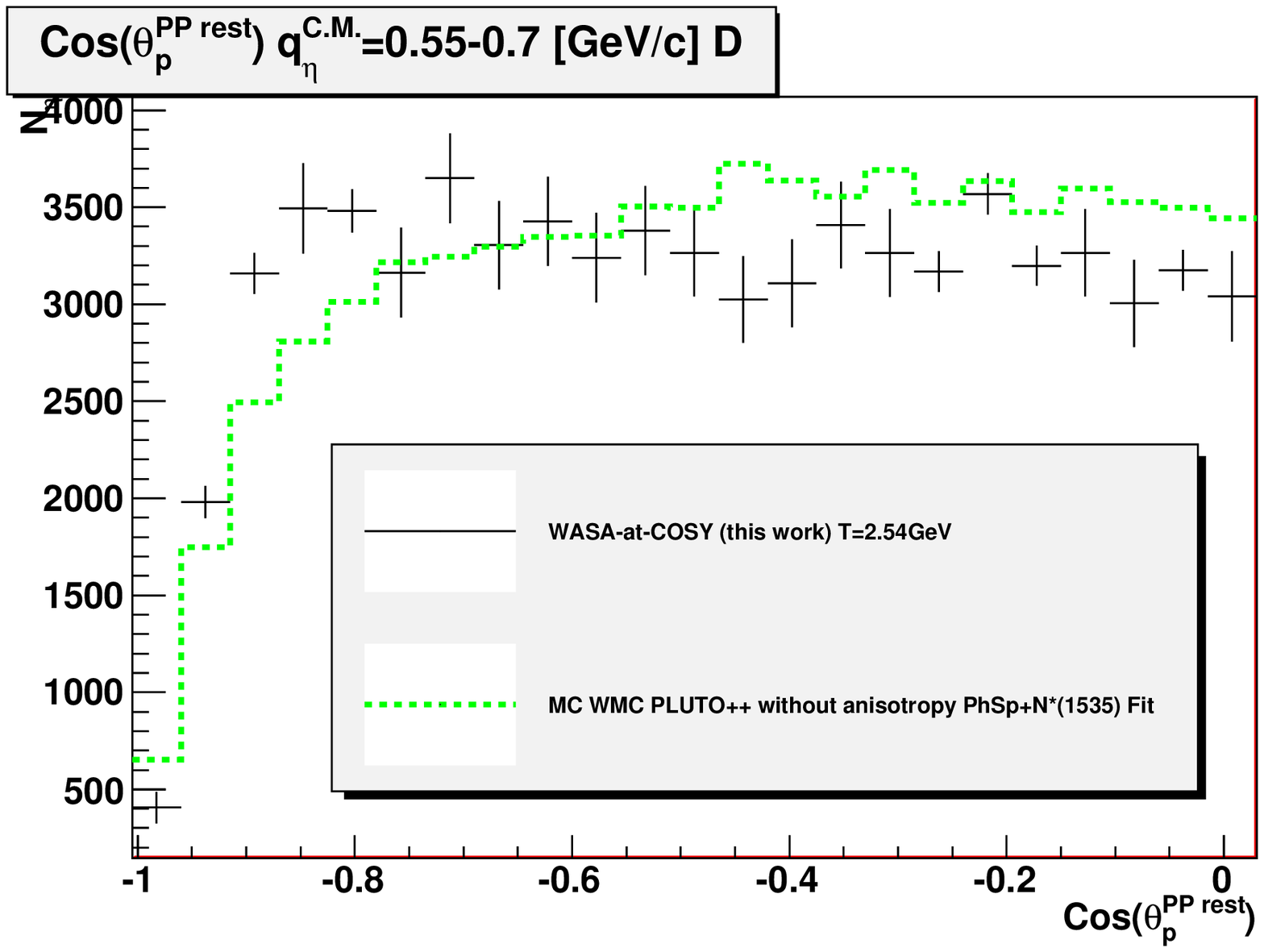}} \label{fig:EtaPPMomDepND}}
}
\caption{$\cos(\theta_{p^{pp}})$ distribution for different momenta of the $\eta$ in CM system $q_{\eta}^{CM}$.
Experimental data (black marker), Monte-Carlo simulation without any proton angle anisotropy (green line)}
\label{fig:EtaPPMomDepN}
\end{sidewaysfigure}

\begin{sidewaysfigure}

\centering
{
\subfigure[$q_{\eta}^{CM}=0.45-0.475\mathrm{~GeV/c}$]{\fbox{\includegraphics[width=0.45\textwidth]{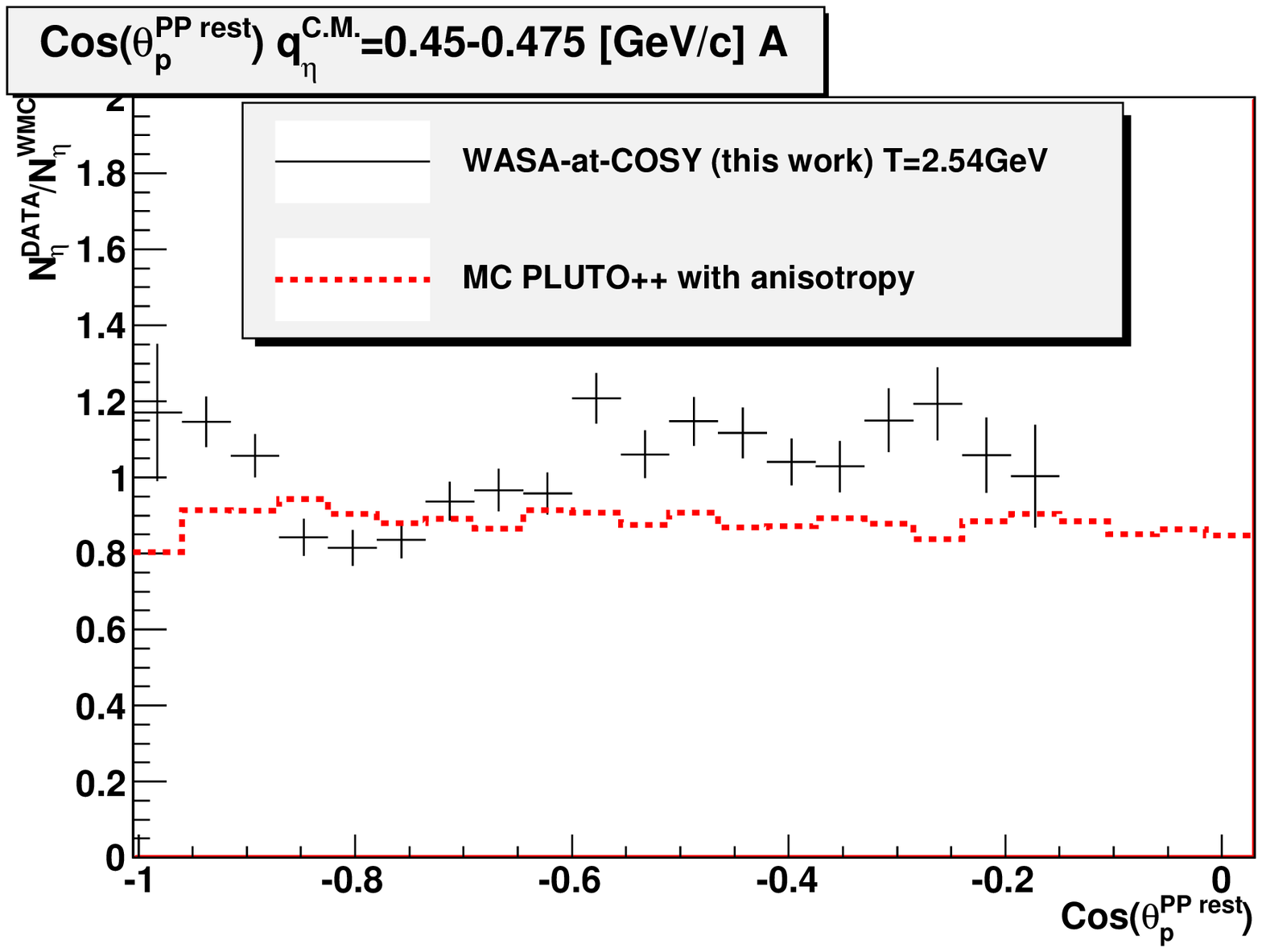}} \label{fig:EtaPPMomDiffNA}}\quad
\subfigure[$q_{\eta}^{CM}=0.475-0.5\mathrm{~GeV/c}$]{\fbox{\includegraphics[width=0.45\textwidth]{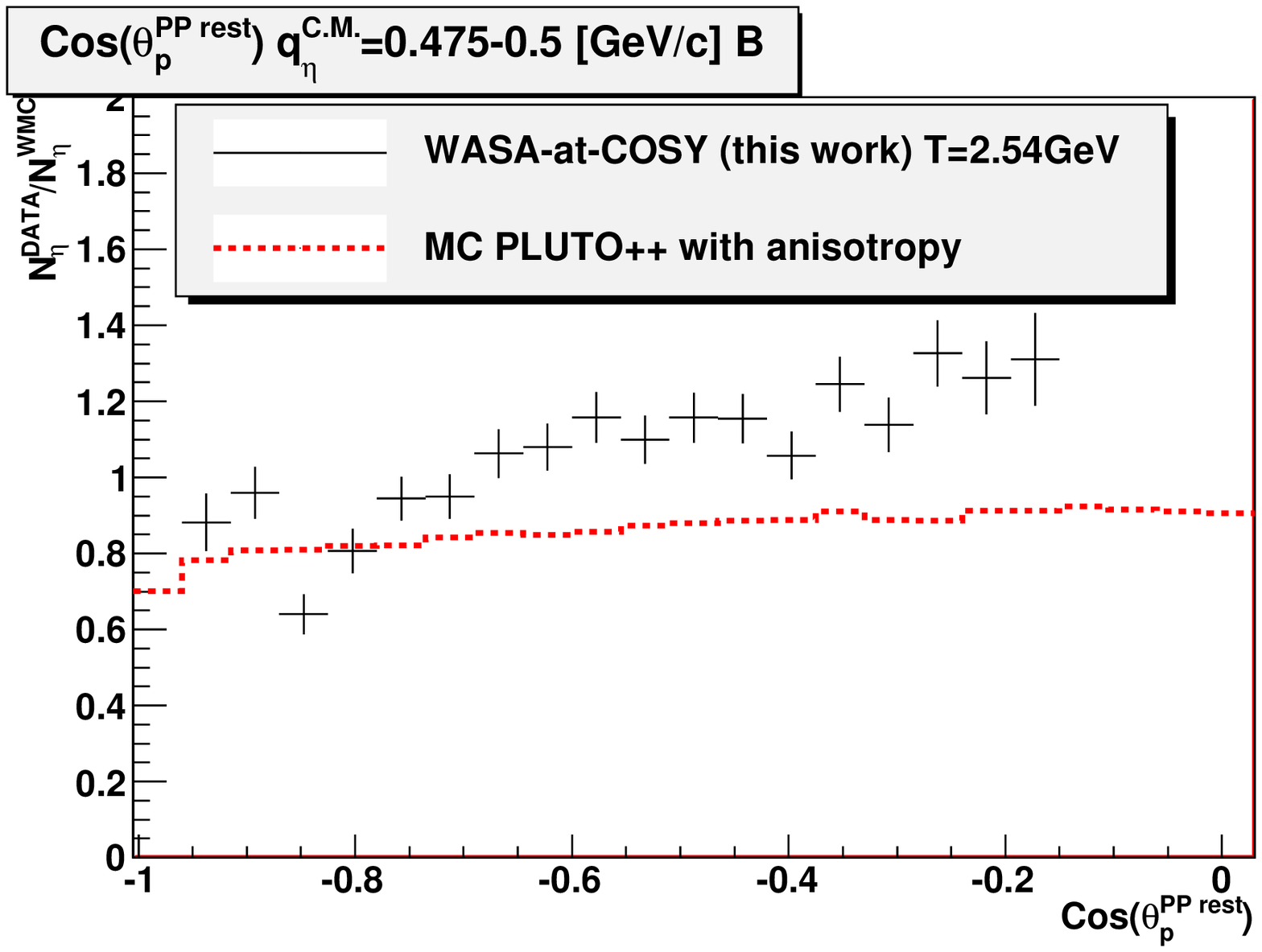}} \label{fig:EtaPPMomDiffNB}}\\
\subfigure[$q_{\eta}^{CM}=0.5-0.55\mathrm{~GeV/c}$]{\fbox{\includegraphics[width=0.45\textwidth]{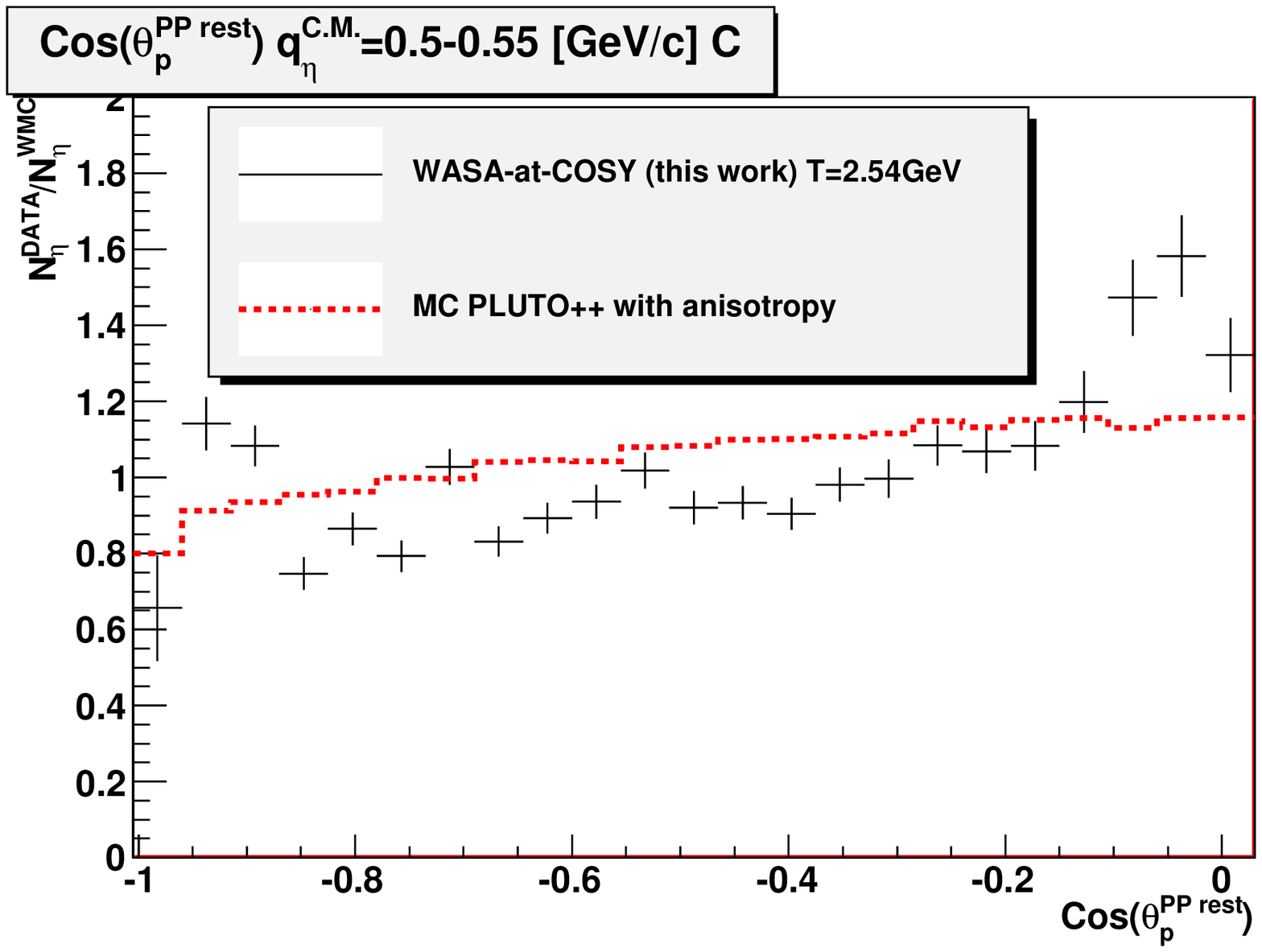}} \label{fig:EtaPPMomDiffNC}}\quad
\subfigure[$q_{\eta}^{CM}=0.55-0.7\mathrm{~GeV/c}$]{\fbox{\includegraphics[width=0.45\textwidth]{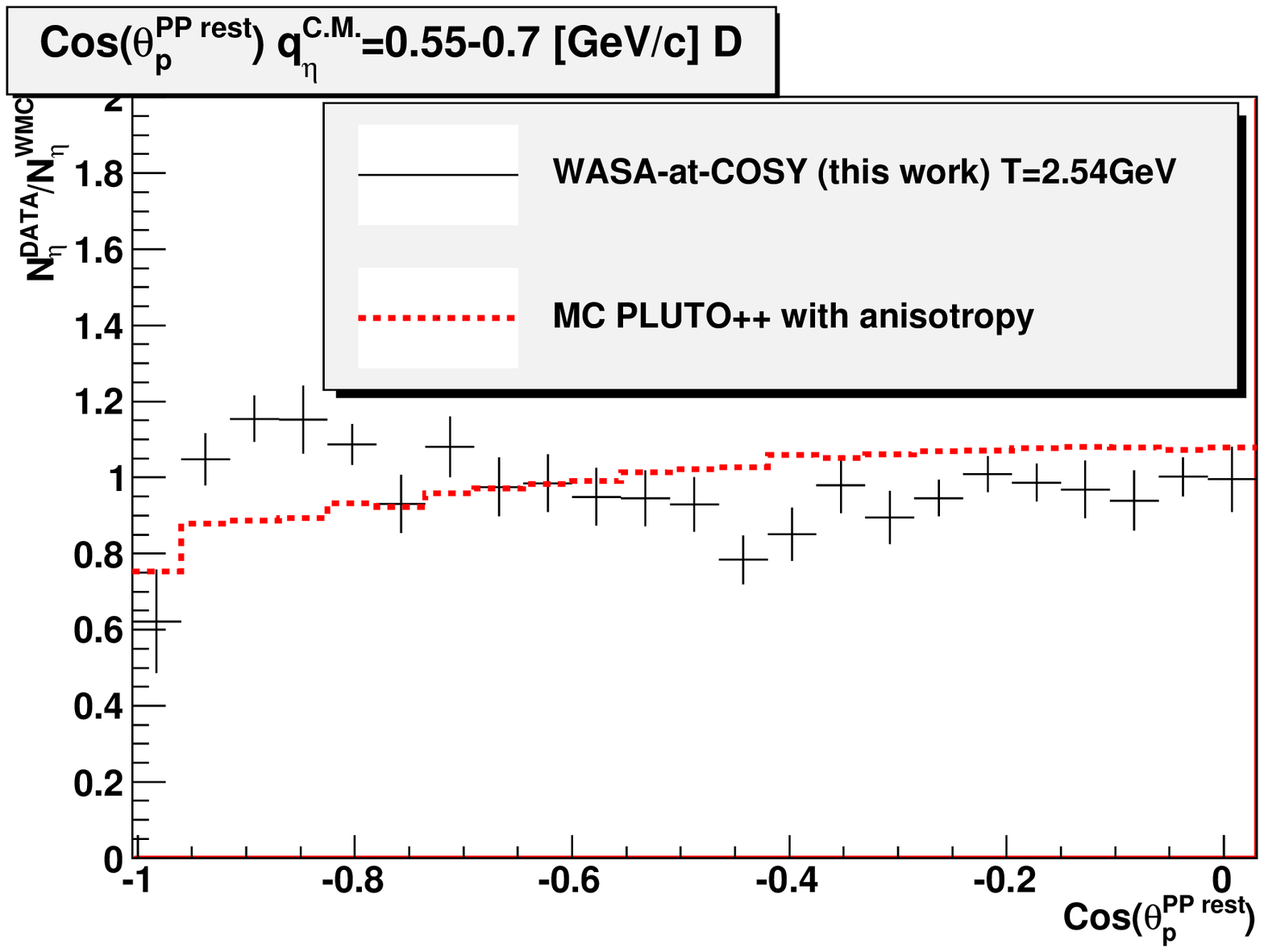}} \label{fig:EtaPPMomDiffND}}
}
\caption{The extracted anisotropy (experimental data divided by the Monte-Carlo simulation of the $\eta$ meson production \myImgRef{EtaIMEtaPFitresultNew2} without any proton angle anisotropy) versus $\cos(\theta_{p^{pp}})$ for different momenta of the $\eta$ in CM system $q_{\eta}^{CM}$.
Experimental data (black marker), Monte-Carlo simulation with anisotropy (arbitrary normalization - to guide the eye) - PLUTO++ parametrization (red line)}
\label{fig:EtaPPMomDiffN}
\end{sidewaysfigure}

Next also the $\cos(\theta_{p^{pp}})$ was studied for the same four different regions of the $q_{\eta}^{CM}$.
To obtain the background free $\cos(\theta_{p^{pp}})$
the missing mass of the two protons was plotted against the $\cos(\theta_{p^{pp}})$. Then the background was fitted outside the $\eta$ meson
peak by the second order polynomial and subtracted from the data, this was done bin by bin.
The experimental data were compared with a Monte-Carlo simulation (phase space and $N^{*}(1535)$ excitation)
 without any assumed proton angle anisotropy (Fig.~\ref{fig:EtaPPMomDepN}).It is seen that for four different ranges in the $\eta$ momentum
the shape of Monte-Carlo and experimental data changes differently. One see here also the momentum dependence of the angular distribution.
To extract this effect and correct for acceptance and efficiency bias, the experimental data were divided by the Monte-Carlo simulation (Fig.~\ref{fig:EtaPPMomDiffN}).
The distribution was compared with a Monte-Carlo simulation with proton angle anisotropy model in PLUTO++ \cite{Pluto} , which is based on \cite{DISTO}.

The extracted anisotropy (experimental data divided by the Monte-Carlo simulation of the $\eta$ meson production \myImgRef{EtaIMEtaPFitresultNew2} without any proton angle anisotropy)
 distributions for four different region of the $\eta$ meson momentum in the CM system
were compared between each other and with the Monte-Carlo simulation with a proton angle anisotropy parametrization based on Pluto++ \cite{Pluto} \myImgRef{EtaPPComaprisonNew2}.
The distributions were normalized so that for the $\cos(\theta_{p^{pp}})=-0.44$ all have the value $1$.    
The effect is not so prominent as in case of the $\cos(\theta_{\eta^{CM}})$. But still one sees a systematic change of the angular distribution
with the $\eta$ momentum.

\myFrameFigure{EtaPPComaprisonNew2}{The extracted anisotropy (experimental data divided by the Monte-Carlo simulation of the $\eta$ meson production \myImgRef{EtaIMEtaPFitresultNew2} without any proton angle anisotropy) as a function of $\cos(\theta_{p^{pp}})$ for four different
 regions of the $\eta$ momentum in the CM system, compared with the Monte-Carlo simulation with a proton angle anisotropy parametrization
 from Pluto++. The normalization of the histograms is described in text.}{Experimental data, $\cos(\theta_{p^{pp}})$ comparison.}

\bigskip
Concluding, one observes the momentum dependence of the angular distribution, for energy $T=2.54\mathrm{GeV}$, which was measured for the first time,
 for the available part of the phase space (Eq.~\ref{eq:EtaAcceptanceCut}). 
The strongest effect is seen in the $\cos(\theta_{\eta^{CM}})$ distribution, which is different 
from the available $\eta$ angle anisotropy model used by PLUTO++ \cite{Pluto} (Integrated over all $\eta$ momentum), which is based on \cite{DISTO} \myImgRef{EtaCMDiffComparison2New2}.  
Since the distribution changes from almost flat to the curved one, when the momentum of the $\eta$ increases, this indicates
that the different partial waves contribute \cite{PDG2008, QMSchiff}.
 
The angular distributions of the $\eta$ meson are available as tables of numbers in Appendix~\ref{appendix:DataTables}.
\newpage ~
\thispagestyle{empty}
\emptydoublepage
\newpage

\section{Summary and Conclusions}\label{sec:SumConclusions}
\thispagestyle{plain}

For the first time the prompt $pp \rightarrow pp 3\pi^{0}$ reaction channel was measured at incident proton momentum of $3.35\mathrm{~GeV/c}$
 with the WASA-at-COSY detector setup (see Section~\ref{sec:setup} on page~\pageref{sec:setup}).
All the final state particles i.e. two protons and three pions, were identified and their four momenta reconstructed
 from the signals in the detectors (see Section~\ref{sec:AnalysisExperimental} on page~\pageref{sec:AnalysisExperimental})
-- around one million of the $pp \rightarrow pp 3\pi^{0}$ clean events were obtained.

First, it was observed that the experimental data could not be described by the Monte-Carlo $pp \rightarrow pp 3\pi^{0}$
 assuming homogeneously and isotropically populated phase space
 (see Fig.~\ref{image_MMpp02PhSp.eps} on page \pageref{image_MMpp02PhSp.eps} and Fig.~\ref{image_ComparisonPhSp.eps} on page \pageref{image_ComparisonPhSp.eps})  
  
Later, the dynamics of the reaction was studied by the missing mass of the two protons $MM_{pp}$ dependent Dalitz and Nyborg plot analysis (see Section~\ref{sec:PP3pi0reaction} on page~\pageref{sec:PP3pi0reaction}) 
together with the proposed kinematic calculations by Monte-Carlo model based on simultaneous excitation of two baryon resonances $\Delta(1232)$ and $N^{*}(1440)$ and their decays
leading to the $pp3\pi^{0}$ final state (Eq.~\ref{eq:Model3pi0First} on page \pageref{eq:Model3pi0First}).
The model assumes two decay branches of the $N^{*}(1440)$, the direct decay $N^{*}(1440) \rightarrow N \pi \pi$
 and the sequential decay $N^{*}(1440) \rightarrow \Delta(1232) \pi \rightarrow N \pi \pi$ and it is based on the kinematic calculations by PLUTO++ event generator \cite{Pluto} (see Appendix~\ref{appendix:wmc} on page \pageref{appendix:wmc}).

\myTable{
\footnotesize
\begin{adjustwidth}{-2cm}{-2cm}
\begin{tabular}{|c|c|c|c|}
\hline
This Work & PDG  & CELSIUS-WASA  & Bonn-Gatchina  \\
(Eq.~\ref{eq:RatioRoper} on page \pageref{eq:RatioRoper}) & \cite{PDG2008} &\cite{SkorodkoRoper} &(at $m_{N^{*}(1440)}=1436\mathrm{~MeV/c^{2}}$) \cite{BonnRoper} \\
\hline
\hline
 $0.039 \pm 0.011 (stat.)  \pm 0.008 (sys.)$ & $0.166-0.5$ & $1.0 \pm 0.1$& $1.20 \pm 0.11$\\ 
\hline
\end{tabular}
\end{adjustwidth}
}{Comparison of the ratio $R=\Gamma(N^{*}(1440) \rightarrow N \pi \pi )/\Gamma(N^{*}(1440) \rightarrow \Delta(1232) \pi \rightarrow N \pi \pi)$
 of the partial decay widths for the decay of the Roper resonance $N^{*}(1440 )$. The ratio is presented for the Roper resonance mass of $1440\mathrm{~MeV/c^{2}}$.}{tab:SumN1440}

The fraction of those two decays was extracted from the experimental data by comparing event populations on Dalitz and Nyborg plots using the chi-square method (see Section~\ref{par:modelparam} on page~\pageref{par:modelparam}).     
The resulting value of the ratio $R=\Gamma(N^{*}(1440) \rightarrow N \pi \pi )/\Gamma(N^{*}(1440) \rightarrow \Delta(1232) \pi \rightarrow N \pi \pi)=0.039 \pm 0.011 (stat.)  \pm 0.008 (sys.)$ (Eq.~\ref{eq:RatioRoper} on page \pageref{eq:RatioRoper})
 together with comparison to the existing data is presented in Table~\ref{tab:SumN1440}.
The obtained value differs from the other experimental measurements \cite{PDG2008, SkorodkoRoper,BonnRoper} significantly.
The closest is the estimation from the PDG~\cite{PDG2008} $R=0.166-0.5$ ($\sim 7$~standard~deviations difference), which is based on the Partial Wave Analysis \cite{PWA} of many data sets.
The Bonn-Gatchina value \cite{BonnRoper} (based on the Partial Wave Analysis \cite{PWA}) as well as the CELSIUS-WASA \cite{SkorodkoRoper}
 (based on the amplitude considerations of the $pp \rightarrow pp 2\pi^{0}$ reaction \cite{ValenciaModel,SkorodkoThesis}) give value $R\approx 1$ ($\sim 96$~standard~deviations difference and $\sim 9$~standard~deviations including errors of Bonn-Gatchina and CELSIUS-WASA). 
The difference between the results might be also interpreted as complex multi-quark structure of the $N^{*}(1440)$ resonance -- ``breathing mode of the nucleon'' \cite{HPMorschBreathing}.  
This complex structure may lead to a different behavior depending on the production mechanism.  
Nevertheless, for the first time one measures this ratio $R$ in a direct way -- by comparing experimental data with the Monte-Carlo simulations; 
not by the extraction from the Partial Wave Analysis \cite{PDG2008,BonnRoper} or as a result of the amplitude analysis \cite{SkorodkoRoper}, which is an indirect method.   
It is also seen from the studies presented in this work that the $N^{*}(1440) \rightarrow \Delta(1232) \pi \rightarrow N \pi \pi$ sequential decay is a leading mode of the $3\pi^{0}$ production (constitutes $\sim 95\%$ Eq.~\ref{eq:3pi0Model}).

To describe the event population in $MM_{pp}$ (proton-proton missing mass) the missing mass population function was introduced, extracted
 from the experimental data (Eq.\ref{eq:3pi0Model} on page \pageref{eq:3pi0Model}). 
It was shown that the model describes the data significantly better than the homogeneously and isotropically populated phase space.

\noindent Studying the kinematics of the reaction, two possible explanations of the origin of $f(MM_{pp})$~(Eq.\ref{eq:3pi0Model}) -- the missing mass population function were considered.

\begin{itemize}
 \item The possibility of $\Delta(1232) N^{*}(1440)$ interaction via One Boson Exchange (OBE) (see Section~\ref{par:DNinteraction} on page \pageref{par:DNinteraction}), which was excluded due to completely
 different behavior of the $f(MM_{pp})$ as a function of the average difference of the four momentum than in OBE \cite{OBE1,OBE2,OBE3,OBE4,OBE5,OBE6,OBE7,OBELast};

 \item Next, it was shown that the $MM_{pp}$ is very sensitive to the structure of the spectral line shape of the $N^{*}(1440)$ (see Section~\ref{par:Nlineshape} on page \pageref{par:Nlineshape}).
The proposed modification of the spectral line mainly of the $N^{*}(1440)\rightarrow \pi^{0}\Delta(1232)$ substitutes the explicitly added $f(MM_{pp})$;
since the main effect influencing the description of the data is due to the modification of the $N^{*}(1440)\rightarrow\pi^{0}\Delta(1232)$ line shape (it is the
leading mode of $3\pi^{0}$ production $\sim 95\%$; see Eq.~\ref{eq:3pi0Model}). 
In particular the proposed modification of this spectral line was found to be very similar to the Breit-Wigner distribution.
This might indicate that the proposed by PLUTO++ \cite{Pluto} modification of the $N^{*}(1440)$ spectral line caused by decay to $\Delta(1232)$ (unstable hadron) is not so strong as proposed
 and seems to be not necessary in case of the reaction considered in this work. Nevertheless more detailed studies of this effect are needed.
For instance, it remains inconclusive whether the spectral line shape of $N^{*}(1440)$ remains the same in case of $N^{*}(1440) \rightarrow p\pi^{0}\pi^{0}$
since the contribution of this branch consists of only $\sim 5\%$ (see Eq.~\ref{eq:3pi0Model}).  
 
\end{itemize}

\noindent It would be very interesting task to see how the proposed in this work spectral line of the $N^{*}(1440)\rightarrow \pi^{0}\Delta(1232)$ works
 for the case of other reactions by high statistics experiments. 

\noindent The possibility of molecule or bound state creation of $\Delta(1232)N^{*}(1440)$ system
as well as the excitations of the quark-gluon degrees of freedom was not excluded.
It would require to consider the dynamical microscopic model of the reaction which is missing
(anyway such models are always parameter dependent).        

\medskip
The detailed validation of the Monte-Carlo developed Model (Eq.~\ref{eq:3pi0Model}) was performed (see Section~\ref{subsec:MCModelValidation} on page~\ref{subsec:MCModelValidation}).
 
\noindent To check the consistency of the model with experimental data, 
the model was compared with the experimental data for all considered spectra i.e. Dalitz and Nyborg plots for different missing mass ranges
-- the statistical analysis was performed (see Section~\ref{par:ModelConsistency} on page \pageref{par:ModelConsistency}).
It was shown that the Monte Carlo developed model describes $\sim80\%$ of the experimental data within the statistical errors of one standard deviation.
It was also shown that around $100\%$ of the experimental data points
is described by the proposed model withing the statistical error of $\sim 2.5$ standard deviations.
Concluding the developed model fully describes the data within the statistical precision of data and model.   
Five body $pp \rightarrow pp 3\pi^{0}$ reaction is fully described by an initially two body process.

\noindent Verification of a possibility of other processes contribution to the Monte Carlo developed model (Eq.~\ref{eq:3pi0Model}) was performed
using Monte-Carlo simulation of $pp \rightarrow pp 3\pi^{0}$ reaction (based on homogeneously and isotropically populated phase space)
 to mimic other processes (see Section~\ref{par:ModelOtherProc} on page~\pageref{par:ModelOtherProc}).
The chi-square method was used for this task.
Concluding, the contribution of other processes estimated here falls to value $\sim 2\%$.

\medskip
The cross section for the $pp \rightarrow pp 3\pi^{0}$ reaction channel, using known cross section of $pp \rightarrow pp \eta(3\pi^{0})$ reaction as a normalization,
 was extracted (see Section~\ref{subsec:CrossSection} on page \pageref{subsec:CrossSection}):

\begin{equation}
\sigma_{pp \rightarrow pp 3\pi^{0}} = 123 \pm 1 (stat.) \pm 8 (sys.) \pm 19 (norm.)~\mu b ~.\nonumber
\end{equation}

\noindent The result was compared with the available data \cite{ETACS01} and models for $pp \rightarrow pp 3\pi^{0}$ cross sections \cite{3pi0CSmodel1, 3pi0CSmodel2, 3pi0CSmodel3}, (Fig.~\ref{image_CS3pi0MyPoint} on page~\pageref{image_CS3pi0MyPoint}).
The data confirm the cross section scaling model based on Delof Final State Interaction \cite{3pi0CSmodel1}. 

\noindent One can predict also the cross section for the $pp \rightarrow pp \pi^{+}\pi^{-}\pi^{0}$ reaction assuming that the reaction follows also via simultaneous excitation of the
$\Delta(1232)$ and $N^{*}(1440)$ baryon resonances, see Table~\ref{tab:ReacScenario} on page \pageref{tab:ReacScenario}. 
The cross section for the  reaction is predicted to be $\sigma_{pp \rightarrow pp \pi^{+}\pi^{-}\pi^{0}} = 861 \pm 147 \mu b$. 

\medskip
Using the developed model (Eq.~\ref{eq:3pi0Model}) the acceptance and efficiency correction of the missing mass of two
 protons dependent Dalitz and Nyborg plots was done (see Section~\ref{subsec:AccEff} on page \pageref{subsec:AccEff}).
The acceptance and efficiency corrected Dalitz and Nyborg plots are available as tables of numbers in Appendix~\ref{appendix:DataTables}.

\bigskip
The $pp \rightarrow pp \eta$ reaction was measured parallely via $\eta$ meson decay into three neutral pions.
Around $200k$ events after background subtraction was available (see Section~\ref{sec:etaprod} on page~\pageref{sec:etaprod}).

First, the experimental accessibility of the phase space was studied.
One finds that the acceptance for this reaction was limited.
It could be expressed in the $\eta$ meson momentum in the Center of Mass frame $q_{\eta}^{CM}$ and cosine of the scattering angle of the $\eta$ meson
 in the CM frame $\cos(\theta_{\eta^{CM}})$ as follows:

\begin{eqnarray}
 &q_{\eta}^{CM}&=0.45-0.7~GeV/c\nonumber \\
&\cos(\theta_{\eta^{CM}})&=-1.0-0.0~.\nonumber
\end{eqnarray}

\medskip
Next, the production mechanism was investigated (see Section~\ref{par:etaprod} on page~\pageref{par:etaprod}).
Two dominant production mechanisms were considered: the resonant production (via excitation of $N^{*}(1535)$) and the non resonant production.
The two scenarios were simulated by the Monte-Carlo and fitted to the background subtracted experimental data distribution of invariant mass of proton-$\eta$ system (see Fig.~\ref{image_EtaIMEtaPFitresultNew2} on page~\pageref{image_EtaIMEtaPFitresultNew2}). 
The contribution of the $N^{*}(1535)$ resonance in the production mechanism was obtained and compared with existing experimental data, see Table~\ref{tab:SumEtaN1535}.
One finds that when the beam kinetic energy $T$ increases the $N^{*}(1535)$ resonance contribution decreases and from $T=2.54\mathrm{~GeV}$ stabilizes at value $\sim 43\%$.   
One can try to compare the result obtained in this work with the closest value at $T=2.5\mathrm{~GeV}$~\cite{DISTO,HadesProc}. 
Between these two measurements the $T$ changes by $40\mathrm{~MeV}$ and excess energy $Q$ by $13\mathrm{~MeV}$ in this range the value changes
 by $\sim 4$~standard~deviations, which is a significant difference.

\myTable{
\footnotesize
\begin{adjustwidth}{-2cm}{-2cm}
\begin{tabular}{|c|c|c|c|c|c|}
\hline
 Beam Kinetic Energy $T$     & $2.15\mathrm{~GeV}$  & $2.5\mathrm{~GeV}$ & $2.54\mathrm{~GeV}$ & $2.85\mathrm{~GeV}$ & $3.5\mathrm{~GeV}$ \\
Excess Energy $Q$ & $325\mathrm{~MeV}$& $442\mathrm{~MeV}$ &$455\mathrm{~MeV}$ &$554\mathrm{~MeV}$ &$752\mathrm{~MeV}$ \\  
 & DISTO & DISTO & This Work & DISTO & HADES \\
                   & \cite{DISTO, HadesProc} & \cite{DISTO,HadesProc} & (Eq.~\ref{eq:MyN1535contrib}) on page \pageref{eq:MyN1535contrib} & \cite{DISTO,HadesProc} & \cite{HadesData}\\ 
\hline
\hline
$N^{*}(1535)$ Contribution                  &$59\%$& $51\%$ & $43.4 \pm 0.8(stat.) \pm 2.0(sys.)\% $ & $43\%$ & $41\%$ \\ 
\hline
\end{tabular}
\end{adjustwidth}
}{Contribution of the $N^{*}(1535)$ in the $pp \rightarrow pp \eta$ production mechanism.}{tab:SumEtaN1535}

\medskip
Later, the angular distributions of the $\eta$ meson in the CM system were studied as well as the proton angular distributions in the proton-proton rest frame.
Both the distributions were studied for four ranges of the $\eta$ meson momentum in the Center of Mass frame $q_{\eta}^{CM}$ (see Section~\ref{par:etaDistr} on page~\pageref{par:etaDistr}). 
For the first time, one observes the momentum dependence of the angular distributions. The strongest effect is seen in the $\cos(\theta_{\eta^{CM}})$ distribution, which is different 
from the $\eta$ angle anisotropy predicted by PLUTO++ \cite{Pluto} based on \cite{DISTO} (Fig.~\ref{image_EtaCMDiffComparison2New2} on page~\pageref{image_EtaCMDiffComparison2New2}).  
Since the distribution changes from almost flat to the curved one, when the momentum of the $\eta$ increases, this indicates
that the different partial waves can contribute \cite{PDG2008, QMSchiff}. It seems that to describe the angular distributions for the three highest ranges of $q_{\eta}^{CM}$ at least p wave is important.
The appropriate model would be needed as well as the high statistics experimental measurements for different beam energy covering full phase space are necessary
to study this effect in details.  

\noindent The angular distributions of the $\eta$ meson are available as tables of numbers in Appendix~\ref{appendix:DataTables}.

\bigskip

Concluding, the multipion reactions in nucleon-nucleon collisions can be used as a precision tool to directly access the properties of the baryon resonances
 by using proposed in this work methods (one would name it multipion spectroscopy):
\begin{itemize}
 \item the baryon resonances are identified by their unique decays topology on the missing mass of two protons $MM_{pp}$ dependent Dalitz and Nyborg plots
-- the strength of the different resonances decays and branching ratios could be precisely extracted (see Section~\ref{subsec:3pi0MCmodel} on page \pageref{subsec:3pi0MCmodel}),

 \item the missing mass of the two protons $MM_{pp}$ distribution is sensitive to the baryon resonance spectral line shape and to the interaction between the baryons (see Section~\ref{par:fMMppOrigin} on page \pageref{par:fMMppOrigin}).
\end{itemize}

\noindent Here also variable other than $MM_{pp}$ might be considered depending on the reaction details. 

\medskip

It would be very valuable to investigate the properties of prompt $pp \rightarrow pp 3\pi^{0}$ reaction for many different energy regions to see the influence of the different baryon resonances contributions
 as well as the  $pp \rightarrow pp \pi^{+}\pi^{-}\pi^{0}$ reaction to confirm the predictions for the cross section and to study the dynamics of this reaction.
Also never measured prompt $pp \rightarrow pp 4\pi$ and $pp \rightarrow pp 5\pi$ reactions both for charged and neutral pions in final state could be very interesting object for studies to understand these processes;
since e.g. the $pp \rightarrow pp 4\pi$ could proceed via simultaneous excitation of two $N^{*}(1440)$ resonances and in the $pp \rightarrow pp 5\pi$ case both the $N^{*}(1440)$ and higher baryon resonance might be involved.

Besides, there exist no dynamical microscopic model for the prompt pion productions in nucleon-nucleon collisions like $3\pi, 4\pi, 5\pi$ productions in contrast to the $NN \rightarrow NN \pi \pi$ reactions 
where complete dynamical microscopic model based on the excitations and decays of various baryon resonances exists \cite{ValenciaModel}.
The results presented in this work (available as tables of numbers in Appendix~\ref{appendix:DataTables}) could be used as an input for testing such a model in future.
Due to the high energy needed to excite $3\pi, 4\pi, 5\pi$, it might be more plausible to use for the future model the microscopic approaches based on the Quantum Chromo Dynamics (QCD) (which takes into account excitations of quark-gluon degrees of freedom) \cite{QCD01,QCD02,HSD0,HSD}
rather then common  existing effective microscopic models \cite{OBE1,OBE2,OBE3,OBE4,OBE5,OBE6,OBE7,OBELast} ( which mimic interaction by exchange of various light mesons like $\pi, \eta, \rho, \omega$
and are more applicable for lower energies \cite{OBElimits}) and which might be very difficult theoretically. 

Full Partial Wave Analysis \cite{PWA} of the resulting data can also be performed in order to get the more details of the reaction mechanism,
 again it forms a formidable task. 

Other approaches to the analysis and visualization of the multidimensional data phase space of prompt multipion productions and model comparisons could be also considered.
One may think about the generalization of the Dalitz Plot to five-particle Dalitz Plot (five dimensional pentahedron representation) and to even more-particle Dalitz Plot in the way how it was proposed and successfully used to visualize atomic break-up processes by using Four-particle Dalitz Plot (four dimensional tetrahedron representation)
which visualize the multidimensional data phase space \cite{4Ddalitz,4Ddalitz2}.

\noindent Another idea would be to use the SOM -- Self-Organizing Map \cite{SOM} or the GTM -- The Generative Topographic Mapping \cite{GMT} techniques which visualize the multidimensional data phase space as a two-dimensional plot \cite{GMT1}.
The Andrews Curves \cite{AndrewsCurves} and their extensions \cite{AndrewsCurvesGen} which represent a multidimensional data points as an orthogonal curves could be also taken into account.   
The methods like discussed above are mathematically complicated and may require time-demanding supercomputing.

\bigskip
\noindent The multipion reactions seems to be promising field of the future scientific exploitations particularly in the area of the baryon resonances spectroscopy.

\newpage ~
\thispagestyle{empty}
\emptydoublepage
\newpage
\begin{appendices}

\newpage~
\thispagestyle{empty}
\emptydoublepage
\section{\mbox{Kinematics of five particle} \mbox{phase space}}\label{appendix:Kine5part}
\thispagestyle{plain}
\renewcommand{\thetable}{\Alph{section}.\arabic{table}}
\renewcommand{\theequation}{\Alph{section}.\arabic{equation}}
\renewcommand{\thefigure}{\Alph{section}.\arabic{figure}}
\setcounter{table}{0}
\setcounter{equation}{0}
\setcounter{figure}{0}

One would like to express the event distribution (Eq.~\ref{eq:ProbPhSp4} on page \pageref{eq:ProbPhSp4}) in the following variables:

%
%
%

{
\begin{eqnarray}
M^{2}_{12} &=& \left( E_{1}+E_{2} \right)^{2}+\left( \overrightarrow{p_{1}} + \overrightarrow{p_{2}} \right)^{2} = \left( E_{1}+E_{2} \right)^{2}+p_{12}^{2}\\
M^{2}_{25} &=& s-2 \sqrt{s}\left(E_3+E_4\right)-2 E_1 \left(\sqrt{s}+E_2-E_4\right)+\\
            &-&\left(E_2-E_4\right) \left(E_2+2 E_3+E_4\right)+M^{2}_{123}\nonumber\\
M^{2}_{34} &=& M^{2}_{125}-s+2 \sqrt{s} \left(E_3+E_4\right)\\
M^{2}_{1235} &=& s - 2\sqrt{s}E_{4} + m^{2}_{4}\\
M^{2}_{123} &=& \left(E_{1} + E_{2} + E_{3}\right)^{2}+p_{123}^{2}\\
M^{2}_{45} &=& \left(\sqrt{s} - E_{1}-E_{2}-E_{3}\right)^{2}+p^{2}_{123} 
\end{eqnarray}
} 
and the inverse relations are

{
\footnotesize
\begin{eqnarray}
 E_{1} &=&\frac{s- M^{2}_{34}+ M^{2}_{125}}{2 \sqrt{s}}+\label{eq:NewVariablesE1}\\
   &-& \frac{\sqrt{4 s M^{2}_{25}+\left(s- M^{2}_{45}\right)^{2}-2 M^{2}_{123} \left(s+M^{2}_{45}\right)+M^{4}_{123}}}{2 \sqrt{s}}\nonumber\\
E_{2} &=& \frac{\sqrt{4 s M^{2}_{25}+\left(s-M^{2}_{123}\right)^2-2 M^{2}_{45} \left(s+M^{2}_{123}\right)+M^{4}_{45}}}{2 \sqrt{s}}+\label{eq:NewVariablesE2}\\
      &+& \frac{m_4^2-M^{2}_{45}+M^{2}_{123}-M^{2}_{1235}}{2 \sqrt{s}}\nonumber\\
E_{3} &=& \frac{M^{2}_{34}-m_4^2-M^{2}_{125}+M^{2}_{1235}}{2 \sqrt{s}}\label{eq:NewVariablesE3}\\
E_{4} &=& \frac{s+m_4^2-M^{2}_{1235}}{2 \sqrt{s}}\label{eq:NewVariablesE4}\\
p_{12} &=& \frac{1}{2 \sqrt{s}}\Biggl( 2 m_4^2\left(s-M^{2}_{34}-M^{2}_{45}+M^{2}_{123}+M^{2}_{125}-M^{2}_{1235}\right)+\label{eq:NewVariablesp12} \\
       &+& \left(s-M^{2}_{34}-M^{2}_{45}+M^{2}_{123}+M^{2}_{125}-M^{2}_{1235}\right)^2-4 s M^{2}_{12}+m_4^4 \biggr)^{1/2}\nonumber \\
p_{123} &=& \frac{\sqrt{-2 M^{2}_{45}
   \left(s+M^{2}_{123}\right)+\left(s-M^{2}_{123}\right)^2+M^{4}_{45}}}{2\sqrt{s}}\label{eq:NewVariablesp123}
\end{eqnarray}
}

For the definitions of the symbols see Section~\ref{sec:DefinitionOfVariables} on page \pageref{sec:DefinitionOfVariables}.\\
The Jacobian

\begin{equation}
 Jac = \frac{\partial\left(E_{1}E_{2}E_{3}E_{4}p_{12}p_{123} \right)}{\partial \left( M^{2}_{12}M^{2}_{25}M^{2}_{34}M^{2}_{1235}M^{2}_{123}M^{2}_{45} \right)} = \frac{1}{8 \sqrt{B} \sqrt{C} \sqrt{D + E}}
\label{eq:jacobian}
\end{equation}

where 
{
\footnotesize
\begin{eqnarray}
B &=& 4 s M^{2}_{25}+\left(s-M^{2}_{45}\right)^{2}-2 M^{2}_{123}\left(s+M^{2}_{45}\right)+M^{4}_{123}\\
C &=& -2 M^{2}_{45}\left(s+M^{2}_{123}\right)+\left(s-M^{2}_{123}\right)^{2}+M^{4}_{45}\\
D &=& 2 m_{4}^{2}\left(s-M^{2}_{34}-M^{2}_{45}+M^{2}_{123}+M^{2}_{125}-M^{2}_{1235}\right)\\
E &=& \left(s-M^{2}_{34}-M^{2}_{45}+M^{2}_{123}+M^{2}_{125}-M^{2}_{1235}\right)^{2}-4 s M^{2}_{12}+m_{4}^{4}
\end{eqnarray}
}

one can rewrite (Eq.~\ref{eq:jacobian}):

\begin{equation}
 Jac = \left[ 32 s \left(s - 2E_{1}\sqrt{s} -M^{2}_{34} + M^{2}_{125}\right) p_{12} p_{123} \right]^{-1}
\label{eq:jacobianNew}
\end{equation}
 
with $E_{1},p_{12}, p_{123} $ given by (Eq.~\ref{eq:NewVariablesE1},~\ref{eq:NewVariablesp12},~\ref{eq:NewVariablesp123}).

\bigskip
Now one can now rewrite the event distribution (Eq.~\ref{eq:ProbPhSp4}) in the invariant masses
{
\begin{equation}
 d^{6}\mathcal{P} = \frac{\pi^{4} dM^{2}_{12}dM^{2}_{25}dM^{2}_{34}dM^{2}_{1235}dM^{2}_{123}dM^{2}_{45}}{32 s \left(s - 2E_{1}\sqrt{s} -M^{2}_{34} + M^{2}_{125}\right) p_{12} p_{123}} \left|\mathcal{M}\right|^{2}
\label{eq:ProbPhSp5}
\end{equation}
}

The boundaries of the physical region in $E_{1},E_{2},E_{3},E_{4},p_{12},p_{123}$ variables are
\begin{eqnarray}
 \left[ \sqrt{s} - \left( E_{1}+E_{2}+E_{3} \right)-E_{4}\right]^{2} &=& m^{2}_{5} + \left(p_{123} \pm p_{4} \right)^{2}\label{eq:bound1}\\
p^{2}_{12} &=& \left(p_{1} \pm p_{2}\right)^{2}\label{eq:bound2}\\
p^{2}_{123} &=& \left(p_{12} \pm p_{3}\right)^{2}\label{eq:bound3}
\end{eqnarray}

Which could be transformed into new variables by \newline the relations (Eq.~\ref{eq:NewVariablesE1},~\ref{eq:NewVariablesE2},~\ref{eq:NewVariablesE3},~\ref{eq:NewVariablesE4},~\ref{eq:NewVariablesp12},~\ref{eq:NewVariablesp123}). 

\bigskip

In order to obtain any two dimensional event distributions one has to integrate (Eq.~\ref{eq:ProbPhSp5}) over four variables with the boundary conditions (Eq.~\ref{eq:bound1},~\ref{eq:bound2},~\ref{eq:bound3}).

\newpage ~
\thispagestyle{empty}
\emptydoublepage
\newpage

\section{\mbox{WASA-at-COSY} \mbox{Detector Calibration}}\label{appendix:DetectorCalibration}
\thispagestyle{plain}
\renewcommand{\thetable}{\Alph{section}.\arabic{table}}
\renewcommand{\theequation}{\Alph{section}.\arabic{equation}}
\renewcommand{\thefigure}{\Alph{section}.\arabic{figure}}
\setcounter{table}{0}
\setcounter{equation}{0}
\setcounter{figure}{0}

The Calibration of the detectors is a conversion from electronic channels of ADC(Analog To Digital Converter) or TDC(Time To Digital Converter) 
i.e. arbitrary units of energy and time, 
which are measured by the detector electronics, to the physical units like GeV and ns.
The calibration procedure is different for different detector type, since the principle of measurements are based on other physics phenomena.
The WASA at Cosy detector setup section~\ref{subsec:detector} is built from three different detector types i.e. Plastic Scintillators, Straw 
Tube Detectors and Electromagnetic Calorimeter. The appropriate calibration procedure for them is described below.

\subsection*{Plastic Scintillators}
To obtain calibration of the plastic scintillators several effects have to be taken into account which are causing the non linear conversion.
One uses fast protons from the proton-proton elastic scattering reaction which are close to the minimum ionization having specific constant energy loss in detector elements.

\myFrameSmallFigurer{FRHCalib0}{The light output non uniformity check. ADC signal times $\cos(\theta)$ as a function of the scattering angle $\theta$ of one element of the FTH detector for the elastic scattered protons. Red curve shows the fit of the polynomial of the third degree \cite{CFRThesis}.}{The light output non uniformity check.}

First the possible non uniformity of light collection efficiency by scintillator has to be checked, which depends on geometrical shape of scintillators.
The ADC signal times $\cos(\theta)$ as a function of the scattering angle $\theta$ in detector element is checked \myImgRef{FRHCalib0}. 
The deviations from straight line indicate the non uniformity, the fit to the data is performed to correct for this effect.

\myFrameFigurer{FRHCalib}{Left Figure, ADC signal corrected for non uniformity of light collection efficiency for the two subsequent layers of the 
FRH detector for the elastic scattered protons, characteristic points marked. 
On the right Figure, energy deposit for the characteristic points, as marked on left Figure, for the Monte-Carlo simulation in GeV versus ADC value. The fit (solid line)
 to the correlation describes the calibration function. The linear function marked as a dotted line \cite{VlasovThesis}. }{Calibration of the FRH detector.}

After the corrections for non uniformity of light collection efficiency for each detector element the corrected 
ADC signal (energy loss) for two subsequent detector layers is plotted (dE-E plot) (\ref{image_FRHCalib}~left) and compared with the same dependence for the Monte-Carlo simulation
for the indicated points in the plot which correspond to:

\begin{description}
 \item[(0)] Zero Point
 \item[(1)] Minimum Ionizing Point
 \item[(3)] Punch-through Point - the kinetic energy of the particle is larger then the stopping power of the detector
 \item[(2),(4)] Equilibrium Points - the energy loss of the particle punching through the current layer (2) is as large as the energy loss of the particle stopped in the current layer (4)
 \item[(5)] Maximum Deposit Point
 \end{description}   

Later for the indicated points the energy deposit in one layer is compared to the energy deposit from the Monte-Carlo simulation, resulting as a correlation plot (\ref{image_FRHCalib}~right).
Next the fit to the correlation plot is performed to get the conversion from the ADC energy deposit (light output) to energy deposit in energy units.
As it is seen the relation is not linear. The nonlinearities can rise from the nonlinearities of the photomultiplier tubes as well as the quenching effect in scintillator.

\subsection*{Straw Tube Detectors}
The Calibration of the Straw Tube Detectors like MDC and FPC is essential to achieve high spatial resolution with these detectors.
In addition to the positions of the anode wires the drift time is measured which is then converted using calibration function to the
drift distance which is understood as a closest approach of the particle trajectory to the nearest anode wire of the straw tube.
The time to distance relation (the calibration) depends on the magnetic field in which the detectors are, the gas mixture in the detectors and the voltage applied to the anode wires.
It has to be found for each change of these conditions.
To derive the calibration the following assumptions has be fulfilled:
\begin{itemize}
 \item the signals in the detectors are consequence of physical particle tracks - no noise
 \item the straw tubes are $100\%$ efficient
 \item the straw tubes are homogeneously irradiated
\end{itemize}

\begin{equation}
  \dfrac{dn}{dr} = \dfrac{N_{Tot}}{R_{Tube}} = const. 
\label{eqAppE:HomoCond} 
\end{equation}

where $n$~-~number of events, $r$~-~distance measured from anode wire, $N_{Tot}$~-~total number of events registered by straw tube, $R_{Tube}$~-~the radius of the straw tube.

Now one can write the drift velocity $v(t)$~:

\begin{equation}
 v(t) = \dfrac{dr}{dt} = \dfrac{dn}{dt}\dfrac{dr}{dn} = \dfrac{R_{Tube}}{N_{Tot}} \dfrac{dn}{dt}
\label{eqAppE:DriftV}   
\end{equation}

To get the time to distance relation (the calibration) one integrates \myEqRef{eqAppE:DriftV} and gets:

\begin{equation}
 D(t) = R_{Tube} \dfrac{\int_{T_{0}}^{t} n(t)dt}{\int_{T_{0}}^{T_{max}} n(t)dt} =  R_{Tube} \dfrac{\int_{T_{0}}^{t} n(t)dt}{N_{Tot}} 
\label{eqAppE:Calib} 
\end{equation}

where $T_{0}$~-~starting point of the drift time measurement, $T_{max}$~-~maximal drift time
It is essential to precisely determine the $T_{0}$ which is also the time reference of the individual TDCs.
To eliminate the trigger time and jitter one uses the relative time between the straws and nearest plastic scintillator.
For the FPC the FTH detector is used and for the MDC the PSB detector respectively.

The process of the MDC calibration is shown \myImgRef{DriftTime}, the gas mixture used in the straw tubes $Ar(80\%)+C_{2}H_{6}(20\%)$ causes the linearity of the time to distance relation - the calibration.  
\newline
\myFrameFigure{DriftTime}{The Calibration of the MDC, the gas mixture used ($Ar(80\%)+C_{2}H_{6}(20\%)$). Left: the drift time spectra, Center: the integrated drift time distribution (red vertical lines indicate integration range, after identification the maximal drift range - blue horizontal lines), Right: the time to distance relation used as a calibration  \cite{CFRThesis}}{The Calibration of the MDC.}

\newpage
\subsection*{Electromagnetic Calorimeter}
The first step in the calibration procedure of the electromagnetic Calorimeter was to obtain the preliminary calibration constants by measuring the response of 
each individual crystal to photons coming from radioactive source \cite{BRJmaster}.
Using these calibration as a first step, for each data taking period the two photon decays of the mesons $\pi^{0}, \eta$ are used to obtain the actual
 set of calibration constants.

For the experimental data the events with the two neutral tracks in Central Detector were selected (see Appendix~\ref{appendix:TrackReconstructionCD}), regarded as photons, and invariant mass of them was computed:

\begin{eqnarray}
IM_{\gamma_{1}\gamma_{2}} & = & \sqrt{(E_{1}^{\gamma}+ E_{2}^{\gamma})^{2} - (\overrightarrow{p}_{1}^{\gamma} + \overrightarrow{p}_{2}^{\gamma})^{2} }\\
& = & \sqrt{2 k_{1} k_{2} E_{1}^{\gamma} E_{2}^{\gamma} (1- cos(\theta_{1,2}^{\gamma}) ) } \nonumber
 \label{eqAppE:IMgg}  
\end{eqnarray}

where $E_{1}^{\gamma}, E_{2}^{\gamma}$~-~energies of the photons,$\overrightarrow{p}_{1}^{\gamma}, \overrightarrow{p}_{2}^{\gamma}$~-~momenta vectors of the photons,
 $cos(\theta_{1,2}^{\gamma})$~-~opening angle between the photons, $k_{1}, k_{2}$~-~new calibration factor for the photons.

For each central crystal element of the two neutral clusters in calorimeter, defined as the element with the highest energy deposit in cluster,
 the invariant mass of the two photons $IM_{\gamma_{1}\gamma_{2}}$ was assigned.
It is assumed that the central crystal element of the cluster has the highest impact on the invariant mass and the invariant mass
shift is associated with it.  

To correct for the shift in invariant mass, the deviation from the nominal meson mass of $\pi^{0}$ or $\eta$ is used to determine the new calibration
 factor for the photon energies for each crystal:

\begin{equation}
 \left(E_{1,2}^{\gamma}\right)_{New} = k_{1,2} E_{1,2}^{\gamma} = \dfrac{M^{2}_{\pi^{0}, \eta}}{IM_{\gamma_{1}\gamma_{2}}^{2}} E_{1,2}^{\gamma}
\end{equation}

where $\left(E_{1,2}^{\gamma}\right)_{New}$~-~individually calibrated energies of the photons, $M_{\pi^{0}}= 134.978~\mathrm{MeV/c^{2}}$, $M_{\eta}= 547.8~\mathrm{MeV/c^{2}}$ from PDG~\cite{PDG2008}.
\newline
This procedure is repeated iteratively for each crystal until the correct meson mass position is reached and the result remains stable.

After the iterations also the additional correction is used to improve the result and avoid overcompensation.
The individually calibrated energies of the photons ($\left(E_{1,2}^{\gamma}\right)_{New}$) are corrected for the average position of the invariant mass:

\begin{equation}
 \left(E_{1,2}^{\gamma}\right)_{Avg} = k_{1,2}^{Avg} \left(E_{1,2}^{\gamma}\right)_{New} = \dfrac{ \left(IM^{2}_{ \gamma_{1}\gamma_{2} }\right)^{Avg} }{IM_{\gamma_{1}\gamma_{2}}^{2}}\left(E_{1,2}^{\gamma}\right)_{New}
\end{equation}     
where $\left(E_{1,2}^{\gamma}\right)_{Avg}$~-~average corrected energies of the photons and $\left(IM^{2}_{ \gamma_{1}\gamma_{2} }\right)^{Avg}$~-~invariant mass average over all crystals. 

The following way of the calibration implicitly corrects for the border crystal effects and shower leakages.

\newpage~
\thispagestyle{empty}
\emptydoublepage


\section{\mbox{The~WASA-at-COSY} \mbox{Monte-Carlo} Simulation}\label{appendix:wmc}
\thispagestyle{plain}

\renewcommand{\thetable}{\Alph{section}.\arabic{table}}
\renewcommand{\theequation}{\Alph{section}.\arabic{equation}}
\renewcommand{\thefigure}{\Alph{section}.\arabic{figure}}
\setcounter{table}{0}
\setcounter{equation}{0}
\setcounter{figure}{0}

It is a standard approach in the nuclear and high energy physics to have a full Monte-Carlo simulation of the experimental setup.
The reasons for that are as follows:
\begin{itemize}
\item one needs to cross check the reconstruction procedure
\item it is necessary to get the error parametrization for the Kinematic Fit 
\item one needs to estimate the total reconstruction efficiency
(geometrical acceptance, reconstruction efficiency) for later data correction
\item it could be also used to determine the detector resolution for the indirectly and directly measured observables
\end{itemize}

To do all of the above tasks a Monte-Carlo simulation should fulfill two very important conditions:

\begin{itemize}
\item The virtual detector in Monte-Carlo simulation should be a reflection of the physical one, 
as close as possible. It has to mimic the performances and status of all the components as it was present during the experiment

\item The kinematics of the simulated events has to reflect the event kinematics during the experiment, one has to know the true physical
mechanism of the reaction or at least one has to mimic it by the model as accurate as possible.

\end{itemize}

One needs excellent tools to fulfill these conditions.

The \textbf{GEANT3} (\textbf{Ge}ometry \textbf{an}d \textbf{T}racking) program from CERN \cite{Geant31} is used for simulation of
the physical processes of particles interactions with detector medium in which the whole geometry of WASA-at-COSY detector is virtualized.
That includes the active materials as particle detectors themselves as well as passive one e.g. flanches, supports, air around detector etc.
As an output from GEANT one gets the detector response for the kinematic configuration (4-vectors of the particles).

For the generation of the kinematic configuration of the reaction one uses \textbf{Pluto++} Monte-Carlo event generator \cite{Pluto} version $5.31$.  
In Pluto++ in addition to generation of homogeneously and isotropically populated phase space many realistic models of
 the reaction mechanism are implemented e.g. the production of $\eta$ meson via $N^{*}(1535)$ with angular anisotropy
 or decays of baryon resonances like $\Delta(1232), N^{*}(1440)$.
In the hadronic interactions at low energies for the unstable hadrons (like baryon resonances) one has to take into account the
deviations from the fixed-width Breit-Wigner distribution.
For realistic simulations this is done in \textbf{Pluto++} by introducing the mass dependent width $\Gamma^{Tot}(m)$ being a function of the partial widths $\Gamma^{k}(m)$.   
The mass of the resonance $m$ is sampled from the relativistic Breit-Wigner distribution $h(m)$ with
an appropriate static pole mass $M_{R}$ Table~\ref{tab:wmcMassWidth} of the resonance and the mass dependent width $\Gamma^{Tot}(m)$ (Eq.~\ref{eq:RelBreitWigner},~\ref{eq:RelBreitWigner2}).

\begin{eqnarray}
 h(m) &=& A\frac{m^{2}\Gamma^{Tot.}(m)}{ \left(M^{2}_{R} - m^{2}\right)^{2} + m^{2}\left(\Gamma^{Tot.}(m)\right)^{2}}\label{eq:RelBreitWigner}\\
\Gamma^{Tot}(m) &=& \sum_{k}^{N}\Gamma^{k}(m)\label{eq:RelBreitWigner2}
\end{eqnarray}

with $N$ number of decay modes of the resonance. The constant $A$ is chosen that the integral of the $h(m)$ is equal to $1$.

\myTable{
\begin{tabular}{|c||c|c|}
\hline
Baryon Resonance name & Pole Mass $M_{R}$  & Static Width\\
&$\mathrm{[MeV/c^{2}}]$ &$\mathrm{[MeV/c^{2}}]$ \\
\hline
\hline
$\Delta(1232)P_{33}$ & $1232$ & $120$ \\
\hline
$N^{*}(1440)P_{11}$ & $1440$& $350$ \\
\hline
$N^{*}(1535)S_{11}$& $1535$ & $150$ \\ 
\hline
\end{tabular}
}{Pole Mass and Static Width the selected baryon resonances used in Pluto++ event generator.}{tab:wmcMassWidth}

Also the decays of the resonances to stable and unstable particles are considered by \textbf{Pluto++}. 
In case of the decays of the resonances to the unstable particles spectral line shape of the resonance
is modified by the unstable particle spectral line shape.  
In \textbf{Pluto++} such a cases are threated explicitly giving a possibility to calculate the realistic spectral functions.   

All in all these effects changes the spectral line shape of the resonances from the relativistic Breit-Wigner distribution shape.

\begin{sidewaysfigure}[t!bp]
\centering
{
\subfigure[$\Delta(1232)$ line shape, where $N^{*}(1440) \rightarrow p \pi^{0}\pi^{0}$\newline (stable particles)]{\fbox{\includegraphics[width=0.45\textwidth]{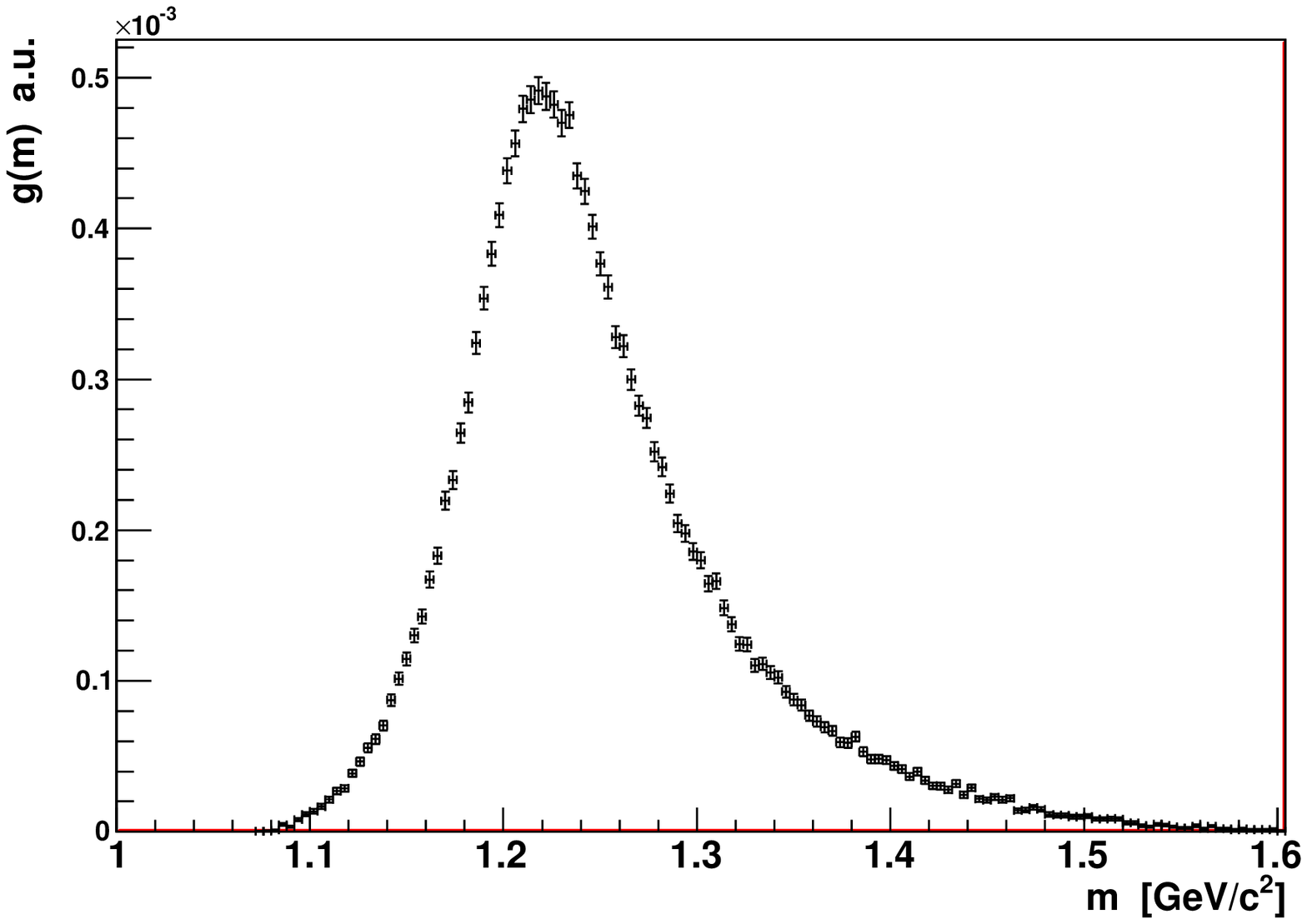}} \label{fig:D1232N1440lineshapeStableD}}\quad
\subfigure[$N^{*}(1440)$ line shape, where $N^{*}(1440) \rightarrow p \pi^{0}\pi^{0}$\newline (stable particles)]{\fbox{\includegraphics[width=0.45\textwidth]{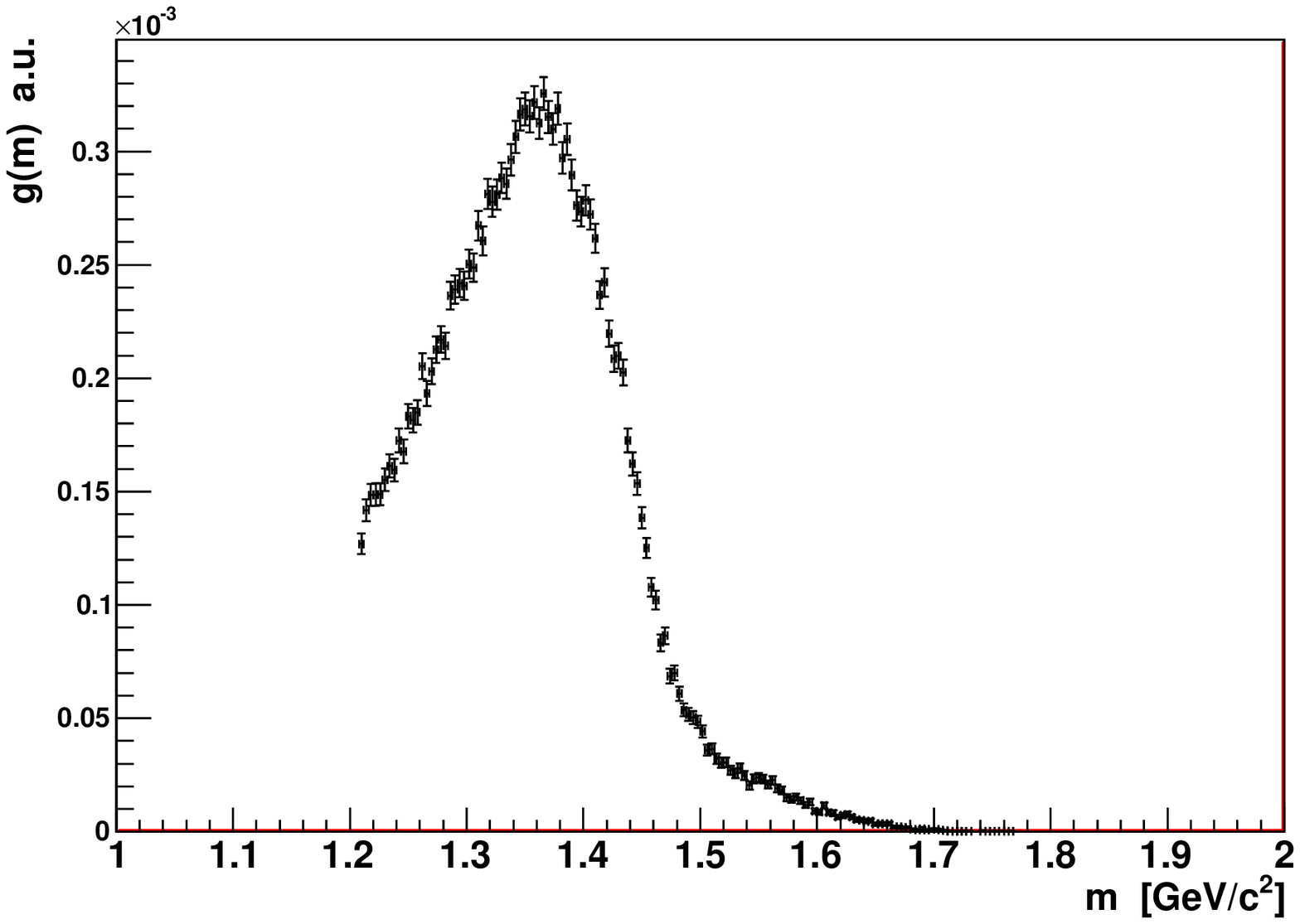}} \label{fig:D1232N1440lineshapeStableN}}\\
\subfigure[$\Delta(1232)$ line shape, where $N^{*}(1440) \rightarrow \pi^{0} \Delta(1232)$\newline (unstable particle)]{\fbox{\includegraphics[width=0.45\textwidth]{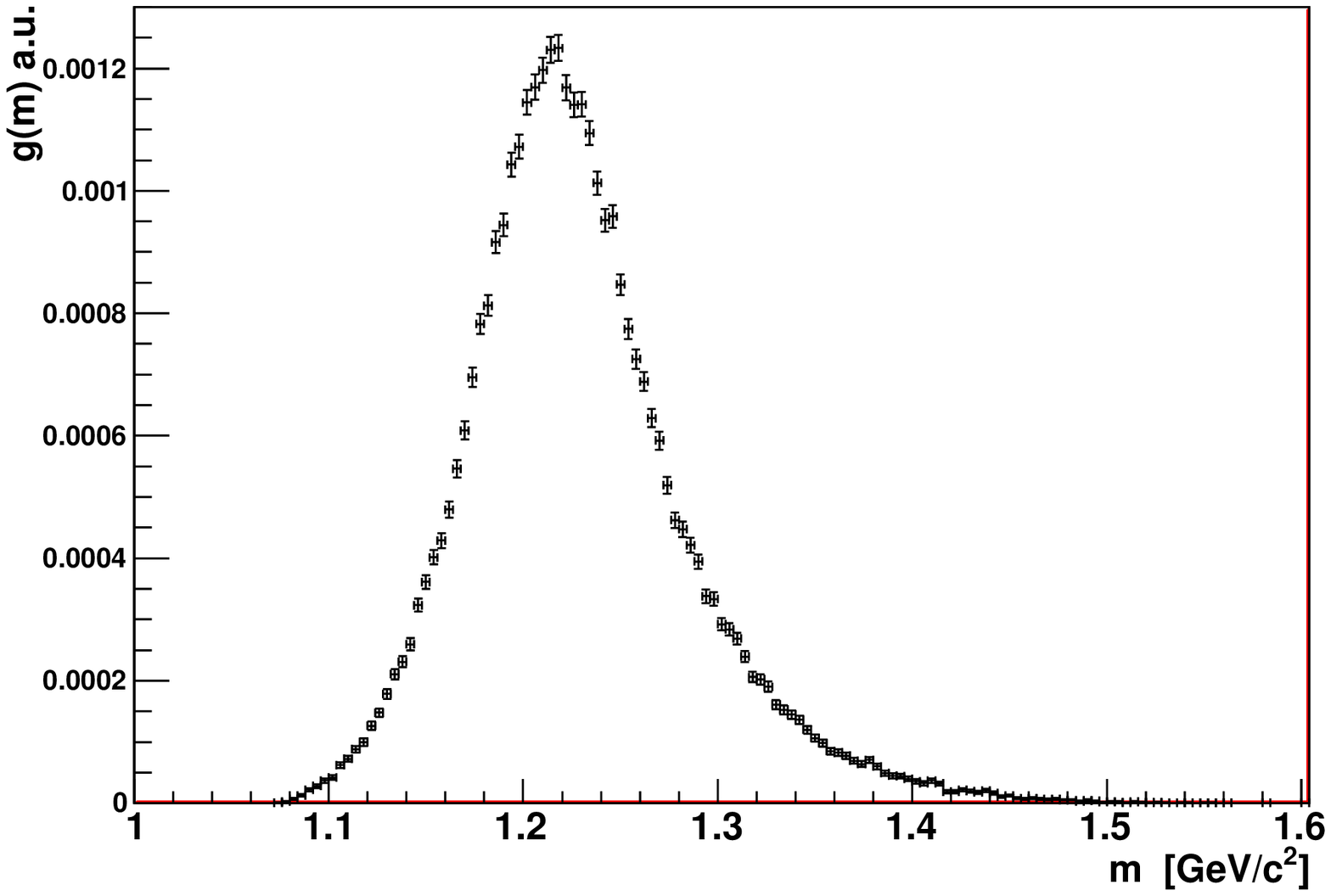}} \label{fig:D1232N1440lineshapeUnStableD}}\quad
\subfigure[$N^{*}(1440)$ line shape, where $N^{*}(1440) \rightarrow \pi^{0} \Delta(1232)$\newline (unstable particle)]{\fbox{\includegraphics[width=0.45\textwidth]{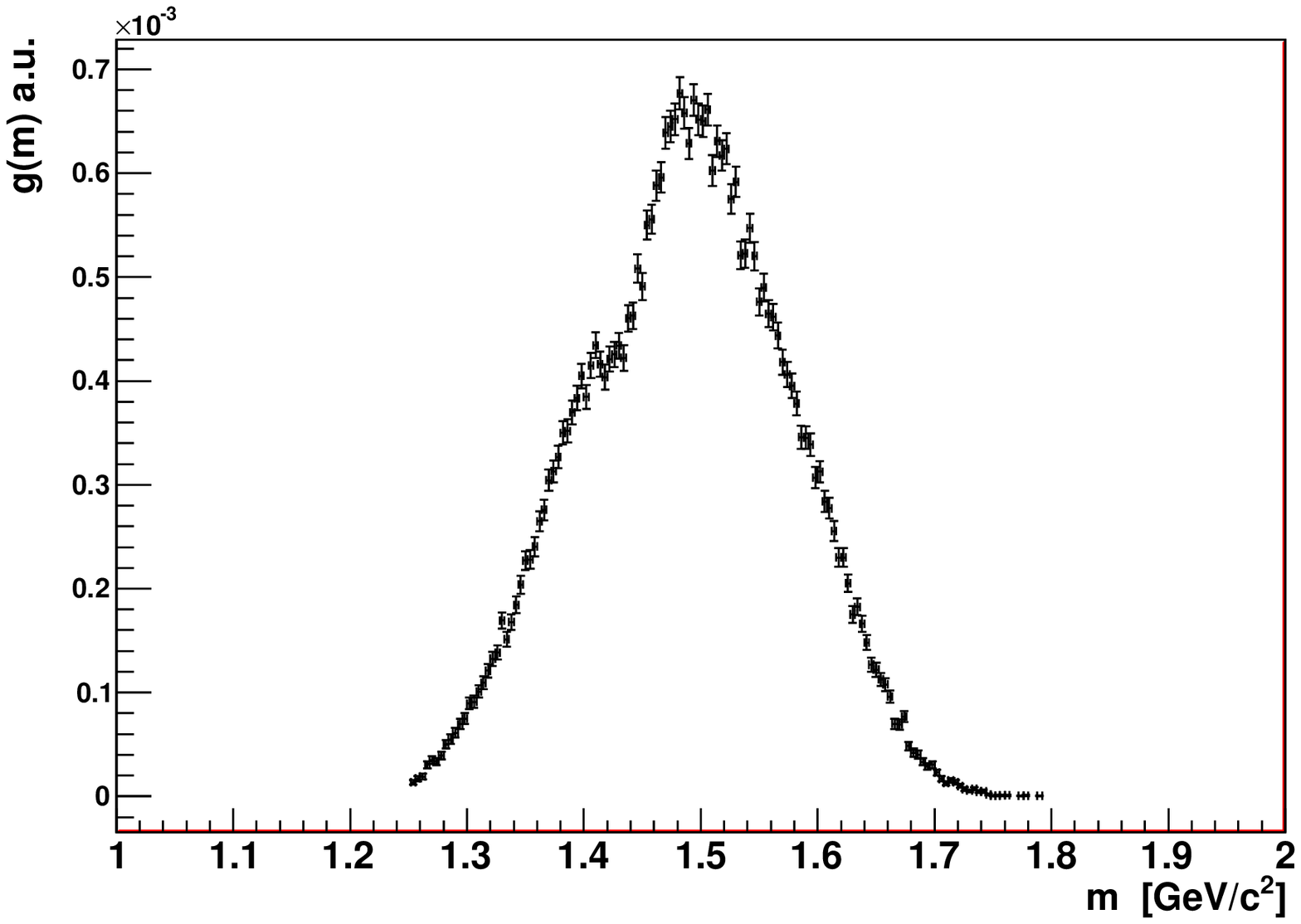}} \label{fig:D1232N1440lineshapeUnStableN}}
}
\caption{Realistic effective spectral line shape $g(m)$ of $\Delta(1232)$ and $N^{*}(1440)$
in the $pp \rightarrow \Delta(1232) N^{*}(1440)$ reaction at incident proton momentum of $3.350\mathrm{~GeV/c^{2}}$.
The $\Delta(1232)$ decays into $p\pi^{0}$ (stable particles), the $N^{*}(1440)$ decays into $p\pi^{0}\pi^{0}$ (stable particles) or into $\pi^{0}\Delta(1232)$ (unstable particle),
when later $\Delta(1232)$ decays into $p\pi^{0}$. 
The strong difference between the $N^{*}(1440)$ spectral line shape in case of decays into stable and unstable particles visible.
In case of $N^{*}(1440)$ the $p\pi^{0}\pi^{0}$ threshold is seen on the left.
Calculations by \textbf{Pluto++} (via Monte-Carlo method).  
}
\label{fig:D1232N1440lineshape}
\end{sidewaysfigure}

Realistic effective spectral line shapes $g(m)$ for $\Delta(1232)$, $N^{*}(1440)$ and $N^{*}(1535)$ in the reactions $pp \rightarrow p N^{*}(1535) \rightarrow p p \eta(3\pi^{0})$
and $pp \rightarrow \Delta(1232) N^{*}(1440)$ at incident proton momentum of $3.350\mathrm{~GeV/c^{2}}$
 calculated by \textbf{Pluto++} are presented in (Fig.~\ref{image_PlutoSpectralpNeta_Nmass.eps} and Fig.~\ref{fig:D1232N1440lineshape}). 
The $\Delta(1232)$ decays into $p\pi^{0}$ (stable particles), the $N^{*}(1440)$ decays into $p\pi^{0}\pi^{0}$ (stable particles) or into $p\Delta(1232)$ (unstable particle),
when later $\Delta(1232)$ decays into $p\pi^{0}$.
The strong difference between the $N^{*}(1440)$ spectral line shape is seen in case of decays into stable (Fig.~\ref{fig:D1232N1440lineshapeStableN})
 and unstable particles (Fig.~\ref{fig:D1232N1440lineshapeUnStableN}).   

The realistic effective spectral line shape of the resonances dependents on the internal properties of the resonances as well as on their decay products. 
The detail informations can be found \cite{Pluto}. 

\myFrameFigure{PlutoSpectralpNeta_Nmass.eps}{Realistic effective spectral line shape of $N^{*}(1535)$
in the $pp \rightarrow p N^{*}(1535) \rightarrow p p \eta(3\pi^{0})$ reaction at incident proton momentum of $3.350\mathrm{~GeV/c^{2}}$.
The threshold for $N^{*}(1535) \rightarrow p \eta$ is seen on the left.
Calculations by \textbf{Pluto++} (via Monte-Carlo method).}{N1535 line shape}

\newpage~
\thispagestyle{empty}
\emptydoublepage


\section{Track Reconstruction \mbox{in WASA-at-COSY}}\label{appendix:TrackReconstruction}
\thispagestyle{plain}
\renewcommand{\thetable}{\Alph{section}.\arabic{table}}
\renewcommand{\theequation}{\Alph{section}.\arabic{equation}}
\renewcommand{\thefigure}{\Alph{section}.\arabic{figure}}
\setcounter{table}{0}
\setcounter{equation}{0}
\setcounter{figure}{0}
The track reconstruction i.e. the reconstruction of particle trajectory from the detector information is the main step of the analysis.
When particle flies through the detector medium it interacts with single detector elements,  it ``hits the detector'' and one is talking about \textit{the hit}.
Then one can combine the hits related to the one particle in one detector to a group which we call \textit{the cluster}.
Later the clusters from different detectors related to one particle are combine to the particle trajectory called \textit{the track},
 which is the goal of the track reconstruction. 
These assignments are done by several different computer algorithms optimized for the given detector type.
Due to the structure of the WASA-at-COSY detector setup (Section~\ref{subsec:detector} on page \pageref{subsec:detector}) we have two types of tracks, described below.

\subsection*{The Forward Detector Tracks}
\label{appendix:TrackReconstructionFD}
These are the tracks in Forward Detector of the WASA-at-COSY identified as charged particles (usually protons).

The creation of tracks begins with the creation of the clusters from the hits in the layers of Forward Detector components,
 here only time coincidence is used - the time of the cluster is calculated as an average time of contributing hits. 
Now the creation of the track from clusters begins. First the FTH clusters in all three
 layers of FTH detector are checked for time coincidence and geometrical overlap forming uniquely defined pixel in FTH.
The time of the track is set to the average time of the clusters forming FTH pixel. 
The position of the pixel sets the coordinate of the track (polar $\theta$ and azimuthal $\phi$ angle ) in coordinate system of the WASA-at-COSY
with origin at beam-target overlap point. Later this information is refined by projecting the FTH pixel to the FPC planes, in each layer of the four FPC
 the straw tube is selected which is the closest one to the FTH pixel and then the crossing point of these tubes is calculated which gives the corrected new coordinate of the track.

By using this method with FPC detector, the so called binary mode, the angular resolution improves by factor two giving
 $FWHM_{\theta}\sim0.2~\mathrm{[deg]}$ and $FWHM_{\phi}\sim3~\mathrm{[deg]}$ (mainly due to the high granularity of the FPC).     

In the last step the information from clusters from the FWC, FTH and FVH detectors (The FRI is not used) is incorporated into the track by comparing time coincidence and azimuthal
 angle overlap between the track and those clusters. 

\subsection*{The Central Detector Tracks}
\label{appendix:TrackReconstructionCD}
These are the tracks in Central Detector of the WASA-at-COSY identified as neutral or charged particles (i.e. photons, electrons,pions, protons ).

The Central Detector consists of three completely different detector components MDC, PSB, SEC which can contribute to the track creation and track type.

The same as for the FD tracks procedure begins with a creation of the clusters from hits in these three detector components,
 if a contribution of the detector in track was justified.
The tracks with MDC and/or PSB cluster are charged tracks identified with a charged particle.
 The absence of MDC and/or PSB cluster (the veto condition)
 determines the neutral particle identified with a photon. To distinguish between the charged and neutral track one needs only one from two detector information (MDC or PSB).

The PSB is a plastic scintillator and it gives very fast binary information if there was a charged particle. 
If one is interested in more details like particle momentum one needs to to use the MDC in the track building (MDC is a drift chamber). 
Since in this work the goal is to detect the photons from the pion decays it is enough to use the contributions of PSB and SEC detectors
 in track building process for distinguishing between neutral and charged tracks. 
Also from the point of view of computing time the PSB decision information is around one order of magnitude faster then the MDC information.
For $pp \rightarrow pp 3\pi^{0} \rightarrow pp 6\gamma$ reaction only tracks with PSB and SEC detector contribution were considered.
The details about the MDC reconstruction could be found in \cite{MDCreco1}.

First the cluster in PSB is build from overlapping elements if the elements are in time coincidence, then the azimuthal angle of the cluster is calculated
as an average of the elements. The polar angle is fixed for different parts of the PSB detector, for forward $30\mathrm{[deg]}$, central $90\mathrm{[deg]}$ and backward $140\mathrm{[deg]}$.
The time of the cluster is calculated as an average of the time for elements and the energy deposit as a maximum from the elements.

Next the cluster is SEC detector is found. Since the photons hitting SEC detector produce electromagnetic showers, their transversal development 
depends on the Moliere Radius which exceeds the size of the crystal so one photon hitting the crystal can develop to other crystals and the cluster building has to combine it.
The cluster building works iteratively, the neighboring elements are checked for time coincidence (maximum $50\mathrm{~ns}$) and minimum energy deposit of $2\mathrm{~MeV}$ and combined to the cluster.
The center of the cluster is taken as a contributing crystal with highest energy deposit. The time is taken from the cluster center.
The energy deposit of the cluster is the sum of all contributing elements. The position of the cluster $\overrightarrow{R}$ is calculated as a weighted average of the positions of the contributing crystals $\overrightarrow{r}_{i}$

\begin{equation}
 \overrightarrow{R} = \dfrac{\sum_{i}w_{i}\overrightarrow{r}_{i}}{\sum_{i}w_{i}}
\end{equation}   
where

\begin{equation}
 w_{i} = MAX\left\lbrace 0; w_{0} + \ln\left( \frac{Edep_{i}}{\sum_{i}Edep_{i}}\right)  \right\rbrace 
\end{equation}

here $w_{0}=5$ and $Edep_{i}$~-~energy deposit of the $i-th$ crystal contributing to the cluster.

Now the PSB cluster and SEC cluster information is combined to the track by checking the angular overlapping between the clusters and time coincidence.
The neutral tracks (identified with photons) are those without assigned PSB cluster (the veto condition).

\newpage~
\thispagestyle{empty}
\emptydoublepage
\section{The Kinematic Fit}\label{appendix:kfit}
\thispagestyle{plain}
\renewcommand{\thetable}{\Alph{section}.\arabic{table}}
\renewcommand{\theequation}{\Alph{section}.\arabic{equation}}
\renewcommand{\thefigure}{\Alph{section}.\arabic{figure}}
\setcounter{table}{0}
\setcounter{equation}{0}
\setcounter{figure}{0}

\paragraph{The idea of the kinematic fit\\}

The experimentalist deals with the measurements $m_{i}$ which are always biased by the measurements specific
uncertainly $\sigma_{m}^{i}$ (i.e. accuracy of the detectors, reconstruction accuracy etc.). All this informations
has to be taken into account in the data analysis, so that in the final results all those effects are compensated.
One way of handling that problem is the kinematic fitting\cite{kfitCrennell, kfitCampbell}.
The kinematic fitting procedure is a data transformation technique which takes into account the information
about the errors $\sigma_{m}^{i}$ of the measured quantities $m_{i}$ to compute the most probable value $f_{i}$ of
the true unknown value $t_{i}$, the estimator of the true value 
\begin{equation}
f_{i}=T(\{m_{j},\sigma_{m}^{j}\})
\label{eqAppA:Estimator}                    
\end{equation}

Considering reaction
\begin{equation}
\underbrace{A+B}_{\mathbb{P}_{in}}\rightarrow \underbrace{1+2+3+4+\ldots+k}_{\mathbb{P}_{out}}
\label{eqAppA:reaction}                    
\end{equation}

where $\mathbb{P}_{in}=\mathbb{P}_{A}+\mathbb{P}_{B}$~-~4-momentum vector of the input channel\\
$\mathbb{P}_{out}=\mathbb{P}_{1}+\mathbb{P}_{2}+\mathbb{P}_{3}+\mathbb{P}_{4}+\ldots$~-~4-momentum vector of the output channel

from the conservation of energy and momentum the relation has to be fulfilled
\begin{equation}
\mathbf{C}=\mathbb{P}_{in}-\mathbb{P}_{out}=\mathbb{O}
\label{eqAppA:KinematicalCondition}
\end{equation}

and one knows that the number of independent variables to describe this reaction is

\begin{equation}
3k-4
\label{eqAppA:NumberIndependant}
\end{equation}
where $k$~-~number of final state particles \cite{byckling}.

Usually the initial state $\mathbb{P}_{in}$ is well known and is fixed(it is not a purpose of the fit).
What is measured, are the the final state particles $i=1, 2, 3, \ldots, k$.
The following parameters of the final state particles are measured: their Energy($E_{i}$), polar angle ($\theta_{i}$) an azimuthal angle ($\phi_{i}$)
which define clearly the particle. Lets call such a parameters for one given particle $m_{i}$, of course each measurement has its uncertainly $\sigma_{m}^{i}$, where $i$ denotes the final state particles.     
Now define a vector of all measurements $\mathbf{M}=[m_{i}]$ and a covariance matrix for the measurements $\mathbf{V}=[cov(p,l)]$(where $p,l=1, 2, 3, \ldots, k$) on the diagonal of this matrix are $(\sigma_{m}^{i})^2$,
in case of no correlations between the measurements the matrix is diagonal.
What we want from this information is the set of new variables which will "replace`` the measurements.
They have to take into account all the uncertainties $\mathbf{V}$. Such a new vector will be called $\mathbf{F}=[f_{i}]$.
To realize the task to find the estimator $\mathbf{F}$ \myEqRef{eqAppA:Estimator} one can use the Least Square Method(The Chi-square Method).
So first lets build the the Chi-square functional:
\begin{equation}
\mathbf{H(F)} = \left( \mathbf{F} - \mathbf{M} \right) \mathbf{V}^{-1} \left( \mathbf{F} - \mathbf{M} \right)^{T}~.
\label{eqAppA:Chi2fuctional}                    
\end{equation}

To incorporate the kinematic condition \myEqRef{eqAppA:KinematicalCondition} to the functional one uses the Lagrange Multipliers,
now the functional has the form:

\begin{equation}
\mathbf{H(F)} = \left( \mathbf{F} - \mathbf{M} \right) \mathbf{V}^{-1} \left( \mathbf{F} - \mathbf{M} \right)^{T} + \mathbf{\lambda} \mathbf{C}
\label{eqAppA:Chi2fuctional2}                    
\end{equation}

here $\mathbf{\lambda}$ denotes the Lagrange multiplier vector. 
In general in addition to the kinematic condition \myEqRef{eqAppA:KinematicalCondition} one can request other condition as for example:
that two final state particles come from a intermediate resonance, lets call these extra conditions $\mathbf{D}$ and rewriting functional as

\begin{equation}
\mathbf{H(F)} = \left( \mathbf{F} - \mathbf{M} \right) \mathbf{V}^{-1} \left( \mathbf{F} - \mathbf{M} \right)^{T} + \mathbf{\lambda}_{C}\mathbf{C}+\mathbf{\lambda}_{D}\mathbf{D} 
\label{eqAppA:Chi2fuctionalFinal}                    
\end{equation}

The solution for $\mathbf{F}$ one gets for $\mathbf{H(F)}$ having minimum.
This corresponds to well known mathematical problem of minimization of the functional with extra boundary conditions.

The minimal $\mathbf{H(F)}=\chi^2_{Fit}$, in case of Gaussian $\mathbf{V}$, is distributed as a $\chi^2_{NDF}$ distribution with Number Degrees of Freedom $NDF=4+N_{D}$ \cite{kfitFett}, $N_{D}$~-~number of extra constraints $\mathbf{D}$.
In general not all of three variables ($E_{i}, \theta_{i}, \phi_{i}$) for each final state particles have to be measured to identified completely the reaction \myEqRef{eqAppA:reaction}, \myEqRef{eqAppA:NumberIndependant} tells us about it.
The unmeasured variables could be retrieved from the available information, by solving \myEqRef{eqAppA:Chi2fuctionalFinal}.
Denoting $N_{u}$~-~number of unmeasured variables, the $NDF$ of $\chi^2_{Fit}$ distribution changes now to $NDF=4+N_{D}-N_{u}$. 
Knowing the Probability Distribution Function of $\chi^2_{Fit}$ one can test the deviation from theoretical $\chi^2_{NDF}$, 
it is easily done by introducing the Probability of the Fit (Complementary Cumulative Distribution Function or survival function):
\begin{equation}
Prob_{Fit}(\chi^2_{Fit}) = \int_{\chi^2_{Fit}}^{\infty}\chi^2_{NDF}(k)dk = 1 - F(\chi^2_{Fit},NDF) 
\label{eqAppA:Probability}                    
\end{equation}
where $F(\chi^2_{Fit},NDF)$ is the Cumulative Distribution Function (CDF). 
Since the $Prob_{Fit}$ is a type of Cumulative Function it should have a flat distribution on range $0-1$.

One uses numerical methods to find a solution of \myEqRef{eqAppA:Chi2fuctionalFinal} for given $\mathbf{M,V}$ \cite{Lyons,KFitBhabha,kfitHondt} 

\newpage
\paragraph{Example of the kinematic fit\\}

As an Example of kinematic fit lets consider the following reaction:

\begin{equation}
p p \rightarrow pp \pi^{0} \pi^{0} \pi^{0} \rightarrow pp \gamma \gamma \gamma \gamma \gamma \gamma
\end{equation}
One has $8$ particles in the final state so the number of independent variables to describe this reaction is $3*8-4 = 20$.
If we would measure for each of the particles Energy ($E$), polar angle ($\theta$) and azimuthal angle ($\phi$) then we would have
$8*3 = 24$ variables, the information would be redundant.

Lets consider that we measure $E, \theta, \phi$ for photons and only $\theta, \phi$ for the protons, then we have still $6*8+2*2 = 22 $ independent variables,
that is still enough to fully describe our reaction. Having the following situation we can now do the kinematic fitting, one has number of unmeasured variables $N_{u} =2$
($E$ for two protons - will come as a result of the fit) and additional number constraints $N_{D} = 3$ (the mass constraints for the photons which should give three $\pi^{0}$).
The Number Degrees of Freedom $NDF$ for the $\chi^2_{Fit}$ fit is $NDF=4+N_{D}-N_{u} = 4 + 3 -2 = 5$ and if the errors are distributed as Gaussian, the $\chi^{2}_{Fit}$
should be distributed as a $\chi^{2}_{NDF}$ distribution with $NDF=5$.

\myFrameSmallFigure{Chi2GausSmear}{Example of the $\chi^{2}_{Fit}$ distribution for the kinematic fitting for the reaction $p p \rightarrow pp 3\pi^{0} \rightarrow pp 6\gamma$ for the Gaussian distributions of the errors.
 The black histogram corresponds to the Monte-Carlo simulation, the red line denotes theoretical $\chi^{2}$.
Number of events is shown on vertical axis.
}{Example of the $\chi^{2}_{Fit}$ distribution for the Kinematic Fitting - Gaussian Errors}

\myFrameSmallFigure{ProbGausSmear}{Example of the $Prob_{Fit}$ distribution for the kinematic fitting for the reaction $p p \rightarrow pp 3\pi^{0} \rightarrow pp 6\gamma$ for the Gaussian distributions of the errors, the Monte-Carlo Simulation.
Number of events is shown on vertical axis.
}{Example of the $Prob_{Fit}$ distribution for the Kinematic Fitting - Gaussian Errors}

In the \myImgRef{Chi2GausSmear} is presented the $\chi^{2}_{Fit}$ distribution for the Kinematic Fit
for the above case for the Monte-Carlo simulation assuming homogeneous and isotropic populated Phase Space with Gaussian smearing used for the variables.
One sees that the $\chi^{2}_{Fit}$ agrees with a theoretical $\chi^{2}_{NDF}$ distribution with $NDF=5$. Also the $Prob_{Fit}$ distribution is presented
in \myImgRef{ProbGausSmear}. It is flat, it confirms ones more that $\chi^{2}_{Fit}$ agrees with a theoretical $\chi^{2}_{NDF}$ and that the errors of the variables are Gaussian.

\myFrameSmallFigure{Chi2BreitWSmear}{Example of the $\chi^{2}_{Fit}$ distribution for the kinematic fitting for the reaction $p p \rightarrow pp 3\pi^{0} \rightarrow pp 6\gamma$ for the Non Gaussian distributions of the errors. The black histogram corresponds to the Monte-Carlo simulation, the red line denotes theoretical $\chi^{2}$.
Number of events is shown on vertical axis.
}{Example of the $\chi^{2}_{Fit}$ distribution for the Kinematic Fitting - Non Gaussian Errors}

\myFrameSmallFigure{ProbBreitWSmear}{Example of the $Prob_{Fit}$ distribution for the kinematic fitting for the reaction $p p \rightarrow pp 3\pi^{0} \rightarrow pp 6\gamma$ for the Non Gaussian distributions of the errors, the Monte-Carlo Simulation.
Number of events is shown on vertical axis.
}{Example of the $Prob_{Fit}$ distribution for the Kinematic Fitting - Non Gaussian Errors}

\myFrameSmallFigure{ProbComp2.eps}{Example of the $Prob_{Fit}$ distribution for the kinematic fitting for the reaction hypothesis $p p \rightarrow pp 3\pi^{0} \rightarrow pp 6\gamma$
the Monte-Carlo simulation of the $p p \rightarrow pp 4\pi^{0} \rightarrow pp 8\gamma$, $p p \rightarrow pp 5\pi^{0} \rightarrow pp 10\gamma$, $p p \rightarrow pp \eta` \rightarrow pp 3\pi^{0} \rightarrow pp 6\gamma$ reaction.
The same amount of each reaction type was simulated. }{Example of the $Prob_{Fit}$ distribution.}

When the errors are not Gaussian distributed, or the reaction hypothesis is not true (e.g. background reaction contamination) the $\chi^{2}_{Fit}$ 
does not agrees with a theoretical $\chi^{2}_{NDF}$ 
distribution see \myImgRef{Chi2BreitWSmear}. This is also reflected on the $Prob_{Fit}$ which is not homogeneously distributed anymore, 
the population of small probability values is higher \myImgRef{ProbBreitWSmear} in case of non Gaussian errors. 
The $Prob_{Fit}$ is slightly non homogeneous with very prominent increase at small probability \myImgRef{ProbComp2.eps} in case the reaction hypothesis is not true. 
To deal with this problem one selects events for 
which the probability $Prob_{Fit}$ for the signal reaction is in good agreement with flat distribution, in this example that would be $Prob_{Fit}>0.2$.
Of course the bigger the $Prob_{Fit}$ threshold for selection the ''cleaner`` will be the data sample but the statistics will reduce - one has to find always the compromise.

In addition to the probability function checks one can construct the "Residual spectra'' for fit:
\begin{equation}
 residual_{i} = m_{i}-f_{i}
\end{equation}
in case of Gaussian errors and no correlations between the measured variables (i.e. the $\mathbf{V}$ matrix diagonal)
the $residual_{i}$ variable should be distributed as Gaussian distribution with mean $0$ and standard deviation $\sigma_{residual}^{i}$ given by the relation \cite{KFITByron}:

\begin{equation}
 \sigma_{residual}^{i} = \sqrt{ (\sigma_{m}^{i})^{2} - (\sigma_{f}^{i})^{2} }
\end{equation}
 which may still "puzzle many users`` \cite{KFITEadie}.

One can also define the "Pulls'' distribution:
\begin{equation}
 pull_{i} = \dfrac{residual_{i}}{\sigma_{residual}^{i}}
\end{equation}
the $pull_{i}$ variable should be distributed as standard Gaussian distribution with mean $0$ and sigma $1$.


\newpage
\paragraph{Illustration of the kinematic fit - simple example\\}

To illustrate the idea of the kinematic fit and how it works look at some simple example.
Lets assume that we have measured $x=0.5$ with the error $\sigma_{x}=1$ and $y=0.5$ with the error $\sigma_{y}=1$ and we know that the relation
\begin{equation}
 xy=1
\end{equation}
should hold for the variables (as constraint; it corresponds to the \myEqRef{eqAppA:KinematicalCondition}).

\myFrameSmallFigure{KFitExample}{Illustration of kinematic fitting. The measurement is marker, lines correspond to the constraint. Two candidates for solution are marked $P1$ and $P2$. The solution of the problem is $P1$.}{Illustration of Kinematic Fitting}

Now we want to combine this information together, as in the kinematic fit. One constructs the function to minimize:

\begin{equation}
 h(x^{*}, y^{*}) = \left( \frac{x^{*}-x}{\sigma_{x}}\right)^{2} + \left( \frac{y^{*}-y}{\sigma_{y}}\right)^{2}
\label{eqAppA:EXPminfunction}
\end{equation}

with the constraint:

\begin{equation}
 x^{*}y^{*}=1
\label{eqAppA:EXPcondition}
\end{equation}

where $x^{*},y^{*}$ correspond to the most probable value which we are looking for.

One has to find the minimum of (Eq.~\ref{eqAppA:EXPminfunction}) with condition (Eq.~\ref{eqAppA:EXPcondition}).
To solve this problem one uses the Lagrange multipliers method.
One constructs the functional to minimize:
\begin{equation}
 H(x^{*}, y^{*}) = h(x^{*}, y^{*}) + \lambda\left(  x^{*}y^{*}-1\right) 
\label{eqAppA:EXPfuctional}
\end{equation}

here $\lambda$ is Lagrange multiplier.

One gets the equations:

\begin{eqnarray}
 \frac{\partial H(x^{*}, y^{*})}{\partial x^{*}} &=& 0 \\
\frac{\partial H(x^{*}, y^{*})}{\partial y^{*}} &=& 0 \nonumber \\
x^{*}y^{*}-1 & = & 0 \nonumber
\end{eqnarray}

There are two candidates for solution of these equations \myImgRef{KFitExample}:

\begin{enumerate}
 \item $x^{*} = 1$, $y^{*}=1$, $\lambda=-1$ $(P1)$
\item $x^{*} = -1$, $y^{*}=-1$, $\lambda=-3$ $(P2)$
\end{enumerate}

Since $H(P1)<H(P2)$, so the $P1$ is the solution of the problem.


\newpage ~
\thispagestyle{empty}
\emptydoublepage
\section{\mbox{Bayesian~Likelihood} \mbox{energy~reconstruction}}\label{appendix:Bayes}
\thispagestyle{plain}
\renewcommand{\thetable}{\Alph{section}.\arabic{table}}
\renewcommand{\theequation}{\Alph{section}.\arabic{equation}}
\renewcommand{\thefigure}{\Alph{section}.\arabic{figure}}
\setcounter{table}{0}
\setcounter{equation}{0}
\setcounter{figure}{0}



\paragraph{The idea of the method\\}

The FRH dE-E telescope of the WASA~at~COSY detector (Section \ref{subsubsec:fd} on page \pageref{subsubsec:fd}) enables the reconstruction of kinetic energy of the particles in FD by the dE-E method. This is done in ``standard'' way using parametrization of $E_{dep}-E_{kin}$ relations depending on the particle type, the detector plane where the particle was stopped, the scattering angle, and the number of layers used for energy reconstruction.
The method using full multidimensional information from the telescope, taking into account all unique correlations simultaneously - not iteratively - of the particle energy loss in five layers of FRH, was missing. Such a method is presented below.

Generalizing the problem: Knowing the energy deposit of the particle in five layers of FRH one wants to find  the estimator of the kinetic energy of the particle $T(E_{kin})$. There are two general methods of searching for the estimators of the true values:

\begin{itemize}
 \item Least Squares Estimation \textbf{LSE} (first described by Johann Carl Friedrich Gauss) - the estimators of that method have no general asymptotic properties, if measurements are Gaussian distributed then the estimators are unbiased and most efficient \cite{LSEbook}. 

 \item Maximum Likelihood Estimation \textbf{MLE} (first described by Sir Ronald Aylmer Fisher) - the estimators of that method are asymptotically optimal, independently on the distribution of the measurements, the estimators are asymptotically unbiased and asymptotically most efficient \cite{MLEbook}.
\end{itemize}

The \textbf{MLE} method is the most general one with the optimal performances of the estimators so it was chosen. The estimator of the kinetic energy of the particle $T(E_{kin})$ is the one which maximizes the likelihood function $L$ which corresponds to minimization of $l=-\log(L)$.
The likelihood for the dE-E telescope is defined as the conditional probability $f$ that the energy of the particle $E$ is equal to the true kinetic energy $E_{kin}$ under the condition that one has the energy deposits in layers of FRH telescope $E_{dep}^{1}, E_{dep}^{2}, E_{dep}^{3}, E_{dep}^{4}, E_{dep}^{5}$ i.e.
\begin{equation}
 L(E)=f\left( E=E_{kin}/E_{dep}^{1}, E_{dep}^{2}, E_{dep}^{3}, E_{dep}^{4}, E_{dep}^{5} \right)
\end{equation}

Now using Bayes theorem \cite{BayesianResoning} we can write

\begin{eqnarray}
L(E) &=& f\left( E=E_{kin}/E_{dep}^{1}, E_{dep}^{2}, E_{dep}^{3}, E_{dep}^{4}, E_{dep}^{5} \right) \\ 
&=&\frac{f\left( E_{dep}^{1}, E_{dep}^{2}, E_{dep}^{3}, E_{dep}^{4}, E_{dep}^{5} / E=E_{kin} \right)}{\int f\left( E_{dep}^{1}, E_{dep}^{2}, E_{dep}^{3}, E_{dep}^{4}, E_{dep}^{5} / E=E_{kin} \right)dE_{kin}}\nonumber
\end{eqnarray}

The only thing which one has to derive is the probability function \\ $f\left( E_{dep}^{1}, E_{dep}^{2}, E_{dep}^{3}, E_{dep}^{4}, E_{dep}^{5} / E=E_{kin} \right)$ which is five dimensional probability function in energy deposits conditioned by energy (six dimensional function), to simplify the problem one may assume the factorization:
\begin{eqnarray}
L(E) &=&f\left( E_{dep}^{1}, E_{dep}^{2}, E_{dep}^{3}, E_{dep}^{4}, E_{dep}^{5} / E=E_{kin}\right) \\  
&=&f_{1}*f_{2}*f_{3}*f_{4}*f_{5}\nonumber
\label{eq:AppALikelihood}
\end{eqnarray}

where $f_{i}$ are independent and defined as

\begin{eqnarray}
f_{1} &=& f\left( E_{dep}^{1} / E_{1}=E_{kin}\right) \\
f_{2} &=& f\left( E_{dep}^{2} / E_{2}=E_{kin}-E_{dep}^{1}\right) \\
f_{3} &=& f\left( E_{dep}^{3} / E_{3}=E_{kin}-E_{dep}^{1}-E_{dep}^{2}\right) \\
f_{4} &=& f\left( E_{dep}^{4} / E_{4}=E_{kin}-E_{dep}^{1}-E_{dep}^{2}-E_{dep}^{3}\right) \\
f_{5} &=& f\left( E_{dep}^{5} / E_{5}=E_{kin}-E_{dep}^{1}-E_{dep}^{2}-E_{dep}^{3}-E_{dep}^{4}\right)
\end{eqnarray}

The $f_{i}$ functions were derived using Monte-Carlo GEANT3 simulation for the single particle tracks in FRH telescope for the given kinetic energy. Example of the derived probability function for protons is on \myImgRef{f3}. 

\myFrameSmallFigure{f3}{$f_{3}$ probability function derived using Monte-Carlo GEANT3 simulated single protons tracks in FRH telescope.
On vertical axis energy loss in third layer of FRH in $GeV$ (here denoted as $Edep3$). On horizontal axis the difference between the kinetic
 energy and energy loss in first and second layer of FRH in $GeV$ (here denoted $E3$)
}{$f_{3}$ probability function derived using Monte-Carlo GEANT3 simulated single protons tracks in FRH telescope}

\newpage
\paragraph{The reconstruction properties\\}

Using above defined likelihood function $L(E)$ (Eq.~\ref{eq:AppALikelihood}) one may now reconstruct the kinetic energy of the particles. The reconstruction of the protons kinetic energy is in the most common interest of WASA-at-COSY collaboration, due to interest of studying the decays of the $\eta '$ meson \cite{EtaPrimeJany, EtaPrimeDuniecJany}.

Using Monte-Carlo GEANT3 simulation of the WASA-at-COSY detector, the single proton tracks illuminating FRH detector are generated with the kinetic energy from $50\mathrm{~MeV}$ to $2\mathrm{~GeV}$. The kinetic energy was reconstructed using proposed method. On the \myImgRef{resLlog5} the relative resolution for the reconstructed kinetic energy $(E_{true}-E_{rec})/E_{true}$ as a function of the minimized likelihood $l=-\log(L)$ is presented, as one can see when the likelihood $l$ is getting larger(the degree of belief\footnote{term taken from \cite{BayesianResoning}} is getting smaller) the relative resolution becomes worse. Cutting on the likelihood value rejects the events with the bad resolution i.e. events witch have low degree of belief.

Since the likelihood function was derived using Monte-Carlo with some final statistical sample of the data, to control this systematical effect,
 in addition to the likelihood $L$ one may calculate the relative error of the minimized likelihood $L_{err}=\Delta L/L$, using standard error propagation method for (Eq.~\ref{eq:AppALikelihood}). 

The relative resolution for the reconstructed kinetic energy as a function of the error of the minimized likelihood $L_{err}$ is shown \myImgRef{resLlog5Err}. 
One sees that the energy resolution of the values above $0.25$ of the $L_{err}$ is very bad. Performing a cut on the $L_{err}$ can suppress 
this systematical effect on $L$.              

\myFrameSmallFigure{resLlog5}{Relative resolution for the reconstructed kinetic energy versus the minimized Likelihood function $l$ (here denoted as $Llog5$) for the Monte-Carlo single proton tracks.}{Relative resolution for the reconstructed kinetic energy versus the minimized Likelihood function $l$ for the Monte-Carlo single proton tracks}

\myFrameSmallFigure{resLlog5Err}{Relative resolution for the reconstructed kinetic energy versus the error of the minimized likelihood function $L_{err}$ (here denoted as $LErr5$) for the Monte-Carlo single proton tracks.}{Relative resolution for the reconstructed kinetic energy versus the error of the minimized Likelihood function $L_{err}$ for the Monte-Carlo single proton tracks}

\myFrameSmallFigure{etrueLlog5}{True kinetic energy in GeV (i.e. assumed in the Monte-Carlo simulations, here denoted as $Etrue$) versus the likelihood $l$ for the Monte-Carlo single proton tracks.}{True kinetic energy versus the Likelihood $l$ for the Monte-Carlo single proton tracks}

When one plots true kinetic energy versus the Likelihood $l$ \myImgRef{etrueLlog5} one immediately sees that these variables are directly correlated, so performing a cut on the Likelihood $l$ one can select a certain range of the true kinetic energy of the particle.

Having now good established Likelihood reconstruction, one may think of doing particle identification i.e. one wants
 to test the hypothesis of the particle type. Lets assume that we would like to distinguish protons from charged pions knowing the energy deposits in FRH.
 The straightforward way is to compare the probability function $f$ conditioned by the particle hypothesis:

\begin{eqnarray}
 R_{PID} &=& \frac{f\left( E=E_{kin}^{Proton}, \mathrm{Proton} /E_{dep}^{1}, E_{dep}^{2}, E_{dep}^{3}, E_{dep}^{4}, E_{dep}^{5} \right)}{f\left( E=E_{kin}^{Pion}, \mathrm{Pion} /E_{dep}^{1}, E_{dep}^{2}, E_{dep}^{3}, E_{dep}^{4}, E_{dep}^{5} \right)} \\
& = & \frac{L(E^{Proton}, \mathrm{Proton})}{L(E^{Pion}, \mathrm{Pion})}\nonumber
\label{eq:AppARPID}
\end{eqnarray}

which is exactly equal, from the definition, to the ratio of the likelihood function for proton and pion (PID stays for Particle Identification).

\myFrameHugeFigure{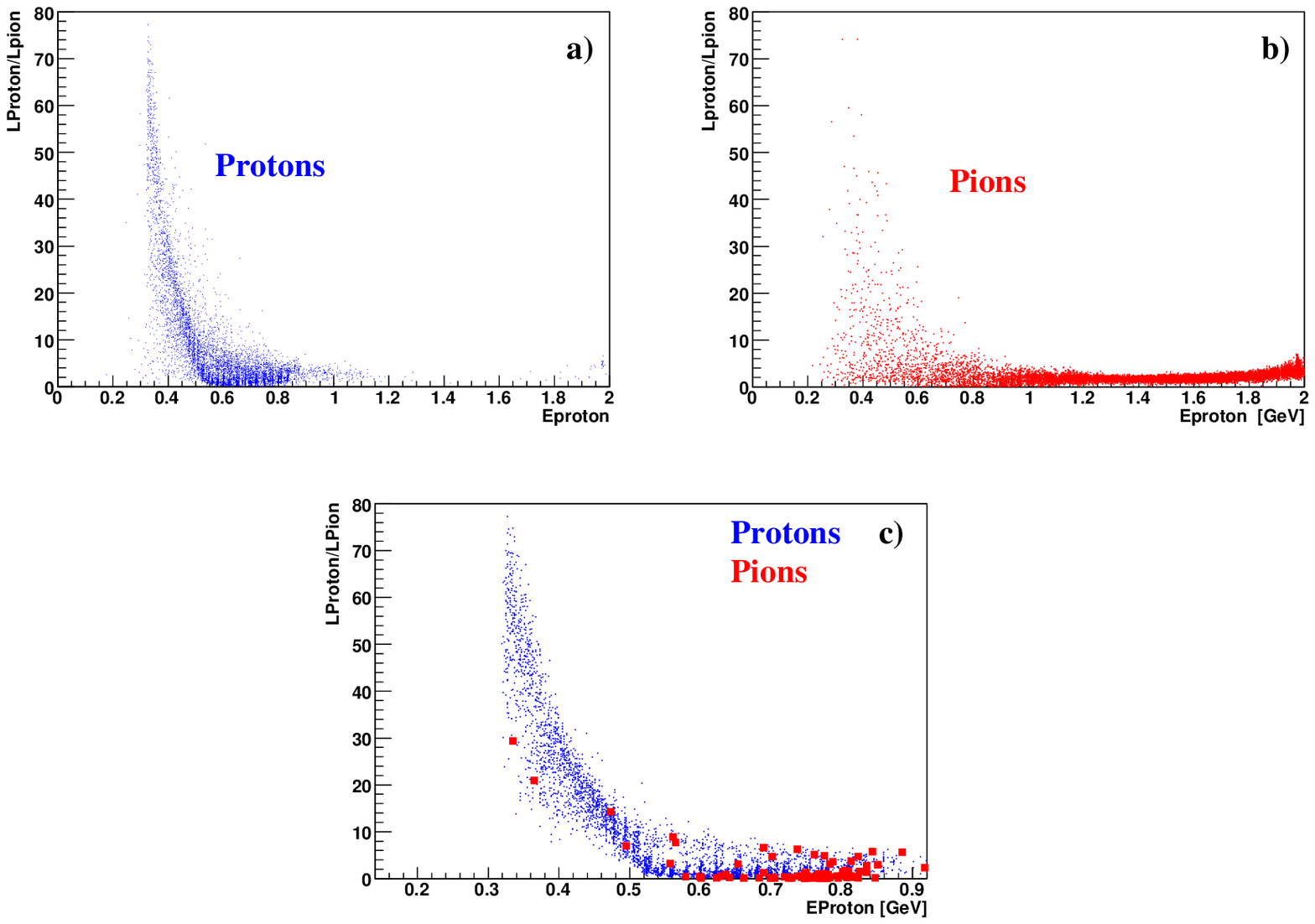}{Particle Identification using likelihood function. $R_{PID}$ ration (Eq.~\ref{eq:AppARPID}) versus the reconstructed kinetic energy in GeV assuming proton as a particle for the single Monte-Carlo tracks of a) protons, b) charged pions, c) protons(blue) and pions(red) with $l^{proton}<4.4$ and $L_{err}^{proton}<0.25$ condition.}{Particle Identification using Likelihood function.}

Using the (Eq.~\ref{eq:AppARPID}) the PID (\ref{image_PID}a,b). Restricting ourself to the particular range of the Likelihood and its error (\ref{image_PID}c) one can get a good separation between protons, and pions up to $\sim 700\mathrm{MeV}$.


It is essential to know, as mentioned above, how the reconstruction works for the protons coming from the $pp \rightarrow pp \eta'$ production at incident proton energy $T=2.54\mathrm{~GeV}$. These protons have the kinetic energy range of $300-800\mathrm{~MeV}$. Once more the single proton tracks are generated using Monte-Carlo GEANT3 simulation with energy range up to $2\mathrm{~GeV}$ later the cut on reconstructed kinetic energy was performed $E_{rec}=300-800\mathrm{~MeV}$.  

\myFrameSmallFigure{Rec1}{True kinetic energy in GeV (i.e. assumed in the Monte-Carlo simulations, here denoted as $Etrue$) versus reconstructed kinetic energy in GeV using:a) standard reconstruction b) Likelihood based for $l<4.4$ and $L_{err}<0.25$; for the Monte-Carlo single proton tracks with $E_{rec}=300-800\mathrm{~MeV}$.}{True Kinetic energy versus reconstructed kinetic energy using}

\myFrameSmallFigure{Rec2}{Relative kinetic energy resolution for:a) standard reconstruction b) Likelihood based for $l<4.4$ and $L_{err}<0.25$, as a function of true kinetic energy in GeV (i.e. assumed in the Monte-Carlo simulations, here denoted as $Etrue$); for the Monte-Carlo single proton tracks with $E_{rec}=300-800\mathrm{~MeV}$.}{Relative kinetic energy resolution}

The quality of the reconstruction for both reconstruction methods is shown (Figs.~\ref{image_Rec1},~\ref{image_Rec2}). The benefits of the likelihood method are seen on \myImgRef{Rec3}: the wrongly reconstructed particles are cut out without loosing efficiency. For the conditions ($l<4.4$ and $L_{err}<0.25$) the reconstruction efficiency of $0.57$ was reached.

\myFrameSmallFigure{Rec3}{Energy resolution $E_{true}-E_{rec}$ in GeV for the reconstructed kinetic energy using:(blue) standard reconstruction, (red) Likelihood based for $l<4.4$ and $L_{err}<0.25$; for the Monte-Carlo single proton tracks with $E_{rec}=300-800\mathrm{~MeV}$.}{Energy resolution $E_{true}-E_{rec}$ for the reconstructed kinetic energy}

\newpage
\paragraph{Verification of the method\\}

Using Monte-Carlo simulation the $pp \rightarrow pp \eta'$ production at incident proton energy $T=2.54\mathrm{~GeV}$
 was generated, and the missing mass of the two final state protons was reconstructed using standard and likelihood one method of kinetic energy reconstruction \myImgRef{MMlog}.
 Using the Likelihood based reconstruction one gets sharper $\eta'$ peak.

The $pp\rightarrow pp \eta$ WASA-at-COSY data from April 2007 at incident proton energy $T=1.4\mathrm{~GeV}$ were analyzed. 
The comparison of the missing mass of the two protons for standard and Likelihood method is shown \myImgRef{Rec4}. 
For the Likelihood method the $\eta$ signal is much more pronounced and the multipion background is suppressed.  

One concludes that the developed Bayesian likelihood kinetic energy reconstruction method had a better performances in all tests then the standard reconstruction
 method since it uses the full multidimensional information simultaneously. The method could be easily generalized for the full FD telescope, 
since it is based on the likelihood, also future detectors could be treated the same way like planned TOF and DIRC detector.
 
Nevertheless further developments are required to fully adopt and optimize the method for the experimental data,
 like derivation of the likelihood function using real experimental detector response. 

\myFrameSmallFigure{MMlog}{Missing mass of two protons, Monte-Carlo simulation of $pp \rightarrow pp \eta'$, reconstructed using: (blue) standard reconstruction, (red) Likelihood based for $l<4.4$ and $L_{err}<0.25$.}{Missing mass of two protons}

\myFrameSmallFigure{Rec4}{Missing Mass of the two protons with a cut on proton reconstructed energy $E_{rec}=150-500\mathrm{~MeV}$ for the WASA-at-COSY, experimental data $pp\rightarrow pp \eta$ at incident proton energy $T=1.4\mathrm{~GeV}$ using a) standard, b) Likelihood one for $l<4.4$ and $L_{err}<0.25$, proton kinetic energy reconstruction.}{Missing Mass of the two protons with a cut}

\newpage ~
\thispagestyle{empty}
\emptydoublepage

\section{Data Tables -- Results}\label{appendix:DataTables}
\thispagestyle{plain}
\renewcommand{\thetable}{\Alph{section}.\arabic{table}}
\renewcommand{\theequation}{\Alph{section}.\arabic{equation}}
\renewcommand{\thefigure}{\Alph{section}.\arabic{figure}}
\setcounter{table}{0}
\setcounter{equation}{0}
\setcounter{figure}{0}
%


\myTable{
\begin{tabular}{|c|}
\hline
\includegraphics[height=0.55\textheight, width=\textwidth]{Dal1A.epsi}\\  
\hline
\end{tabular}
}{Acceptance and efficiency corrected Dalitz Plot $ppX$,\newline for $MM_{pp}=0.4-0.5\mathrm{~GeV/c^{2}}$ (Fig.~\ref{image_Dalitz1AccCor2D}).
The errors of invariant masses are determined by selected bin sizes and chosen as a half of the bin size.
Additional global uncertainties of the absolute normalization of $19\%$ have to be included.
Fully expandable version of the table is available in the attached electronic version of the thesis.
}{tab:DataDal1A}

\myTable{
\begin{tabular}{|c|}
\hline
\includegraphics[height=0.85\textheight, width=\textwidth]{Dal1B.epsi}\\  
\hline
\end{tabular}
}{Acceptance and efficiency corrected Dalitz Plot $ppX$,\newline for $MM_{pp}=0.6-0.7\mathrm{~GeV/c^{2}}$ (Fig.~\ref{image_Dalitz1AccCor2D}).
The errors of invariant masses are determined by selected bin sizes and chosen as a half of the bin size.
Additional global uncertainties of the absolute normalization of $19\%$ have to be included.
Fully expandable version of the table is available in the attached electronic version of the thesis.
}{tab:DataDal1B}

\myTable{
\begin{tabular}{|c|}
\hline
\includegraphics[height=0.85\textheight, width=\textwidth]{Dal1C.epsi}\\  
\hline
\end{tabular}
}{Acceptance and efficiency corrected Dalitz Plot $ppX$,\newline for $MM_{pp}=0.7-0.8\mathrm{~GeV/c^{2}}$ (Fig.~\ref{image_Dalitz1AccCor2D}).
The errors of invariant masses are determined by selected bin sizes and chosen as a half of the bin size.
Additional global uncertainties of the absolute normalization of $19\%$ have to be included.
Fully expandable version of the table is available in the attached electronic version of the thesis.
}{tab:DataDal1C}

\myTable{
\begin{tabular}{|c|}
\hline
\includegraphics[height=0.85\textheight, width=\textwidth]{Dal1D.epsi}\\  
\hline
\end{tabular}
}{Acceptance and efficiency corrected Dalitz Plot $ppX$,\newline for $MM_{pp}=0.8-0.9\mathrm{~GeV/c^{2}}$ (Fig.~\ref{image_Dalitz1AccCor2D}).
The errors of invariant masses are determined by selected bin sizes and chosen as a half of the bin size.
Additional global uncertainties of the absolute normalization of $19\%$ have to be included.
Fully expandable version of the table is available in the attached electronic version of the thesis.
}{tab:DataDal1D}

\myTable{
\begin{tabular}{|c|}
\hline
\includegraphics[height=0.85\textheight, width=\textwidth]{Dal1E.epsi}\\  
\hline
\end{tabular}
}{Acceptance and efficiency corrected Dalitz Plot $ppX$,\newline for $MM_{pp}=0.9-1.0\mathrm{~GeV/c^{2}}$ (Fig.~\ref{image_Dalitz1AccCor2D}).
The errors of invariant masses are determined by selected bin sizes and chosen as a half of the bin size.
Additional global uncertainties of the absolute normalization of $19\%$ have to be included.
Fully expandable version of the table is available in the attached electronic version of the thesis.
}{tab:DataDal1E}

\myTable{
\begin{tabular}{|c|}
\hline
\includegraphics[height=0.85\textheight, width=\textwidth]{Dal2A.epsi}\\  
\hline
\end{tabular}
}{Acceptance and efficiency corrected Dalitz Plot $3\pi^{0}$,\newline for $MM_{pp}=0.4-0.5\mathrm{~GeV/c^{2}}$ (Fig.~\ref{image_Dalitz2AccCor2D}).
The errors of invariant masses are determined by selected bin sizes and chosen as a half of the bin size.
Additional global uncertainties of the absolute normalization of $19\%$ have to be included.
Fully expandable version of the table is available in the attached electronic version of the thesis.
}{tab:DataDal2A}

\myTable{
\begin{tabular}{|c|}
\hline
\includegraphics[height=0.85\textheight, width=\textwidth]{Dal2B.epsi}\\  
\hline
\end{tabular}
}{Acceptance and efficiency corrected Dalitz Plot $3\pi^{0}$,\newline for $MM_{pp}=0.6-0.7\mathrm{~GeV/c^{2}}$ (Fig.~\ref{image_Dalitz2AccCor2D}).
The errors of invariant masses are determined by selected bin sizes and chosen as a half of the bin size.
Additional global uncertainties of the absolute normalization of $19\%$ have to be included.
Fully expandable version of the table is available in the attached electronic version of the thesis.
}{tab:DataDal2B}

\myTable{
\begin{tabular}{|c|}
\hline
\includegraphics[height=0.85\textheight, width=\textwidth]{Dal2C.epsi}\\  
\hline
\end{tabular}
}{Acceptance and efficiency corrected Dalitz Plot $3\pi^{0}$,\newline for $MM_{pp}=0.7-0.8\mathrm{~GeV/c^{2}}$ (Fig.~\ref{image_Dalitz2AccCor2D}).
The errors of invariant masses are determined by selected bin sizes and chosen as a half of the bin size.
Additional global uncertainties of the absolute normalization of $19\%$ have to be included.
Fully expandable version of the table is available in the attached electronic version of the thesis.
}{tab:DataDal2C}

\myTable{
\begin{tabular}{|c|}
\hline
\includegraphics[height=0.85\textheight, width=\textwidth]{Dal2D.epsi}\\  
\hline
\end{tabular}
}{Acceptance and efficiency corrected Dalitz Plot $3\pi^{0}$,\newline for $MM_{pp}=0.8-0.9\mathrm{~GeV/c^{2}}$ (Fig.~\ref{image_Dalitz2AccCor2D}).
The errors of invariant masses are determined by selected bin sizes and chosen as a half of the bin size.
Additional global uncertainties of the absolute normalization of $19\%$ have to be included.
Fully expandable version of the table is available in the attached electronic version of the thesis.
}{tab:DataDal2D}

\myTable{
\begin{tabular}{|c|}
\hline
\includegraphics[height=0.85\textheight, width=\textwidth]{Dal2E.epsi}\\  
\hline
\end{tabular}
}{Acceptance and efficiency corrected Dalitz Plot $3\pi^{0}$,\newline for $MM_{pp}=0.9-1.0\mathrm{~GeV/c^{2}}$ (Fig.~\ref{image_Dalitz2AccCor2D}).
The errors of invariant masses are determined by selected bin sizes and chosen as a half of the bin size.
Additional global uncertainties of the absolute normalization of $19\%$ have to be included.
Fully expandable version of the table is available in the attached electronic version of the thesis.
}{tab:DataDal2E}

\myTable{
\begin{tabular}{|c|}
\hline
\includegraphics[height=0.85\textheight, width=\textwidth]{NybA.epsi}\\  
\hline
\end{tabular}
}{Acceptance and efficiency corrected Nyborg Plot,\newline for $MM_{pp}=0.4-0.5\mathrm{~GeV/c^{2}}$ (Fig.~\ref{image_Dalitz3AccCor2D}).
The errors of invariant masses are determined by selected bin sizes and chosen as a half of the bin size.
Additional global uncertainties of the absolute normalization of $19\%$ have to be included.
Fully expandable version of the table is available in the attached electronic version of the thesis.
}{tab:DataDal3A}

\myTable{
\begin{tabular}{|c|}
\hline
\includegraphics[height=0.85\textheight, width=\textwidth]{NybB.epsi}\\  
\hline
\end{tabular}
}{Acceptance and efficiency corrected Nyborg Plot,\newline for $MM_{pp}=0.6-0.7\mathrm{~GeV/c^{2}}$ (Fig.~\ref{image_Dalitz3AccCor2D}).
The errors of invariant masses are determined by selected bin sizes and chosen as a half of the bin size.
Additional global uncertainties of the absolute normalization of $19\%$ have to be included.
Fully expandable version of the table is available in the attached electronic version of the thesis.
}{tab:DataDal3B}

\myTable{
\begin{tabular}{|c|}
\hline
\includegraphics[height=0.85\textheight, width=\textwidth]{NybC.epsi}\\  
\hline
\end{tabular}
}{Acceptance and efficiency corrected Nyborg Plot,\newline for $MM_{pp}=0.7-0.8\mathrm{~GeV/c^{2}}$ (Fig.~\ref{image_Dalitz3AccCor2D}).
The errors of invariant masses are determined by selected bin sizes and chosen as a half of the bin size.
Additional global uncertainties of the absolute normalization of $19\%$ have to be included.
Fully expandable version of the table is available in the attached electronic version of the thesis.
}{tab:DataDal3C}

\myTable{
\begin{tabular}{|c|}
\hline
\includegraphics[height=0.85\textheight, width=\textwidth]{NybD.epsi}\\  
\hline
\end{tabular}
}{Acceptance and efficiency corrected Nyborg Plot,\newline for $MM_{pp}=0.8-0.9\mathrm{~GeV/c^{2}}$ (Fig.~\ref{image_Dalitz3AccCor2D}).
The errors of invariant masses are determined by selected bin sizes and chosen as a half of the bin size.
Additional global uncertainties of the absolute normalization of $19\%$ have to be included.
Fully expandable version of the table is available in the attached electronic version of the thesis.
}{tab:DataDal3D}

\myTable{
\begin{tabular}{|c|}
\hline
\includegraphics[height=0.85\textheight, width=\textwidth]{NybE.epsi}\\  
\hline
\end{tabular}
}{Acceptance and efficiency corrected Nyborg Plot,\newline for $MM_{pp}=0.9-1.0\mathrm{~GeV/c^{2}}$ (Fig.~\ref{image_Dalitz3AccCor2D}).
The errors of invariant masses are determined by selected bin sizes and chosen as a half of the bin size.
Additional global uncertainties of the absolute normalization of $19\%$ have to be included.
Fully expandable version of the table is available in the attached electronic version of the thesis.
}{tab:DataDal3E}

\myTable{
\footnotesize
\begin{tabular}{|c|c|c|c|}
\hline
 $\cos(\theta_{\eta^{CM}})$ & Error of $\cos(\theta_{\eta^{CM}})$ & anisotropy [a.u.] & Error of anisotropy [a.u.] \\  
\hline
\hline
$-0.983$	&	$0.023$	&	$0.937$	&	$0.048$	\\
\hline
$-0.938$	&	$0.023$	&	$1.039$	&	$0.043$	\\
\hline
$-0.893$	&	$0.023$	&	$1.009$	&	$0.041$	\\
\hline
$-0.848$	&	$0.023$	&	$1.010$	&	$0.041$	\\
\hline
$-0.803$	&	$0.023$	&	$0.949$	&	$0.042$	\\
\hline
$-0.758$	&	$0.023$	&	$1.077$	&	$0.049$	\\
\hline
$-0.713$	&	$0.023$	&	$0.968$	&	$0.047$	\\
\hline
$-0.668$	&	$0.023$	&	$1.000$	&	$0.056$	\\
\hline
$-0.623$	&	$0.023$	&	$1.007$	&	$0.063$	\\
\hline
$-0.578$	&	$0.023$	&	$1.001$	&	$0.067$	\\
\hline
$-0.533$	&	$0.023$	&	$0.812$	&	$0.065$	\\
\hline
$-0.488$	&	$0.023$	&	$1.054$	&	$0.093$	\\
\hline
$-0.443$	&	$0.023$	&	$0.810$	&	$0.087$	\\
\hline
$-0.398$	&	$0.023$	&	$0.88$	&	$0.11$	\\
\hline
\end{tabular}
}{Angular distribution of the $\eta$-meson in the CM system for the $\eta$-meson momentum in the CM system $q_{\eta}^{CM}=0.45-0.475\mathrm{~GeV/c}$ (Fig.~\ref{image_EtaCMDiffComparison2New2}).
}{tab:DataEtaA}

\myTable{
\footnotesize
\begin{tabular}{|c|c|c|c|}
\hline
 $\cos(\theta_{\eta^{CM}})$ & Error of $\cos(\theta_{\eta^{CM}})$ & anisotropy [a.u.] & Error of anisotropy [a.u.] \\  
\hline
\hline
$-0.983$	&	$0.023$	&	$1.332$	&	$0.058$	\\
\hline
$-0.938$	&	$0.023$	&	$1.159$	&	$0.048$	\\
\hline
$-0.893$	&	$0.023$	&	$1.062$	&	$0.043$	\\
\hline
$-0.848$	&	$0.023$	&	$1.017$	&	$0.042$	\\
\hline
$-0.803$	&	$0.023$	&	$1.134$	&	$0.049$	\\
\hline
$-0.758$	&	$0.023$	&	$1.151$	&	$0.051$	\\
\hline
$-0.713$	&	$0.023$	&	$1.155$	&	$0.056$	\\
\hline
$-0.668$	&	$0.023$	&	$1.000$	&	$0.056$	\\
\hline
$-0.623$	&	$0.023$	&	$1.054$	&	$0.066$	\\
\hline
$-0.578$	&	$0.023$	&	$1.110$	&	$0.073$	\\
\hline
$-0.533$	&	$0.023$	&	$0.978$	&	$0.078$	\\
\hline
$-0.488$	&	$0.023$	&	$1.042$	&	$0.093$	\\
\hline
$-0.443$	&	$0.023$	&	$0.780$	&	$0.078$	\\
\hline
$-0.398$	&	$0.023$	&	$0.89$	&	$0.10$	\\
\hline
\end{tabular}
}{Angular distribution of the $\eta$-meson in the CM system for the $\eta$-meson momentum in the CM system $q_{\eta}^{CM}=0.475-0.5\mathrm{~GeV/c}$ (Fig.~\ref{image_EtaCMDiffComparison2New2}).
}{tab:DataEtaB}

\newpage

\myTable{
\tiny
\begin{tabular}{|c|c|c|c|}
\hline
 $\cos(\theta_{\eta^{CM}})$ & Error of $\cos(\theta_{\eta^{CM}})$ & anisotropy [a.u.] & Error of anisotropy [a.u.] \\  
\hline
\hline
$-0.983$	&	$0.023$	&	$1.383$	&	$0.043$	\\
\hline
$-0.938$	&	$0.023$	&	$1.189$	&	$0.038$	\\
\hline
$-0.893$	&	$0.023$	&	$1.097$	&	$0.035$	\\
\hline
$-0.848$	&	$0.023$	&	$1.168$	&	$0.038$	\\
\hline
$-0.803$	&	$0.023$	&	$1.118$	&	$0.039$	\\
\hline
$-0.758$	&	$0.023$	&	$1.091$	&	$0.041$	\\
\hline
$-0.713$	&	$0.023$	&	$1.040$	&	$0.042$	\\
\hline
$-0.668$	&	$0.023$	&	$1.000$	&	$0.044$	\\
\hline
$-0.623$	&	$0.023$	&	$1.065$	&	$0.050$	\\
\hline
$-0.578$	&	$0.023$	&	$0.920$	&	$0.053$	\\
\hline
$-0.533$	&	$0.023$	&	$0.923$	&	$0.060$	\\
\hline
$-0.488$	&	$0.023$	&	$0.922$	&	$0.071$	\\
\hline
$-0.443$	&	$0.023$	&	$1.229$	&	$0.097$	\\
\hline
$-0.398$	&	$0.023$	&	$0.740$	&	$0.079$	\\
\hline
$-0.353$	&	$0.023$	&	$0.95$	&	$0.11$	\\
\hline
$-0.308$	&	$0.023$	&	$1.09$	&	$0.14$	\\
\hline
\end{tabular}
}{Angular distribution of the $\eta$-meson in the CM system for the $\eta$-meson momentum in the CM system $q_{\eta}^{CM}=0.5-0.55\mathrm{~GeV/c}$ (Fig.~\ref{image_EtaCMDiffComparison2New2}).
}{tab:DataEtaC}

\myTable{
\scriptsize
\begin{tabular}{|c|c|c|c|}
\hline
 $\cos(\theta_{\eta^{CM}})$ & Error of $\cos(\theta_{\eta^{CM}})$ & anisotropy [a.u.] & Error of anisotropy [a.u.] \\  
\hline
\hline
$-0.983$	&	$0.023$	&	$1.374$	&	$0.044$	\\
\hline
$-0.938$	&	$0.023$	&	$1.274$	&	$0.038$	\\
\hline
$-0.893$	&	$0.023$	&	$1.290$	&	$0.039$	\\
\hline
$-0.848$	&	$0.023$	&	$1.204$	&	$0.038$	\\
\hline
$-0.803$	&	$0.023$	&	$1.106$	&	$0.037$	\\
\hline
$-0.758$	&	$0.023$	&	$1.037$	&	$0.037$	\\
\hline
$-0.713$	&	$0.023$	&	$1.073$	&	$0.043$	\\
\hline
$-0.668$	&	$0.023$	&	$1.000$	&	$0.044$	\\
\hline
$-0.623$	&	$0.023$	&	$0.900$	&	$0.044$	\\
\hline
$-0.578$	&	$0.023$	&	$0.963$	&	$0.051$	\\
\hline
$-0.533$	&	$0.023$	&	$0.886$	&	$0.057$	\\
\hline
$-0.488$	&	$0.023$	&	$0.852$	&	$0.062$	\\
\hline
$-0.443$	&	$0.023$	&	$0.825$	&	$0.067$	\\
\hline
$-0.398$	&	$0.023$	&	$0.898$	&	$0.093$	\\
\hline
$-0.353$	&	$0.023$	&	$1.00$	&	$0.12$	\\
\hline
$-0.308$	&	$0.023$	&	$0.89$	&	$0.13$	\\
\hline
$-0.263$	&	$0.023$	&	$0.81$	&	$0.14$	\\
\hline
$-0.218$	&	$0.023$	&	$0.75$	&	$0.15$	\\
\hline
\end{tabular}
}{Angular distribution of the $\eta$-meson in the CM system for the $\eta$-meson momentum in the CM system $q_{\eta}^{CM}=0.55-0.7\mathrm{~GeV/c}$ (Fig.~\ref{image_EtaCMDiffComparison2New2}).
}{tab:DataEtaD}

\newpage
\myTable{
\tiny
\begin{tabular}{|c|c|c|c|}
\hline
 $\cos(\theta_{p^{pp}})$ & Error of $\cos(\theta_{p^{pp}})$ & anisotropy [a.u.] & Error of anisotropy [a.u.] \\  
\hline
\hline
$-0.983$	&	$0.023$	&	$1.05$	&	$0.16$	\\
\hline
$-0.938$	&	$0.023$	&	$1.026$	&	$0.060$	\\
\hline
$-0.893$	&	$0.023$	&	$0.946$	&	$0.051$	\\
\hline
$-0.848$	&	$0.023$	&	$0.753$	&	$0.044$	\\
\hline
$-0.803$	&	$0.023$	&	$0.729$	&	$0.042$	\\
\hline
$-0.758$	&	$0.023$	&	$0.748$	&	$0.044$	\\
\hline
$-0.713$	&	$0.023$	&	$0.838$	&	$0.046$	\\
\hline
$-0.668$	&	$0.023$	&	$0.865$	&	$0.050$	\\
\hline
$-0.623$	&	$0.023$	&	$0.857$	&	$0.049$	\\
\hline
$-0.578$	&	$0.023$	&	$1.081$	&	$0.060$	\\
\hline
$-0.533$	&	$0.023$	&	$0.949$	&	$0.056$	\\
\hline
$-0.488$	&	$0.023$	&	$1.027$	&	$0.057$	\\
\hline
$-0.443$	&	$0.023$	&	$1.000$	&	$0.060$	\\
\hline
$-0.398$	&	$0.023$	&	$0.931$	&	$0.056$	\\
\hline
$-0.353$	&	$0.023$	&	$0.920$	&	$0.060$	\\
\hline
$-0.308$	&	$0.023$	&	$1.029$	&	$0.075$	\\
\hline
$-0.263$	&	$0.023$	&	$1.068$	&	$0.085$	\\
\hline
$-0.218$	&	$0.023$	&	$0.947$	&	$0.089$	\\
\hline
$-0.173$	&	$0.023$	&	$0.90$	&	$0.12$	\\
\hline
\end{tabular}
}{Angular distribution of the proton in the proton-proton rest frame for the $\eta$-meson momentum in the CM system $q_{\eta}^{CM}=0.45-0.475\mathrm{~GeV/c}$ (Fig.~\ref{image_EtaPPComaprisonNew2}).
}{tab:DataPA}

\myTable{
\tiny
\begin{tabular}{|c|c|c|c|}
\hline
 $\cos(\theta_{p^{pp}})$ & Error of $\cos(\theta_{p^{pp}})$ & anisotropy [a.u.] & Error of anisotropy [a.u.] \\  
\hline
\hline
$-0.938$	&	$0.023$	&	$0.764$	&	$0.066$	\\
\hline
$-0.893$	&	$0.023$	&	$0.831$	&	$0.059$	\\
\hline
$-0.848$	&	$0.023$	&	$0.555$	&	$0.046$	\\
\hline
$-0.803$	&	$0.023$	&	$0.699$	&	$0.051$	\\
\hline
$-0.758$	&	$0.023$	&	$0.818$	&	$0.050$	\\
\hline
$-0.713$	&	$0.023$	&	$0.823$	&	$0.050$	\\
\hline
$-0.668$	&	$0.023$	&	$0.921$	&	$0.055$	\\
\hline
$-0.623$	&	$0.023$	&	$0.936$	&	$0.053$	\\
\hline
$-0.578$	&	$0.023$	&	$1.003$	&	$0.058$	\\
\hline
$-0.533$	&	$0.023$	&	$0.952$	&	$0.055$	\\
\hline
$-0.488$	&	$0.023$	&	$1.003$	&	$0.057$	\\
\hline
$-0.443$	&	$0.023$	&	$1.000$	&	$0.056$	\\
\hline
$-0.398$	&	$0.023$	&	$0.916$	&	$0.054$	\\
\hline
$-0.353$	&	$0.023$	&	$1.079$	&	$0.063$	\\
\hline
$-0.308$	&	$0.023$	&	$0.986$	&	$0.062$	\\
\hline
$-0.263$	&	$0.023$	&	$1.149$	&	$0.075$	\\
\hline
$-0.218$	&	$0.023$	&	$1.093$	&	$0.083$	\\
\hline
$-0.173$	&	$0.023$	&	$1.13$	&	$0.11$	\\
\hline
\end{tabular}
}{Angular distribution of the proton in the proton-proton rest frame for the $\eta$-meson momentum in the CM system $q_{\eta}^{CM}=0.475-0.5\mathrm{~GeV/c}$ (Fig.~\ref{image_EtaPPComaprisonNew2}).
}{tab:DataPB}

\newpage

\myTable{
\footnotesize
\begin{tabular}{|c|c|c|c|}
\hline
 $\cos(\theta_{p^{pp}})$ & Error of $\cos(\theta_{p^{pp}})$ & anisotropy [a.u.] & Error of anisotropy [a.u.] \\  
\hline
\hline
$-0.983$	&	$0.023$	&	$0.70$	&	$0.15$	\\
\hline
$-0.938$	&	$0.023$	&	$1.223$	&	$0.075$	\\
\hline
$-0.893$	&	$0.023$	&	$1.159$	&	$0.057$	\\
\hline
$-0.848$	&	$0.023$	&	$0.800$	&	$0.047$	\\
\hline
$-0.803$	&	$0.023$	&	$0.926$	&	$0.047$	\\
\hline
$-0.758$	&	$0.023$	&	$0.849$	&	$0.044$	\\
\hline
$-0.713$	&	$0.023$	&	$1.101$	&	$0.051$	\\
\hline
$-0.668$	&	$0.023$	&	$0.891$	&	$0.043$	\\
\hline
$-0.623$	&	$0.023$	&	$0.956$	&	$0.044$	\\
\hline
$-0.578$	&	$0.023$	&	$1.002$	&	$0.048$	\\
\hline
$-0.533$	&	$0.023$	&	$1.090$	&	$0.050$	\\
\hline
$-0.488$	&	$0.023$	&	$0.986$	&	$0.047$	\\
\hline
$-0.443$	&	$0.023$	&	$1.000$	&	$0.046$	\\
\hline
$-0.398$	&	$0.023$	&	$0.969$	&	$0.045$	\\
\hline
$-0.353$	&	$0.023$	&	$1.051$	&	$0.048$	\\
\hline
$-0.308$	&	$0.023$	&	$1.068$	&	$0.054$	\\
\hline
$-0.263$	&	$0.023$	&	$1.161$	&	$0.057$	\\
\hline
$-0.218$	&	$0.023$	&	$1.144$	&	$0.061$	\\
\hline
$-0.173$	&	$0.023$	&	$1.160$	&	$0.070$	\\
\hline
$-0.128$	&	$0.023$	&	$1.284$	&	$0.087$	\\
\hline
$-0.083$	&	$0.023$	&	$1.58$	&	$0.11$	\\
\hline
$-0.038$	&	$0.023$	&	$1.69$	&	$0.11$	\\
\hline
\end{tabular}
}{Angular distribution of the proton in the proton-proton rest frame for the $\eta$-meson momentum in the CM system $q_{\eta}^{CM}=0.5-0.55\mathrm{~GeV/c}$ (Fig.~\ref{image_EtaPPComaprisonNew2}).
}{tab:DataPC}

\newpage

\myTable{
\footnotesize
\begin{tabular}{|c|c|c|c|}
\hline
 $\cos(\theta_{p^{pp}})$ & Error of $\cos(\theta_{p^{pp}})$ & anisotropy [a.u.] & Error of anisotropy [a.u.] \\  
\hline
\hline
$-0.983$	&	$0.023$	&	$0.79$	&	$0.17$	\\
\hline
$-0.938$	&	$0.023$	&	$1.337$	&	$0.087$	\\
\hline
$-0.893$	&	$0.023$	&	$1.473$	&	$0.078$	\\
\hline
$-0.848$	&	$0.023$	&	$1.47$	&	$0.11$	\\
\hline
$-0.803$	&	$0.023$	&	$1.387$	&	$0.069$	\\
\hline
$-0.758$	&	$0.023$	&	$1.188$	&	$0.097$	\\
\hline
$-0.713$	&	$0.023$	&	$1.38$	&	$0.10$	\\
\hline
$-0.668$	&	$0.023$	&	$1.245$	&	$0.098$	\\
\hline
$-0.623$	&	$0.023$	&	$1.257$	&	$0.096$	\\
\hline
$-0.578$	&	$0.023$	&	$1.212$	&	$0.097$	\\
\hline
$-0.533$	&	$0.023$	&	$1.206$	&	$0.093$	\\
\hline
$-0.488$	&	$0.023$	&	$1.185$	&	$0.092$	\\
\hline
$-0.443$	&	$0.023$	&	$1.000$	&	$0.082$	\\
\hline
$-0.398$	&	$0.023$	&	$1.086$	&	$0.089$	\\
\hline
$-0.353$	&	$0.023$	&	$1.250$	&	$0.094$	\\
\hline
$-0.308$	&	$0.023$	&	$1.142$	&	$0.090$	\\
\hline
$-0.263$	&	$0.023$	&	$1.207$	&	$0.061$	\\
\hline
$-0.218$	&	$0.023$	&	$1.288$	&	$0.061$	\\
\hline
$-0.173$	&	$0.023$	&	$1.259$	&	$0.064$	\\
\hline
$-0.128$	&	$0.023$	&	$1.237$	&	$0.097$	\\
\hline
$-0.083$	&	$0.023$	&	$1.20$	&	$0.10$	\\
\hline
$-0.038$	&	$0.023$	&	$1.278$	&	$0.065$	\\
\hline
\end{tabular}
}{Angular distribution of the proton in the proton-proton rest frame for the $\eta$-meson momentum in the CM system $q_{\eta}^{CM}=0.55-0.7\mathrm{~GeV/c}$ (Fig.~\ref{image_EtaPPComaprisonNew2}).
}{tab:DataPD}

\end{appendices}





\newpage ~
\thispagestyle{empty}
\emptydoublepage

\phantomsection
\addcontentsline{toc}{section}{References}
\thispagestyle{plain}

\newpage ~
\thispagestyle{empty}
\emptydoublepage

\section*{Acknowledgments}
\addcontentsline{toc}{section}{Acknowledgments}
\rhead{Acknowledgments}
\lhead{Acknowledgments}
\thispagestyle{plain}

\large
I would like to thank all people that helped me to create this dissertation and without whom it wouldn't have been possible. 

\bigskip

First of all I would like to express my enormous gratitude to my supervisor Prof. Zbigniew Rudy for first introduction to the secrets of data analysis,
 for the time spent on many discussions, for his guidance, encouragement, support and patience.

\medskip

I am also very grateful to Dr. Volker Hejny from  \mbox{FZ-J\"ulich}, for showing me the aspects of the advanced data analysis and for the guidance through the analysis steps.
\medskip

I am also very grateful to Prof. Bugus{\l}aw Kamys for allowing me to prepare this dissertation in the Nuclear Physics Department of the Jagiellonian University
 and for his advices, support and discussions. 
\medskip

I would like also to thank Prof. Hans Str\"oher for possibility of stay in \mbox{Forschungszentrum-J\"ulich} and for giving me the opportunity to work with WASA~at~COSY collaboration.
\medskip

I want to express my appreciation to Prof. Lucjan Jarczyk for many interesting ideas and fruitful discussions.
\medskip

I also thank all colleagues from WASA~at~COSY collaboration,  
specially: Dr. Andrzej Kupsc, Dr. hab. Susan Schadmand, Dr. Magnus Wolke, Dr. hab. Frank Goldenbaum, Dr. Christian Pauly and Dr. Christoph Redmer. 

\medskip

I thank all my colleagues from the IKP Forschungszentrum-J\"ulich and from the Nuclear Physics Department of the Jagiellonian University
 for the pleasant atmosphere of daily work.

\medskip

I also want to express my gratitude to my beloved wife for sharing daily life and fascination to physics with me.

\newpage ~
\thispagestyle{empty}
\emptydoublepage

\newpage ~
\thispagestyle{empty}
\emptydoublepage


\end{document}